\documentclass[epsf,12pt]{iopart}
\usepackage{iopams}
\usepackage{here} 
\usepackage{epsf}  
\usepackage{graphicx} 
\usepackage{dcolumn}% 
\newcommand{\beq}{\begin{equation}}
\newcommand{\eeq}{\end{equation}}

\begin{document}

\title[Explosion Mechanism of Core-Collapse Supernovae]
{Explosion Mechanism, Neutrino Burst, and Gravitational Wave in Core-Collapse Supernovae}

\author{Kei Kotake, Katsuhiko Sato\dag and Keitaro Takahashi\ddag
\footnote[3]{
To whom correspondence should be addressed
(kkotake@heap.phys.waseda.ac.jp, or ktakahas@Princeton.EDU)}
}
\address{ \ Science \& Engineering, Waseda University, 3-4-1 Okubo,
Shinjuku, Tokyo, 169-8555, Japan}
\address{\dag\ Department of Physics, School of Science, the University of
Tokyo, 7-3-1 Hongo, Bunkyo-ku, Tokyo 113-0033, Japan}
\address{\ddag Department of Physics, Princeton University, Princeton,
NJ 08544, U.S.A.}

\begin{abstract}
Core-collapse supernovae are among the most energetic explosions in the 
universe marking the catastrophic end of massive stars.
%Core-collapse supernovae are one of the most important issues in
%astrophysics. 
In spite of rigorous studies for several decades, we still
don't understand the explosion mechanism completely. Since they 
are related to 
many astrophysical phenomena such as nucleosynthesis,
gamma-ray bursts and acceleration of cosmic rays, understanding of their 
physics has been of wide interest to the astrophysical community.

In this article, we review recent progress in the study of core-collapse
supernovae focusing on the explosion mechanism, supernova neutrinos, and
 the gravitational waves. As for the explosion mechanism, we present a
 review paying particular attention to the roles 
of multidimensional aspects, such as
 convection, rotation, and magnetic fields, on the neutrino heating
 mechanism.
%, which has been most promising to produce successful explosions.  
Next, we discuss supernova neutrinos,
which is a powerful tool to probe not only deep inside of the supernovae
but also intrinsic properties of neutrinos. 
For this purpose, it is 
necessary to understand neutrino oscillation which has been established
recently by a lot of experiments. Gravitational astronomy is now also 
becoming reality. 
 We present an extensive review on the physical foundations and
 the emission mechanism of gravitational waves in detail, and discuss the possibility of their detections.
\end{abstract}

\tableofcontents
\clearpage

\section{Overview \label{intro}}

Core-collapse supernovae are among the most energetic explosions in the 
universe. They mark the catastrophic end of stars more massive than
8 solar masses leaving behind compact remnants such as neutron stars
or stellar mass black holes. Noteworthy, they have been thought to be
extremely important astrophysical objects and thus have been of
wide interest to the astrophysical community. The nucleosynthesis
in these massive stars, and their subsequent explosions, are responsible
for most of the heavy element enrichment in our galaxy. So naturally,
any attempt to address human origins must begin with an understanding
of core-collapse supernovae.

At the moment of explosion, most of the binding energy of the core is released
as neutrinos. These neutrinos, which we call them as {\it supernova
neutrinos} in the following, are temporarily confined in the core and
escape to the outer region by diffusion. Thus supernova neutrinos will have
valuable information of deep inside of the core. In fact, the detection
of neutrinos from SN1987A paved the way for the {\it Neutrino Astronomy},
which is an alternative to the conventional astronomy by electromagnetic waves.
Even though neutrino events from SN1987A were just two dozens, they have been
studied extensively and allowed us to have a confidence that the basic
picture of core-collapse supernova is correct.

Here it is worth mentioning that supernova neutrinos have attracted not only
astrophysicist but also particle physicist. This is because supernova neutrinos
are also useful to probe intrinsic properties of neutrinos as well as
supernova dynamics. Conventionally they are used to set constraints on
neutrino mass, lifetime and electric charge etc. More recent development
involves neutrino oscillation, which have been established experimentally
in the last decade. Neutrino oscillation on supernova neutrinos is important
in two ways. First, since neutrino oscillation changes the event spectra,
we cannot obtain the information on the physical state of the core without
a consideration of neutrino oscillation. Second, since supernova has
a distinct feature as a neutrino source compared with other sources
such as the sun, atmosphere, accelerator and reactor, it also acts as
a laboratory for neutrino oscillation.

Supernova is now about to start even another astronomy, {\it Gravitational-Wave
Astronomy}. In fact, core-collapse supernovae have been supposed to be one of
the most plausible sources of gravitational waves. Currently a lot of
long-baseline laser interferometers such as GEO600, LIGO, TAMA300 and VIRGO
are running and preparing for the first direct observation, by which
the prediction by Einstein's theory of General Relativity can be confirmed.

Astrophysicists have been long puzzled by the origins of the gamma-ray bursts
since their accidental discovery in the late sixties. Some recent observations
imply that the long-duration gamma-ray bursts are associated with core-collapse
supernovae. In a theoretical point of view, the gamma-ray bursts are considered
to be accompanied by the failed core-collapse supernovae, in which not
the neutron star but the black hole is left behind. It is one of the most
exciting issue to understand how the failed core-collapse supernovae can
produce the observed properties of the gamma-ray bursts.

In order to obtain the understanding of these astrophysical phenomena
related to core-collapse supernovae and the properties of neutrino and
gravitational-wave emissions, it is indispensable to understand the explosion
mechanism of core-collapse supernovae. However one still cannot tell
it exactly albeit with the elaborate efforts during this 40 years. At present,
detections of neutrinos and gravitational waves from nearby core-collapse
supernovae are becoming reality. Since neutrinos and gravitational waves can be
the only window that enables us to see directly the innermost part of
core-collapse supernovae, their information is expected to help us to understand
the explosion mechanism itself.  Under the circumstances, the mutual
understanding of the explosion mechanism, the supernova neutrinos, and
the gravitational waves, which we will review in turn in this article,
will be important.

The plan of this article is as follows. We begin by a brief description
of the standard scenario of core-collapse supernovae in section
\ref{supernova_theory}. In section \ref{section:nu_osc}, we give
a tool to discuss supernova neutrinos and their observation,
neutrino oscillation. Although neutrino oscillation is thought to
have only a negligible effect on the dynamics of supernova, it is
necessary when we try to interpret observed neutrinos and extract
information of supernova from them. Then supernova neutrinos and their
neutrino oscillation are elaborately reviewed in section
\ref{section:SN_nu_osc}. With respect to the study of the explosion mechanism, 
good progress in the multi-dimensional models has been made recently.
We review these studies in section \ref{exp_mecha}. 
Finally, gravitational waves in core-collapse supernovae are reviewed 
in section \ref{GW_sec}, in which we pay a particular attention 
to the predicted characteristics of gravitational waves and their 
detectability for the currently running and planning laser
interferometers.  So far a number of excellent reviews already 
exist on various topics in this article. This article goes beyond such 
reviews to cover more the state-of-the-art investigations.

\clearpage
% check ok
\section{Supernova Theory \label{supernova_theory}}
\subsection{The Fate of Massive Star \label{fate}}
In a historical view point, {\it supernovae} owe their name to astronomers Walter Baarde and Fritz
Zwicky, who already in the early 1930's realized that these objects
show a sudden bursts in luminosity that slowly decays, similar to common
novae, but much more luminous and rare \cite{label1}.
Their high luminosities comparable to the integrated light of their host 
galaxies and their broad spectral lines led them to conclude that
supernovae were very energetic explosions produced at the death of the
massive star.
What is amazing is that they suggested that a
supernova derive their tremendous energy from gravitational collapse, in
particular that the inner part of the star collapses to a neutron star.
Although much observational and theoretical
progress have been made since then, and many physical principles and
important details 
%for the explosion mechanism 
have been identified, the
basic picture of the early 1930's still holds nowadays. 
To begin with, we review the current understanding of the fate of the
massive stars in the following.

\begin{figure}[H]
\begin{center}
\epsfxsize=12 cm
\rotatebox[origin=c]{-90}{
\epsfbox{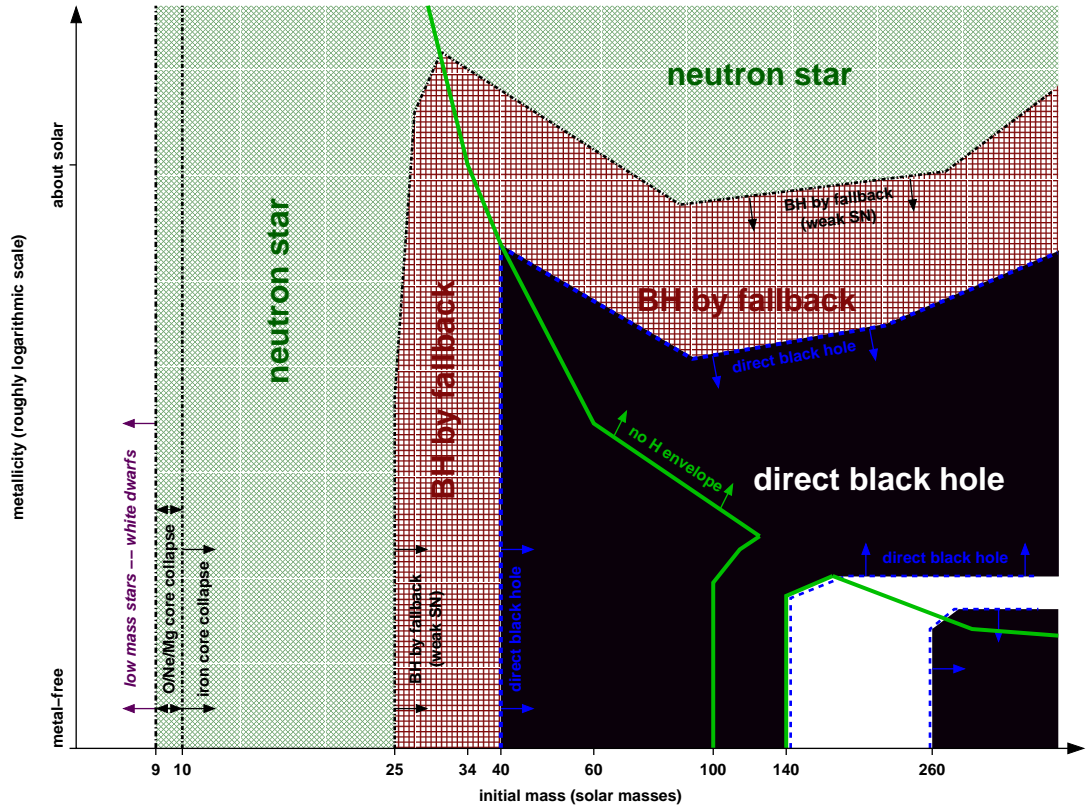}}
\caption{Remnants of massive single stars as a function of initial
 metalicity and initial mass. In the regions above the thick green line
 (for the higher initial metalicity), 
the hydrogen envelope is stripped during its evolution  due to the
 active mass loss processes. The dashed blue line indicates the border
 of the regime of direct black hole formation. The white strip near the
 right lower corner indicates the occurrence of the pair-instability
 supernovae. In the white region on the left side at lower mass, the
 stellar cores do not collapse and end their lives as white dwarfs. This
 figure is taken from Heger et al (2003) \cite{hegershinka}.}
\label{heger_fate}
\end{center}
\end{figure}

%To begin with, we review the current understanding of the stellar
%evolutions of the massive single stars in the next section.
The fate of a single massive stars, that is
to say, whether the remnant formed after stellar collapse will be 
a neutron star or a black hole, is 
mainly determined by its mass at birth 
and by the history of its mass loss during its evolution.  
The mass loss is expected to be 
crucially affected by the initial metalicity of the star, because the
mass loss rate by the stellar winds is sensitive to the photon opacity, which
is determined by the metalicity. 
The stars with high initial metalicity
have more mass loss, and thus, have smaller helium cores and hydrogen
envelopes during its evolution. Stellar collapse of such stars tends to 
lead to the formation of a neutron star, while for the lower metalicity
stars, a black hole \cite{hegershinka}. Figure \ref{heger_fate}
illustrates how the remnants of massive stars will be as a function of
the initial mass and the metalicity (this figure is taken from
\cite{hegershinka}). From the figure, stellar collapse of the stars with
the initial masses above $\sim 9 M_{\odot}$ and below  $\sim 25
M_{\odot}$ lead to the formation of neutron
stars. Above $25
M_{\odot}$,  black holes are expected to be formed
 either by fall-back of matter after the weak explosion (below
 $40M_{\odot}$) or directly if the
 stellar core is too massive to produce the outgoing shock wave (above
 $40 M_{\odot}$). Given a fixed initial mass above $40 M_{\odot}$, the
 stars with smaller initial metalicity tend to form a black hole
 directly due to the more heavier core as a result of the less
 mass-loss activities during evolution. 

Recently, the fate of massive stars has been paid considerable attention.
This is mainly due to the accumulating observations that the death of
massive stars and supernova-like events are associated with the
long-duration gamma-ray bursts (GRBs) (see, for example,
\cite{lazzati}). The fact that accompanying supernovae are in general
more energetic (they are frequently referred to as ``hypernovae'' in the
literature) than the canonical core-collapse supernovae is another
reason for this frenzy \cite{mazzali}.
According to the most widely accepted theoretical models, 
it is believed that  a black hole/an accretion disk system
supported by the sufficient angular momentum is required \cite{macfad}. 
In addition to the rapid rotation, 
the strong magnetic fields, as high as $10^{15}\sim 10^{16}$ G 
in the central regions are also pointed out to be helpful 
for producing the GRBs. In order to determine the progenitor of the
gamma-ray bursts, stellar rotation and magnetic fields should be
taken into consideration. Such investigation has just begun \cite{heger04}. In addition, 
the astrophysical details of the geometry or 
environment of the black hole/accretion system are currently hidden 
from us both observationally and computationally. Although these are
open questions now, this situation may change in the near future with the
development of gravitational-wave and neutrino observatories 
and more sophisticated astrophysical simulation 
capabilities (see \cite{piran} for a review). 
 
Very massive stars between $140M_{\odot} \le M \le 260
 M_{\odot}$ with the smaller initial metalicity are considered
 to become unstable to the electron-positron pair-instability ($\gamma
 \gamma \rightarrow e^{+}e^{-}$) during 
its evolution, which lead
 to the complete disruption of the star. Recently, explosions of 
metal-poor stars have been paid great attention because such
 stars are related to the first stars (the so-called Population III
 stars) to form in the universe.
So far two hyper metal-poor stars, HE0107-5240 \cite{christlieb} and 
HE1327-2326 \cite{frebel}, whose
 metalicity is smaller than $1/100000$ of the sun,
have been discovered. They provided crucial clues 
to the star formation history \cite{schneider} and the synthesis 
of chemical elements \cite{umeda,iwamoto} in the early universe. 
Furthermore, neutrino emissions and 
gravitational waves from such stars are one of the most exciting 
research issues. 

In this review, we focus on the ordinary supernova which lead to the
neutron star formation ($\sim 9 M_{\odot} \le M \le 25 M_{\odot}$ with
the solar metalicity). As will be explained below, the most promising
scenario of the explosion mechanism of such stars are the
neutrino-heating explosion. After we shortly refer to the current status of the
presupernova models in section \ref{pre} (for details, see,
\cite{wooshinka,hegershinka}),  we explain the
scenario in section \ref{standardscenario}.

\subsection{Presupernova Models \label{pre}}
\begin{figure}[H]
\begin{center}
\epsfxsize=16 cm
\epsfbox{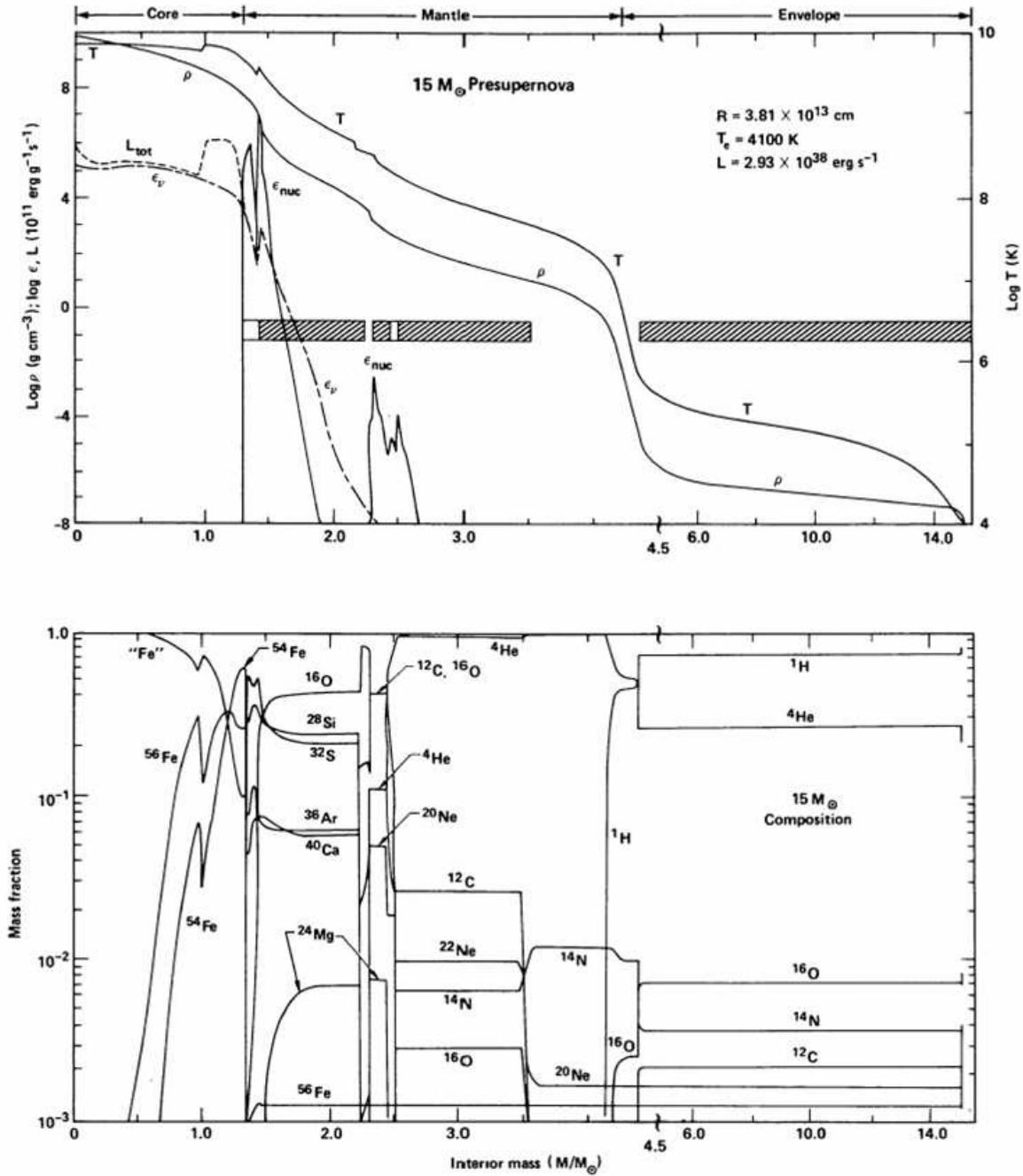}
\caption{Structure of the precollapse star (Woosley and Weaver's 15
 $M_{\odot}$ model taken from \cite{woosrevshinka}). In the upper panel,  the temperature and
 density profiles are given. $L_{\rm tot}$, $\epsilon_{\nu}$, and 
$\epsilon_{\rm nuc}$, represents 
the total energy loss and the contribution from the neutrino emission
 and from the nuclear-energy generation,
 respectively. In the lower panel, the composition profile is given.}
\label{woos_shinka}
\end{center}
\end{figure}
In Figure \ref{woos_shinka}, an example of the precollapse stellar model by Woosley and
Weaver (1995) \cite{ww:95}, which has been often employed as an initial condition of
 core-collapse simulations, is shown.
The iron core is surrounded by shells of lighter elements (the bottom
panel of Figure \ref{woos_shinka}). This is called
onion-skin structure. The size of iron core is the order of $10^{9}$ cm
while the stellar radius is larger than $10^{13}$ cm. At the core and
the surrounding shell, the density decreases steeply and hence the
dynamical timescale of the core  (:$\tau_{\rm dyn} \sim
(G\bar{\rho})^{-1/2}$ with $G$ and $\bar{\rho}$ being the
      gravitational constant and the average density) is much shorter than that of the
envelope (see the upper panel of Figure \ref{woos_shinka}).
That is, the dynamics of the iron core is not affected by the envelope.
Therefore we focus on the core hearafter for a while.

The late evolutional stage of massive stars are strongly affected by 
weak interactions. In fact, it can be seen from the upper panel of 
Figure \ref{woos_shinka} that
the dominant energy loss process in the iron core is the neutrino
emissions (see $L_{\rm tot}$, $\epsilon_{\nu}$, and
$\epsilon_{\rm nuc}$ in the panel).  
The generated neutrinos, which are well transparent for densities
$\bar{\rho} \le 10^{11}~\rm{g}~\rm{cm}^{-3}$, escape the star
carrying away energy and thus cooling the star. Due to the weak
interactions, namely electron capture and beta decay, not only the core 
entropy $s$, but also the electron fraction $Y_e$, which is the electron to
baryon ratio, changes. Since the mass of the presupernova core 
can be approximately expressed by the effective Chandrasekar mass \cite{chandra,timmes}, 
\begin{equation}
M_{\rm Ch} = 5.83 \bar{Y}_e^2 \Bigl[1 + \Bigl(\frac{\bar{s}_e}{\pi \bar{Y}_e}\Bigr)^2\Bigr]M_{\odot},
\label{chand}
\end{equation}
with $\bar{Y}_e$ and $\bar{s}_e$ being the average values of electron
 fraction and electronic entropy per baryon in the core,  
 the weak interaction rates play an important
role of determining the 
core mass. Putting the typical values of $\bar{Y}_e = 0.45, \bar{s}_e =
 0.52$ in 
a $15 M_{\odot}$ star into Eq. (\ref{chand}), one has
an Chandrasekhar mass of $1.34 M_{\odot}$ which is close to the core
mass obtained by the stellar evolution calculation
 (see Figure \ref{woos_shinka}).

\begin{figure}[H]
\begin{center}
\epsfxsize=7 cm
\epsfbox{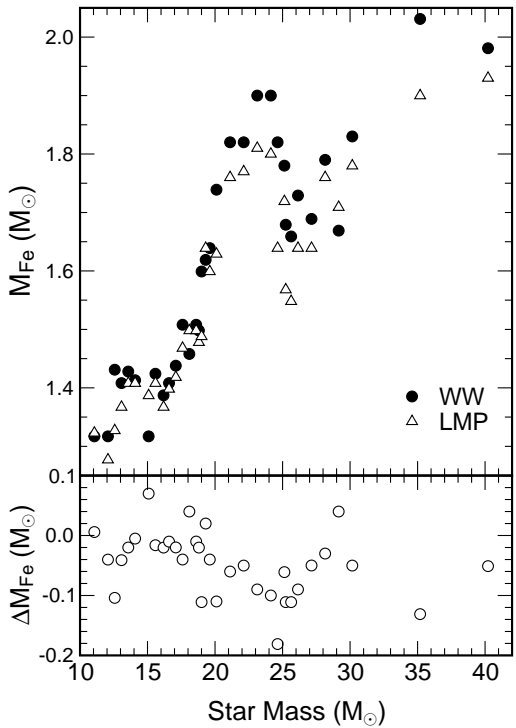}
\end{center}
\caption{Comparison of the values of the iron core
masses for $11 - 40 M_{\odot}$ stars 
between the Woosley and Weaver's models (WW) using the FFN
interaction rates and the ones using the shell model weak interaction
rates by Langanke and Mart\'{i}nez-Pinedo (LMP)
 \cite{lmp1,lmp2}. $\Delta M_{\rm Fe}$ represents the difference of the
 masses between the two computations. This figure is taken from \cite{langankerev}.}
\label{LMP}
\end{figure}

So far presupernova models have been constructed by
employing the weak interaction rates by Fuller, Fowler and Newman (FFN)
 \cite{fuller1,fuller2,fuller3} for electron-capture rates with an 
older set of beta decay rates.
%Figure \ref{woos_shinka}, albeit routinely used in supernova
%simulations, was constructed by
%employing the weak interaction rates by Fuller, Fowler and Newman (FFN)
% \cite{fuller1,fuller2,fuller3} for electron-capture rates with an 
%older set of beta decay rates. 
As well known, the electron capture and 
its inverse are dominated by Fermi and Gamow-Tellar transitions. A
correct description of the Gamow-Tellar transitions is difficult because it
requires to solve the many-body problem in the nuclear structure.
Due to the restricted available
experimental data in the mid 1980's, the tabulations of FFN could not
fully describe the Gamow-Taylor distributions in nuclei. 
This has been practicable recently by the new-shell model calculation by
 Langanke and Mart\'{i}nez-Pinedo (\cite{lmp1,lmp2}, see \cite{langankerev} for review).
According to Heger {\it et al} \cite{heger_prl,heger_weak}, who 
studied the effect of the shell model
 rates on presupernova models by repeating the calculations of Woosley
 and Weaver (1995) \cite{ww:95} while 
fixing the other stellar physics, the iron core mass is 
found to be reduced about up to $0.2 M_{\odot}$ than the ones in Woosley
and Weaver's 
computations (see Figure \ref{LMP}).

%\begin{figure}[H]
%\begin{center}
%\epsfxsize=12 cm
%\rotatebox[origin=c]{-90}{
%\epsfbox{figure_kotake/hegerF1.ps}}
%\end{center}
%\caption{Internal structure of a 22 $M_{\odot}$ star of solar metalicity
% as a function of time (logarithm of time till core
% collapse). Convection plays an important role in the envelope outside
% the helium burning shell, but also in shells during oxygen and silicon
% burning. This figure is taken from \cite{}.}
%\end{figure}

\subsection{Standard Scenario of Core-collapse Supernova Explosion
\label{standardscenario}}
In the following, we shall briefly outline
 the modern picture of the explosion mechanism of core-collapse
supernovae (see, also \cite{betherev,suzuki} for reviews). 

\subsubsection{onset of infall}
In the late-time iron core of a massive star, 
the pressure, which supports it against the
core's own gravitational force, is dominated by a  
degenerate gas of relativistic electrons,
\begin{equation}
P_{\rm e} = \frac{1}{4}(3 \pi^2)^{1/3}
\Bigl(\frac{\rho}{m_{u}}\Bigr)^{4/3}Y_{\rm e}^{4/3},
\label{degene_press}
\end{equation} 
where $Y_{\rm e} \equiv n_{e^{-}} - e_{e^{+}}$ is electron fraction per 
baryon, $m_u$ is the atomic mass unit, and $\rho$ is the density.
At the typical core densities and
temperatures ($\rho \sim 10^{10}~{\rm g}~{\rm cm}^{-3}$ and
$T \sim 10^{10}~{\rm K}$), the electron capture on Fe nuclei occurs via
\begin{equation}
{}^{56}{\rm Fe} + e^{-} \rightarrow {}^{56}{\rm Mn} + \nu_{e},
\end{equation}
because the Fermi energy of electrons,
\begin{eqnarray}
\mu_{e} & = & (3 \pi^{2} n_{e})^{\frac{1}{3}} \hbar c \\
& = & 11.1 {\rm MeV} \left(\frac{\rho Y_{e}}{10^{10} {\rm g}~{\rm cm}^{-3}}
\right)^{\frac{1}{3}}
\end{eqnarray}
exceeds the mass difference between the nuclei, namely, 
$m_{\rm Mn} - m_{\rm Fe} = 3.7 {\rm MeV}$. Decrease of the 
electron fraction results in the reduction of the pressure support and the core begins to
collapse. Note that neutrinos escape freely from the core before the central
density for $\rho \lesssim 10^{11} {\rm g}~{\rm cm}^{-3}$ as will be
mentioned in the next subsection.

The onset of core-collapse can be also understood by the fact that the 
\footnote[2]{Strictly speaking, this adiabatic index is
a pressure-averaged adiabatic. See for details, section III - {\it (f)} in
\cite{bruenn85}.} adiabatic index:
\begin{equation}
\Gamma_{S} \equiv \left(\frac{\partial \ln p}{\partial \ln \rho}\right)_{S},
\label{adia_def}
\end{equation}
is lowered below $4/3$, which is the instability
condition against the radial perturbation of a spherical star \cite{chandra}. 
From Eq. (\ref{degene_press}), the adiabatic index becomes
\begin{eqnarray}
\Gamma_{S} & = & 
\left.\frac{\partial \ln{P_{e}}}{\partial \ln{\rho}}\right|_{S}
= \left.\frac{\partial \ln{P_{e}}}{\partial \ln{\rho}}\right|_{S,Y_{e}}
+ \left.\frac{\partial \ln{Y_{e}}}{\partial \ln{\rho}}\right|_{S}
\left.\frac{\partial \ln{P_{e}}}{\partial \ln{Y_{e}}}\right|_{S} 
+ \left.\frac{\partial \ln{S}}{\partial \ln{\rho}}\right|_{Y_e}
\left.\frac{\partial \ln{P_{e}}}{\partial \ln{S}}\right|_{Y_e}
\label{final_term}
\\
& = & \frac{4}{3}\left( 1+
\left.\frac{\partial \ln{Y_{e}}}{\partial \ln{\rho}}\right|_{S}
\right),
\end{eqnarray}
where the final term on Eq.(\ref{final_term}) is set to zero, because
the collapse proceeds almost adiabatically. Progression of electron
capture implies negative $\left.\frac{\partial \ln{Y_{e}}}{\partial
\ln{\rho}}\right|_{S}$ which makes $\Gamma_{S}$ less than $4/3$.

Furthermore, the endothermic photodissociation of iron nuclei,
\begin{equation}
\gamma + ^{56} {\rm Fe} \rightarrow 13\alpha + 4 n -124.4 {\rm MeV}  
\label{photoreaction}
\end{equation}
occurs for the temperature $T \ge 5\times 10^{9}$ K, which leads to the
reduction of the thermal pressure support. In addition, the internal energy
produced by the core contraction is exhausted by this reaction. Both of
them promote the core collapse. 

Since the degenerate pressure of relativistic electrons in finite
temperature can be expressed as, 
\begin{equation}
P_e = \frac{\mu_e^4}{12 \pi^2 c^3 \hbar^3}\Biggl[ 1 + \frac{2}{3}
\Bigl(\frac{S_e}{\pi Y_e}\Bigr)^2\Biggr],
\end{equation}
the adiabatic index in Eq. (\ref{adia_def}) becomes,
\begin{equation}
\Gamma_S \approx \frac{4}{3}\Bigl[ 1 + \frac{\alpha^2}{1 + 2/3 \alpha^2} 
\left.\frac{\partial \ln{S_{e}}}{\partial \ln{\rho}}\right|_{Y_e} \Bigr],
\label{photo}
\end{equation}
where $S_e = \pi^2 Y_e k_{\rm B} T/ \mu_e$ is the electron entropy with
$k_{\rm B}$ being the Boltzmann constant \cite{baron}. The electron
entropy decreases with the central density during infall phase because
the photodissociation proceeds by the loss of the thermal energy 
of electrons. Hence, $\left.\frac{\partial \ln{S_{e}}}{\partial
\ln{\rho}}\right|_{Y_e}$ in Eq. (\ref{photo}) becomes negative, by which 
the core is shown to be destabilized by the reaction. It is noted that 
the entropy transfer from electron to nucleon occurs during the core
collapsing phase because the reduction of
electron entropy leads to the increase of the entropy of nucleons, while
conserving the total entropy \cite{baron}.

\subsubsection{neutrino trapping}
  After the onset of gravitational collapse, the core
proceeds to contract under the pull of the self gravitational
      force, unnoticed by the rest of the outer part of the star, 
      on a free-fall time scale, which is of the order of 
      $\tau \sim (G\bar{\rho})^{-1/2} \sim 0.04~{\rm sec}~(\bar{\rho}/1 \times
      10^{10} ~{\rm g}~{\rm cm}^{-3})^{-1/2}$ with the average 
density of the core of $\sim 10^{10}~{\rm g}~{\rm cm}^{-3}$. 
When the central
      densities exceed $10^{11} - 10^{12}~{\rm g}~{\rm cm}^{-3} $, 
      electron neutrinos, which can escape freely from the core at first,
      begins to be trapped in the core, because the timescale for 
       electron neutrino
      diffusion from the core becomes longer than that for the
      timescale of the core-collapse. This is the so-called  ``neutrino
      trapping'' , which plays very important roles in supernova
      physics \cite{Satotrap1,Satotrap2,fried}.

      During the core collapses, only electron neutrinos ($\nu_e$) are
 produced copiously by electron captures. Since the wavelength of
 neutrinos with the typical energy of $E_{\nu}$,
\begin{equation}
\lambda \approx 20~{\rm fm}\Bigl(\frac{E_{\nu}}{10~{\rm MeV}}\Bigr)^{-1},
\end{equation}
is longer than the size of the
 nuclei of $_{26} ^{56} {\rm Fe}$, 
\begin{equation}
r_{\rm nuc} \sim 1.2 A^{1/3}~{\rm fm} \approx 5 \left(\frac{A}{56}\right)^{\frac{1}{3}} {\rm fm},
\end{equation}
neutrinos are scattered coherently off $A$ nucleons in the
 nucleus, by which the cross section ($\nu_e + A \rightarrow \nu_e + A$)
 becomes roughly $A^2$ times of the cross section of each scattering of nucleons
 ($\nu_e + n,p \rightarrow \nu_e + n,p$). Thus the coherent scattering
 of neutrinos is the dominant opacity
 source for the neutrinos during the infall phase.

 The mean free path determined by the coherent scattering of the
 neutrinos on the iron nuclei is estimated to be,
\begin{eqnarray}
\lambda_{\nu} &=& \frac{1}{\sigma_{A}~n_{A}},
\label{eq:mean_free_path}
\end{eqnarray}
where $n_A = \rho/(A m_{u})$ is the number density of nuclei and 
$\sigma_A$ is the cross 
section of the coherent scattering in the leading order (see the
detailed one in \cite{thomp_detailed}),
\begin{equation}
\sigma_{A} = \frac{1}{16}\sigma_0 \Bigl(\frac{E_{\nu}}{m_{\rm e}c^2}\Bigr)^2
A^2\Bigl[1 - \frac{Z}{A} + (4 \sin^{2}\theta_{w} - 1)\frac{Z}{A}\Bigr]^2, 
\label{eqcoh}
\end{equation}
where $\sigma_0 = 4G_{F}^2(m_{\rm e} c^2)^2/(\pi (\hbar c)^4) = 1.705
\times 10^{-44}~{\rm cm}^2$ is a
convenient reference cross section of weak interactions, $G_F$ and $\theta_{w}$ is the Fermi coupling constant and the
Weinberg angle.  
The mean electron-neutrino energy in the iron ($_{26} ^{56} {\rm Fe}$) core can be estimated as follows,
\beq
E_{\nu} \approx \frac{5}{6} \mu_{e}
= \frac{5}{6} 
\left( 3 \pi^{2} \frac{\rho Y_{e}}{m_{u}} \right)^{\frac{1}{3}} \hbar c
\approx 10.3 {\rm MeV} \left(\frac{\rho}{3 \times 10^{10} {\rm g} \; 
{\rm cm}^{-3}}\right)^{\frac{1}{3}}
\left(\frac{Y_{e}}{26/56}\right)^{\frac{1}{3}}.
\label{eq:e_nu}
\eeq
 Introducing Eqs. (\ref{eqcoh}) and (\ref{eq:e_nu}) to
 Eq. (\ref{eq:mean_free_path}), the mean free path in Eq. (\ref{eq:mean_free_path}) becomes
\begin{eqnarray}
\lambda_{\nu} &=& \frac{1}{\sigma_{A}~n_{A}} \nonumber \\
&=& 8.5 \times 10^{6}~{\rm cm}~\left(\frac{\rho}{3 \times 10^{10} {\rm g} {\rm cm}^{-3}}\right)^{-1}
\left(\frac{A}{56}\right)^{-1}
\left(\frac{E_{\nu}}{10.3~{\rm MeV}}\right)^{-2},
\\ \label{middest}
&\approx& 10^{7} {\rm cm}
\left(\frac{\rho}{3 \times 10^{10} {\rm g} {\rm cm}^{-3}}\right)^{-5/3}
\left(\frac{A}{56}\right)^{-1}
\left(\frac{Y_{e}}{26/56}\right)^{-\frac{2}{3}}.
\label{mean_free_path}
\end{eqnarray}
Since the mean free path becomes smaller than the size of the core, 
\beq
R_{\rm core} \approx \Bigl(\frac{3M_{\rm core}}{4 \pi \rho}\Bigr)^{1/3}
\approx 2.7 \times 10^{7} {\rm cm} 
\left(\frac{\rho}{3 \times 10^{10} {\rm g} {\rm cm}^{-3}}\right)^{-\frac{1}{3}}
\left(\frac{Y_{e}}{26/56}\right)^{\frac{2}{3}},
\eeq
as the central density increases (note $\lambda_{\nu} \propto
\rho^{-5/3}$, while $R_{\rm core}\propto \rho^{-1/3}$), neutrinos cannot
escape freely from the core. This suggests that there is a
characteristic surface determining the escape or trapping of neutrinos in
the core. The radial position of the neutrino sphere is usually defined
as the surface where the neutrino ``optical'' depth,
\begin{equation}
\tau(r,E_{\nu}) = \int^{\infty}_{r} \frac{dr}{\lambda_{\nu}},
\end{equation}
becomes $2/3$. The neutrino sphere is the
effective radiating surface for neutrinos, in analogy with the
``photosphere'' of normal light emitting surface. It is noted that its
position differs from neutrino species to species and is dependent on
the neutrino energy. The neutrino sphere, which we are now discussing, is of
the electron neutrinos defined by their mean energy.

Introducing Eq. (\ref{middest}) to
the above equation, one obtains
\beq
\tau(r,E_{\nu}) \propto E^{2}_{\nu} \int^{\infty}_{r}
\rho(r) A(r) dr.
\eeq
Taking the distribution of the density, 
which can be approximated by 
\begin{equation}
\rho(r) = H \frac{1}{r^{3}}~~(~{\rm with}~ H=3\times10^{31})
\label{eq:core_density_profile}
\end{equation}
during the collapsing phase \cite{betherev}, and taking the typical values of $A = 56$
 at the central density of $\rho = 10^{12}~{\rm g}~{\rm
cm}^{-3}$, the optical depth becomes
\beq
\tau(r,E_{\nu}) \approx 6.1 
\left(\frac{E_{\nu}}{10 {\rm MeV}}\right)^{2}
\left(\frac{\rho(r)}{10^{12} {\rm g} \; {\rm cm}^{-3}}\right)^
{\frac{2}{3}}.
\eeq
Thus the typical radial position and the density of the neutrino sphere
 ($\tau(R_{\nu},E_{\nu})$ = 2/3) becomes 
\beq
R_{\nu} \approx 1.0 \times 10^{7}~{\rm cm}
\left(\frac{E_{\nu}}{10 {\rm MeV}}\right) ,
\label{eq:nu_sphere_radius}
\eeq
\beq
\rho(R_{\nu}(E_{\nu})) = 3.6 \times 10^{10}  {\rm g} \; {\rm cm}^{-3}
\left(\frac{E_{\nu}}{10 {\rm MeV}}\right)^{-3}.
\eeq
Neutrinos produced at $R > R_{\nu}$ can freely escape
from the core, while neutrinos produced inside propagates outwards by a
random-walk induced by the coherent scattering. The diffusion time for
neutrinos to diffuse out from the core of size $R$, can
be estimated as, 
\beq
t_{\rm diff} = \frac{3 R^{2}}{c\lambda_{\nu}}\approx 2.3 \times 10^{-1} {\rm sec} 
\left(\frac{\rho}{10^{12} {\rm g} \; {\rm cm}^{-3}}\right). 
\eeq
Since the dynamical timescale of the core,
\beq
t_{\rm dyn} = 4 \times 10^{-3} {\rm sec} 
\left(\frac{\rho}{10^{12} {\rm g} \; {\rm cm}^{-3}}\right)^{-\frac{1}{2}}, 
\eeq  
is shorter for the core density of $10^{11} \sim 10^{12} {\rm g} \; {\rm
cm}^{-3}$. This also means that neutrinos cannot freely escape from the core
and trapped in the core. After the neutrino trapping, the lepton
fraction ($Y_{l} = Y_{e} + Y_{\nu}$), where $Y_{\nu} = n_{\nu_e} -n_
{\bar{\nu}_{e}}$  is the electron-type
neutrino fraction per baryon, is kept nearly constant during the
collapse stage. Furthermore, electron neutrinos also
      become degenerate like electrons and the $\beta-$ equilibrium is 
      established between 
       $e^{-} + p \rightarrow n + \nu_{e}$ and its inverse. 
       After the achievement of the $\beta-$ equilibrium, the entropy is
       conserved and the collapse proceeds adiabatically.
\subsubsection{homologous collapse}
\begin{figure}
\begin{center}
\epsfxsize=7 cm
\epsfbox{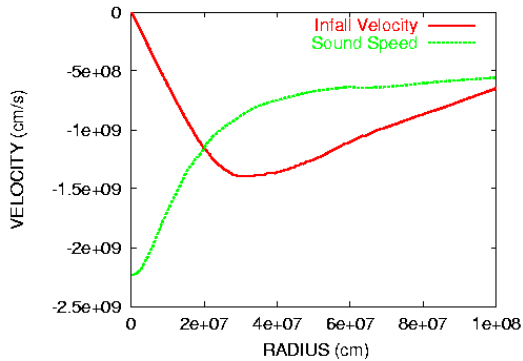}
\end{center}
\caption{Infall velocity and sound velocity versus radius at the central
 density of $10^{12}~{\rm g}~{\rm cm}^{-3}$ of a 15 $M_{\odot}$
 progenitor model. The region inside and outside the sonic
 point ($R \approx 200$ km, at which the two curves cross) roughly corresponds to the inner core and the outer
 core, respectively.}
\label{homocola}
\end{figure}

The collapsing core consists of two parts: the 
(homologously collapsing) inner core and the (supersonically infalling)
 outer core. This structure is clearly seen in Figure
\ref{homocola}.
Matter inside the sonic point (the point in the star where the sound speed
equals the magnitude of the infall velocity) stays in communication and
collapses homologously (velocity roughly proportional to
radius). On the
other hand, the material outside the sonic point falls in quasi-free
fall with velocity proportional to the inverse square of the
radius.  Beautiful analytic studies have been done on this phase
of collapse by \cite{weber,yahil_lattimer}, who predict that the size of the homologous core is
roughly the Chandrasekhar mass (see Eq. \ref{chand}). For a typical value of $Y_e$ in the inner core, the mass of
the inner core can be estimated $M_{\rm
ch} \approx 0.5 - 0.8 M_{\odot}$, which is in good agreement with the
mass of the inner core obtained in a realistic numerical simulation \cite{bruenn85}.
  
%\begin{figure}[hbt]
%\begin{center}
%\includegraphics[width=0.5\textwidth]{figure_kotake/r_vs_cs_and_v.eps}
%\end{center}
%\vspace{-0.3cm}
%\caption{Infall velocity and sound velocity versus radius at the central
% density of $10^{12}~{\rm g}~{\rm cm}^{-3}$ of the 15 $M_{\odot}$ star. The region inside and outside the sonic
% point ($R \approx 200$ km, at which the two curves cross) roughly corresponds to the inner core and the outer
% core, respectively.}
%\label{homocola}
%\end{figure}

\subsubsection{core bounce}
\begin{figure}
\begin{center}
\epsfxsize=10 cm
\epsfbox{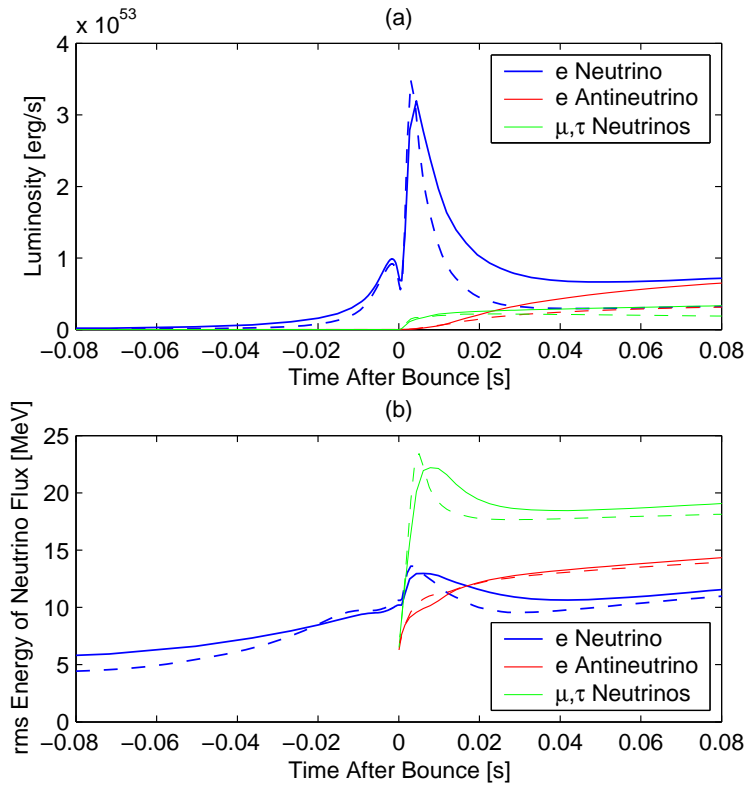}
\end{center}
\caption{The luminosities and root-mean-square energies of the neutrinos
 as a function of time. The results of the 13 $M_{\odot}$ model are
 drawn with dashed lines and the results of the 40 $M_{\odot}$ model
 with solid lines. 
%A thick line belongs to the electron neutrino, a line width medium
%width to the electron anitineutrino, and a thin line to the $\mu$ and
% $\tau$ neutrinos. 
The time is measure from the core bounce. This figure is taken from \cite{Liebendoerfer_et_al_04}.}
\label{lieb_neutro}
\end{figure}

When nuclear densities are reached in the
collapsing core ($\rho_{\rm c} \sim 3 \times 10^{14} {\rm
g}~{\rm cm}^{-3}$), repulsive nuclear forces
halt the collapse of the inner core driving a shock wave
into the outer core.
As the shock propagates into the outer core with dissociating nuclei
into free nucleons, the electron capture process 
$e^{-} + p \rightarrow n + \nu_{e}$ generates a huge amount 
of electron neutrinos just behind the shock.
Before the shock arrives at the neutrino sphere, these electron
neutrinos cannot escape in the hydrodynamical scale. Because
these regions are opaque to the final state electron neutrinos and they are 
effectively trapped because their diffusion time is much longer than
      that for the shock propagation.  As the shock 
waves move out in outer radius and pass through the neutrino sphere, the previously trapped electron neutrinos decouple
      from the matter and propagate ahead of the shock waves.
This sudden liberation of electron neutrinos is called the
neutronization burst (or ``breakout'' burst) (see the top panel of
Figure \ref{lieb_neutro}).
The duration of the neutronization burst is the
timescale of the shock propagation and, hence, less than $\sim 20~{\rm msec}$. 
While the peak luminosity exceeds $10^{53}~{\rm erg}~{\rm s}^{-1}$, the
total energy emitted in the neutronization burst is only of the order of
$10^{51}~{\rm erg}$ due to the short duration timescale.
 This electron-neutrino
 breakout signal is expected to be detected from the Galactic supernova in
      modern neutrino detectors such as SuperKamiokande and the Sudbury
      Neutrino Observatory (see section \ref{section:SN_nu_osc}). 
      
 The breakout of the 
     electron neutrinos is almost simultaneous with the appearance of
      the other neutrino species. In the hot post-bounce region, the
      electron degeneracy is not high and relativistic positrons are
      also created thermally leading the production of 
      anti-electron neutrinos
      ($\bar{\nu}_{e}'~s$) via reaction $e^{+} + n \rightarrow p + \bar{\nu}_{e}$.
      Mu- and tau- neutrinos are also produced in this epoch by the
      electron-positron annihilation ($e^{+}e^{-} \rightarrow 
      \nu_{\mu,\tau}\bar{\nu}_{\mu,\tau}$), nucleon-nucleon bremsstrahlung 
      ($N N^{'} \rightarrow N N^{'}\nu_{\mu,\tau}\bar{\nu}_{\mu,\tau}$), and
      neutrino/anti-neutrino annihilation 
      ($\nu_{e}\bar{\nu}_{e} \rightarrow \nu_{\mu,\tau}
     \bar{\nu}_{\mu.\tau}$) (see the bottom panel of Figure
     \ref{lieb_neutro}). Note that  
      each process listed above also contributes for $\nu_{e}$ and 
      $\bar{\nu}_{e}$ neutrinos, the production of these neutrinos are
      predominantly determined by the charged-current interactions,
      $e^{-} + p \rightarrow n + \nu_{e}$ and $e^{+} + n \rightarrow p 
      + \bar{\nu}_{e}$.

      At a radius of $100~{\rm km} \sim 200~{\rm km}$, the shock
      generated by core bounce stalls as a result of the following two 
      effects. First, as the shock propagates outwards, it dissociates
      the infalling iron-peak nuclei into free nucleons, thus giving up 
      $\sim 8.8~{\rm MeV}$ per nucleon in binding energy. Second, and
      most importantly, as the shock dissociates nuclei into free
      nucleons, electron capture on the newly-liberated protons to
      produce electron-neutrinos in the reaction 
      $e^{-} + p \rightarrow n + \nu_{e}$. 

\begin{figure}
\begin{center}
\epsfxsize=5. cm
\epsfbox{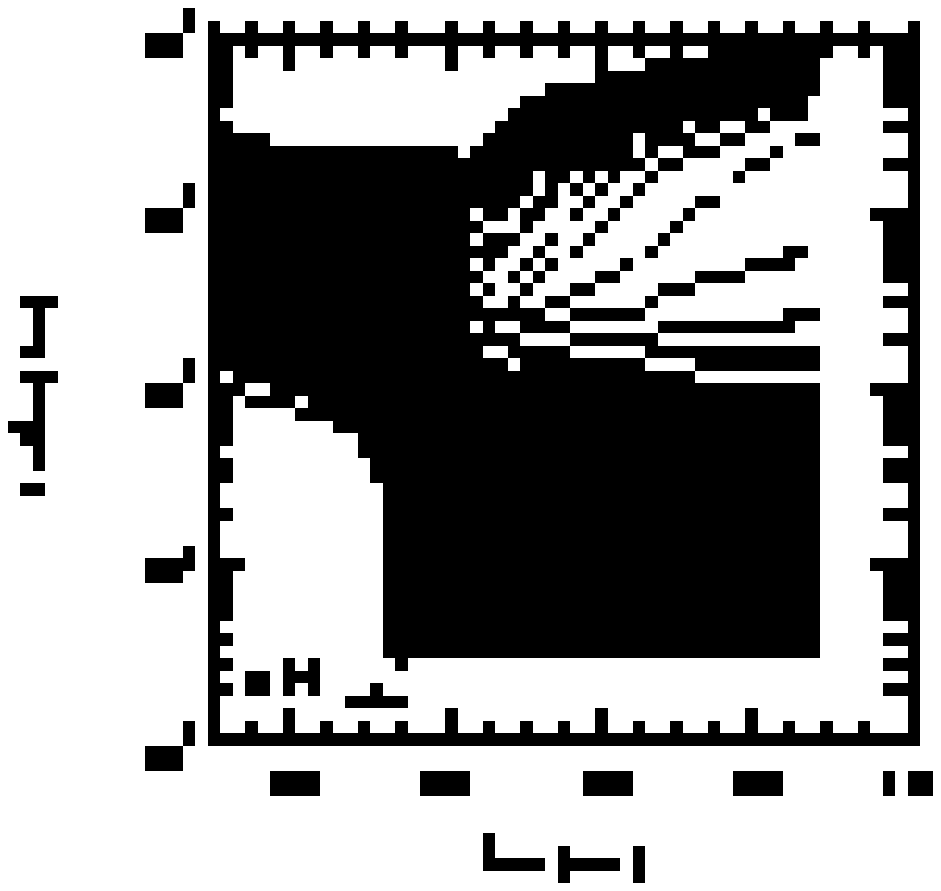}
\epsfxsize=5. cm
\epsfbox{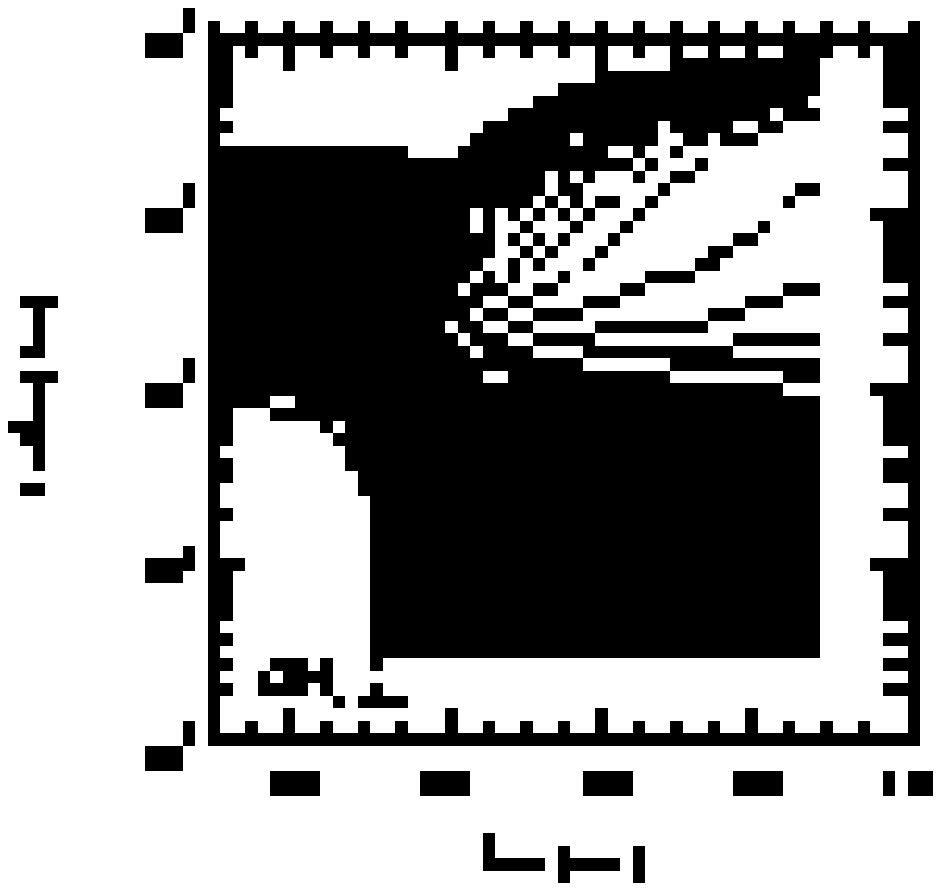}
\epsfxsize=5. cm
\epsfbox{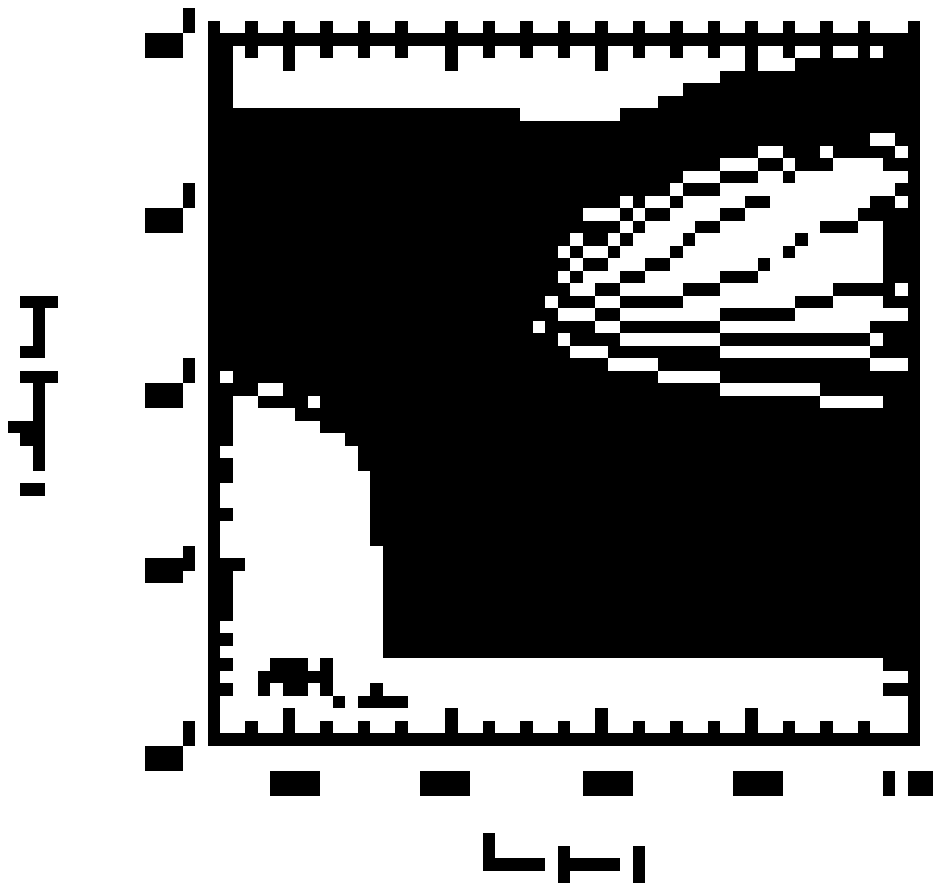}
\epsfxsize=5. cm
\epsfbox{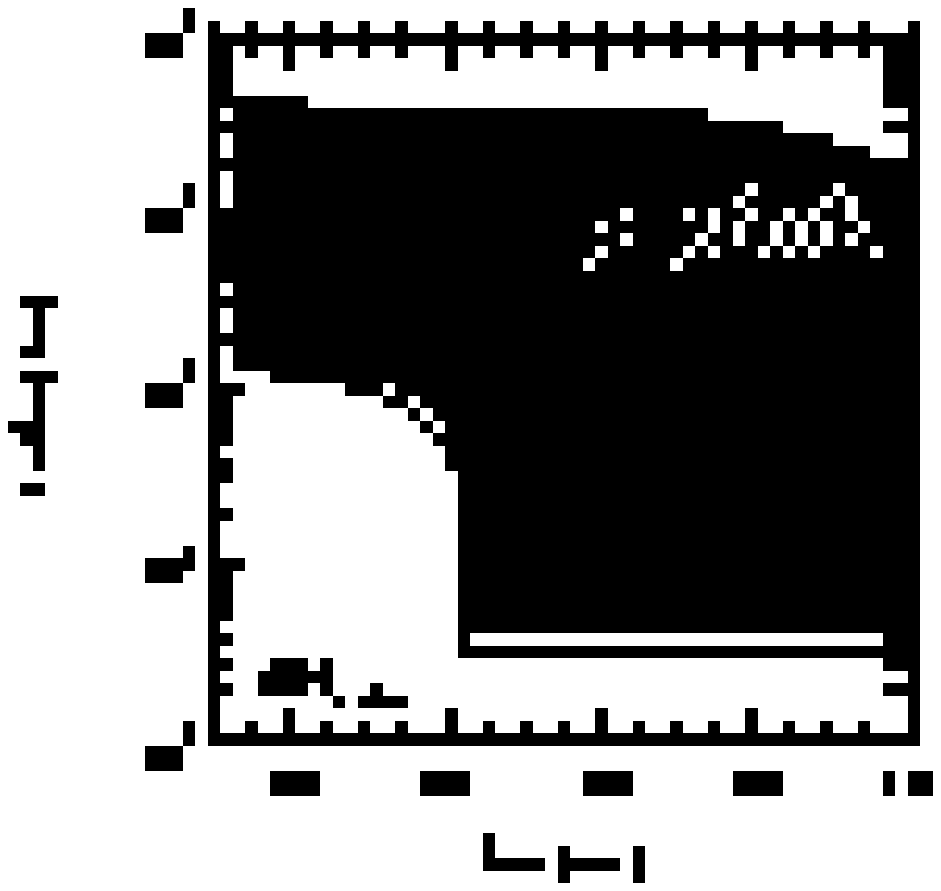}
\epsfxsize=5. cm
\epsfbox{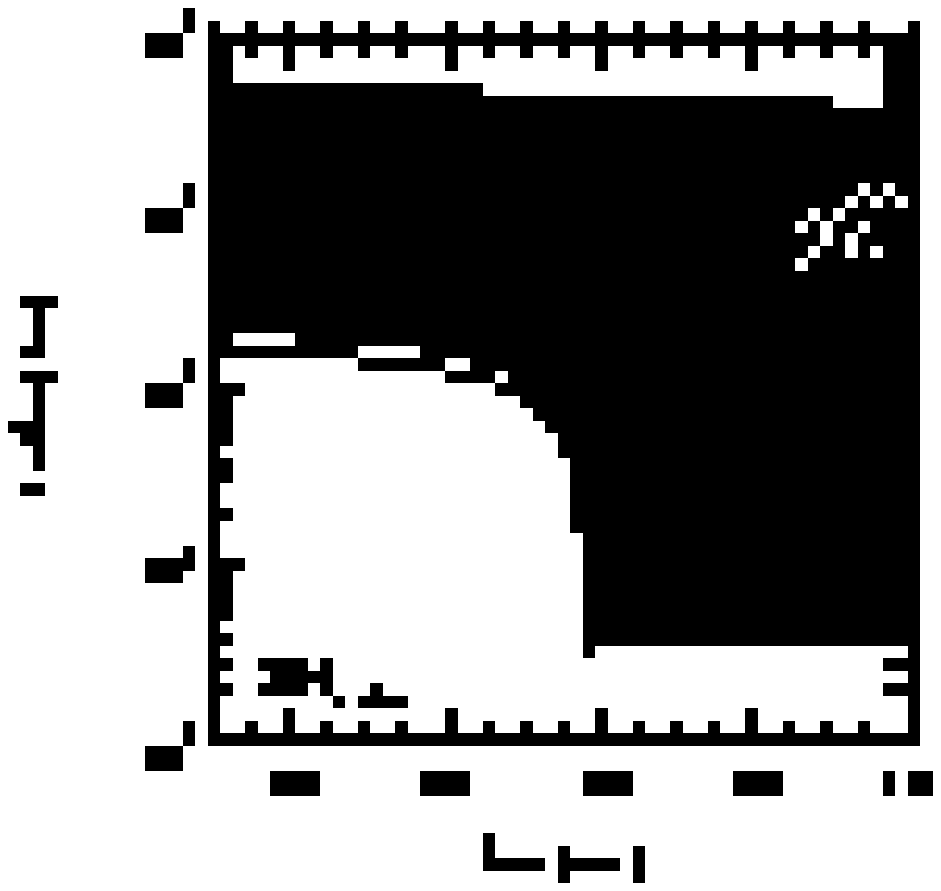}
\end{center}
\caption{Success or failure of the prompt explosion mechanism
 (courtesy of K. Sumiyoshi and see also \cite{sumi,sumi_prep}). Each
 panel shows the trajectories of mass mesh in radius as a function of
 time for the different progenitor models. The number in the panels
 represents the mass of the progenitor models of $11, 12, 15, 18, 20 M_{\odot}$ by \cite{ww:95} with the corresponding iron masses of $1.32, 1.32, 1.32, 1.46, 1.74 M_{\odot}$. It can be
 seen that the prompt explosion mechanism works in the case of the relatively
 smaller progenitor mass with the smaller iron core.      
\label{prompt}
}
\end{figure}

      Only for very special combinations of physical parameters, such as
      the stellar model of the progenitor or the incompressibility of
      nuclear matter, resulting in an extraordinary smaller cores,
      the so-called prompt explosion might work \cite{hill1,arnet1,hillnomo,baron}, in which the shock wave
      at core bounce propagates through the outer core to produce
      explosions without the shock-stall (see Figure \ref{prompt}).

\begin{figure}
\begin{center}
\epsfxsize=5in
\epsfbox{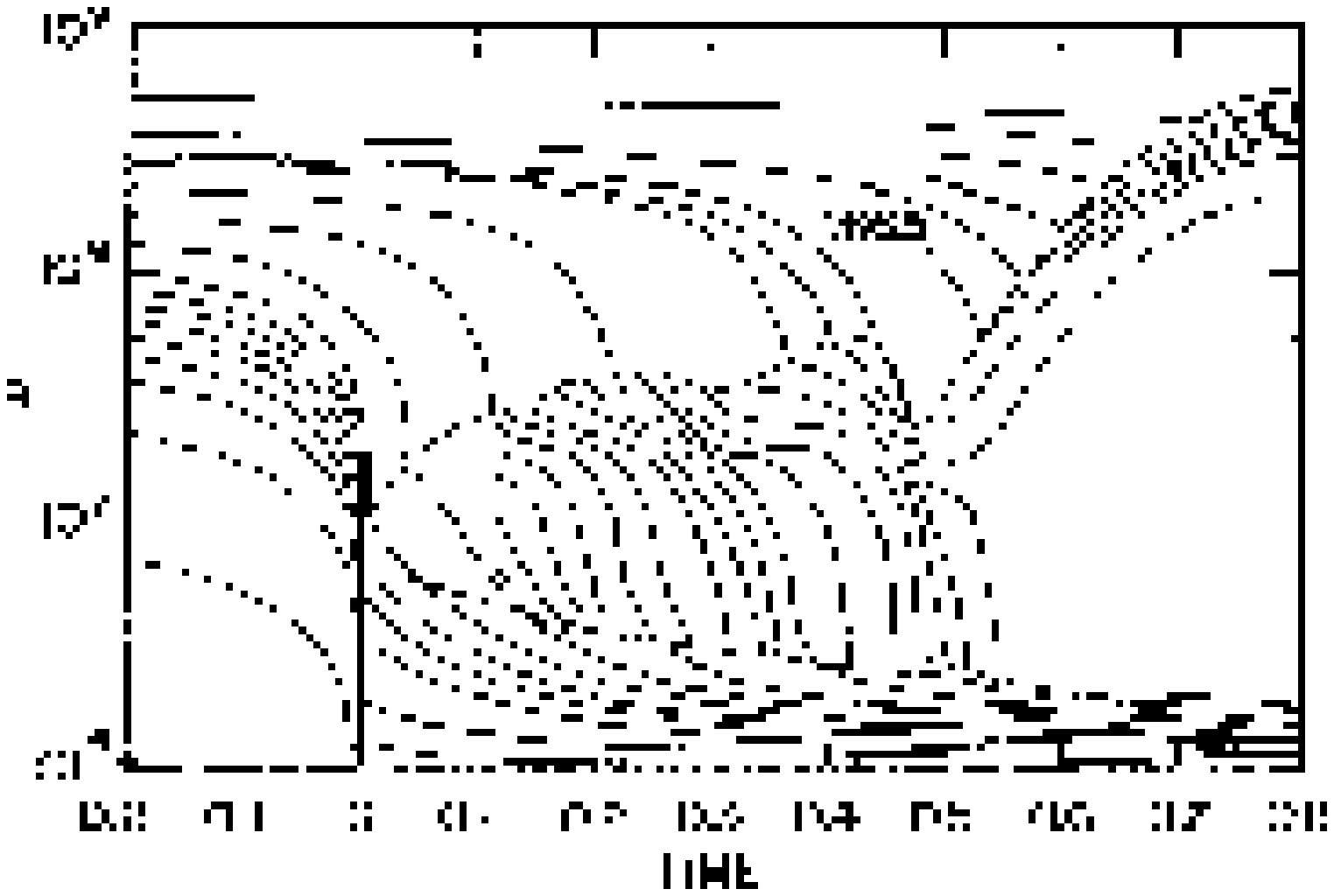}
\end{center}
\caption{Successful of delayed explosion taken from \cite{wilson1985}.
The $x$ and $y$ axis represents the time in unit of second measured from core bounce  and the radius from the stellar center in unit of centi-meter, respectively.  
Lines are trajectories of selected mass zones.
The dashed line represents the shock front. 1.665 $M_{\odot}$ in the figure shows the mass point which is expelled outwards by the second shock due to the neutrino heating. As a result,  the shock wave once weakened at $\sim 500$ msec revives and then successfully 
propagates to the surface of the iron core.
\label{fig:delayed_explosion}
}
\end{figure}
\subsubsection{delayed explosion}
      Only several milliseconds after the shock-stall, a quasi-hydrostatic equilibrium is maintained between the 
      newly-born protoneutron star (with radius of $\sim 50 - 80$ km)
      and the stalled accretion shock (with radius of $\sim 100 - 200$ km). The core of the protoneutron star
      is hot and dense, producing high-energy neutrinos of all species. 
      If the energy transfer from the neutrinos to the material near the
      stalled shock is large enough, the stalled shock can be
      revived to produce the successful explosion. This neutrino-heating
      mechanism was discovered from the numerical simulations by Wilson
      \cite{wilson1985} (see Figure \ref{fig:delayed_explosion}). It is interesting to note that already in 1960's, Colgate and White proposed that neutrino heating was
      essential for producing the explosions \cite{colgate}. 
      The amount of gravitational binding energy ($E_{\rm grav}$)
      released is huge,
\begin{equation}
E_{\rm kin} << E_{\rm grav} \sim \frac{3}{5}\frac{G M_{\rm NS}^2}{R} \sim 3 \times 10^{53}
\Bigl(\frac{M_{\rm NS}}{1.4 M_{\odot}}\Bigr)^2 \Bigl(\frac{R_{\rm NS}}{10 {\rm km}}\Bigr)^{-1}~{\rm erg},
\end{equation}
in contrast to the kinetic energy of
      canonical observed supernovae ($E_{\rm kin} \sim 10^{51}$ erg),
where $M_{\rm NS}$ and $R_{\rm NS}$ are the typical mass and the radius
       of a neutron star.
      Therefore in order to produce the explosions by the
      neutrino-heating mechanism, a small fraction
      ($\sim 1 \%$) of the binding energy should be transfered, via
      neutrinos, to the
      mantle above the
      protoneutron star that is ejected as the supernova.

\begin{figure}
\begin{center}
\epsfxsize=10 cm
\epsfbox{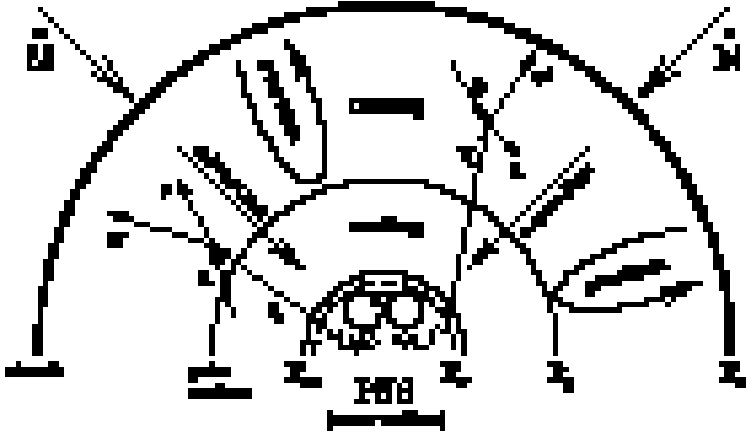}
\end{center}
\caption{Sketch of the stellar core during the shock revival
 phase. $R_{\nu}$ is the radius of neutrino sphere, from which neutrinos
 are emitted freely, $R_{\rm ns}$ is the radius of the protoneutron
 star, $R_{g}$ is the radius (see text) and $R_{s}$ is the radius of the
 stalled shock. The shock expansion is impeded by mass infall to the
 shock front at $R_{s}$ by the mass infall at a rate $\dot{M}$. This figure is taken from Janka (2001) \cite{jankashock}.}
\label{janka_fig}
\end{figure}

For the better understanding of the mechanism, we give an
 order-of-magnitude estimation according to \cite{betherev,jankashock}. 
 Let us assume the situation that a neutrino sphere is formed at
 radius of $R_{\nu}$ and from there, the isotropic neutrino with
 luminosity of $L_{\nu}$ is emitted (see Figure \ref{janka_fig}).
 Then the neutrino heating rate of nucleons
 via the reactions, $n+\nu_e \rightarrow e^{-} + p$ and $p+\bar{\nu_e}
 \rightarrow e^{+} + n$ at a radius $R$ ($R_{\nu} < R < R_{\rm s}$) can
 be estimated as, 
\begin{eqnarray}
 Q_{\nu}^{+} &\cong&  \frac{L_{\nu}~\sigma(\epsilon_{\nu}) Y_{N}}{4 \pi R^2} 
 \nonumber \\ 
 &\sim& 53 \Bigl(\frac{L_{\nu}}{10^{52}~{\rm erg}~{\rm s}^{-1}}\Bigr)
\Bigl(\frac{\epsilon_{\nu}}{15~{\rm MeV}}\Bigr)^2 
\Bigl(\frac{R}{150~{\rm km}}\Bigr)^{-2}
\Bigl(\frac{Y_{N}}{1.0}\Bigr)~
\Bigl[\frac{\rm MeV}{{\rm s}\cdot{\rm nucleon}}\Bigr],
\label{abs}
\end{eqnarray}
where $L_{\nu}$ is a
typical neutrino luminosity, $\epsilon_{\nu}$ is the mean energy of
neutrinos, $\sigma(\epsilon_{\nu})$ is the cross section of the above
absorption processes. Here we take $Y_{N}$, the sum of the fraction of free
nucleon and protons, to be $1$ because nuclei are nearly fully dissociated into free
nucleons after the passage of the shock waves. Outside the stalled
shock, on the other hand, the above heating rates are
suppressed because of the absence of the free nucleons.
Note that each value assumed
in the above estimation is taken from the recent 1D core-collapse
simulation \cite{Lieben01}. 
On the other hand, the gravitational binding energy per baryon at a
radius of $R$ can be given as follows:
\begin{equation}
- \frac{GM_{NS}m_u}{R} = - 8.7 \Bigl(\frac{M_{\rm NS}}{1.4M_{\odot}}\Bigr)
\Bigl(\frac{R}{150 {\rm km}}\Bigr)^{-1}~[{\rm MeV/nucleon}],
\label{heat_bethe}
\end{equation}
where we take $M_{NS}$ to be a typical mass scale of $1.4 M_{\odot}$. Comparing the neutrino heating rate (r.h.s. of Eq. (\ref{abs})) 
with the binding energy (r.h.s. of Eq. (\ref{heat_bethe})), one can see
that the neutrino heating can give the matter enough energy to be
      expelled from the core in 0.16 second. In realistic situations, 
      the cooling of matter occurs 
      simultaneously via the very inverse process of the heating
      reactions, which  delays the timescale of the
      explosion up to $\sim 1~{\rm sec}$
      \cite{wilson1985}. These timescales are
       much longer
      than those of the prompt explosion
      mechanism ($O(10~{\rm msec})$). Thus the 
      neutrino-driven mechanism is sometimes called as the delayed explosion
      mechanism. 

\begin{figure}[H]
\begin{center}
\epsfxsize=16 cm
\epsfbox{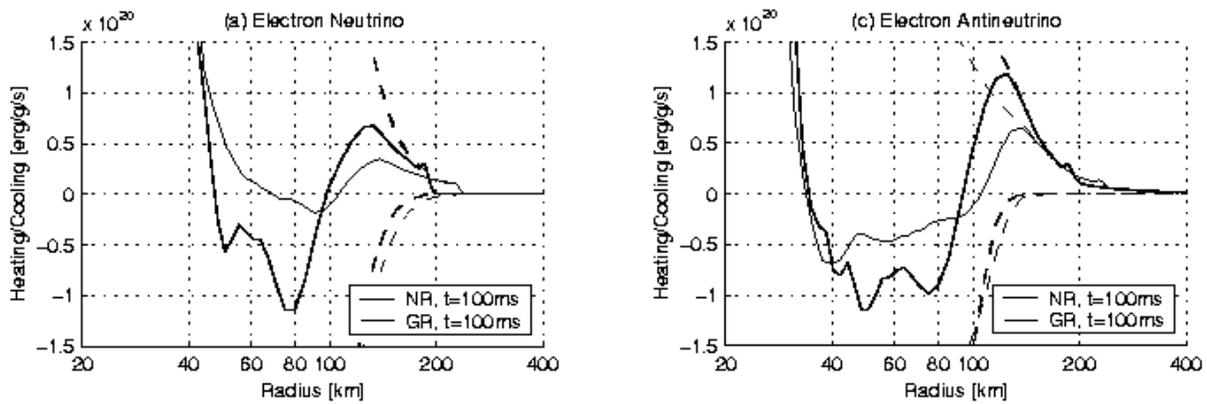}
\end{center}
\caption{Heating and cooling rates by electron (left) and anti-electron 
 (right) neutrinos at 100 msec after core bounce taken from Liebend\"{o}refer et
 al. 2001 \cite{Lieben01}. At the moment, the neutrino sphere defined by
 its mean energy and the stalled shock is located at $R_{\nu} \sim 80$ km and
 $R_{s} \sim 200$ km, respectively. In both panels, dashed lines with positive
 values, dashed lines with negative lines, and solid lines represent 
the heating rates, cooling rates, and the net rates, respectively. From
 the figures, it is shown that the location where the net heating
 changes sign defines the gain radius $R_{g}\sim 80 \sim 100$ km. Lines
 labeled by ``NR'' or ``GR'' indicate that the results are obtained by
 the Newtonian or general relativistic simulations. }
\label{lieben_gain}
\end{figure}

Noteworthy, a characteristic radial position, which is the so-called gain radius, 
in which the neutrino heating and
      cooling balances and above which the neutrino heating dominates over
      the neutrino cooling, are formed after the shock-stall
      \cite{bethegain}. In the following, we estimate the position of
      the gain-radius by an order-of-magnitude estimation. In addition
      to the neutrino heating rate (Eq. (\ref{abs})), the
      neutrino cooling rate of nucleons is given as follows:
\begin{equation}
Q_{\nu}^{-} = - \sigma(T) a^{'}c T^4,
\end{equation}
where $T$ is the temperature of the material, $\sigma(T)$ is the
corresponding neutrino absorption cross section,  $a^{'} = 7/16 \times
1.37\cdot 10^{26}~{\rm erg}~{\rm cm}^{-3}~{\rm MeV}^4 $ is the
radiation density constant of neutrinos, and $c$ is the speed of
light. Since we assume for simplicity that the distribution function of 
neutrino is the fermi distribution with a vanishing chemical
potential, then $a^{'} T^4$ represents the energy density of neutrinos
which yields to a black body radiation.
Here we write $L_{\nu}$ in
equation (\ref{abs}) as follows,
\begin{equation}
L_{\nu} = \pi R_{\nu}^2 a^{'} c T_{\nu}^4,
\end{equation}
assuming again that the neutrinos from the neutrino sphere are a Fermi
distribution of the temperature $T_{\nu}$ of the neutrino sphere.
Noting that $\sigma(T)/\sigma(T_{\nu}) = (T/T_{\nu})^2$, 
the net heating rate can be written, 
\begin{equation}
Q_{\rm tot} = Q_{\nu}^{+} - Q_{\nu}^{-} = Q_{\nu}^{+}
\Biggl[1 - \Bigl(\frac{2R}{R_{\nu}}\Bigr)^2\Bigl(\frac{T}{T_{\nu}}\Bigr)^6
\Biggr].
\end{equation}
Using the simple power law relation,
\begin{equation}
T = T_{\rm s}\frac{R_s}{R},
\end{equation}
which yields a good approximation in the radiation dominated atmosphere 
$R_{\nu}< R < R_{\rm s}$ \cite{jankashock}, the position of the gain radius $R_{g}$
becomes
\begin{equation}
R_{g} = \sqrt{\frac{(2R_s)^3}{R_{\nu}}\Bigl(\frac{T_s}{T_{\nu}}\Bigr)^3}.
\end{equation} 
Taking data obtained from a state-of-the-art numerical simulations 
\cite{Lieben01},
namely, $T_{\nu} = 4.8 {\rm MeV}, T_{s} = 1{\rm MeV}, R_{s}= 200{\rm
km}, R_{\nu} = 80{\rm km}$, the gain radius becomes $85$ km, which is
in good agreement with the position numerically obtained by the
corresponding simulations (see Figure \ref{lieben_gain}). 

      The extent of the region of the net neutrino heating and the magnitude
      of the net neutrino energy deposition are responsible for producing
      the successful explosions and dependent crucially on the
      neutrino energy density and the flux outside the
      neutrino sphere, in which the neutrino semi-transparently
      couples to the matter. Thus the accurate treatment of neutrino 
      transport is an important task in order to address the success or
      failure of the supernova explosions in numerical simulations.

%Importantly, a robust conclusion of detailed
%      numerical models is that directly above the electron
%      neutrino sphere $(R_{\nu_{e}} \sim 80~{\rm km})$ is a region of net
%      neutrino cooling. At what is so-called the gain radius $(R_{g})$,
%      heating balances cooling  $(R_{g} \sim 100~{\rm km})$. Treating
%      the problem stationary, $R_{g}$ must also correspond to a region
%      with a maximum entropy because $T v~ds/dr|_{R_{g}} \sim H - C = 0$,
%      where $s$ is the specific entropy and $T, v$ are the postshock
%      temperature and velocity, respectively. Only between $R_{g}$ and
%      the radius of the shock ($R_{s} \sim$ 200 km) is a region of net
%      neutrino heating.

If the neutrino heating mechanism works sufficiently to revive the
      stalled shock wave,  the shock wave goes into the stellar
      envelope and finally blows off. This is observed as a supernova
      after the shock breaks out the photosphere. Unlike the case in the
      iron core, the photodissociation and the energy loss due to
      neutrinos is negligible in the stellar envelope and the binding
      energy is small, the shock wave successfully explodes the whole
      star. The propagation time of the shock wave depends on the
      stellar radius and is in the range from several hours to days.

 Here we shall mention that there is another type of supernovae, driven
 by a quite different physical mechanism.
Supernovae Type Ia characterized by the absence of hydrogen lines 
in their spectra are thought to be caused by a thermonuclear explosion 
of a white dwarf that is completely disrupted in this event (for a
 review of SN Ia explosion models see \cite{hillenie}). 
Since the luminosities at the 
explosions of Type Ia supernovae are almost constant, they are 
good candidates to determine extragalactic distances and to measure the 
basic cosmological parameters. We will not consider them in this
review. Supernovae we pay attention to in this thesis are the so-called
 Type II, Type Ib and
Ic, (for the
 recent observational classifications of supernovae, see \cite{hamuy} for example.)
%(Type Ib: the
%absence of silicon lines and the presence of strong helium lines, Type
%Ic: without the silicon lines and with/without weak helium lines in their
%spectra, respectively).
 For convenience, we have used the common name 
``core-collapse supernovae'' for supernovae of Types II and Ib/c.

\clearpage

%%%%%%%%%%%%%%%%%%%%%%%%%%%%%%%%%%%%%%%%%%%%%%%%%%%%%%%%%%%%%%%%%%%%%%%%%%%%%%%%%%%%%%%%%%%%%%%
%%%%%%%%%%%%%%%%%%%%%%%%%%%%%%%%%%%%%%%%%%%%%%%%%%%%%%%%%%%%%%%%%%%%%%%%%%%%%%%%%%%%%%%%%%%%%%%
\section{Neutrino Oscillation\label{section:nu_osc}}
%%%%%%%%%%%%%%%%%%%%%%%%%%%%%%%%%%%%%%%%%%%%%%%%%%%%%%%%%%%%%%%%%%%%%%%%%%%%%%%%%%%%%%%%%%%%%%%
%%%%%%%%%%%%%%%%%%%%%%%%%%%%%%%%%%%%%%%%%%%%%%%%%%%%%%%%%%%%%%%%%%%%%%%%%%%%%%%%%%%%%%%%%%%%%%%
In this section, we give
a fundamental tool to discuss supernova neutrinos and their observation,
neutrino oscillation. These topics have been seldom reviewed systematically so
far. Starting from the physical foundation,
we give an elaborate description of the neutrino oscillation, which has
been established
recently by a lot of experiments. Based on this, 
neutrino oscillations in supernovae are reviewed in the next section.  
%%%%%%%%%%%%%%%%%%%%%%%%%%%%%%%%%%%%%%%%%%%%%%%%%%%%%%%%%%%%%%%%%%%%%%%%%%%%%%%%%%%%%%%%%%%%%%%
\subsection{Overview}
%%%%%%%%%%%%%%%%%%%%%%%%%%%%%%%%%%%%%%%%%%%%%%%%%%%%%%%%%%%%%%%%%%%%%%%%%%%%%%%%%%%%%%%%%%%%%%%
 
So far, we know three types of neutrino, $\nu_{e}, \nu_{\mu}$ and $\nu_{\tau}$.
These are partners of the corresponding charged leptons, electron, muon and tauon, respectively,
and produced via charged current interactions such as $\beta$-decay and decays of muons and tauons.
Thus, $\nu_{e}, \nu_{\mu}$ and $\nu_{\tau}$ are called {\it flavor (weak) eigenstates}, which mean
the eigenstates of the weak interaction. On the other hand, we can also consider eigenstates of
their free Hamiltonian. They are called {\it mass eigenstates} denoted as $\nu_{i} (i = 1, 2, 3)$
and have definite masses $m_{i}$.

These types of eigenstates come from essentially different physical concept so that they
do not necessarily coincide with each other. In fact, this is the case with the quark sector:
Flavor eigenstates are linear combinations of mass eigenstates which are determined by
a unitary matrix called Cabbibo-Kobayashi-Maskawa matrix. Then, like the quark sector,
it is natural to consider that the leptons are also mixing.

The lepton mixing means that, for example, $\nu_{e}$, which is produced by $\beta$-decay,
is a linear combination of some mass eigenstates $\nu_{i}$. More generally, neutrinos are
always produced and detected in flavor eigenstates, which are not eigenstates of
the propagation Hamiltonian. This mismatch leads to neutrino oscillation.

Neutrino oscillation can be roughly understood as follows.
Expressing the wave function of the neutrino by plane wave, each mass eigenstate evolves
as $\exp{[i(E_{i} t - \vec{k}_{i} \cdot \vec{x})]}$, where $E_{i}$ and $\vec{k}_{i}$ are energy
and momentum of the mass eigenstate $\nu_{i}$. Because different masses $m_{i}^{2}$ lead to
different dispersion relations $E_{i}^{2} = k_{i}^{2} + m_{i}^{2}$, phase differences between
the wave functions would appear as the neutrino evolves. Thus time-evolved wave function of a flavor
eigenstate is no longer the original linear combination of mass eigenstates, which means that there is
a probability that the neutrino is detected as a different flavor from the original flavor.

%%%%%%%%%%%%%%%%%%%%%%%%%%%%%%%%%%%%%%%%%%%%%%%%%%%%%%%%%%%%%%%%%%%%%%%%%%%%%%%%%%%%%%%%%%%%%%%
\subsection{Vacuum Oscillation \label{subsection:vacuum}}
%%%%%%%%%%%%%%%%%%%%%%%%%%%%%%%%%%%%%%%%%%%%%%%%%%%%%%%%%%%%%%%%%%%%%%%%%%%%%%%%%%%%%%%%%%%%%%%

Let us start with the Klein-Gordon equation neglecting the spin degree of freedom of neutrino,
which is not important unless neutrino has large magnetic dipole moment. The equations of motion
of the mass eigenstates in vacuum are
\begin{equation}
\left( \frac{\partial^{2}}{\partial t^{2}} - \nabla^{2} + M^{2} \right) \Psi^{\rm (m)} = 0,
\end{equation}
where $\Psi^{\rm (m)} = (\nu_{1},\nu_{2},\nu_{3})$ is a wave-function vector of the mass eigenstates
and $M = {\rm diag}(m_{1},m_{2},m_{3})$ is the mass matrix. Let us expand the wave function as
\begin{equation}
\Psi^{\rm (m)}(t,\vec{x}) = e^{-iEt} \Psi_{E}^{\rm (m)}(\vec{x}),
\end{equation}
where we assumed all the mass eigenstates have the same energy $E$. Although this assumption is
not physically appropriate, the results below are not affected if the neutrino is ultra-relativistic.
Then the equations of motion become
\begin{equation}
\left( -E^{2} - \nabla^{2} + M^{2} \right) \Psi_{E}^{\rm (m)} = 0
\end{equation}
If the neutrino is ultra-relativistic $( E \sim k_{i} \gg m_{i})$, we have
\begin{equation}
-E^{2} - \nabla^{2} = - (E+i\nabla) (E-i\nabla) \approx -(E+i\nabla) 2E,
\end{equation}
which leads to
\begin{equation}
i \frac{\partial}{\partial z} \Psi_{E}^{\rm (m)}
= - \left( E - \frac{M^{2}}{2E} \right) \Psi_{E}^{\rm (m)},
\label{eq:EOM}
\end{equation}
where we set the direction of motion to $z$ direction. The first term on the r.h.s. of
Eq. (\ref{eq:EOM}) is irrelevant for neutrino oscillation because it just contributes to
overall phase and will be neglected from now on.

On the other hand, flavor eigenstates can be written as linear combinations of
mass eigenstates as,
\begin{equation}
\Psi_{E}^{\rm (f)} = U \Psi_{E}^{\rm (m)},
\end{equation}
where $U$ is the mixing matrix, which is also referred to as the Maki-Nakagawa-Sakata (MNS) matrix
\cite{MakiNakagawaSakata62}. This matrix corresponds to the Kobayashi-Maskawa matrix in the quark sector
and often parameterized as,
\begin{equation}
U =
\left( \begin{array}{ccc}
c_{12} c_{13} & s_{12} c_{13} & s_{13} e^{-i \delta} \\
-c_{23} s_{12} - c_{12} s_{23} s_{13} e^{i\delta} & 
c_{12} c_{23} - s_{12} s_{23} s_{13} e^{i\delta} & c_{13} s_{23} \\
s_{12} s_{23} - s_{13} e^{i\delta} & 
- c_{12} s_{23} - c_{23} s_{12} s_{13} e^{i\delta} & c_{13} c_{23}
\end{array} \right)
\label{eq:mixing_matrix}
\end{equation}
where $s_{ij} \equiv \sin{\theta_{ij}}$ and $c_{ij} \equiv \cos{\theta_{ij}}$,
$\theta_{ij} (ij=1,2,3)$ are mixing angles and $\delta$ is $CP$ phase.
In terms of the flavor eigenstates, the equations of motion (\ref{eq:EOM}) are expressed as,
\begin{equation}
i \frac{\partial}{\partial z} \Psi_{E}^{\rm (f)} = \frac{U M^{2} U^{\dag}}{2E} \Psi_{E}^{\rm (f)}.
\label{eq:EOM2}
\end{equation}
Here it should be noted that the mass matrix for flavor eigenstates, $U M U^{\dag}$, is not
diagonal in general.

As a simple example, let us assume that there are only two neutrino species,
$\nu_{e}$ and $\nu_{\mu}$. Then the mixing matrix can be written as,
\begin{equation}
U =
\left( \begin{array}{cc}
\cos{\theta} & \sin{\theta} \\
-\sin{\theta} & \cos{\theta}
\end{array} \right)
\end{equation}
and the equations of motion reduce to
\begin{equation}
i \frac{\partial}{\partial z} \left( \begin{array}{c} \nu_{1} \\ \nu_{2} \end{array} \right)
=
\left( \begin{array}{cc}
\frac{m_{1}^{2}}{2E} & 0 \\
0 & \frac{m_{2}^{2}}{2E}
\end{array} \right)
\left( \begin{array}{c} \nu_{1} \\ \nu_{2} \end{array} \right)
\end{equation}
for mass eigenstates and
\begin{equation}
i \frac{\partial}{\partial z} \left( \begin{array}{c} \nu_{e} \\ \nu_{\mu} \end{array} \right)
= \frac{\Delta m^{2}}{4 E}
\left( \begin{array}{cc}
- \cos{2 \theta} & \sin{2 \theta} \\
\sin{2 \theta} & \cos{2 \theta}
\end{array} \right)
\left( \begin{array}{c} \nu_{e} \\ \nu_{\mu} \end{array} \right)
\label{eq:EOM_2f}
\end{equation}
for flavor eigenstates. Here $\Delta m^{2} \equiv m_{2}^{2} - m_{1}^{2}$ and we again neglected
a term proportional to identity matrix. Then consider a neutrino which is purely electron-type
at first. Noting that electron-type neutrino can be written in terms of mass eigenstates as,
\begin{equation}
|\nu_{e}\rangle = \cos{\theta} |\nu_{1}\rangle + \sin{\theta} |\nu_{2}\rangle,
\end{equation}
the neutrino evolves according to,
\begin{equation}
|\nu(z)\rangle = \exp{\left(-i\frac{m_{1}^{2}}{2E}z\right)} \cos{\theta} |\nu_{1}\rangle
                 + \exp{\left(-i\frac{m_{2}^{2}}{2E}z\right)} \sin{\theta} |\nu_{2}\rangle.
\end{equation}
Multiplying $\langle \nu_{e}|$, we obtain the probability that this state is electron type,
\begin{equation}
P_{\nu_{e} \rightarrow \nu_{e}}(z) = \left| \langle \nu_{e} | \nu(z) \rangle \right|^{2}
= 1 - \sin^{2}{2\theta} \sin^{2}{\left( \pi \frac{z}{\ell_{\rm osc}} \right)},
\label{eq:conversion_probability}
\end{equation}
where
\begin{equation}
\ell_{\rm osc} \equiv \frac{4 \pi E}{\Delta m^{2}}
= 2.48 \times 10^{7} {\rm cm} \left( \frac{E}{1{\rm MeV}} \right)
                              \left( \frac{10^{-5} {\rm eV}}{\Delta m^{2}} \right)
\label{eq:osc_length}
\end{equation}
is called the {\it oscillation length}. It is easy to show that
\begin{eqnarray}
&& P_{\nu_{\mu} \rightarrow \nu_{e}}(z) = P_{\nu_{e} \rightarrow \nu_{\mu}}(z) \\
&& P_{\nu_{e} \rightarrow \nu_{e}}(z) = P_{\nu_{\mu} \rightarrow \nu_{\mu}}(z)
   = 1 - P_{\nu_{e} \rightarrow \nu_{\mu}}(z)
\end{eqnarray}
as expected by unitarity.

\begin{figure}[t]
\begin{center}
\epsfbox{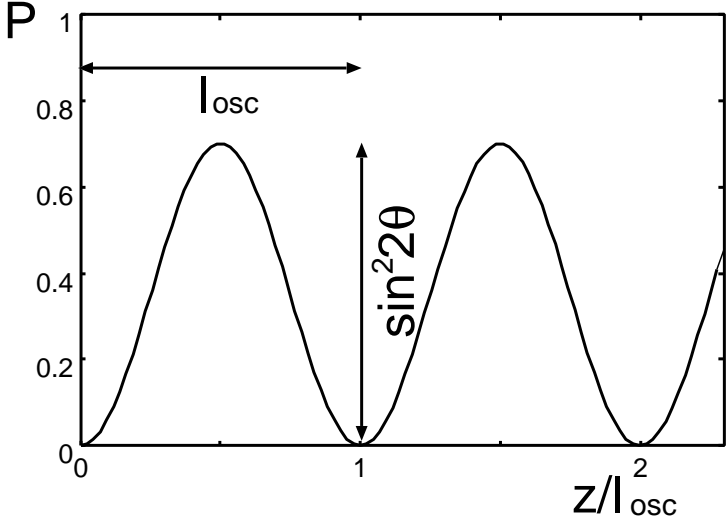}
\end{center}
\vspace{-0.5cm}
\caption{Conversion probability $P_{\nu_{e} \rightarrow \nu_{\mu}}(z)$ as a function of
propagation distance $z$ normalized by the oscillation length $\ell_{\rm osc}$
\label{fig:conv_prob_vac}}
\end{figure}

The probability $P_{\nu_{e} \rightarrow \nu_{\mu}}(z)$ as a function of propagation distance $z$
is plotted in Fig. \ref{fig:conv_prob_vac}. It oscillates with respect to $z$ and the wave length
is the oscillation length $\ell_{\rm osc}$. Here it will be worth noting that the oscillation length
depends on the neutrino energy and the mass difference of the two involved mass eigenstates as is
seen in (\ref{eq:osc_length}). The amplitude is determined by the mixing angle and is the largest
when $\theta = \pi/4$. Thus even if a neutrino is produced as an electron-type neutrino, there is
non-zero probability that it is detected as a muon-type neutrino if there is a mixing between the
two neutrino flavors. It is this phenomenon which is known as the neutrino oscillation.

Let us consider a more general case with many neutrino species. A neutrino state with a flavor
$\alpha$ can be written as a linear combination of the mass eigenstates,
\begin{equation}
|\nu_{\alpha}\rangle = \sum_{i} U^{*}_{\alpha i} |\nu_{i}\rangle.
\end{equation}
Then the evolution of a neutrino which is originally $\nu_{\alpha}$ is
\begin{equation}
|\nu(z)\rangle = \sum_{i} U^{*}_{\alpha i} \exp{\left(-i\frac{m_{i}^{2}}{2E}z\right)}
                 |\nu_{i}\rangle,
\end{equation}
and the probability $P_{\nu_{\alpha} \rightarrow \nu_{\beta}}(z)$ is
\begin{equation}
P_{\nu_{\alpha} \rightarrow \nu_{\beta}}(z)
= \sum_{i} |U_{\alpha i} U_{\beta i}|^{2}
  + 2 {\rm Re} \sum_{i>j} U^{*}_{\alpha i} U_{\beta i} U_{\alpha j} U^{*}_{\beta j}
                          \exp{\left(-i\frac{\Delta m_{ij}^{2}}{2E}z\right)}
\end{equation}
where $\Delta m_{ij}^{2} \equiv m_{i}^{2} - m_{j}^{2}$.

\paragraph{averages}

\begin{figure}[hbt]
\begin{center}
  \epsfxsize=8 cm
 \epsfbox{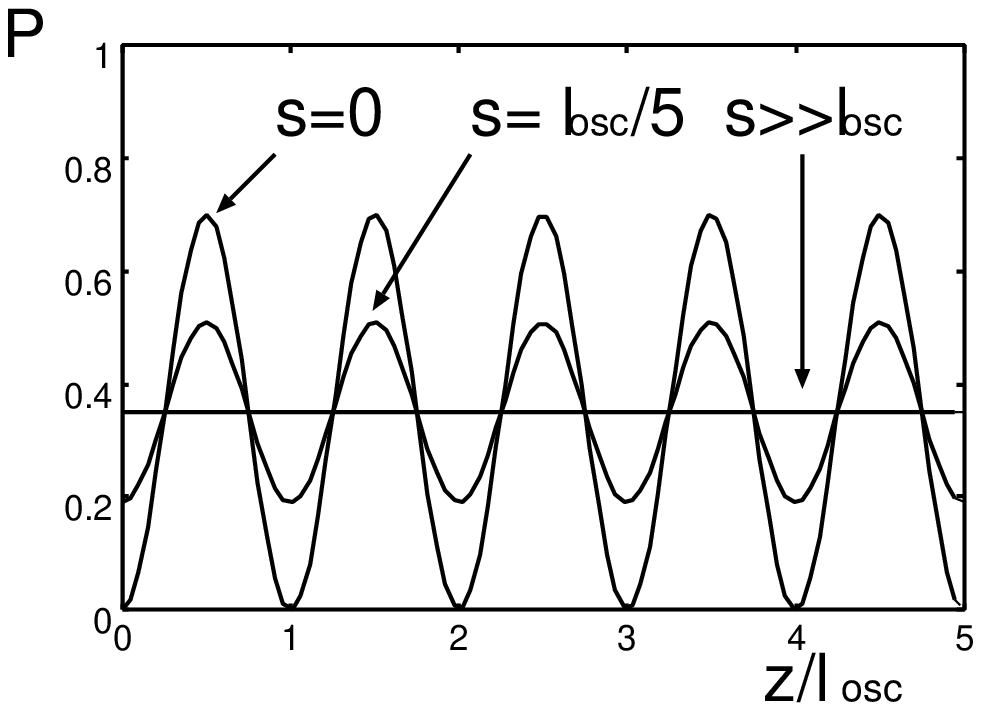}
 \caption{Conversion probability $P_{\nu_{e} \rightarrow \nu_{\mu}}(z)$ averaged
 by taking the finite size of a source into account.
 \label{fig:ave_source}}
%\hspace{0.5cm}  
  \epsfxsize=8 cm
 \epsfbox{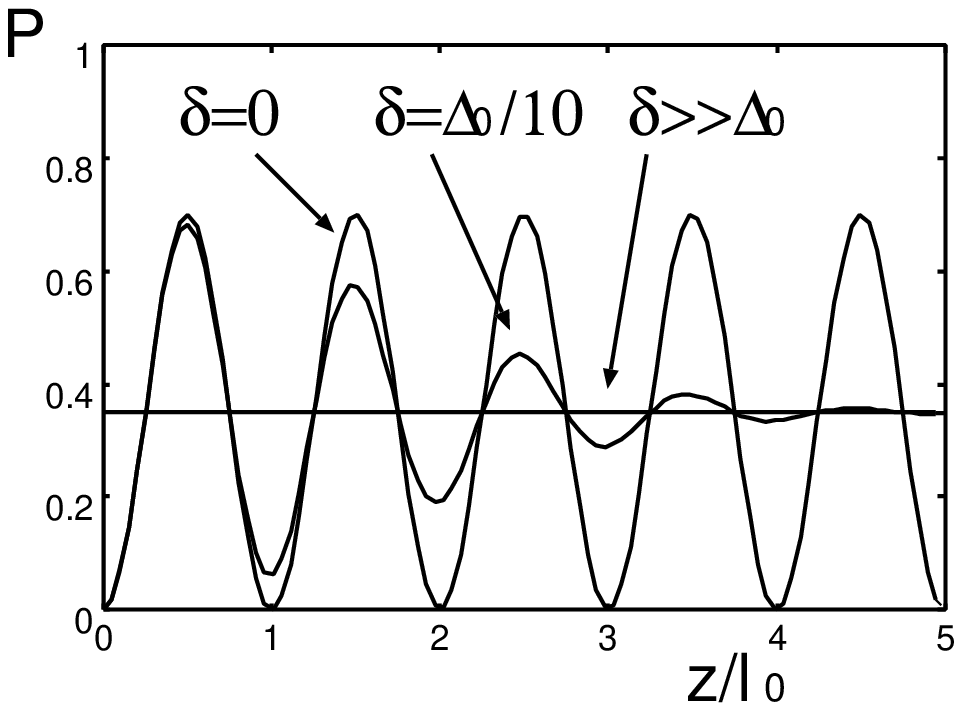}
 \caption{Conversion probability $P_{\nu_{e} \rightarrow \nu_{\mu}}(z)$ averaged
 by a Gaussian energy distribution.
 \label{fig:ave_energy}}
\end{center}
\end{figure}

Finally we consider two averages of the conversion probability concerned with
the finite size of a source and a finite energy width.

A neutrino source is, in general, not point-like and has a finite size. For example,
in the sun, there is a spherical neutrino source with a radius $\approx 10^{10} {\rm cm}$
\cite{BahcallText}. In this case, the finite size will average the phase of the oscillation
of the conversion probability. Denoting the source distribution as $f(z_{0})$, the conversion
probability is given by
\begin{equation}
P_{\nu_{e} \rightarrow \nu_{\mu}}(z) = 
\sin^{2}{2\theta} \int dz_{0} f(z_{0}) 
\sin^{2}{\left(\frac{\pi (z - z_{0})}{l_{\rm osc}}\right)},
\end{equation}
where $z$ is the distance between the center of the source and the detector.
If we assume a Gaussian distribution with a width of $s$, that is,
\begin{equation}
f(z_{0}) = \frac{1}{s \sqrt{2 \pi}} e^{-\frac{z_{0}^{2}}{2 s^{2}}},
\end{equation}
the conversion probability is computed as,
\begin{equation}
P_{\nu_{e} \rightarrow \nu_{\mu}}(z) = 
\frac{1}{2} \sin^{2}{2\theta}
\left[ 1 - e^{-\frac{2 \pi^{2} s^{2}}{l_{\rm osc}^{2}}}
\cos{\left(\frac{2 \pi z}{l_{\rm osc}}\right)} \right],
\end{equation}
and is shown in Fig. \ref{fig:ave_source}. As one can see, if a source has a finite size,
the amplitude of the probability oscillation become small while the average value remains unchanged.
In the limit of $s \rightarrow 0$, it reduces to Eq. (\ref{eq:conversion_probability}) and
the oscillation is completely smoothed for $s \gg \ell_{\rm osc}$.

Next, let us consider a finite energy width. If a source has a finite energy width
we must also average the conversion probability by the energy spectrum because
the oscillation length (\ref{eq:osc_length}) depends on neutrino energy:
\begin{eqnarray}
&& P_{\nu_{e} \rightarrow \nu_{\mu}}(z) =
   \sin^{2}{2\theta} \int dE g(E)
   \sin^{2}\left(\frac{\Delta(E) z}{2}\right), \\
&& \Delta(E) \equiv \frac{2\pi}{l_{\rm osc}(E)} =
   \frac{\Delta m^{2}}{2E},
\end{eqnarray}
where $g(E)$ is the energy spectrum of neutrinos. As a simple example, we consider
a spectrum with Gaussian $\Delta(E)$:
\begin{equation}
g(E) = \frac{1}{\delta \sqrt{2\pi}} 
       e^{-\frac{(\Delta - \Delta_{0})^{2}}{2 \delta^{2}}},
~~~
\Delta_{0} \equiv \frac{2\pi}{l_{0}(E)} 
= \frac{\Delta m^{2}}{2 E_{0}},
\end{equation}
where $\delta$ is the width and $E_{0}$ is the central energy. Here it should be noted that
this distribution is not Gaussian with respect to neutrino energy. Then we have
an averaged conversion probability,
\begin{equation}
P_{\nu_{e} \rightarrow \nu_{\mu}}(z) =
\frac{1}{2} \sin^{2}{2\theta}
\left[ 1 - e^{-\frac{\delta^{2} z^{2}}{2}}
\cos{\left(\frac{2 \pi z}{l_{0}}\right)} \right].
\end{equation}
This is plotted in Fig. \ref{fig:ave_energy}. Although the oscillation is smoothed
as in the case of the finite-size source, the behavior is different between the two cases.
Since difference in energy leads to difference in oscillation length, the phase difference of
two neutrinos with different energies increases as they propagate a long distance.
Therefore, the conversion probability will cease to oscillate in the end regardless of
the magnitude of the energy width.

%%%%%%%%%%%%%%%%%%%%%%%%%%%%%%%%%%%%%%%%%%%%%%%%%%%%%%%%%%%%%%%%%%%%%%%%%%%%%%%%%%%%%%%%%%%%%%%
\subsection{Oscillation in Matter}
%%%%%%%%%%%%%%%%%%%%%%%%%%%%%%%%%%%%%%%%%%%%%%%%%%%%%%%%%%%%%%%%%%%%%%%%%%%%%%%%%%%%%%%%%%%%%%%

The behavior of the neutrino oscillation changes if the neutrino propagates in the presence
of matter, not in vacuum. Due to the interaction with matter, neutrino gains effective mass,
which modifies the dispersion relation. If the interaction is flavor-dependent, like that
with electrons, the change of the dispersion relation is also flavor-dependent. Remembering
that the neutrino oscillation in vacuum is induced by different dispersion relations due to
different masses, it is easy to imagine that further change in dispersion relations will change
the behavior of the neutrino oscillation. This effect, the MSW effect, was first pointed out
by Wolfenstein \cite{Wolfenstein78,Wolfenstein79} and discussed in detail by Mikheyev and Smirnov
\cite{MikheyevSmirnov85,MikheyevSmirnov86a,MikheyevSmirnov86b}.
As will be discussed below, if matter is homogeneous, the situation is essentially the same
as the vacuum oscillation with effective mixing angles determined by matter density and
the original mixing angles.

What is interesting and important is a case with varying density. In fact, the MSW effect
with varying density gave the solution to the long-standing solar neutrino problem
\cite{SKsolar04,SNO05} and will also play a important role in supernovae.

%%%%%%%%%%%%%%%%%%%%%%%%%%%%%%%%%%%%%%
\subsubsection{constant density \label{subsubsection:constant_density}}
%%%%%%%%%%%%%%%%%%%%%%%%%%%%%%%%%%%%%%

At low energies only the elastic forward scattering is important and it can be described by
the refraction index $n_{\rm ref}$. In terms of the forward scattering amplitude $f(E)$,
the refraction index is written as
\begin{equation}
n_{\rm ref} = 1 + \frac{2 \pi n}{E^{2}} f(E),
\end{equation}
where $n$ is the target density. Then we have the dispersion relation in matter as
\begin{equation}
(n_{\rm ref} E)^{2} = k^{2} + m^{2}.
\end{equation}
If we rewrite this dispersion relation as
\begin{equation}
(E-V_{\rm eff})^{2} = k^{2} + m^{2},
\end{equation}
we obtain the effective potential $V_{\rm eff}$ as
\begin{equation}
V_{\rm eff} = (1-n_{\rm ref}) E = - \frac{2 \pi n}{E} f(E).
\end{equation}

On the other hand, low-energy effective Hamiltonian for weak interaction between a neutrino
and a target fermion is
\begin{equation}
H_{\rm int}
= \frac{G_{F}}{\sqrt{2}} \bar{\psi}_{f} \gamma_{\mu} (C_{V}-C_{A}\gamma_{5}) \psi_{f}
  \bar{\psi}_{\nu} \gamma^{\mu} (1-\gamma_{5}) \psi_{\nu}
\end{equation}
where $\psi_{f}$ is the target fermion field, $\psi_{\nu}$ is the neutrino.
Here the coupling constant $G_{F}$ is the Fermi constant,
\begin{equation}
G_{F} = 1.2 \times 10^{-5} {\rm GeV}^{-2} = 9 \times 10^{-38} {\rm eV} ~ {\rm cm}^{3},
\end{equation}
and $C_{V}$ and $C_{A}$ are vector weak charge and axial-vector weak charge, respectively, which
depend on the species of the target. For the neutral-current interactions, charges are shown
in Table \ref{table:coupling_const} and for the charged-current interaction, $C_{V}=C_{A}=1$.

\begin{table}
\begin{center}
\begin{tabular}{cccc} \hline
fermion  & neutrino & $C_{V}$  & $C_{A}$  \\ \hline
electron & $\nu_{e}$ & $1/2+2\sin^{2}{\theta_{\rm W}}$  & $1/2$ \\
         & $\nu_{\mu,\tau}$ & $-1/2+2\sin^{2}{\theta_{\rm W}}$ & $-1/2$  \\
proton   & $\nu_{e,\mu,\tau}$ & $1/2-2\sin^{2}{\theta_{\rm W}}$ & $1.37/2$  \\ 
neutron  & $\nu_{e,\mu,\tau}$ & $-1/2$  & $-1.15/2$  \\ 
neutrino($\nu_{\alpha}$) & $\nu_{\alpha}$  & $1$  & $1$  \\ 
         & $\nu_{\beta \not= \alpha}$  & $1/2$  & $1/2$  \\ \hline
\end{tabular}
\caption{Effective weak coupling constant for neutral-current interactions.
Here $\theta_{\rm W}$ is the Weinberg angle $\sin^{2}{\theta_{\rm W}} \approx 0.23$.
\label{table:coupling_const}}
\end{center}
\end{table}

Using the Hamiltonian and coupling constant, the forward scattering amplitude, refraction index
and effective potential are computed as,
\begin{eqnarray}
&& f(E) = \mp C'_{V} G_{F} \frac{E}{2\sqrt{2}\pi}, \\
&& n_{\rm ref} = 1 \mp C'_{V} G_{F} \frac{n_{f} - n_{\bar{f}}}{\sqrt{2}E}, \\
&& V_{\rm eff} = \pm C'_{V} G_{F} \frac{n_{f} - n_{\bar{f}}}{\sqrt{2}}
\equiv   C'_{V} G_{F} n_{B} \frac{Y_{f}}{\sqrt{2}},
\end{eqnarray}
for neutrino and anti-neutrino, respectively. Here $n_{B}$ is the baryon density, $n_{f}$ and
$n_{\bar{f}}$ are fermion and anti-fermion number density,
$Y_{f} \equiv (n_{f} - n_{\bar{f}})/n_{B}$ is the fermion number fraction per baryon and
\begin{equation}
C'_{V} = \left\{ \begin{array}{ll}
C_{V} & (f \not= \nu) \\
2 C_{V} & (f = \nu) 
\end{array} \right. .
\end{equation}
Assuming charge neutrality ($Y_{e} = Y_{p}$), we have
\begin{equation}
V_{\rm eff} = \pm \sqrt{2} G_{F} n_{B} \times
\left\{ \begin{array}{ll}
( -\frac{1}{2}Y_{n} + Y_{e} + 2 Y_{\nu_{e}}) & ({\rm for} \;\; \nu_{e}) \\
( -\frac{1}{2}Y_{n} + Y_{\nu_{e}}) & ({\rm for} \;\; \nu_{\mu,\tau}) 
\end{array} \right.
\label{eq:effective_potential}
\end{equation}

With this effective potential, the wave equation (\ref{eq:EOM2}) is modified as
\begin{equation}
i \frac{\partial}{\partial z} \Psi_{E}^{\rm (f)} =
\left[ A + \frac{U M^{2} U^{\dag}}{2E} \right] \Psi_{E}^{\rm (f)},
\label{eq:EOM3}
\end{equation}
where $A$ is the mass matrix representing the contribution from interactions with matter.
Neglecting the background neutrino, we have
\begin{equation}
A = \frac{G_{F} n_{B}}{\sqrt{2}}
\left(\begin{array}{ccc}
3 Y_{e} - 1 & 0 & 0 \\
0 & Y_{e} - 1 & 0 \\
0 & 0 & Y_{e} - 1
\end{array}\right).
\end{equation}
Here we used $Y_{n} = 1 - Y_{p} = 1 - Y_{e}$ but the contribution from neutrons is not important
for neutrino oscillation because it is proportional to identity matrix. This reflects the fact
that the interaction with neutrons is via the neutral-current interaction which occurs equally to
all flavors.

\begin{figure}[hbt]
\begin{center}
\epsfxsize = 12 cm
\epsfbox{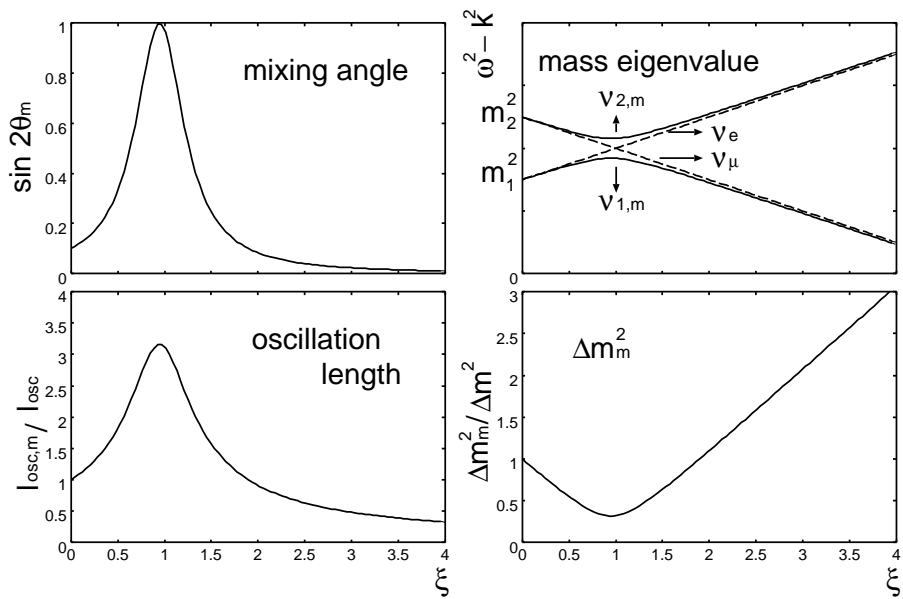}
\end{center}
\caption{Behaviors of various parameters in matter as functions of the dimensionless density
parameter $\xi$. The mixing angle in vacuum is set as $\sin^{2}{2\theta}=0.1$.
\label{fig:matter_parameter}
}
\end{figure}

Again, let us consider the two-flavor case. The wave equation in matter (\ref{eq:EOM3}) can
be rewritten as
\begin{equation}
i \frac{\partial}{\partial z} \left( \begin{array}{c} \nu_{e} \\ \nu_{\mu} \end{array} \right)
= \frac{\Delta m_{m}^{2}}{4 E}
\left( \begin{array}{cc}
- \cos{2 \theta_{\rm m}} & \sin{2 \theta_{\rm m}} \\
\sin{2 \theta_{\rm m}} & \cos{2 \theta_{\rm m}}
\end{array} \right)
\left( \begin{array}{c} \nu_{e} \\ \nu_{\mu} \end{array} \right),
\label{eq:EOM_2f_m}
\end{equation}
up to terms proportional to identity matrix.
This is exactly the same form as the vacuum case (\ref{eq:EOM_2f}) with modified parameters,
\begin{eqnarray}
&& \sin{2\theta_{\rm m}} = \frac{\sin{2\theta}}{\sqrt{(\xi-\cos{2\theta})^{2} + \sin^{2}{2\theta}}}, \\
&& \Delta m_{\rm m}^{2} = \Delta m^{2} \sqrt{(\xi-\cos{2\theta})^{2} + \sin^{2}{2\theta}},
\end{eqnarray}
where $\xi$ is the dimensionless density parameter:
\begin{eqnarray}
\xi &=& \frac{2\sqrt{2}G_{F}n_{B}E}{\Delta m^{2}} \nonumber \\
    &=& 1.53 \times 10^{-2} \left( \frac{Y_{e} \rho}{1 {\rm g} ~ {\rm cm}^{-3}} \right)
        \left( \frac{E}{1 {\rm MeV}} \right)
        \left( \frac{10^{-5} {\rm eV}^{2}}{\Delta m^{2}} \right).
\label{eq:density-parameter}
\end{eqnarray}
The oscillation length in matter is also defined in the same way,
\begin{equation}
\ell_{\rm osc,m}
\equiv \frac{4 \pi E}{\Delta m_{\rm m}^{2}}
= \frac{\sin{2\theta_{\rm m}}}{\sin{2\theta}} \ell_{\rm osc}
= \frac{\ell_{\rm osc}}{\sqrt{(\xi-\cos{2\theta})^{2} + \sin^{2}{2\theta}}}.
\label{eq:osc-length_matter}
\end{equation}
Thus neutrino oscillation occurs with modified mixing angle $\theta_{\rm m}$ and oscillation
length $\ell_{\rm osc, m}$. Mass eigenvalues can be obtained by diagonalizing (\ref{eq:EOM3}) as
\begin{equation}
m_{\rm m}^{2} =
\frac{m^{2}_{1} + m^{2}_{2}}{2}
+  \frac{\Delta m^{2}}{2}
   \left[ (2Y_{e} - 1) \xi
          \pm \sqrt{(\xi-\cos{2\theta})^{2} + \sin^{2}{2\theta}}
   \right].
\end{equation}
In Fig. \ref{fig:matter_parameter}, behaviors of various parameters in matter as functions of
the dimensionless density parameter $\xi$ are shown.

When $\xi = \cos{2\theta}$, the mixing angle in matter, $\theta_{\rm m}$, becomes maximum ($\pi/4$).
This is called {\it resonance} and the resonance condition
can be rewritten as
\begin{equation}
\rho_{\rm res} = 
1.3 \times 10^{2} {\rm g} ~ {\rm cm}^{-3}
\cos{2\theta} \left( \frac{0.5}{Y_{e}} \right) 
\left( \frac{1 {\rm MeV}}{E} \right) 
\left( \frac{\Delta m^{2}}{10^{-5} {\rm eV}^{2}} \right).
\label{eq:res_condition}
\end{equation}
At the resonance density, the matter oscillation length and the mass-squared difference
become maximum and minimum, respectively:
\begin{equation}
(l_{\rm osc,m})_{\rm res} = \frac{l_{\rm osc}}{\sin{2\theta}}, ~~~~
(\Delta m^{2}_{\rm m})_{\rm res} = \Delta m^{2} \sin{2\theta}.
\label{eq:resonance-length}
\end{equation}
On the other hand, in the case of anti-neutrino, the signature of the matter effect
is different so that there is no resonance. However, we will discuss possible resonance
in anti-neutrino sector later.

%%%%%%%%%%%%%%%%%%%%%%%%%%%%%%%%%%%%%%
\subsubsection{varying density \label{subsubsection:varying_density}}
%%%%%%%%%%%%%%%%%%%%%%%%%%%%%%%%%%%%%%

The solar neutrinos are produced at the center with the density about $150 {\rm g} ~ {\rm cm}^{-3}$,
and escape outward into vacuum. In this case and many more cases in astrophysical systems including supernovae,
the neutrinos propagates in an inhomogeneous medium and the neutrino oscillation becomes much more
complicated. As we saw in section \ref{subsubsection:constant_density}, flavor eigenstates and
mass eigenstates can be related by the effective mixing angles in matter as
\begin{equation}
\left(
\begin{array}{c} \nu_{e} \\ \nu_{\mu} \\ \nu_{\tau} \end{array}
\right) = U(\theta_{i, {\rm m}})
\left(
\begin{array}{c} \nu_{1,{\rm m}} \\ \nu_{2,{\rm m}} \\ \nu_{3,{\rm m}} 
\end{array}
\right).
\end{equation}
In an inhomogeneous matter, the mixing angles $\theta_{i, {\rm m}}$ are functions of $z$.
Due to this dependence of the mixing angles on $z$, the wave equations cannot be solved analytically
in general.

To see this, consider a two-flavor case. The wave equations were given in (\ref{eq:EOM_2f_m}):
\begin{equation}
i \frac{\partial}{\partial z} \left( \begin{array}{c} \nu_{e} \\ \nu_{\mu} \end{array} \right)
= \frac{\Delta m_{m}^{2}}{4 E}
\left( \begin{array}{cc}
- \cos{2 \theta_{\rm m}} & \sin{2 \theta_{\rm m}} \\
\sin{2 \theta_{\rm m}} & \cos{2 \theta_{\rm m}}
\end{array} \right)
\left( \begin{array}{c} \nu_{e} \\ \nu_{\mu} \end{array} \right).
\label{eq:EOM_2_m2}
\end{equation}
If the mixing matrix $U$ is constant,  we can diagonalize the equations by multiplying $U$
to the both sides of (\ref{eq:EOM_2_m2}). However, if the mixing matrix $U$ depends on $z$,
the derivative operator $\partial/\partial z$ and $U$ do not commute so that the l.h.s. does
not result in a simple form, although the r.h.s. is diagonalized:
\begin{equation}
i\frac{\partial}{\partial z}
\left( \begin{array}{c} \nu_{1, {\rm m}} \\ \nu_{2, {\rm m}} \end{array} \right)
+ i U \left(\frac{\partial}{\partial z} U^{\dag} \right)
  \left( \begin{array}{c} \nu_{1, {\rm m}} \\ \nu_{2, {\rm m}} \end{array} \right)
= \frac{1}{2 \omega}
\left(\begin{array}{cc}
m^{2}_{1,{\rm m}} & 0 \\
0 & m^{2}_{2,{\rm m}} 
\end{array}\right)
\left( \begin{array}{c} \nu_{1, {\rm m}} \\ \nu_{2, {\rm m}} \end{array} \right).
\end{equation}
Thus the equations are not effectively diagonalized and can be written as,
\begin{equation}
i\frac{\partial}{\partial z}
\left( \begin{array}{c} \nu_{1, {\rm m}} \\ \nu_{2, {\rm m}} \end{array} \right)
= \left(\begin{array}{cc}
- \frac{\Delta m^{2}_{\rm m}}{2\omega} & -i \frac{\partial \theta_{\rm m}}{\partial z} \\
i \frac{\partial \theta_{\rm m}}{\partial z} & \frac{\Delta m^{2}_{\rm m}}{2\omega} 
\end{array}\right)
\left( \begin{array}{c} \nu_{1, {\rm m}} \\ \nu_{2, {\rm m}} \end{array} \right).
\end{equation}
This shows that even the mass eigenstates are mixing if the density is inhomogeneous and
the magnitude of the mixing depends on $\partial \theta_{\rm m}/\partial z$. 

Let us first consider the mixing of the mass eigenstates qualitatively. If the diagonal component is
much larger than the off-diagonal component everywhere, that is,
\begin{equation}
\left| \frac{\partial \theta_{\rm m}}{\partial z} \right| \ll \frac{\Delta m^{2}_{\rm m}}{2\omega},
\label{eq:ad_condition}
\end{equation}
the mass eigenstates will propagate without mixing. In other words, the heavier state remains heavier
and the lighter state remains lighter. There is no energy jump and this case can be said to be
"adiabatic". Contrastingly, if the condition (\ref{eq:ad_condition}) is not satisfied,
the heavier state can change to the lighter state and vice versa, which is a "non-adiabatic" case.
The non-adiabaticity is largest where the change of the mixing angle is rapid, which is expected to be
around the resonance point as can be expected from Fig. \ref{fig:matter_parameter}.

\begin{figure}[hbt]
\begin{center}
\epsfbox{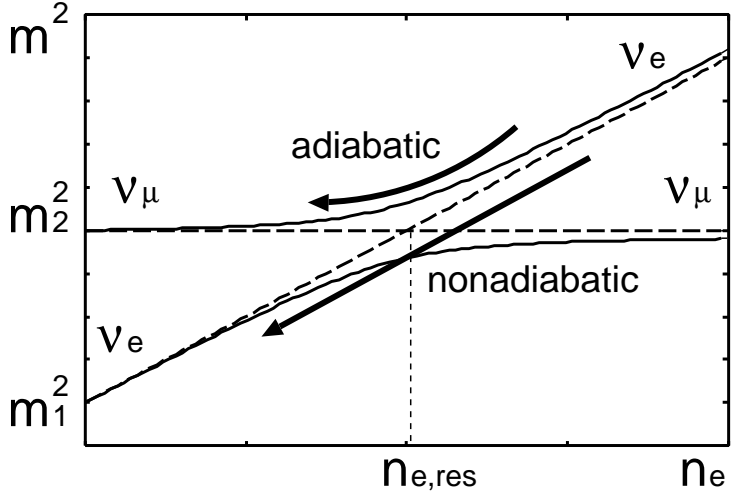}
\end{center}
\caption{Schematic view of two-flavor resonance in an inhomogeneous medium.
The dashed lines show the masses of the mass eigenstates and the solid lines show the effective
masses of the flavor eigenstates. Contribution from the neutral-current interactions are subtracted.
\label{fig:2_resonance}
}
\end{figure}

Fig. \ref{fig:2_resonance} will be helpful to understand the oscillation in an inhomogeneous matter.
Assume that the vacuum mixing angle is so small that the lighter state is almost $\nu_{e}$, that is,
$\nu_{e}$ is effectively "ligher" than $\nu_{\mu}$ in vacuum. In a dense region, on the other hand,
$\nu_{e}$ is effectively "heavier" than $\nu_{\mu}$. The two flavors have the same "mass" at the
resonance point. Then let us consider a case where a $\nu_{e}$ is produced at a dense region and escape
into vacuum as the solar neutrinos. If the resonance is adiabatic, the heavier state will remain
heavier, which means that a $\nu_{e}$ produced at the dense region will emerge as a $\nu_{\mu}$.
In contrast, if the resonance is non-adiabatic, a $\nu_{e}$ will remain a $\nu_{e}$.

Thus, the survival probability of $\nu_{e}$ depends largely on the adiabaticity of the resonance.
The importance of the resonance in the solar neutrino problem was first pointed out by Mikheyev
and Smirnov \cite{MikheyevSmirnov85}. It was Bethe who found the essence of the MSW effect
in an inhomogeneous medium to be the level crossing of the flavor eigenstates \cite{Bethe86}.

Let us discuss more quantitatively. The adiabatic condition (\ref{eq:ad_condition}) at the resonance
point can be rewritten as,
\begin{equation}
1 \ll \frac{\Delta m^{2}}
{E \left| \frac{\partial \ln{n_{e}}}{\partial z} \right|_{\rm res}} 
\sin{2\theta} \tan{2\theta}.
\end{equation}
The adiabaticity parameter $\gamma$ is defined as
\begin{equation}
\gamma \equiv \frac{\Delta m^{2}}
{E \left| \frac{\partial \ln{n_{e}}}{\partial z} \right|_{\rm res}} 
\sin{2\theta} \tan{2\theta},
\label{eq:adiabaticity}
\end{equation}
so that the adiabaticity condition is,
\begin{equation}
\gamma \gg 1.
\label{eq:ad_condition2}
\end{equation}
What we want to know is the conversion probability of the matter eigenstates,
$P_{\nu_{1,{\rm m}} \rightarrow \nu_{2,{\rm m}}}$.
For a general profile of matter density, this cannot be obtained analytically. But for
some special cases, analytic expression is known \cite{KuoPantaleone89}. 
If we write the probability as,
\begin{equation}
P_{\nu_{1,{\rm m}} \rightarrow \nu_{2,{\rm m}}}
 = \frac{\exp{[-\frac{\pi \gamma F}{2}]} - 
         \exp{[-\frac{\pi \gamma F}{2 \sin^{2}{\theta}}]}}
        {1 - \exp{[-\frac{\pi \gamma F}{2 \sin^{2}{\theta}}]}},
\end{equation}
the factor $F$ is given by,
\begin{equation}
F = \left\{ \begin{array}{ll}
1 & (n_{e} \propto z) \\
\frac{(1-\tan^{2}{\theta})^{2}}{(1+\tan^{2}{\theta})^{2}} 
& (n_{e} \propto z^{-1}) \\
1-\tan^{2}{\theta} & (n_{e} \propto e^{-z/z_{0}})
\end{array} \right. .
\end{equation}
Note that $P_{\nu_{1,{\rm m}} \rightarrow \nu_{2,{\rm m}}} \sim 1$ when $\gamma \gg 1$ and
$P_{\nu_{1,{\rm m}} \rightarrow \nu_{2,{\rm m}}} \sim 0$ when $\gamma \ll 1$ in any cases,
as expected.

If we obtain the conversion probability $P_{\nu_{1,{\rm m}} \rightarrow \nu_{2,{\rm m}}}$
in some way, we can compute, for example, the survival probability of
$\nu_{e}$, $P_{\nu_{e} \rightarrow \nu_{e}}$.
When a $\nu_{e}$ is produced at the center of the sun, the probabilities that it is
$\nu_{1,{\rm m}}$ and $\nu_{2,{\rm m}}$ are $\cos^{2}{\theta_{\rm c}}$ and $\sin^{2}{\theta_{\rm c}}$,
respectively, where $\theta_{\rm c}$ is the mixing angle at the center. First, if there is
no conversion between $\nu_{1,{\rm m}}$ and $\nu_{2,{\rm m}}$, that is, if 
$P_{\nu_{1,{\rm m}} \rightarrow \nu_{2,{\rm m}}} = 0$,
\begin{eqnarray}
P_{\nu_{e} \rightarrow \nu_{e}} & = &
\cos^{2}{\theta_{\rm c}}\cos^{2}{\theta}
+\sin^{2}{\theta_{\rm c}}\sin^{2}{\theta} \nonumber \\
& = & \frac{1+\cos{2\theta_{\rm c}}\cos{2\theta}}{2}.
\end{eqnarray}
On the other hand, when $P_{\nu_{1,{\rm m}} \rightarrow \nu_{2,{\rm m}}}$ is non-zero,
\begin{eqnarray}
P_{\nu_{e} \rightarrow \nu_{e}} & = &
\biggl[ (1-P) \cos^{2}{\theta_{\rm c}} + P \sin^{2}{\theta_{\rm c}} \biggr]
\cos^{2}{\theta}
+ \biggl[P \cos^{2}{\theta_{\rm c}} + (1-P) \sin^{2}{\theta_{\rm c}} \biggr]
  \sin^{2}{\theta} 
\nonumber \\
& = & \frac{1+\cos{2\theta_{\rm c}}\cos{2\theta}}{2} - P \cos{2\theta_{\rm c}}\cos{2\theta},
\end{eqnarray}
where $P \equiv P_{\nu_{1,{\rm m}} \rightarrow \nu_{2,{\rm m}}}$.

In most cases, the survival probability is determined by the adiabaticity parameter $\gamma$.
Because $\gamma$ depends on the neutrino parameters, mixing angles and mass differences, and
density profile of matter, neutrino observation from various systems will allow us to investigate them.

If the matter density is much larger than the solar case, the two-flavor analysis is invalid and
we have to take three flavors into account. This is exactly what we do later to consider
neutrino oscillation in supernova. For a three flavor case, there are two resonance points.
Although the situation will become more complicated, the essence is the same as the two-flavor case,
the adiabaticity of the resonance.

%%%%%%%%%%%%%%%%%%%%%%%%%%%%%%%%%%%%%%%%%%%%%%%%%%%%%%%%%%%%%%%%%%%%%%%%%%%%%%%%%%%%%%%%%%%%%%%
\subsection{Experiment of Neutrino Oscillation \label{subsection:ex_nu-osc}}
%%%%%%%%%%%%%%%%%%%%%%%%%%%%%%%%%%%%%%%%%%%%%%%%%%%%%%%%%%%%%%%%%%%%%%%%%%%%%%%%%%%%%%%%%%%%%%%

Because the neutrino oscillation is a phenomenon beyond the standard model of particle physics,
many experiments have been conducted to verify it in various systems. One of the attractive
features of neutrino oscillation experiment is that it does not need high-energy accelerator.

In this section, we review neutrino oscillation experiments starting from general remarks
about the experiments.

%%%%%%%%%%%%%%%%%%%%%%%%%%%%%%%%%%%%%%
\subsubsection{general remarks \label{subsubsection:general_remarks}}
%%%%%%%%%%%%%%%%%%%%%%%%%%%%%%%%%%%%%%

The basic of neutrino oscillation experiment is to observe neutrinos from a known source.
Although all flavors except sterile neutrino can be ideally detected, $\nu_{e}$ and $\bar{\nu}_{e}$
are easier to detect than other flavors so that they are often used as signal. In this respect,
neutrino oscillation experiment can be classified into two types. One is called
"appearance experiment", in which, for example, we detect neutrinos from a $\nu_{\mu}$ source.
If we observe even a single event of $\nu_{e}$, this is an evidence of flavor conversion.
Another type is called "disappearance experiment", in which we observe $\nu_{e}$s ($\bar{\nu}_{e}$s)
from a $\nu_{e}$ ($\bar{\nu}_{e}$) source with known flux. If we observe less number of $\nu_{e}$s,
this can also be an evidence of flavor conversion.

Each type has its own advantage and disadvantage. By the appearance experiment, we cannot reject
neutrino oscillation phenomenon even if we did not observe the signal. This is because
$\nu_{\mu}$ might have changed into $\nu_{\tau}$, not $\nu_{e}$. Contrastingly, the disappearance
experiment can tell whether neutrino oscillation occurred or not, if only $\nu_{e}$s were converted
into any type of neutrino. However, we cannot know the oscillation channel, that is, which flavor
$\nu_{e}$s were converted to. In this respect, the appearance experiment can probe a selected channel
of neutrino oscillation.

Many kinds of experiments have been done so far and each has different parameter region
it can probe. Here we discuss how the neutrino parameters can be probed.
More concretely, let us consider what we could know if there was no signal in an appearance experiment.
For simplicity, we consider just two-flavor oscillation in vacuum.
As we saw in section \ref{subsection:vacuum}, conversion probability $P_{\nu_{\mu} \rightarrow \nu_{e}}$
as a function of the propagation distance is,
\begin{equation}
P_{\nu_{\mu} \rightarrow \nu_{e}}(z)
= \sin^{2}{2\theta} \sin^{2}\left(\frac{\pi z}{l_{\rm osc}}\right)
\end{equation}
If we detect no $\nu_{e}$ signal, it means that the conversion probability is smaller than
a certain value $\delta$ which is determined by the noise level of the experiment.
When the baseline $L$ is much smaller than the oscillation length $\ell_{\rm osc}$,
it is written as,
\begin{equation}
P_{\nu_{\mu} \rightarrow \nu_{e}}(L) \approx 
\sin^{2}{2\theta} \left(\frac{\pi L}{l_{\rm osc}}\right)^{2}
< \delta.
\end{equation}
Substituting the definition of $\ell_{\rm osc}$ (\ref{eq:osc_length}), this reduces to
\begin{equation}
\Delta m^{2} \sin{2\theta} < \frac{E \sqrt{\delta}}{L}.
\end{equation}
On the other hand, when $L \gg \ell_{\rm osc}$, finite energy width of the neutrino beam
will average the oscillation of the conversion probability so that the no signal means,
\begin{equation}
P_{\nu_{\mu} \rightarrow \nu_{e}}(L)
= \frac{1}{2} \sin^{2}{2\theta} < \delta,
\end{equation}
which reduces to,
\begin{equation}
\sin{2\theta} < \sqrt{2 \delta}.
\end{equation}
Note that we cannot obtain information about $\Delta m^{2}$ in this case.

Thus if we want to probe small $\Delta m^{2}$, experiments with small $E/L$ are advantageous.
Various systems with characteristic neutrino energy, baseline and possible $\Delta m^{2}$ which
can be probed are shown in Table \ref{table:probe_msd}. Analysis of an disappearance experiment
can be done essentially in the same way.

\begin{table}[t]
\caption{Probe of squared mass difference by various experiments.
\label{table:probe_msd}}
\begin{center}
\begin{tabular}{cccc} 
source & energy $E ({\rm MeV})$ & baseline $L ({\rm m})$ & 
$\Delta m^{2} ({\rm eV}^{2})$ \\ \hline
accelerator & $10^{3} \sim 10^{5}$ & $10^{2} \sim 10^{6}$ & $10^{-3} \sim 10^{2}$ \\
reactor     & $1 \sim 2$           & $10 \sim 10^{5}$   & $10^{-5} \sim 10^{-1}$ \\ 
atmosphere  & $\sim 10^{3}$      & $10^{5} \sim 10^{7}$ & $10^{-2} \sim 10^{-4}$ \\ 
Sun         & $\sim 1$           & $\sim 10^{11}$       & $\sim 10^{-11}$ \\
\hline 
\end{tabular}
\end{center}
\end{table}

%%%%%%%%%%%%%%%%%%%%%%%%%%%%%%%%%%%%%%
\subsubsection{accelerator experiment \label{subsubsection:accelerator}}
%%%%%%%%%%%%%%%%%%%%%%%%%%%%%%%%%%%%%%

Accelerator experiment is the most popular experiment of neutrino oscillation.
One of the advantage of accelerator experiment is that we can control the neutrino source
while Sun, atmosphere and supernovae are uncontrolled and rather unknown sources.
Although there have been a lot of accelerator experiments so far, the basic concept
is similar as we will review below.

First, protons are accelerated and collided with target nuclei to produce $\pi^{\pm}$:
\begin{equation}
p + N \rightarrow \pi^{+} \; , \; \pi^{-}, \cdots.
\end{equation}
Then $\pi^{+}$s or $\pi^{-}$s are absorbed and the others decay to produce neutrinos.
For example,
\begin{eqnarray}
&& \pi^{+} \rightarrow \mu^{+} + \nu_{\mu}, \\
&& \mu^{+} \rightarrow e^{+} + \nu_{e} + \bar{\nu}_{\mu}
\end{eqnarray}
In this way, if $\pi^{-}$s are absorbed completely, we have a neutrino beam which consists
of $\nu_{e}$, $\nu_{\mu}$ and $\bar{\nu}_{\mu}$. Thus, if we detect $\bar{\nu}_{e}$s
in this beam, we can confirm neutrino oscillation $\bar{\nu}_{\mu} \rightarrow \bar{\nu}_{e}$.
In fact, $\pi^{-}$s can not be absorbed completely and decay like,
\begin{eqnarray}
&& \pi^{-} \rightarrow \mu^{-} + \bar{\nu}_{\mu}, \\
&& \mu^{-} \rightarrow e^{-} + \bar{\nu}_{e} + \nu_{\mu}.
\end{eqnarray}
Consequently, some $\bar{\nu}_{e}$s will be produced and they become one of main noises
in this kind of experiment.

\begin{table}[t]
\caption{Accelerator experiments
\label{table:accelerator}}
\begin{center}
\begin{tabular}{ccccll} 
experiment & baseline  & neutrino energy & detector & status & reference\\ \hline
CCFR   & $\sim 1{\rm km}$ & $30 - 500{\rm GeV}$
& 690 ton target calorimeter & completed &\cite{CCFR95,CCFR97} \\
NuTeV  & $1.4{\rm km}$  & $\sim 100{\rm GeV}$
& 690 ton target calorimeter & completed & \cite{CCFR95,NuTeV02} \\
KARMEN & $17.5{\rm m}$ & $\sim 50{\rm MeV}$  
& $56{\rm ton}$ liquid scintillator & completed & \cite{KARMEN98,KARMEN02} \\
LSND   & $30{\rm m}$   & $\sim 50{\rm MeV}$  
& $167{\rm ton}$ liquid scintillator & completed & \cite{LSND01a,LSND01b,LSND01c} \\ 
NOMAD  & $625{\rm m}$   & $\sim 50{\rm GeV}$
& 2.7 ton drift chambers & completed & \cite{NOMAD01,NOMAD03} \\
MiniBooNE & $490{\rm m}$ & $0.1 \sim 2{\rm GeV}$ 
& $445{\rm ton}$ mineral oil & ongoing & \cite{MiniBooNE00,MiniBooNE04} \\ 
K2K    & $250{\rm km}$ & $\sim 1{\rm GeV}$ 
& $22500{\rm ton}$ water (SK) & ongoing &\cite{K2K01,K2K03,K2K04} \\ 
          &              &                  & & & and \cite{K2K05}
\end{tabular}
\end{center}
\end{table}
\begin{figure}[hbt]
\begin{center}
 \begin{minipage}{6cm}
 \begin{center}
 \includegraphics[width=1\textwidth]{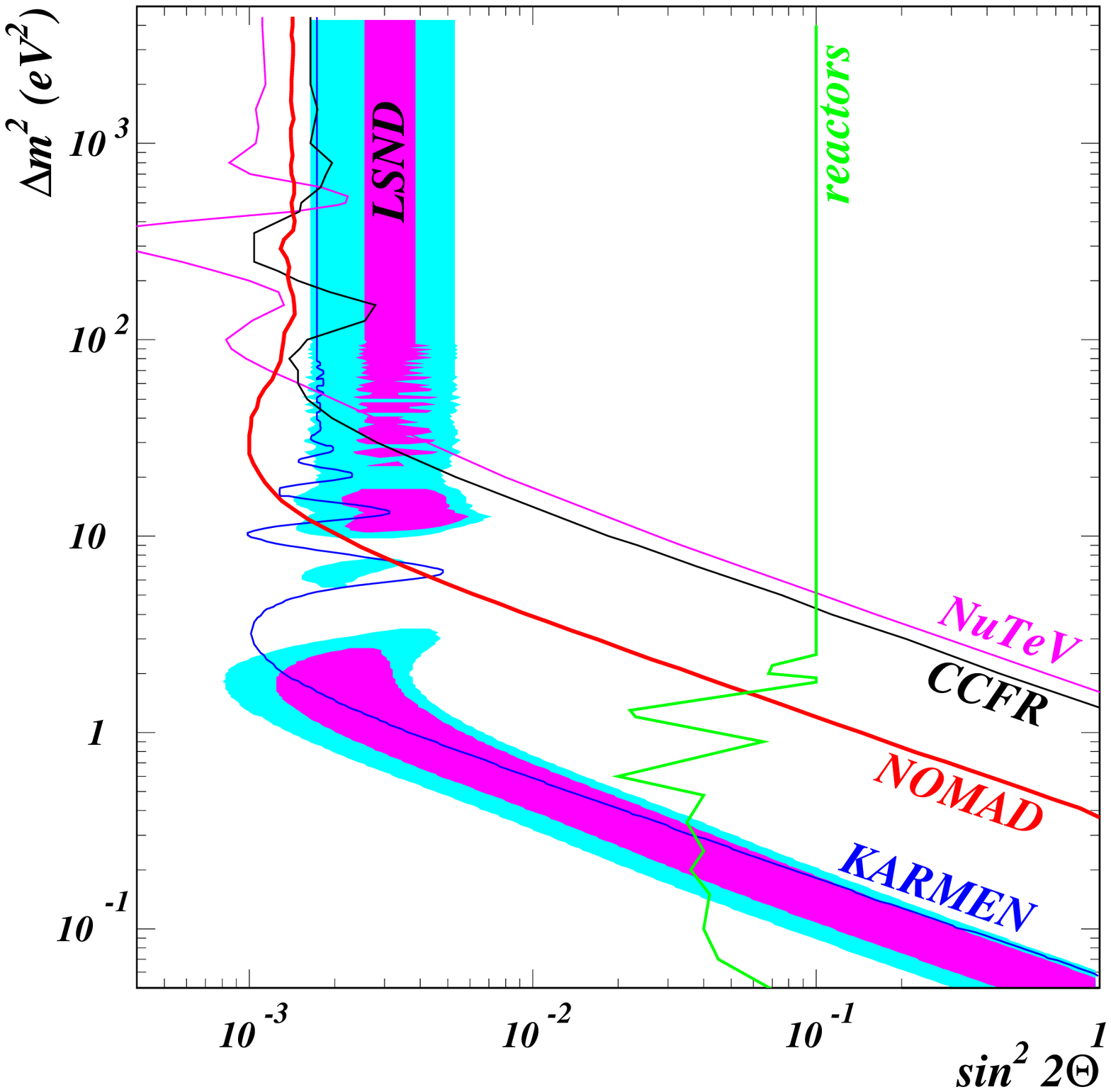}
 \vspace{-0.5cm}
 \caption{Parameter regions allowed by the LSND observation (purple, blue). Exclusion
 curves from various other experiments are shown as well, with the region on the right
 side excluded \cite{NOMAD03}.
 \label{fig:LSND}}
 \end{center}
 \end{minipage}
\hspace{0.5cm}  
 \begin{minipage}{6cm}
 \begin{center}
 \includegraphics[width=1\textwidth]{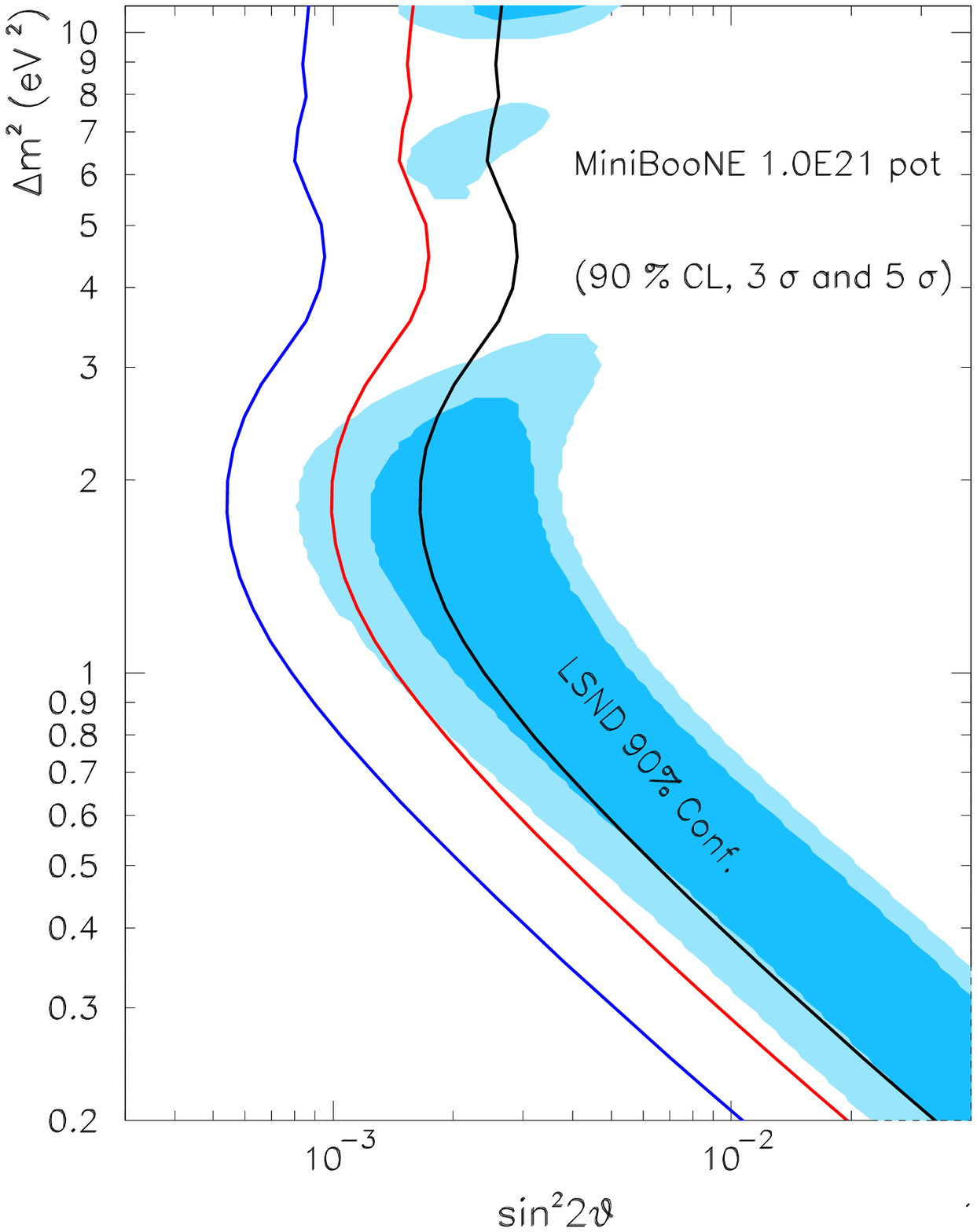}
 \vspace{-0.5cm}
 \caption{Expected excluded regions in case of a non-observation of signal in MiniBooNE
 \cite{MiniBooNE04}.
 \label{fig:MiniBooNE}}
 \end{center}
 \end{minipage}
\end{center}
\end{figure}

\begin{figure}[hbt]
\begin{center}
 \begin{minipage}{7.0cm}
 \begin{center}
 \includegraphics[width=1\textwidth]{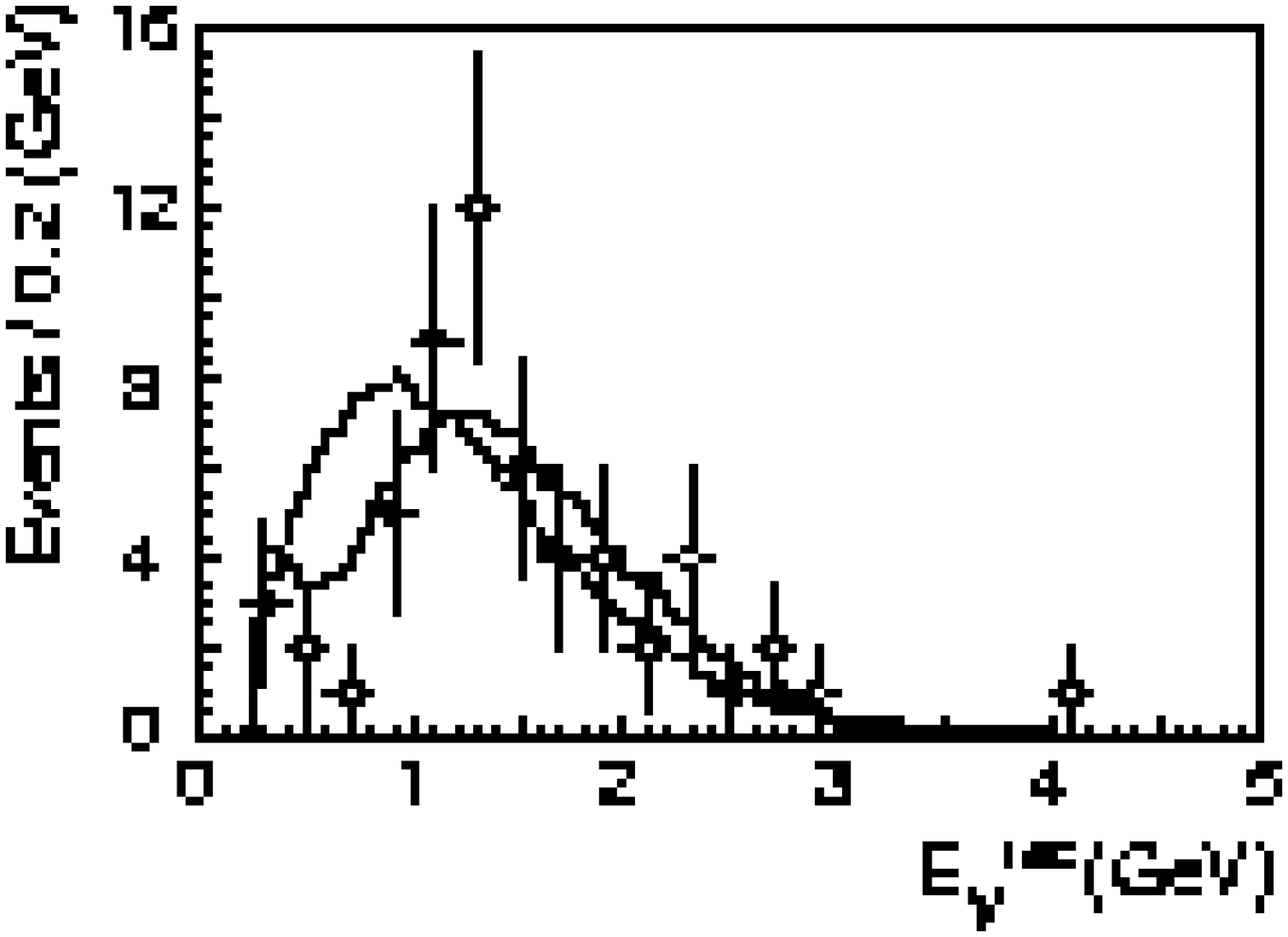}
 \vspace{-0.5cm}
 \caption{The reconstructed energy spectrum for the $\nu_{\mu}$-like sample.
 Points with error bars are data. The solid line is the best fit spectrum and the dashed line
 is the expected spectrum without oscillation \cite{K2K05}.
 \label{fig:K2K-data}}
 \end{center}
 \end{minipage}
\hspace{0.5cm}  
 \begin{minipage}{7.0cm}
 \begin{center}
 \includegraphics[width=1\textwidth]{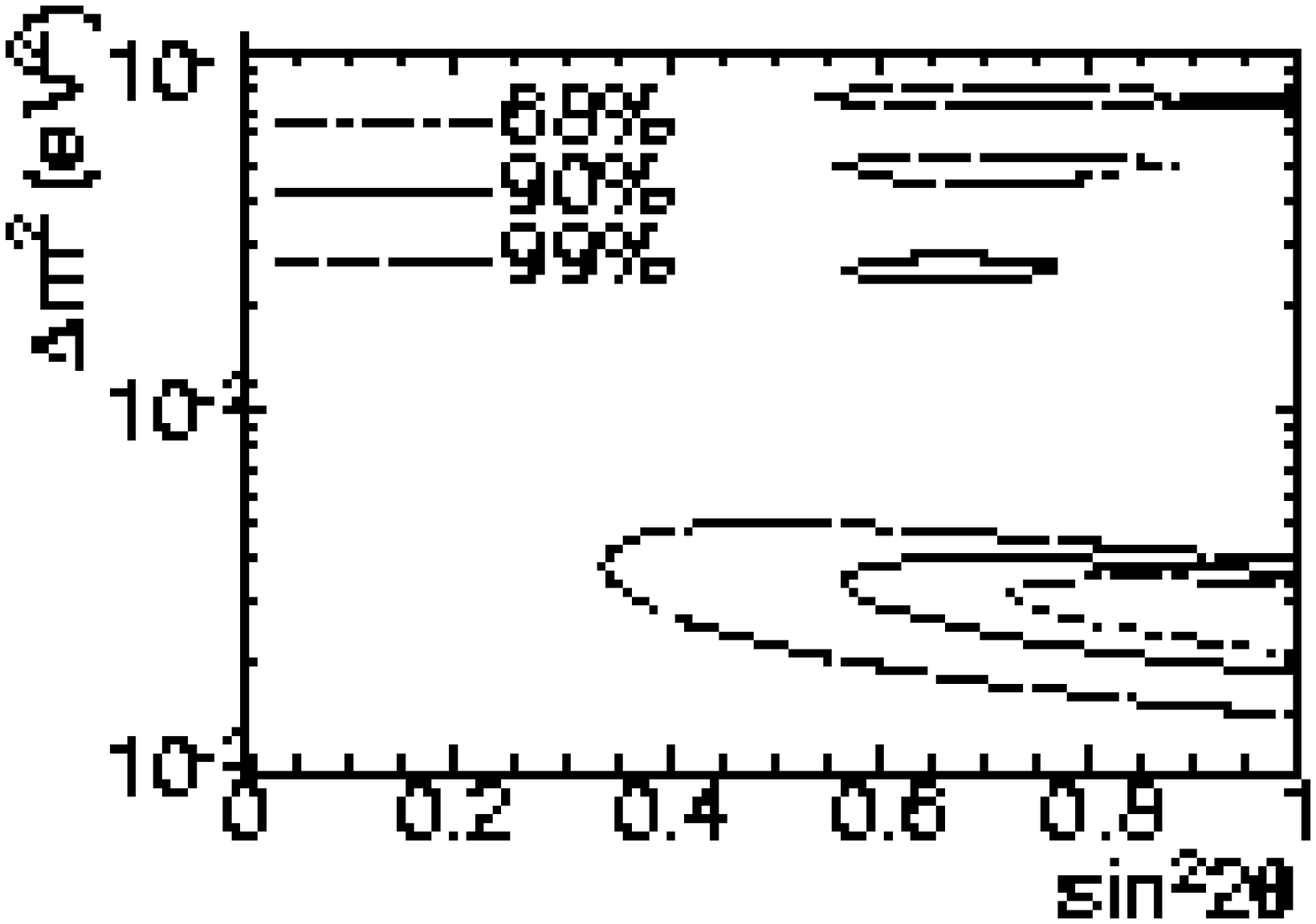}
 \vspace{-0.5cm}
 \caption{Allowed regions of oscillation parameters for $\nu_{\mu} \rightarrow \nu_{\tau}$.
 Dashed, solid and dot-dashed lines are 68.4 \%, 90 \% and 99 \% C.L. contours, respectively
 \cite{K2K05}.
 \label{fig:K2K_constraint}}
 \end{center}
 \end{minipage}
\end{center}
\end{figure}

In this experiment, we can control the amount of the produced neutrinos. If there are
$\bar{\nu}_{e}$ events we can compute the conversion probability
$P_{\bar{\nu}_{\mu} \rightarrow \bar{\nu}_{e}}$. Or, we can obtain an upper limit for
the conversion probability if no signal was obtained. Anyway, we can extract information 
about neutrino parameters as we saw in section \ref{subsubsection:general_remarks}.

The current and past accelerator experiments are classified into long-baseline experiments
and short-baseline experiments. Of course, the latter is technically easier so that early
experiments have rather short baselines. However, implication from the recent solar and
atmospheric neutrino observations has made it necessary to probe very small $\Delta m^{2}$
by long-baseline experiments. Features of some of the current and past accelerators are
shown in Table \ref{table:accelerator}. Among them, CCFR, NuTeV, KARMEN and NOMAD had
no signal for neutrino oscillation and obtained constraints on oscillation parameters
which are shown in Fig. \ref{fig:LSND}.

The LSND experiment performed at Los Alamos observed $87.9 \pm 22.4 \pm 6.0$ excess events
in the $\bar{\nu}_{\mu} \rightarrow \bar{\nu}_{e}$ appearance channel \cite{LSND01c}.
This signal corresponds to a transition probability of $P = (0.264 \pm 0.067 \pm 0.045) \% $,
which is $\sim 3.3 \sigma$ away from zero. If we interpret it in terms of two-flavor
oscillation, parameter regions shown in Fig. \ref{fig:LSND} are allowed. Although most of
the allowed regions are excluded by other experiments, there are still some surviving regions
with $\Delta m_{\rm LSND}^{2} \approx 0.1-1 {\rm eV}^{2}$. This remaining regions are
expected to be confirmed or denied by the MiniBooNE experiment \cite{MiniBooNE00,MiniBooNE04}
in the near future. Fig. \ref{fig:MiniBooNE} shows expected excluded regions in case of
a non-observation of signal in MiniBooNE.

The K2K experiment \cite{K2K01,K2K03,K2K04,K2K05}, the KEK to Kamioka long-baseline neutrino
oscillation experiment, is an accelerator based project with 250 km baseline which is much
longer than those of the past experiments. This long baseline make it possible to explore
neutrino oscillation in the same $\Delta m^{2}$ region as atmospheric neutrinos.
In \cite{K2K05}, five-year data with 57 $\nu_{\mu}$ candidates was reported and their energy
distribution is shown in Fig. \ref{fig:K2K-data}. The distortion of energy spectrum,
which signals neutrino oscillation, is clearly seen in Fig. \ref{fig:K2K-data} and
the probability that the result would be observed without neutrino oscillation is
$0.0050 \% (4.0 \sigma)$. Fig. \ref{fig:K2K_constraint} shows a two-flavor neutrino
oscillation analysis with $\nu_{\mu}$ disappearance. The best fit point is,
\begin{equation}
\sin^{2}{2 \theta} = 1.0, ~~~ \Delta m^{2} = 2.8 \times 10^{-3} {\rm eV}^{2}.
\end{equation}

%%%%%%%%%%%%%%%%%%%%%%%%%%%%%%%%%%%%%%
\subsubsection{reactor experiment \label{subsubsection:reactor}}
%%%%%%%%%%%%%%%%%%%%%%%%%%%%%%%%%%%%%%

\begin{table}[t]
\caption{Reactor experiments
\label{table:reactor}}
\begin{center}
\begin{tabular}{ccccc} 
experiment       & baseline            & status    & reference\\ \hline
Bugey (France)   & $15,40,95 {\rm m}$  & completed & \cite{Achkar95} \\
CHOOZ (France)   & $1 {\rm km}$        & completed & \cite{CHOOZ98,CHOOZ99}  \\ 
Palo Verde (USA) & $750 {\rm m}$       & completed & \cite{Boehm00a,Boehm00b} \\ 
KamLAND (Japan)  & $100 {\rm km} \sim$ & running   & \cite{KamLAND03,KamLAND04,KamLAND05} \\
\hline
\end{tabular}
\end{center}
\end{table}

In nuclear reactors, $\bar{\nu}_{e}$s are isotropically emitted by $\beta$-decay of neutron-rich
nuclei. Reactor experiment of neutrino oscillation is detecting these $\bar{\nu}_{e}$s and
seeing if there is a deficit compared with the expected flux. The flux and spectrum of
$\bar{\nu}_{e}$ are determined by the power of the reactor and abundance of 
${}^{235}$U, ${}^{238}$U, ${}^{239}$Pu and ${}^{241}$Pu. Because reactor neutrinos have
relatively low energy, they are well suited in exploring the region of small $\Delta m^{2}$ at
modest baselines. For example, to explore the parameter $\Delta m^{2}$ down to
$10^{-3} {\rm eV}^{2}$ a reactor experiment with energy around 5 MeV requires a baseline of
$L = 1$ km, while an accelerator experiment with $E = 5$ GeV would require $L = 1,000$ km.

\begin{figure}[htb]
\begin{center}
 \begin{minipage}{6cm}
 \begin{center}
 \includegraphics[width=1\textwidth]{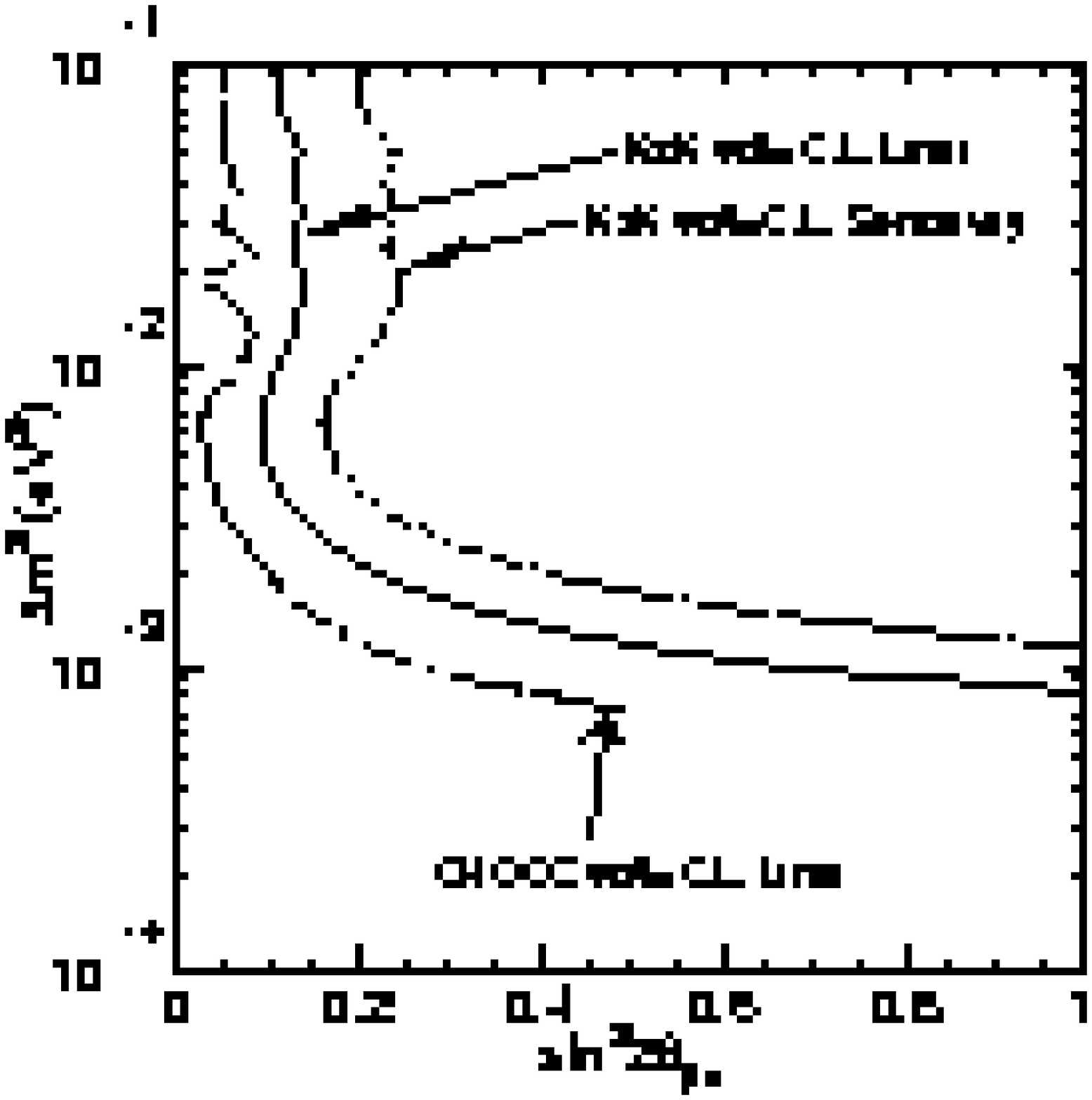}
 \vspace{-1.5cm}
 \caption{Constraint to conversion $\bar{\nu}_{e} \rightarrow \bar{\nu}_{x}$ from CHOOZ experiment
 \cite{CHOOZ99} and that of conversion $\bar{\nu}_{\mu} \rightarrow \bar{\nu}_{e}$ from K2K experiment
 \cite{K2K04}.
 \label{fig:CHOOZ-K2K}}
 \end{center}
 \end{minipage}
\hspace{0.5cm}  
 \begin{minipage}{6cm}
 \begin{center}
 \includegraphics[width=1\textwidth]{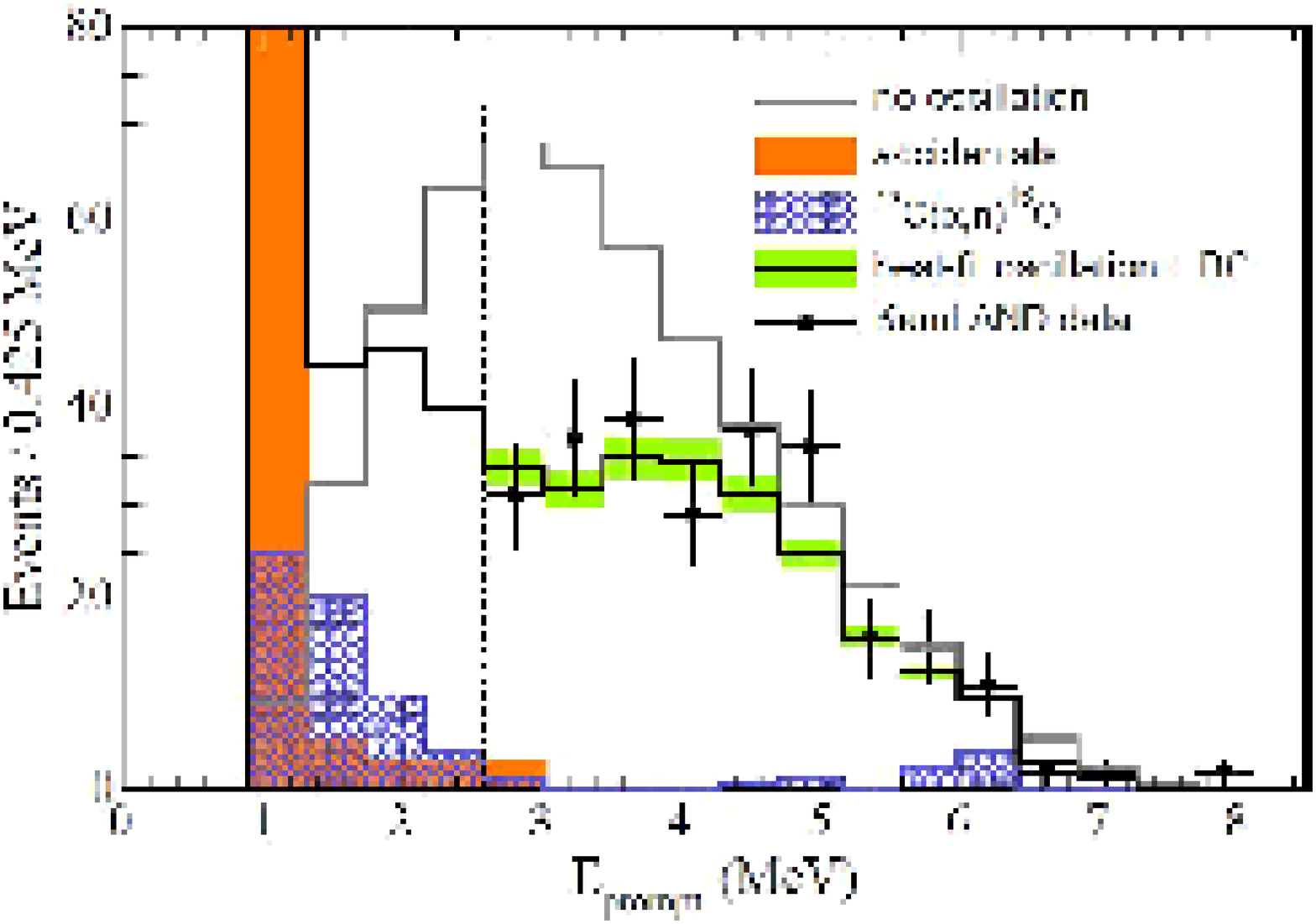}
 \vspace{-0.8cm}
 \caption{Prompt event energy spectrum of $\bar{\nu}_{e}$ candidate events with associated
 background spectra. The shaded band indicates the systematic error in the best-fit reactor
 spectrum above 2.6 MeV \cite{KamLAND05}.
 \label{fig:KamLAND-data}}
 \end{center}
 \end{minipage}
\end{center}
\vspace{-0.5cm}
\begin{center}
\includegraphics[width=0.85\textwidth]{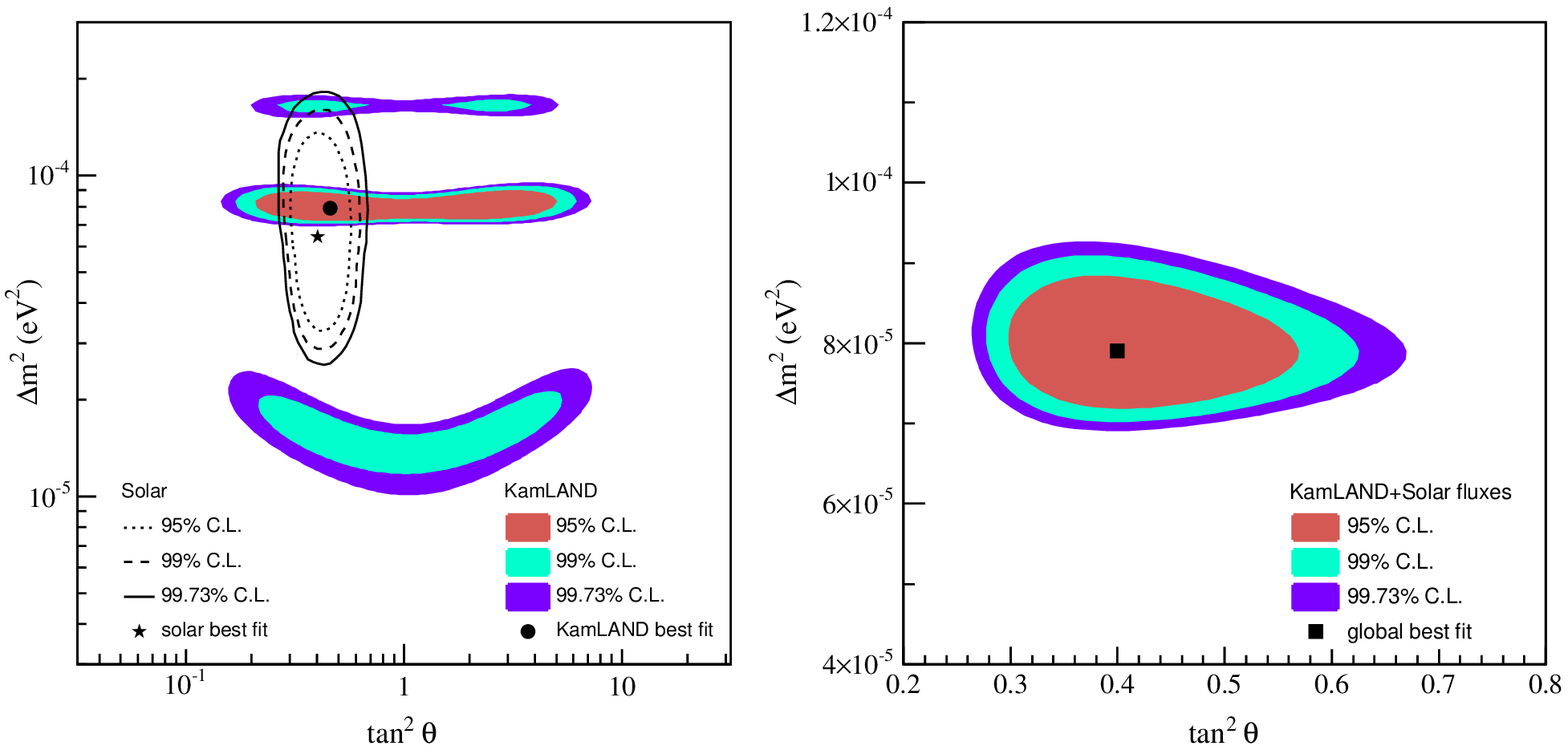}
\end{center}
\vspace{-0.8cm}
\caption{(a) Neutrino oscillation parameter allowed region from KamLAND anti-neutrino data
(shaded regions) and solar neutrino experiments (lines) which we discuss later \cite{KamLAND05}.
(b) Result of a combined two-neutrino oscillation analysis of KamLAND and the observed solar
neutrino fluxes.
\label{fig:KamLAND_constraint}}
\end{figure}

The current and past reactor experiments are shown in Table \ref{table:reactor}.
Among these, CHOOZ \cite{CHOOZ99} gives the strongest constraint on the mixing angle
of $\bar{\nu}_{e} \rightarrow \bar{\nu}_{x}$ for $\Delta m^{2} > 10^{-3} {\rm eV}^{2}$
(Fig. \ref{fig:CHOOZ-K2K}).

Inspired by the recent development of solar neutrino observation, a long-base line experiment
called KamLAND \cite{KamLAND03,KamLAND04,KamLAND05} was constructed where there was once
the Kamiokande detector. KamLAND consists of 1000ton liquid scintillator and its primary purpose
is to confirm the solution to the solar neutrino problem, which will be discussed later.
There are 16 reactors with distances about 100 $\sim$ 1000km from KamLAND so that
the $\bar{\nu}_{e}$ flux at KamLAND is sufficiently large to probe neutrino oscillation
Because of the long baseline and detectability of low-energy neutrinos, KamLAND can probe
much smaller $\Delta m^{2}$ compared with the past experiments.

Fig. \ref{fig:KamLAND-data} shows the prompt event energy spectrum of $\bar{\nu}_{e}$ candidate
events with associated background spectra \cite{KamLAND05}. They observed 258 $\bar{\nu}_{e}$
candidate events with energies above 3.4 MeV compared to 365.2 events expected in the absence
of neutrino oscillation. Accounting for 17.8 expected background events, the statistical
significance for reactor $\bar{\nu}_{e}$ disappearance is 99.998\%. Also the observed energy
spectrum disagrees with the expected spectral shape in the absence of neutrino oscillation at 99.6\%
significance and prefers the distortion expected from $\bar{\nu}_{e}$ oscillation effects
rather than those from neutrino decay and decoherence. A two-neutrino oscillation analysis of
the KamLAND data gives
\begin{equation}
\Delta m^{2} = 7.9^{+0.6}_{-0.5} \times 10^{-5} {\rm eV}^{2},
\end{equation}
as shown in Fig. \ref{fig:KamLAND_constraint}. A global analysis of data from KamLAND and
solar neutrino experiments, which will be discussed later, yields
\begin{equation}
\Delta m^{2} = 7.9^{+0.6}_{-0.5} \times 10 ^{-5} {\rm eV}^{2},
\tan^{2}{\theta} = 0.40^{+0.10}_{-0.07}.
\end{equation}

%%%%%%%%%%%%%%%%%%%%%%%%%%%%%%%%%%%%%%
\subsubsection{atmospheric neutrino}
%%%%%%%%%%%%%%%%%%%%%%%%%%%%%%%%%%%%%%

\begin{figure}[hbt]
\begin{center}
\includegraphics[width=0.6\textwidth]{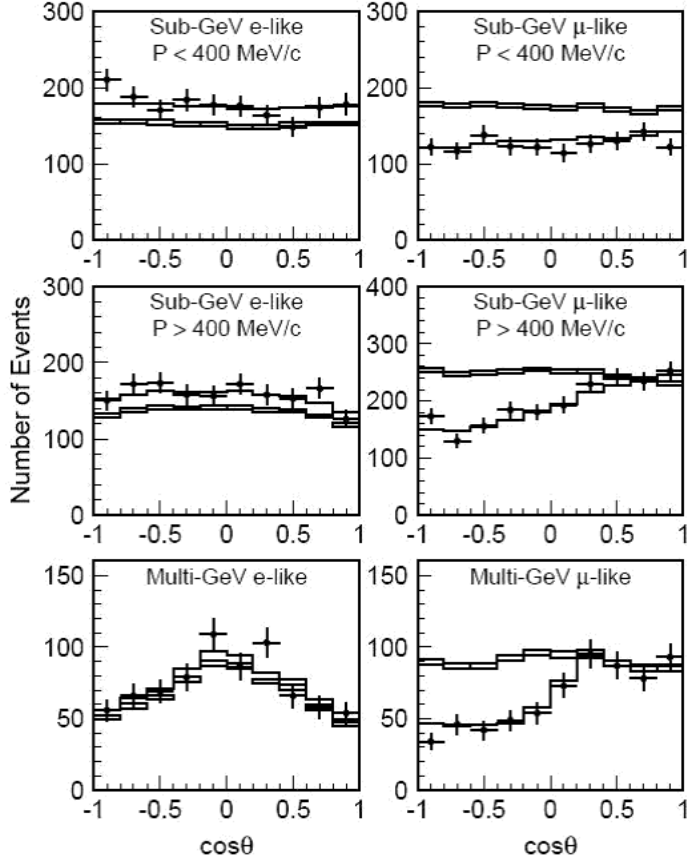}
\end{center}
\vspace{-0.3cm}
\caption{The zenith angle distribution for fully-contained 1-ring events \cite{SKatm05}.
The points show the data, box histograms show the non-oscillated Monte Carlo events and
the lines show the best-fit expectations for $\nu_{\mu} \leftrightarrow \nu_{\tau}$ oscillations
with $\sin^{2}{2 \theta} = 1.00$ and $\Delta m^{2} = 2.1 \times 10^{-3} {\rm eV}^{2}$.
The height of the boxes shows the statistical error of the Monte Carlo.
\label{fig:SK_atm_obs}}
\end{figure}
When cosmic rays enter the atmosphere, they collide with the atmospheric nuclei to produce
a lot of mesons (mostly pions):
\begin{equation}
N_{\rm CR} + N_{\rm air} \rightarrow \pi^{\pm}, \pi^{0}, \cdots .
\end{equation}
The pions decay and the decay product, such muons, further decay to produce neutrinos:
\begin{equation}
\pi^{\pm} \rightarrow \mu^{\pm} + \nu_{\mu}(\bar{\nu}_{\mu}), ~~~
\mu^{\pm} \rightarrow e^{\pm} + \nu_{e}(\bar{\nu}_{e}) + 
\bar{\nu}_{\mu}(\nu_{\mu})
\label{eq:atmospheric_nu}
\end{equation}

When cosmic rays enter the atmosphere, they collide with the atmospheric nuclei to produce
a lot of mesons (mostly pions):
\begin{equation}
N_{\rm CR} + N_{\rm air} \rightarrow \pi^{\pm}, \pi^{0}, \cdots .
\end{equation}
The pions decay and the decay product, such as muons, further decay to produce neutrinos:
\begin{equation}
\pi^{\pm} \rightarrow \mu^{\pm} + \nu_{\mu}(\bar{\nu}_{\mu}), ~~~
\mu^{\pm} \rightarrow e^{\pm} + \nu_{e}(\bar{\nu}_{e}) + 
\bar{\nu}_{\mu}(\nu_{\mu})
\label{eq:atmospheric_nu}
\end{equation}
These neutrinos are called the {\it atmospheric neutrinos} and they provided the first strong
indication for neutrino oscillation. Their energy range and path length varies from 0.1 to 10 GeV
and from 10 to 10,000 km, respectively, which indicates that atmospheric neutrinos can
provide an opportunity for oscillation studies over a wide range of energies and distances.
From (\ref{eq:atmospheric_nu}), we simply expect that the flux ratio
$(\nu_{\mu} + \bar{\nu}_{\mu})/(\nu_{e} + \bar{\nu}_{e}) = 2$.
This is roughly correct, even though the ratio depends on a lot of factors such as neutrino energy
and zenith angle of the incoming neutrino if detailed decay processes and geometric effects
are taken into account \cite{Battistoni03,Honda01,Honda04}.
However, the results from many experiments showed that this ratio was
about unity. This was once called the atmospheric neutrino problem, which is now interpreted
successfully in terms of neutrino oscillation. In fact, the SK was the first to prove neutrino
oscillation phenomenon by its observation of the atmospheric neutrino with a high accuracy.

\begin{figure}[hbt]
\begin{center}
\includegraphics[width=0.6\textwidth]{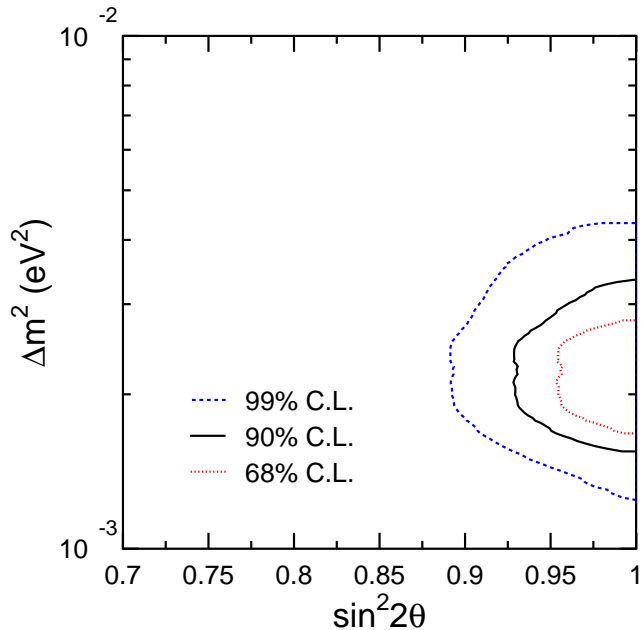}
\end{center}
\vspace{-1.2cm}
\caption{Allowed oscillation parameters for $\nu_{\mu} \leftrightarrow \nu_{\tau}$ oscillations
from the atmospheric neutrino observation at SK \cite{SKatm05}. Three contours correspond to
the 68\% (dotted line), 90\% (solid line) and 99\% (dashed line) C.L. allowed regions.
\label{fig:atm_constraint}}
\end{figure}

It was reliable identification of $\nu_{\mu}$ that allowed SK to prove neutrino oscillation.
Because the SK detector is huge (diameter $\sim$ 39m and height $\sim$ 42m), $\nu_{\mu}$s
with energy $\sim 1$ GeV can be identified as fully contained events, for which all
neutrino-induced interactions occur in the detector and neutrino energies and directions
can be accurately obtained. The results from SK are shown in Fig. \ref{fig:SK_atm_obs}.
As can be seen, $\nu_{e}$ flux is consistent with the theoretical calculation while
there is a deficit in $\nu_{\mu}$ flux. It indicates that $\nu_{\mu} \leftrightarrow \nu_{\tau}$
oscillation would be solution to the atmospheric neutrino problem.
In Fig. \ref{fig:atm_constraint}, allowed oscillation parameters for
$\nu_{\mu} \leftrightarrow \nu_{\tau}$ oscillations are shown. The best fit values are
\begin{equation}
\Delta m^{2}_{\rm atm} = 2.1 \times 10^{-3} {\rm eV}^{2}, ~~~
\sin^{2}{\theta_{\rm atm}} = 1.00.
\end{equation}
These results are confirmed by other experiments such as Soudan 2 \cite{Soudan03}
and MACRO \cite{MACRO01,MACRO03} (see also \cite{GiacomelliGiorgini05}).

As we saw in Eq. (\ref{eq:conversion_probability}), the conversion and survival probabilities
depends on $L/E$, where $L$ and $E$ are neutrino path length and energy, and oscillates
with respect to $L/E$. This behavior was confirmed by the atmospheric neutrino observation at SK
\cite{SKatm04}. This means that it rejected other possibilities such as neutrino decay
and decoherence which had different dependence on $L/E$ and proved that solution to
the atmospheric neutrino problem was truly neutrino oscillation.

%%%%%%%%%%%%%%%%%%%%%%%%%%%%%%%%%%%%%%
\subsubsection{solar neutrino \label{subsubsection:solar}}
%%%%%%%%%%%%%%%%%%%%%%%%%%%%%%%%%%%%%%

\begin{figure}[hbt]
\begin{center}
\rotatebox{270}{\includegraphics[width=0.44\textwidth]{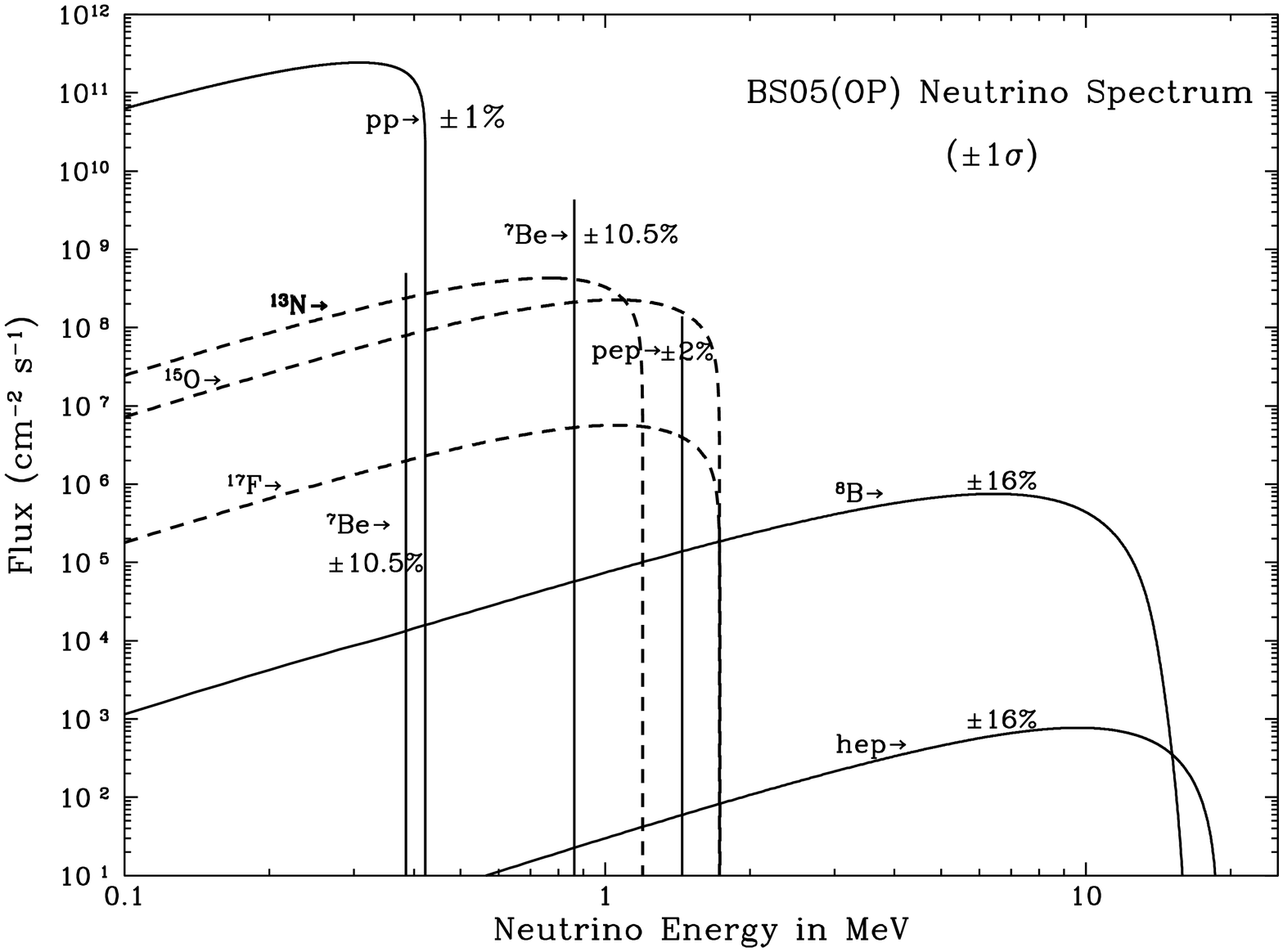}}
\vspace{-0.2cm}
\caption{Solar neutrino flux from various nuclear reactions \cite{BahcallSerenelliBasu05}.
\label{fig:solar-nu_flux}}
\vspace{-0.5cm}
\rotatebox{270}{\includegraphics[width=0.48\textwidth]{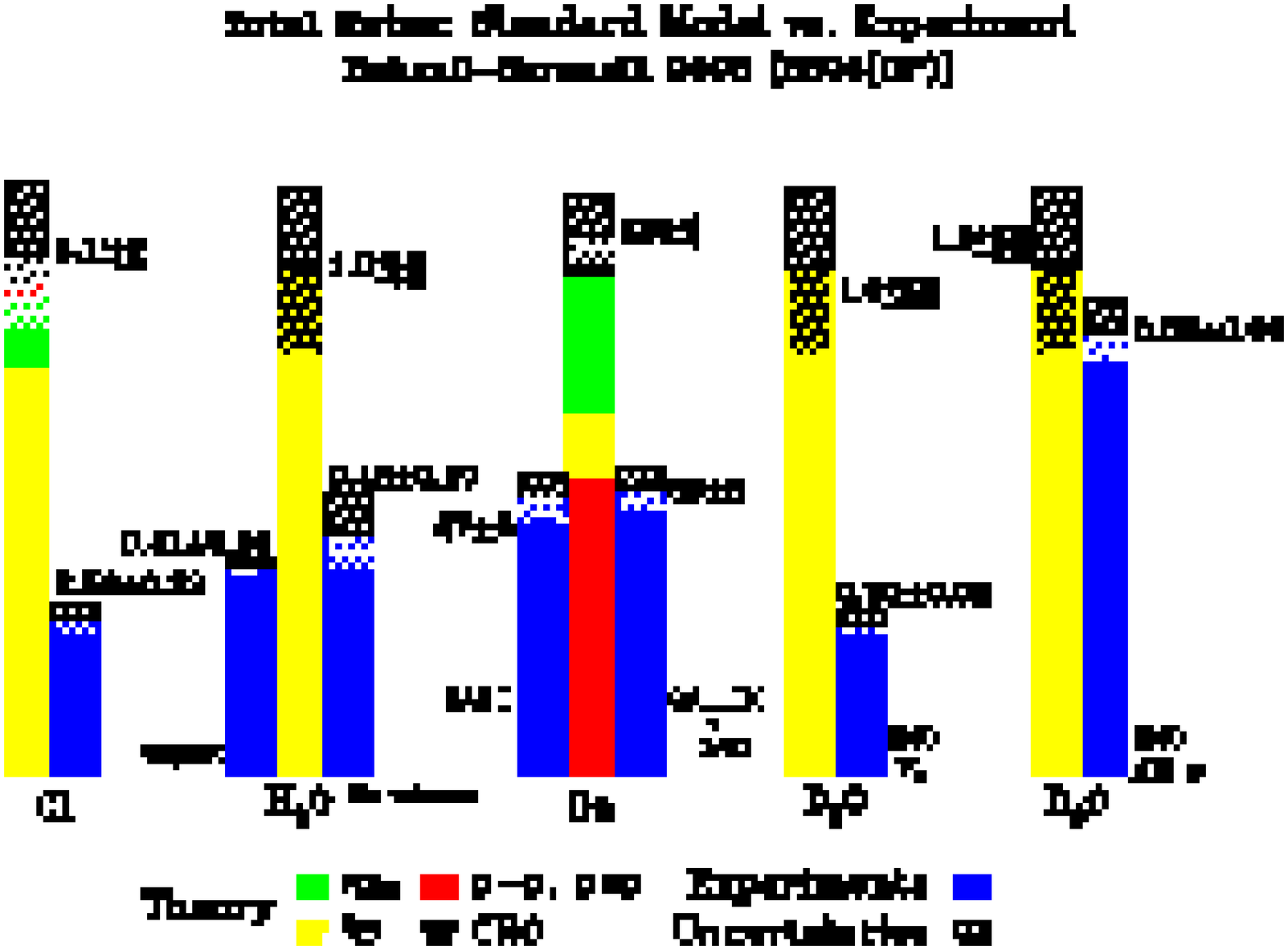}}
\vspace{-0.2cm}
\caption{Observed and predicted fluxes of solar neutrino \cite{BahcallHP}.
\label{fig:solar-nu_flux_obs}}
\end{center}
\end{figure}

In the central region of the sun, $\nu_{e}$s are continuously produced by
the hydrogen fusion,
\begin{equation}
4p \rightarrow {}^{4}{\rm He} + 2 e^{+} + 2 \nu_{e} + 26.731 {\rm MeV}.
\end{equation}
These neutrinos are called {\it solar neutrinos}. The first experiment of the solar neutrino
observation was the chlorine experiment at Homestake by R. Davis and his collaborators
in the 1960's \cite{Davis68,Cleveland98}. On the other hand, the theoretical study of
the solar neutrino was pioneered by J. N. Bahcall. The neutrino production rate in the sun
has been calculated based on the standard solar model, which reproduces the current state
of the sun by following the evolution of a main-sequence star with solar metalicity and mass.
The solar neutrino flux calculated by the current standard solar model
\cite{BahcallBasu05,BahcallSerenelliBasu05} is shown in Fig. \ref{fig:solar-nu_flux}. 

However, the fluxes observed by several detectors have been substantially smaller than
the predicted flux (Fig. \ref{fig:solar-nu_flux_obs}). This is so called
{\it solar neutrino problem} and has been studied for several decades
(for reviews, see \cite{Goswami03,Bahcall04}, and a text book by Bahcall \cite{BahcallText}).
Historically, the solar neutrino problem was attributed to incompleteness of the solar model
and/or unknown neutrino property including neutrino oscillation. With the improvement of both
the solar model and observation, especially at SuperKamiokande
\cite{SKsolar98,SKsolar99,SKsolar01,SKsolar04}, it is now common to think that the uncertainties
of the solar model cannot solely explain the gap between the observations and prediction.

The critical observation has been conducted by Sudbury Neutrino Observatory (SNO)
\cite{SNO01,SNO02,SNO04,SNO05}, which clearly showed that electron neutrinos are converted
to the other flavors. The SNO was designed primarily to search for a clear indication of
neutrino flavor conversion for solar neutrinos without relying on solar model calculations.
Its significant feature is the use of 1000 tons of heavy water which allows the distinction
between the following three signal:
\begin{eqnarray}
&& \nu_{e} + d \rightarrow p + p + e^{-}, \label{eq:d_CC} \\
&& \nu + d \rightarrow p + n + \nu, \label{eq:d_NC} \\
&& \nu + e^{-} \rightarrow \nu + e^{-}. \label{eq:e_NC}
\end{eqnarray}
Here the reaction (\ref{eq:d_CC}) occurs through the charged current interaction
and is relevant only to $\nu_{e}$, while the other two reactions are through the
neutral current interaction and sensitive to all flavors. It should be noted that
SK can identify only the electron scattering event (\ref{eq:e_NC}) for energies of
the solar neutrinos ($< 10 {\rm MeV}$), although its volume is much larger than
that of SNO.

\begin{figure}[hbt]
\begin{center}
 \begin{minipage}{7.0cm}
 \begin{center}
 \includegraphics[width=1.1\textwidth]{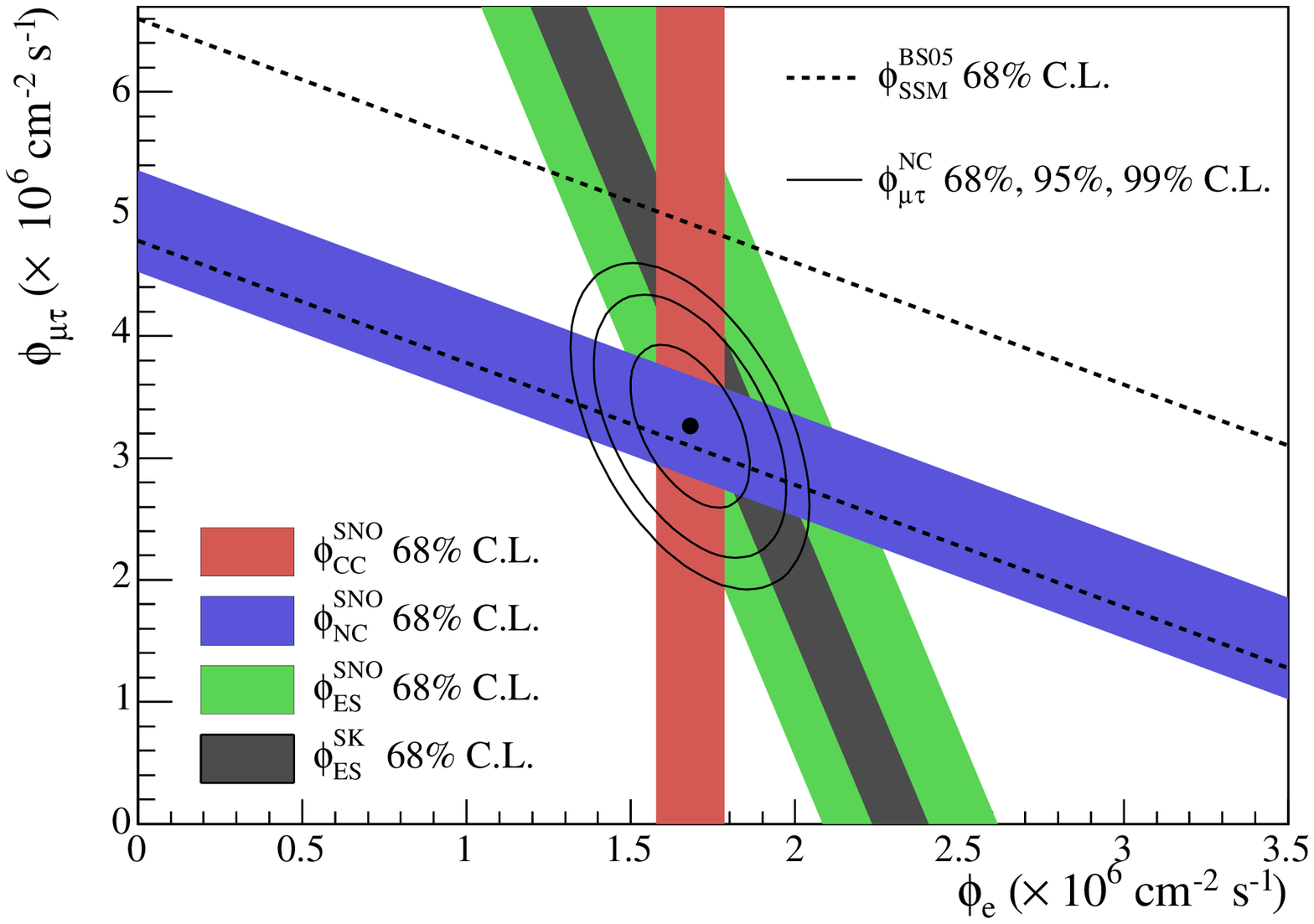}
 \caption{Flux of $\mu+\tau$ neutrinos versus flux of electron neutrinos obtained
 from SNO \cite{SNO05} and SK \cite{SKsolar02}. The total ${}^{8}$Be solar neutrino flux
 predicted by the Standard Solar Model \cite{BahcallSerenelliBasu05} is shown as
 dashed lines. The point represents $\phi_e$ from the CC flux and $\phi_{\mu\tau}$
 from the NC-CC difference with 68\%, 95\%, and 99\% C.L. contours included.
 \label{fig:SNO_obs}}
 \end{center}
 \end{minipage}
\hspace{0.5cm}  
 \begin{minipage}{6.0cm}
 \begin{center}
 \includegraphics[width=1.0\textwidth]{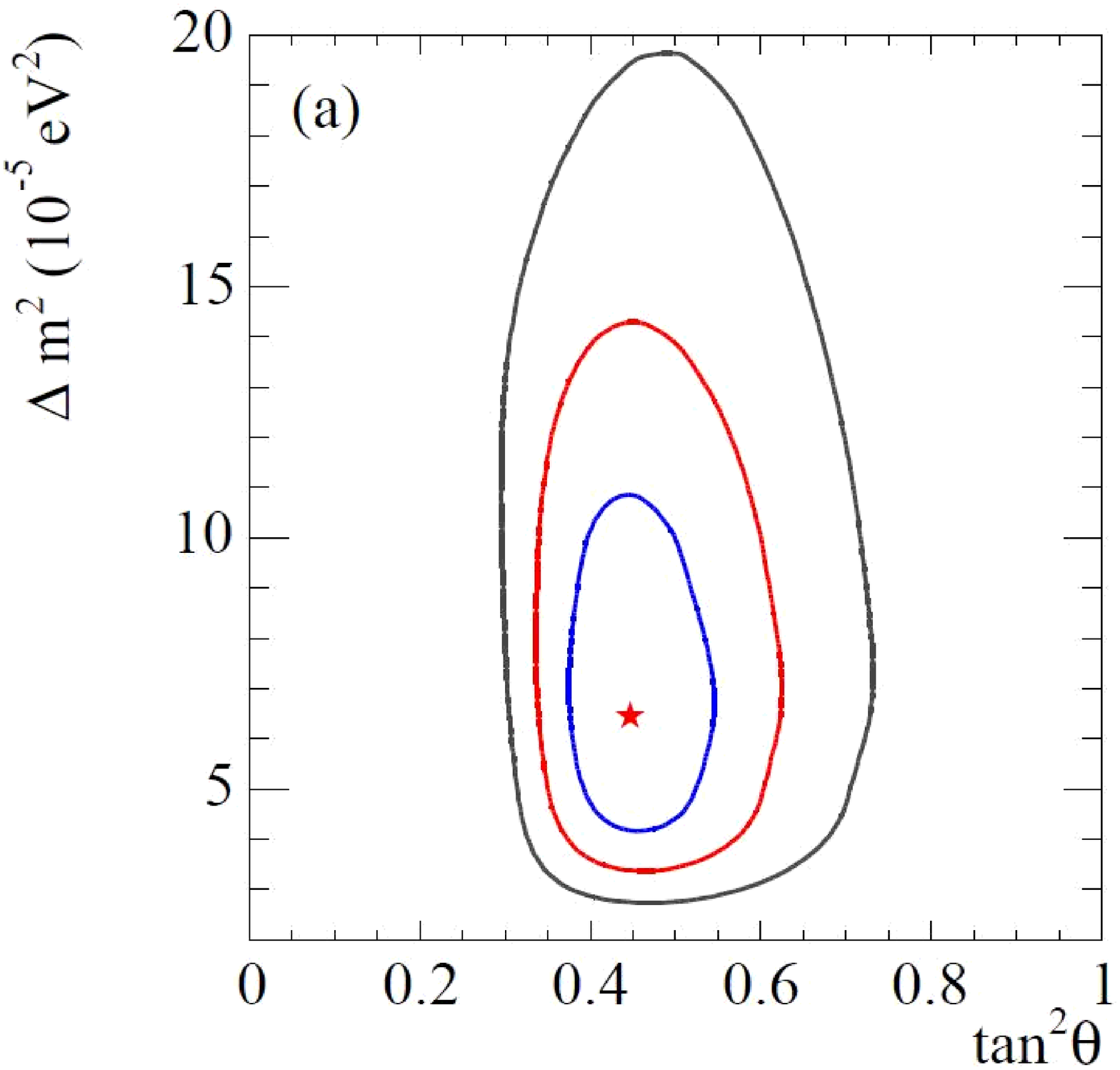}
 \caption{Global neutrino oscillation analysis using solar neutrino data \cite{SNO05}.
 The solar neutrino data includes SNO data, the rate measurements from the Cl, SAGE,
 Gallex/GNO, and SK-I zenith spectra.
 \label{fig:solar_constraint}}
 \end{center}
 \end{minipage}
\end{center}
\end{figure}

Fig. \ref{fig:SNO_obs} shows the fluxes of $\mu+\tau$ neutrinos and electron neutrinos
obtained from SNO \cite{SNO05} and SK \cite{SKsolar02}. Combining the signals from
the three channels, it clearly shows that there is non-zero flux of $\nu_{\mu}$ and
$\nu_{\tau}$, which is a strong evidence of neutrino flavor conversion. If we interpret
these data by 2-flavor neutrino oscillation $\nu_{e} \leftrightarrow \nu_{x}$,
we obtain constraint on the mixing angle and mass-squared difference as
in Fig. \ref{fig:solar_constraint} with the best fit:
\begin{equation}
\Delta m^{2}_{\rm solar} = 6.5^{+4.4}_{-2.3} \times 10^{-5} {\rm eV}^{2}, ~~~
\tan^{2}{\theta_{\rm solar}} = 0.45^{+0.09}_{-0.08}.
\end{equation}
Also it should be noted that, as can be seen in Fig. \ref{fig:SNO_obs}, the prediction of
the standard solar model is, at least concerning the high-energy neutrinos, confirmed by
the observation.

%%%%%%%%%%%%%%%%%%%%%%%%%%%%%%%%%%%%%%
\subsubsection{current status of neutrino parameters \label{subsubsection:current_status}}
%%%%%%%%%%%%%%%%%%%%%%%%%%%%%%%%%%%%%%

\begin{figure}[hbt]
\begin{center}
\includegraphics[width=1\textwidth]{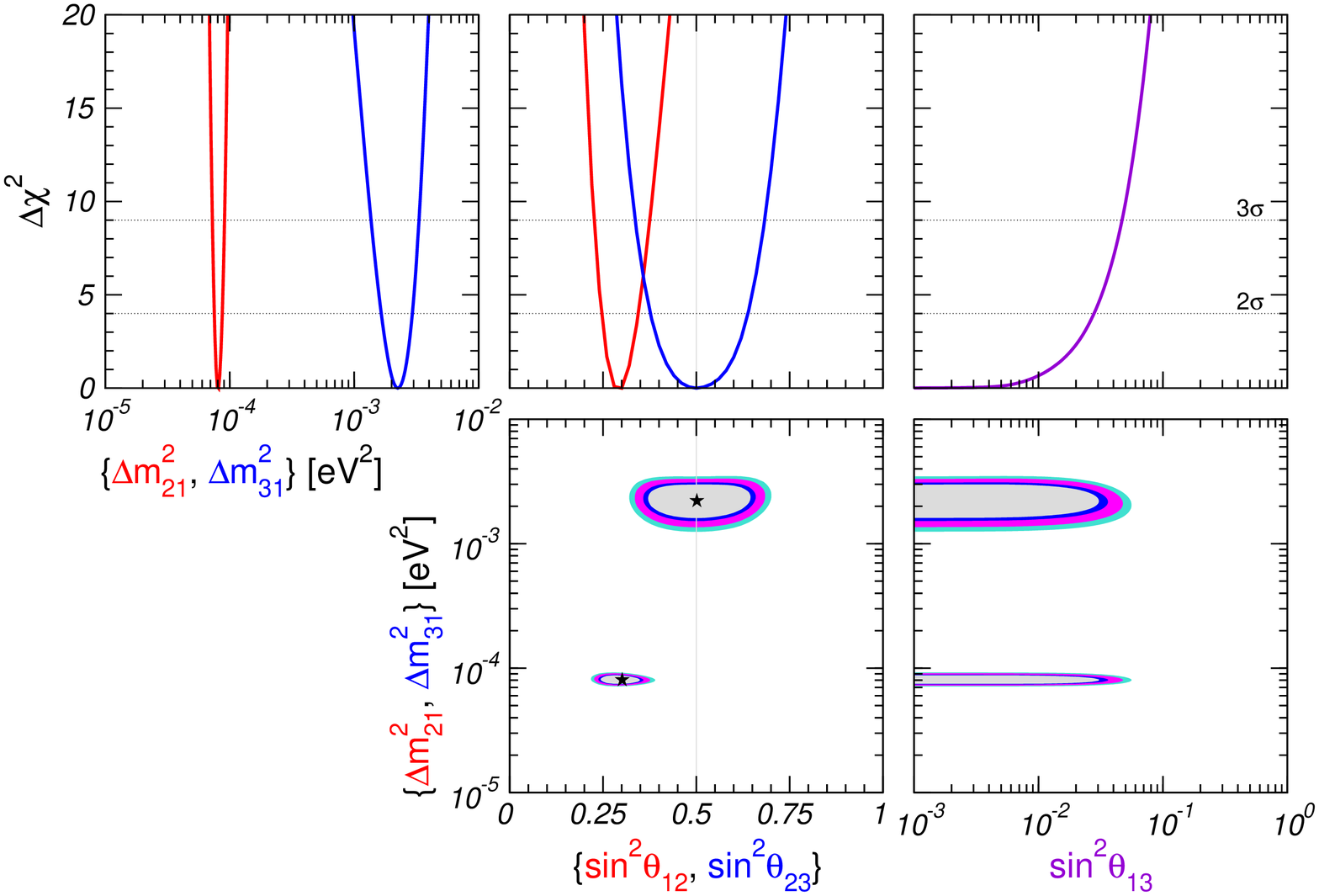}
\vspace{-0.5cm}
\caption{Projections of the allowed regions from the global oscillation data at 90\%, 95\%, 99\%,
and 3$\sigma$ C.L. for 2 d.o.f. for various parameter combinations. Also shown is $\Delta \chi^2$
as a function of the oscillation parameters
$\sin^{2}{\theta_{12}}, \sin^{2}{\theta_{23}}, \sin^{2}{\theta_{13}},
\Delta m^{2}_{21}, \Delta m^{2}_{31}$, minimized with respect to all undisplayed parameters
\cite{Maltoni04}.
\label{fig:osc-parameters}}
\vspace{0.5cm}
\includegraphics[width=0.6\textwidth]{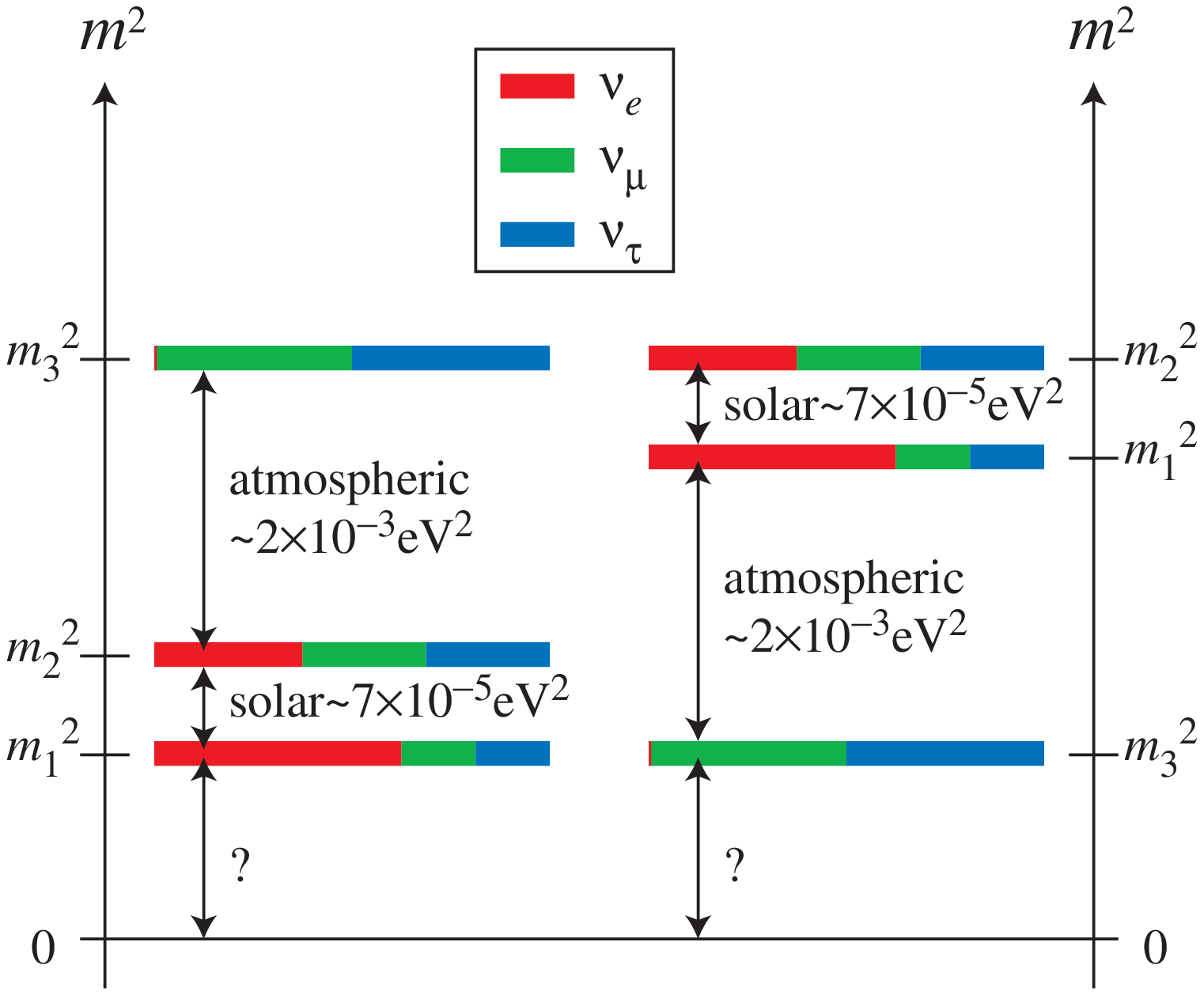}
\vspace{-0.3cm}
\caption{Possible configurations of neutrino mass states as suggested by oscillations:
the normal (left) and inverted (right) hierarchy.
The flavor composition is shown as well \cite{FeruglioStrumiaVissani02,FeruglioStrumiaVissani03}.
\label{fig:mass-spectrum}}
\end{center}
\end{figure}

Here we summarize the information of neutrino oscillation parameters obtained so far.
Combined analysis based on all the data from various neutrino oscillation experiments has been
done by many authors \cite{Fogli03,Gonzalez03,Maltoni03,Maltoni04,StrumiaVissani05}.
Basically three-flavor neutrino oscillation can explain all the data reasonably except
the LSND data discussed in section \ref{subsubsection:accelerator}. Maltoni et al.
\cite{Maltoni04} performed a general fit to the global data in the five-dimensional parameter
space ($\theta_{12}, \theta_{23}, \theta_{13}, \Delta m_{12}^{2}$ and $\Delta m_{13}^{2}$),
and showed projections onto various one- or two-dimensional subspaces. They gave,
\begin{itemize}
\item $\sin^{2}{\theta_{12}} \approx 0.30$, mostly from solar neutrino data
\item $\sin^{2}{\theta_{23}} \approx 0.50$, mostly from atmospheric neutrino data
\item $\sin^{2}{\theta_{13}} \leq 0.30$, mostly from atmospheric neutrino and CHOOZ data
\item $\Delta m_{12}^{2} \approx 8.1 \times 10^{-5} {\rm eV}^{-2}$, mostly from KamLAND data
\item $|\Delta m_{13}^{2}| \approx 2.2 \times 10^{-3} {\rm eV}^{-2}$,
      mostly from atmospheric neutrino data
\end{itemize}
and allowed regions of the parameters are summarized in Table \ref{table:osc-parameters} and
Fig. \ref{fig:osc-parameters}. We can see that most of the parameters are known with high
accuracies.

Contrastingly, we have rather poor information on some of key neutrino parameters.
First, only a loose upper bound has been obtained for $\sin^{2}{\theta_{13}}$.
As we will discuss in section \ref{section:nu-osc_SN}, this parameter acts an important role
in supernova neutrino oscillation.

The next is the signature of $\Delta m_{13}^{2}$. There are two mass schemes according to
the signature (Fig. \ref{fig:mass-spectrum}). One is called {\it normal hierarchy} with
$m_{1}^{2} \approx m_{2}^{2} \ll m_{3}^{2}$ and another is called {\it inverted hierarchy}
with $m_{3}^{2} \ll m_{1}^{2} \approx m_{2}^{2}$. The mass scheme is also crucial when
we consider neutrino oscillation in supernova.

The value of CP violation parameter (see Eq. (\ref{eq:mixing_matrix})) is not known either.
Although it would have small impact on supernova neutrino oscillation so that we have
neglected it, it is important in considering the structure and origin of neutrino masses.
Also it will be important particle-theoretically whether $\theta_{23}$ is maximal ($\pi/4$)
or not and whether $\theta_{13}$ is exactly zero or just small.

As we saw in section \ref{subsubsection:accelerator}, the LSND experiment gave us
an implication of $\bar{\nu}_{\mu} \rightarrow \bar{\nu}_{e}$ oscillation with
$\Delta m_{\rm LSND}^{2} \approx 1 {\rm eV}^{2}$. However, noting that
$\Delta m_{\rm LSND}^{2} \gg \Delta m_{\rm atm}^{2} \gg \Delta m_{\rm solar}^{2}$, it is easy
to see that three-flavor oscillation scheme discussed above cannot explain the LSND data.
This is because we have only two independent mass-squared differences with three flavors and
they are completely determined by the solar, atmospheric, reactor and accelerator experiments.
If the LSND results are confirmed by another experiment like MiniBooNE, we will need some
new physics beyond the standard three-flavor neutrino oscillation. One possibility is
to add an extra neutrino which mixes with standard neutrinos. It must not have a charge
of weak interaction because the LEP experiments imply that there are no very light degrees
of freedom which couple to Z-boson \cite{LEP03}. Thus the extra neutrino must be {\it sterile}.
We will not discuss sterile neutrino further in this review. For further study about sterile
neutrino, see \cite{Cirelli04,Strumia04} and references therein.

\begin{table}[t] 
\begin{center}
\caption{Best-fit values, 2$\sigma$, 3$\sigma$, and 4$\sigma$ intervals (1 d.o.f.) for
the three-flavor neutrino oscillation parameters from global data \cite{Maltoni04}.
\label{table:osc-parameters}}
\begin{tabular}{|l|c|c|c|c|}
        \hline
        parameter & best fit & 2$\sigma$ & 3$\sigma$ & 4$\sigma$
        \\
        \hline
        $\Delta m^2_{21}\: [10^{-5} {\rm eV}^{2}]$
        & 8.1   & 7.5--8.7 & 7.2--9.1 & 7.0--9.4\\
        $\Delta m^2_{31}\: [10^{-3}{\rm eV}^{2}]$
        & 2.2   & 1.7--2.9 & 1.4--3.3 & 1.1--3.7\\
        $\sin^2\theta_{12}$
        & 0.30  & 0.25--0.34 & 0.23--0.38 & 0.21--0.41\\
        $\sin^2\theta_{23}$
        & 0.50  & 0.38--0.64 & 0.34--0.68 & 0.30--0.72 \\
        $\sin^2\theta_{13}$
        & 0.000 &  $\leq$ 0.028 & $\leq$ 0.047  & $\leq$ 0.068 \\
        \hline
\end{tabular}
\end{center}
\end{table}

%%%%%%%%%%%%%%%%%%%%%%%%%%%%%%%%%%%%%%
%\subsubsection{future precision neutrino-oscillation experiments}
%%%%%%%%%%%%%%%%%%%%%%%%%%%%%%%%%%%%%%

%\begin{table}[t]
%\caption{Future accelerator experiments
%\label{table:accelerator}}
%\begin{center}
%\begin{tabular}{cccccc} 
%experiment & baseline  & neutrino energy & detector & status & reference\\ \hline
%MINOS  & $750{\rm km}$ & $1 \sim 25{\rm GeV}$ 
%& $3300{\rm ton}$ scintillator & $2005 \sim$ & \cite{} \\ 
%\end{tabular}
%\end{center}
%\end{table}

\clearpage

%%%%%%%%%%%%%%%%%%%%%%%%%%%%%%%%%%%%%%%%%%%%%%%%%%%%%%%%%%%%%%%%%%%%%%%%%%%%%%%%%%%%%%%%%%%%%%%
%%%%%%%%%%%%%%%%%%%%%%%%%%%%%%%%%%%%%%%%%%%%%%%%%%%%%%%%%%%%%%%%%%%%%%%%%%%%%%%%%%%%%%%%%%%%%%%
\section{Neutrino Oscillation in Supernova \label{section:SN_nu_osc}}
%%%%%%%%%%%%%%%%%%%%%%%%%%%%%%%%%%%%%%%%%%%%%%%%%%%%%%%%%%%%%%%%%%%%%%%%%%%%%%%%%%%%%%%%%%%%%%%
%%%%%%%%%%%%%%%%%%%%%%%%%%%%%%%%%%%%%%%%%%%%%%%%%%%%%%%%%%%%%%%%%%%%%%%%%%%%%%%%%%%%%%%%%%%%%%%

%%%%%%%%%%%%%%%%%%%%%%%%%%%%%%%%%%%%%%%%%%%%%%%%%%%%%%%%%%%%%%%%%%%%%%%%%%%%%%%%%%%%%%%%%%%%%%%
\subsection{Overview}
%%%%%%%%%%%%%%%%%%%%%%%%%%%%%%%%%%%%%%%%%%%%%%%%%%%%%%%%%%%%%%%%%%%%%%%%%%%%%%%%%%%%%%%%%%%%%%%

As we saw in section \ref{supernova_theory}, core-collapse supernovae are powerful sources
of neutrinos with total energies about $10^{53}$ erg. Since neutrinos are considered
to dominate the dynamics of supernova, they reflect the physical state of deep inside
of the supernova, which cannot be seen by electromagnetic waves. Neutrinos are emitted
by the core and pass through the mantle and envelope of the progenitor star.
Since the interactions between matter and neutrinos are extremely weak, one may
expect that neutrinos bring no information about the mantle and envelope.
In fact, they do bring the information through neutrino oscillation because resonant
oscillation discussed in section \ref{subsubsection:varying_density} depends on
the density profile around the resonance point. Thus neutrinos are also a useful tool
to probe the outer structure of supernova, including propagation of shock waves.

On the other hand, supernova has been attracting attention of particle physicist, too,
because it has some striking features as a neutrino source. As we discussed
in section \ref{subsection:ex_nu-osc}, there have been a lot of neutrino oscillation
experiment, which allowed us to know many important parameters such as mixing angles
and mass-squared differences. However, there are still some unknown parameters
and physical structure of neutrinos which are difficult to probe by the conventional
approaches. In this situation, supernova has been expected to give us information
on fundamental properties of neutrinos which cannot be obtained from other sources.

%%%%%%%%%%%%%%%%%%%%%%%%%%%%%%%%%%%%%%%%%%%%%%%%%%%%%%%%%%%%%%%%%%%%%%%%%%%%%%%%%%%%%%%%%%%%%%%
\subsection{Supernova Neutrino\label{subsection:SN-neu}}
%%%%%%%%%%%%%%%%%%%%%%%%%%%%%%%%%%%%%%%%%%%%%%%%%%%%%%%%%%%%%%%%%%%%%%%%%%%%%%%%%%%%%%%%%%%%%%%

Here we review the basic properties of neutrinos emitted during various phases from the onset
of the gravitational collapse to the explosion. Supernova is roughly
a blackbody source for neutrinos of all flavors with a temperature of several MeV. What is
important, in the context of neutrino oscillation, is that each flavor has a different
temperature, flux and its time evolution. The differences are significant especially among
$\nu_{e}$, $\bar{\nu}_{e}$ and the other flavors denoted $\nu_{x}$. Although the quantitative
understanding of the differences are not fully established in the current numerical simulation,
we can still have qualitative predictions and some quantitative predictions.

%%%%%%%%%%%%%%%%%%%%%%%%%%%%%%%%%%%%%%
\subsubsection{neutrino emission during various phases}
%%%%%%%%%%%%%%%%%%%%%%%%%%%%%%%%%%%%%%

Here let us follow again the supernova processes discussed in section \ref{supernova_theory}
focusing on neutrinos. First of all, the core collapse is induced by electron capture,
\begin{equation}
e^{-} + A \rightarrow \nu_{e} + A',
\end{equation}
which produces $\nu_{e}$s. They can escape freely from the core because the core is optically
thin for the neutrinos in the early stage of the collapse. However, the luminosity is
negligible compared with the later phases.

As the core density increases, the mean free path of $\nu_{e}$, $\lambda_{\nu}$, becomes smaller
due to the coherent scattering with nuclei, $\nu_{e} A \rightarrow \nu_{e}$. Neutrinosphere
is formed when the mean free path $\lambda_{\nu}$ becomes smaller than the core size, $R$.
Further, if the diffusion timescale of $\nu_{e}$,
\begin{equation}
t_{\rm diff} \approx \frac{3 R^{2}}{c \lambda_{\nu}},
\end{equation}
is larger than the dynamical timescale of the core,
\begin{equation}
t_{\rm dyn} \approx \frac{1}{\sqrt{G \rho}},
\end{equation}
$\nu_{e}$s cannot escape from the core during the collapse, that is, $\nu_{e}$s are trapped.
When neutrinos are trapped and become degenerate, the average neutrino energy increases
and the core become optically-thicker because the cross section of the coherent scattering
increases as $\sigma_{\rm coh} \propto E_{\nu}^{2}$. Since low-energy neutrinos can escape
easily from the core, most of neutrinos emitted during the collapse phase have relatively
low energy ($< 30 {\rm MeV}$).

\begin{figure}[hbt]
\begin{center}
 \begin{minipage}{7cm}
 \begin{center}
 \includegraphics[width=1.0\textwidth]{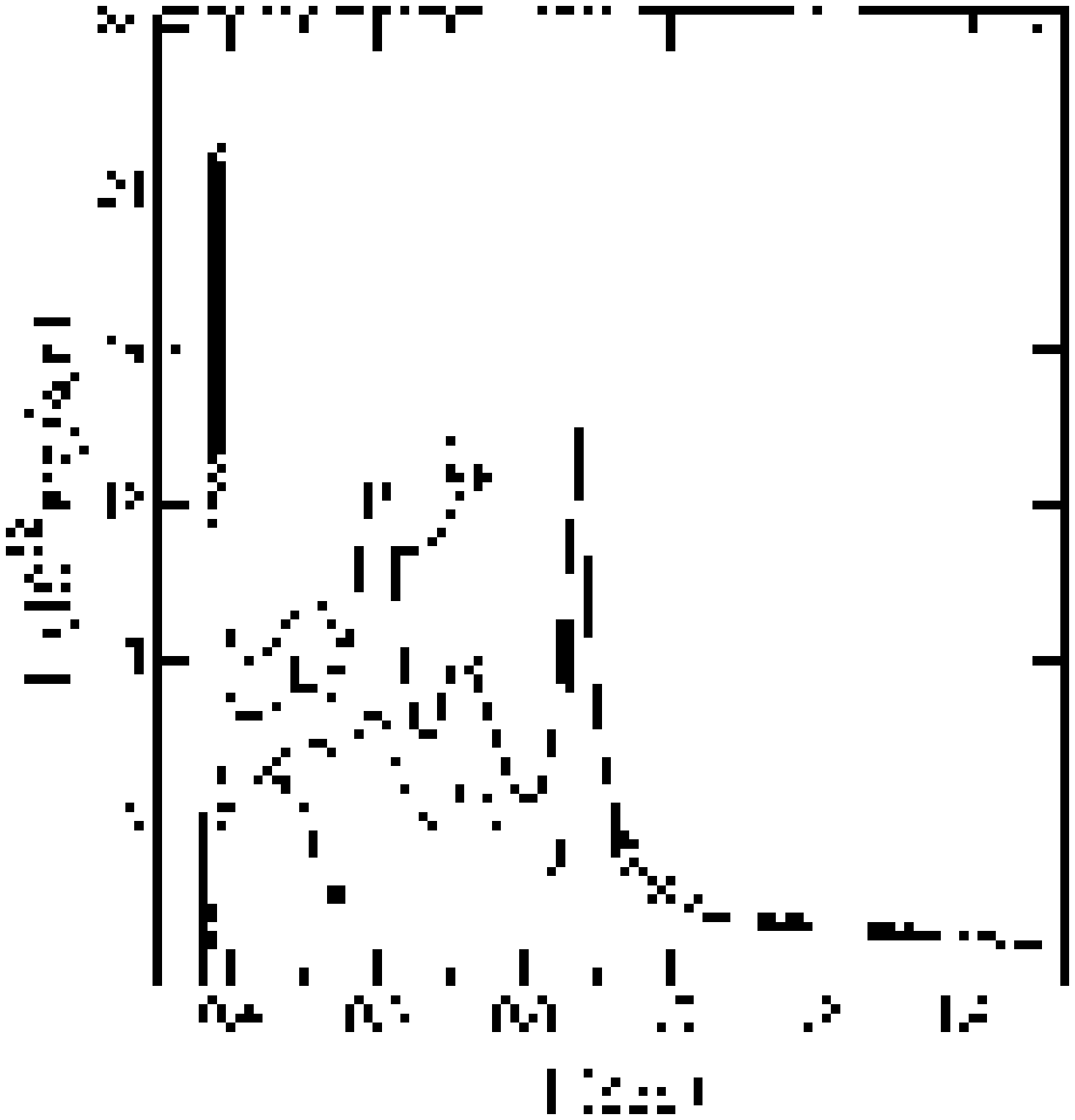}
 \caption{Time evolution of neutrino luminosities calculated by a numerical model
 of a $25 M_{\odot}$ progenitor star with a $2.05 M_{\odot}$ core \cite{MayleWilsonSchramm87}.
 The neutronization peak is drawn to be half the actual value. Here $\nu, \bar{\nu}$ and
 $\mu$ denote $\nu_{e}, \bar{\nu}_{e}$ and $\nu_{x}$, respectively.
 \label{fig:nu_luminosity_delay}}
 \end{center}
 \end{minipage}
\hspace{0.5cm}  
 \begin{minipage}{7cm}
 \begin{center}
 \includegraphics[width=1.1\textwidth]{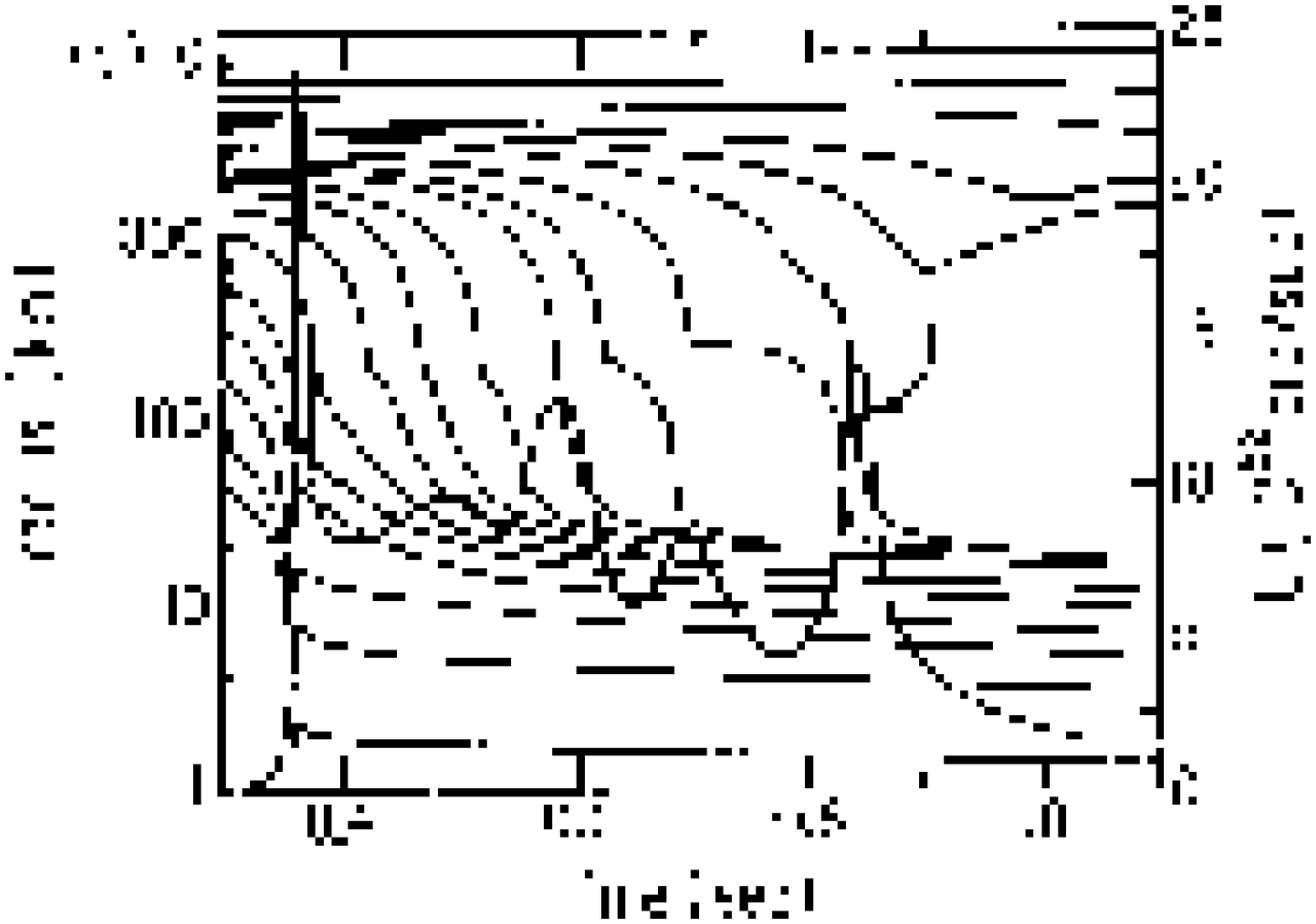}
 \caption{Electron neutrino luminosity and motion of mass shells based on the same model
 as the one in Fig. \ref{fig:nu_luminosity_delay} \cite{MayleWilsonSchramm87}.
 The neutrino luminosities increase when outer core matter accretes onto the protoneutron star.
 \label{fig:luminosity-mass_shell}}
 \end{center}
 \end{minipage}
\end{center}
\end{figure}

The collapse of the inner core stops when the central density exceeds the nucleus density
and a shock wave stands between the inner core and the outer core falling with
a super-sonic velocity. It should be noted that the shock wave stands in a region with
a high density ($\approx 10^{12} \sim 10^{14} {\rm g/cc}$), which is much deeper than
where the neutrinosphere is formed ($\approx 10^{10} \sim 10^{12} {\rm g/cc}$).

In the shocked region, nuclei are decomposed into free nucleons. Because the cross section
of the coherent scattering is proportional to the square of mass number
($\sigma_{\rm coh} \propto A^{2}$), $\nu_{e}$ can freely propagate in the shocked region.
Besides, the cross section of electron capture is much larger for free proton
than nuclei. Thus a lot of $\nu_{e}$s are emitted like a burst while the shock wave propagate
through the core. This process, {\it neutronization burst}, works for about 10 msec and
the emitted neutrino energy is estimated as,
\begin{eqnarray}
&& {\rm peak ~ luminosity} \sim 10^{53} {\rm erg} ~ {\rm s}^{-1}, \\
&& {\rm total ~ energy} \sim 10^{51} {\rm erg}.
\end{eqnarray}

Some fraction of the shocked outer core accretes onto the protoneutron star, where the
gravitational energy is converted into thermal energy. Through thermal processes like
$\gamma + \gamma \rightarrow e^{+} + e^{-}$, positrons are produced and through processes
like,
\begin{equation}
e^{+} + n \rightarrow \bar{\nu}_{e} + p, ~~~ e^{+} + e^{-} \rightarrow \nu + \bar{\nu},
\end{equation}
in addition to the electron capture, neutrinos of all flavors are produced. This accretion
phase continue for $O(10)$ msec for the prompt explosion and $O(1)$ sec for the delayed
explosion.

Finally, the protoneutron star cools and deleptonizes to form a neutron star. In this process,
thermal neutrinos of all flavors are emitted with a timescale $O(10)$ sec, which is the
timescale of the neutrino diffusion. Dominant production process of the neutrinos depends
on the temperature: pair annihilation of electrons and positrons
$e^{-} + e^{+} \rightarrow \nu + \bar{\nu}$ for relatively high temperatures and
nucleon bremsstrahlung $N + N' \rightarrow N + N' + \nu + \bar{\nu}$ for low energies.

The total energy of neutrinos emitted in the cooling phase of the protoneutron star is
roughly the same as the binding energy of the neutron star,
\begin{equation}
E_{\rm NS} \approx \frac{G M_{\rm NS}^{2}}{R_{\rm NS}}
\approx 3 \times 10^{53} {\rm erg}
        \left( \frac{M_{\rm NS}}{M_{\odot}} \right)^{2}
        \left( \frac{10 {\rm km}}{R_{\rm NS}} \right).
\end{equation}
About $99 \% $ of the energy is emitted as neutrinos.

Fig. \ref{fig:nu_luminosity_delay} shows time evolution of neutrino luminosities
calculated by a simulation of the Livermore group \cite{MayleWilsonSchramm87}.
This is based on a numerical progenitor model with mass $25 M_{\odot}$
and a $2.05 M_{\odot}$ core. The time evolution of $\nu_{e}$
luminosity is superimposed on the motion of mass shells in Fig. \ref{fig:luminosity-mass_shell}.
The first $\nu_{e}$ peak is the neutronization burst, whose amplitude is drawn to be half
the actual value in the figure. The next $O(1)$ sec is the matter accretion phase.
The luminosities of $\nu_{e}$ and $\bar{\nu}_{e}$ are greater than that of $\nu_{x}$
because there are additional contributions to $\nu_{e}$ and $\bar{\nu}_{e}$ luminosities
from the charged current interaction of the pair-annihilation of $e^{+} e^{-}$.
As can be seen in Fig. \ref{fig:luminosity-mass_shell}, the neutrino luminosities increase
when outer core matter accretes onto the protoneutron star. The final is the cooling phase
of the protoneutron star, during which neutrino luminosities decrease exponentially with
the neutrino diffusion timescale, $O(10)$ sec.

To summarize, neutrino emission from a supernova can divided into three phases.
Because different mechanisms work during the three phases, they have different timescales
and neutrino luminosities.

%%%%%%%%%%%%%%%%%%%%%%%%%%%%%%%%%%%%%%
\subsubsection{average energy}
%%%%%%%%%%%%%%%%%%%%%%%%%%%%%%%%%%%%%%
\begin{figure}[hbt]
\begin{center}
\includegraphics[width=0.5\textwidth]{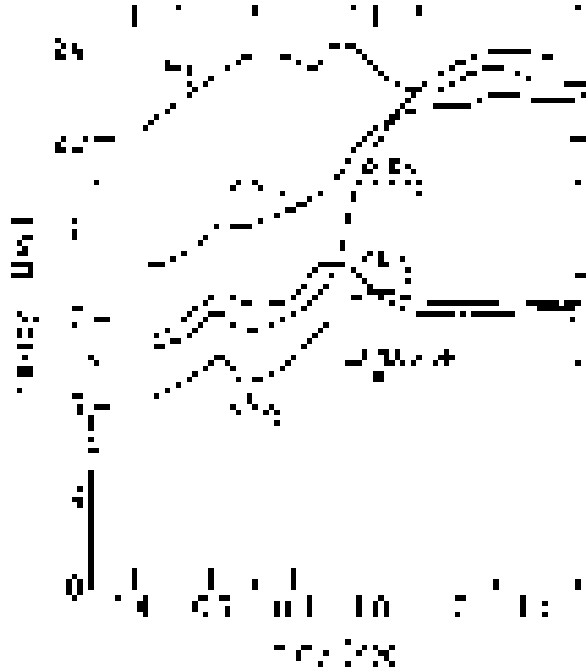}
\end{center}
\vspace{-0.3cm}
\caption{Time evolution of average neutrino energies \cite{MayleWilsonSchramm87}.
Here $\nu, \bar{\nu}$ and $\mu$ denote $\nu_{e}, \bar{\nu}_{e}$ and $\nu_{x}$, respectively,
and the subscripts S and C stand for the calculation method. As neutronization of
the protoneutron star proceeds, the difference between $\bar{\nu}_{e}$ and $\nu_{x}$ energies
decreases.
\label{fig:nu_average_energy}}
\end{figure}

Average energy of emitted neutrinos reflects the temperature of matter around the
neutrinosphere. Interactions between neutrinos and matter are sufficiently strong inside
the neutrinosphere so that thermal equilibrium is realized there. Since the temperature
is lower in the outer region, neutrino average energy becomes lower as the radius of the
neutrinosphere is larger. Then the problem is what determines the radius of the neutrinosphere.
Basically, it is determined by the strength of interactions between neutrinos and matter.

Interactions between neutrinos and matter are,
\begin{eqnarray}
&& \nu_{e} + n \leftrightarrow e^{-} + p, \label{eq:nue-n} \\
&& \bar{\nu}_{e} + p \leftrightarrow e^{+} + n, \label{eq:anue-p} \\
&& \nu + e^{\pm} \leftrightarrow \nu + e^{\pm}, \label{eq:nu-e} \\
&& \nu + N \leftrightarrow \nu + N. \label{eq:nu-N}
\end{eqnarray}
Here it should be noted that the reactions (\ref{eq:nue-n}) and (\ref{eq:anue-p})
are relevant only to $\nu_{e}$ and $\bar{\nu}_{e}$, respectively. Furthermore,
although all flavors interact with matter through the reaction (\ref{eq:nu-e}),
interactions for $\nu_{e}$ and $\bar{\nu}_{e}$ are contributed from both the neutral and charged
current, while that for $\nu_{x}$ is contributed only from the neutral current.
Interaction (\ref{eq:nu-N}) occurs equally to all flavors. Therefore, interactions of
$\nu_{e}$ and $\bar{\nu}_{e}$ are stronger than those of $\nu_{x}$. Because there are
more neutrons than protons in the protoneutron star, $\nu_{e}$ couples stronger to matter
than $\bar{\nu}_{e}$. Thus, it is expected that average energies of neutrinos have
the following inequality:
\begin{equation}
\langle E_{\nu_{e}} \rangle < \langle E_{\bar{\nu}_{e}} \rangle
< \langle E_{\nu_{x}} \rangle.
\label{eq:nu-energy-hierarchy}
\end{equation}
Although this hierarchy would be a robust prediction of the current supernova theory,
it is highly difficult to estimate the differences of the average energies without
detailed numerical simulations.

Fig. \ref{fig:nu_average_energy} shows time evolution of average neutrino energies obtained
from the Livermore simulation \cite{MayleWilsonSchramm87}. We can see the hierarchy of
neutrino average energies, Eq. (\ref{eq:nu-energy-hierarchy}). The difference between
$\bar{\nu}_{e}$ and $\nu_{x}$ energies decreases in time because number of protons decreases
as neutronization of the protoneutron star proceeds.

The differences of average energies are important particularly for neutrino oscillation.
As an extreme case, neutrino oscillation does not affect neutrino spectra at all
if all flavors have exactly the same energy spectrum. However, prediction of
neutrino spectra by numerical simulation is highly sensitive and model-dependent
although the qualitative feature, Eq. (\ref{eq:nu-energy-hierarchy}), is confirmed
by a lot of simulations. Simulations by the Livermore group
\cite{MayleWilsonSchramm87,WilsonMayleWoosleyWeaver86} predict relatively large differences
of average energies, while simulations of protoneutron star cooling by
Suzuki predict much smaller differences \cite{Suzuki93,SumiyoshiSuzukiToki95}.

%%%%%%%%%%%%%%%%%%%%%%%%%%%%%%%%%%%%%%
\subsubsection{energy spectrum}
%%%%%%%%%%%%%%%%%%%%%%%%%%%%%%%%%%%%%%

The position of the neutrinosphere is determined by the strength of the interactions between
neutrinos and matter. However, since the cross sections of the interactions depend on
neutrino energy, the neutrinosphere has a finite width even for one flavor. Therefore,
the energy spectra of neutrinos are not simple blackbodies. Because neutrinos with lower energies
interact relatively weakly with matter, their neutrinospheres have smaller radii compared to
those of high-energy neutrinos. As a result, the energy spectrum has a pinched shape compared
to the Fermi-Dirac distribution.

\begin{figure}[hbt]
\begin{center}
 \begin{minipage}{7cm}
 \begin{center}
 \includegraphics[width=1\textwidth]{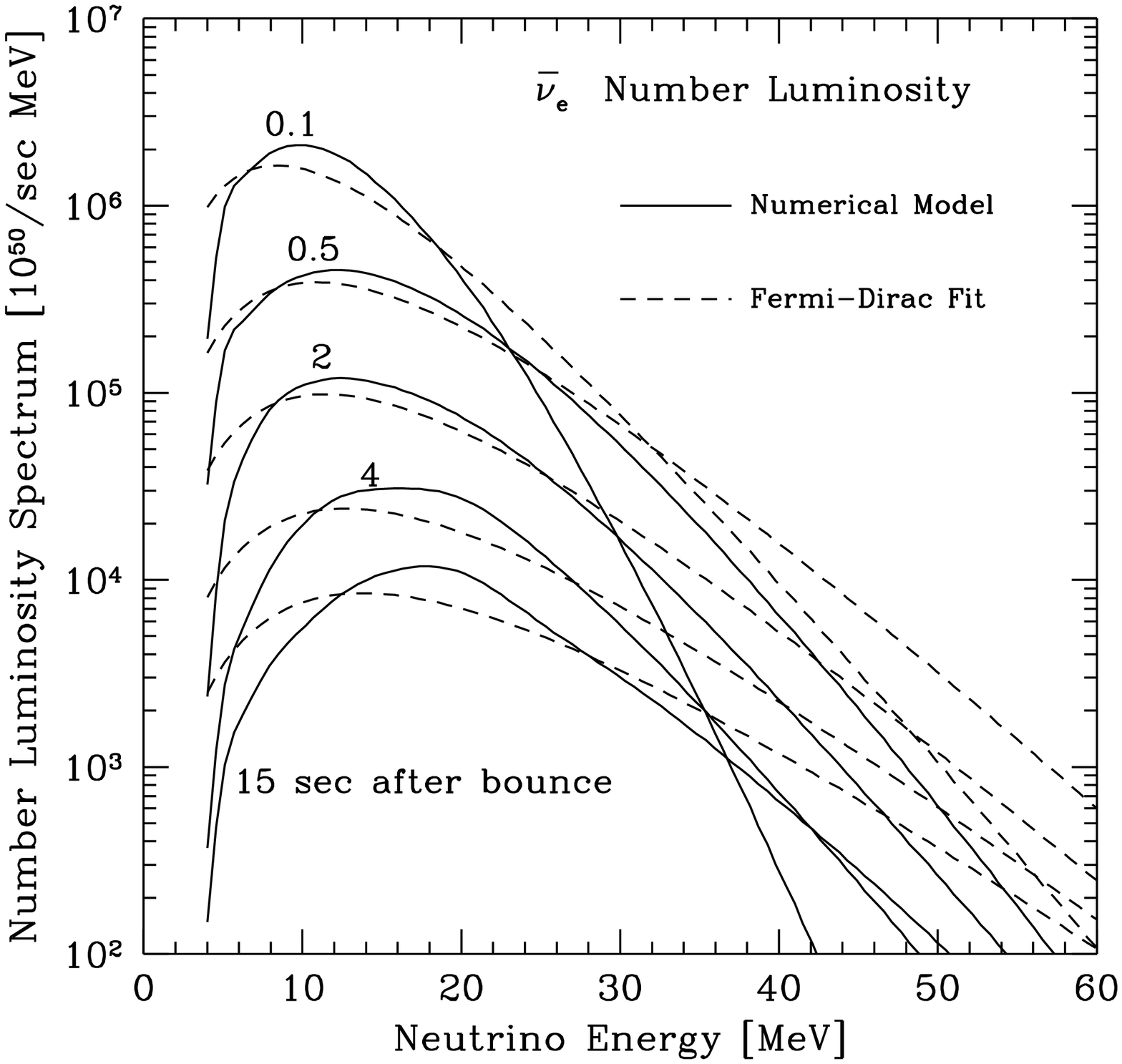}
 \caption{Energy spectra of $\bar{\nu}_{e}$ at different times and the Fermi-Dirac distribution
 with the same average energies \cite{TotaniSatoDalhedWilson98}. The chemical potentials
 of the Fermi-Dirac distributions are set to zero.
 \label{fig:anue_spe}}
 \end{center}
 \end{minipage}
\hspace{0.5cm}  
 \begin{minipage}{7cm}
 \begin{center}
 \includegraphics[width=1\textwidth]{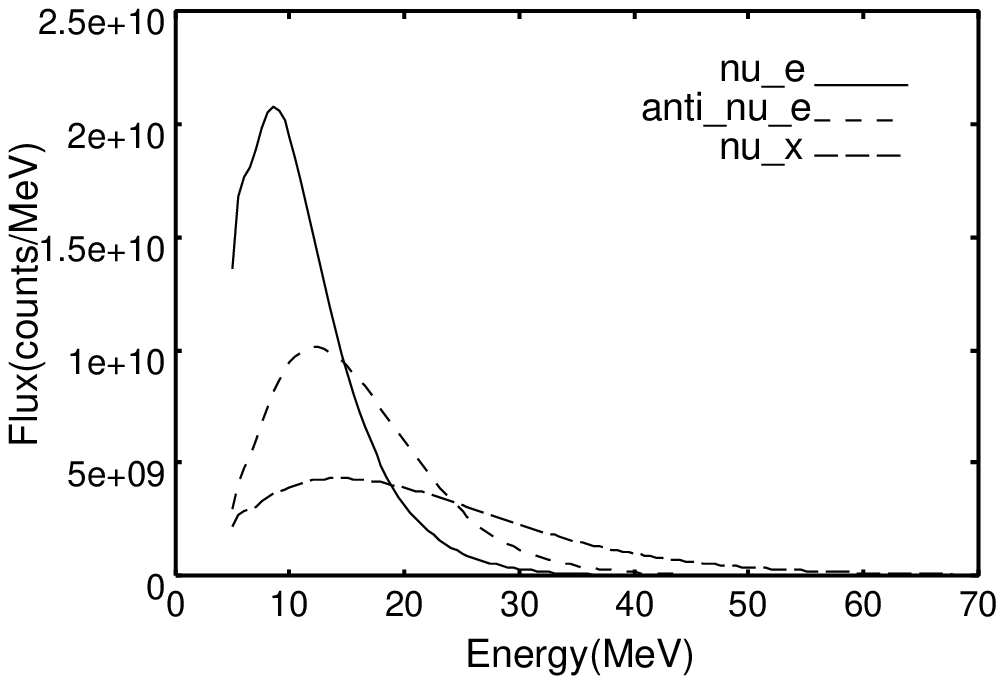}
 \caption{Time-integrated energy spectra \cite{TotaniSatoDalhedWilson98}.
 \label{fig:original_spe}}
 \end{center}
 \end{minipage}
\end{center}
\end{figure}

Fig. \ref{fig:anue_spe} shows energy spectra of $\bar{\nu}_{e}$ at different times and
the Fermi-Dirac distribution with the same average energies \cite{TotaniSatoDalhedWilson98}.
We can see the pinched Fermi-Dirac distribution at each time.
Time-integrated energy spectra are shown in Fig. \ref{fig:integrated_spe}.

In the literature, the neutrino spectrum is sometimes parameterized in several forms.
One popular way to parametrize is called a "pinched" Fermi-Dirac spectrum
(e.g. \cite{LunardiniSmirnov01}),
\begin{equation}
F^{0}_{\alpha}(E)
= \frac{L_{\alpha}}{F(\eta_{\alpha}) T_{\alpha}^{4}}
  \frac{E^{2}}{e^{E/T_{\alpha} - \eta_{\alpha}} + 1},
\end{equation}
where $L_{\alpha}$ and $T_{\alpha}$ are the luminosity and effective temperature
of $\nu_{\alpha}$, respectively, and $\eta_{\alpha}$ is a dimensionless pinching parameter.
The normalization factor $F(\eta_{\alpha})$ is
\begin{equation}
F(\eta_{\alpha}) \equiv \int^{\infty}_{0} \frac{x^{3}}{e^{x-\eta_{\alpha}} + 1},
\end{equation}
where $F(0) = 7 \pi^{4} / 120 \sim 5.68$. Their typical values obtained from numerical
simulations are,
\begin{eqnarray}
&& \langle E_{\bar{e}} \rangle = (14-22) {\rm MeV}, ~~~
   \frac{\langle E_{x} \rangle}{\langle E_{\bar{e}} \rangle} = (1.1-1.6), ~~~
   \frac{\langle E_{e} \rangle}{\langle E_{\bar{e}} \rangle} = (0.5-0.8), \\
&& \frac{L_{e}}{L_{x}} = (0.5-2), ~~~
   \frac{L_{\bar{e}}}{L_{x}} = (0.5-2), \\
&& \eta_{e} = (0-3), ~~~
   \eta_{\bar{e}} = (0-3), ~~~
   \eta_{x} = (0-2).
\end{eqnarray}
Note that the average energy depends on both $T_{\alpha}$ and $\eta_{\alpha}$,
and for $\eta_{\alpha} = 0$ we have $\langle E_{\alpha} \rangle \sim 3.15 T_{\alpha}$.

On the other hand, Keil et al. suggested the following form \cite{Keil03,KeilRaffeltJanka03},
\begin{equation}
F^{0}_{\alpha}(E)
= \frac{L_{\alpha}}{\langle E_{\alpha} \rangle}
  \frac{\beta_{\alpha}^{\beta_{\alpha}}}{\Gamma(\beta_{\alpha})}
  \left( \frac{E}{\langle E_{\alpha} \rangle} \right)^{\beta_{\alpha}-1}
  \exp{\left( - \beta_{\alpha} \frac{E}{\langle E_{\alpha} \rangle} \right)},
\label{eq:spectrum-Keil}
\end{equation}
where $L_{\alpha}$ and $E_{\alpha}$ denote the flux normalization and average energy,
respectively, and  $\beta_{\alpha}$ is a dimensionless parameter that relates to
the width of the neutrino spectrum and typically takes on values $3.5-6$. It should be noted
that these quantities are dependent on both the flavor and time. Spectra obtained from
numerical simulations are well fitted by
\begin{eqnarray}
&& \langle E_{e} \rangle = 12 {\rm MeV}, ~~~
   \langle E_{\bar{e}} \rangle = 15 {\rm MeV}, ~~~
   \langle E_{x} \rangle = 24 {\rm MeV}, \\
&& \frac{L_{e}}{L_{x}} = 2.0, ~~~
   \frac{L_{\bar{e}}}{L_{x}} = 1.6,
\end{eqnarray}
for the ones by the Livermore group and
\begin{eqnarray}
&& \langle E_{e} \rangle = 12 {\rm MeV}, ~~~
   \langle E_{\bar{e}} \rangle = 15 {\rm MeV}, ~~~
   \langle E_{x} \rangle = (15-18) {\rm MeV}, \\
&& \frac{L_{e}}{L_{x}} = (0.5-0.8), ~~~
   \frac{L_{\bar{e}}}{L_{x}} = (0.5-0.8),
\end{eqnarray}
for the ones by the Garching group which will be mentioned in the next section.

%%%%%%%%%%%%%%%%%%%%%%%%%%%%%%%%%%%%%%
\subsubsection{recent developments}
%%%%%%%%%%%%%%%%%%%%%%%%%%%%%%%%%%%%%%

\begin{figure}[hbt]
\begin{center}
\epsfbox{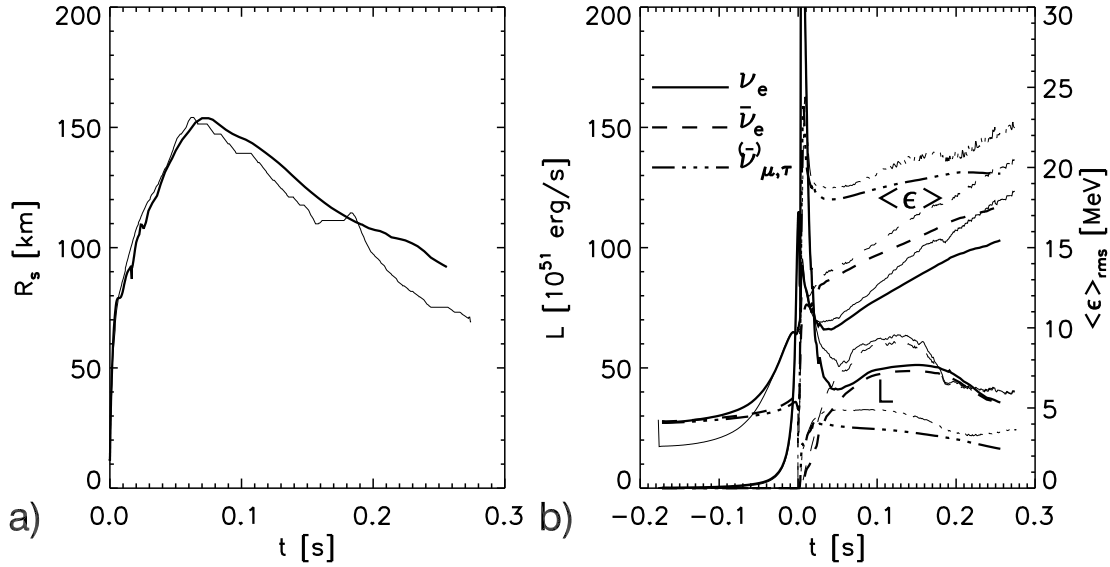}
\end{center}
\vspace{-0.3cm}
\caption{(a) Position of the accretion front as a function of time.
(b) Neutrino luminosities and rms energies as functions of time. In both figures, thick and
thin lines show result from AGILE-BOLTZTRAN and VERTEX, respectively. The differences
in the neutrino results are mainly indirect consequences of the approximate treatment
of general relativity in the VERTEX simulation.
\cite{Lieben03}
\label{fig:SN-model_updated}}
\end{figure}

Around neutrinosphere and shock front, neutrinos strongly couple the dynamics of different
layers on short propagation time scales so that neither diffusion nor free streaming
is a good approximation in this region. An accurate treatment of the neutrino transport
and neutrino-matter interactions therefore are important not only for obtaining reliable
neutrino spectra but also for following the dynamics of supernova correctly. It requires
to solve energy- and angle-dependent Boltzmann transport equation, which is an extremely
tough job. However, recent growing computer capability has made it possible to solve
the Boltzmann equation in consistent with hydrodynamics
\cite{RamppJanka00,Liebendorfer01a,Lieben01,RamppJanka02,ThompsonBurrowsPinto03,Lieben03}.
The result of simulations of several groups agree that spherically symmetric models with
standard microphysical input fail to explode by the neutrino-driven mechanism.

In \cite{Lieben03}, Liebend\"orfer et al. gave a direct and detailed comparison
between two independent codes of neutrino radiation-hydrodynamics, AGILE-BOLTZTRAN of
the Oak Ridge-Basel group and VERTEX of the Garching group.
Fig. \ref{fig:SN-model_updated} shows (a) the position of the accretion front as a function
of time, and (b) the neutrino luminosities and rms energies as functions of time.
The two codes are reasonably consistent as to the position of the accretion front.
The differences in the neutrino results are mainly indirect consequences of the approximate
treatment of general relativity in the VERTEX simulation. It can be seen that differences of
rms neutrino energies are,
\begin{equation}
\langle E_{\bar{\nu}_{e}} \rangle - \langle E_{\nu_{e}} \rangle \sim 2-3 {\rm MeV}, ~~~
\langle E_{\nu_{x}} \rangle - \langle E_{\bar{\nu}_{e}} \rangle \sim 3-4 {\rm MeV}.
\end{equation}
However, since recent sophisticated simulations have not succeeded in explosion,
it is conservative to consider that we do not have definite quantitative predictions
about neutrino spectra.

Despite of the advent of the sophisticated simulations, traditional ones by
the Livermore group are still useful for neutrino oscillation study of supernova neutrinos.
This is because the latter has a great advantage that it covers the full evolution from
the collapse over the explosion to the Kelvin-Helmholtz cooling phase of the newly formed
neutron star, while the former covers typically at most 1 sec.

%%%%%%%%%%%%%%%%%%%%%%%%%%%%%%%%%%%%%%%%%%%%%%%%%%%%%%%%%%%%%%%%%%%%%%%%%%%%%%%%%%%%%%%%%%%%%%%
\subsection{Neutrino Oscillation}
%%%%%%%%%%%%%%%%%%%%%%%%%%%%%%%%%%%%%%%%%%%%%%%%%%%%%%%%%%%%%%%%%%%%%%%%%%%%%%%%%%%%%%%%%%%%%%%

%%%%%%%%%%%%%%%%%%%%%%%%%%%%%%%%%%%%%%
\subsubsection{overview}
%%%%%%%%%%%%%%%%%%%%%%%%%%%%%%%%%%%%%%

Observation of supernova neutrinos can give us information on deep inside of supernova which
cannot be seen by electromagnetic waves. However, in general, neutrinos do not reach
the earth as they were produced at the core due to neutrino oscillation. As in the case
of the solar neutrino, resonant neutrino oscillation occurs in the star. In the current case,
however, there are two resonance points involving three generations of neutrino
because neutrinos of all three flavors are produced in the supernova and the core density
is sufficiently high.

A key point in the three-generation resonance is the mass hierarchy of neutrino.
Because there are three generation, there are two mass differences, which are obtained
by several experiments as
\begin{eqnarray}
&& \Delta m_{12}^{2} \approx 8 \times 10^{-5} {\rm eV}^{2}, \nonumber \\
&& \Delta m_{13}^{2} \approx \Delta m_{23}^{2} \approx 2 \times 10^{-3} {\rm eV}^{2}.
\label{eq:Delta_m}
\end{eqnarray}
As was discussed in section \ref{subsubsection:current_status}, there are two ways to order
these two mass difference: one is normal hierarchy, for which $m_{1}^{2} \approx m_{2}^{2} \ll m_{3}^{2}$,
and another is inverted hierarchy, for which $m_{3}^{2} \ll m_{1}^{2} \approx m_{2}^{2}$.

\begin{figure}[hbt]
\begin{center}
 \includegraphics[width=0.45\textwidth]{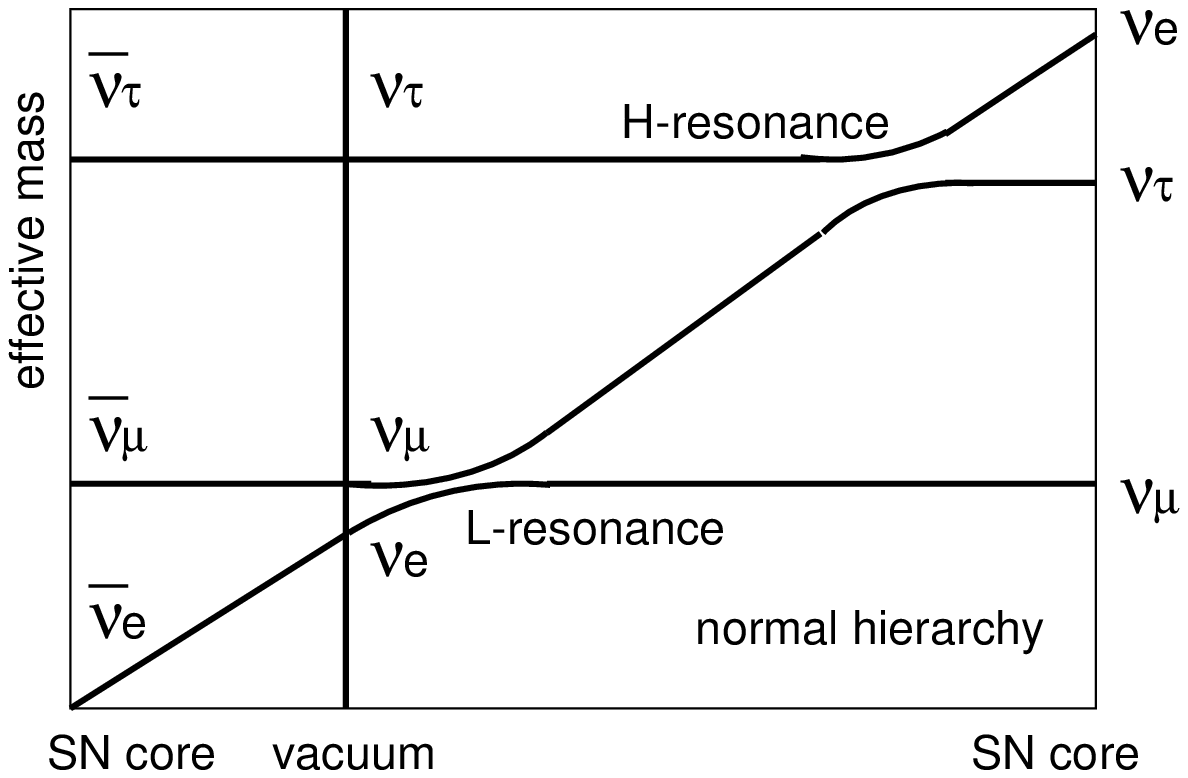}
 \includegraphics[width=0.45\textwidth]{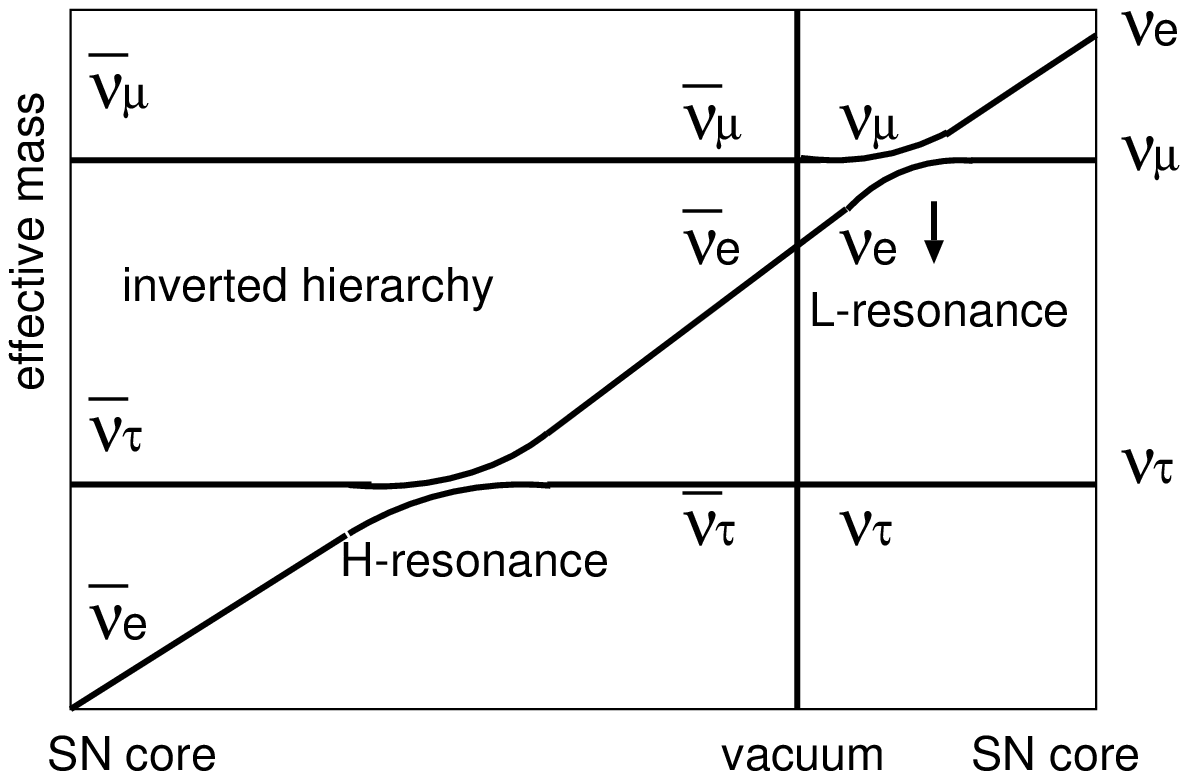}
 \caption{Resonance schemes for normal mass hierarchy (left) and inverted hierarchy (right).
\label{fig:resonance_mass-hierarchy}} 
\end{center}
\end{figure}

The two resonance points are called {\it H-resonance} which occurs at denser region
and {\it L-resonance} which occurs at less dense region. Because the resonance density
is proportional to the mass difference of the two involved mass eigenstates
(see Eq. (\ref{eq:res_condition})), H-resonance and L-resonance correspond to
$\Delta m_{13}^{2}$ and $\Delta m_{12}^{2}$, respectively. Likewise, the involved mixing
angles are $\theta_{13}$ and $\theta_{12}$ for H- and L-resonance, respectively.
Thus, adiabaticities at the two resonance points depend on the following parameters:
\begin{eqnarray}
&& {\rm H-resonance ~ at ~ higher ~ density} \rightarrow \Delta m_{13}^{2}, \theta_{13}, \\
&& {\rm L-resonance ~ at ~ lower ~ density} \rightarrow \Delta m_{12}^{2}, \theta_{12}.
\end{eqnarray}
The situation is quite different for the two mass hierarchies. If the mass hierarchy is normal,
there are two resonances in the neutrino sector, while there is one resonance in both
the neutrino and anti-neutrino sector for the inverted mass hierarchy. The schematic views of
the two situations are shown in Fig. \ref{fig:resonance_mass-hierarchy}.
The vertical line corresponds to vacuum and (anti-)neutrino sector is right (left) hand side
of it. Here it should be noted that $\nu_{e}$ and $\bar{\nu}_{e}$ become effectively heavier
and lighter, respectively, in matter if neutral-current contributions to the effective mass
are subtracted. As can be seen in Fig. \ref{fig:resonance_mass-hierarchy}, L-resonance always
occurs in the neutrino sector, while H-resonance occurs in the neutrino sector for
the normal hierarchy and in the anti-neutrino sector for the inverted hierarchy.

Neutrinos produced at the core have energy spectra which reflect the physical state
of the core. Then neutrino oscillation changes the spectra according to the adiabaticities
of the resonances. Since the adiabaticities depend on the density profile of the progenitor
star and neutrino oscillation parameters, neutrino spectra observed at the earth are
determined by:
\begin{itemize}
\item spectra at the core,
\item density profile of the progenitor star,
\item neutrino oscillation parameters (mixing angles, mass differences and mass hierarchy).
\end{itemize}
If we observe supernova neutrinos, a mixture of these information will be obtained.
The former two of the three directly reflect the explosion mechanism of supernova,
of which we have the basic picture as we discussed in section \ref{supernova_theory}.
However, since we have not succeeded in reproduce an explosion with a numerical simulation
and we have not observed supernova neutrinos besides the small number of events (19 events)
from SN1987A, it would be conservative to think that our understanding of the explosion mechanism,
especially quantitative understanding, is incomplete. Thus, if we want to extract
information on neutrino oscillation parameters, we have to conduct an analysis based
on qualitative features of supernova neutrinos such as,
\begin{itemize}
\item $\nu_{e}$ emission by the neutronization burst,
\item differences in the average energies and fluxed between flavors.
\end{itemize}
However, as we discuss later, number of the neutronization burst events would not be
statistically sufficient unless supernova occurs several persec from the earth.
Anyway, supernova is a unique source to probe neutrino oscillation parameters
as is implied by the comparison to other neutrino sources
(Table \ref{table:neutrino_sources}). Supernova neutrinos have quite different features
from those of other sources so that useful information which cannot be obtained
from other experiments is expected to be obtained.

\begin{table}[t]
\caption{Comparison of some neutrino sources.
\label{table:neutrino_sources}}
\begin{center}
\begin{tabular}{|c|c|c|c|} \hline
neutrino source & solar          & atmospheric      & supernova         \\ \hline
production site & solar center   & atmosphere       & supernova core    \\ \hline
flavor          & ${\nu}_{e}$    & $\nu_{e}, \nu_{\mu}, \bar{\nu}_{e}, \bar{\nu}_{\mu}$ 
                & ${\nu}_{e}, {\bar{\nu}}_{e}, {\nu}_{x}$               \\ \hline
energy          & $\sim 10$ MeV  & $\sim 1$ GeV     & $10 \sim 70$ MeV  \\ \hline
resonance       & once           & none             & twice             \\ \hline
distance        & 1AU            & 10km             & 10kpc             \\ 
                &                & $\sim 10^{4}$km  & (Galactic center) \\ \hline
\end{tabular}
\end{center}
\end{table}

On the other hand, if we want information on supernova itself, we have to subtract the effect
of neutrino oscillation. Some of the neutrino oscillation parameter, such as 2 mass
differences, $\theta_{12}$ and $\theta_{23}$, have been obtained with excellent accuracies
by recent experiments as we saw in section \ref{subsubsection:current_status}. However,
only a loose constraint has been obtained on $\theta_{13}$ and we have no idea on
the mass hierarchy. Therefore it is not straightforward to obtain the intrinsic property
of supernova neutrinos.

%%%%%%%%%%%%%%%%%%%%%%%%%%%%%%%%%%%%%%
\subsubsection{resonance points}
%%%%%%%%%%%%%%%%%%%%%%%%%%%%%%%%%%%%%%

\begin{figure}[hbt]
\begin{center}
 \begin{minipage}{7.0cm}
 \begin{center}
 \includegraphics[width=1.0\textwidth]{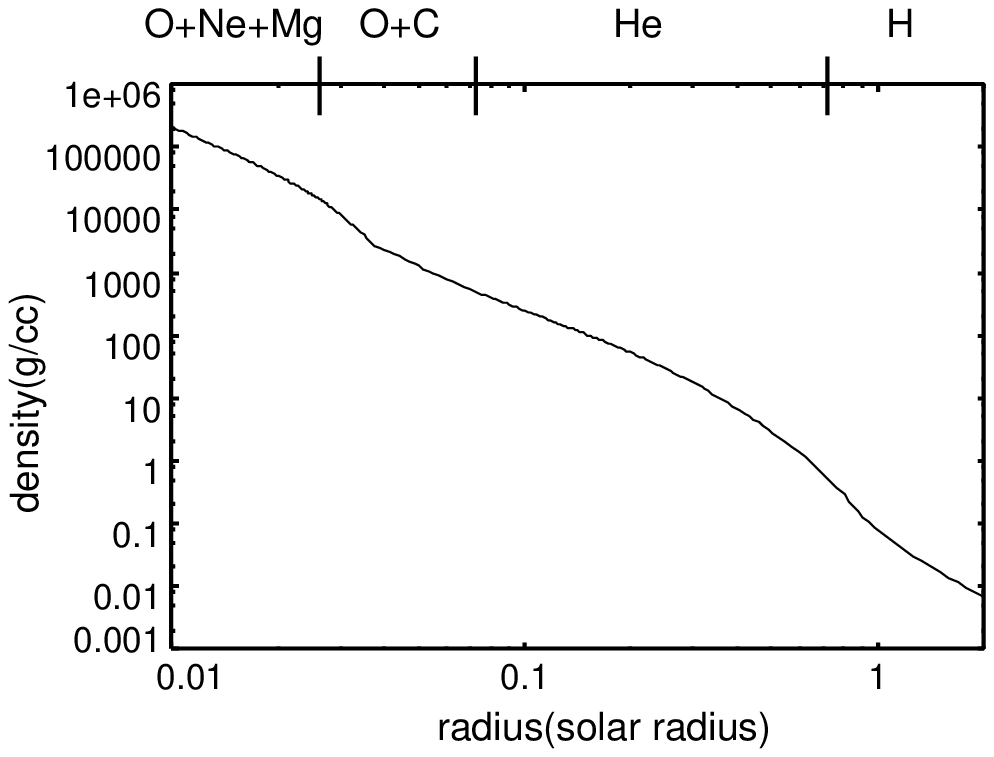}
 \caption{Density profile of a progenitor star with $15 M_{\odot}$ just before
          core collapse \cite{WoosleyWeaver95,SSC}. Also shown corresponding layers.
 \label{fig:density_progenitor}}
 \end{center}
 \end{minipage}
\hspace{0.5cm}  
 \begin{minipage}{6.85cm}
 \begin{center}
 \includegraphics[width=1.0\textwidth]{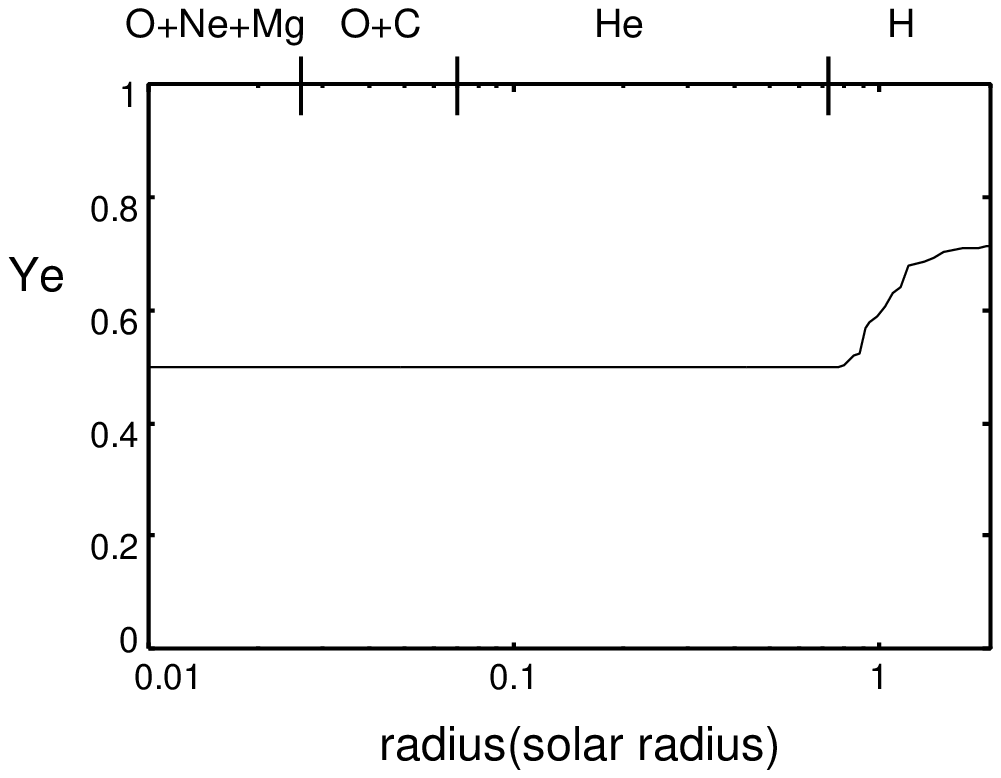}
 \caption{Electron-fraction profile of a progenitor star with $15 M_{\odot}$ just before
          core collapse \cite{WoosleyWeaver95,SSC}.
 \label{fig:Ye_progenitor}}
 \end{center}
 \end{minipage}
\end{center}
\end{figure}

As we saw in Eq. (\ref{eq:res_condition}), the resonance density is,
\begin{equation}
\rho_{\rm res} = 1.3 \times 10^{6} {\rm g} ~ {\rm cm}^{-3}
\cos{2\theta} \left(\frac{0.5}{Y_{e}}\right) 
\left(\frac{10 {\rm MeV}}{E_{\nu}}\right) 
\left(\frac{\Delta m^{2}}{1 {\rm eV}^{2}}\right),
\end{equation}
from which, knowing the approximate values of the mass differences Eq. (\ref{eq:Delta_m}), 
the densities of H- and L-resonance regions are obtained as,
\begin{eqnarray}
&& \rho_{\rm H} \approx (1-10) \times 10^{3} ~ {\rm g} ~ {\rm cm}^{-3}, \\
&& \rho_{\rm L} \approx (20-200) ~ {\rm g} ~ {\rm cm}^{-3}.
\end{eqnarray}
We show a density profile of a progenitor star with mass $15 M_{\odot}$ just before
the core collapse in Fig. \ref{fig:density_progenitor}, which is based on a numerical model
by Woosley and Weaver \cite{WoosleyWeaver95,SSC}. Electron fraction is also shown in
Fig. \ref{fig:Ye_progenitor} for reference. As the figure shows, the resonance regions
are far from the core:
\begin{eqnarray}
&& r_{\rm H} \approx (0.05-0.1) R_{\odot}, \\
&& r_{\rm L} \approx (0.1-0.2) R_{\odot}.
\end{eqnarray}
Therefore, it is conventional to assume that the dynamics of supernova
is not affected by neutrino oscillation. Also it has often been assumed that the shock wave
does not affect the structure of the region where the resonances occur so that static
models of progenitor star have often been used to analyze neutrino oscillation.
In fact, as we discuss later, shock wave reaches the resonance region in $O(1)$ sec and
can change the adiabaticity of the resonances.

%%%%%%%%%%%%%%%%%%%%%%%%%%%%%%%%%%%%%%
\subsubsection{conversion in supernova}
%%%%%%%%%%%%%%%%%%%%%%%%%%%%%%%%%%%%%%

Here we give a general discussion on flavor conversion in supernova following \cite{DigheSmirnov00}.
Our purpose here is to express neutrino fluxes at the surface of the star in terms of the original
fluxes. Let us denote the original flux of flavor $\alpha$ at the core as $F^{0}_{\alpha}$.
Because $\nu_{\mu}, \nu_{\tau}, \bar{\nu}_{\mu}$ and $\bar{\nu}_{\tau}$ interact
with matter almost equally in supernova, their fluxes can also be considered to be equal and we define
\begin{equation}
F^{0}_{x} \equiv F^{0}_{\mu} =F^{0}_{\tau} =F^{0}_{\bar{\mu}} = F^{0}_{\bar{\tau}} 
\end{equation}
as the non-electron neutrino flux. For possible difference in $\nu_{\mu}$ and $\nu_{\tau}$
fluxes, see \cite{AkhmedovLunardiniSmirnov02}.

Although three-generation resonance is complicated in general, the hierarchy of the two
mass-squared differences and the smallness of $U_{e3}$ simplify it substantially so that
the two resonances can be considered to be two two-generation resonances
\cite{KuoPantaleone88,KuoPantaleone89,MikheyevSmirnov89}:
\begin{equation}
P^{\rm total}_{\nu_{e} \rightarrow \nu_{e}}
= P^{\rm H}_{\nu_{e} \rightarrow \nu_{e}}
  \times P^{\rm L}_{\nu_{e} \rightarrow \nu_{e}}.
\end{equation}

\paragraph{neutrino sector}

For extremely high densities as in supernova core ($\rho \gg \rho_{\rm H}, \rho_{\rm L}$),
all mixings are suppressed so that the flavor eigenstates coincide with the eigenstates
in the medium:
\begin{equation}
\nu_{3,{\rm m}} = \nu_{e}, ~~~ \nu_{2,{\rm m}} = \nu_{\tau'}, ~~~ \nu_{1,{\rm m}} = \nu_{\mu'}.
\end{equation}
Correspondingly, the original fluxes of the eigenstates in medium are written as,
\begin{equation}
F^{0}_{3,{\rm m}} = F^{0}_{e}, ~~~
F^{0}_{2,{\rm m}} = F^{0}_{\tau'} = F^{0}_{x}, ~~~
F^{0}_{1,{\rm m}} = F^{0}_{\mu'} = F^{0}_{x}.
\end{equation}
We now want to calculate the fluxes of mass eigenstates at the surface of the star assuming
normal hierarchy. They will be written in terms of the original fluxes
$F^{0}_{\alpha}, (\alpha = e, \mu, \tau)$ and flip probabilities at the H- and L-resonances
$P_{\rm H}$ and $P_{\rm L}$. First let us consider the fate of $F^{0}_{e}$. At the H-resonance,
a fraction $P_{\rm H}$ of $F^{0}_{e}$ flip to lighter state
$\nu_{2,{\rm m}}$ and $(1 - P_{\rm H})$ remain to be $\nu_{3,{\rm m}}$. Among the $\nu_{2,{\rm m}}$
with flux $P_{\rm H} F^{0}_{e}$, a fraction $P_{\rm L}$ flip to the lightest state $\nu_{1,{\rm m}}$
and $(1 - P_{\rm L})$ remain to be $\nu_{2,{\rm m}}$. As a consequence,
\begin{equation}
F^{0}_{e} \Longrightarrow
\left\{ \begin{array}{l}
F_{1} : P_{\rm L} P_{\rm H} F^{0}_{e} \\
F_{2} : (1 - P_{\rm L}) P_{\rm H} F^{0}_{e} \\
F_{3} : (1 - P_{\rm H}) F^{0}_{e}
\end{array} \right. .
\end{equation}
In the same way, $F_{i}$ are contributed from $\nu_{\mu'}$ and $\nu_{\tau'}$ as,
\begin{equation}
F^{0}_{\mu'} \Longrightarrow
\left\{ \begin{array}{l}
F_{1} : (1 - P_{\rm L}) F^{0}_{\mu'} \\
F_{2} : P_{\rm L} F^{0}_{\mu'} \\
F_{3} : 0
\end{array} \right.
, ~~~
F^{0}_{\tau'} \Longrightarrow
\left\{ \begin{array}{l}
F_{1} : P_{\rm L} (1 - P_{\rm H}) F^{0}_{\tau'} \\
F_{2} : (1 - P_{\rm L}) (1 - P_{\rm H}) F^{0}_{\tau'} \\
F_{3} : P_{\rm H} F^{0}_{\tau'}
\end{array} \right. .
\end{equation}
Summing all the contributions, we obtain
\begin{eqnarray}
&& F_{1} = P_{\rm L} P_{\rm H} F^{0}_{e} + (1 - P_{\rm L} P_{\rm H}) F^{0}_{x}, \\
&& F_{2} = (1 - P_{\rm L}) P_{\rm H} F^{0}_{e}
           + (1 - P_{\rm H} + P_{\rm L} P_{\rm H}) F^{0}_{x}, \\
&& F_{3} = (1 - P_{\rm H}) F^{0}_{e} + P_{\rm H} F^{0}_{x},
\end{eqnarray}
which are rewritten as,
\begin{equation}
F_{i} = a_{i} F^{0}_{e} + (1 - a_{i}) F^{0}_{x},
\label{eq:flux_i_e0}
\end{equation}
where
\begin{equation}
a_{1} = P_{\rm L} P_{\rm H}, ~~~
a_{2} = (1 - P_{\rm L}) P_{\rm H}, ~~~
a_{3} = 1 - P_{\rm H}.
\end{equation}
Because the mass eigenstates are the eigenstates of the Hamiltonian in vacuum,
they propagate independently to the earth. Further, their coherence is lost
on the way to the earth so that the neutrinos arrive at the surface of the earth
as incoherent fluxes of the mass eigenstates.

Taking into account the neutrino mixing, the net flux of $\nu_{e}$ is, upto
the geometrical factor $1/4 \pi L^{2}$,
\begin{eqnarray}
F_{e} &=& \sum_{i} \left| U_{ei} \right|^{2} F_{i} \nonumber \\
      &=& F^{0}_{e} \sum_{i} \left| U_{ei} \right|^{2} a_{i}
          + F^{0}_{x} \left( 1 - \sum_{i} \left| U_{ei} \right|^{2} a_{i} \right) \nonumber \\
      &=& p F^{0}_{e} + (1-p) F^{0}_{x},
\label{eq:flux_e_f0}
\end{eqnarray}
where we used the unitarity condition $\sum_{i} \left| U_{ei} \right|^{2} = 1$ and
the "total survival probability of $\nu_{e}$", $p$, is defined as,
\begin{eqnarray}
p &\equiv& \left| U_{ei} \right|^{2} a_{i} \nonumber \\
  &=& \left| U_{e1} \right|^{2} P_{\rm L} P_{\rm H}
      + \left| U_{e2} \right|^{2} (1 - P_{\rm L}) P_{\rm H}
      + \left| U_{e3} \right|^{2} (1 - P_{\rm H}).
\label{eq:survival-p_e}
\end{eqnarray}
Since the total flux is conserved, that is,
\begin{equation}
F^{0}_{e} + 2 F^{0}_{x} = F_{e} + 2 F_{x},
\end{equation}
we obtain the net flux of $\nu_{x}$ as
\begin{equation}
F_{\mu} + F_{\tau} = 2 F_{x} = (1-p) F^{0}_{e} + (1+p) F^{0}_{x}.
\end{equation}
Thus the fluxes at the surface of the earth can be expressed by the original fluxes
and the survival probability $p$. Here it should be noted that not only the fluxes
but also the survival probability depend on the neutrino energy.

So far, we have assumed the normal hierarchy. In fact, the case with the inverted hierarchy
reduces the same expressions with $P_{\rm H} = 1$.

\paragraph{anti-neutrino sector}

Next we consider the antineutrino sector. Again, the flavor eigenstates coincide with
the eigenstates in the medium at the core with sufficiently high densities:
\begin{equation}
\bar{\nu}_{3,{\rm m}} = \bar{\nu}_{\tau'}, ~~~
\bar{\nu}_{2,{\rm m}} = \bar{\nu}_{\mu'}, ~~~
\bar{\nu}_{1,{\rm m}} = \bar{\nu}_{e},
\end{equation}
so that the original fluxes of the eigenstates in the medium are given by,
\begin{equation}
F^{0}_{\bar{3},{\rm m}} = F^{0}_{\bar{\tau}'}, ~~~
F^{0}_{\bar{2},{\rm m}} = F^{0}_{\bar{\mu}'} = F^{0}_{x}, ~~~
F^{0}_{\bar{1},{\rm m}} = F^{0}_{\bar{e}} = F^{0}_{x}.
\end{equation}
Because the small mixing angle $\theta_{e3}$ is further suppressed in the medium,
the $\bar{\nu}_{e} \leftrightarrow \bar{\nu}_{3}$ transitions are negligible, Also,
the state $\bar{\nu}_{3}$ is so far from the level crossing that it propagates
adiabatically, that is, $\bar{\nu}_{\tau}' \rightarrow \bar{\nu}_{3}$. The other
two states can flip through H-resonance for the inverted hierarchy. (For a possible
L-resonance of anti-neutrino sector, see \cite{SmirnovSpergelBahcall94}.)
In the same way as the neutrino sector, we obtain,
\begin{eqnarray}
&& F_{\bar{e}} = \bar{p} F^{0}_{\bar{e}} + (1 - \bar{p}) F^{0}_{x}, \\
&& F_{\bar{\mu}} + F_{\bar{\tau}} = 2 F_{x}
   = (1 - \bar{p}) F^{0}_{\bar{e}} + (1 + \bar{p}) F^{0}_{x}.
\end{eqnarray}
where
\begin{equation}
\bar{p} = \left| U_{e1} \right|^{2} \bar{P}_{\rm H}
          + \left| U_{e3} \right|^{2} (1 - \bar{P}_{\rm H}),
\end{equation}
is the effective survival probability of $\bar{\nu}_{e}$. For the normal hierarchy,
$\bar{P}_{\rm H} = 1$.

\paragraph{summary}

Summarizing the considerations above, the neutrino fluxes can be written as,
\begin{equation}
\left( \begin{array}{c}
F_{e} \\ F_{\bar{e}} \\ 4 F_{x}
\end{array} \right)
=
\left( \begin{array}{ccc}
p & 0 & 1 - p \\
0 & \bar{p} & 1 - \bar{p} \\
1 - p & 1 - \bar{p} & 2 + p + \bar{p}
\end{array} \right)
\left( \begin{array}{c}
F^{0}_{e} \\ F^{0}_{\bar{e}} \\ F^{0}_{x}
\end{array} \right)
\label{eq:flux-mix}
\end{equation}
with survival probabilities,
\begin{eqnarray}
&& p = \left| U_{e1} \right|^{2} P_{\rm L} P_{\rm H}
       + \left| U_{e2} \right|^{2} (1 - P_{\rm L}) P_{\rm H}
       + \left| U_{e3} \right|^{2} (1 - P_{\rm H}), \\
&& \bar{p} = \left| U_{e1} \right|^{2} \bar{P}_{\rm H}
             + \left| U_{e3} \right|^{2} (1 - \bar{P}_{\rm H}),
\label{eq:effective_survival-prob}
\end{eqnarray}
which depend on neutrino energy, neutrino oscillation parameters and mass scheme.

%%%%%%%%%%%%%%%%%%%%%%%%%%%%%%%%%%%%%%
\subsubsection{survival probabilities \label{subsubsection:survival_probability}}
%%%%%%%%%%%%%%%%%%%%%%%%%%%%%%%%%%%%%%

In the previous section, we obtained an general expression of neutrino fluxes at the surface
of the earth in terms of original fluxes and survival probabilities, which can be expressed by
the conversion probabilities at the two resonances and oscillation parameters.

The survival probabilities $p$ and $\bar{p}$ can be calculated from the following
Schr\"odinger-like equation,
\begin{equation}
i \frac{d}{dt}
\left( \begin{array}{c} \nu_{e} \\ \nu_{\mu} \\ \nu_{\tau} \end{array} \right)
= H(t)
  \left( \begin{array}{c} \nu_{e} \\ \nu_{\mu} \\ \nu_{\tau} \end{array}\right),
\label{eq:Schrodinger}
\end{equation}
where the effective Hamiltonian is given by
\begin{equation}
H(t) = U 
\left(\begin{array}{ccc}
0 & 0 & 0\\
0 & \Delta m^{2}_{12} / 2E & 0\\
0 & 0 & \Delta m^{2}_{13} / 2E
\end{array}\right) U^{-1}
+ 
\left(\begin{array}{ccc}
A(t) & 0 & 0\\
0 & 0 & 0\\
0 & 0 & 0
\end{array}\right),
\end{equation}
and matter function $A(t)$ is given by
\begin{equation}
A(t) = \sqrt{2} G_{F} n_{e}(t).
\end{equation}
When the two resonances are perfectly adiabatic or non-adiabatic, the situation is simple.
Let us first estimate the adiabaticies of the two resonances.
If we assume the density profile of the progenitor star as
\begin{equation}
\rho (r) = \rho_{0} r^{-n},
\end{equation}
the adiabaticity parameter $\gamma$, defined in Eq. (\ref{eq:adiabaticity}), is written as,
\begin{equation}
\gamma
= \frac{1}{2 n} \frac{\sin^{2}{2 \theta}}{(\cos{2 \theta})^{1+1/n}}
\left( \frac{\Delta m^{2}}{E} \right)^{1-1/n}
\left( \frac{2 \sqrt{2} G_{F} Y_{e} \rho_{0}}{m_{p}} \right)^{1/n}.
\end{equation}
Adopting an approximate value $n=-3$ in the mantle, we have
\begin{eqnarray}
\gamma
&=& 9 \times 10^{3} \frac{\sin^{2}{2 \theta_{13}}}{(\cos{2 \theta_{13}})^{4/3}}
\left( \frac{\Delta m_{13}^{2}}{2 \times 10^{-3} {\rm eV}^{2}} \right)^{2/3}
\left( \frac{10 {\rm MeV}}{E} \right)^{2/3} ~~~ ({\rm for ~ H-resonance}) \nonumber \\
& &
\label{eq:adiabaticity-H} \\
&=& 10^{3} \frac{\sin^{2}{2 \theta_{12}}}{(\cos{2 \theta_{12}})^{4/3}}
\left( \frac{\Delta m_{12}^{2}}{8 \times 10^{-5} {\rm eV}^{2}} \right)^{2/3}
\left( \frac{10 {\rm MeV}}{E} \right)^{2/3},  ~~~ \nonumber \\ 
& & ({\rm for ~ L-resonance})
\label{eq:adiabaticity-L}
\end{eqnarray}
Substituting the mixing angle obtained from solar neutrino observation into
Eq. (\ref{eq:adiabaticity-L}), it is seen that the L-resonance is perfectly adiabatic for
an energy range of supernova neutrino ($E = 1 - 100 {\rm eV}$). On the other hand,
for the current constraint on $\theta_{13}$, the H-resonance can be either perfectly adiabatic
or perfectly non-adiabatic.

\begin{figure}[hbt]
\begin{center}
\includegraphics[width=1\textwidth]{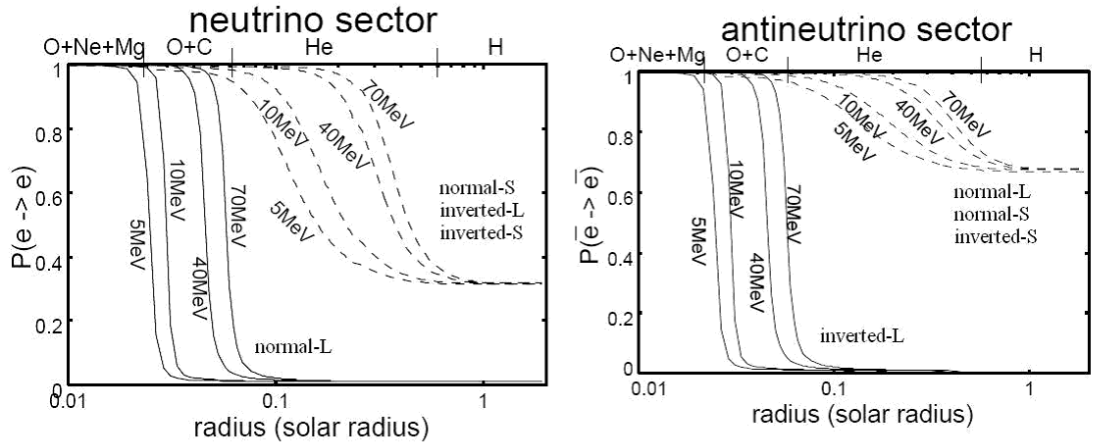}
\end{center}
\vspace{-0.8cm}
\caption{Survival probabilities of $\nu_{e}$ (left) and $\bar{\nu}_{e}$ (right)
in the progenitor star \cite{KT03}. Survival probabilities depend drastically on
the mass scheme (normal or inverted) and the value of $\theta_{13}$ (Large or Small).
When the resonances are sufficiently adiabatic or non-adiabatic, difference in
neutrino energy leads to just difference of the position of resonance.
\label{fig:conv_prob}}
\end{figure}

Takahashi et al. followed the evolution of survival probabilities,
$P_{\nu_{e} \rightarrow \nu_{e}}$ and $P_{\bar{\nu}_{e} \rightarrow \bar{\nu}_{e}}$,
in \cite{KT01,KT03}. Their calculations were performed by solving the Schr\"odinger
equation (\ref{eq:Schrodinger}) numerically along the density profile of the progenitor star
given in Fig. \ref{fig:density_progenitor}. Fig. \ref{fig:conv_prob} shows the evolution of
the survival probabilities with sufficiently large (L, $\sin^{2}{2 \theta_{13}} > 10^{-3}$) and
small (S, $\sin^{2}{2 \theta_{13}} < 10^{-5}$) values of $\theta_{13}$ \cite{KT03}.
Here "normal-L" means a model with sufficiently large $\theta_{13}$ and normal mass hierarchy.
As can be seen the survival probabilities depend drastically on the mass scheme and the value
of $\theta_{13}$. It is also seen that difference in neutrino energy leads to just difference
of the position of resonance because the resonances are sufficiently adiabatic or non-adiabatic.
Here it should be noted that mass hierarchy is not important when $\theta_{13}$ is sufficiently
small because, in this case, the H-resonance is perfectly non-adiabatic, which is equivalent
to the absence of the H-resonance.

%%%%%%%%%%%%%%%%%%%%%%%%%%%%%%%%%%%%%%
\subsubsection{neutrino spectra}
%%%%%%%%%%%%%%%%%%%%%%%%%%%%%%%%%%%%%%

\begin{figure}[hbt]
\begin{center}
 \begin{minipage}{7.0cm}
% \begin{center}
% \epsfbox{figure_ktaro/original_spe.eps}
\includegraphics[width=1\textwidth]{figure_ktaro/original_spe.eps}
 \caption{Original neutrino spectra \cite{TotaniSatoDalhedWilson98} used in \cite{KT03}.
 \label{fig:original_spe}}
% \end{center}
 \end{minipage}
\hspace{0.5cm}  
 \begin{minipage}{7.0cm}
% \begin{center}
% \epsfbox{figure_ktaro/original_time.eps}
\includegraphics[width=1\textwidth]{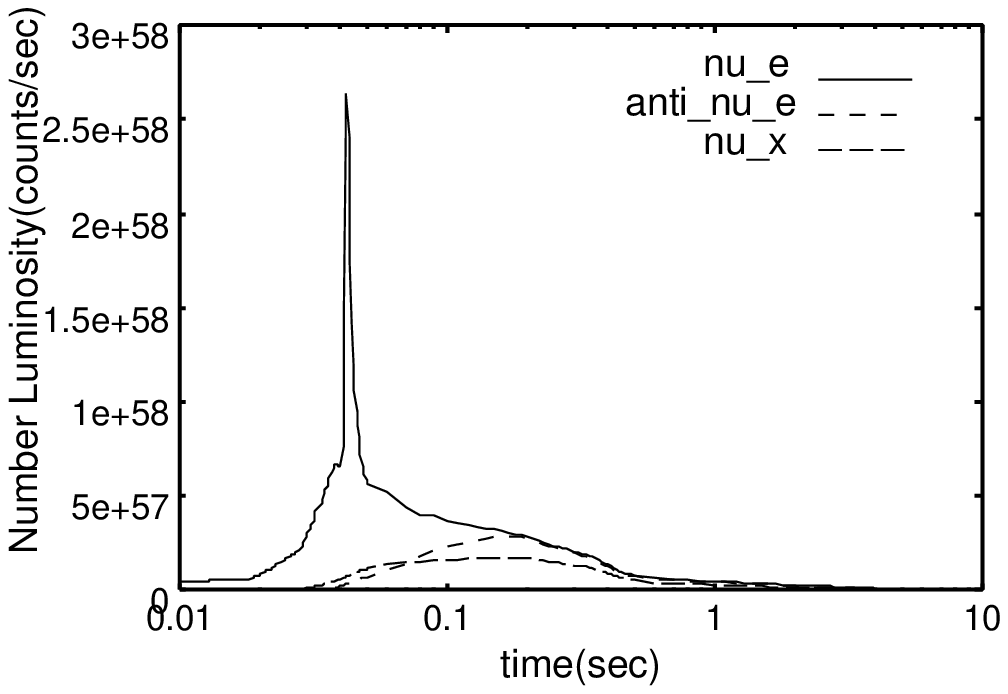}
 \caption{Original evolution of neutrino flux \cite{TotaniSatoDalhedWilson98} used in \cite{KT03}.
 \label{fig:original_time}}
% \end{center}
 \end{minipage}
\end{center}
\end{figure}

\begin{figure}[hbt]
\begin{center}
\epsfxsize = 16 cm
\epsfbox{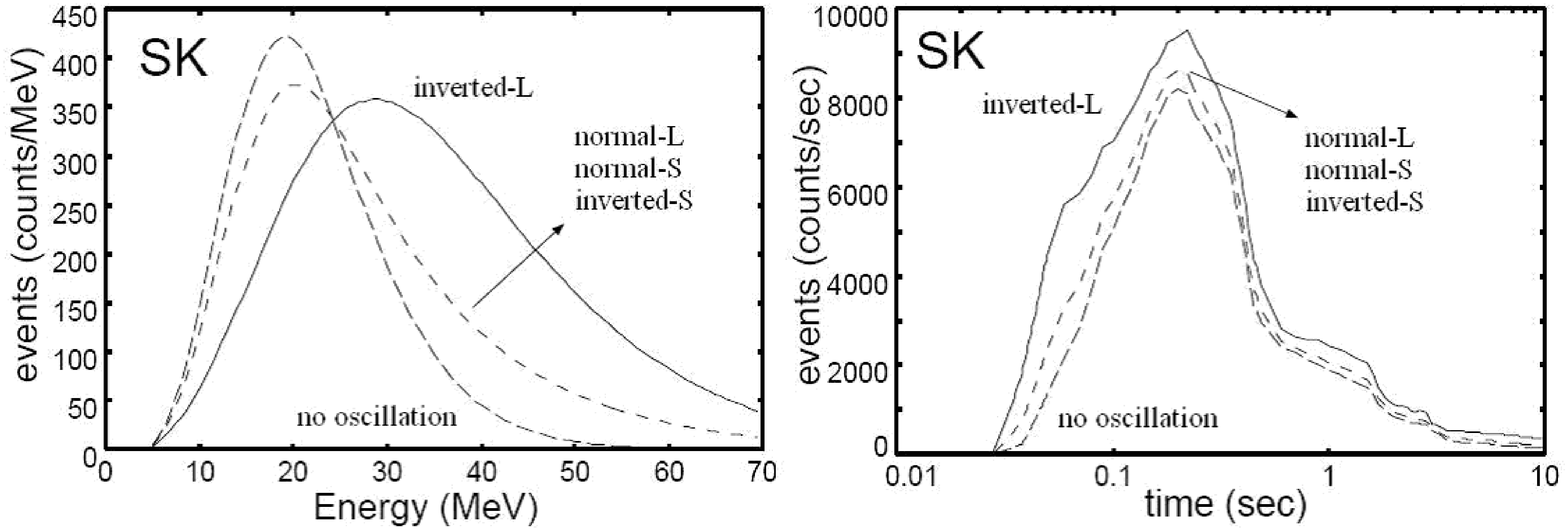}
\end{center}
\vspace{-0.8cm}
\caption{Time-integrated energy spectra (left) and the time evolution of the number of neutrino
events (right) at SK. Only $\bar{\nu}_{e} p$ CC interaction, which is the dominant event at SK,
is taken into account \cite{KT03}.
\label{fig:SK_SN-event}}
\vspace{0.2cm}
\begin{center}
\epsfxsize = 16 cm
\epsfbox{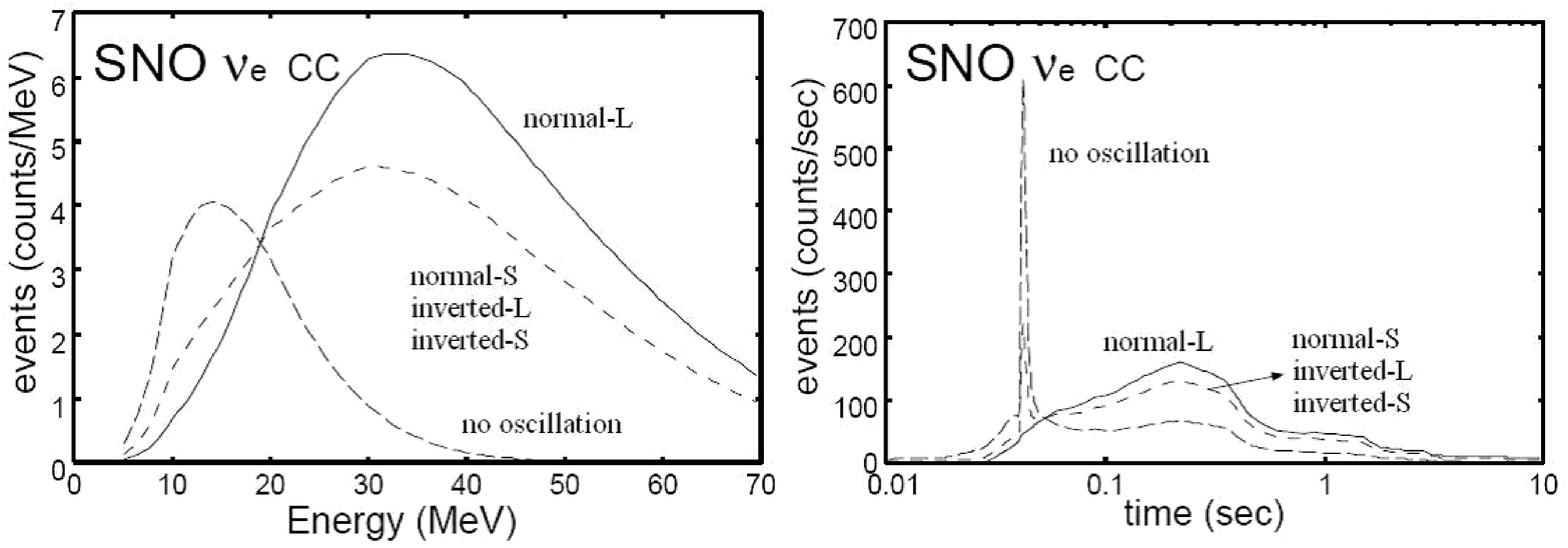}
\end{center}
\vspace{-0.8cm}
\caption{Time-integrated energy spectra (left) and the time evolution of the number of $\nu_{e} d$
events (right) at SNO. Neutronization burst is suppressed for normal-S, inverted-L and inverted-S,
and absent for normal-L \cite{KT03}.
\label{fig:SNO_SN-event}}
\vspace{0.2cm}
\begin{center}
\epsfxsize = 16 cm
\epsfbox{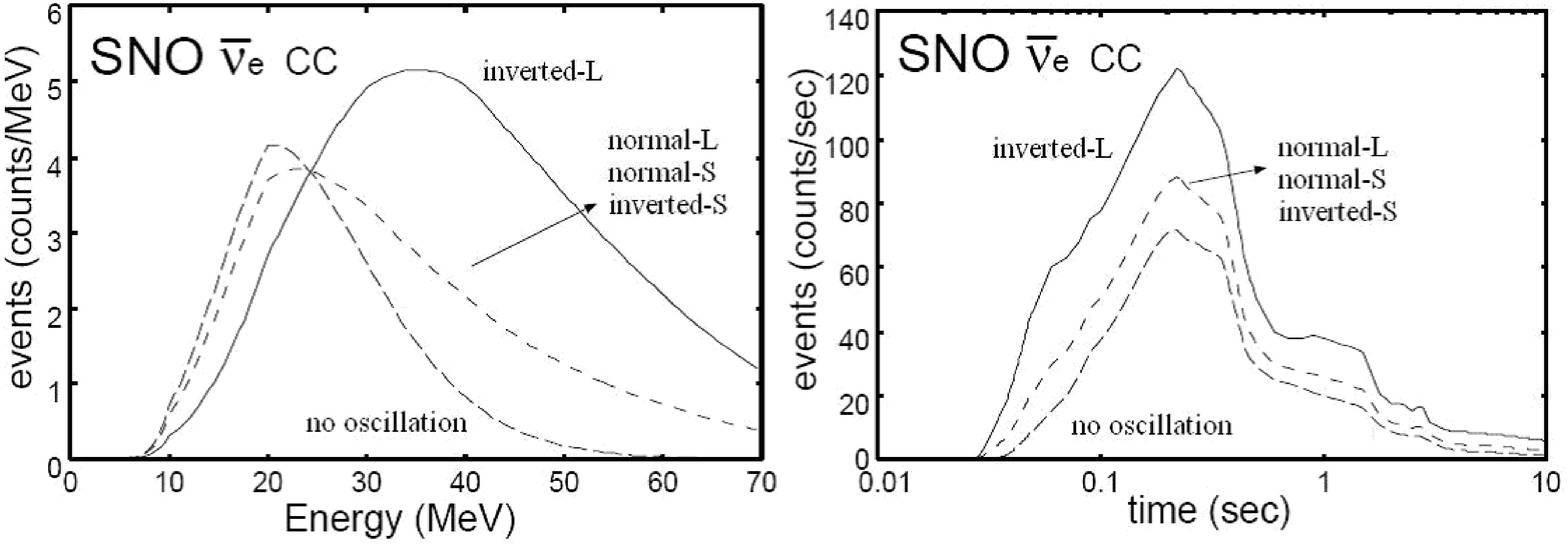}
\end{center}
\vspace{-0.8cm}
\caption{Time-integrated energy spectra (left) and the time evolution of the number of
$\bar{\nu}_{e} d$ events (right) at SNO \cite{KT03}.
\label{fig:SNO_SN-event_anti}}
\end{figure}

With survival probabilities obtained in section \ref{subsubsection:survival_probability},
information of detector, which will be discussed in section \ref{subsection:detector}, and
original neutrino fluxes, we can calculate number of events at the detector. In \cite{KT03},
original neutrino fluxes calculated by a realistic model of a collapse-driven supernova by
the Lawrence Livermore group \cite{WilsonMayleWoosleyWeaver86,TotaniSatoDalhedWilson98}
was used and is shown in Figs. \ref{fig:original_spe} and \ref{fig:original_time}.
Figs. \ref{fig:SK_SN-event}, \ref{fig:SNO_SN-event} and \ref{fig:SNO_SN-event_anti} show
time-integrated energy spectra and time evolution of the number of neutrino events at
SK and SNO. Event number of each interaction is also shown in Tables \ref{table:SK_SN-event}
and \ref{table:SNO_SN-event}. The distance between the earth and the supernova was set to 10 kpc,
which corresponds to the galactic center. As in section \ref{subsubsection:survival_probability},
"normal" and "inverted" represent the mass hierarchy and "L" and "S" mean that $\theta_{13}$
is enough large and small for the H-resonance to be perfectly adiabatic and non-adiabatic,
respectively.

With a supernova at 10 kpc, SK will have an enormous number of events, which is mostly
$\bar{\nu}_{e} p$ events, and make a statistical study possible. SNO will have much smaller
events than SK but it can count the number of $\nu_{e}$s. Thus it is expected that we can
obtain many useful information by combining data from the two detectors. However, at both
detectors, event number of neutronization burst will be rather small.

\begin{table}
\caption{Number of events at SK \label{table:event_SK}
\label{table:SK_SN-event}}
\begin{center}
\begin{tabular}{c|cc|cc|c}
hierarchy & normal & & inverted & & \\
$\theta_{13}$ & large & small & large & small & no osc \\ \hline
$\bar{\nu}_{e}p$ & 9459 & 9427 & 12269 & 9441 & 8036 \\
$\nu_{e} e^{-}$ & 186 & 171 & 171 & 171 & 132 \\
$\bar{\nu}_{e} e^{-}$ & 46 & 46 & 56 & 46 & 42 \\
$\nu_{x} e^{-}$ & 98 & 98 & 77 & 98 \\
$O \nu_{e}$ & 297 & 214 & 297 & 214 & 31 \\
$O \bar{\nu}_{e}$ & 160 & 158 & 296 & 159 & 92 \\
\hline
total & 10245 & 10114 & 13084 & 10129 & 8441  \\ 
burst & 15.7 & 16.7 & 20.1 & 16.7 & 12.4
\end{tabular}
\end{center}
\end{table}

\begin{table}
\caption{Number of events (CC) at SNO \label{table:event_SNO}
\label{table:SNO_SN-event}}
\begin{center}
\begin{tabular}{c|cc|cc|c}
hierarchy & normal & & inverted & & \\
$\theta_{13}$ & large & small & large & small & no osc \\ \hline
$\nu_{e} d {\rm (CC)}$ & 237 & 185 & 185 & 185 & 68 \\ 
$\bar{\nu}_{e} d {\rm (CC)}$ & 118 & 117 & 190 & 117 & 82 \\
\hline 
total & 355 & 302 & 375 & 302 & 150 \\
burst & 0.6 & 1.1 & 1.1 & 1.1 & 2.1
\end{tabular}
\end{center}
\end{table}

\begin{figure}[hbt]
\begin{center}
\epsfxsize = 16 cm
\epsfbox{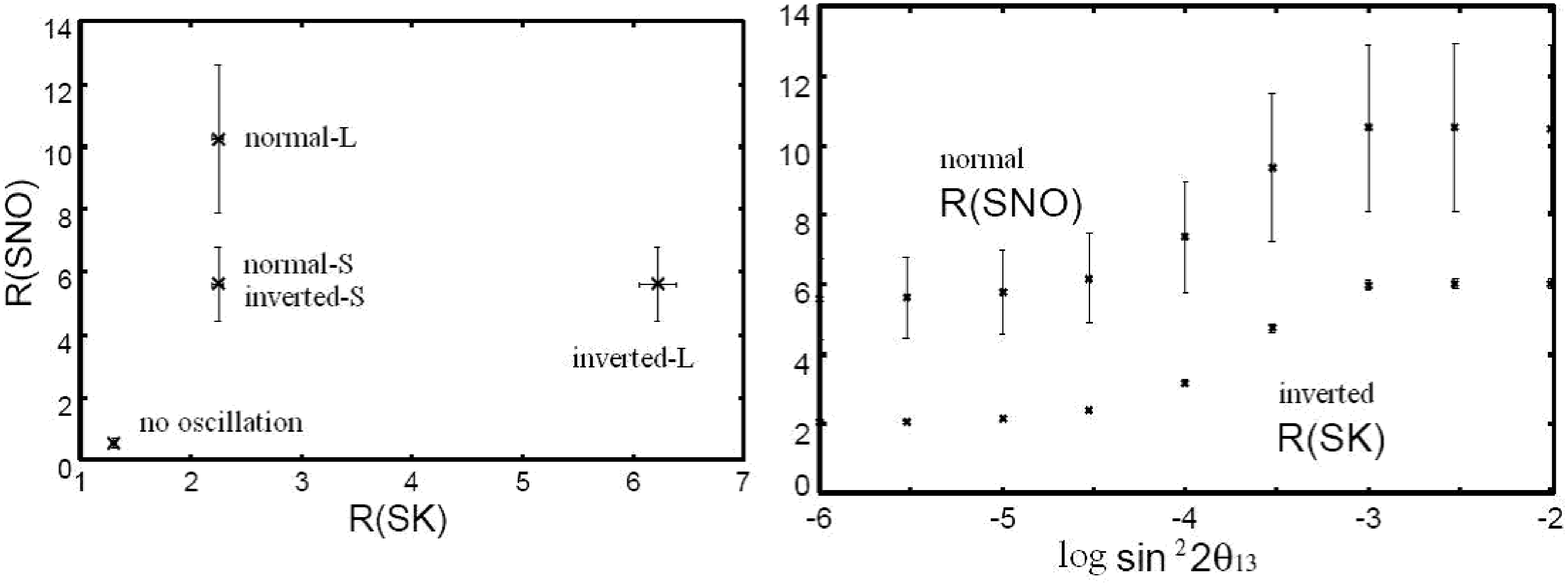}
\end{center}
\vspace{-0.8cm}
\caption{(left) Ratios of high-energy to low-energy events at SK and SNO.
(right) Dependence of the ratios on $\sin^{2}{\theta_{13}}$. In both figures,
only $\nu_{e} d$ events are taken into account for $R(SNO)$ and error bars represent
the statistical errors only \cite{KT03}.
\label{fig:SN-event-ratio}}
\end{figure}

In general, neutrino oscillation makes the $\nu_{e}$ and $\bar{\nu}_{e}$ spectra harder,
since the original average energies of $\nu_{e}$ and $\bar{\nu}_{e}$ are smaller than that of
$\nu_{x}$. In other words, neutrino oscillation produces high energy $\nu_{e}$ and $\bar{\nu}_{e}$
from $\nu_{x}$. As a result, the number of high-energy events increases and that of low-energy events
decreases. The boundary between high energy and low energy is around 20 MeV. Note that the amounts
of these increase and decrease depend on the adiabaticity parameters, and therefore the neutrino
oscillation parameters, as can be seen in Figs. \ref{fig:SK_SN-event}, \ref{fig:SNO_SN-event}
and \ref{fig:SNO_SN-event_anti}. This feature can be used as a criterion for the magnitude of
the neutrino oscillation effects. In \cite{KT03}, the following simple ratios were introduced
as a criterion:
\begin{equation}
R =
\frac{{\rm number \; of \; events \; at} \;
      20 {\rm MeV} < E_{\nu} < 70 {\rm MeV}}
     {{\rm number \; of \; events \; at} \; 5 {\rm MeV} < E_{\nu} < 20 {\rm MeV}}.
\end{equation}
The ratios at SK and SNO are plotted in the left of Fig. \ref{fig:SN-event-ratio}. In this figure,
only $\nu_{e} d$ CC events are taken into account for $R_{\rm SNO}$ and error bars represent
the statistical errors only. The ratios $R_{\rm SK}$ and $R_{\rm SNO}$ can be considered
as estimators of neutrino conversion at neutrino sector and anti-neutrino sector, respectively.
As can be seen, the ratios will give a reasonable implication for the mass hierarchy and the value
of $\theta_{13}$, especially SK is expected to give valuable information on $\theta_{13}$ with
its large event number in case of inverted hierarchy. However the mass hierarchy cannot be
distinguished if $\theta_{13}$ is very small. Note that the $\nu_{e}$ flux and the $\bar{\nu}_{e}$
flux contain essentially different information about the neutrino oscillation parameters.
For example, inverted-L and inverted-S are distinguishable from $\bar{\nu}_{e}$ events,
but they are not distinguishable from $\nu_{e}$ events.

To this point, we have considered only two extreme cases with perfectly adiabatic and
non-adiabatic H-resonances. It is also interesting to investigate the intermediate
cases. In the right of Fig. \ref{fig:SN-event-ratio} the $\theta_{13}$ dependences
of $R_{\rm SK}$ and $R_{\rm SNO}$ are plotted. Here it should be note that $R_{\rm SK}$
and $R_{\rm SNO}$ vary only in the cases of inverted and normal hierarchies, respectively,
as expected from Figs. \ref{fig:SK_SN-event}, \ref{fig:SNO_SN-event} and
\ref{fig:SNO_SN-event_anti}. In the case of the normal hierarchy, it would be difficult
to determine the value of $\theta_{13}$, due to the small event number at SNO and large
statistical errors, but it will give a hint whether it is very large or very small.
In the case of the inverted hierarchy, the overlap of the error bars is small even in
the intermediate cases. If $\theta_{13}$ is rather large ($\sin^{2}{2 \theta_{13}} > 10^{-3}$),
these rations will give useful information of the mass hierarchy.

%%%%%%%%%%%%%%%%%%%%%%%%%%%%%%%%%%%%%%
\subsubsection{Earth effects}
%%%%%%%%%%%%%%%%%%%%%%%%%%%%%%%%%%%%%%

\begin{figure}[hbt]
\begin{center}
 \begin{minipage}{7.0cm}
 \begin{center}
 \epsfbox{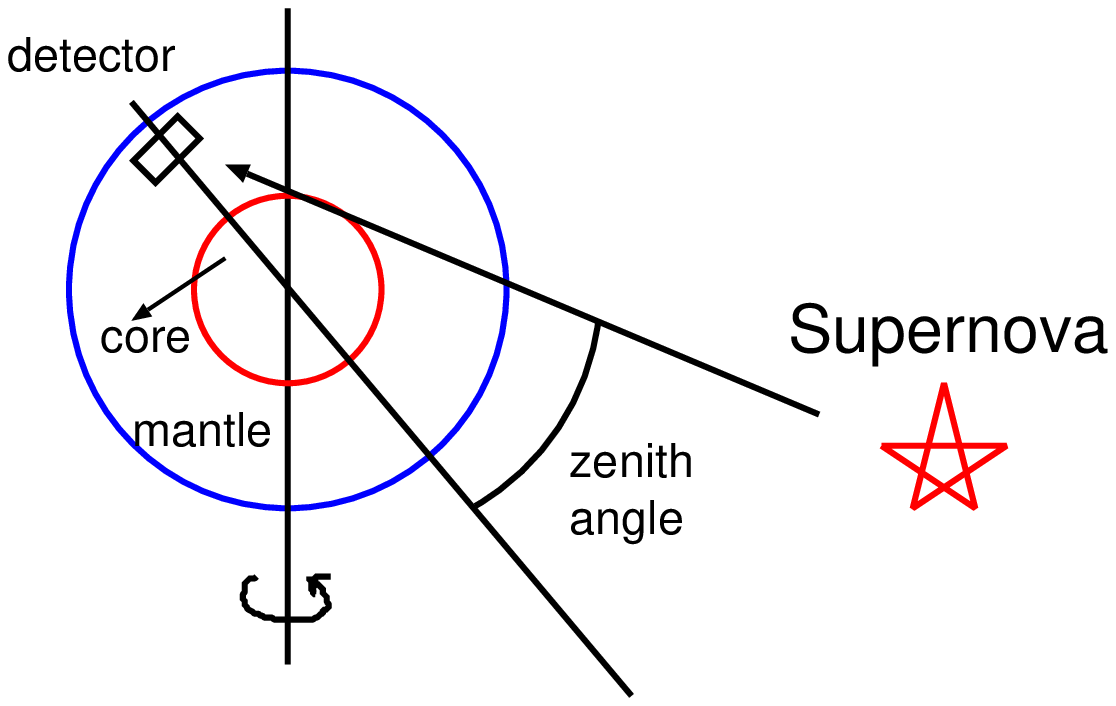}
 \caption{Neutrinos passing through the earth before reaching a detector.
 \label{fig:configuration}}
 \end{center}
 \end{minipage}
\vspace{0.5cm}  
 \begin{minipage}{9.0cm}
 \begin{center}
 \epsfbox{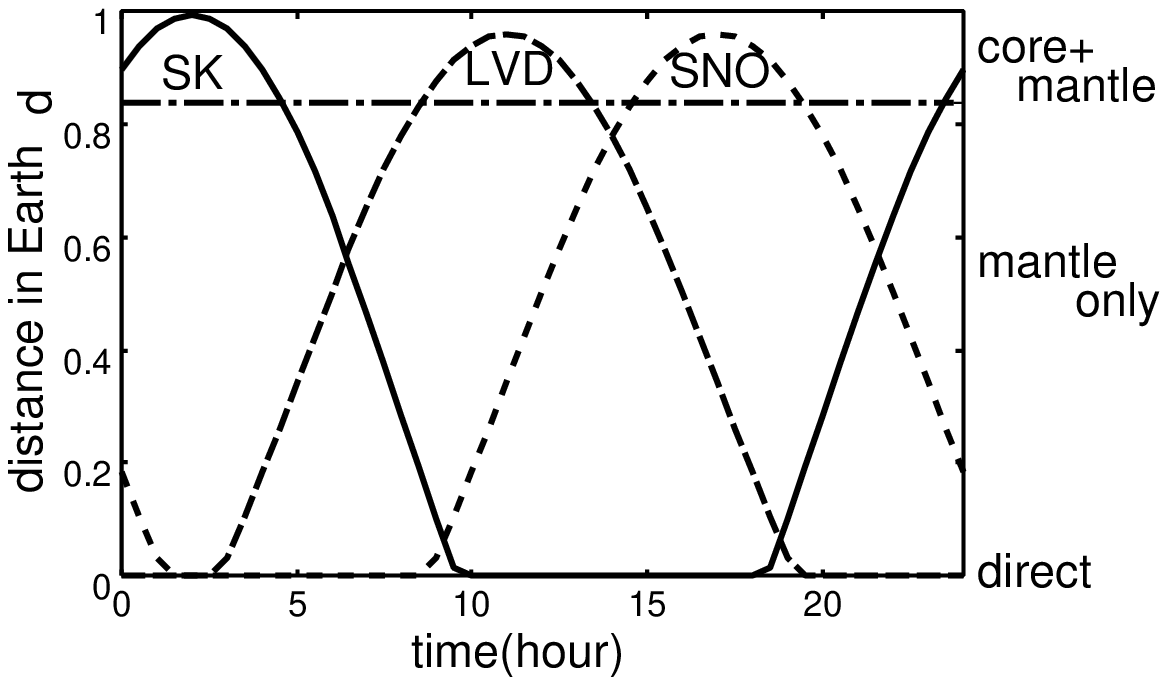}
 \caption{Path lengths in the earth of neutrinos from the galactic center to
 various detectors as functions of the time of the day. The distance in the earth
 is normalized by the earth diameter.
 \label{fig:distance_earth}}
 \end{center}
 \end{minipage}
\end{center}
\end{figure}

We have discussed neutrino oscillation in supernova and its effects on neutrino signal
at detectors. Here we will consider matter effect inside the earth. Actually, neutrinos can
pass through the earth before reaching the detector depending on the configuration of
the supernova, the earth and the detector (Fig. \ref{fig:configuration}). 
Fig. \ref{fig:distance_earth} shows path lengths in the earth of neutrinos from the galactic
center to various detectors as functions of the time of the day. The distance in the earth
is normalized by the earth diameter. As one can see, at any time of the day, at least
one detector among SK (Japan), SNO (Canada) and LVD (Italy) observes neutrinos which have
experienced the earth matter.

The earth matter can have extra effects on the supernova neutrino spectra and
observation/non-observation of them can give us further information on neutrino
oscillation parameters and original neutrino fluxes. First we give a general discussion
on the earth matter effect following again \cite{DigheSmirnov00}.

\begin{figure}[hbt]
\begin{center}
 \begin{minipage}{7.0cm}
 \begin{center}
\epsfxsize = 6.5 cm
 \epsfbox{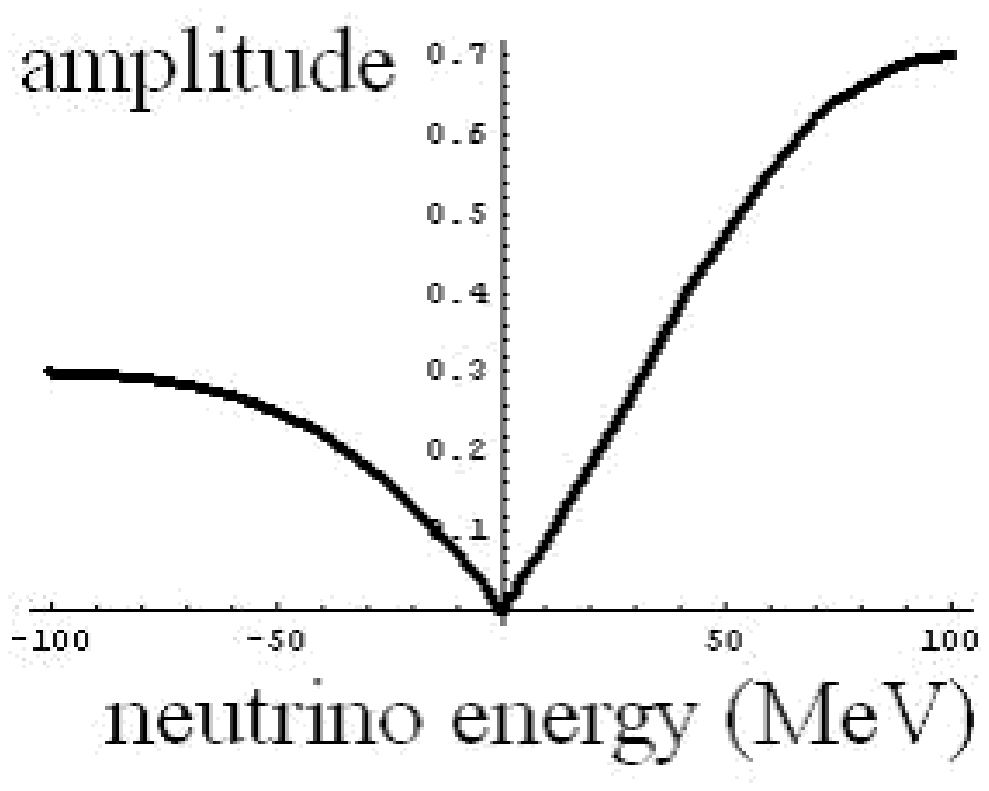}
 \caption{Amplitude of the earth matter effect
 $\xi \sin^{2}{2 \theta_{12}}/[(\xi - \cos{2 \theta_{12}})^{2} + \sin^{2}{2 \theta_{12}}]$
 as a function of $\xi$. Here we set $\sin^{2}{\theta_{12}} = 0.3$. Anti-neutrino sector
 is shown in the negative energy region.
 \label{fig:matter-factor}}
 \end{center}
 \end{minipage}
\hspace{0.5cm}  
 \begin{minipage}{7.0cm}
 \begin{center}
\epsfxsize = 6.5 cm
 \epsfbox{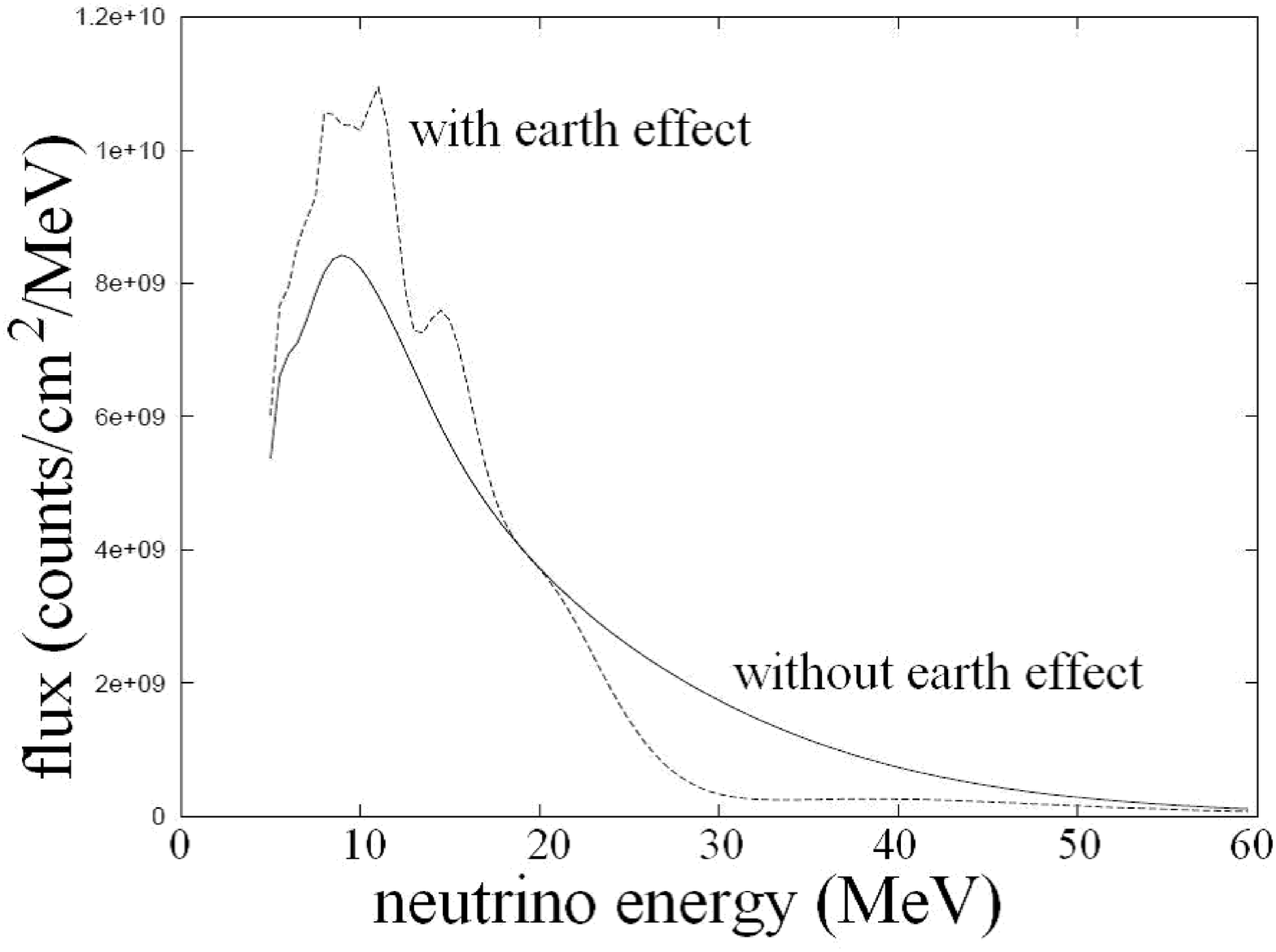}
 \caption{Fluxes of $\nu_{e}$ with the earth effect (dashed line) and without the earth
 effect (solid). Here it is assumed that $\nu_{e}$s propagate 4,000 km in matter with
 the density of the earth core.
 \label{fig:flux_nu-e_earth}}
 \end{center}
 \end{minipage}
\end{center}
\end{figure}

First, we consider the neutrino sector assuming the normal hierarchy. But it is equivalent
for the inverted hierarchy if we set $P_{\rm H} = 0$ and also for the anti-neutrino sector
with the normal and inverted hierarchy if we set $P_{\rm H} = P_{\rm L} = 0$ and
$P_{\rm L} = 0$, respectively. Flux of $\nu_{e}$ at the surface of the earth, $F_{e}$,
is written as Eq. (\ref{eq:flux_e_f0}),
\begin{equation}
F_{e} = \sum_{i} |U_{ie}|^{2} F_{i} = p F^{0}_{e} + (1 - p) F^{0}_{x},
\end{equation}
where $F_{i}$ is the flux of $i$-th mass eigenstate $\nu_{i}$ and $p$ is the survival
probability of $\nu_{e}$ defined in Eq. (\ref{eq:survival-p_e}). Now let $P_{ie}$ be
the probability that a $\nu_{i}$ entering the earth reaches the detector as a $\nu_{e}$.
Then the flux of $\nu_{e}$ at the detector, $F^{\rm D}_{e}$, is,
\begin{equation}
F^{\rm D}_{e} = \sum_{i} P_{ie} F_{i}.
\end{equation}
Rewriting $F_{i}$s by the original fluxes $F^{0}_{\alpha}$ using Eq. (\ref{eq:flux_i_e0}),
we obtain,
\begin{eqnarray}
F^{\rm D}_{e}
&=& F^{0}_{e} \sum_{i} a_{i} P_{ie}
    + F^{0}_{x} \left( 1 - \sum_{i} a_{i} P_{ie} \right) \nonumber \\
&=& p^{\rm D} F^{0}_{e} + (1 - p^{\rm D}) F^{0}_{x},
\end{eqnarray}
where we have used the unitarity condition $\sum_{i} P_{ie} = 1$ in the first line
and,
\begin{equation}
p^{\rm D} \equiv \sum_{i} a_{i} P_{ie},
\end{equation}
is the effective survival probability of $\nu_{e}$ at the detector.
The earth effect can be expressed by the difference between $F_{e}$ and $F^{\rm D}_{e}$:
\begin{equation}
F^{\rm D}_{e} - F_{e} = (p^{\rm D} - p) \left( F^{0}_{e} - F^{0}_{x} \right).
\end{equation}
The difference of the survival probabilities are computed as,
\begin{eqnarray}
p^{\rm D} - p
&=& \sum_{i} a_{i} (P_{ie} - |U_{ie}|^{2}) \nonumber \\
&=& (1 - 2P_{\rm L}) P_{\rm H} (P_{2e} - |U_{2e}|^{2})
    + (1 - P_{\rm H} - P_{\rm L} P_{\rm H}) (P_{3e} - |U_{3e}|^{2}),
\end{eqnarray}
where $\sum_{i} P_{ie} =1$ was used. Because the earth density is rather low and the mixing
angle $\theta_{13}$ is known to be small, we can neglect the second term. Thus we have,
\begin{equation}
F^{\rm D}_{e} - F_{e} =  (1 - 2P_{\rm L}) P_{\rm H} (P_{2e} - |U_{2e}|^{2})
                         \left( F^{0}_{e} - F^{0}_{x} \right).
\end{equation}
Now let us evaluate the factor $(P_{2e} - |U_{2e}|^{2})$. For simplicity, we assume
a constant density and two-flavor oscillation. Remembering that the flavor eigenstates
can be written by both vacuum mass eigenstates and eigenstates in matter, that is,
\begin{equation}
\left(\begin{array}{c} \nu_{e} \\ \nu_{\mu} \end{array}\right)
= U \left(\begin{array}{c} \nu_{1} \\ \nu_{2} \end{array}\right)
= U_{\rm m} \left(\begin{array}{c} \nu_{1, {\rm m}} \\ \nu_{2, {\rm m}} 
\end{array}\right),
\end{equation}
where $U$ and $U_{\rm m}$ are mixing matrix in vacuum and in matter, respectively, we have
\begin{eqnarray}
\left(\begin{array}{c} \nu_{1} \\ \nu_{2} \end{array}\right) & = &
U^{-1} U_{\rm m} \left(\begin{array}{c} \nu_{1, {\rm m}} \\ \nu_{2, {\rm m}} 
\end{array}\right) \nonumber \\
& = & 
\left(\begin{array}{cc}
\cos{(\theta_{12,{\rm m}} - \theta_{12})} & \sin{(\theta_{12,{\rm m}} - \theta_{12})} \\
- \sin{(\theta_{12,{\rm m}} - \theta_{12})} & \cos{(\theta_{12,{\rm m}} - \theta_{12})}
\end{array}\right)
\left(\begin{array}{c} \nu_{1, {\rm m}} \\ \nu_{2, {\rm m}} 
\end{array}\right).
\end{eqnarray}
Now consider a neutrino which is initially a $\nu_{2}$ and enter the earth.
The wave function evolves as,
\begin{eqnarray}
\nu_{2}(z)
&=& - \sin{(\theta_{12,{\rm m}} - \theta_{12})}
      \exp{\left( -i \frac{m_{1,{\rm m}}^{2}}{2 E} z \right)} \nu_{1,{\rm m}} \nonumber \\
& &    + \cos{(\theta_{12,{\rm m}} - \theta_{12})}
      \exp{\left( -i \frac{m_{2,{\rm m}}^{2}}{2 E} z \right)} \nu_{2,{\rm m}} \nonumber \\
&=& - \exp{\left( -i \frac{m_{1,{\rm m}}^{2}}{2 E} z \right)}\times \nonumber \\
& &      \left[ \sin{(\theta_{12,{\rm m}} - \theta_{12})} \cos{\theta_{12,{\rm m}}}
             - \cos{(\theta_{12,{\rm m}} - \theta_{12})} \sin{\theta_{12,{\rm m}}}
               \exp{\left( -i \frac{\Delta m_{\rm m}^{2}}{2 E} z \right)}
      \right] \nu_{e} \nonumber \\
& & + \exp{\left( -i \frac{m_{1,{\rm m}}^{2}}{2 E} z \right)}\times \nonumber \\ & &      \left[ \sin{(\theta_{12,{\rm m}} - \theta_{12})} \sin{\theta_{12,{\rm m}}}
             + \cos{(\theta_{12,{\rm m}} - \theta_{12})} \cos{\theta_{12,{\rm m}}}
               \exp{\left( -i \frac{\Delta m_{\rm m}^{2}}{2 E} z \right)}
      \right] \nu_{\mu}. \nonumber \\
\end{eqnarray}
Then $P_{2e}$, the probability that a $\nu_{2}$ entering the earth is observed
as a $\nu_{e}$ is,
\begin{equation}
P_{2e} = \sin^{2}{\theta_{12}}
         - \frac{\xi \sin^{2}{2 \theta_{12}}}
                {(\xi - \cos{2 \theta_{12}})^{2} + \sin^{2}{2 \theta_{12}}}
           \sin^{2}{\left( \frac{\pi z}{\ell_{\rm osc,m}} \right)},
\end{equation}
where $\xi$ is the dimensionless density parameter defined in Eq. (\ref{eq:density-parameter}),
\begin{eqnarray}
\xi &=& \frac{2\sqrt{2}G_{F}n_{B}E}{\Delta m^{2}} \nonumber \\
    &=& 9.6 \times 10^{-2} \left( \frac{Y_{e} \rho}{5 {\rm g} ~ {\rm cm}^{-3}} \right)
        \left( \frac{E}{10 {\rm MeV}} \right)
        \left( \frac{8 \times 10^{-5} {\rm eV}^{2}}{\Delta m^{2}} \right),
\end{eqnarray}
and $\ell_{\rm osc,m}$ is the oscillation length in matter defined
in Eq. (\ref{eq:osc-length_matter}),
\begin{eqnarray}
\ell_{\rm osc,m}
&=& \frac{\ell_{\rm osc}}{\sqrt{(\xi-\cos{2\theta})^{2} + \sin^{2}{2\theta}}} \nonumber \\
&=& 3.1 \times 10^{2} {\rm km}
    \frac{1}{\sqrt{(\xi-\cos{2\theta})^{2} + \sin^{2}{2\theta}}}
    \left( \frac{E}{10 {\rm MeV}} \right)\times \nonumber \\
& &    \left( \frac{8 \times 10^{-5} {\rm eV}}{\Delta m^{2}} \right).
\end{eqnarray}
Eventually, we have the final expression:
\begin{eqnarray}
F^{\rm D}_{e} - F_{e}
&=&  (1 - 2P_{\rm L}) P_{\rm H}
   \frac{\xi \sin^{2}{2 \theta_{12}}}{(\xi - \cos{2 \theta_{12}})^{2} + \sin^{2}{2 \theta_{12}}}\times\nonumber \\ 
& &   \sin^{2}{\left( \frac{\pi z}{\ell_{\rm osc,m}} \right)}
   \left( F^{0}_{x} - F^{0}_{e} \right).
\label{eq:earth-effect}
\end{eqnarray}
We can easily understand this expression. First of all, there must be a difference between
the fluxes of $\nu_{e}$ and $\nu_{x}$ in order for neutrino oscillation to have any effects.
Since flux of $\nu_{e}$ is larger and smaller than that of $\nu_{x}$ at lower and higher energies,
respectively, the earth effect is expected to change sign at some critical energy where
$F_{e}^{0}(E_{\rm crit}) = F_{x}^{0}(E_{\rm crit})$. This critical energy is about 20 MeV
in the simulation of the Livermore group.

Then there must also be a difference between
the fluxes of $\nu_{1}$ and $\nu_{2}$ at the surface of the earth because we have neglected
the contribution from the third generation. This fact is reflected in the factor
$(1 - 2P_{\rm L}) P_{\rm H}$: if $P_{\rm H}=0$, both $\nu_{1}$ and $\nu_{2}$ have
the original spectrum of $\nu_{x}$ and if $P_{\rm H} = 1/2$, both $\nu_{1}$ and $\nu_{2}$
have the same mixture of the original spectra of $\nu_{e}$ and $\nu_{x}$.
Therefore it is expected that if $\theta_{13}$ is so large ($\sin^{2}{2 \theta_{13}} > 10^{-3}$)
that the H-resonance is perfectly adiabatic, there is no earth matter effect on
$\nu_{e}$ and $\bar{\nu}_{e}$ spectra for the normal and inverted hierarchy, respectively.
Therefore, existence of the earth effect is determined by $\theta_{13}$ and the mass hierarchy:
\begin{itemize}
\item small $\theta_{13}$ $\rightarrow$ $\nu_{e}$ and $\bar{\nu}_{e}$
\item large $\theta_{13}$ and normal hierarchy $\rightarrow$ $\bar{\nu}_{e}$
\item large $\theta_{13}$ and inverted hierarchy $\rightarrow$ $\nu_{e}$
\end{itemize}
Thus observation/non-observation of the earth effects will give further information
on neutrino parameters and mass hierarchy.

Finally, the remaining factor represents the magnitude of the earth matter effect which
oscillates with an amplitude
$|\xi \sin^{2}{2 \theta_{12}}/[(\xi - \cos{2 \theta_{12}})^{2} + \sin^{2}{2 \theta_{12}}]|$
and with an oscillation length $\ell_{\rm osc,m}$. The amplitude is plotted in
Fig. \ref{fig:matter-factor} as a function of $\xi$. Anti-neutrino sector is shown the negative
energy region. As one can see, the earth matter effect will be larger for higher energies.
Comparing the neutrino and anti-neutrino sector, it is expected that the earth effect
in the neutrino sector is larger than that in the anti-neutrino sector because
both the above amplitude and the difference of neutrino spectrum compared with $\nu_{x}$
are larger for $\nu_{e}$ than $\bar{\nu}_{e}$.

In Fig. \ref{fig:flux_nu-e_earth}, fluxes of $\nu_{e}$ with and without the earth effect are shown.
Here it is assumed that $\nu_{e}$s propagate 4,000 km in matter with the density of the earth core.
At low and high energies, $\nu_{e}$ flux increases and decreases, respectively, due to
the earth effect. As is expected, the earth effect vanishes at the critical energy.
Note also that there is a modulation in the spectrum, which is a unique feature of the earth effect
and comes from the factor $\sin^{2}{(\pi z/\ell_{\rm osc,m})}$ in Eq. (\ref{eq:earth-effect}).
This feature will be further discussed in the next section.

%%%%%%%%%%%%%%%%%%%%%%%%%%%%%%%%%%%%%%
\subsubsection{detection and implication from earth effect}
%%%%%%%%%%%%%%%%%%%%%%%%%%%%%%%%%%%%%%
\begin{figure}[hbt]
\begin{center}
\epsfbox{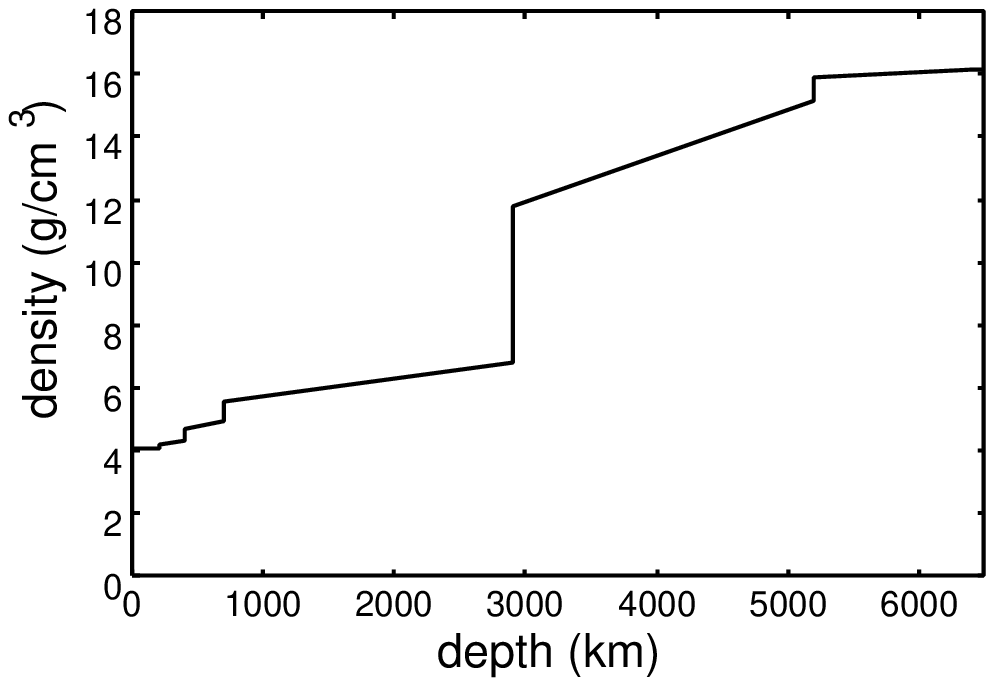}
\end{center}
\vspace{-0.8cm}
\caption{Density profile of the earth \cite{DziewonskiAnderson81}.
\label{fig:earth_density}}
\end{figure}

\begin{figure}[hbt]
\begin{center}
 \begin{minipage}{7.0cm}
 \begin{center}
\epsfxsize = 6cm
 \epsfbox{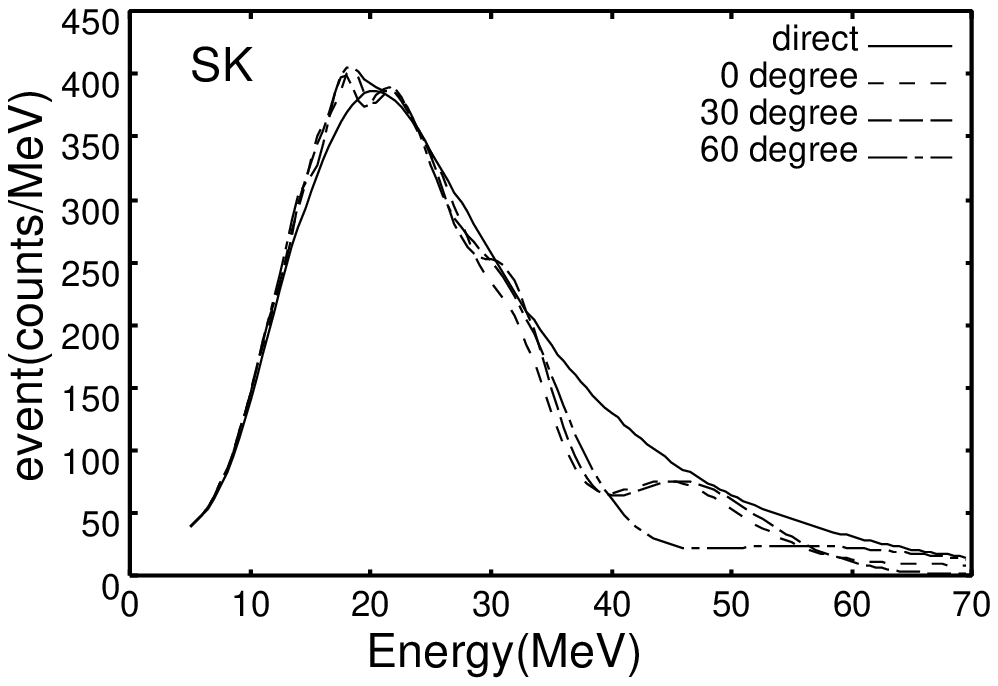}
 \vspace{0.5cm}
 \end{center}
 \end{minipage}
\hspace{0.5cm}  
 \begin{minipage}{7.0cm}
 \begin{center}
\epsfxsize = 6cm
 \epsfbox{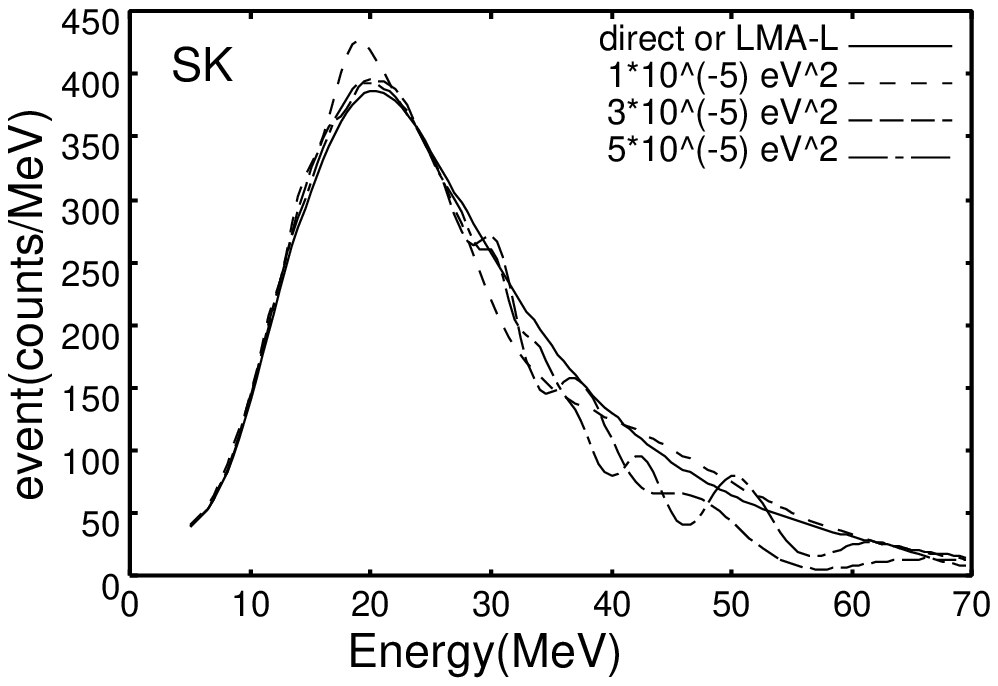}
 \vspace{0.5cm}
 \end{center}
 \end{minipage}
\end{center}
\vspace{0.5cm}
\begin{center}
 \begin{minipage}{7.0cm}
 \begin{center}
\epsfxsize = 6cm
 \epsfbox{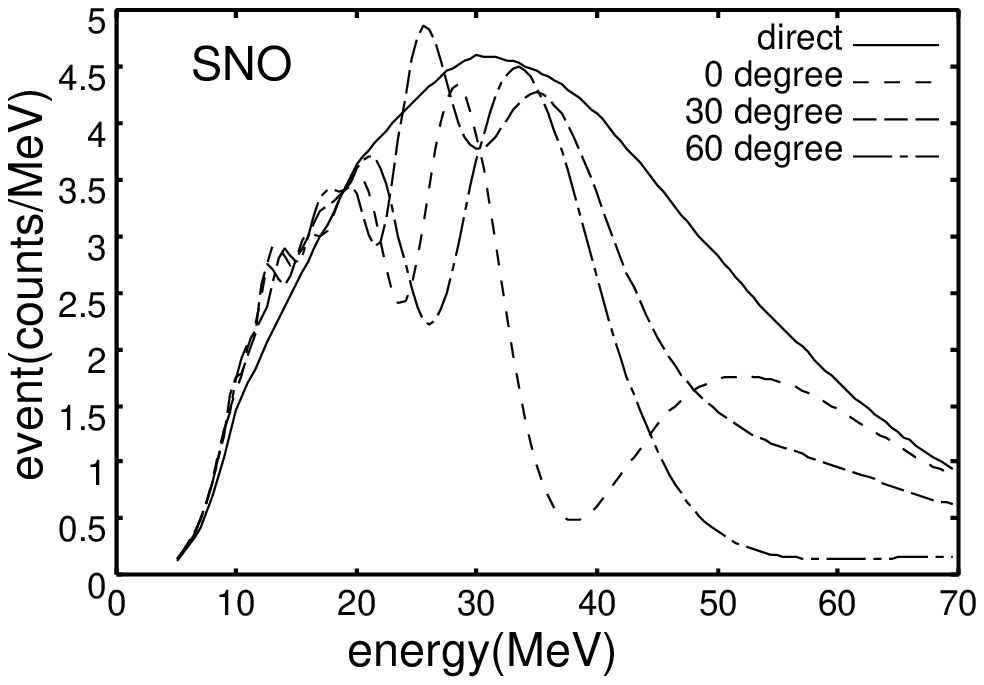}
 \vspace{0.5cm}
 \end{center}
 \end{minipage}
\hspace{0.5cm}  
 \begin{minipage}{7.0cm}
 \begin{center}
\epsfxsize = 6cm
 \epsfbox{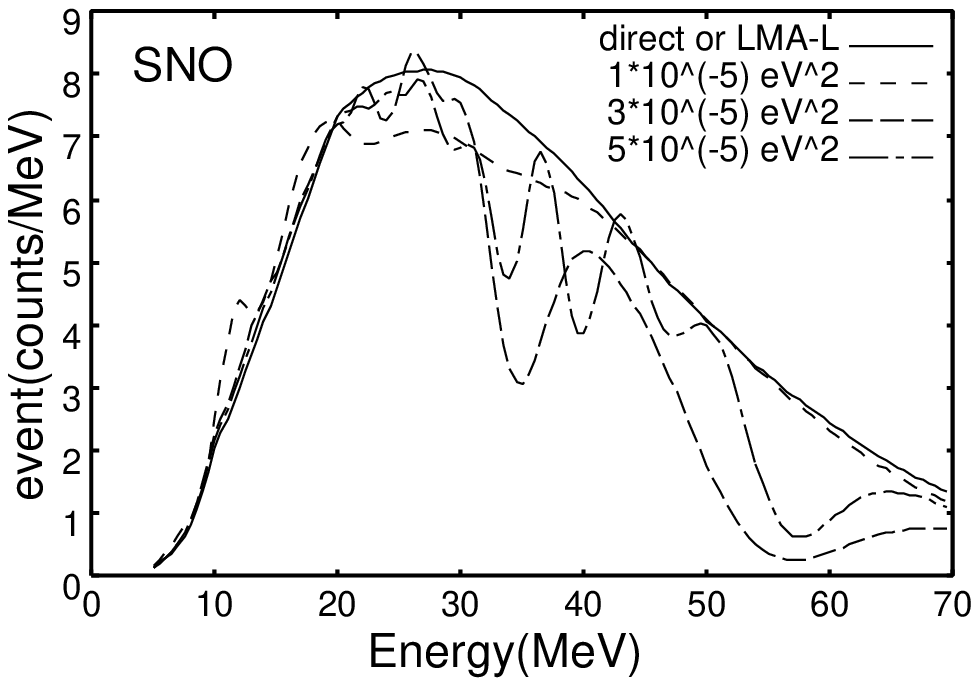}
 \vspace{0.5cm}
 \end{center}
 \end{minipage}
\end{center}
\vspace{-0.5cm}
\caption{Event spectra at SK and SNO with earth effect with some values of zenith angle
and $\Delta m_{12}^{2}$ \cite{KTearth02}. In the left row, $\Delta m_{12}^{2}$ is fixed
to $2 \times 10^{-5} {\rm eV}^{2}$ and, in the right row, zenith angle is fixed to zero.
Only charged-current events are considered in the SNO spectra. The mixing angle $\theta_{13}$
is taken sufficiently small that the H-resonance is perfectly non-adiabatic.
\label{fig:earth_angle_mass}}
\end{figure}

Here we discuss the earth effects more in detail. Because the oscillation length is
almost the same order of the earth radius, the earth effects depend highly on
the path length of neutrinos inside the earth. The path length is determined by
the position of detector on the earth, the direction of supernova and the time of
the day. As we saw in Fig. \ref{fig:distance_earth}, at least one detector among SK,
SNO and LVD will observe neutrinos with the earth effect and at least one detector
will observe neutrinos without the earth effect at anytime if a supernova occurs
at the galactic center.

If we can see a supernova optically, its direction can be determined accurately.
However, it is likely that the light from supernova is obscured by interstellar dust
if the supernova occurs around the galactic center. Methods of direction determination
by neutrinos are suggested and studied by several authors. One strategy is to use
electron scattering events \cite{BeacomFogel99}. As we will see in section
\ref{subsection:detector}, scattered electron in electron scattering event reflects
the direction of the incident neutrino. In \cite{AndoSato02}, the accuracy of the direction
determination was estimated to be about 9 degree using expected event data at SuperKamiokande.
Much better accuracy about $0.6^{o}$ is expected if a megaton water Cherenkov detector
is available \cite{Tomas03}. Another strategy is triangulation using time delay of
neutrino signals at different neutrino detectors \cite{Burrows92}. However, given
the expected statistics, this method was shown to be not so effective \cite{BeacomVogel99}.

As we discussed in the previous section, $\theta_{13}$ and the mass hierarchy determines
the existence of the earth effects. Further, since the oscillation length depends on
$\Delta m_{12}^{2}$, the spectral modulation due to the earth effects will be highly
dependent on the value of $\Delta m_{12}^{2}$. In \cite{KTearth02}, the dependence of
the earth effects on $\Delta m_{12}^{2}$ and the path length of neutrinos inside the earth
was studied in detail. The computational method is essentially the same as that without
the earth effects: the Schr\"odinger equation (\ref{eq:Schrodinger}) needs to be solved
along the density profile of the earth. The standard density profile of the earth is
shown in Fig. \ref{fig:earth_density} \cite{DziewonskiAnderson81}.

Fig. \ref{fig:earth_angle_mass} shows event spectra at SK and SNO with earth effect
with some values of zenith angle and $\Delta m_{12}^{2}$ \cite{KTearth02}. In the left row,
$\Delta m_{12}^{2}$ is fixed to $2 \times 10^{-5} {\rm eV}^{2}$ and, in the right row,
zenith angle is fixed to zero. It is seen that the spectral shape varies with the zenith angle
and $\Delta m_{12}^{2}$. Since the oscillation length is shorter for larger $\Delta m_{12}^{2}$,
the frequency of spectral modulation with respect to neutrino energy is larger for
larger $\Delta m_{12}^{2}$. Thus, two detectors located at separate cites observe neutrinos
with different energy spectra because their path lengths in the earth are different in general.
Also, one can see that the earth effect at SK is smaller than that at SNO. This is because
the dominant events at SK are $\bar{\nu}_{e}$ while both $\nu_{e}$ and $\bar{\nu}_{e}$
contribute to the signal at SNO.

Thus the detection of the earth effect will give us information on $\theta_{13}$,
the mass hierarchy and $\Delta m_{12}^{2}$. There are basically two strategies for detecting
the earth effect. One is to compare the neutrino fluxes at two or more sites and another is
to identify the spectral modulation mentioned above. Because the earth effect is a rather small
effect, the former strategy needs large detectors. On the other hand, the latter strategy needs
only one detector although it must have a sufficient energy resolution to identify
the spectral modulation.
\begin{figure}[hbt]
\begin{center}
 \begin{minipage}{7.cm}
 \begin{center}
\epsfxsize = 6cm
 \epsfbox{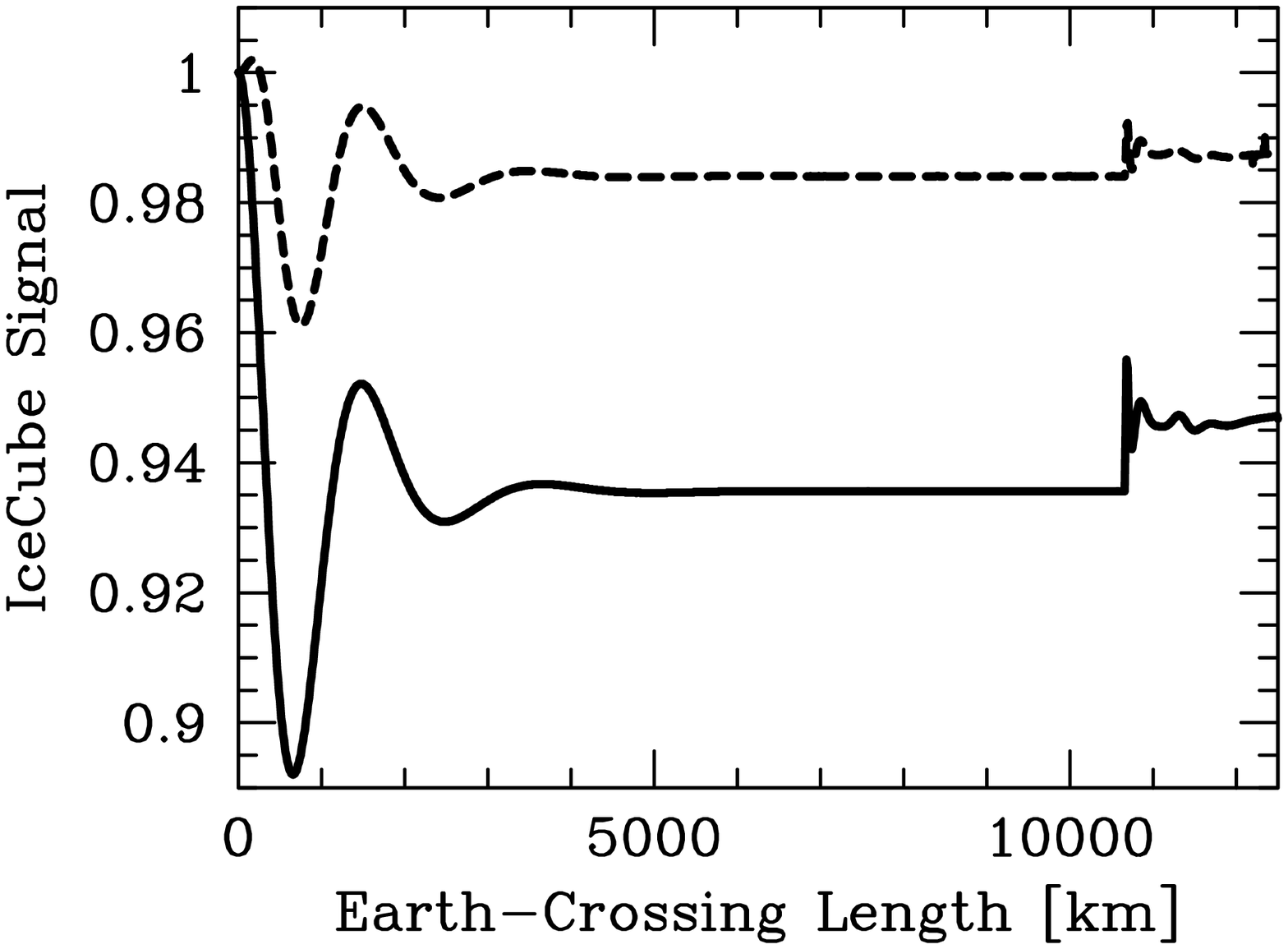}
 \caption{Variation of the expected IceCube signal with neutrino earth-crossing length
 \cite{DigheKeilRaffelt03a}. The signal is normalized to unity when no earth effect is present
 ($L=0$). The solid and dashed lines correspond to the accretion phase and cooling phase,
 respectively.
 \label{fig:earth-effect_IceCube}}
 \end{center}
 \end{minipage}
\hspace{0.5cm}  
 \begin{minipage}{7.cm}
 \begin{center}
\epsfxsize = 6cm
 \epsfbox{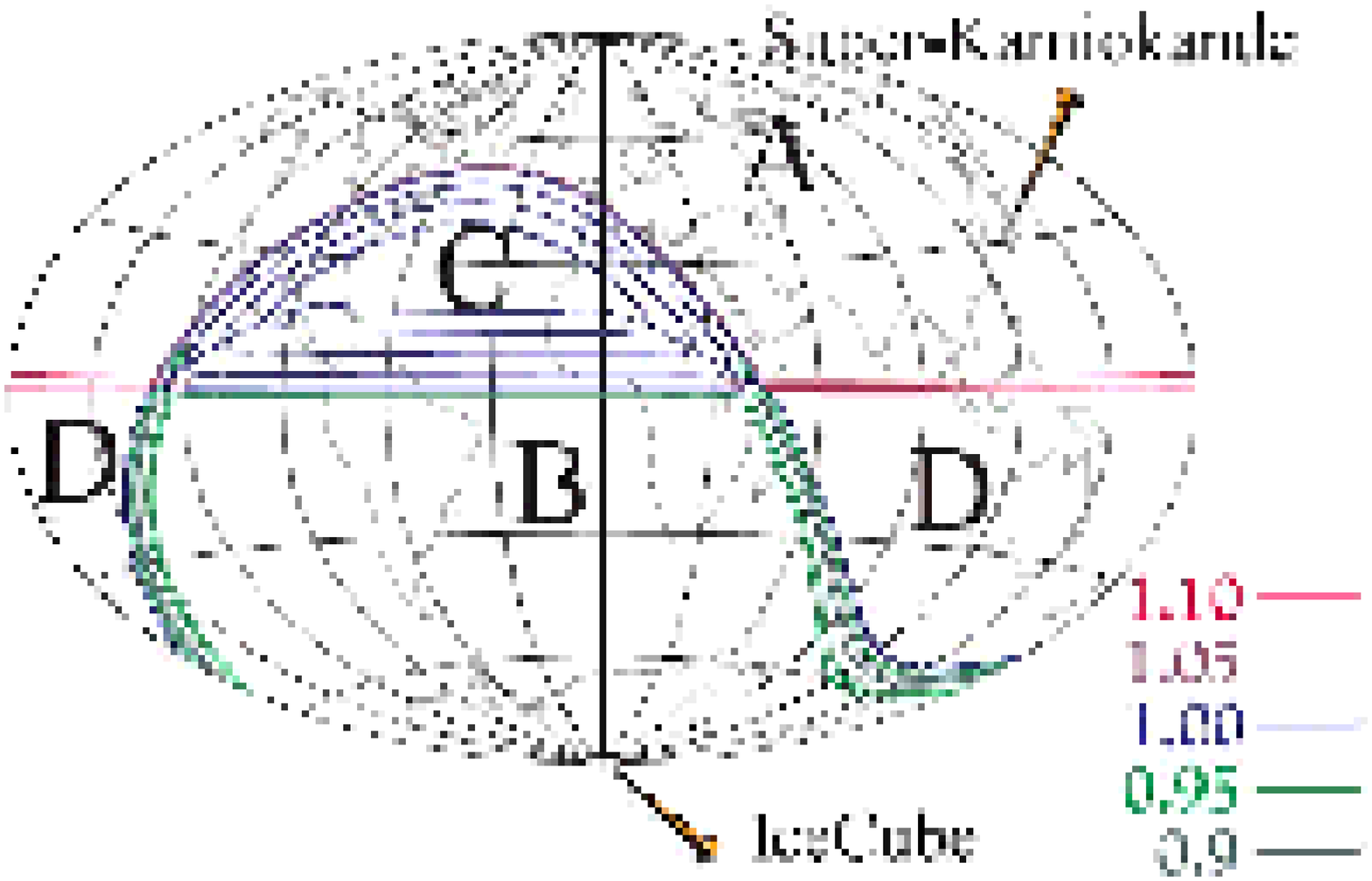}
 \caption{Contours of $N_{\rm SK}/N_{\rm IC}$ on the map of the sky projected on the earth
 \cite{DigheKeilRaffelt03a}
 \label{fig:earth-effect_sky}}.
 \end{center}
 \end{minipage}
\end{center}
\end{figure}

\begin{table}[t]
\begin{center}
\caption{Regions in Fig. \ref{fig:earth-effect_sky} for the earth effect
in IceCube and Super-Kamiokande.
\label{table:earth-effect_sky}}
\begin{tabular}{@{}lllll}
\hline
Region&Sky fraction&
\multicolumn{2}{l}{Neutrinos come from}&$N_{\rm SK}/N_{\rm IC}$\\
&&IceCube&Super-K&\\
\hline
A&0.35&below&above&1.070\\
B&0.35&above&below&0.935\\
C&0.15&below&below&Fluctuations around 1 \\
D&0.15&above&above&1\\
\hline
\end{tabular}
\end{center}
\end{table}

Comparison of neutrino flux at two detector was considered in \cite{DigheKeilRaffelt03a}.
This method is simple but the detectors must be enough large to accomplish statistically
significant detection of the earth effect. They considered IceCube and SuperKamiokande,
both of which are Cherenkov detectors. Because the dominant events come from $\bar{\nu}_{e}$
at both the detectors, the earth effects are relatively small.
Fig. \ref{fig:earth-effect_IceCube} shows the variation of the expected IceCube signal
with neutrino earth-crossing length. The signal is normalized to unity when no earth effect
is present ($L=0$). The solid and dashed lines correspond to the accretion phase and cooling phase,
respectively. Here they used neutrino spectra with,
\begin{equation}
\langle E_{\bar{\nu}_{e}}^{0} \rangle = 15 {\rm MeV}, ~~~
\langle E_{\nu_{x}}^{0} \rangle = 17 {\rm MeV}, ~~~
\frac{F_{\bar{e}}^{0}}{F_{x}^{0}} = 1.5,
\end{equation}
for the accretion phase and,
\begin{equation}
\langle E_{\bar{\nu}_{e}}^{0} \rangle = 15 {\rm MeV}, ~~~
\langle E_{\nu_{x}}^{0} \rangle = 18 {\rm MeV}, ~~~
\frac{F_{\bar{e}}^{0}}{F_{x}^{0}} = 0.8,
\end{equation}
for the cooling phase, and neutrino mixing parameters they used
$\Delta m_{12}^{2} = 6 \times 10^{-5} {\rm eV}^{2}$ and $\sin^{2}{2 \theta_{12}} = 0.9$.

Denoting the number of Cherenkov photon at IceCube as $N_{\rm IC}$ and the equivalent IceCube
signal measured by SK as $N_{\rm SK}$, the contour of the ratio $N_{\rm SK}/N_{\rm IC}$
is shown in Fig. \ref{fig:earth-effect_sky}. The sky can be divided into four regions
according to the direction of neutrinos at the two detectors (Table \ref{table:earth-effect_sky}).
As one can see, deviation of the ratio $N_{\rm SK}/N_{\rm IC}$ from unity is significant
when either IceCube or SK observes neutrinos directly and the other observes neutrinos
from below and the deviation is typically 0.07 in such cases. This deviation is rather large
because, if a supernova explodes at 10 kpc from the earth, the statistical precision for
the total neutrino energy deposition is about $0.2 \%$ for the IceCube and $1 \%$ for the SK.
Further, such ideal cases is realized for about $70 \%$ of the sky. Thus the geographical position
of IceCube with respect to SK at a latitude of $36.4^{o}$ is well-suited for the detection of
the earth effect through a combination of the signals.

The second possibility, making use of the spectral modulation, was pursued 
by Dighe et al. \cite{DigheKeilRaffelt03b,DigheKachelriessRaffeltTomas04}.
Their basic idea is to Fourier-transform the "inverse-energy" spectrum of the signal.
As we saw in the previous section, the spectral modulation due to the earth effect
comes from the following factor in Eq. (\ref{eq:earth-effect}),
\begin{equation}
F^{\rm D}_{e} - F_{e} \propto
\sin^{2}{\left( \frac{\Delta m_{12,{\rm m}}^{2} z}{4 E_{\nu}} \right)}.
\label{eq:earth_osc}
\end{equation}
Therefore the earth effect will have a clear peak in the power spectrum if it is
Fourier-transformed with respect to $E^{-1}$. The position of the peak is determined
by a factor $\Delta m_{12,{\rm m}}^{2} z$, that is, $\Delta m_{12}^{2}$, the density
of the region which neutrinos propagate and the path length. It is important to note that
the peak position is not affected by the primary neutrino spectra so that the value of
$\Delta m_{12}^{2}$ can be determined accurately from the peak position independently of
supernova model if we know the neutrino path length inside the earth.

To illustrate the effectiveness of their method, they consider a 32 kton scintillator detector
and a megaton water Cherenkov detector. The major difference between
the scintillation and the water Cherenkov detector is that the energy resolution
of the scintillation detector is roughly six times better than that of the Cherenkov detector.
Therefore even the volume of the SuperKamiokande will not be sufficient to identify
the earth effect and much larger detector, HyperKamiokande (HK), is needed.

\begin{figure}[hbt]
\vspace{-1cm}
\begin{center}
\epsfxsize = 8 cm
\epsfbox{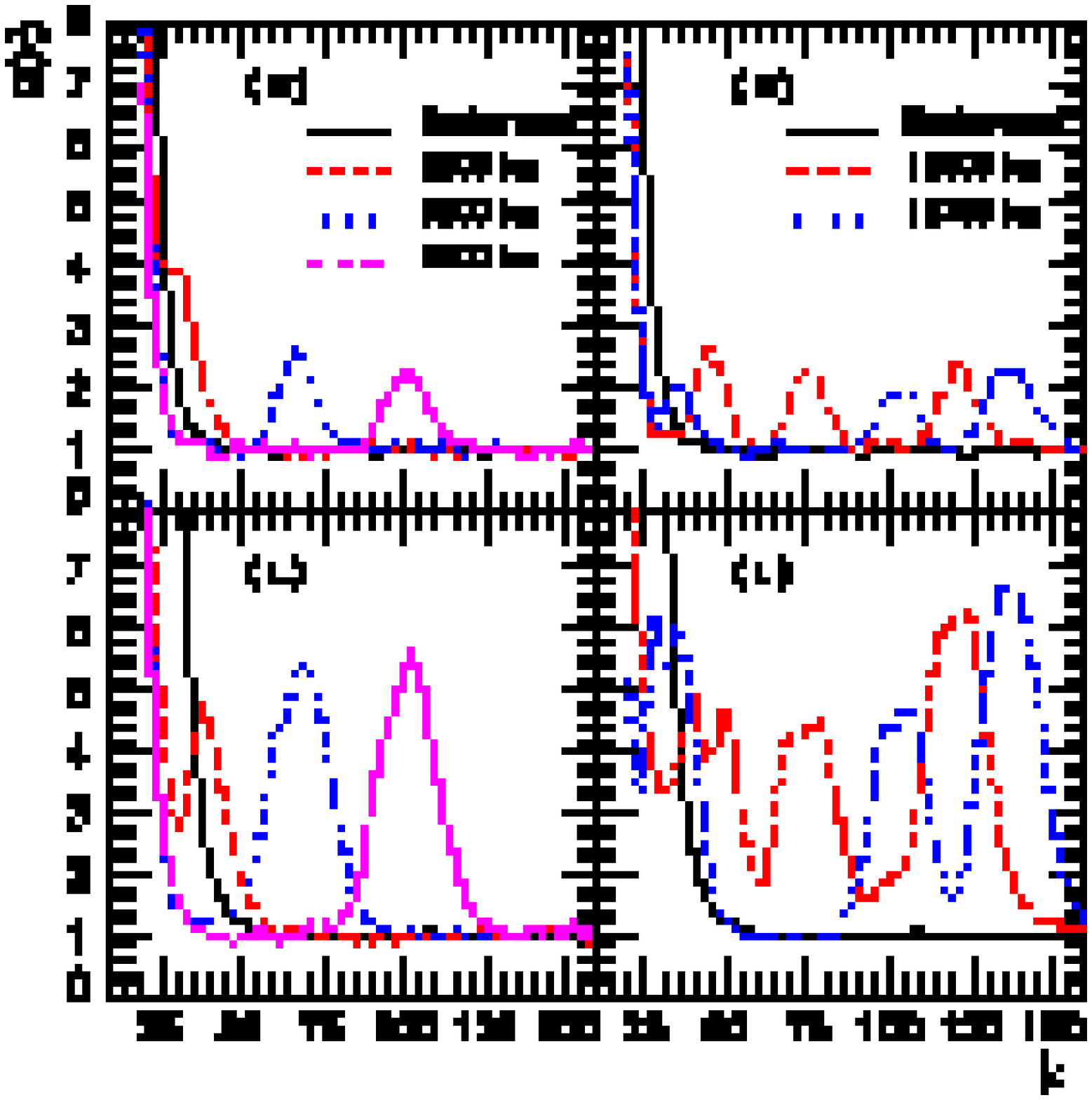}
\end{center}
\vspace{-1.2cm}
\caption{Averaged power spectra in the case of a large scintillator detector
for different SN models, Garching (G) and Livermore (L), and distances traveled through
the earth \cite{DigheKachelriessRaffeltTomas04}.
\label{fig:power-spectrum_SC}}
\vspace{-1cm}
\begin{center}
\epsfxsize = 8 cm
\epsfbox{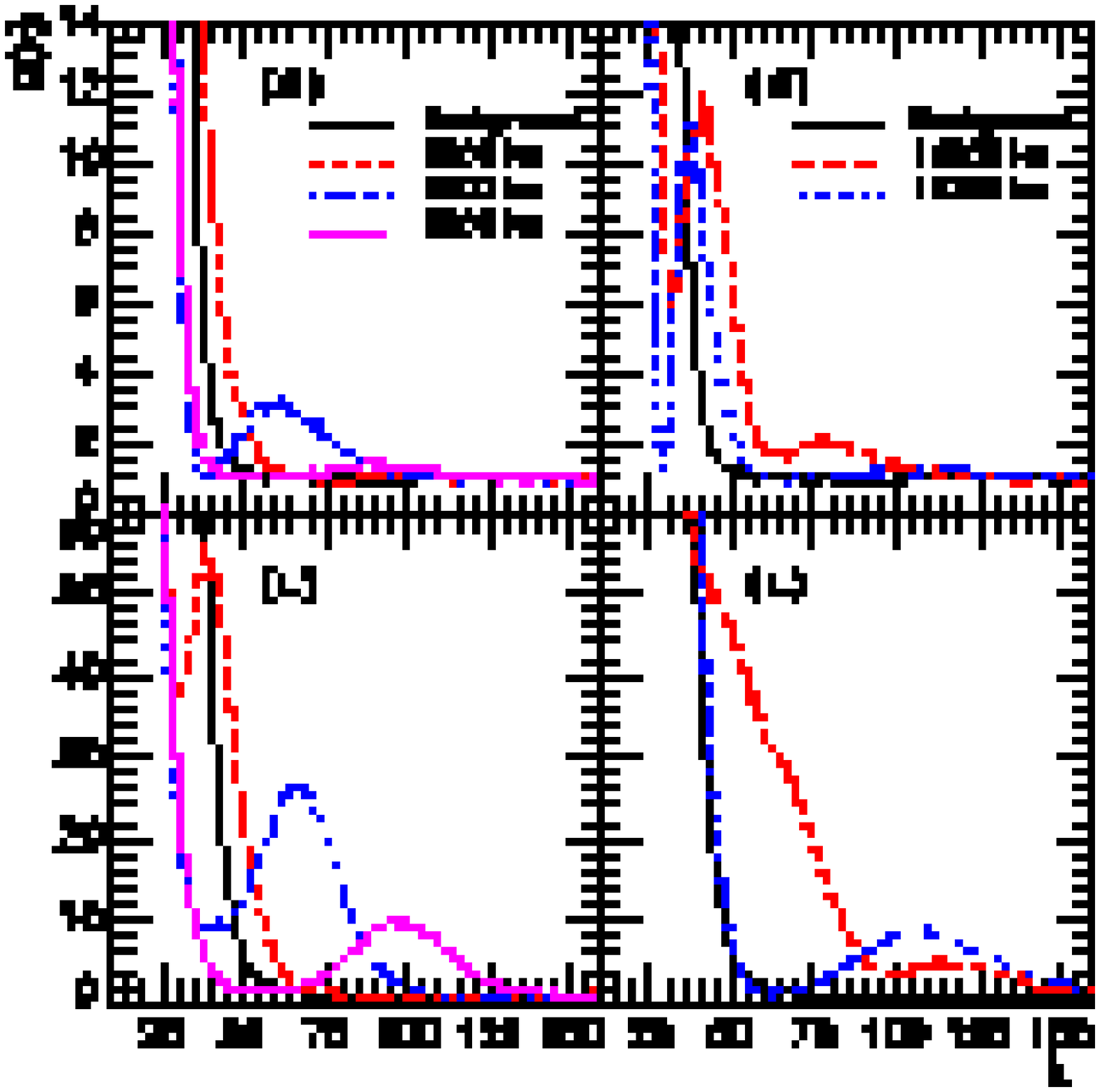}
\end{center}
\vspace{-1.2cm}
\caption{Same as the Fig. \ref{fig:power-spectrum_SC} but for the case of HK
\cite{DigheKachelriessRaffeltTomas04}.
\label{fig:power-spectrum_HK}}
\end{figure}

Figs. \ref{fig:power-spectrum_SC} and \ref{fig:power-spectrum_HK} show expected
power spectra at the scintillator detector and HK, respectively, for different SN models,
Garching (G) and Livermore (L), and distances traveled through the earth. Supernova is
assumed to be 10 kpc from the earth and the power spectra are averaged over 1000 SN simulations. 
The left rows show the power spectra of neutrinos which propagate only in the earth mantle
while the right rows correspond to those which propagate the earth core as well as the mantle.

A clear peak is seen in each figure in the left rows, corresponding to the assumed values
of $\Delta m_{12}^{2}$, the mantle density, the average neutrino energy and the path length.
The peak width reflects the finite energy resolution of the detectors and weak energy dependence
of Eq. (\ref{eq:earth-effect}) besides the factor Eq. (\ref{eq:earth_osc}).
The large power at small $k$ is the dominant contribution from the energy dependence
of the neutrino spectra without the earth effect. On the other hand, if neutrinos propagate
in both the mantle and core, multiple peaks appear as in the right rows of the figures
corresponding to their densities and path lengths in them.

The Livermore model was used as a representative model of conventional simulations
which predict relatively large average-energy differences between flavors. On the other hand,
the Garching model, which treats neutrinos more carefully, predicts rather small average-energy
differences between flavors. As we saw in the previous section, the earth effects as well as
the effects of neutrino oscillation in supernova are larger for a case with larger
average-energy differences. However, as seen in the figure, the peak positions are independent
of the supernova models. 

\clearpage

\begin{figure}[hbt]
\vspace{-0.5cm}
\begin{center}
 \begin{minipage}{7.0cm}
 \begin{center}
\epsfxsize = 6 cm
 \epsfbox{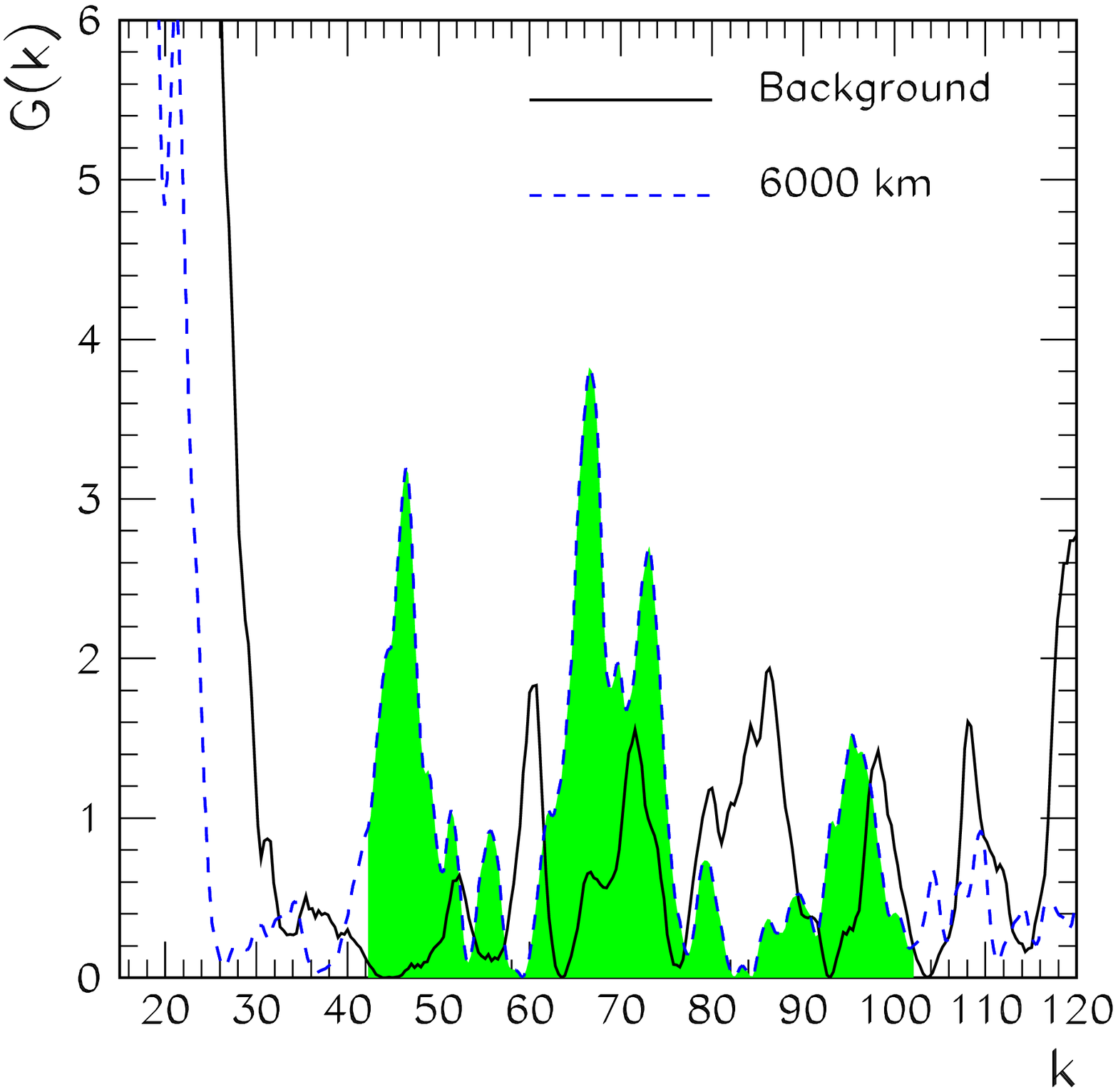}
 \end{center}
 \end{minipage}
\hspace{0.5cm}  
 \begin{minipage}{7.0cm}
 \begin{center}
\epsfxsize = 6 cm
 \epsfbox{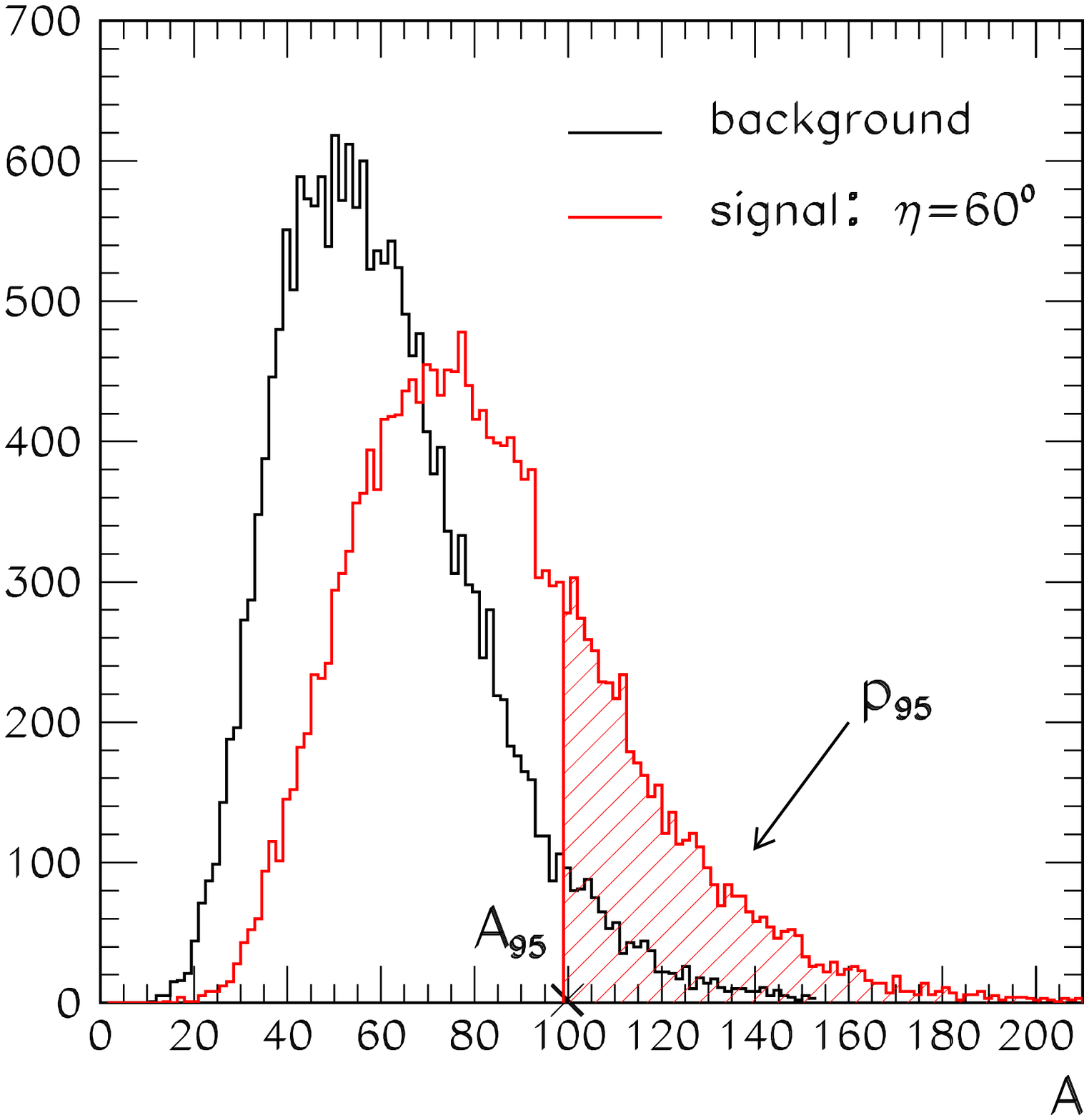}
 \end{center}
 \end{minipage}
\end{center}
\vspace{-0.7cm}
\caption{Left: Realistic spectrum from a single simulation. Right: Area distribution of
the background (black) and the signal (red) obtained for a 32 kton scintillator detector
and Garching model for $\eta=60$ \cite{DigheKachelriessRaffeltTomas04}.
\label{fig:area-distribution}}
\vspace{-0.8cm}
\begin{center}
 \begin{minipage}{7.cm}
 \begin{center}
\epsfxsize = 6 cm
 \epsfbox{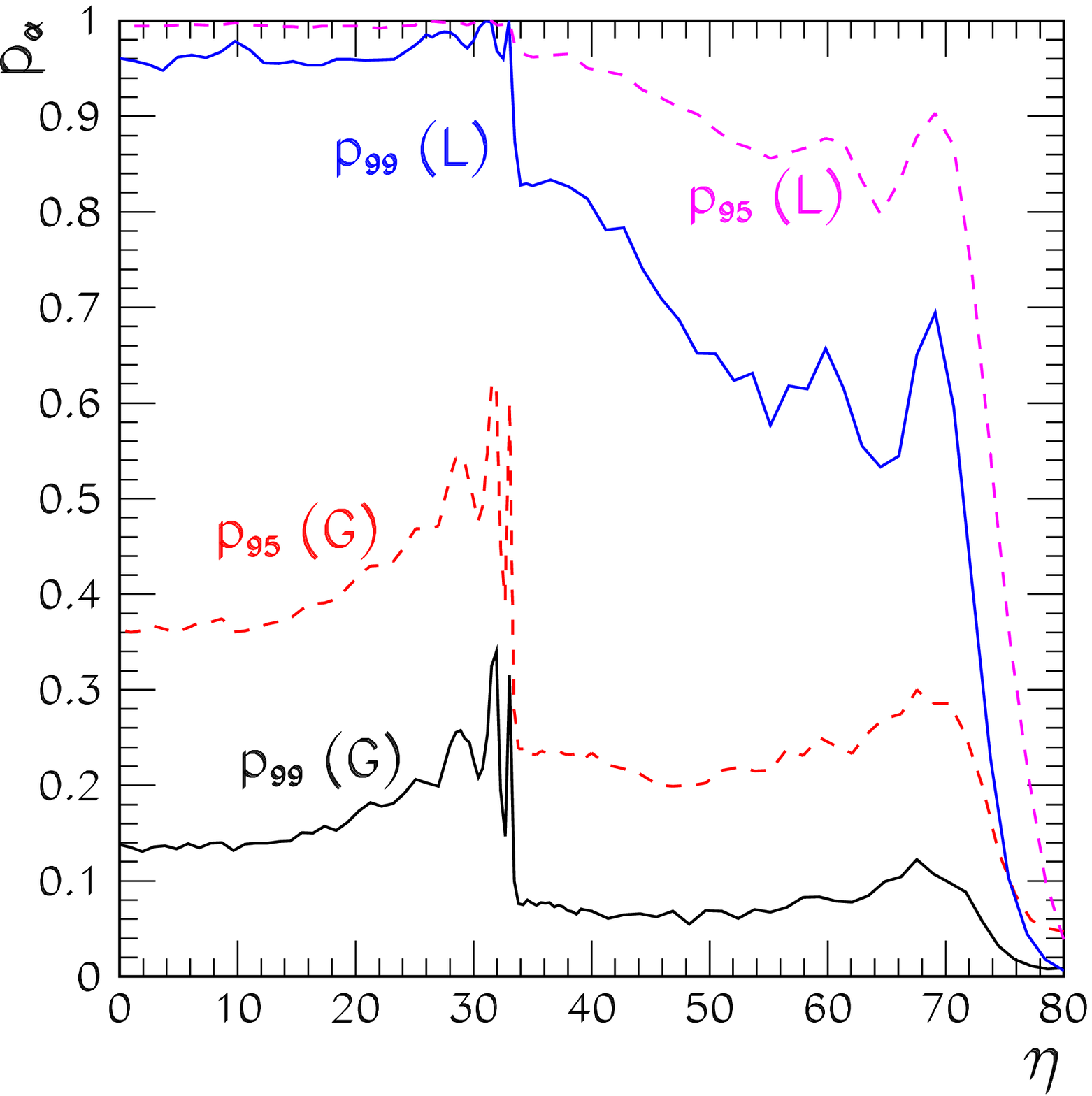}
 \end{center}
 \end{minipage}
\hspace{0.5cm}  
 \begin{minipage}{7.cm}
 \begin{center}
\epsfxsize = 6 cm
 \epsfbox{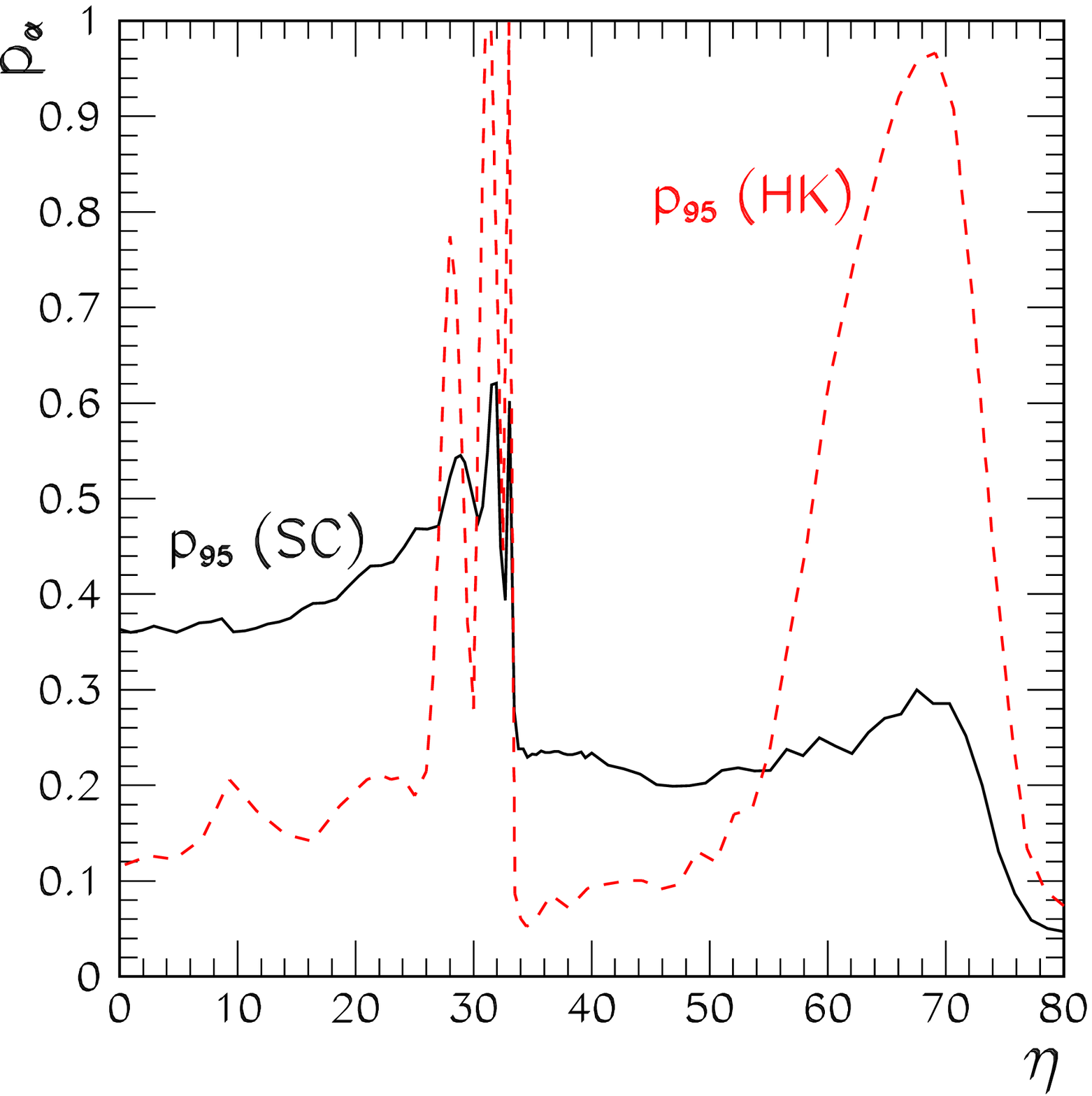}
 \end{center}
 \end{minipage}
\end{center}
\vspace{-0.7cm}
\caption{Left: Comparison of $p_{95}$ and $p_{99}$ for the Garching (G) and Livermore (L)
SN models in a 32 kton scintillator detector. Right: Comparison of $p_{95}$ in
this large scintillator detector (SC) and in the case of a megaton water Cherenkov (HK),
for the Garching model. From \cite{DigheKachelriessRaffeltTomas04}.
\label{fig:p95}}
\end{figure}

To quantify the effectiveness of their method, they introduced an algorithm to identify
the peaks in the real world with the presence of background fluctuations. It is based on
the integration of the area around the expected position of the peak. Once we know
the neutrino path, we can calculate the position of the peaks. It is more robust to
consider the area around the expected position of the peak than to look for the maximum
in the height of the power spectrum, as shown in the left of Fig. \ref{fig:area-distribution}.
They considered, for a single peak case, the interval of integration $k_{\rm peak} \pm \Delta k$
with $\Delta k = 30$, which is roughly the expected width of the peak. For a multiple peak case,
they measure the area $k = 40-160$.

Then we must consider the statistical significance of the result obtained. For this purpose, 
they compared the value of the measured area with the distribution of the area in the case
of no earth effect. The right of Fig. \ref{fig:area-distribution} shows the area distribution
of Monte Carlo simulations with and without the earth effect. Here $A_{95} \sim 100$ denote
the area corresponding to $95 \%$ C.L. detection of the earth effect. The problem is
the probability that we have larger area than $A_{95}$, which is denoted by $p_{95}$.
This probability depends on the distance traveled by the neutrinos through the earth,
which is in turn determined by the location of the supernova in the sky.
In Fig. \ref{fig:p95}, the probability as a function of the nadir angle $\eta$ of supernova
is plotted. The passage through the core corresponds to $\eta < 33^{o}$. The left is
comparison of $p_{95}$ and $p_{99}$ for the Garching and Livermore SN models in a 32 kton
scintillator detector. The right is comparison of $p_{95}$ in this large scintillator detector
(SC) and in the case of a megaton water Cherenkov (HK), for the Garching model.
As can be seen, this method is very effective if the average-energy differences of the original
neutrino spectra are as large as predicted by the Livermore simulation, while the effectiveness
decreases by half with the Garching simulation.

%%%%%%%%%%%%%%%%%%%%%%%%%%%%%%%%%%%%%%
\subsubsection{shock wave and neutrino oscillation}
%%%%%%%%%%%%%%%%%%%%%%%%%%%%%%%%%%%%%%

\begin{figure}[hbt]
\begin{center}
 \begin{minipage}{7.0cm}
 \begin{center}
\epsfxsize = 6cm
 \epsfbox{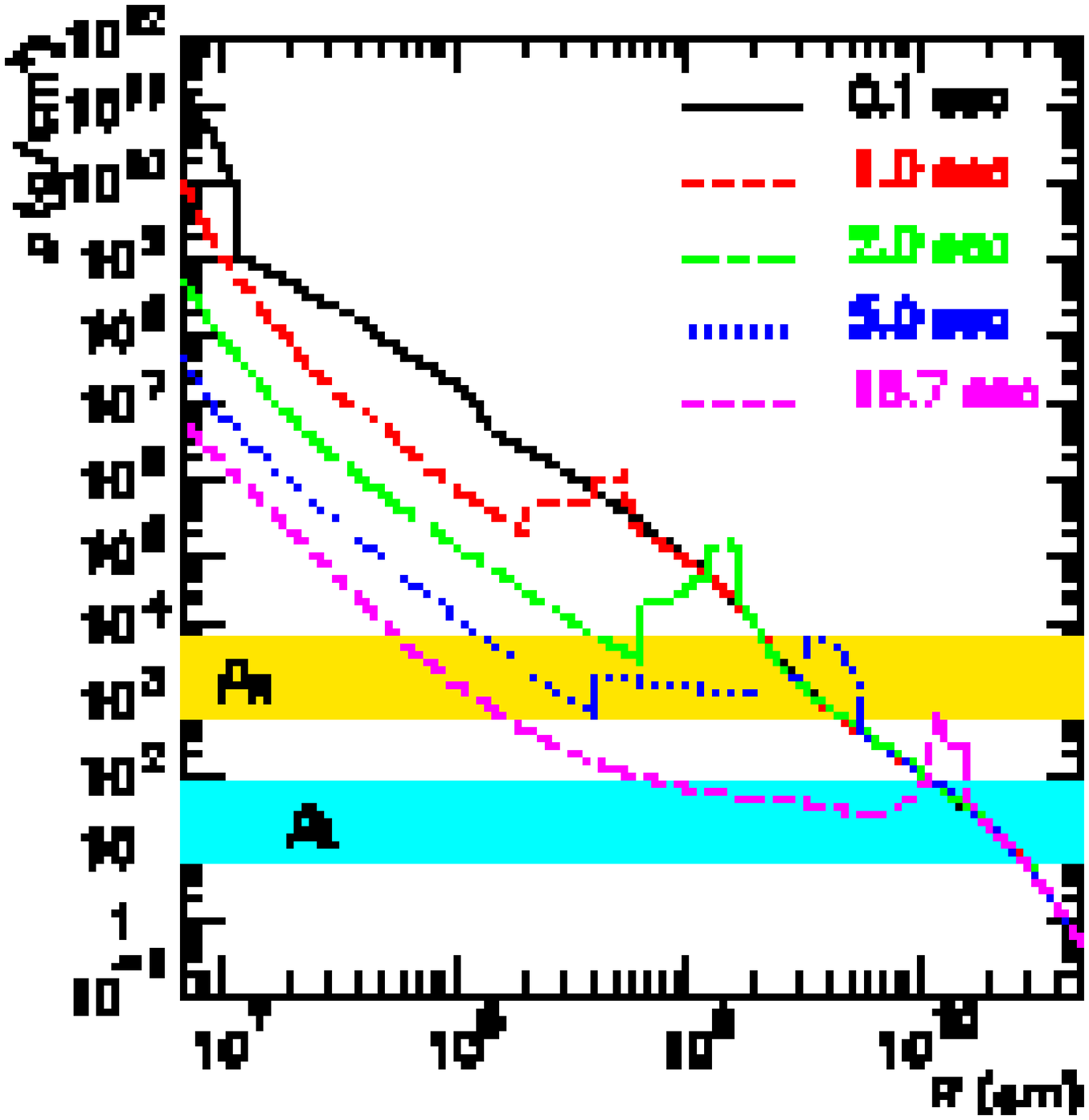}
 \end{center}
 \end{minipage}
\hspace{0.5cm}  
 \begin{minipage}{7.0cm}
 \begin{center}
\epsfxsize = 6cm
 \epsfbox{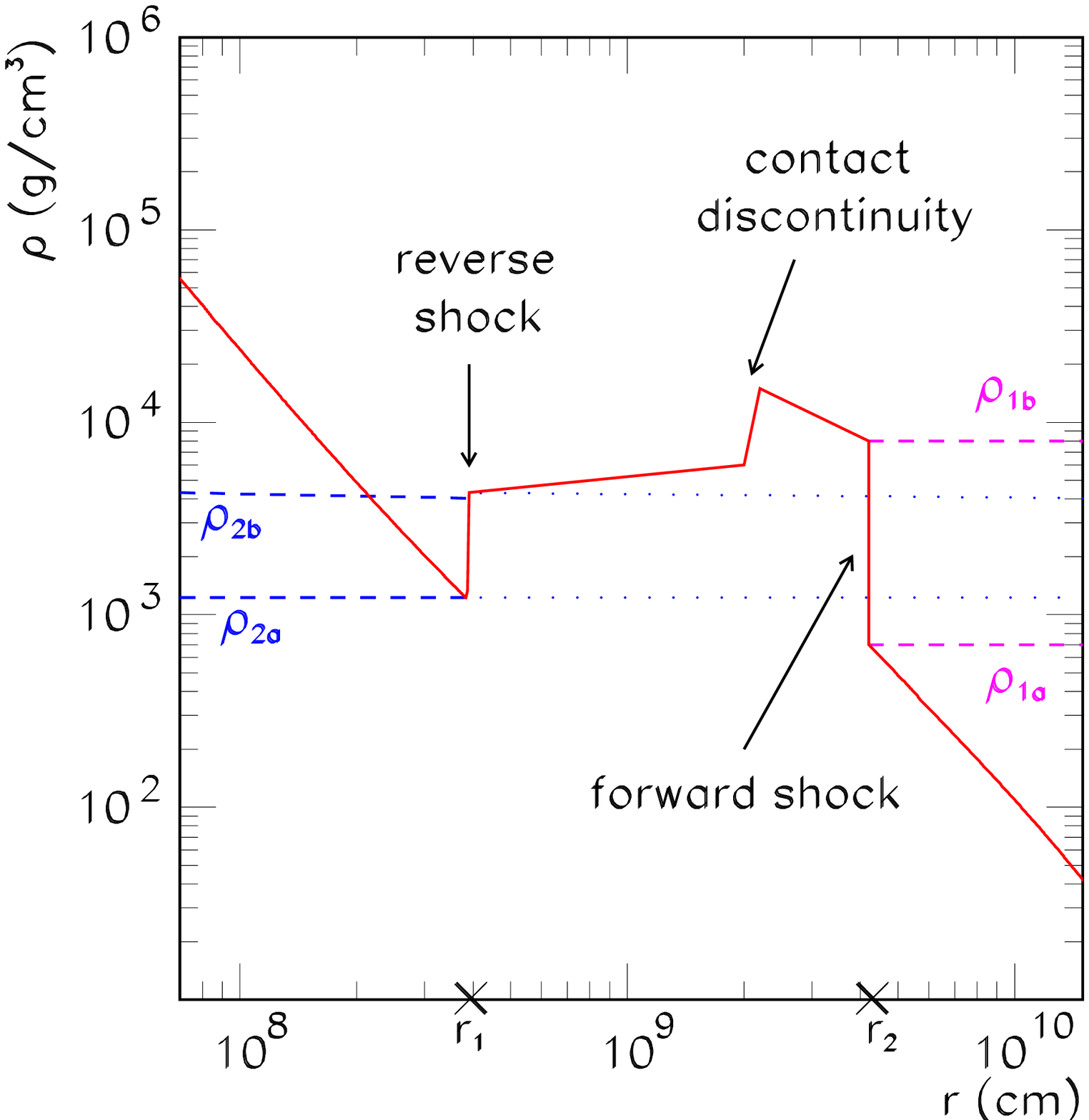}
 \end{center}
 \end{minipage}
\end{center}
\vspace{-0.5cm}
\caption{Left: Propagation of forward and reverse shocks with densities of H- and L-resonance
regions \cite{Tomas04}. The width of the density bands reflects the energy range of
supernova neutrinos. Right: Schematic density profile in the presence of a forward and
reverse shock wave with a contact discontinuity between them. Four key densities are
denoted as $\rho_{\rm 1b}, \rho_{\rm 2b}, \rho_{\rm 2a}$ and $\rho_{\rm 1a}$ from
dense to sparse region, which correspond to the edges of the forward and reverse shocks.
\label{fig:shock-profile}}
\end{figure}

So far we have assumed the structure of supernova envelope to be static during
$\sim 10 {\rm sec}$ of neutrino emission. Actually, as pointed out by Schirato and Fuller
\cite{SchiratoFuller02} and studied further in
\cite{KTshock03,LunardiniSmirnov03,Gil-BotellaRubbia03,Tomas04,Fogli05}, the shock wave
produced at the bounce reaches the resonance regions in several seconds
(left of Fig. \ref{fig:shock-profile}) and can affect the adiabaticities of the resonances.
This fact implies the possibility to probe the propagation of the shock wave by neutrino
observation.

Before we discuss neutrino oscillation, let us summarize basic features of shock wave.
While the shock sweeps the infalling matter, the gain region behind the shock is
accelerated by the neutrinos from the neutron star to form the neutrino-driven wind.
At the interface between this neutrino-driven wind and the shock-accerelated ejecta is formed
a contact discontinuity, which is characterized by a density jump but continuous
velocity and pressure. Even farther behind the forward shock, a reverse shock can appear
due to the collision between the neutrino-driven wind and more slowly moving material.
The forward and reverse shocks are sharp discontinuities where density, pressure and
velocity change on the microscopic (sub-millimeter) scale of the ion mean free path.
The right of Fig. \ref{fig:shock-profile} shows a schematic density profile in the presence of
a forward and reverse shock wave with a contact discontinuity between them.
The existence of the contact discontinuity and reverse shock seems to be a generic feature
in supernova explosion according to recent numerical simulation. However, because the structure
and time evolution of the shock wave depend on the detailed dynamics during the early
stages of the supernova explosion, it is difficult to make an exact prediction.
The situation becomes even more complicated if we consider non-spherical supernova explosion
where violent convective instabilities and large anisotropies make the density structure
chaotic, although the some generic features of the one-dimensional situation are retained.
Here we discuss effects of shock propagation on neutrino oscillation in one-dimensional
supernova following \cite{Tomas04}.

\begin{figure}[hbt]
\begin{center}
 \begin{minipage}{7.0cm}
 \begin{center}
\epsfxsize = 6cm
 \epsfbox{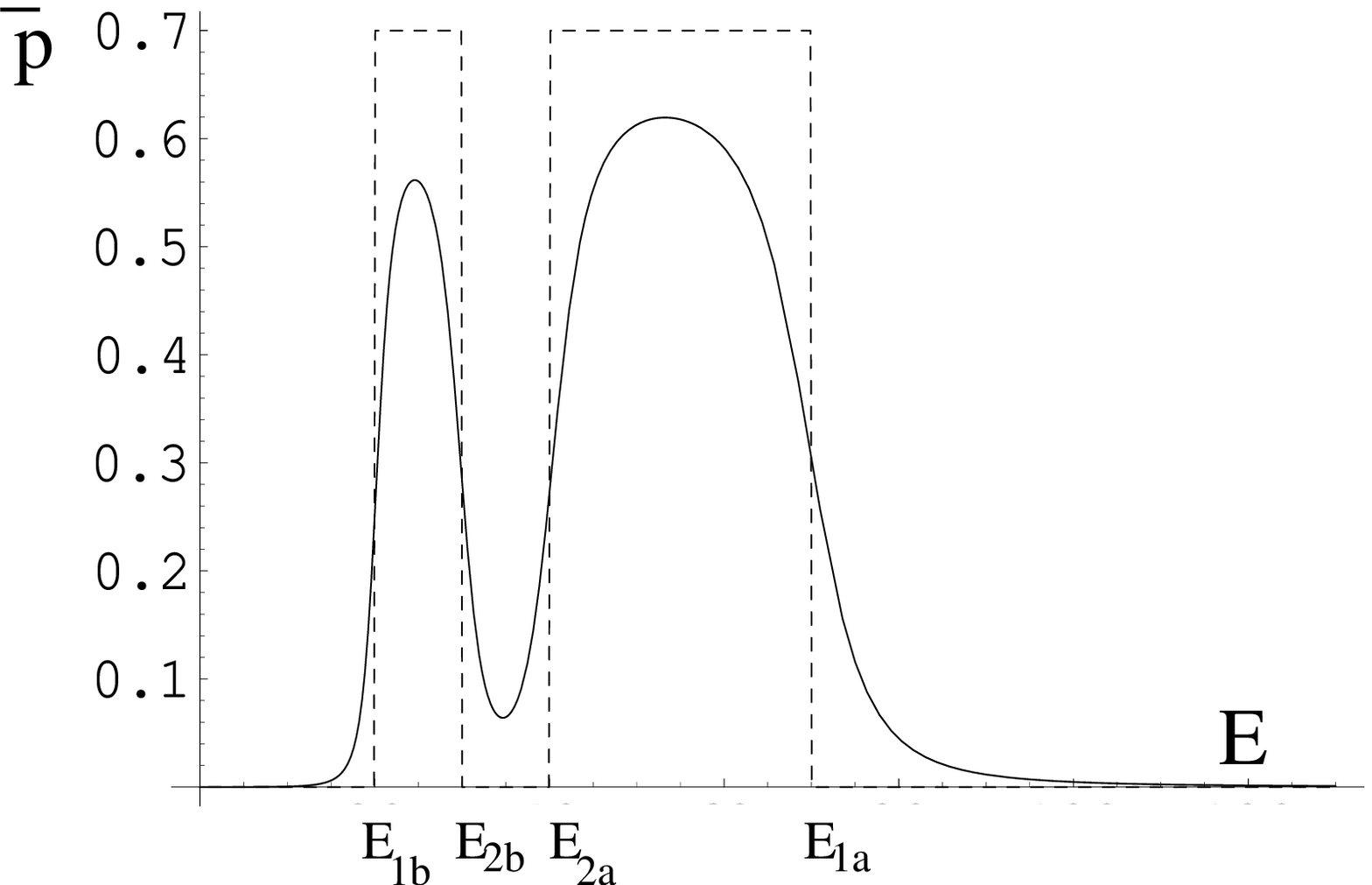}
 \end{center}
 \end{minipage}
\hspace{0.5cm}  
 \begin{minipage}{7.0cm}
 \begin{center}
\epsfxsize = 6cm
 \epsfbox{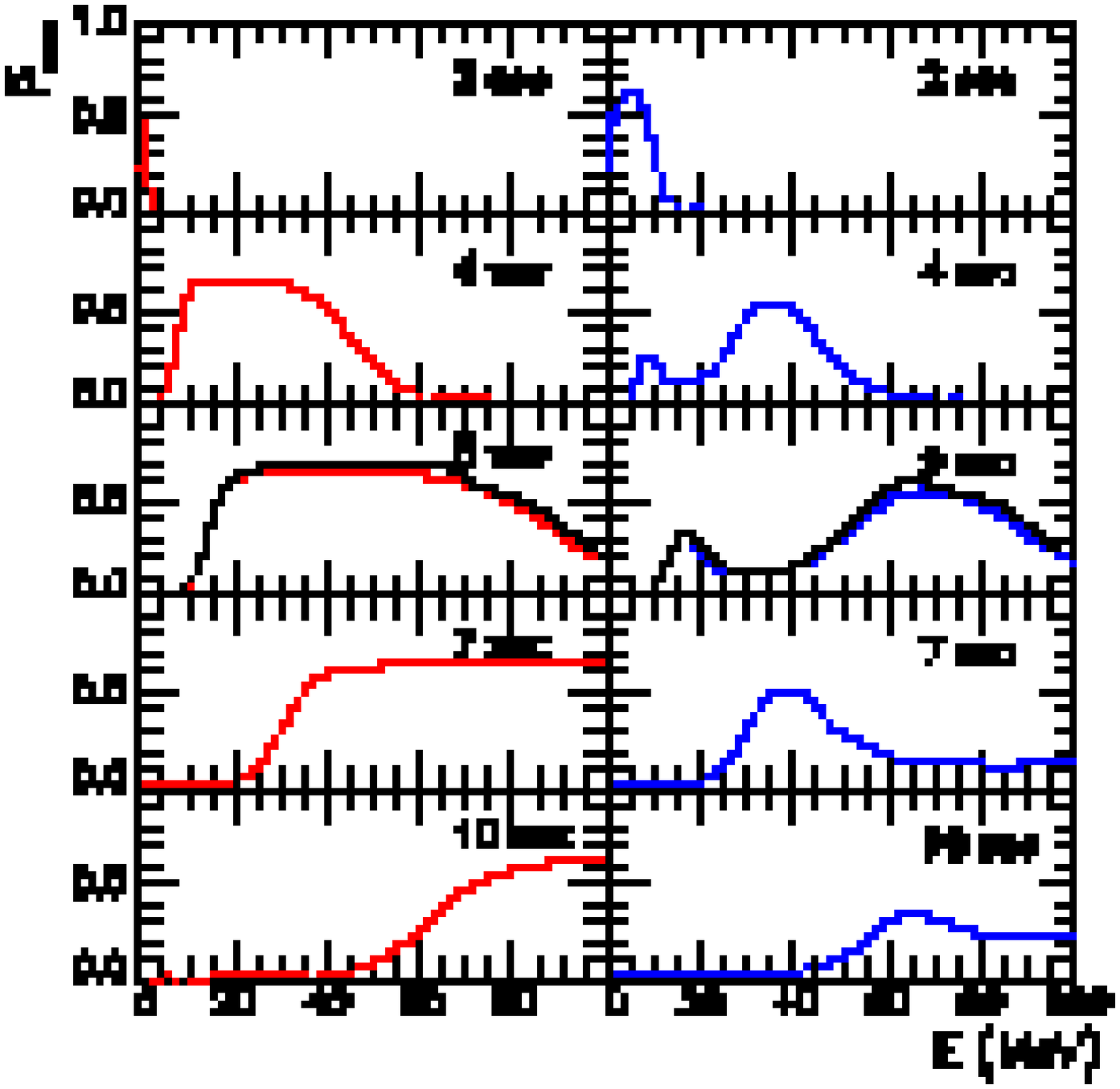}
 \end{center}
 \end{minipage}
\end{center}
\vspace{-0.5cm}
\caption{Left: Survival probability as a function of energy with $\cos^{2}{\theta_{12}}=0.7$.
The dotted line corresponds to the probability based on the simple argument in the text while
the solid line is based on a more realistic analytic evaluation in \cite{Tomas04}.
Right: Survival probability $\bar{p}(E,t)$ as a function of energy at different times averaging
in energies with the energy resolution of Super-Kamiokande: for a profile with
only a forward shock (left) and a profile with forward and reverse shock (right).
At $t=5$ sec, $\bar p(E,t)$ including Earth matter effects for a zenith angle of $62^{o}$
is also shown with a black line. Both are from \cite{Tomas04}.
\label{fig:survival-prob}}
\end{figure}

\begin{figure}[hbt]
\vspace{-0.5cm}
\begin{center}
 \begin{minipage}{7cm}
 \begin{center}
\epsfxsize = 6cm
 \epsfbox{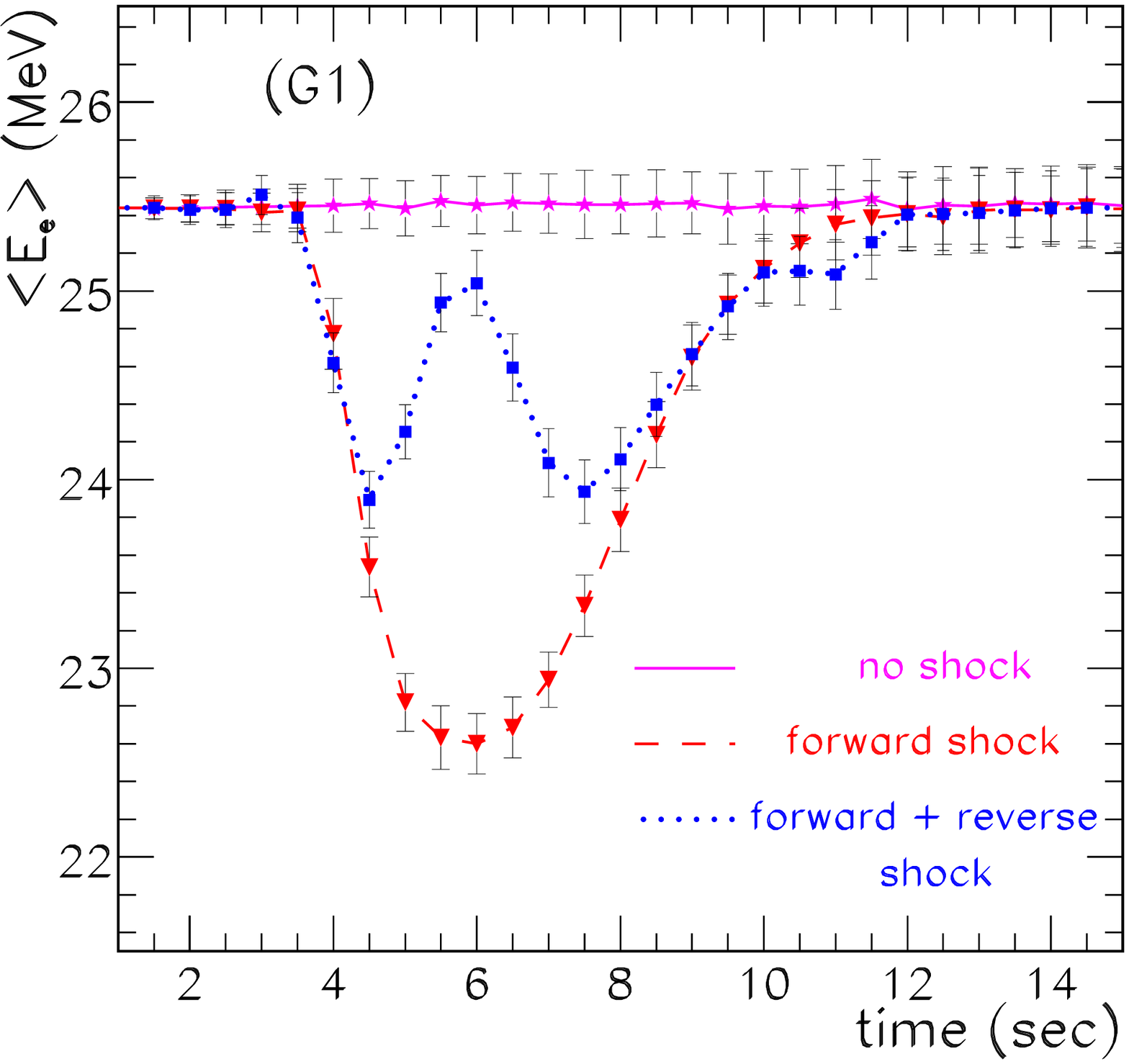}
 \end{center}
 \end{minipage}
\hspace{0.5cm}  
 \begin{minipage}{7cm}
 \begin{center}
\epsfxsize = 6cm
 \epsfbox{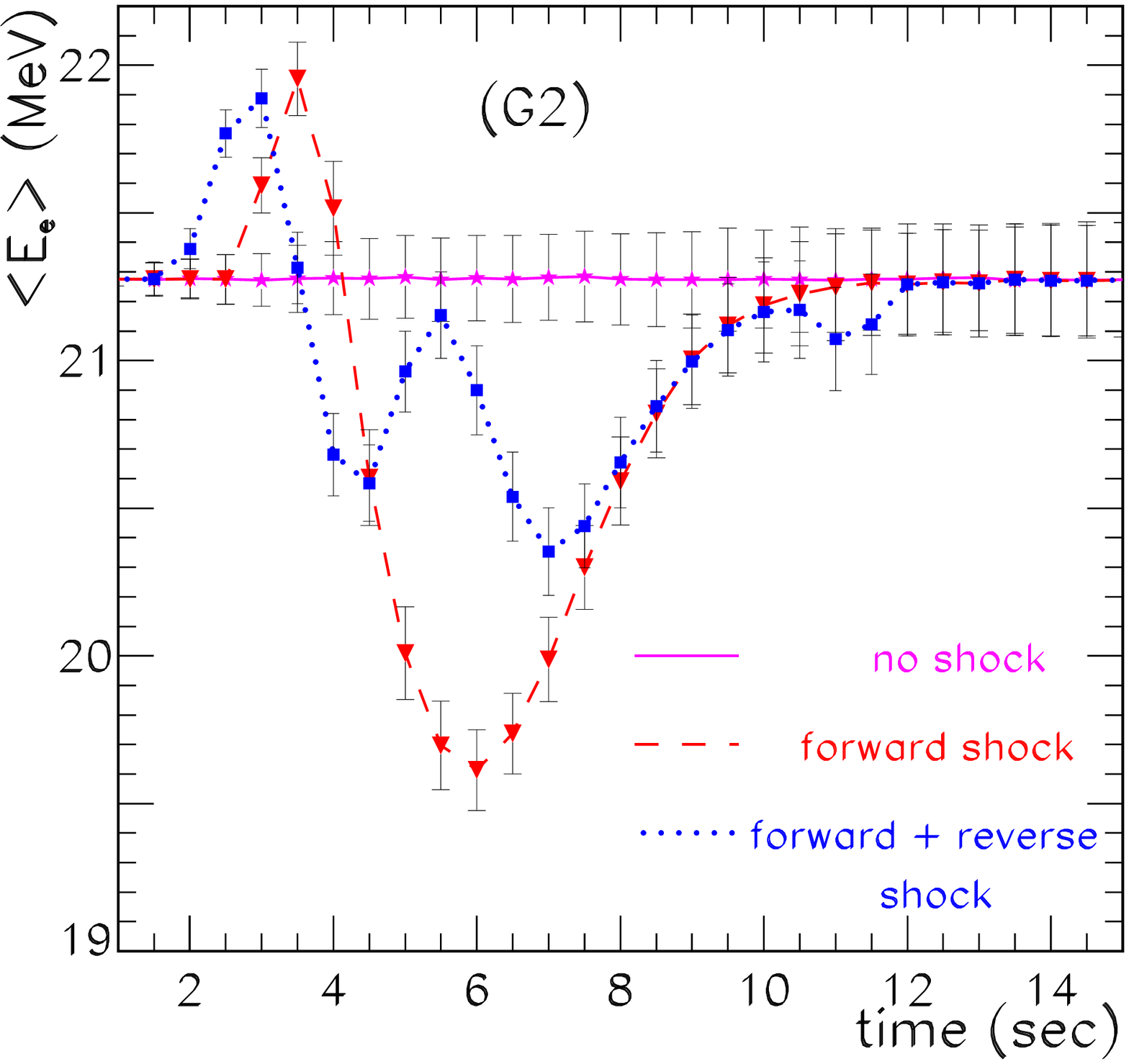}
 \end{center}
 \end{minipage}
\end{center}
\vspace{-1.2cm}
\begin{center}
\epsfxsize = 6cm
 \begin{minipage}{7cm}
 \begin{center}
 \epsfbox{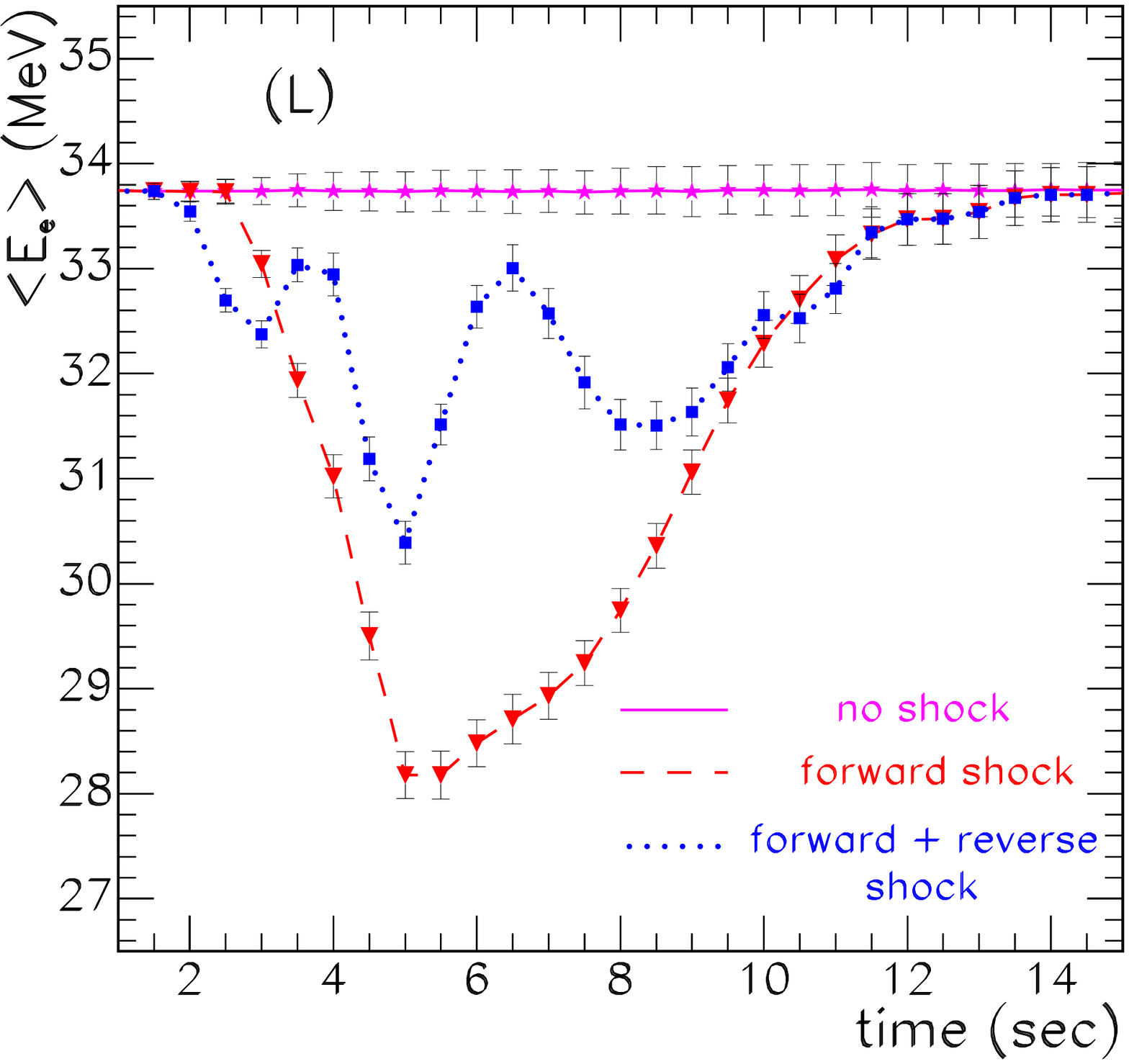}
 \end{center}
 \end{minipage}
\hspace{0.5cm}  
 \begin{minipage}{7cm}
 \begin{center}
\epsfxsize = 6cm
 \epsfbox{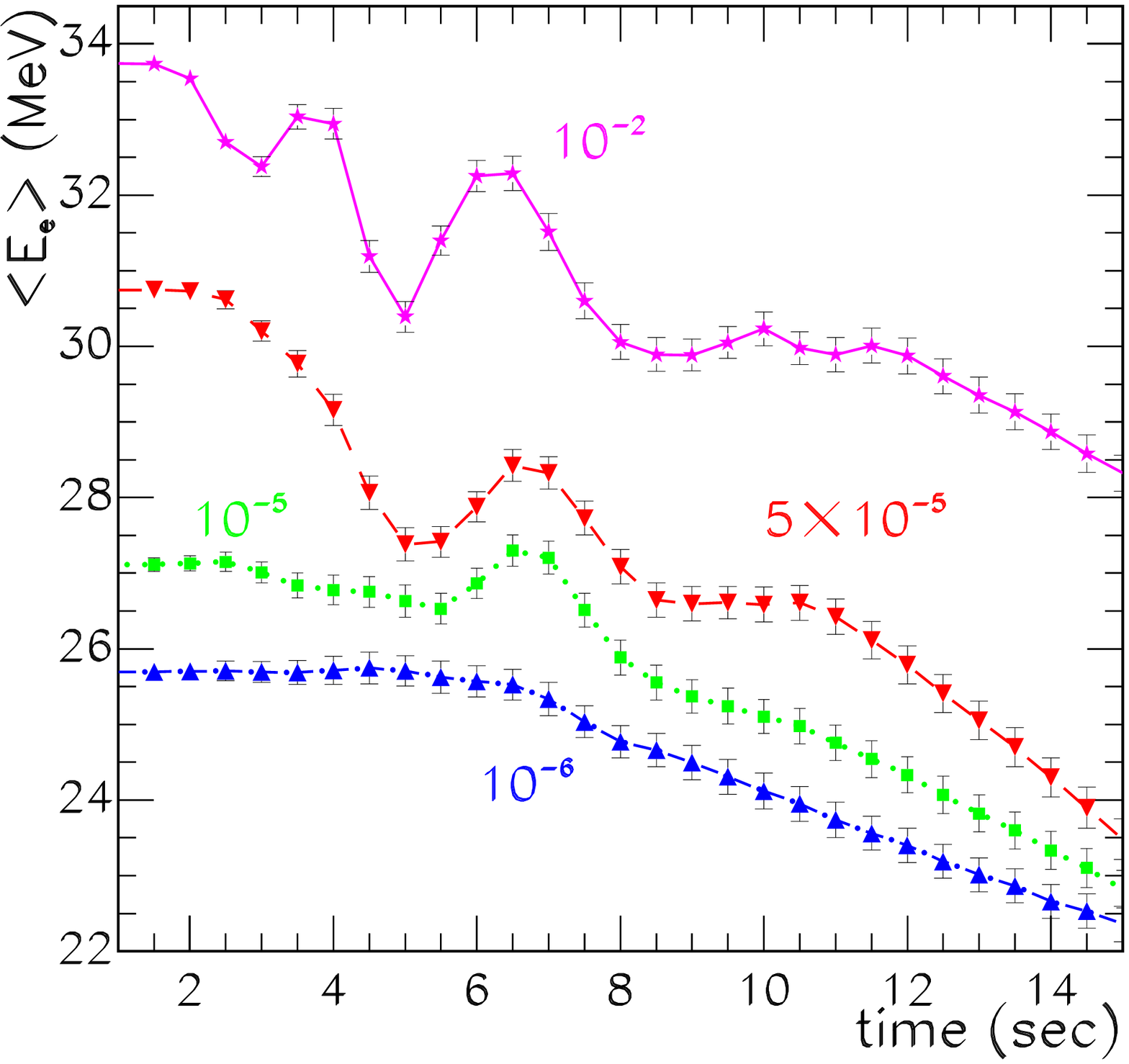}
 \end{center}
 \end{minipage}
\end{center}
\vspace{-0.5cm}
\caption{Upper and lower left: The average energy of $\bar{\nu} p \to n e^{+}$ events
at a megaton water Cherenkov detector binned in time for simulations by the Livermore group (L)
and the Garching group (G1 and G2).
 The average energy is assumed to be static and the error bars represent
$1 \sigma$ errors in any bin. Lower right: Time dependence of $\langle E_{e} \rangle$
for a profile with a forward and reverse shock for several values of $\tan^{2}{\theta_{13}}$
and for model L. Figures from \cite{Tomas04}.
\label{fig:shock_ave-energy}}
\end{figure}

First, let us consider the survival probability of neutrinos which experience the H-resonance,
that is, $\nu_{e}$s and $\bar{\nu}_{e}$ for the normal and inverted hierarchy, respectively.
In this section, we assume the inverted hierarchy and consider $\bar{\nu}_{e}$.
We can obtain a qualitative feature of the survival probability as a function of
neutrino energy from the schematic density profile of the shock shown in the right
of Fig. \ref{fig:shock-profile}. In the figure, four key densities are denoted as
$\rho_{\rm 1b}, \rho_{\rm 2b}, \rho_{\rm 2a}$ and $\rho_{\rm 1a}$ from dense to sparse region.
They correspond to the edges of the forward and reverse shock: the density jumps from
$\rho_{\rm 2a}$ to $\rho_{\rm 2b}$ at the reverse shock and from $\rho_{\rm 1b}$ to
$\rho_{\rm 1a}$ at the forward shock. As we saw in Eq. (\ref{eq:res_condition}),
the resonance density is proportional to the neutrino energy,
\begin{equation}
\rho_{\rm res} = 1.3 \times 10^{3} {\rm g} ~ {\rm cm}^{-3}
\cos{2\theta} \left(\frac{0.5}{Y_{e}}\right) 
\left(\frac{10 {\rm MeV}}{E_{\nu}}\right) 
\left(\frac{\Delta m^{2}}{10^{-3} {\rm eV}^{2}}\right),
\end{equation}
so that there are four energies whose resonance densities are the above four densities,
$E_{\rm 1b}, E_{\rm 2b}, E_{\rm 2a}$ and $E_{\rm 1a}$ from lower to higher energies.
Here it should be noted that the four densities are time-dependent and so are the four
energies.

The essential point in the effect of shock propagation on neutrino oscillation is that
the density gradient at the shocks is so steep that the resonance there tends to be
non-adiabatic even if the mixing angle involved is rather large. Let us assume first
that the resonance is completely non-adiabatic at the shocks and completely adiabatic
elsewhere. The latter means that $\theta_{13}$ is sufficiently large. Then, a $\bar{\nu}_{e}$
with energy $E_{\rm 1b} < E < E_{\rm 2b}$ experiences the H-resonance three times, that is,
behind the reverse shock, between the reverse and forward shock, and at the forward shock,
which are adiabatic, adiabatic and non-adiabatic, respectively. The three resonances are
equivalent to one non-adiabatic resonance. In the same way, we obtain the adiabaticities
of resonances of various neutrino energies,
\begin{itemize}
\item $E < E_{\rm 1b} \Rightarrow$ adiabatic
\item $E_{\rm 1b} < E < E_{\rm 2b} \Rightarrow $
adiabatic, adiabatic, non-adiabatic $\Rightarrow$ effectively non-adiabatic
\item $E_{\rm 2b} < E < E_{\rm 2a} \Rightarrow$
adiabatic, non-adiabatic, non-adiabatic $\Rightarrow$ effectively adiabatic
\item $E_{\rm 2a} < E < E_{\rm 1a} \Rightarrow$ non-adiabatic
\item $E > E_{\rm 1a} \Rightarrow$ adiabatic
\end{itemize}
and the survival probability as a function of neutrino energy become like the left of
Fig. \ref{fig:survival-prob}. As one can see, there are two peaks in the survival probability,
which result, as we will see later, in characteristic features in the time evolution of
observed neutrino average energy and event number. Note that there will be only one
peak without the reverse shock and that the behavior of the survival probability will
become more complicated if the contact discontinuity is so sharp that resonance there
is non-adiabatic. Anyway, since we assumed $\theta_{13}$ to be sufficiently large so that
the survival probability is zero in absence of shock propagation, the effect of the shock
can be thought to be to make the resonance more non-adiabatic effectively.

Because the shock propagate into low-density regions, the key densities decrease in time
and so the key energies increase in time. The right of Fig. \ref{fig:survival-prob}
shows the survival probability $\bar{p}(E,t)$ as a function of energy at different times averaging
in energies with the energy resolution of Super-Kamiokande: for a profile with
only a forward shock (left) and a profile with forward and reverse shock (right).
At $t=5$ sec, $\bar p(E,t)$ including earth matter effects for a zenith angle of $62^{o}$
is also shown with a black line. It can be seen than there is only one peak without
the reverse shock and are two peaks with both the reverse and forward shocks and the peaks
shift to high-energy regions as the shock propagates, as expected.

Upper and lower left of Fig. \ref{fig:shock_ave-energy} show the average energy of
$\bar{\nu_{e}} p \to n e^{+}$ events binned in time for simulations by the Livermore group (L)
and the Garching group (G1 and G2). Magenta, red and blue lines correspond to
a static density profile, a profile with only a forward shock and with forward and
reverse shock, respectively. The error bars are $1 \sigma$ statistical errors assuming
a megaton water Cherenkov detector. As one can see, there are some dips
in the time evolution of the average energy: one dip with only forward shock and double
dips with both forward and reverse shock. This characteristic feature is the direct result
of the dips in the survival probability in Fig. \ref{fig:survival-prob} and can be seen in every
SN model. Lower right of Fig. \ref{fig:shock_ave-energy} is the time dependence of
the average energy for a profile with a forward and reverse shock for several values of
$\tan^{2}{\theta_{13}}$ and for model L. It is seen that the dips are smaller for smaller
values of $\tan^{2}{\theta_{13}}$. This can be understood by remembering that the propagation
of the shock waves through the resonance region tends to make the resonance more non-adiabatic
effectively. When $\tan^{2}{\theta_{13}}$ is small, the H-resonance is non-adiabatic even without
the shock and then the shock has no effect on neutrino oscillation.

\begin{figure}[hbt]
\begin{center}
 \begin{minipage}{7.0cm}
 \begin{center}
\epsfxsize = 6cm
 \epsfbox{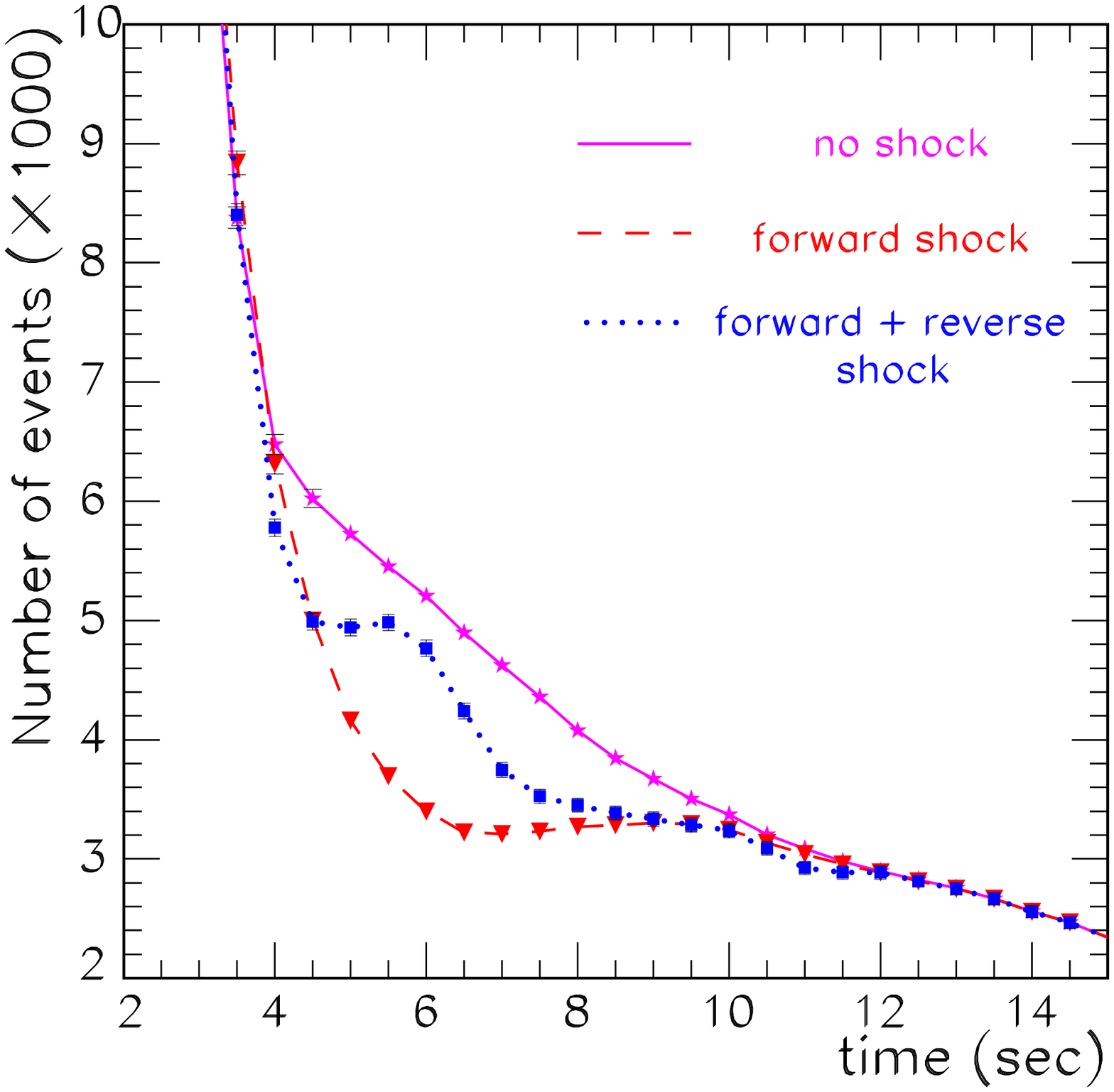}
 \end{center}
 \end{minipage}
\hspace{0.5cm}  
\epsfxsize = 6cm
 \begin{minipage}{7.0cm}
 \begin{center}
 \epsfbox{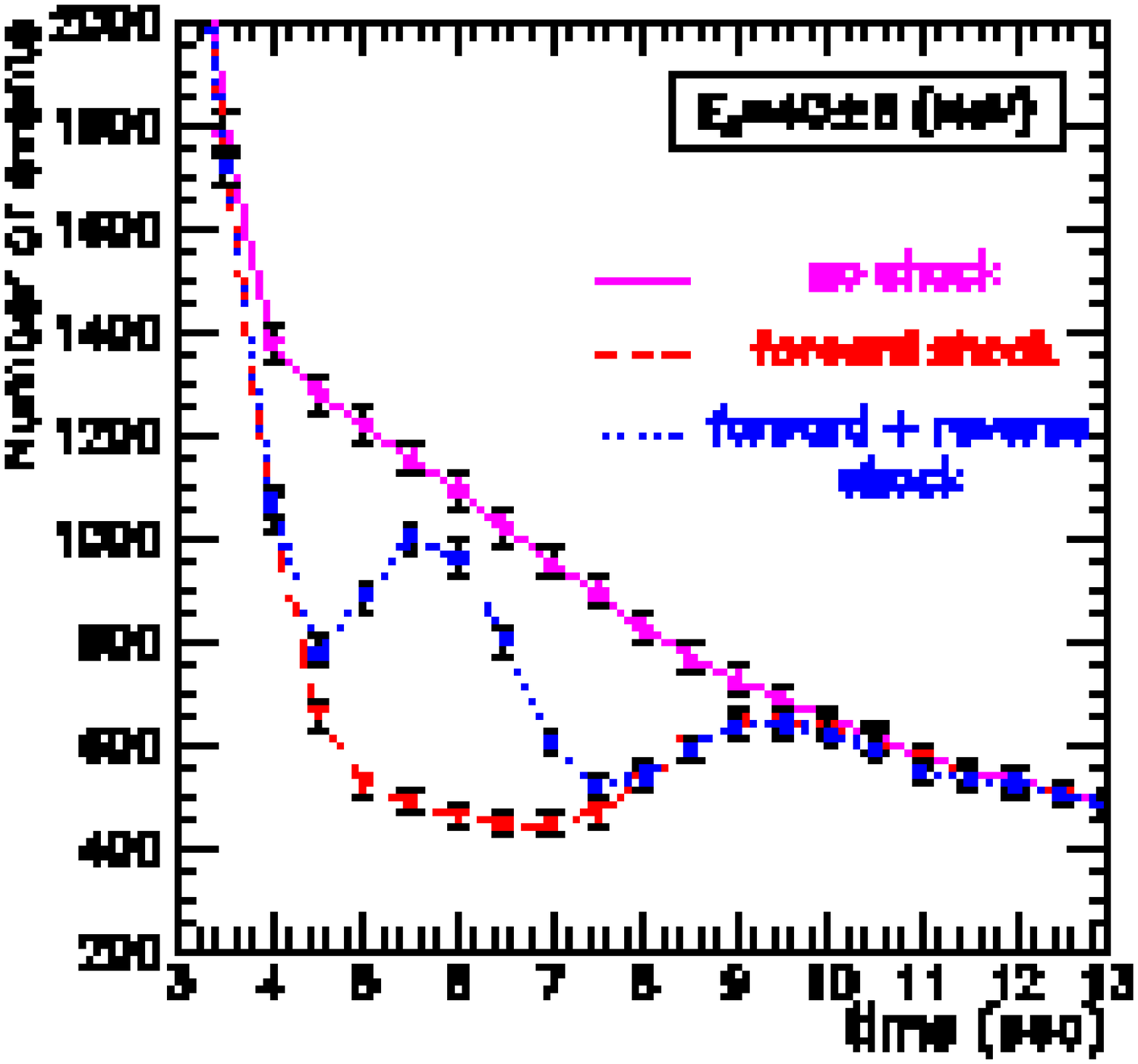}
 \end{center}
 \end{minipage}
\end{center}
\caption{Left: Total number of events as a function of time for a static density profile
and a profile with only forward shock and with both forward and reverse shock.
Right: Same as the left but the energy range is restricted to $E = 40 \pm 5 {\rm MeV}$.
The error bars are statistical errors expected in a megaton water Cherenkov detector.
Both are from \cite{Tomas04}.
\label{fig:shock_event}}
\end{figure}

The imprint of the shock waves can also seen in the time evolution of the event number.
The left of Fig. \ref{fig:shock_event} shows the total number of events as a function of time
for a static density profile and a profile with only forward shock and with both forward and
reverse shock. As one can see, a dip can be seen in the presence of both the forward and
reverse shocks. This is even striking if the energy range is restricted to
$E = 40 \pm 5 {\rm MeV}$ as can be seen in the right of Fig. \ref{fig:shock_event}.

Thus, if $\theta_{13}$ is sufficiently large ($\tan^{2}{\theta_{13}} > 10^{-5}$),
dips appear in the time evolution the neutrino average energy and event number, which would be
relatively model-independent. A megaton water Cherenkov detector, like the HyperKamiokande,
can probe the shock wave propagation efficiently if the mass hierarchy is inverted one
so that the H-resonance occurs at the anti-neutrino sector.

%%%%%%%%%%%%%%%%%%%%%%%%%%%%%%%%%%%%%%
\subsubsection{toward model-independent predictions}
%%%%%%%%%%%%%%%%%%%%%%%%%%%%%%%%%%%%%%
\begin{figure}[hbt]
\begin{center}
 \begin{minipage}{7.0cm}
 \begin{center}
\epsfxsize = 6cm
 \epsfbox{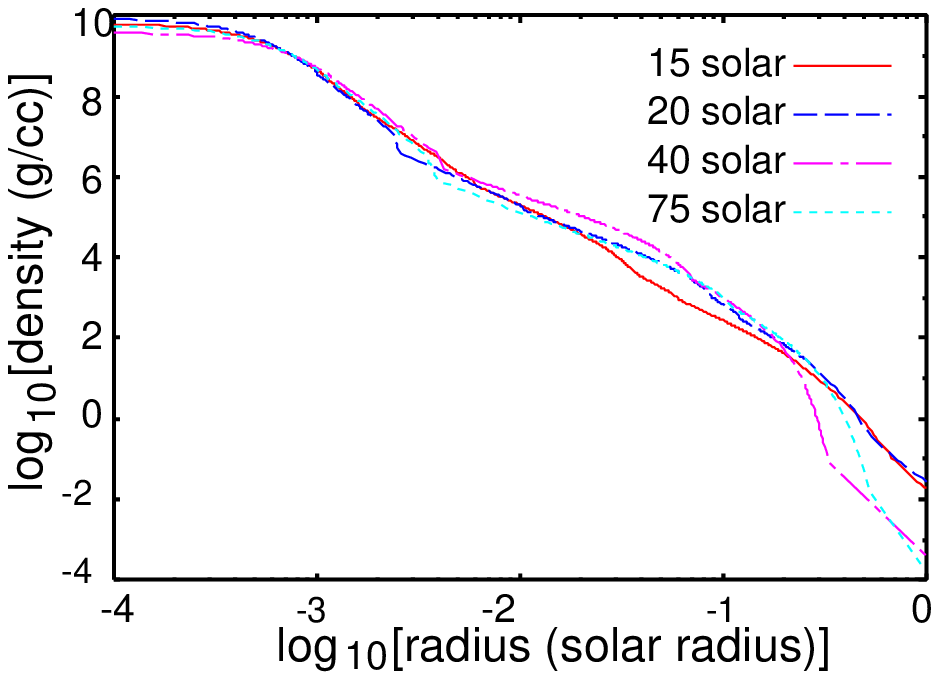}
 \end{center}
 \end{minipage}
\hspace{0.5cm}  
 \begin{minipage}{7.0cm}
 \begin{center}
\epsfxsize = 6cm
 \epsfbox{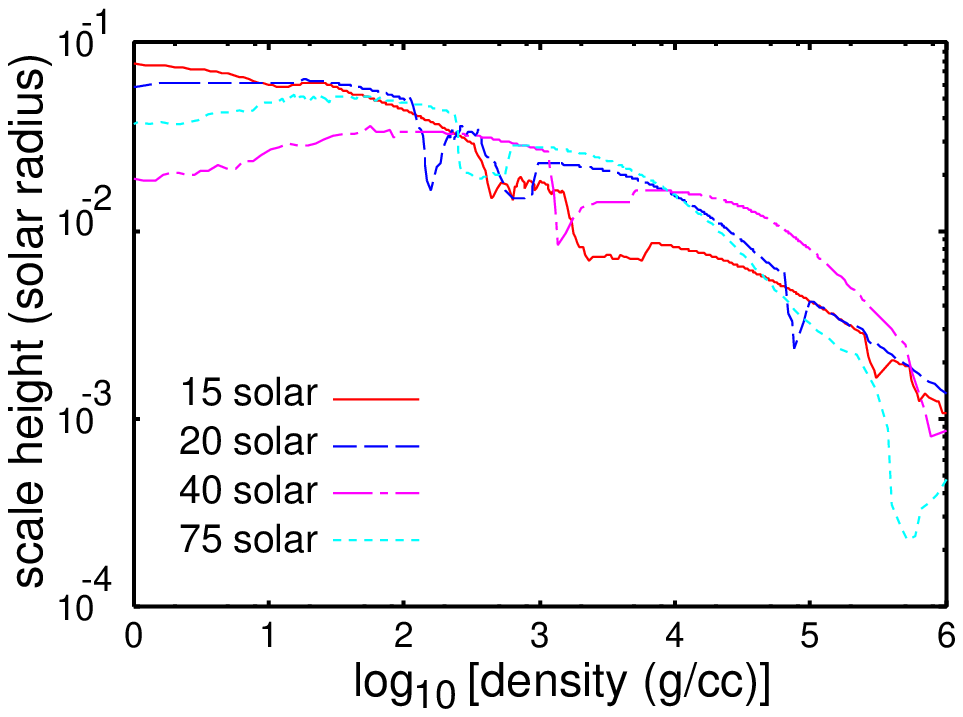}
 \end{center}
 \end{minipage}
\end{center}
\vspace{-0.5cm}
\caption{Density profiles (left) and scale height $n_{e}/|dn_{e}/dr|$ of stars
just before supernova explosion with initial mass $15 M_{\odot}, 20 M_{\odot}, 40 M_{\odot}$
and $75 M_{\odot}$ \cite{KTmass03}. 
\label{fig:mass_density}}
\end{figure}

While type Ia supernovae have rather universal features as far as we observe them,
core-collapse supernovae seem to have a wide variety in luminosity and spectrum.
Since this will reflect the diversity of the presupernova structure, neutrino emission
will also be dependent on the physical state of the progenitor star.
Some important physical quantities about the progenitor star will include its mass,
metalicity, magnetic field and rotation, whose effects on neutrino emission are
still unclear. If we want to extract information about neutrino oscillation parameters,
the effects of these supernova parameters on the neutrino emission and dynamics of
neutrino oscillation must be studied and some model-independent analyses are required.
In \cite{KTmass03}, Takahashi et al. studied the effect of initial mass of a progenitor star
on the neutrino emission and neutrino oscillation. Then some input-physics dependences
of neutrino emission were studied in \cite{Kachelriess05} concentrating on
neutronization-burst signal. Here we review \cite{KTmass03} and summarize \cite{Kachelriess05}
briefly.

The mass of the progenitor star affects the neutrino oscillation signature of
supernova neutrinos through differences in both the mantle and core structures.
The density profile of the progenitor star, especially of the mantle, is important
because it is related to the dynamics of neutrino flavor conversion. On the other hand,
the structure of the iron core at the collapse determines the characteristics of
the neutrino burst, e.g., the average energy and luminosity for each flavor.

Let us first discuss the density profile of a progenitor star. In numerical simulations
of supernova, the initial condition is a star just before the collapse whose structure
is often given by a numerical presupernova model, which is obtained by following the evolution
of a massive star. The evolution of a massive star is significantly affected by mass loss
due to a stellar wind, which we still do not have a definite understanding.
Indeed, the mass loss can become so strong for a star with initial mass more than about
35 $M_{\odot}$ and solar metalicity that the entire hydrogen envelope can be lost prior
to the explosion of the star. It is suggested in \cite{WoosleyHegerWeaver02} that the maximum
in the final mass is about 20 $M_{\odot}$. Thus, a massive star just before the supernova
explosion may have a universal structure independent of the initial mass of the star.
The final density profiles of stars with various initial masses just before the collapse
are shown in the left of Fig. \ref{fig:mass_density} \cite{SSC}. As is expected,
they are similar to each other.

The density profile of the star comes into the adiabaticity parameter as the scale height
$n_{e}/|dn_{e}/dr|$. A smaller scale height, that is, a steeper density profile results
in less adiabatic resonance. The scale heights of stars with various initial masses,
calculated from the density profiles shown in the left of Fig. \ref{fig:mass_density}, are given
in the right of Fig. \ref{fig:mass_density} as functions of the density. The scale heights
vary significantly, but, at the densities relevant for the resonances, differences between
different initial masses are factors of 2 or 3.

\begin{figure}[hbt]
\begin{center}
 \begin{minipage}{7.0cm}
 \begin{center}
\epsfxsize = 6cm
 \epsfbox{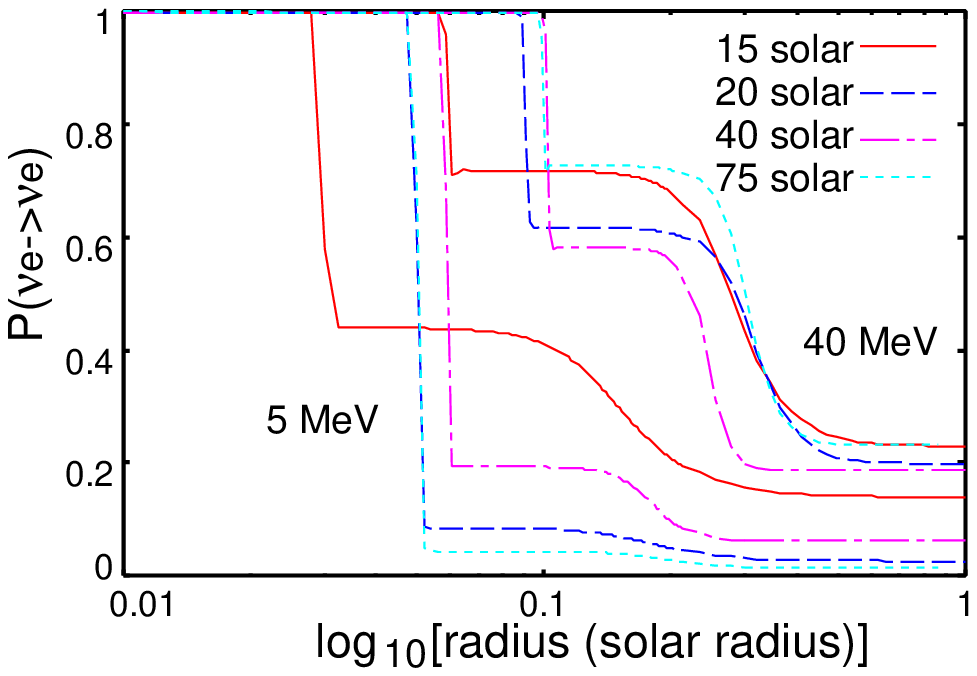}
 \end{center}
 \end{minipage}
\hspace{0.5cm}  
 \begin{minipage}{7.0cm}
 \begin{center}
\epsfxsize = 6cm
 \epsfbox{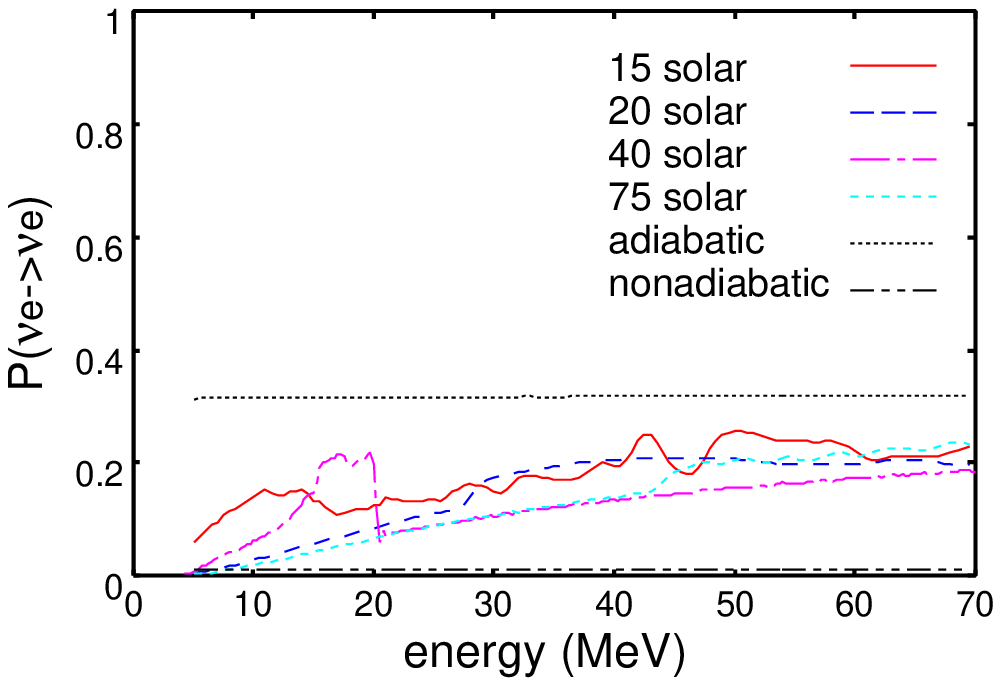}
 \end{center}
 \end{minipage}
\end{center}
\vspace{-0.5cm}
\caption{Left: Evolution of survival probabilities $P(\nu_{e} \rightarrow \nu_{e})$
as functions of radius for different progenitor models. Upper and lower curves correspond
to neutrino energies of 40 MeV and 5 MeV, respectively. Right: Energy dependence of
survival probability $P(\nu_{e} \rightarrow \nu_{e})$ for different progenitor models.
Also shown are the perfectly adiabatic and non-adiabatic cases for the H-resonance.
\label{fig:mass_conv_prob}}
\end{figure}

In the left of Fig. \ref{fig:mass_conv_prob}, the evolution of the survival probability
$P(\nu_{e} \to \nu_{e})$ is shown. The neutrino energies on the plot are 5 MeV and 40 MeV
and we set $\sin^{2}{2 \theta_{13}} = 10^{-4}$. The H-resonance radii and the final
probabilities can be quite different for different progenitors and neutrino energies.
The observationally important quantity is the final probability and this is shown in the right
of Fig. \ref{fig:mass_conv_prob} as a function of the neutrino energy. The differences are
not so small, about $O(10) \%$ at all energies. But if 
$\sin^{2}{2 \theta_{13}}$ is
so large or very small that H-resonance is perfectly adiabatic or non-adiabatic, respectively,
difference in scale height will not affect the neutrino conversion probabilities.

Next we consider the difference in core structure.
As was stated in the above, massive stars experience significant mass loss.
For current empirical mass loss rates, all solar-metalicity stars initially
more massive than about $35\ M_{\odot}$ are thought to become hydrogen-free objects
of roughly $5 M_{\odot}$ at the end of their thermonuclear evolution. The corresponding
upper limit to the mass of the final iron core is about $2 M_{\odot}$ \cite{WoosleyHegerWeaver02}.

As mentioned in subsection \ref{pre}, the mass of the iron core is determined roughly by the Chandrasekhar mass.
For a zero-temperature and constant $Y_{e}$, its Newtonian structure is given by,
\begin{equation}
M_{\rm Ch 0} = 5.83 Y^{2}_{e} M_{\odot}.
\end{equation}
However, there are numerous corrections, some of which are large \cite{TimmesWoosleyWeaver96}.
To the first approximation, the non-zero entropy of the core is important and
\begin{equation}
M_{\rm Ch} \sim M_{\rm Ch 0} 
\left[ 1 + \left( \frac{s_{e}}{\pi Y_{e}} \right)^{2} \right],
\end{equation}
where
\begin{equation} s_{e} = 0.50
\left(
\frac{\rho}{10^{10} {\rm g/cc}} \right)^{-1/3} \left( \frac{Y_{e}}{0.42}
\right)^{2/3} 
\left( \frac{T}{1 {\rm MeV}} \right)
\end{equation}
is the electronic entropy per baryon. More massive stars have higher entropy and contain
larger iron cores on average. However, this general tendency is moderated by the loss
and redistribution of entropy that occurs during the late burning stages. Thus, the mass
of the iron core as a function of the initial mass will be somewhat uncertain in that
a small change in the initial mass results in a large difference in the iron core mass.
According to \cite{WoosleyHegerWeaver02}, the mass of the iron core is
$1.2 (1.4) - 1.6 M_{\odot}$ when the initial mass is between $10 (20) M_{\odot}$ and
$40 M_{\odot}$. This weak dependence of the iron core mass on the ZAMS progenitor mass
leads to a somewhat universal neutrino burst. 

\begin{figure}[hbt]
\vspace{-0.5cm}
\begin{center}
 \begin{minipage}{7.0cm}
 \begin{center}
\epsfxsize = 6cm
 \epsfbox{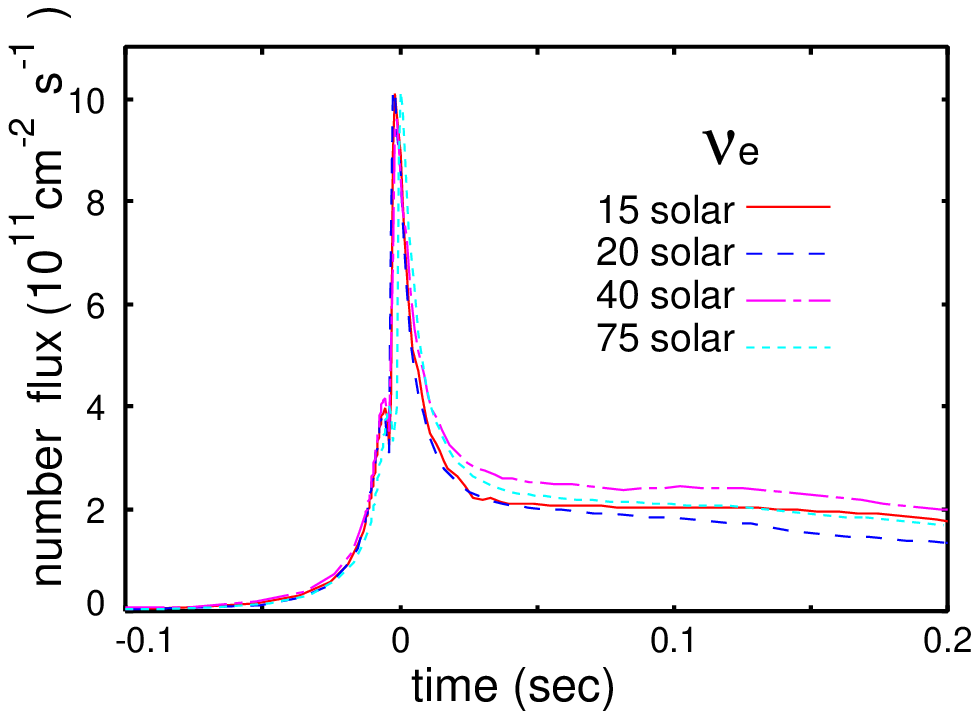}
 \end{center}
 \end{minipage}
\hspace{0.5cm}  
 \begin{minipage}{7.0cm}
 \begin{center}
\epsfxsize = 6cm
 \epsfbox{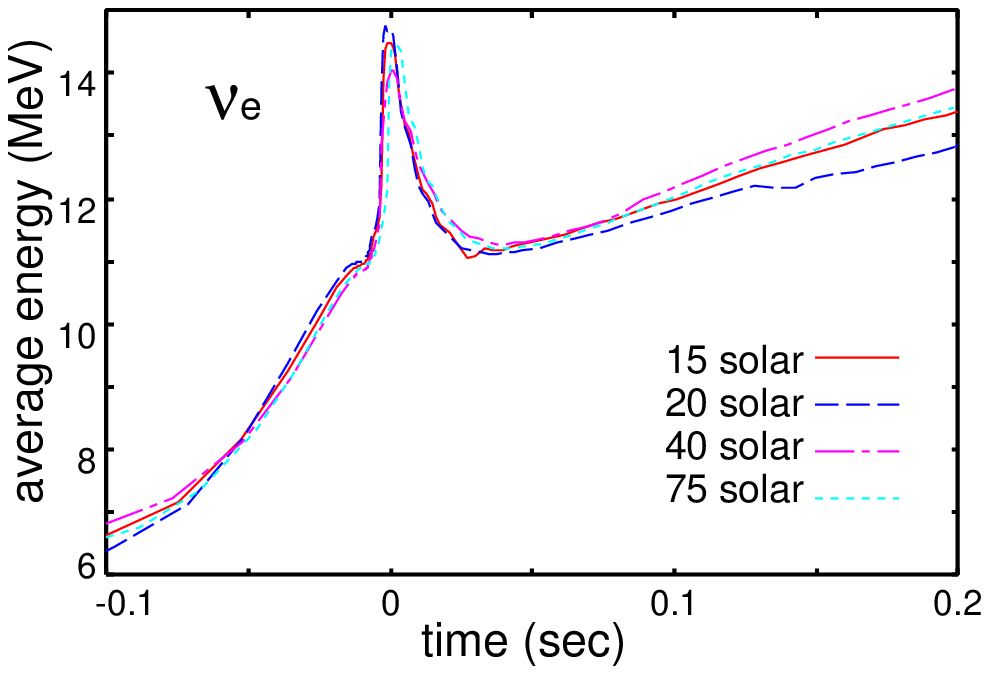}
 \end{center}
 \end{minipage}
\end{center}
\vspace{-0.5cm}
\begin{center}
 \begin{minipage}{7.0cm}
 \begin{center}
\epsfxsize = 6cm
 \epsfbox{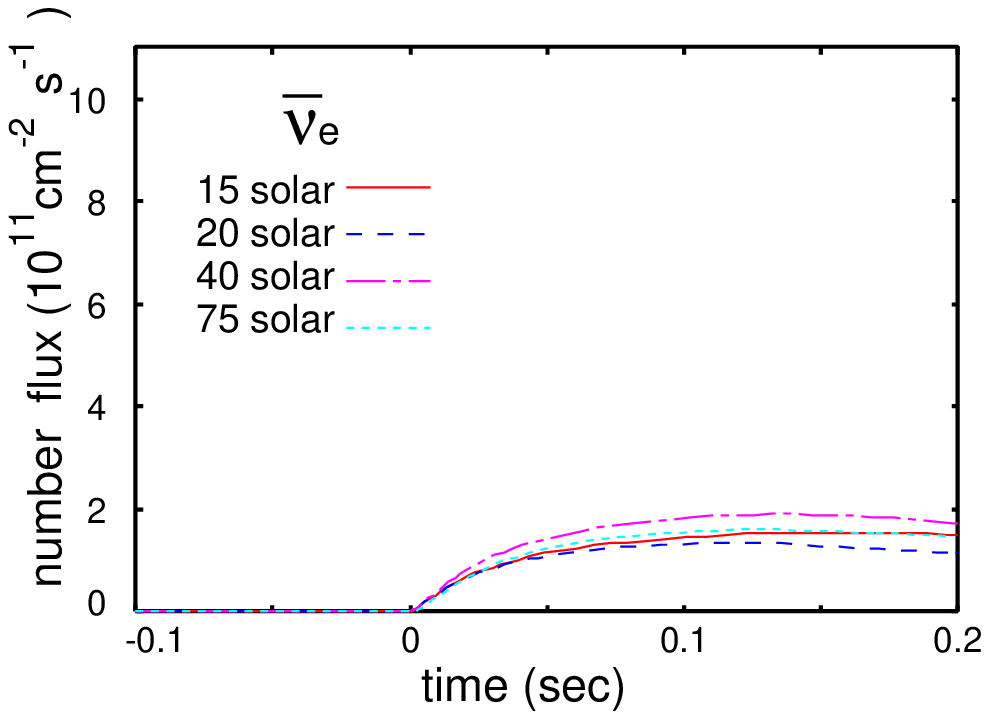}
 \end{center}
 \end{minipage}
\hspace{0.5cm}  
 \begin{minipage}{7.0cm}
 \begin{center}
\epsfxsize = 6cm
 \epsfbox{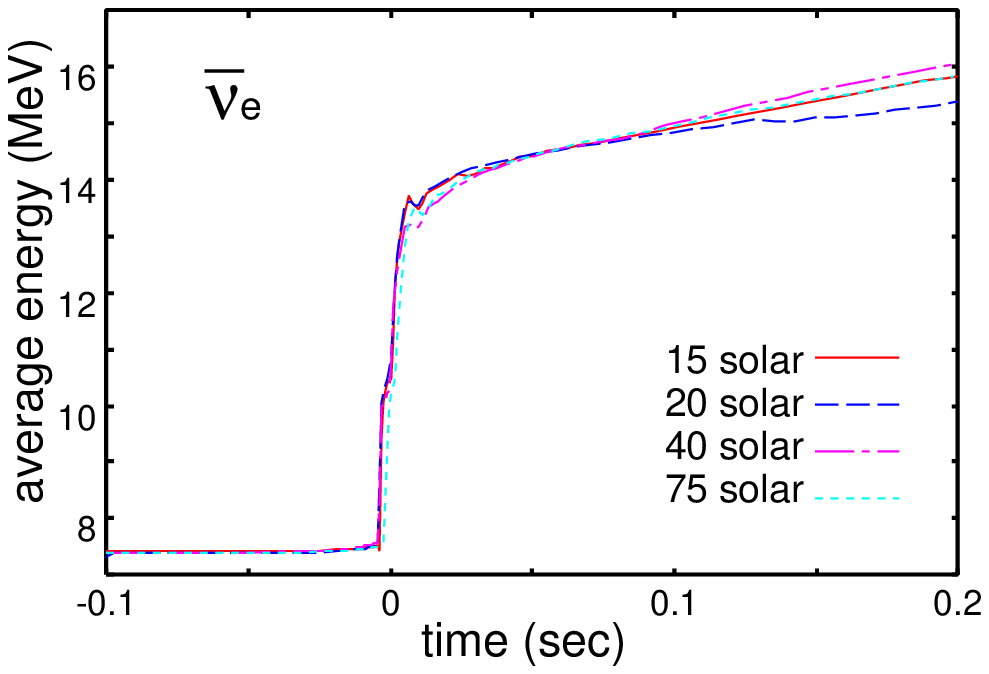}
 \end{center}
 \end{minipage}
\end{center}
\vspace{-0.5cm}
\begin{center}
 \begin{minipage}{7.0cm}
 \begin{center}
\epsfxsize = 6cm
 \epsfbox{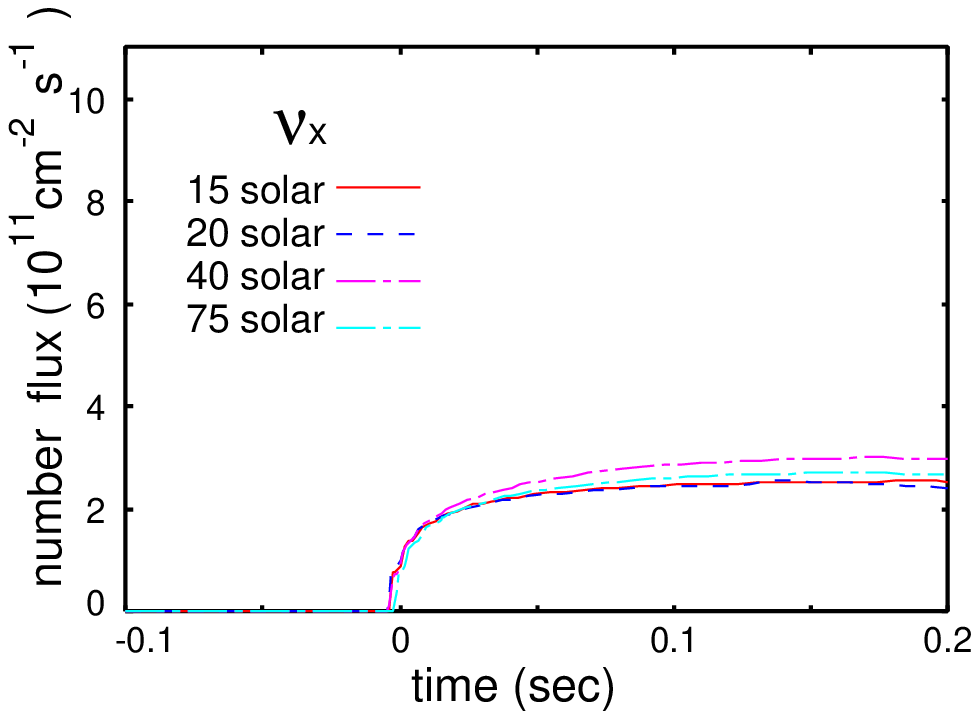}
 \end{center}
 \end{minipage}
\hspace{0.5cm}  
 \begin{minipage}{7.0cm}
 \begin{center}
\epsfxsize = 6cm
 \epsfbox{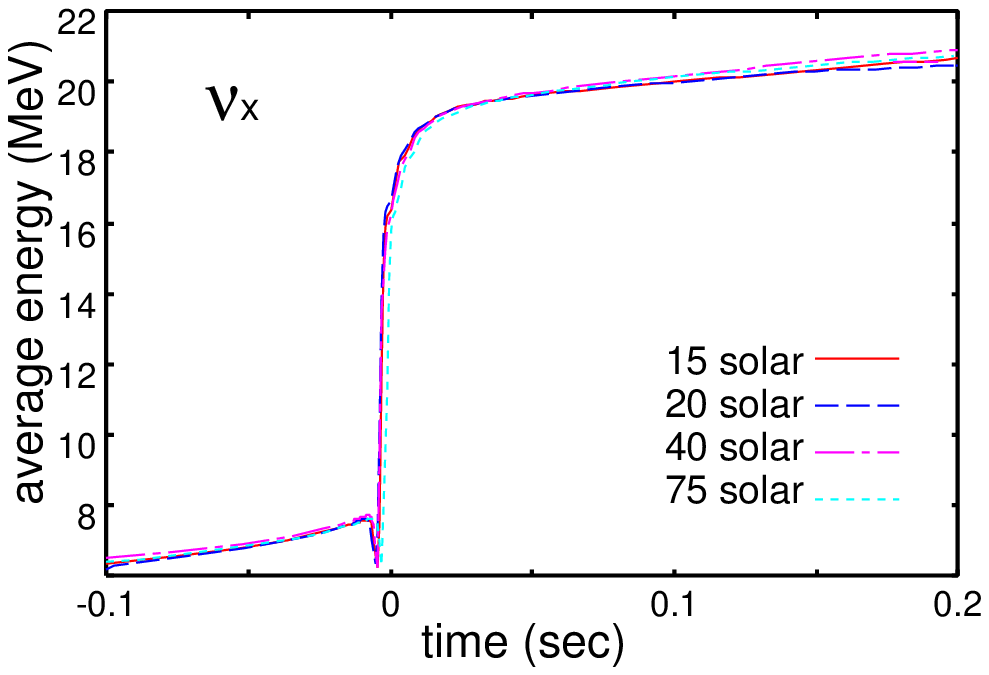}
 \end{center}
 \end{minipage}
\end{center}
\vspace{-0.5cm}
\caption{Left: Evolution of the number flux at the Earth of neutrinos from a supernova
by a progenitor with initial masses $15 M_{\odot}, 20 M_{\odot}, 40 M_{\odot}$
and $75 M_{\odot}$ at a distance of 10 kiloparsecs. Right: Evolution of neutrino average
energies. In all figures, time at the bounce is set to zero.
\label{fig:mass_lum_ave-ene}}
\end{figure}

Fig. \ref{fig:mass_lum_ave-ene} shows the evolution of the average neutrino energy
and number luminosity in the early phase up to 200 milliseconds after bounce. The calculation
is based on dynamical models of core-collapse supernovae in one spatial dimension,
employing a Boltzmann neutrino radiation transport algorithm, coupled to
Newtonian Lagrangean hydrodynamics and a consistent high-density nuclear equation of state.
Details of these simulations are described in \cite{ThompsonBurrowsPinto03}. As can be seen,
the major features of the early neutrino burst are almost independent of the initial mass.

Combined with the discussion on the scale height above, we conclude that the mantle structure
and the features of the neutrino burst depend little on the initial mass of the progenitor star
if $\sin^{2}{2 \theta_{13}} < 10^{-5}$ or $\sin^{2}{2 \theta_{13}} > 10^{-3}$. On one hand,
this means that we can not easily obtain information about the initial mass from observations
of neutrinos during the first 200 milliseconds after bounce. On the other hand, 
this situation is desirable for extracting information about the neutrino parameters.

In \cite{Kachelriess05}, Kachelriess studied dependence of neutronization burst on
input neutrino physics in numerical simulation, as well as progenitor mass. They found
neutronization burst to be relatively independent of the progenitor mass, electron capture
rate and equation of state at high densities. Especially, uncertainties in the number of
neutronization-burst events due to the input physics was estimated to be less than $10 \%$.
This feature can be used not only to probe neutrino parameters such as $\theta_{13}$ and
the mass hierarchy but also to determine the distance to the supernova. The latter is
important because it is likely that the supernova is optically obscured by dust if it occurs
around the galactic center. Their estimation is that distance to the next galactic supernova
is determined with a $6 \%$ error if we have a megaton water Cherenkov detector.

As we have discussed above, effects of some of neutrino parameters and input physics
on supernova neutrinos and their oscillation have been studied so far. Fortunately
to particle physics, the progenitor mass, electron capture rate and equation of state
have relatively small effects on neutrinos and we will not suffer from uncertainties
due to them. However, analyses so far are rather restricted to one-dimensional supernova,
that is, we do not understand multi-dimensional effects such as magnetic field and rotation.
This is because multi-dimensional simulation with sufficiently detailed treatment
of the neutrino physics is still challenging now and much progress in this field
is indispensable.

\clearpage

%%%%%%%%%%%%%%%%%%%%%%%%%%%%%%%%%%%%%%%%%%%%%%%%%%%%%%%%%%%%%%%%%%%%%%%%%%%%%%%%%%%%%%%%%%%%%%%
\subsection{Neutrinos from SN1987A}
%%%%%%%%%%%%%%%%%%%%%%%%%%%%%%%%%%%%%%%%%%%%%%%%%%%%%%%%%%%%%%%%%%%%%%%%%%%%%%%%%%%%%%%%%%%%%%%

On February 23 in 1987, a supernova was found in the Large Magellanic Cloud.
This is the closest supernova from us since a galactic supernova in 17th century.
From various analyses, this supernova, SN1987A, was identified as a type II supernova
whose progenitor star is a blue giant with mass $M \sim 10 M_{\odot}$ at 50 kpc.
Here we review the basic facts of neutrinos from SN1987A and their implication.

%%%%%%%%%%%%%%%%%%%%%%%%%%%%%%%%%%%%%%
\subsubsection{observational facts}
%%%%%%%%%%%%%%%%%%%%%%%%%%%%%%%%%%%%%%

\begin{table}[t]
\caption{Arrival times of neutrinos, energies of prompt electrons and angles between
momenta of neutrinos and corresponding prompt electrons at Kamiokande II and IMB.
\label{table:SN1987A_event}}
\begin{center}
\begin{tabular}{|cccc|cccc|} \hline
KII & & & & IMB & & &  \\ \hline
event & time & energy        & angle        & event & time & energy  & angle         \\
   & (sec)  & (MeV)          & (deg)        &   & (sec) & (MeV)      & (deg)         \\
1  & 0.000  & 20.0 $\pm$ 2.9 & 18  $\pm$ 18 & 1 & 0.000 & 38 $\pm$ 7 & 80  $\pm$ 10  \\
2  & 0.107  & 13.5 $\pm$ 3.2 & 40  $\pm$ 27 & 2 & 0.412 & 37 $\pm$ 7 & 44  $\pm$ 15  \\
3  & 0.303  & 7.5  $\pm$ 2.0 & 108 $\pm$ 32 & 3 & 0.650 & 28 $\pm$ 6 & 56  $\pm$ 20  \\
4  & 0.324  & 9.2  $\pm$ 2.7 & 70  $\pm$ 30 & 4 & 1.141 & 39 $\pm$ 7 & 65  $\pm$ 20  \\
5  & 0.507  & 12.8 $\pm$ 2.9 & 135 $\pm$ 23 & 5 & 1.562 & 36 $\pm$ 9 & 33  $\pm$ 15  \\
6  & 1.541  & 35.4 $\pm$ 8.0 & 32  $\pm$ 16 & 6 & 2.684 & 36 $\pm$ 6 & 52  $\pm$ 10  \\
7  & 1.728  & 21.0 $\pm$ 4.2 & 30  $\pm$ 18 & 7 & 5.010 & 19 $\pm$ 5 & 42  $\pm$ 20  \\
8  & 1.915  & 19.8 $\pm$ 3.2 & 38  $\pm$ 22 & 8 & 5.582 & 22 $\pm$ 5 & 104 $\pm$ 20  \\
9  & 9.219  & 8.6  $\pm$ 2.7 & 122 $\pm$ 30 &   &       &            &               \\
10 & 10.433 & 13.0 $\pm$ 2.6 & 49  $\pm$ 26 &   &       &            &               \\
11 & 12.439 & 8.9  $\pm$ 1.9 & 91  $\pm$ 39 &   &       &            &               \\
\hline
\end{tabular}
\end{center}
\end{table}

\begin{figure}[hbt]
\begin{center}
 \begin{minipage}{8.0cm}
 \begin{center}
\epsfxsize = 6cm
 \epsfbox{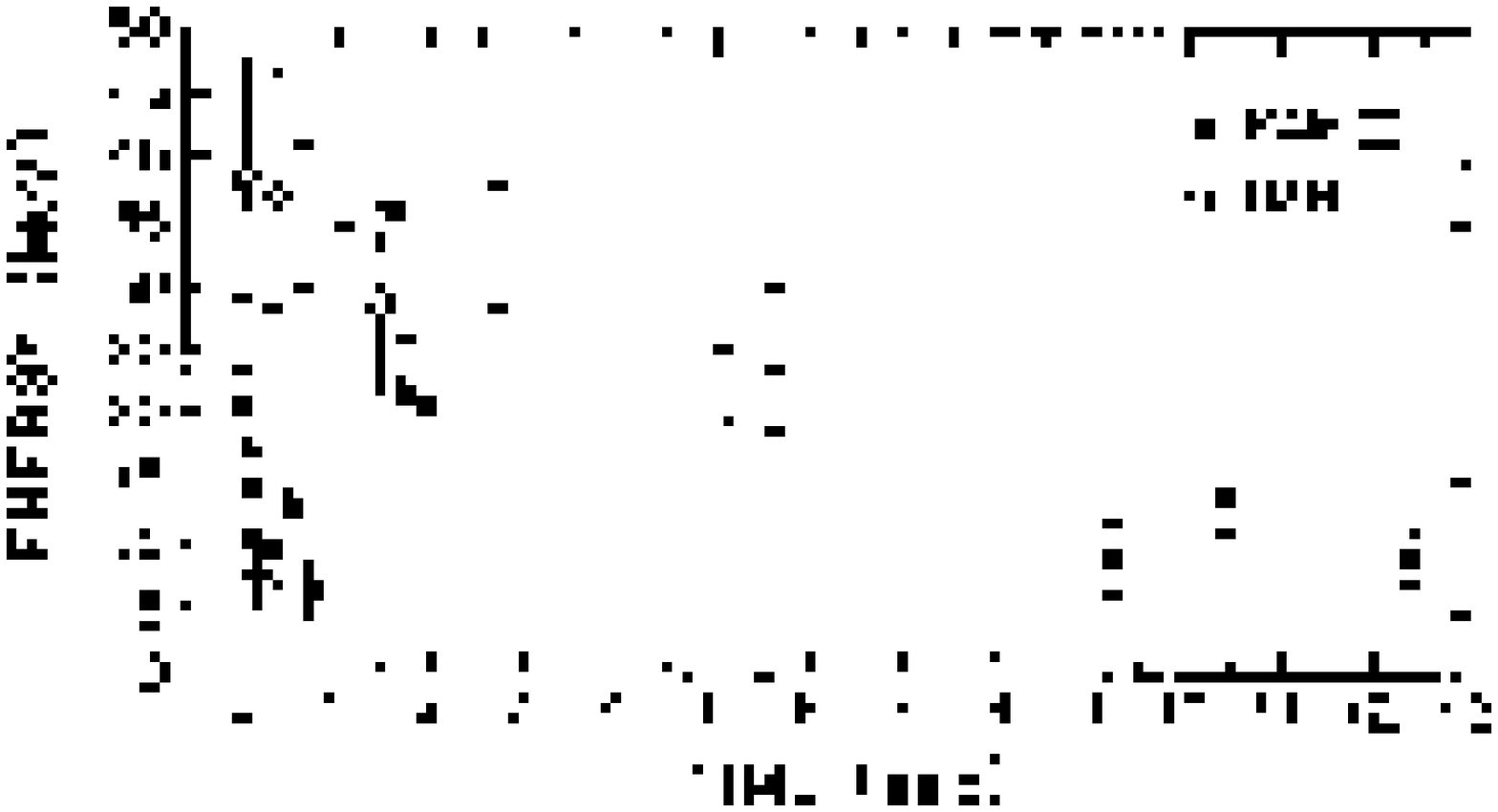}
 \caption{Arrival times and estimated energies of neutrinos observed at
 Kamiokande II and IMB \cite{Hirata88}.
 \label{fig:SN1987A_event}}
 \end{center}
 \end{minipage}
\hspace{0.5cm}  
 \begin{minipage}{6.5cm}
 \begin{center}
\epsfxsize = 6cm
 \epsfbox{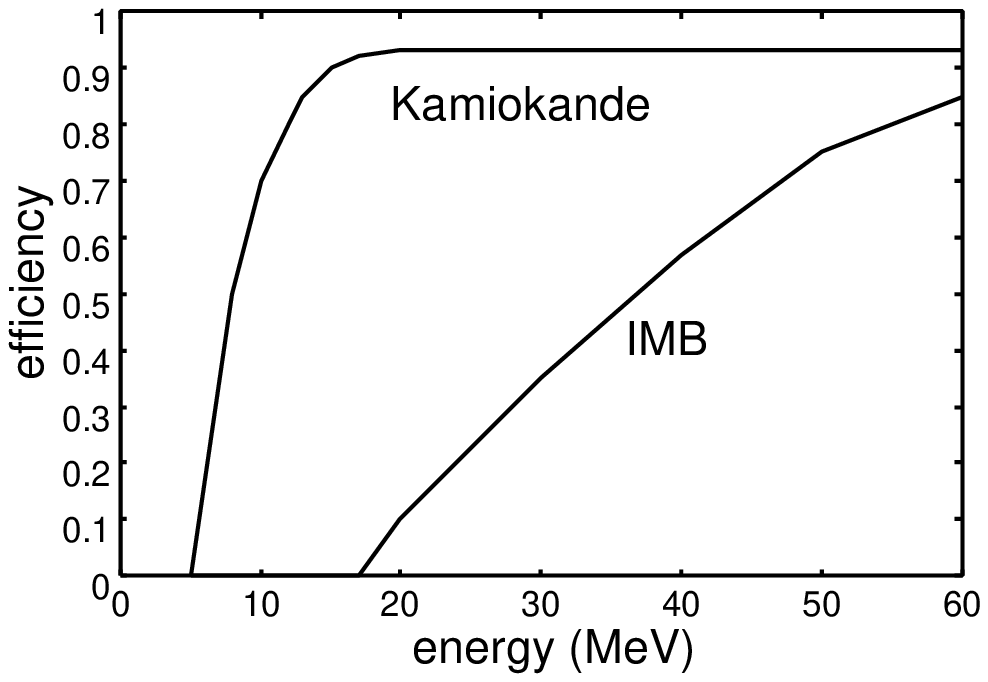}
 \caption{Detection efficiencies at Kamiokande II \cite{Hirata88} and IMB \cite{Bratton88}.
 \label{fig:KII-IMB_efficiency}}
 \end{center}
 \end{minipage}
\end{center}
\end{figure}

Just after the discovery of SN1987A by optical observations, it was expected that neutrinos
from such a close supernova must have been detected. Actually Kamiokande II \cite{Hirata87,Hirata88}
and Irvine-Michigan-Brookhaven detector (IMB) \cite{Bionta87,Bratton88} observed 11 and 8 events,
respectively. (For possible detection of neutrinos at Baksan, see \cite{Alekseev87}.)

Table \ref{table:SN1987A_event} shows the arrival times of neutrinos, energies of
prompt electrons and angles between momenta of neutrinos and corresponding prompt electrons
at Kamiokande II and IMB. Arrival times and estimated energies of neutrinos observed at
the Kamiokande II and IMB are plotted in Fig. \ref{fig:SN1987A_event}. From this figure,
one might think that neutrinos detected at the IMB have higher energies than those detected
at the Kamiokande II. But it is not obvious because we have to consider the difference in detection
efficiencies showed in Fig. \ref{fig:KII-IMB_efficiency}. As one can see, the Kamiokande II
had a high efficiency at more than 20 MeV while the IMB was not effective to detect low-energy
neutrinos. We summarize basic information on the observed neutrinos below.

\paragraph{duration}

The duration of neutrino events is 12.4 sec at the Kamiokande II and 5.6 sec at the IMB.
This is consistent with the diffusion timescale of neutrino, about 10 sec, discussed
in \ref{subsection:SN-neu}. Thus it is confirmed that neutrinos are confined in
the protoneutron star and escape by diffusion.

\paragraph{angular distribution of events}

Since both the Kamiokande II and IMB are water Cherenkov detectors, main events
come from $\bar{\nu}_{e} p \to e^{+} n$ which has about hundred times larger cross section
than that of electron scattering $\nu e^{-} \to \nu e^{-}$. A positron is emitted
isotropically in $\bar{\nu}_{e} p \to e^{+} n$ while an electron scattered by a neutrino
has a forward peak. Angular distribution in Table \ref{table:SN1987A_event} confirms
these arguments.

Let us focus on the first event at the Kamiokande II. In this event the primary electron
is emitted forward. The probability that an electron, which has an isotropic distribution,
is emitted forward inside $20^{o}$ is $3 \%$, which leads to the expectation value of 0.6
for total event number of 19. On the other hand, as we saw in Table \ref{table:SK_SN-event},
event number of electron scattering is about $5 \%$ of that of $\bar{\nu}_{e} p \to e^{+} n$
event. Thus, it is not obvious which reaction the first event came from. If the electron
scattering is the case, $\nu_{e}$ is the most likely for the event because $\nu_{e}$
has the largest cross section in electron scattering.

\paragraph{neutronization burst?}

If the first event at the Kamiokande II was $\nu_{e}$ event, it is possible that
the $\nu_{e}$ is emitted during the neutronization burst. However, the event number
from the neutronization burst was estimated to be about 0.01 in \cite{SatoSuzuki87}.
Thus it is unlikely that the first event at the Kamiokande II was from the
neutronization burst.

\paragraph{neutrino temperature and luminosity}

Neutrinos from supernova have roughly a thermal distribution though they do not exactly.
The effective temperature and luminosity of the {\it observed} $\bar{\nu}_{e}$ were estimated
by several authors \cite{ArafuneFukugita87,JankaHillebrandt89}. They are roughly
consistent with each other and give,
\begin{equation}
T_{\bar{\nu}_{e}} = (3 - 4) {\rm MeV}, ~~~
L_{\bar{\nu}_{e}} = (3 - 6) \times 10^{52} {\rm erg}.
\end{equation}
If we assume that all flavors have the same luminosity, the total neutrino luminosity
is about $3 \times 10^{53} {\rm erg}$. This is about the same as the binding energy
of a neutron star, which indicates that the current supernova theory is roughly correct.

%%%%%%%%%%%%%%%%%%%%%%%%%%%%%%%%%%%%%%
\subsubsection{constraints on neutrino properties}
%%%%%%%%%%%%%%%%%%%%%%%%%%%%%%%%%%%%%%

Observational feature of the neutrinos from SN1987A can be summarized as follows:
\begin{itemize}
\item Duration of the neutrino events is about 12 sec
\item Temperature of $\bar{\nu}_{e}$ is about $(3-4)$ MeV
\item Neutrino total energy is roughly the same as the binding energy of a neutron star
\end{itemize}
We can put constraints on any new physics, unknown processes and exotic particles which
lead to any contradictions with the above observational facts. Here we will discuss topics
related to neutrinos.

\paragraph{mass}

If neutrinos have mass, neutrinos with different energies have different velocities,
\begin{equation}
v_{\nu} \approx 1 - \frac{m_{\nu}^{2}}{2 E_{\nu}^{2}},
\end{equation}
so that the propagation time from the SN1987A to the earth are also different.
Denoting the departure and arrival times as $t_{\rm d}$ and $t_{\rm a}$, respectively,
the propagation time can be written as
\begin{equation}
t_{\rm a} - t_{\rm d} = \frac{D}{v_{\nu}}
\approx D \left(1 + \frac{m_{\nu}^{2}}{2 E_{\nu}^{2}}\right),
\end{equation}
where $D \sim 50 {\rm kpc}$ is the distance between SN1987A and the earth.
Therefore the difference in the propagation times of two neutrinos is,
\begin{equation}
|\Delta t_{\rm a} - \Delta t_{\rm d}| = 
\frac{1}{2} D m_{\nu}^{2} 
\frac{|E_{1}^{2} - E_{2}^{2}|}{E_{1}^{2}E_{2}^{2}},
\end{equation}
where $\Delta t_{\rm a}$ and $\Delta t_{\rm d}$ are the differences in the departure
and arrival times, respectively of the two neutrinos. As is stated above,
$\Delta t_{\rm a} < 12 {\rm sec}$ for any two of the 19 events. To obtain a constraint
on neutrino mass, we need a statistical analysis. Let us first consider a simple case
with the event 3 and 10 at Kamiokande II. In this case $\Delta t_{\rm a} = 10.1 {\rm sec}$,
and then we have
\begin{equation}
|10.1 {\rm sec} - \Delta t_{\rm e}|
= 0.06 {\rm sec} \left(\frac{m_{\nu}}{1 {\rm eV}}\right)^{2}.
\end{equation}
If we assume $\Delta t_{\rm d} \ll 10 {\rm sec}$, we obtain $m_{\nu} < 13 {\rm eV}$.
Statistical studies give slightly weak constraints, $m_{\nu} < (19-30) {\rm eV}$
\cite{ArnettRosner87,BahcallGlashow87,KolbStebbinsTurner87,SatoSuzuki87}

\paragraph{lifetime}
Since the estimated total energies of the emitted neutrinos is roughly the same
as the binding energy of a neutron star, it can be said that most of the neutrinos
did not decay before reaching the earth. Thus, denoting neutrino lifetime as $\tau_{\nu}$,
\begin{equation}
\frac{E_{\nu}}{m_{\nu}} \tau_{\nu} \geq
\frac{D}{c} \approx 5 \times 10^{12} {\rm sec},
\end{equation}
then we have
\begin{equation}
\tau_{\nu} \left(\frac{E_{\nu}}{20 {\rm MeV}}\right) \geq
2.5 \times 10^{5} \left(\frac{m_{\nu}}{1 {\rm eV}}\right) {\rm sec}.
\end{equation}
Note that neutrinos cannot decay without having mass.

\begin{figure}[hbt]
\begin{center}
\epsfbox{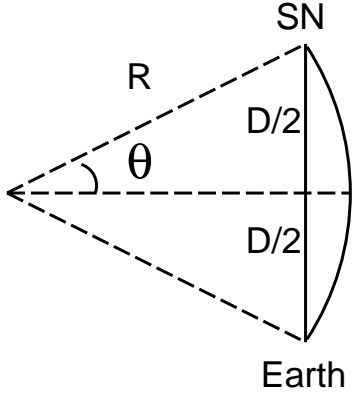}
\end{center}
\vspace{-0.5cm}
\caption{Deflection of path of a charged particle by magnetic field.
\label{fig:path_charge}}
\end{figure}

\paragraph{electric charge}

If neutrinos have electric charge, even if it is extremely small, neutrino trajectory
is deflected by the galactic magnetic field. Then the trajectory of a low-energy neutrino
is longer than that of a high-energy neutrino. Larmor radius of a neutrino with charge $Q_{\nu}$
in a magnetic field $B_{\rm gal}$ is
\begin{equation}
R = \frac{m_{\nu} v}{Q_{\nu} e B_{\rm gal}}.
\end{equation}
From Fig. \ref{fig:path_charge}, the extra distance due to charge is
\begin{equation}
2R \theta - D \approx \frac{R \theta^{3}}{3}
= \frac{D^{3} Q_{\nu}^{2} e^{2} B_{\rm gal}^{2}}{24 m_{\nu}^{2}}
\left(1+\frac{m_{\nu}^{2}}{E_{\nu}^{2}}\right).
\end{equation}
Then time delay of two neutrinos is,
\begin{equation}
\Delta t_{\rm a} = \Delta t_{\rm d} + 
\frac{D^{3} Q_{\nu}^{2} e^{2} B_{\rm gal}^{2}}{24}
\left(\frac{1}{E_{1}^{2}} - \frac{1}{E_{2}^{2}}\right).
\end{equation}
Taking $\Delta t_{\rm d} < 20 {\rm sec}$ and $B_{\rm gal} = 10^{-6} {\rm Gauss}$,
we obtain $Q_{\nu} < 10^{-18}$. A statistical analysis considering the inhomogeneity of
the galactic magnetic field gives $Q_{\nu} < 10^{-17}$ \cite{BahcallSpiegelPress}.

\paragraph{weak equivalence principle}

The difference between arrival times of photons and neutrinos was several hours.
This means that the gravitational constants for them are not so different. Specifically,
\begin{equation}
\frac{G_{\nu} - G_{\gamma}}{G_{\nu} + G_{\gamma}} < 10^{-3},
\end{equation}
is obtained in \cite{Longo88,KraussTremaine88}.

%%%%%%%%%%%%%%%%%%%%%%%%%%%%%%%%%%%%%%
\subsubsection{neutrino oscillation}
%%%%%%%%%%%%%%%%%%%%%%%%%%%%%%%%%%%%%%

Here we discuss neutrino oscillation of neutrinos from SN1987A.
Because most of the observed neutrinos were $\bar{\nu}_{e}$, we concentrate on
the anti-neutrino sector. Then the key point of the neutrino oscillation dynamics is
the adiabaticity of the H-resonance if it exists. If the mass hierarchy is inverted
and $\theta_{13}$ is so large that the H-resonance is perfectly adiabatic, $\bar{\nu}_{e}$ flux
can be written as (see Eqs. (\ref{eq:flux-mix}) and (\ref{eq:effective_survival-prob})),
\begin{eqnarray}
F_{\bar{e}} &=& |U_{e3}|^{2} F_{\bar{e}}^{0} + (1 - |U_{e3}|^{2}) F_{x}^{0} \nonumber \\
&\approx& F_{x}^{0},
\end{eqnarray}
where we used $|U_{e3}|^{2} \ll 1$. Thus the $\bar{\nu}_{e}$ flux observed at the earth
reflects directly the original flux of $\nu_{x}$.

\begin{figure}[hbt]
\begin{center}
\epsfxsize = 8cm
\epsfbox{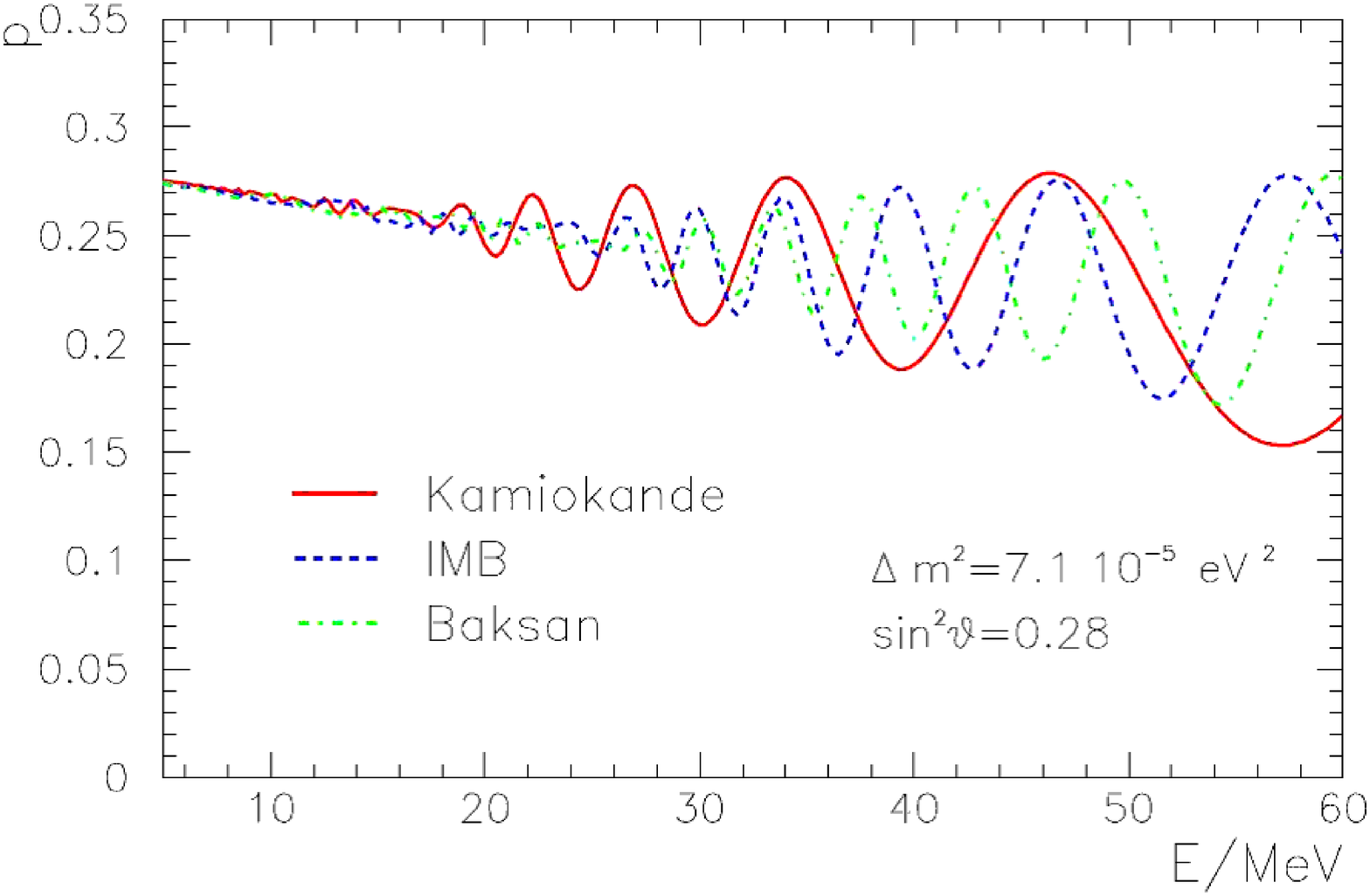}
\end{center}
\vspace{-0.5cm}
\caption{Permutation factor $(1-\bar{p})$ as a function of the neutrino energy
at Kamiokande II, IMB and Baksan \cite{LunardiniSmirnov04}.
\label{fig:permutation}}
\end{figure}

On the other hand, if the mass hierarchy is normal or $\theta_{13}$ is very small,
the situation becomes more complicated, which was studied extensively by Lunardini and Smirnov
\cite{LunardiniSmirnov04}. To interpret the data in terms of neutrino oscillation,
first we have to calculate the survival probability of $\bar{\nu}_{e}, \bar{p}$. It is important
to note that neutrinos detected Kamiokande II and IMB have different survival probabilities
due to the different positions of the two detectors on the earth. Fig. \ref{fig:permutation}
shows the permutation factors $(1-\bar{p})$ as a function of the neutrino energy at Kamiokande II,
IMB and Baksan. Here the H-resonance was assumed to be perfectly non-adiabatic and neutrino
oscillation parameters were set as $\Delta m_{12}^{2} = 7.1 \times 10^{-5} {\rm eV}^{2}$ and
$\sin^{2}{\theta_{12}} = 0.28$. Although the average behavior is the same for all detectors
the phase of oscillation is different. Further, due to the larger distance crossed
by neutrinos inside the earth, for the IMB detector the frequency of the oscillatory curve
in the energy scale is twice as large as the frequency for Kamiokande II.

\begin{figure}[hbt]
\begin{center}
\epsfxsize = 12cm
\epsfbox{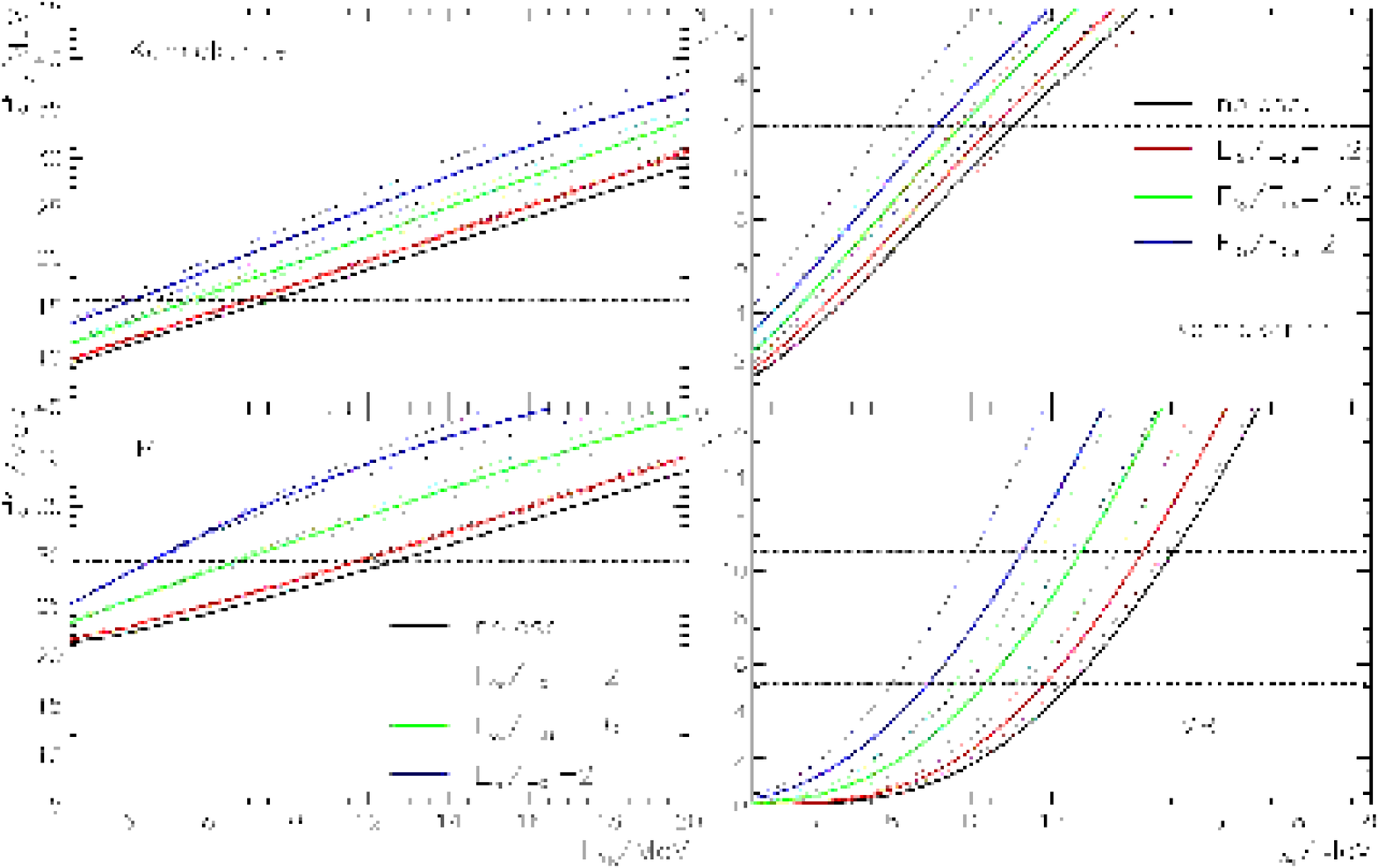}
\end{center}
\vspace{-0.5cm}
\caption{Left: Average energy of positrons, $\bar{\epsilon}^{i}$, in the detectors Kamiokande II
and IMB as a function of the average energy of the original $\bar{\nu}_{e}$, $E_{0\bar{e}}$,
and different values of $E_{0x}/E_{0e}$ and $L_{x}/L_{e}$. 
Right: The predicted numbers of events at Kamiokande II and IMB as a function of $E_{0e}$.
Here $L_{e} = 5.3 \times 10^{52}$ erg was used. In both figures, the horizontal lines represent
the experimental results with the $1\sigma$ error and the solid, dashed and dotted-dashed lines
correspond to $L_{x}/L_{e} = 1, 1.5, 0.667$ respectively. Both figures are from
\cite{LunardiniSmirnov04}.
\label{fig:SN1987A_osc}}
\end{figure}

In \cite{LunardiniSmirnov04}, they took the spectrum parameterization in
Eq. (\ref{eq:spectrum-Keil}). Then, given the parameters
$E_{0e}, E_{0x}, L_{e}, L_{x}, \beta_{e}$ and $\beta_{x}$, we can calculate
the observed average energies and event numbers expected at Kamiokande II and IMB.
Fig. \ref{fig:SN1987A_osc} shows the average energy of positrons (left) and event numbers
(right) at the Kamiokande II and IMB as functions of the average energy of the original
$\bar{\nu}_{e}$, $E_{0\bar{e}}$, and different values of $E_{0x}/E_{0e}$ and $L_{x}/L_{e}$. 
Neutrino oscillation parameters are set to the same value as in Fig. \ref{fig:permutation}.

Let us consider first the no-oscillation case. From the left of Fig. \ref{fig:SN1987A_osc},
we obtain the $\bar{\nu}_{e}$ average energy: from Kamiokande II data,
\begin{equation}
E_{0e}^{\rm K2} = 8.7 \pm 0.9 {\rm MeV},
\end{equation}
and from IMB data,
\begin{equation}
E_{0e}^{\rm IMB} = 14.7 \pm 1.9 {\rm MeV}.
\end{equation}
Thus, the IMB result is more than $3 \sigma$ above the Kamiokande II result. According to
the right of Fig. \ref{fig:SN1987A_osc}, the expected event numbers at Kamiokande II and IMB for
the energies $E_{0e}^{\rm K2}$ and $E_{0e}^{\rm IMB}$ and $L_{e} = 5.3 \times 10^{52} {\rm erg}$
are $N_{e}^{\rm K2} = 7.8 \pm 1.5$ and $N_{e}^{\rm IMB} = 10^{+6}_{-4}$, respectively.
Therefore, to reproduce the observed event numbers at Kamiokande II and IMB,
\begin{equation}
L_{e}^{\rm K2} = 8.2 \times 10^{52} ~ {\rm erg}, ~~~
L_{e}^{\rm IMB} = 4.2 \times 10^{52} ~ {\rm erg},
\end{equation}
are required, respectively. Thus, the IMB signal implies about 2 times higher average
energy and 2 times smaller luminosity in comparison with Kamiokande II.

Neutrino oscillation improves the agreement between the Kamiokande II and IMB.
For $E_{0x}/E_{0e} = 1.6$ and $L_{x}/L_{e} = 1$, we obtain,
\begin{equation}
E_{0e}^{\rm K2} = 6.8 \pm 0.8 ~ {\rm MeV}, ~~~
E_{0e}^{\rm IMB} = 10.3 \pm 1.7 ~ {\rm MeV},
\end{equation}
which now agree in $2 \sigma$ level. Actually, a "concordance" model with
\begin{eqnarray}
&& E_{0e} = 8 {\rm MeV}, ~~~ E_{0x} = 12.8 {\rm MeV}, \nonumber \\
&& L_{e} = L_{x} = 8 \times 10^{52} {\rm erg}, \nonumber \\
&& \beta_{e} = \beta_{x} = 4,
\end{eqnarray}
gives the best fit to the all available data with $\chi^{2} = 11.0$ while
the best-fit no-oscillation model with
\begin{equation}
E_{0e} = 11 {\rm MeV}, ~~~ L_{e} = 5.3 \times 10^{52} {\rm erg}, ~~~ \beta_{e} = 4,
\end{equation}
gives $\chi^{2} = 16.2$. Thus neutrino oscillation leads to a certain improvement of
the global fit of the data. The improvement requires lower average energy of the original
$\bar{\nu}_{e}$ spectrum and larger $\bar{\nu}_{e}$ luminosity. The combination of
smaller average energy and larger luminosity corresponds to a larger radius
of the neutrinosphere: $R_{\rm ns} \propto E_{0e}^{-2} L_{e}^{1/2}$. It follows that
in the concordance model $R_{\rm ns}$ is about 2.4 times larger than in the no-oscillation
model, which gives $R_{\rm ns} = (20-30) {\rm km}$ \cite{LoredoLamb02}.

The concordance model gives rather small average energies of $\bar{\nu}_{e}$ and $\nu_{x}$
compared to those predicted by numerical simulations. The situation becomes even worse
with adiabatic H-resonance, where most of the observed neutrinos were originally $\nu_{x}$s.
Because $\nu_{x}$s are expected to have larger average energy than $\bar{\nu}_{e}$,
$E_{0x}$ must be substantially smaller than that predicted by numerical simulations.
Anyway, since the event numbers of SN1987A neutrinos would be too small to make
a definitive conclusion. We are looking forward to seeing the next galactic supernova.

%%%%%%%%%%%%%%%%%%%%%%%%%%%%%%%%%%%%%%%%%%%%%%%%%%%%%%%%%%%%%%%%%%%%%%%%%%%%%%%%%%%%%%%%%%%%%%%
\subsection{Neutrino detectors \label{subsection:detector}}
%%%%%%%%%%%%%%%%%%%%%%%%%%%%%%%%%%%%%%%%%%%%%%%%%%%%%%%%%%%%%%%%%%%%%%%%%%%%%%%%%%%%%%%%%%%%%%%

It took 25 years to prove the existence of neutrinos experimentally since Pauli
predicted theoretically in 1931. It took another 30 years to use neutrinos as tools in
particle physics and astrophysics. This is because neutrinos have very weak interaction with
other particles. Since the cross section of neutrino reaction is typically $10^{-40} {\rm cm}^{2}$,
mean free path of neutrinos in water is about $10^{16} {\rm cm}$. This is why the current
neutrino detectors are huge in volume. In this section we explain various neutrino detectors
and their detection principles.

%%%%%%%%%%%%%%%%%%%%%%%%%%%%%%%%%%%%%%
\subsubsection{water, heavy water and ice}
%%%%%%%%%%%%%%%%%%%%%%%%%%%%%%%%%%%%%%

Neutrino interactions in water produces a relativistic charged particle which is mostly
an electron or positron. If the velocity of the charged particle is larger than
velocity of light in water, the charged particle emits Cherenkov light. This Cherenkov
light is the signal of water Cherenkov detector. There are many neutrino detectors
which utilize this method, starting from Kamiokande and IMB to the current experiment
SuperKamiokande (SK), Sudbury Neutrino Observatory (SNO) and IceCube (AMANDA). They are based
on the same detection strategy but have different features which come from different target media:
SuperKamiokande, SNO and IceCube use pure water, heavy water and antarctic ice, respectively.

\paragraph{neutrino reactions in water}

\begin{figure}[hbt]
\begin{center}
 \begin{minipage}{7.0cm}
 \begin{center}
 \epsfxsize = 6cm
 \epsfbox{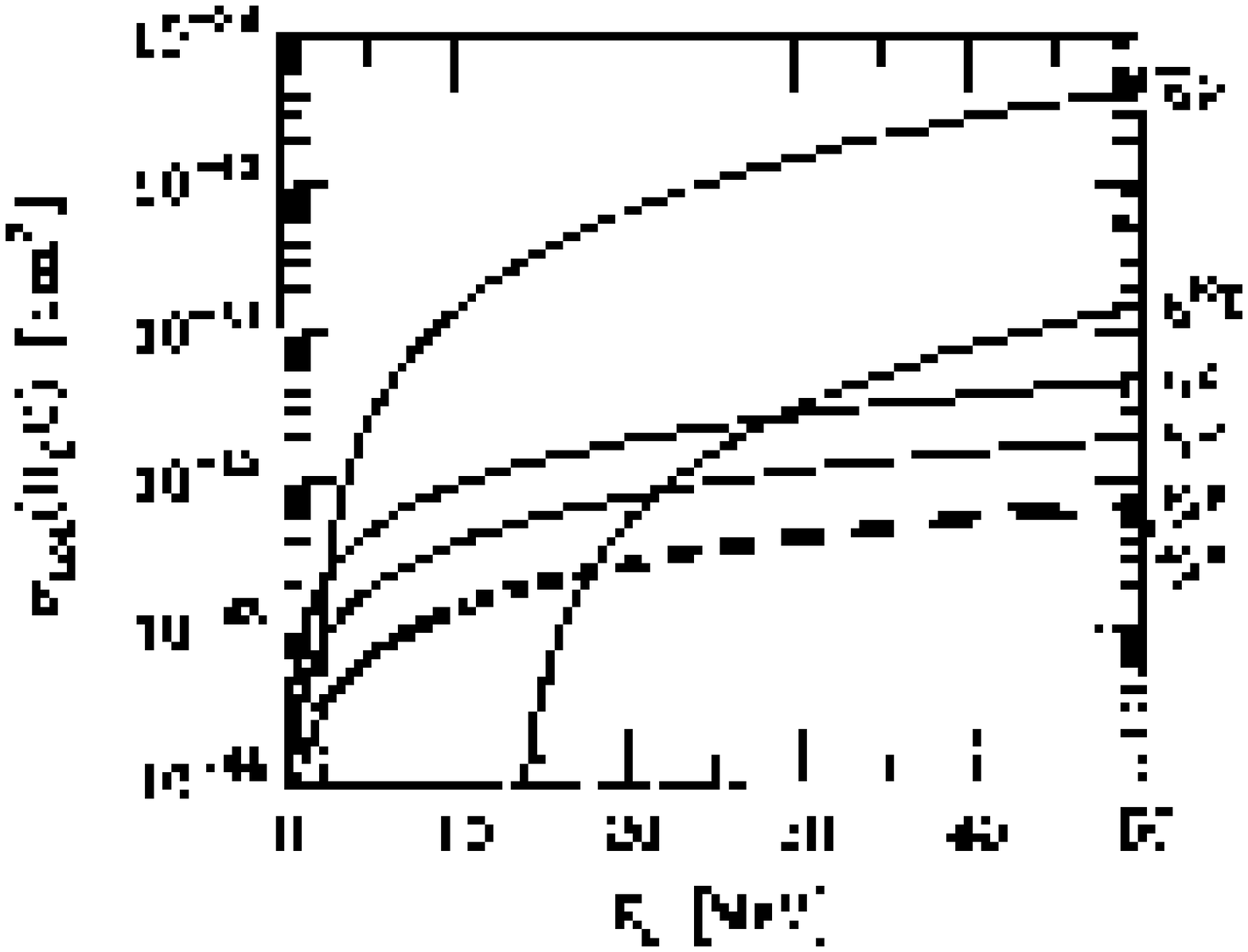}
 \caption{Cross sections of neutrino interactions in water. The dominant interaction
 is $\bar{\nu}_{e} p \rightarrow e^{+} n$ at every energy.
 \label{fig:cross_water}}
 \end{center}
 \end{minipage}
\hspace{0.5cm}  
 \begin{minipage}{7.0cm}
 \begin{center}
\epsfxsize = 6cm
 \epsfbox{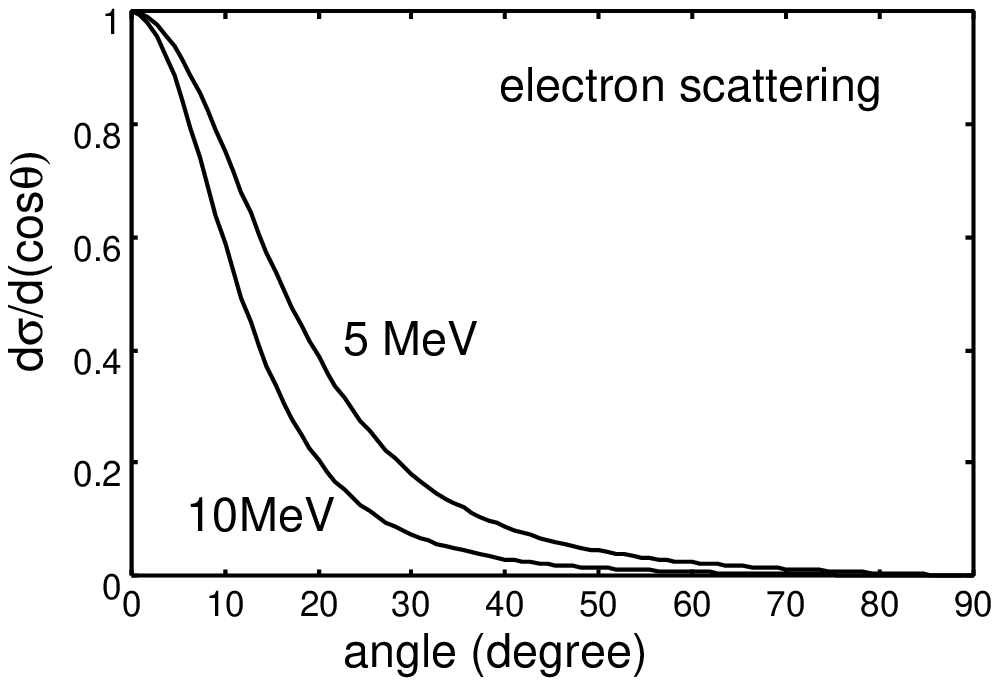}
 \caption{Angular distribution of scatter electron in $\nu_{e} e^{-} \to \nu_{e} e^{-}$
 with $E_{\nu_{e}} = 5$ and 10 MeV.
 \label{fig:ES_angle_degree}}
 \end{center}
 \end{minipage}
\end{center}
\end{figure}

Cross sections of neutrino reactions in water at low energies, $s \leq m_{W}^{2}$,
where $s$ is the center-of-mass energy and $m_{W}$ is the $W$ boson mass, are given by,
\begin{eqnarray}
\sigma(\nu_{e}e \rightarrow \nu_{e}e) & = & 9.33 \times 10^{-44} 
\left(\frac{E_{\nu}}{10 {\rm MeV}} \right) {\rm cm}^{2}, \\
\sigma(\bar{\nu}_{e}e \rightarrow \bar{\nu}_{e}e) & = & 3.88 \times 10^{-44} 
\left(\frac{E_{\nu}}{10 {\rm MeV}} \right) {\rm cm}^{2}, \\
\sigma(\nu_{x}e \rightarrow \nu_{x}e) & = & 1.59 \times 10^{-44} 
\left(\frac{E_{\nu}}{10 {\rm MeV}} \right) {\rm cm}^{2}, \\
\sigma(\bar{\nu}_{x}e \rightarrow \bar{\nu}_{x}e) & = & 1.30 \times 10^{-44} 
\left(\frac{E_{\nu}}{10 {\rm MeV}} \right) {\rm cm}^{2}, \\
\sigma(\nu_{\mu}e \rightarrow \mu\nu_{e} (s \gg m_{\mu}^{2})) & = & 
1 \times 10^{-41} \left(\frac{E_{\nu}}{1 {\rm GeV}} \right) {\rm cm}^{2}, \\
\sigma(\bar{\nu}_{e}p \rightarrow e^{+}n) & = & 9.77 \times 10^{-42} 
\left(\frac{E_{\nu}}{10 {\rm MeV}} \right)^{2} {\rm cm}^{2}, \label{eq:inverse-beta} \\
\sigma(\nu_{e}{}^{16}{\rm O} \rightarrow e^{-}{}^{16}{\rm F}) & = & 
1.1 \times 10^{-42} 
\left(\frac{E_{\nu} - 13 {\rm MeV}}{10 {\rm MeV}} \right)^{2} {\rm cm}^{2}, \\
\sigma(\bar{\nu}_{e}{}^{16}{\rm O} \rightarrow e^{-}{}^{16}{\rm N}) & = & 
1.1 \times 10^{-42} 
\left(\frac{E_{\nu} - 13 {\rm MeV}}{10 {\rm MeV}} \right)^{2} {\rm cm}^{2},
\end{eqnarray}
and shown in Fig. \ref{fig:cross_water}.

Among these, the inverse beta decay (\ref{eq:inverse-beta}) has the largest cross section
at all energies so that most of the events at water Cherenkov come from this reaction.
Because the emitted positron has an almost isotropic distribution for low-energy neutrinos
like supernova neutrinos, it is difficult to know the direction of the incident neutrino.

On the other hand, electron scattering event gives us information of the direction of
the incident neutrino, although the cross section is much smaller than that of the inverse
beta decay. The angle between the scattered electron and the incident neutrino is given by,
\begin{equation}
\cos{\theta} =
\frac{E_{\nu} + m_{e}}{E_{\nu}} \left(\frac{1}{1 + 2m_{e}/E_{e}}\right)^{1/2},
\end{equation}
where $m_{e}$ and $E_{e}$ are the electron mass and electron energy, respectively.
Angular distribution of the scattered electron in $\nu_{e} e^{-} \to \nu_{e} e^{-}$ is
plotted in Fig. \ref{fig:ES_angle_degree}. Here it should be noted that electron scattering
occurs for all the flavors and they cannot be distinguished at low energies ($s \ll m_{\mu}$),
although the cross sections are different. The $\nu_{e} e^{-}$ cross section is larger than
those of the other flavors because electron scattering of $\nu_{e}$ is contributed from
the charged current interaction as well as the neutral current interaction.

\paragraph{SuperKamiokande}

The SuperKamiokande \cite{SKHP} detector is a cylindrical 50,000 ton water Cherenkov detector
located at the Kamioka mine in Japan. It lies 1,000 m underneath the top of Mt. Ikenoyama,
(i.e. 2,700 m water equivalent underground), resulting in a cosmic ray muon rate of 2.2 Hz,
a reduction of 10.5 compared to the rate at the surface. As a supernova neutrino detector,
SK has a fiducial volume of 32,000 ton.The detector is optically separated into two regions,
the inner and outer detectors. The inner detector of the SuperKamiokande-I detector,
which operated from April 1996 to July 2001, was instrumented with 11,146 50-cm diameter
inward facing photomultiplier tubes (PMTs) which provide $40 \%$ photocathode coverage.
This photocathode coverage made it possible to detect low energy electrons
down to $\sim 5$ MeV. Also, SK uses the anti-counter which surrounds the inner detector
detect and remove events due to cosmic-ray muons. The direction of a charged particle is
reconstructed using the directionality of the Cherenkov light. Angular resolution is about
25 degree for a 10 MeV electron. Energy resolution $\sigma$ for low-energy neutrinos is given by,
\begin{equation}
\sigma = 1.5 \left( \frac{E}{10 {\rm MeV}} \right)^{\frac{1}{2}} {\rm MeV}.
\end{equation}
As discussed in section \ref{subsection:ex_nu-osc}, SK has been playing a central role
in the field of neutrino oscillation experiment by observing neutrinos from the sun, atmosphere
and accelerator. It is also expected to give tremendous information on supernova if it occurs
in the future. For a detailed description of SK detector, see \cite{SK99}.

Due to an unfortunate accident in 2001, $60 \%$ of the PMTs were destroyed and the observation
was interrupted for a while. However, the observation was restarted by redistributing
the survived PMTs and SK is expected to resume normal observation with the original
number of PMTs in 2006.

\paragraph{IceCube}

IceCube \cite{IceCubeHP} is an extended experiment of AMANDA \cite{AMANDAHP,AMANDA02} and
consists of an array of 4800 optical modules on 80 strings, regularly spaced by 125 m
in antarctic ${\rm km}^{3}$ ice. It covers an area of approximately $1 {\rm km}^{2}$,
with the optical modules at depths of 1.4 to 2.4 km below surface. Each string carries
60 optical modules, vertically spaced by 17 m. IceCube is primarily designed to observe
high-energy neutrinos ($E > 1 {\rm TeV}$) from astrophysical sources. In order to reach
the large volume needed to detect the expected small fluxes at high energies,
the density of optical modules must be too sparse to measure low-energy neutrinos
such as solar neutrinos. However, it is expected that IceCube can detect
a supernova neutrino burst because the Cherenkov glow of the ice can be identified
as time-correlated noise among all phototubes. The observed quantity is the number
of Cherenkov photons caused by supernova neutrinos as a function of time.
Thus, although IceCube cannot identify individual neutrino, it can measure
neutrino luminosity of supernova. For a detailed description of IceCube, see \cite{IceCube04}.
Supernova neutrino detection at IceCube was studied in detail in
\cite{HalzenJacobsenZas96,DigheKeilRaffelt03a}.

\paragraph{Sudbury Neutrino Observatory}

\begin{figure}[hbt]
\begin{center}
\epsfbox{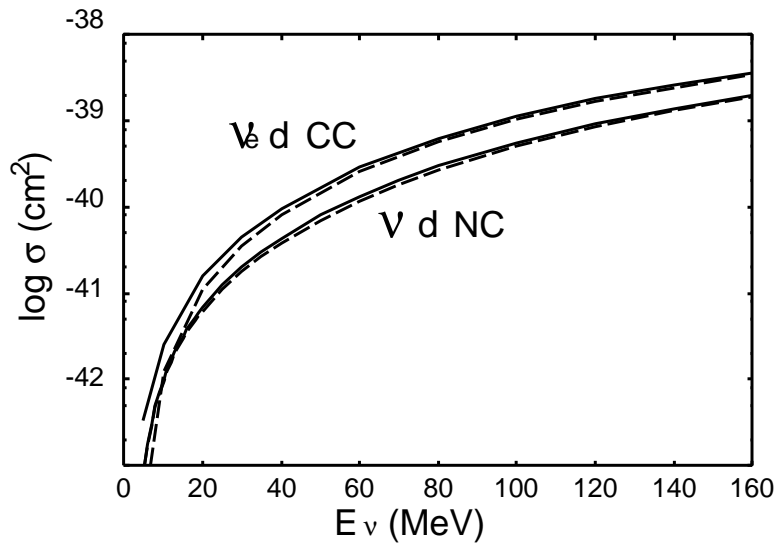}
\end{center}
\vspace{-0.5cm}
\caption{Cross sections of neutrino interactions with deuteron. Solid lines are neutrinos
and dashed lines are antineutrinos.
\label{fig:cross_d}}
\end{figure}

Sudbury Neutrino Observatory (SNO) \cite{SNOHP} is located in a large cavity excavated
at the 2,039 m level (6,000 m water equivalent) in the Creighton mine near Sudbury.
The 1,000 tons of heavy water are contained in an acrylic vessel surrounded by
a light water shield.

SNO's uniqueness is the use of heavy water as its neutrino detection medium.
Neutrinos interact in heavy water in two additional ways not possible in ordinary water.
One is through the charged current interaction:
\begin{eqnarray}
\nu_{e} + d & \rightarrow & e^{-} + p + p ~ (E_{\rm th} = 1.44 {\rm MeV}),
\label{eq:d-CC} \\
\bar{\nu}_{e} + d & \rightarrow & e^{+} + n + n ~ (E_{\rm th} = 4.03 {\rm MeV}),
\end{eqnarray}
whose cross sections are plotted in Fig. \ref{fig:cross_d}.
These neutrino absorption reactions can only happen if the neutrino is an electron neutrino
or electron anti-neutrino at low-energies ($s < m_{\mu}^{2}$). Thus the neutrino absorption
reaction exclusively counts electron neutrinos and electron anti-neutrinos. 

Another is the deuteron breakup reaction:
\begin{equation}
\nu + d \rightarrow \nu + n + p ~ (E_{\rm th} = 2.22 {\rm MeV}).
\end{equation}
This reaction occurs through the neutral current interaction and it occurs with equal probability
for all neutrino flavors. In this reaction, no new charged particle is created and
the free neutron cannot by itself create Cerenkov light. However, after scattering off of
the nuclei in the heavy water it is eventually captured by another deuteron,
creating a tritium nucleus and releasing a high energy $\gamma$ ray. This $\gamma$ ray
then scatters an electron in the heavy water and it is this secondary electron
which creates the Cerenkov light. 

SNO played a critical role in solving the solar neutrino problem with its ability to identify
$\nu_{e}$ and neutral current events, as discussed in section \ref{subsubsection:solar}.
This feature will also have a great impact on observation of supernova neutrinos.
For a detailed description of SNO, see \cite{SNO00}.

%%%%%%%%%%%%%%%%%%%%%%%%%%%%%%%%%%%%%%
\subsubsection{scintillator}
%%%%%%%%%%%%%%%%%%%%%%%%%%%%%%%%%%%%%%

As a high energy particle propagates in medium, it loses energy exciting electrons in the medium.
A part of the deposited energy is emitted as scintillation photons. Scintillator detector
uses the scintillation photons as a signal to detect high energy particles. Here we summarize
basic feature of scintillator detector as a neutrino detector.

As to neutrino detector, liquid hydrocarbon is often used as a medium. In this medium,
neutrinos interact with electrons, protons and carbon nuclei. Dominant contribution
to the events comes from inverse beta decay,
\begin{equation}
\bar{\nu}_{e} + p \rightarrow e^{+} + n,
\end{equation}
where the positron has almost the same energy as that of the incident neutrino.
Then the neutron is absorbed into proton in about $170 \mu {\rm sec}$,
\begin{equation}
n + p \rightarrow d + \gamma_{2.2 {\rm MeV}} ~~~
(E_{\rm th} = 1.80 {\rm MeV}),
\end{equation}
and the emitted photon scatter an electron, which then emits Cherenkov photons. Thus
the positron and the delayed 2.2 MeV photon are the signal of inverse beta decay.

\begin{figure}[hbt]
\begin{center}
\epsfbox{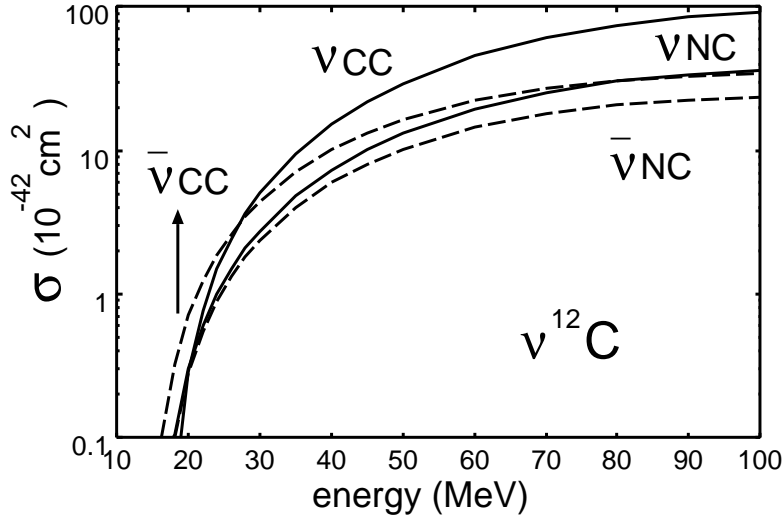}
\end{center}
\vspace{-0.5cm}
\caption{Cross sections of neutrino interactions with ${}^{12}{\rm C}$. Solid lines are neutrinos
and dashed lines are antineutrinos.
\label{fig:cross_C}}
\end{figure}

Since carbon nuclei are abundant in scintillator, the following reactions also contribute
to the events,
\begin{enumerate}
\item neutral current interaction
\begin{equation}
{}^{12}{\rm C} + \nu \rightarrow {}^{12}{\rm C}^{*} + \nu \;\; 
(E_{\rm th} = 15.11 {\rm MeV}),
\end{equation}
\begin{equation}
{}^{12}{\rm C}^{*} \rightarrow {}^{12}{\rm C} + \gamma_{15.11 {\rm MeV}},
\end{equation}
\item $\nu_{e}$ capture by charged current interaction
\begin{eqnarray}
&& {}^{12}{\rm C} + \nu_{e} \rightarrow {}^{12}{\rm N} + e^{-} ~~
   (E_{\rm th} = 17.34 {\rm MeV}), \\
&& {}^{12}{\rm N} \rightarrow {}^{12}{\rm C} + e^{+} + \nu_{e} ~~
   (\tau_{1/2} = 11.00 {\rm msec}),
\end{eqnarray}
\item $\bar{\nu}_{e}$ capture by charged current interaction
\begin{eqnarray}
&& {}^{12}{\rm C} + \bar{\nu}_{e} \rightarrow {}^{12}{\rm B} + e^{+} ~~
   (E_{\rm th} = 14.39 {\rm MeV}), \\
&& {}^{12}{\rm B} \rightarrow {}^{12}{\rm C} + e^{+} + \nu_{e} ~~
   (\tau_{1/2} = 20.20 {\rm msec}).
\end{eqnarray}
\end{enumerate}
The cross section of these reactions are plotted in Fig. \ref{fig:cross_C}.
Because each reaction has its unique signal, we can identify them including
inverse beta decay. This is an advantage of scintillator detector.

\paragraph{KamLAND}

KamLAND (Kamioka Liquid scintillator Anti-Neutrino Detector) is located underneath
Mt. Ikenoyama in Gifu prefecture in central Japan, where Kamiokande was once located.
It is a 1,000 ton liquid scintillator which is composed of $80 \%$ dodecane and $20 \%$
pseudocumene, whose typical composition is ${\rm C}_{n} {\rm H}_{2n}$. The primary purpose
of the KamLAND is to probe the LMA solution of the solar neutrino problem by observing
reactor neutrinos from the entire Japanese nuclear power industry as discussed in
section \ref{subsubsection:reactor}.

Because of low background level, KamLAND is also suitable for observation of supernova
neutrinos. In \cite{Bandyopadhyay03}, supernova neutrino detection at KamLAND was
studied.

\paragraph{LVD}

LVD (Large Volume Detector) is located in the INFN Gran Sasso National Laboratory,
Italy. It consists of an array of 840 liquid scintillator
(${\rm C}_{n} {\rm H}_{2n+2}$ with $\langle n \rangle = 9.6$) counters and the active
scintillator mass is 1,000 ton. The main purpose of the project is detection of supernova
neutrinos. For details of the detector, see \cite{Aglietta92,Selvi03}. Analyses of
supernova neutrinos expected to be observed at LVD were performed in
\cite{LunardiniSmirnov01,KTearth02}.

%%%%%%%%%%%%%%%%%%%%%%%%%%%%%%%%%%%%%%%%%%%%%%%%%%%%%%%%%%%%%%%%%%%%%%%%%%%%%%%%%%%%%%%%%%%%%%%
%\subsection{Future directions}
%%%%%%%%%%%%%%%%%%%%%%%%%%%%%%%%%%%%%%%%%%%%%%%%%%%%%%%%%%%%%%%%%%%%%%%%%%%%%%%%%%%%%%%%%%%%%%%

\clearpage

\section{Explosion Mechanism of Core-Collapse Supernovae \label{exp_mecha}}

\subsection{Status of Spherical Models}
\begin{figure}
\begin{center}
\epsfxsize=10cm
\epsfbox{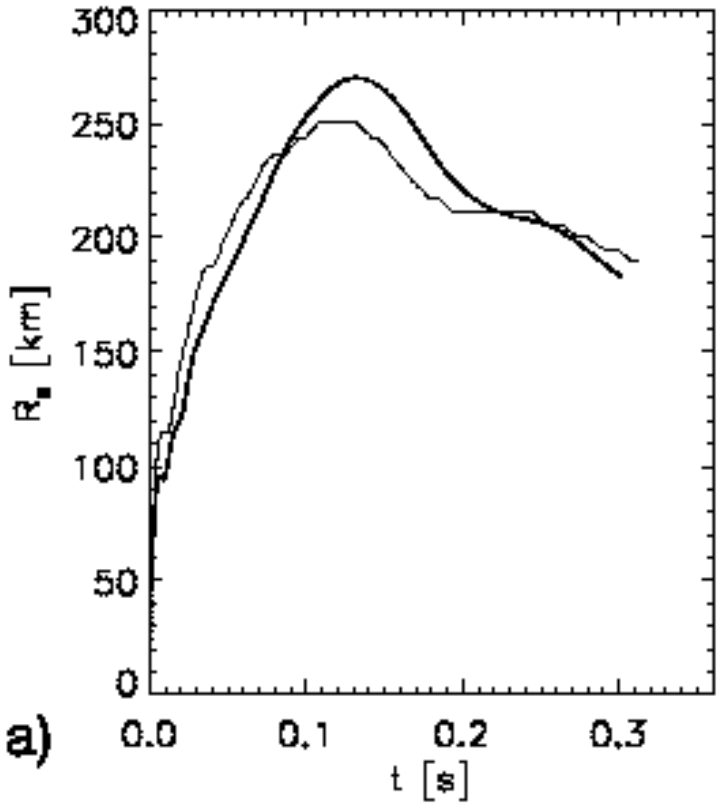}
\end{center}
\caption{Comparison of the radial position of the shock waves as a function of time 
obtained by the two independent groups. The thin and thick lines are
 based on the simulations of a $15 M{\odot}$ progenitor star, performed by \textsc{vertex} code of the Garching group \cite{ramp} and \textsc{agile-boltztran} code of Oak Ridge-Basel group \cite{Liebendoerfer_et_al_04}, respectively. It is shown that spherically symmetric models
with standard microphysical input fail to explode by the delayed,
neutrino-driven mechanism. This figure is taken from Liebend\"{o}rfer {\it et al} \cite{Lieben03}.}
\label{liebenfig}
\end{figure}

Although the gross physical conditions of core-collapse supernovae are
understood as denoted in the section \ref{supernova_theory}, recent 
numerical simulations assuming {\it spherical symmetry}, 
however, with the current input physics
(neutrino interactions and the equations of state of dense matter) and
with/without general relativity,
do not yield successful explosions by the neutrino heating mechanism (see Figure \ref{liebenfig}) \cite{ramp,Lieben01,tomp, Lieben03}.

\begin{figure}
\begin{center}
\epsfxsize= 10cm
\epsfbox{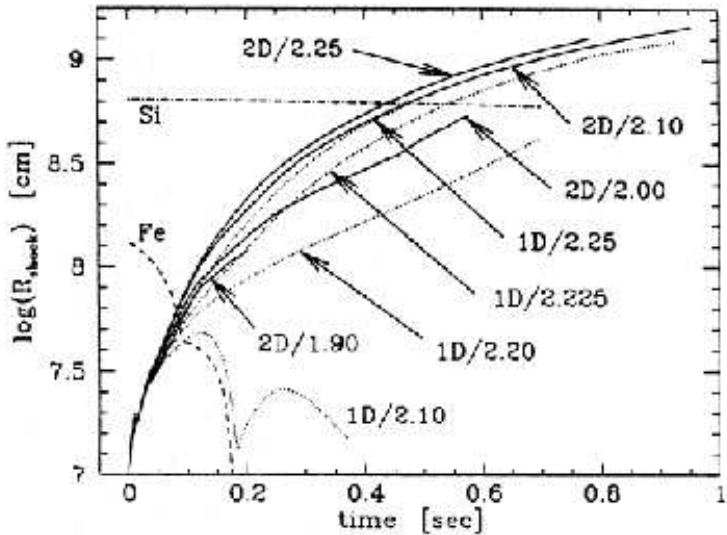}
\end{center}
\caption{Shock positions as a function of time after core bounce taken from 
\cite{jankamueller94}. 1D or 2D represents the one or two dimensional models. The numbers indicate the size of the neutrino luminosities in unit of $10^{52}~\rm{erg}~\rm{s}^{-1}$ injected 
from the proto-neutron star. Given a neutrino luminosity of $2.1\times 10^{52}~\rm{erg}~\rm{s^{-1}}$, it can be seen that the unsuccessful explosion model in the 1D simulation turns to lead the successful explosion in the 2D simulation. Generally, the shock can propagate to the outer regions in the 2D simulations due to the convection in the hot bubble, which boosts the neutrino-heating efficiency.     
\label{shockrevival_janka}
}
\end{figure}

\begin{figure}
\begin{center}
\epsfxsize=10cm
\epsfbox{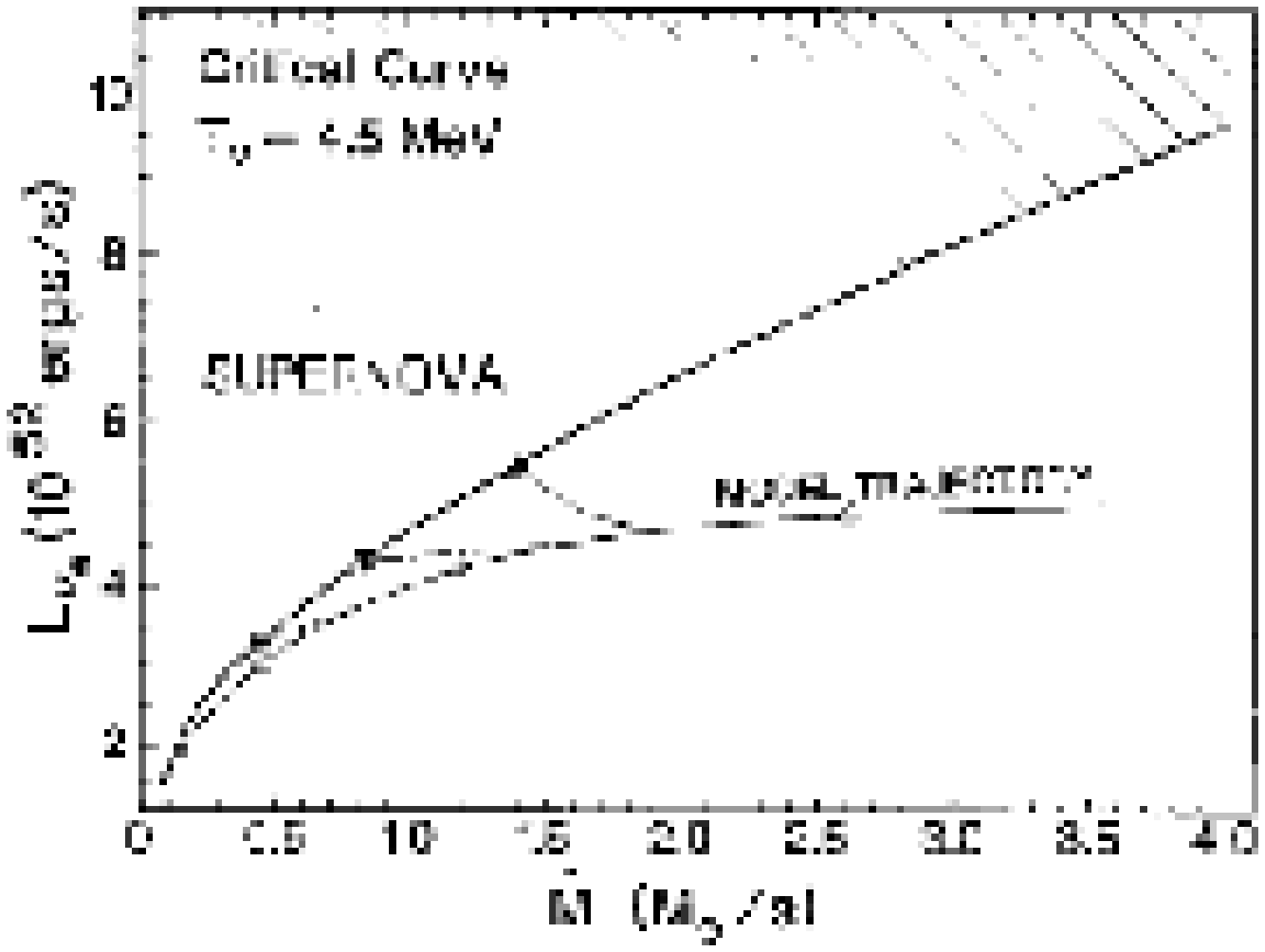}
\end{center}
\caption{Critical curve for the shock-revival as a function of the neutrino luminosity
 ($L_{\nu_e}$ in unit of $10^{52}~{\rm erg}~{\rm s}^{-1}$) emitting from
 the central protoneutron star and the accretion rate ($\dot{M}$ in unit of
 $M_{\odot}$) through the stalled shock. The ``Model Trajectory'' in
 this figure is taken from
 the results obtained in the numerical simulations by Bruenn (1982) \cite{bruenn92}, in which no explosion was
 obtained. If the evolution line of a real core crosses the critical
 curve into the hatched region, a neutrino-driven supernova should begin. 
 This figure is taken from Burrows \& Goshy (1993) \cite{bur93}.}
\label{bg}
\end{figure}

The problem apparently stems from the
weak-interacting natures of neutrinos. The neutrino heating 
occurs only inefficient for the successful explosions in the spherical 
models. 
Whether neutrinos 
succeed in reviving the stalled shock wave depends on the efficiency of
the energy transfer to the postshock layer, which in turn increases with
the neutrino luminosity.
 In fact, it was pointed out that the stalled shock revives leading to
 explosions for otherwise failed explosion models by enhancing the neutrino
luminosity artificially by a few tens percents from the original value
\cite{jankamueller94} (see Figure \ref{shockrevival_janka} and compare
1D/2.10 and 1D/2.225).
It was also found by the study of the static configurations after the 
shock-stagnation that for a given accretion rate there is a critical 
luminosity for the shock revival \cite{bur93} (see Figure \ref{bg}). 
Janka {\it et al.}
reported that the manipulation of
omitting the velocity dependent term ($O(v/c) \sim 10 \%$) in the neutrino transport equations
increased the neutrino energy deposition in the heating region and was
sufficient to covert a failed model into a exploding one \cite{janka04manu}.
These facts suggest that all we have to obtain for the
successful explosions, is the relatively small amount of boost of 
the neutrino luminosity and energy from the values we have obtained 
in the failed explosion models.

In the last couple of years, both numerics and microphysics have been 
developed. The former, in particular, 
has seen major progress \cite{ramp,Lieben01,tomp,Lieben03,buras}. 
Ever since Wilson first proposed the neutrino heating mechanism
\cite{wilson1985}, the accurate treatment of neutrino transport has been an 
important task, however mainly due to the computational intensity, 
some approximations, such as 
the multi-group-flux-limited diffusion approximation as the most
familiar example \cite{bruenn85,bruenn87,mezza98}, have been employed 
in the 1D spherical symmetric
simulations. This situation has changed completely 
lately (see \cite{cardall05} for a complete set of references). A couple of groups \cite{ramp,Lieben01,tomp,Lieben03}, 
have published the state-of-the-art direct solutions of the Boltzmann 
equation for neutrinos, some of them extended even to 2D computations
in the multi-group-flux-limited diffusion approximation
\cite{buras,livne}. Although they have still not found successful 
explosions, the importance of the accurate treatment of neutrino
transfer should be never missed. 

% koko

The microphysics such as neutrino reaction rates and 
equations of state have also been studied in detail. Recently,
shell-model calculations of nuclear properties \cite{langanke04} revealed that the
treatment of the electron capture rates for various nuclei should be
 changed significantly from the previous ones \cite{bruenn85}.
 In almost all the supernova simulations, the electron captures rates on 
nuclei were cut off above a few $\sim 10^{10} ~{\rm g}~{\rm cm}^{-3}$ 
because only the resonant Gamow-Taylor transitions could be treated 
in the average heavy nuclei, which is calculated by the employed
equation of state (EOS)
\cite{bruenn85}. 
\begin{figure}
\begin{center}
\epsfxsize = 10 cm
\epsfbox{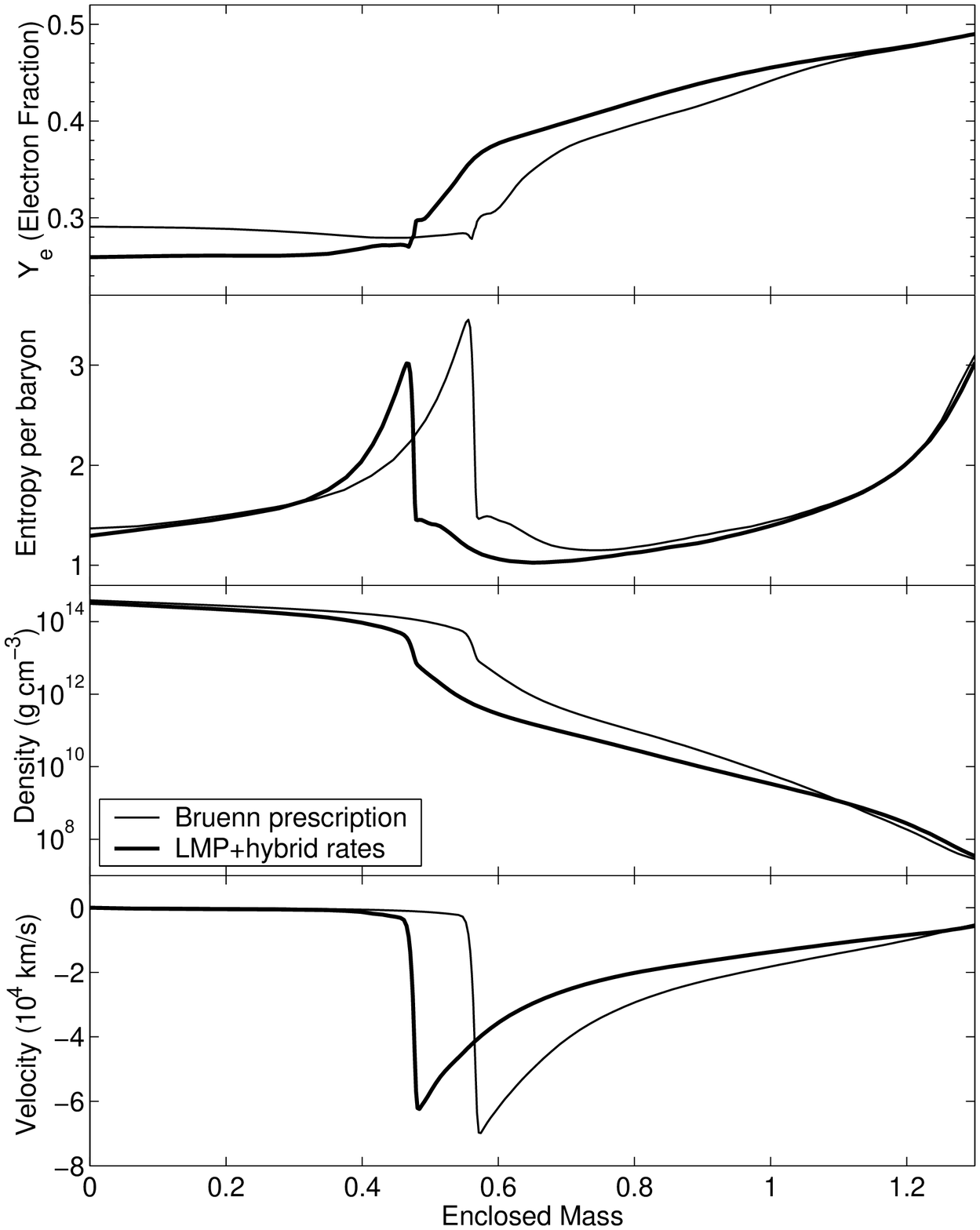}
\caption{The electron fraction, entropy, density and velocity as
 functions of the enclosed mass at the beginning of bounce for a 15
 $M_{\odot}$ model. The thin line is a simulation using the conventional
 capture rates by Bruenn parametrization \cite{bruenn85} while 
the thick line for a simulation
 using the new rates by Langanke Mart\'{i}nez-Pinedo (LMP) 
 \cite{lmp1,lmp2}. This figure is taken from Hix {\it et al.} (2003)
 \cite{hix}. }
\label{prl_langanke}
\end{center}
\end{figure}
Despite of the quantitative change of the
electron capture rates and hence the lepton fraction at the central
portion of the core, subsequent shock propagations were found to show 
no significant change in comparison with the previous studies 
due to cancellation effects \cite{hix} (see Figure \ref{prl_langanke}). 

\begin{figure}
\epsfxsize=8cm
\epsfbox{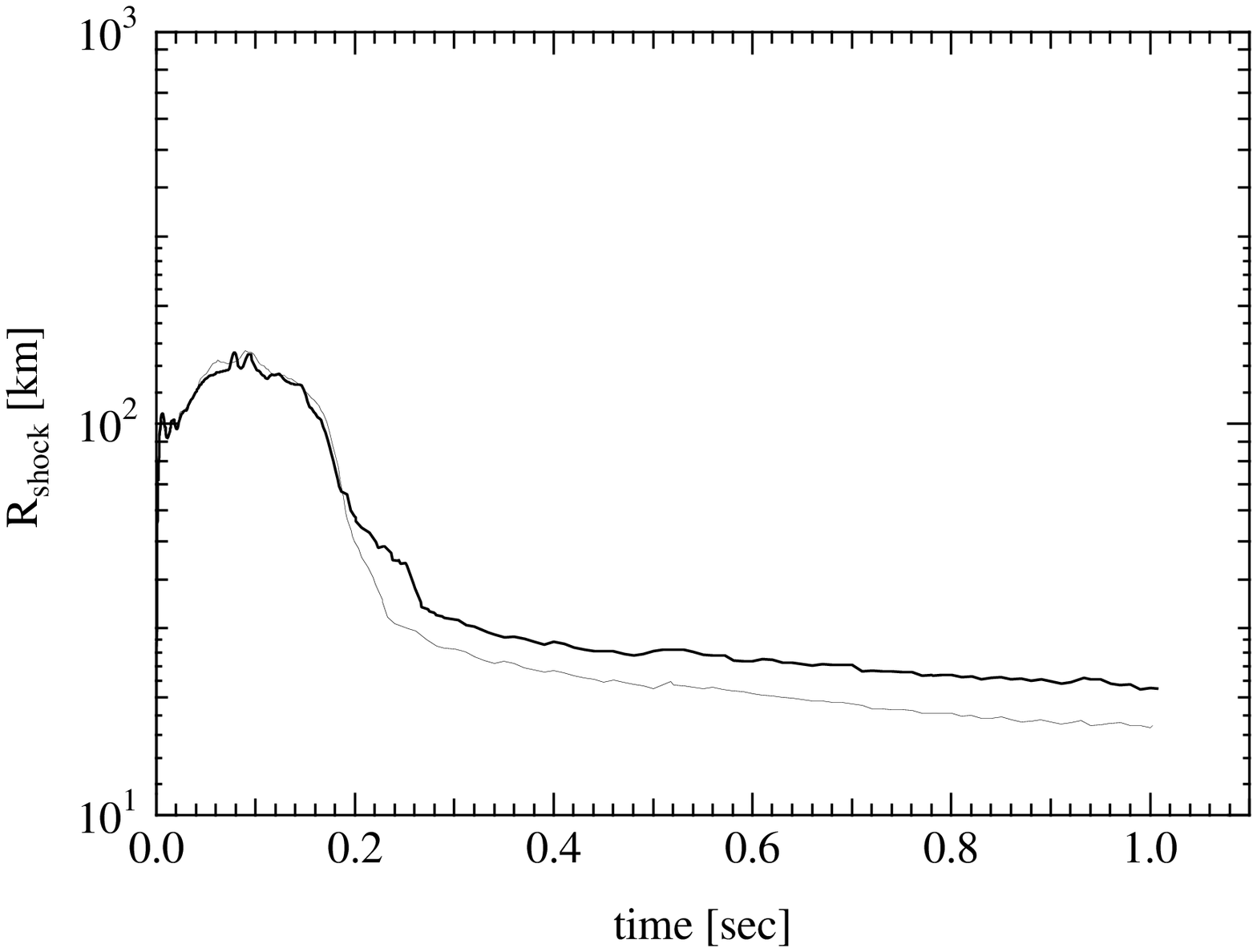}
\epsfxsize=8cm
\epsfbox{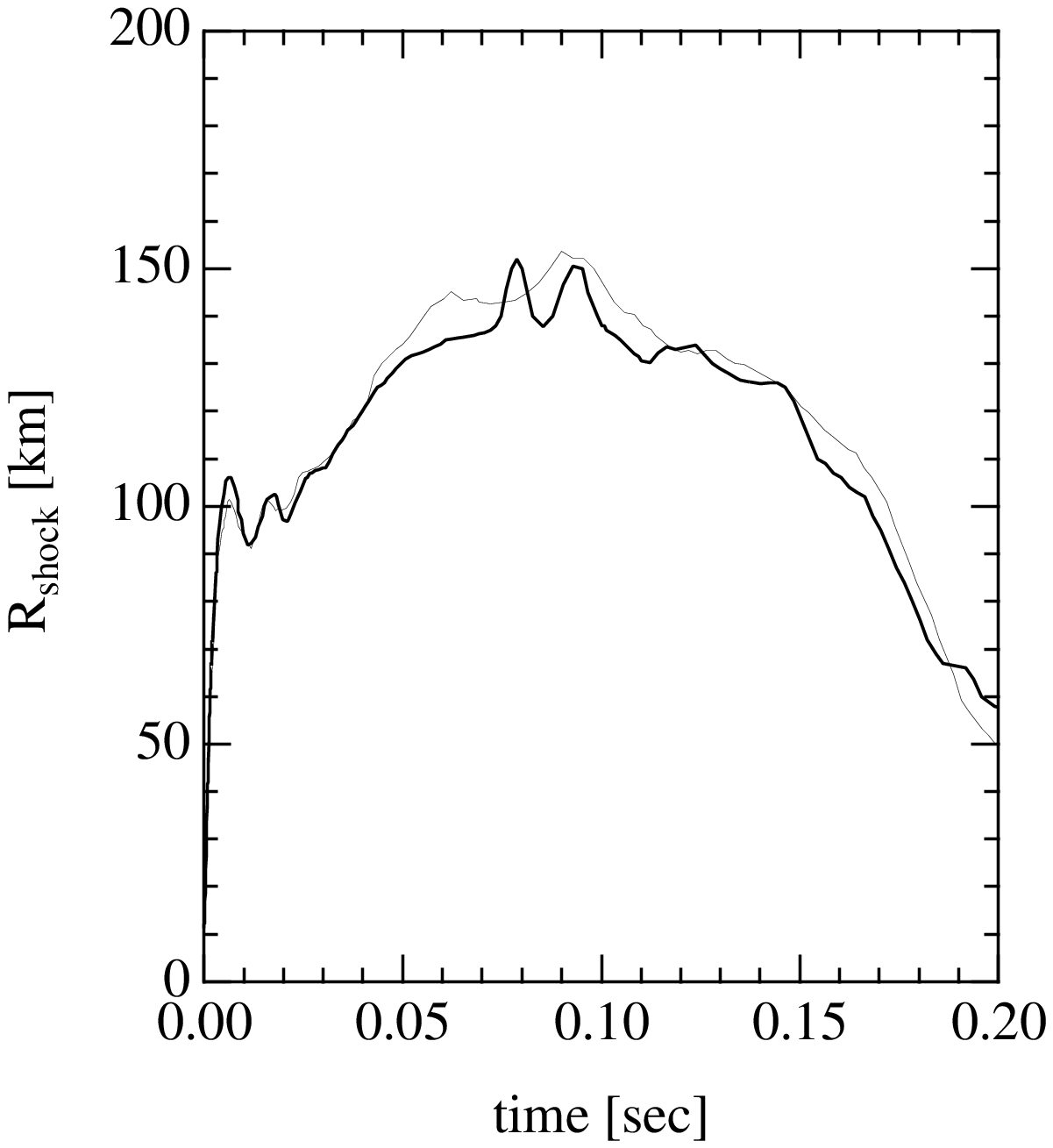}
\caption{Radial positions of shock waves in models employing 
SHEN EOS and Lattimer \& Swesty EOS are 
shown by thick and thin lines, respectively, as a function of time after
 bounce. The evolutions at early and late times are displayed 
in left and right panels, respectively. This figure is taken from
 Sumiyoshi {\it et al} (2005) \cite{sumi_shock}.}
\label{sumi_shock}
\end{figure}

As for the EOS of dense matter, 
 we have at least two kinds of EOS
now available based on different realistic descriptions of nuclear
interactions, namely EOS by Lattimer \& Swesty (LS EOS) \cite{Lat91} and 
EOS by Shen {\it et al} (SHEN EOS) \cite{shen98}. 
Sumiyoshi {\it et al.} performed 1D adiabatic hydrodynamic simulations 
employing the two kinds of equations
of state, respectively, and found that there does appear the difference in the
chemical compositions between the EOS's during the infalling phase, however, which disappears in the later
phases, and hence no significant differences such in the remnant masses and
the explosion energies are obtained \cite{sumi_prep}. Currently they 
performed the general relativistic neutrino radiation hydrodynamic
 core-collapse simulations assuming spherically symmetric and
investigated long-term postbounce evolution after core-bounce employing the
 two equations of state. They found that, for both EOSs, the core does
 not explode and the shock waves stall similarly in the first 100
 milliseconds after core bounce (see Figure \ref{sumi_shock}).

\begin{figure}
\begin{center}
\epsfxsize = 6.5 cm
\epsfbox{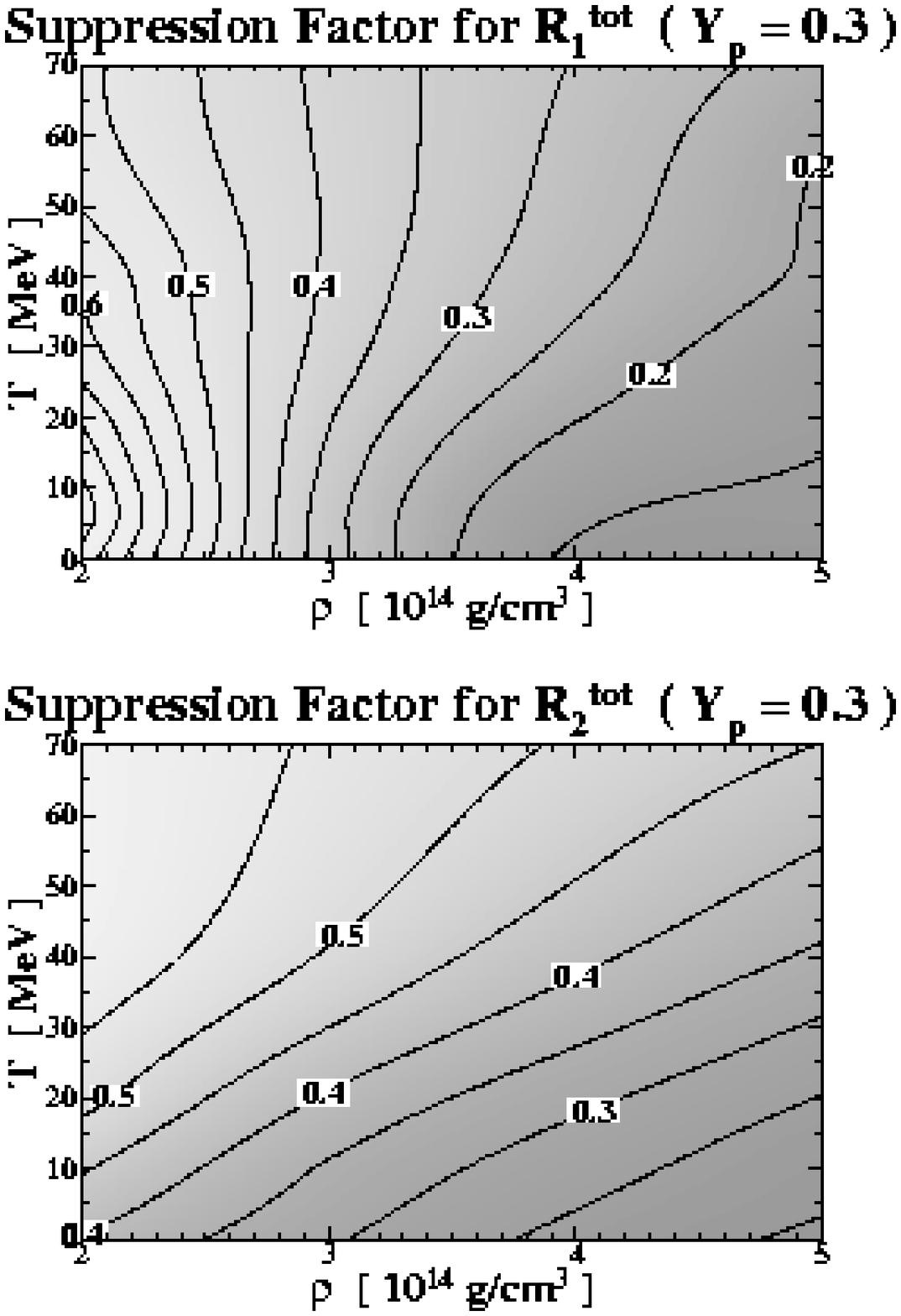}
\epsfxsize = 7 cm
\epsfbox{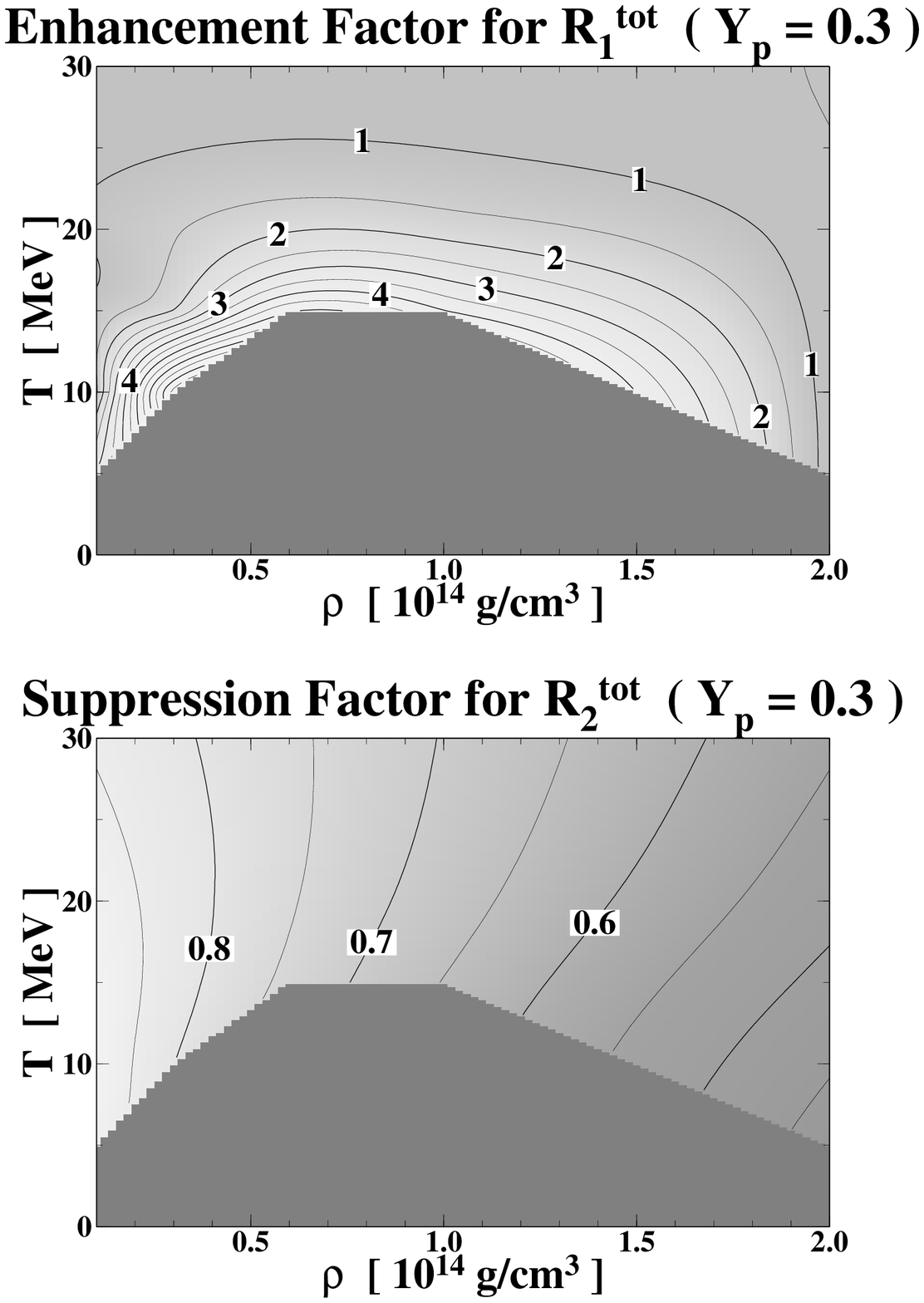}
\caption{Effect of inter-nucleon interactions on the neutrino-nucleon
 scattering rates taken from \cite{yamatoki}. Left and right panels show the
 contours of the suppression and enhancement factors, which is 
the ratio of the scattering rate with to without the corrections.
 The top and bottom panel shows the 
contribution from the density correlation function ($R_1$, vector
 currents) and  the spin-density correlation ($R_2$, axial-vector currents), respectively. 
Note the difference of the density scale in the right and left panels.
It can be seen that the scattering rate is
 suppressed due to the corrections in the high density
 regime ($\rho \ge 10^{14}~{\rm g}~{\rm cm}^{-3}$) (see left panels), 
while the vector current contribution (right top panel) 
is enhanced in the low density region ($\rho \le 10^{14}~{\rm g}~{\rm cm}^{-3}$) and in the vicinity of the liquid-gas
 phase transition regions (the dark regions in the right panels). The
 above features are generally common in case of the other values of
 the proton fraction, $Y_{\rm p}$. See for details Yamada and Toki (2000) \cite{yamatoki}.}
\label{yamatoki_fig}
\end{center}
\end{figure}

\begin{figure}
\begin{center}
\epsfxsize = 7 cm
\epsfbox{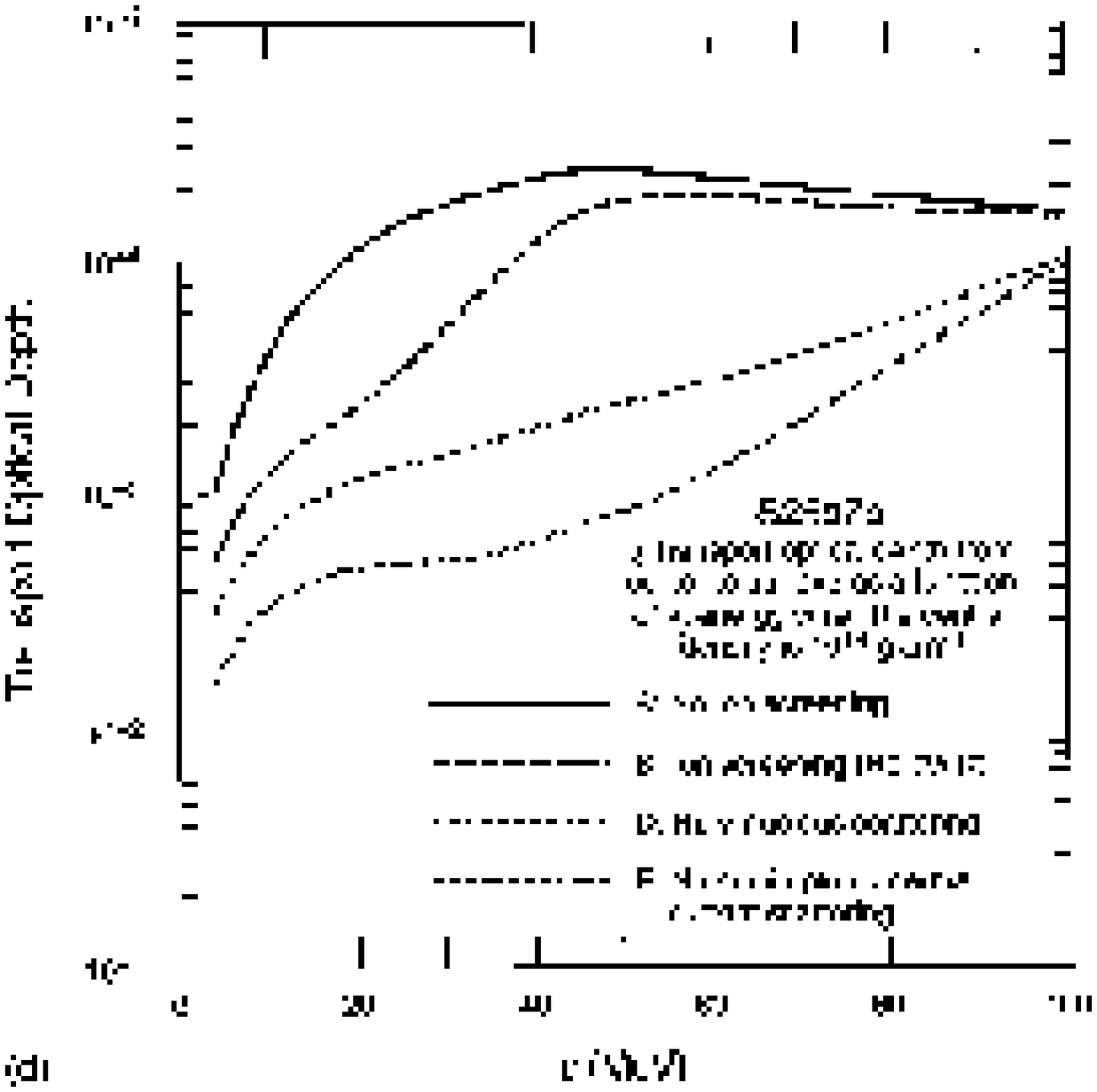}
\epsfxsize = 7 cm
\epsfbox{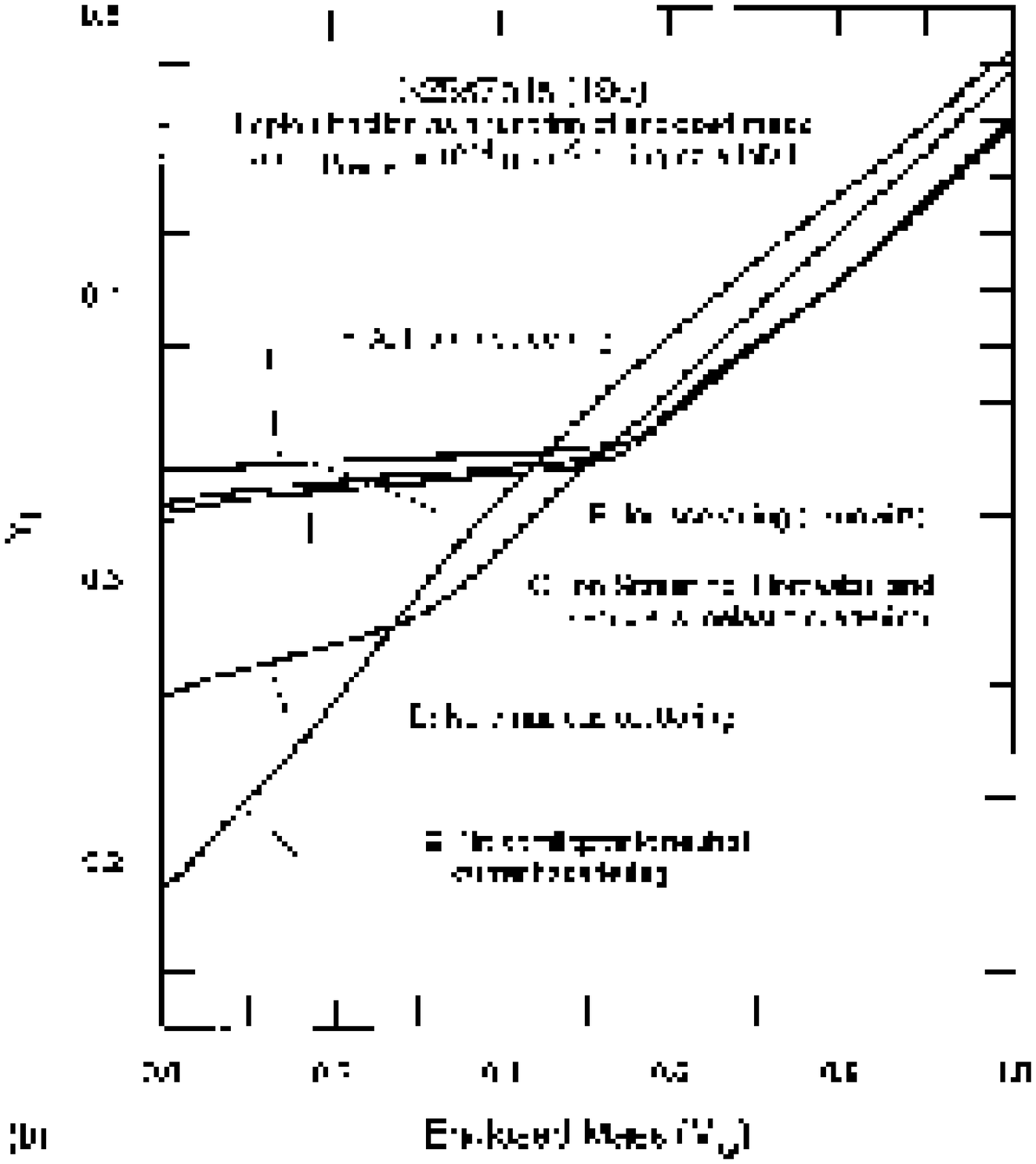}
\caption{Effect of ion-ion correlations on neutrino-nucleus elastic
 scattering during core-collapse of a $25 M_{\odot}$ progenitor star
 taken from \cite{bruenn_ion}. Left
 panel shows the optical depth along a radial path from the stellar
 center to the surface of as a function of
 the neutrino energy when the central density is $10^{14}~{\rm g}~{\rm cm}^{-3}$. Right panel shows the lepton fraction profiles at the same time with the right panel. From the left panel, the optical depth of the lower-energy neutrinos is shown to be lowered due to the screening effect of nucleus  (compare the lines labeled as A and B). On the other hand, this change leads to a smaller change of the lepton fraction $\sim 0.015$ in the central region (right panel). }
\label{ion-ion}
\end{center}
\end{figure}

\begin{figure}
\begin{center}
\epsfxsize = 8 cm
\epsfbox{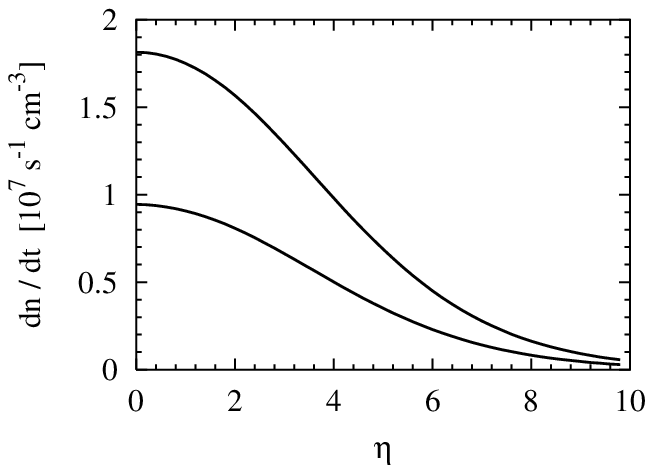}
\caption{The pair-production rates by the process $\nu_e + {\bar \nu}_e \rightarrow \nu_{\mu,\tau} + \bar{\nu}_{\mu,\tau}$ (top line) and 
$e^{+} + e^{-} \rightarrow \nu_{\mu,\tau} + \bar{\nu}_{\mu,\tau}$ (bottom
 line) as a function of the degeneracy parameter of electron-type
 neutrinos ($\eta_{\nu_e}$) and of electrons ($\eta_{e}$), respectively. Here $T = 12$ MeV and $\eta_{\nu_{\mu,\tau}} = 0$
is assumed. The new rate, namely, $\nu_e + {\bar \nu}_e \rightarrow \nu_{\mu,\tau} + \bar{\nu}_{\mu,\tau}$, is shown to dominate over the electron-positron pair reaction, which was considered only the sources for producing $\nu_{\mu,\tau}$ and $\bar{\nu}_{\mu,\tau}$.
This figure is taken from Buras {\it et al.} (2003)
 \cite{burasbrem}. }
\label{buras_new_source}
\end{center}
\end{figure}

As for the neutrino reaction rates, 
the effects of the inter-nucleon many body effects near nuclear density 
have been elaborately studied (see Figure \ref{yamatoki_fig} and
\cite{bursaw1,bursaw2,reddy,yamatoki}.)
Inter-ion correlations work for suppressing the
neutrino-nucleus elastic scattering \cite{horowitz_ion,ito_ion}, however, have been pointed out to 
lead to non noticeable changes of the dynamic during core-collapse (see
Figure \ref{ion-ion}, \cite{bruenn_ion,marek_ion}). 
Relatively small corrections to the standard neutrino interaction
processes \cite{bruenn85} such 
as the detailed reaction kinematics of nucleon thermal motions,
recoil, and weak magnetism contributions \cite{horomag} as well 
as nucleon bremsstrahlung ($NN^{'} \rightleftharpoons
NN^{'} \nu \bar{\nu}$) \cite{hannestad}, 
pair-annihilation/creations among different neutrino flavors
\cite{burasbrem}, such as $\nu_e + \bar{\nu}_e \rightarrow \nu_{\mu,\tau} +
\bar{\nu}_{\mu,\tau}$ (see Figure \ref{buras_new_source}), and the quenching of the axial vector coupling 
constants in dense matter \cite{carter} have been partly or fully
incorporated and their importance has been evaluated in the recent 
1D computations \cite{rampp_buras,tomp,buras}. 
These nuclear corrections effectively work for the suppression of the
opacities leading to the larger neutrino luminosities, which is able to
act in favor of enhancing the explosion \cite{rampp_buras}.     

However even with these sophistications, the successful explosion has not yet
been found for the present \cite{rampp_buras,buras}. We may still missing some
important microphysical processes if we are to obtain the successful
explosion assuming spherical symmetry. The phase transitions with various
possible geometrical structures from isolated
nuclei to uniform nuclear matter (the so-called ``nuclear pasta'') have
been elaborately studied from a nuclear physics point of view
\cite{gensan1,gensan2}. Due to this non-uniformity of the nuclei, the
neutrino opacities are shown to be lowered, which are conventionally
estimated by a single heavy nucleus approximation \cite{horonuc}. 
The inelastic neutrino-nucleus
interaction has been pointed out to play the same important role of
thermalizing neutrinos as effective as the 
neutrino scattering on electrons \cite{bruennhax}.
These microphysical ingredients, which have not been routinely taken
into the modern supernova simulations, should deserve further investigations.

\subsection{Multidimensional Aspects in Core-Collapse Supernovae \label{kare}}
\begin{figure}
\begin{center}
\epsfxsize=4.8 cm
\epsfbox{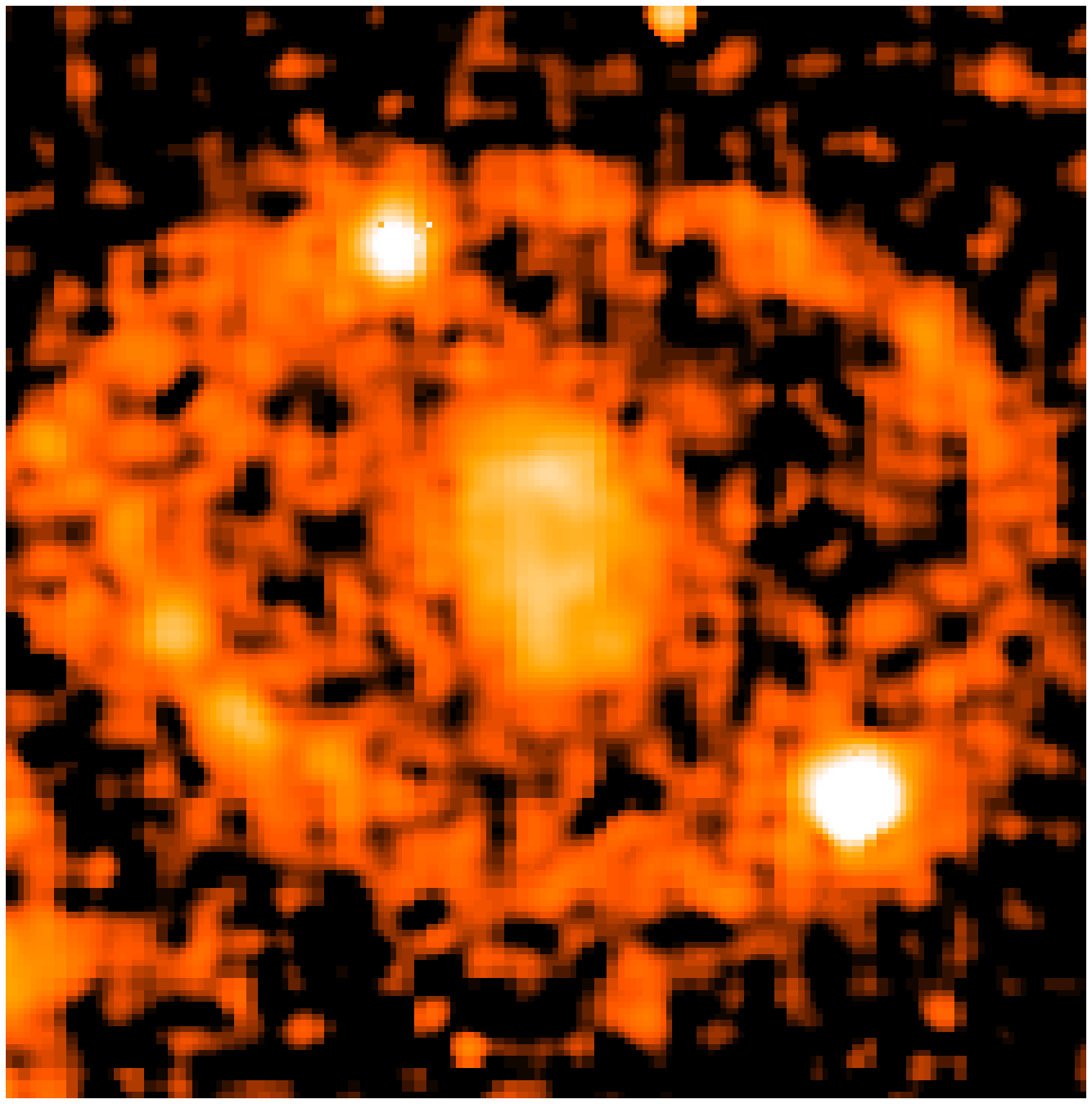}
\epsfxsize=5 cm
\epsfbox{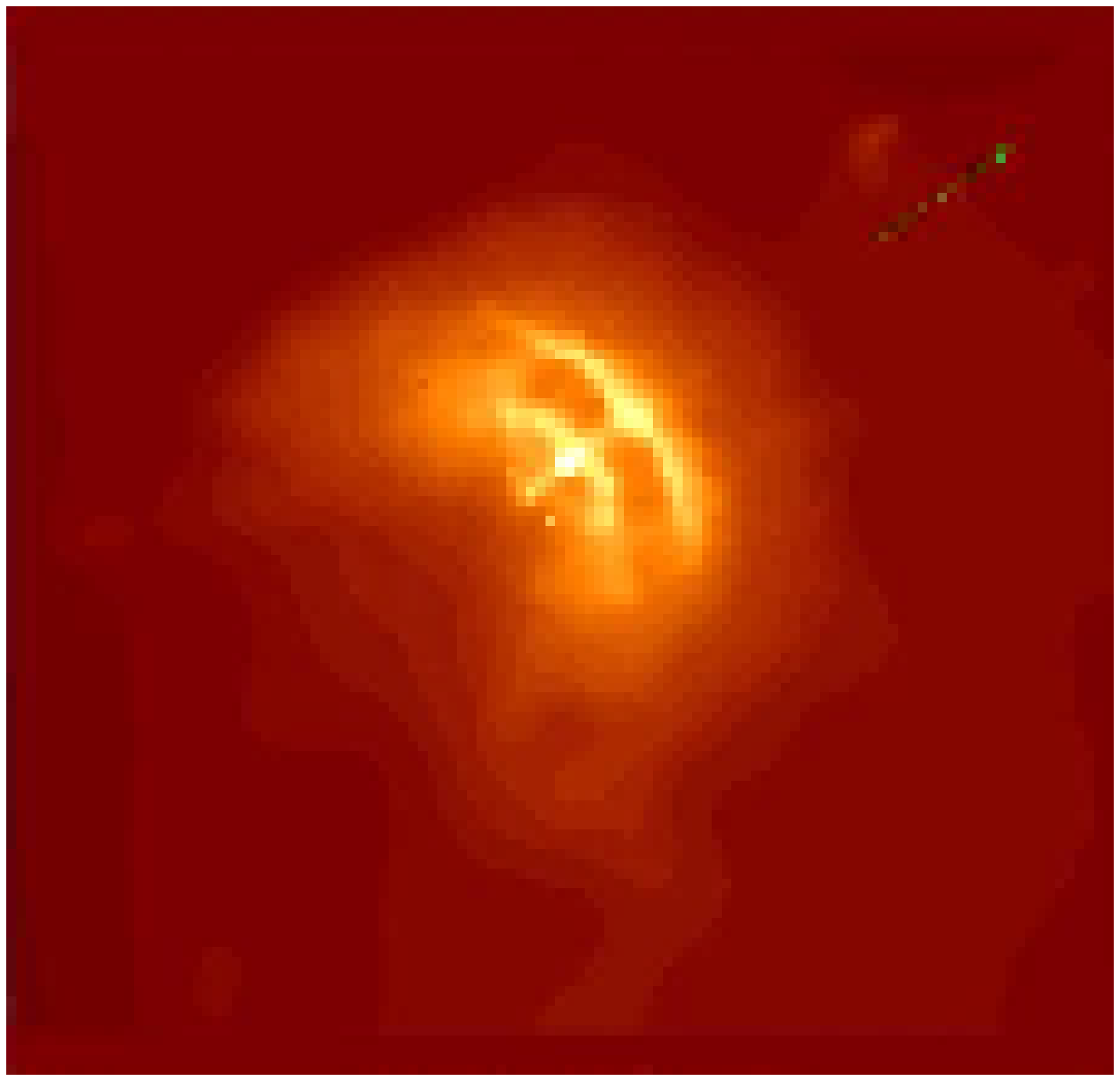}
\epsfxsize=5 cm
\epsfbox{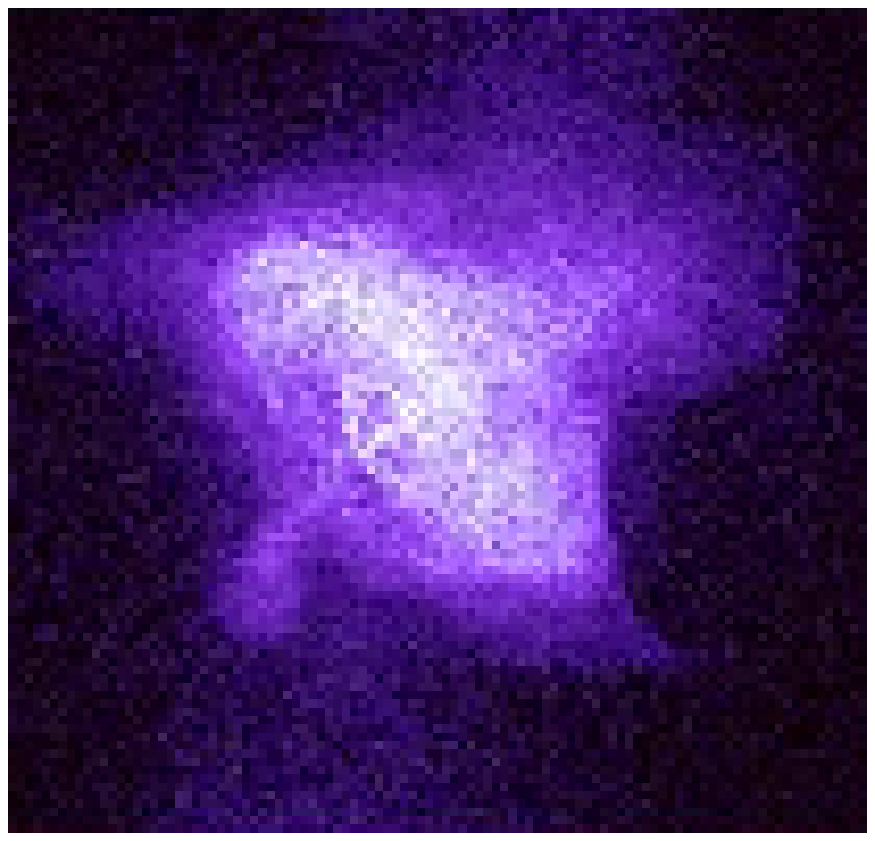}
\end{center}
\caption{Observational evidences for asphericity in core-collapse
 supernovae. Left panel shows the inner debris of the supernova ejecta
 with an axis ratio of $\sim$ 2 (inner red region) and the ring around it produced in the presupernova era (taken from \cite{wang02}). The middle and right panel shows the nebulae of Vela and Crab pulsars, respectively (taken from \cite{pavlov} and \cite{weiss}). 
\label{asphericalobs}
}
\end{figure}

%%%%%%%%%%%%%%%%%%%%%%%%%%%%%%%%%%%%%%%%%%%%%%%%%%%%%%%%%%%
%%%%%%%%%%%%%%%%%%%%%%%%%%%%%%%%%%%%%%%%%%%%%%%%%%%%%%%%%%%
%\subsubsection{Importance of Multi-dimensional Aspects\\}
Ever since SN1987A was observed, most researchers in this field think 
that the dynamics of supernova is aspherical. The following observations of SN1987A have been
attainable because SN 1987A is the nearest
supernova for us and thus have been observed by modern observational 
instruments. For the explanation of the observed shape of the
lightcurve, the unexpectedly early appearance of X, $\gamma$-ray 
emissions, and Doppler features of spectral lines, the existence of
a large-scale mixing between the deep interior and the hydrogen envelope
is suggested (for a recent review, see \cite{nometal94}). 
The expanding debris of SN1987A is directly confirmed to be globally
asymmetric by images of
{\it{Hubble Space Telescope (HST)}} \cite{plait} and its axis roughly
aligns with the small axis of the rings \cite{wang02} (see the left panel of 
Figure \ref{asphericalobs}).
In recent years, it is noted that the 
same features are drawn for other core-collapse supernovae (see
\cite{hoefrev} and references therein). Spectroporalimetry shows that
substantial asymmetry is common in core-collapse supernovae, indicating
bi-polar explosions with axis ratios up to $\sim 2$ 
\cite{wang02,mendez,hoflich,wang96,wang01}. The degree of the asymmetry 
 tends to increase with time when greater depths are observed
 \cite{wang01,leonard}.  Both suggests that a connection of the
asymmetries with the central engine. For core-collapse supernovae with
good time and wavelength coverage, the orientation of the polarization
vector tends to stray constant both in time and with wavelength, which
suggest that there is a global symmetry axis in the ejecta
\cite{wang01,leonard}. Two oppositely directed jets \cite{fesen} 
with the ejected material in a toroidal structure \cite{willi} 
around the center have been observed
in the remnant of Cas A supernova. Young
neutron stars are observed with high space velocities
(typically $300 -400$ km/s \cite{lyne,lorimer}, with highest values greater than $1000$ km/s 
\cite{arzoumanian}), which are most likely imparted to the neutron star
by a kick at the moment of the explosion. Interestingly, the recent X-ray observations have 
shown the correlation between the direction of pulsar motions and the spin axis of their supernovae in Vela and Crab pulsars \cite{helfand,pavlov} (see the middle and right panels of Figure \ref{asphericalobs}). 
All these observational evidences can be 
naturally interpreted as an evidence that the inner portions of the
explosion are strongly aspherical.  

  So far many physical ingredients have been suggested in order to
  produce such asymmetric explosions. We review each
  ingredient one by one in the following. 
\subsection{Roles of Convections and Hydrodynamical Instabilities \label{convec_chap2}}
The convections and hydrodynamical instabilities have been 
long supposed to be responsible for producing the aspherical explosions.
We give a summary of past studies about them, paying
attention to where and why the convections and instabilities are likely
to occur in the supernova cores and their possible roles of producing the
aspherical explosions.
%\begin{itemize}
\subsubsection{convection near and below the protoneutron star }
It has been widely recognized that the convection 
in the protoneutron star (PNS) could play a crucial role in enhancing the 
neutrino luminosities from the PNS.
In fact, Wilson and his collaborators obtained the exploding models \cite{wilsonmayle88,wilsonmayle93}, 
in which a neutron-finger convection, which will be stated below, in a
PNS was assumed, otherwise they did not see the successful
explosions. This illuminates the importance of the convection
in the PNS. 

After the shock wave stalls ($t \sim 10 ~{\rm ms}$ after core bounce), 
the outer parts of the PNS
are convective unstable, because the deleptonization occurring in the
shocked material outside the neutrino sphere produces a negative lepton
gradient and the weakening of the prompt shock wave gives rise to a
negative entropy gradient in the same region. Epstein first proposed
 that these two factors are an sufficient condition for the
so-called prompt convections
to occur \cite{epsteinconv}. In fact, the existence of these instabilities were demonstrated in most hydrodynamic simulations
     (\cite{burro93,buroheyfrex} see, however, \cite{bruennmezza94}). 

\begin{figure}
\begin{center}
\epsfxsize=6.0 cm
\epsfbox{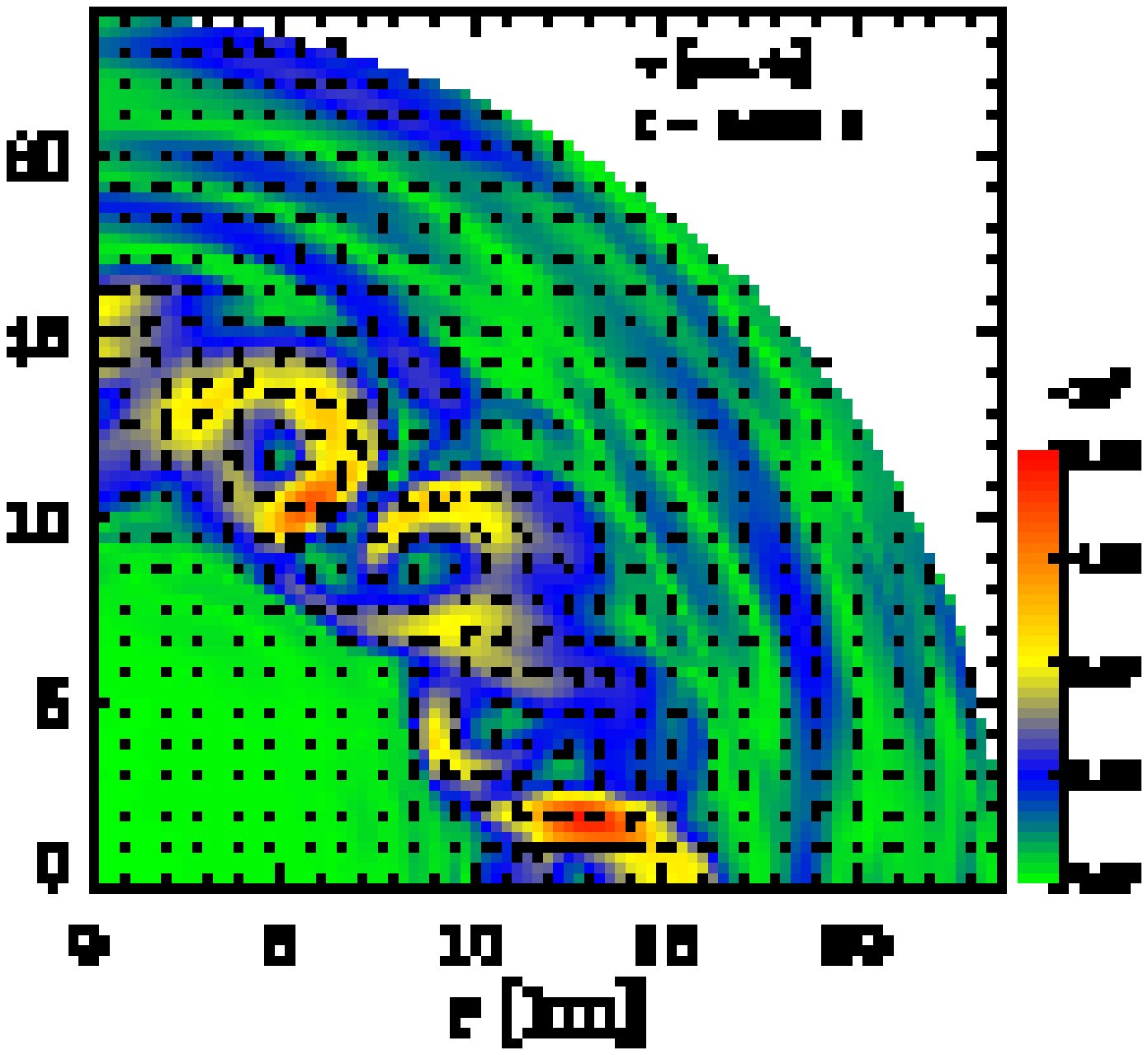}
\epsfxsize=6.0 cm
\epsfbox{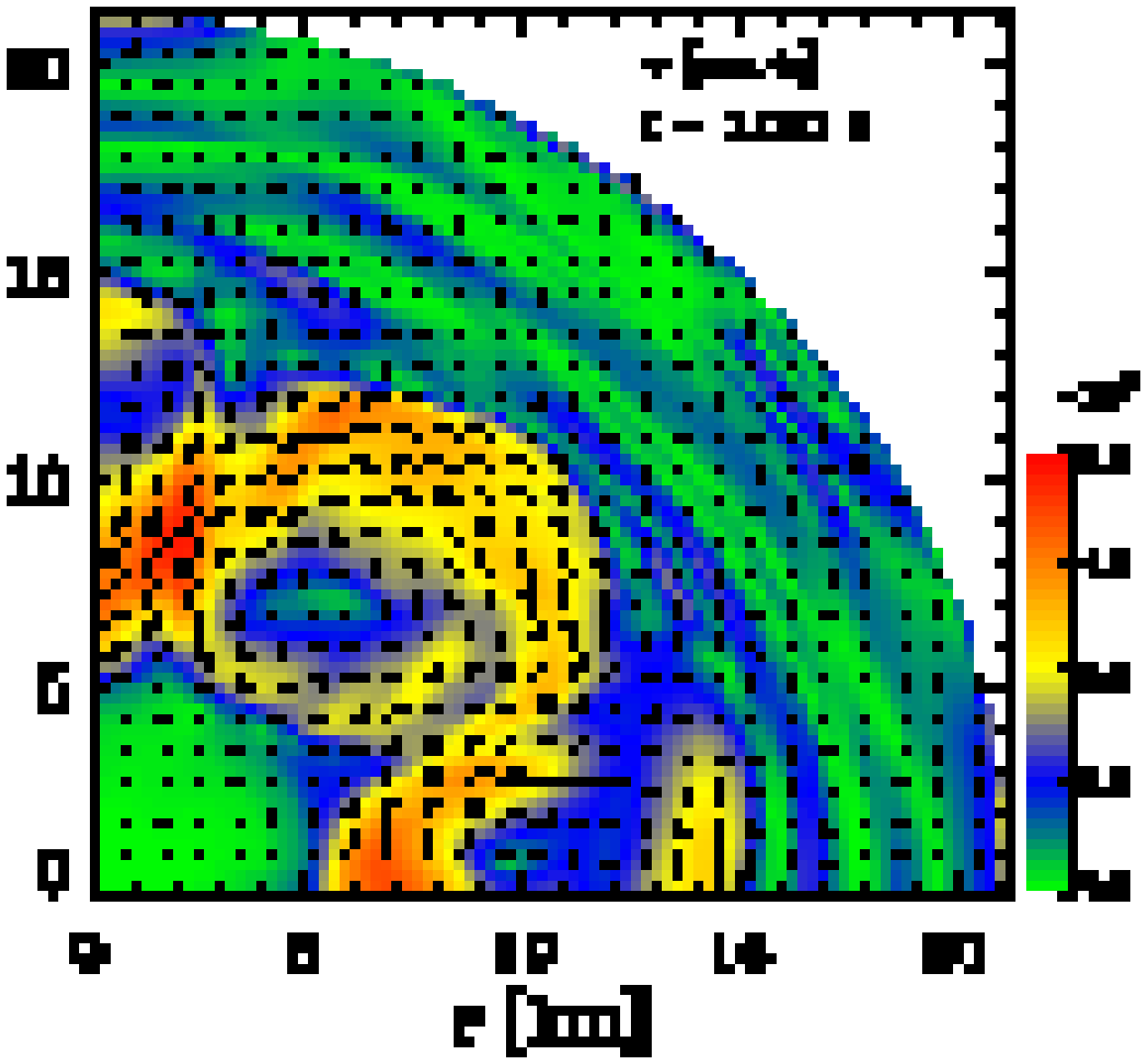}
\epsfxsize=7.0 cm
\epsfbox{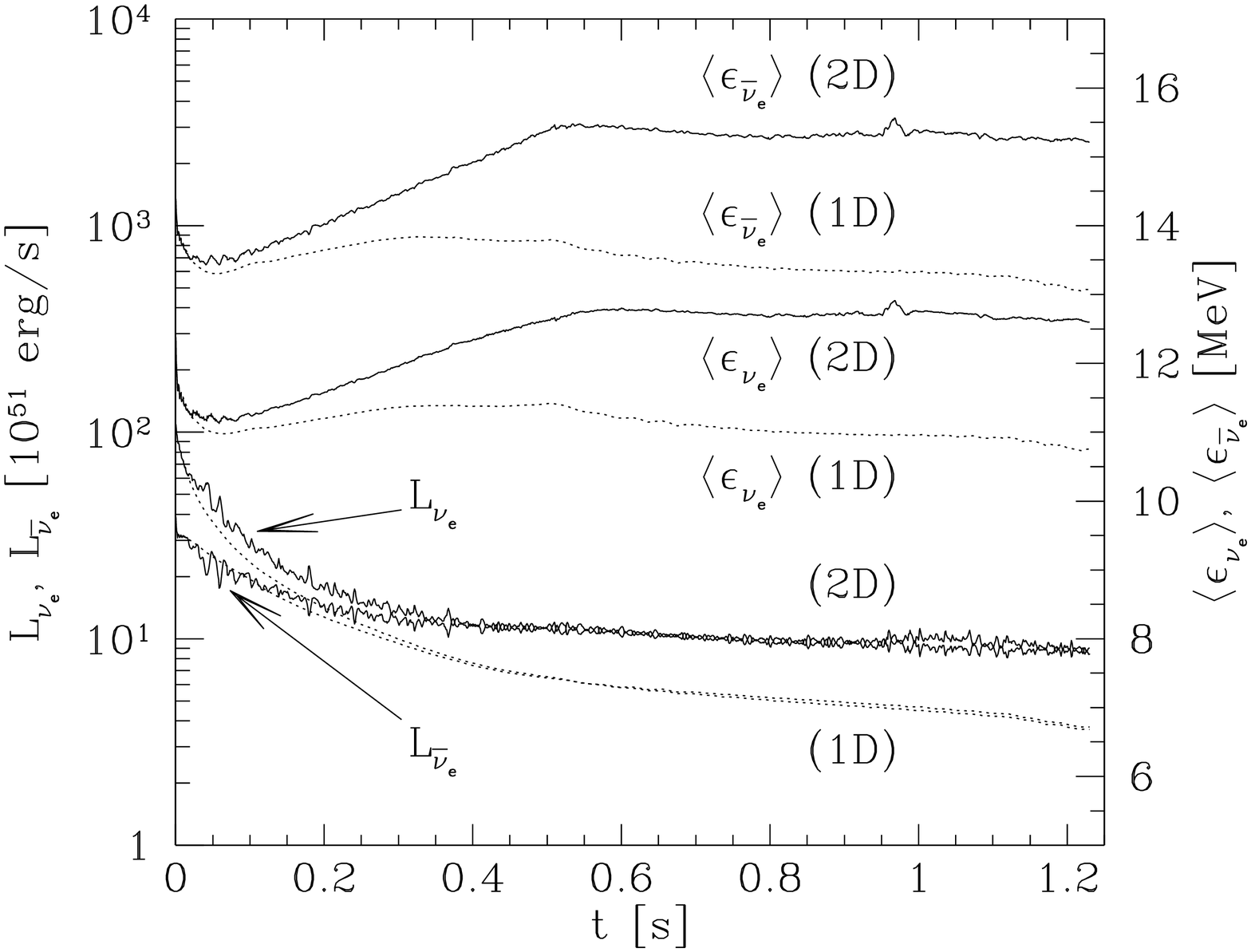}
\end{center}
\caption{Convection inside the protoneutron star. Top left and right panels show the absolute values of the velocity for two instants (about 0.5s (left) and 1 s (right) obtained in the 2D simulations of \cite{keil}.
 The growth of the convective region can be seen. Bottom panel shows the time evolutions of neutrino luminosities $L_{\nu}$ and mean energies of $\nu_{e}$ and $\bar{\nu_{e}}$ for the model in the top panels without (``1D''; dotted) and with convection (``2D'': solid). Significant rise in the neutrino luminosities and energies can be clearly seen. 
Top panels are taken from Janka and Keil
 \cite{jankeil}, and the bottom panel is taken from Keil {\it et al}
 \cite{keil}.}
\label{keil_conve}
\end{figure}

After this prompt convection, the so-called ``Neutron finger''
instability develops inside the PNS \cite{wilsonmayle93}. This instability is 
driven in the presence of a positive entropy gradient and a
negative lepton gradient, both of which are mostly satisfied inside
the PNS. 
In the long-run 2D hydrodynamical
     simulations in the PNS, Keil {\it et al.} found that the neutrino 
     luminosities
     increase $\sim 50 \%$ due to the convection at times later than $200 - 300$ ms after 
     core bounce (see Figure \ref{keil_conve}), which are expected to crucial for reviving the stalled 
     shock wave \cite{keil}. 

\begin{figure}
\begin{center}
\epsfxsize=12.0 cm
\epsfbox{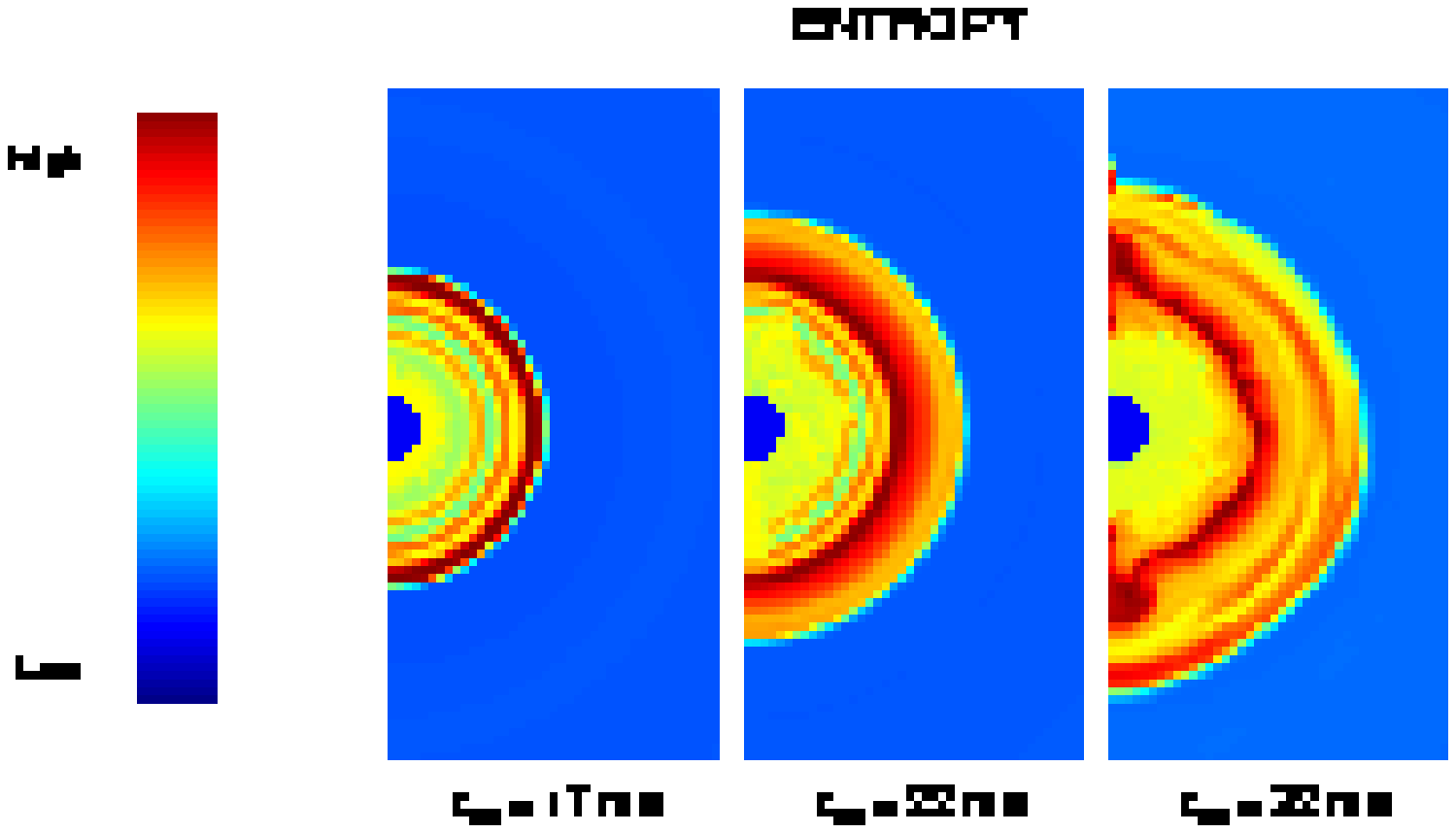}
\epsfxsize=12.0 cm
\epsfbox{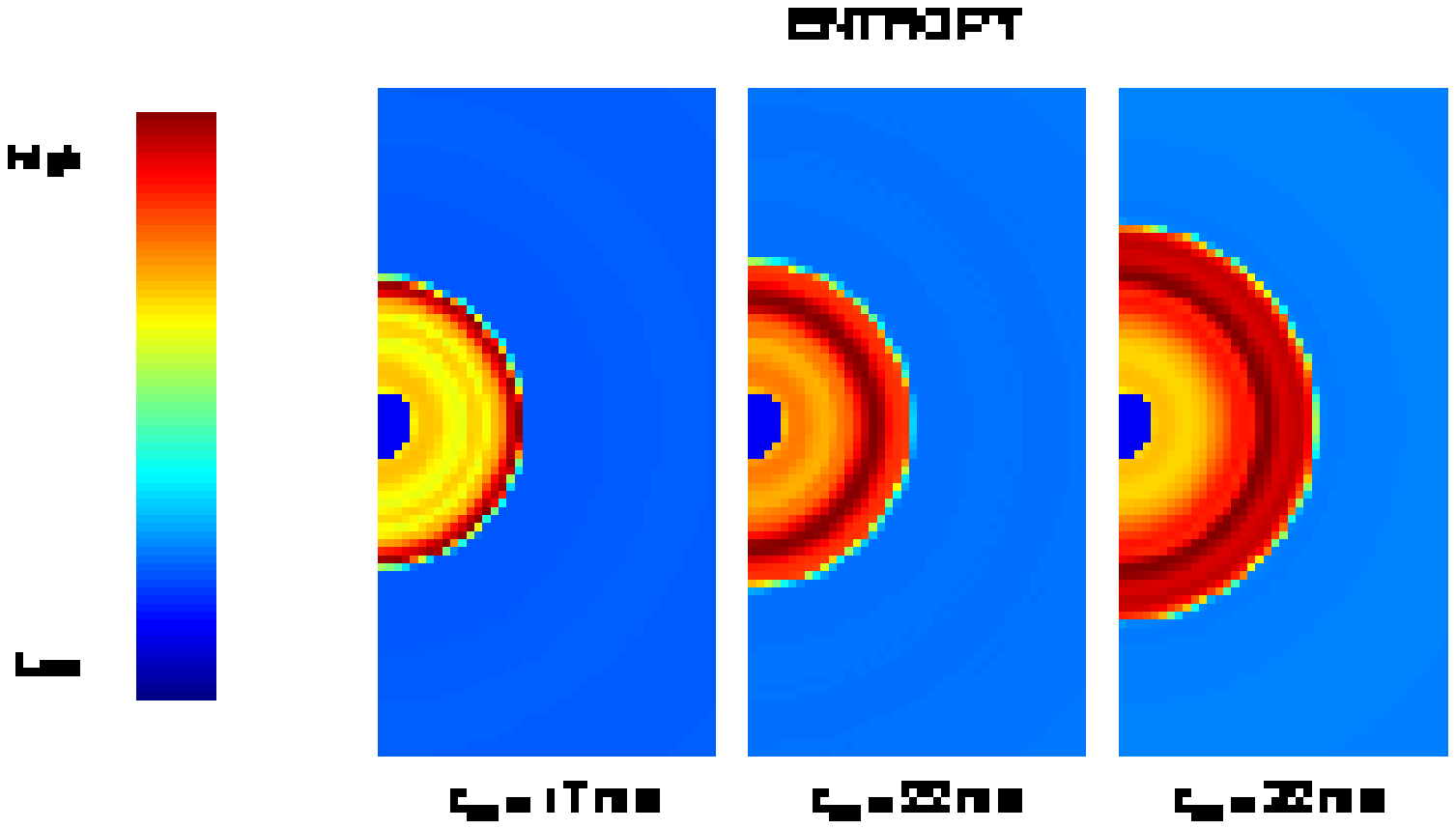}
\end{center}
\caption{Two-dimensional plots showing the entropy distribution of the
 15 $M_{\odot}$ model in a simulation ``without'' neutrino transport
 (top panel) and with neutrino transport (down panel). 
 Taken into account the neutrino transport, albeit assuming spherical
 symmetry, it is seen that the instability in the protoneutron star is damped
 out within a short time.
This figure is taken from Mezzacappa {\it et al} (1998) \cite{mezza1998}.}
\label{mezzacappa}
\end{figure}

It is noted that their neutrino transport was 
     coupled only to the radial directions and thus suppresses the neutrino
     transport in the angular directions, essentially underestimating
     the stabilizing effect of the neutrino transport. On the other
     hand, Mezzacappa
     {\it et al}
     found only mild convective activity in the region near
     the neutrino sphere \cite{mezza1998}. 
     Since the neutrino transport in Mezzacappa
     {\it et al} \cite{mezza1998} was assumed to be
     spherical, it might result in overestimating the stabilizing
     effect \cite{mezza1998} (see Figure \ref{mezzacappa}). 
     The recent two-dimensional numerical simulation shows that the PNS convection really does occur, however, is not
     important for boosting the
     neutrino luminosity, because the
     convectively active
     layer is formed rather deep inside
     the PNS and is surrounded by a convectively stable
     shell \cite{buras}. However, one may argue that the conclusion may be
     still subject to change because their neutrino transport, albeit in the
     state-of-the-art manner, is not fully spatially two-dimensional, and hence
     could not reproduce all of the properties of the convective flows. 

%     self-consistent multidimensional neutrino transport simulations
%     might change the conclusion, the
%     convective activity in the PNS might be nowadays considered to play a
%     little influence for the SN dynamics.  
%While one may argue that
%     the ray-bay-ray neutrino transport fails to reproduce all of the
%     properties of the convective flows,
%    In a recent study by Buras {\it et al} \cite{buras}, they 
%     concluded that the PNS convection is not important for boosting the
%     neutrino luminosity from the PNS.  Remembering a caveat that their ....%  and thus a future
%     self-consistent multidimensional neutrino transport simulations
%     might change the conclusion, the
%     convective activity in the PNS might be nowadays considered to play a
%     little influence for the SN dynamics.  
%\end{itemize}
%\begin{itemize}     
\subsubsection{convection in the hot-bubble regions }
In the hot bubble, in which the neutrino heating dominates over the
     neutrino cooling (the regions between $R_{g}$ and $R_{s}$ in Figure
     \ref{janka_fig}), convections
     are expected to occur by the negative entropy gradient. 
 In fact, a
     dynamical overturn between the hot, neutrino-heated, rising
     materials above gain radius and the cold postshock matter beneath
     has been demonstrated in the two-dimensional
     \cite{hera92,hera,buroheyfrex,jankamueller94,scheck} or
     the three-dimensional \cite{fryer03} numerical simulations. 
     As clearly demonstrated in \cite{jankamueller94}, these convections indeed aid the
     shock revival and can lead to explosions in the case where 
     the spherical models fail. 

     In most of the preceding studies, it is noted that 
the neutrino luminosities from
     the protoneutron star are given 
     by hand at some inner boundary (the so-called lightbulb approximation). 
     In order to boost the neutrino-heating efficiency by the convection
     in the hot bubbles,
     a sufficient deposition of the neutrino luminosity from the hot
     protoneutron star is required, which unfortunately no previously
     published models with the neutrino transport in the whole regions
     have failed to reproduce.
 \begin{figure}
\begin{center}
\epsfxsize=7.5cm
\epsfbox{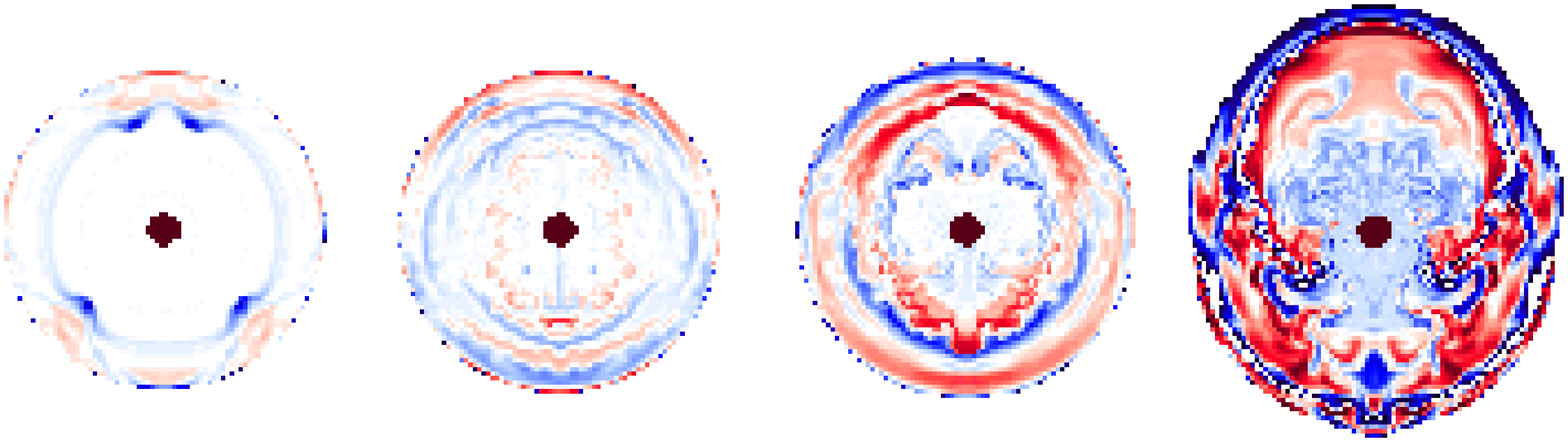}
\epsfxsize=7.5cm
\epsfbox{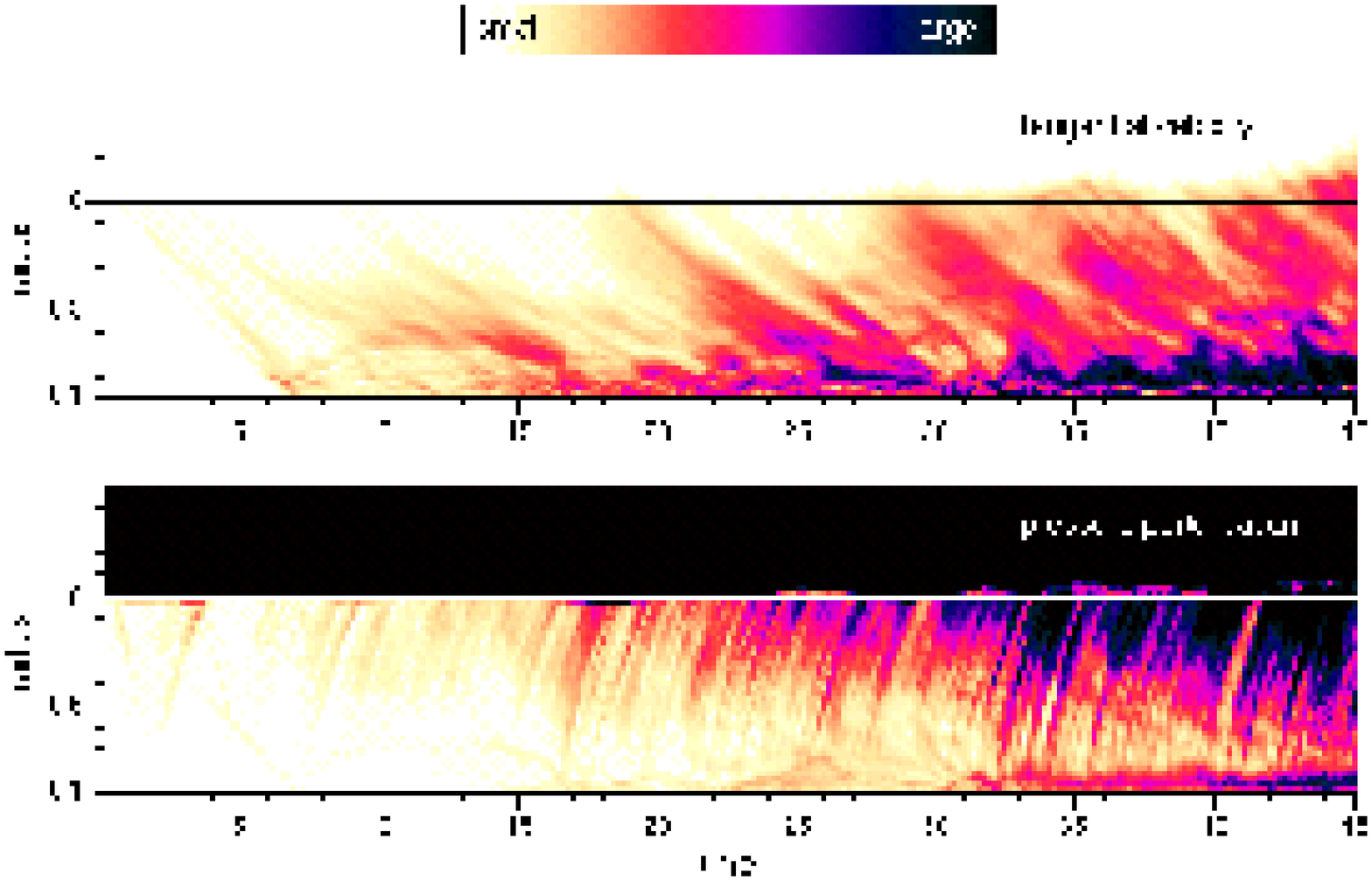}
\end{center}
\caption{Left panel shows the time evolution of the gas entropy (from left to right in this panel), 
 illustrating the growth of the instability induced by the aspherical perturbation, which is added at the stalled shock initially. Right panel shows the time evolution of the two-dimension simulation of the left panel in spacetime diagrams. The color at a given radius and time corresponds
to the angle-average of the absolute value of the
tangential velocity (right top) and absolute value of the deviation 
of pressure from the steady-state solution, $\Delta P/P$ (right bottom).  
The downflow of the tangential velocity is seen to be reflected at the inner 
boundary, producing the pressure perturbation propagating outwards 
(see the stripes in the top and down panels). 
Direct evidence of the vorticity-acoustic cycle are seen in these velocity 
plots. These figures are taken from Blondin {\it et al} \cite{blo03}.}
\label{blondinfigure}
\end{figure}
%\end{itemize}     

%\begin{itemize}
\subsubsection{instability induced by the non-radial oscillation of the stalled shock waves}
     It is recently pointed out that the
     stalled shock waves are subject to low-mode ($l=1~{\rm or}~2$) 
     aspherical oscillations 
     \cite{blo03}. Interestingly, it is not due to the convection
     induced by the negative gradients of entropy or lepton fraction
     but due to the so-called vorticity-acoustic cycle.
     In the cycle, the vorticity perturbations given at
     the stalled shocks propagate inwards and reflected at some inner
     boundary, which is presumably the surface of the protoneutron star,
     producing the acoustic waves, and propagate outwards and  
     create new vorticity perturbations when reaching the stalled
     shock waves. This closed cycle amplifies the growth of the aspherical
     oscillations (see Figure \ref{blondinfigure}). This cycle was first discovered in the context of
     stability analysis of the accretion disk around the black holes \cite{fog1,fog2}.
     In the simulations by Blondin {\it et al.} \cite{blo03}, 
     the oblique shock waves are
     found to feed vorticity in the
     postshock region and lead to growing turbulence.
     Although the model computations by them are
     not so realistic in the sense that no neutrino transport was
     included, which
     should affect the growth of the pressure waves predominantly by
     neutrino cooling, such instability seemingly appears 
     in some recent realistic supernova simulations \cite{jankapro04}.  
%\end{itemize}
Combined with the convections and the hydrodynamical instabilities
 stated above, 
it is expected to have a good chance to obtain a successful explosion,
because the neutrino heating mechanism seems very close to explosion as 
it is in the spherical collapse.
However, the difficulties of
 multi-dimensional treatment of neutrino transport
have hampered the definitive answer to 
this problem. 

Although some smoothed particle hydrodynamic (SPH) 
simulations have found explosions 
induced by the combination of 
neutrino heating and convection in the heating region
\cite{hera,fryer03},
there has been persistent concern with their approximate treatment of 
neutrino transfer. This situation may be changing recently.
Some groups are preparing for full spatially multi-dimensional neutrino transport simulations \cite{buras,livne}.
In these simulations, the dynamics of the whole core is computed with 
a code implemented with a Boltzmann solver. 
But still no successful explosions have been reported so far.
We may have to look for alternatives 
to the convections and the hydrodynamical instabilities in order to get the successful explosions.
\clearpage
\subsection{Roles of Rotation \label{roleofrotation}}
\begin{figure}
\begin{center}
\epsfxsize = 12.5 cm
\epsfbox{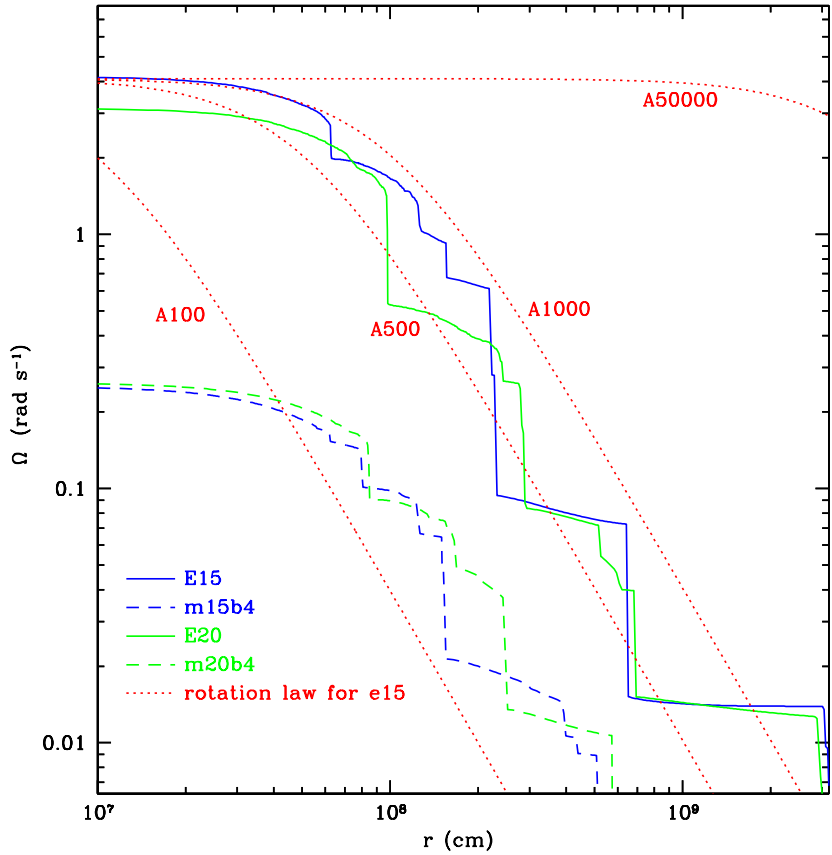}
\end{center}
\caption{Initial angular velocity as a function of the radius 
obtained in the recent stellar evolution calculations \cite{heger00,heger03}. 15 (blue) and 20 (green) indicate the mass of
 the progenitor in unit of the solar mass ($M_{\odot}$). Much smaller
 angular velocities by models m15b4 and m20b4 were evolved taking into
 account of the magnetic fields. The dotted red lines were constructed
 with the rotation law of Eq. (\ref{init_velo}) using the central angular
 velocity $\Omega_{0}$ of E15 ($4~\rm{rad}~{s}^{-1}$). Labeled number in
 the dotted red lines, such as 100, 500, represents the values of
 $R_{0}$ in Eq. (\ref{init_velo}).
 This figure is taken from \cite{ott}. }
\label{initial_ang}
\end{figure}
 In addition to the hydrodynamical instabilities, rotation can
produce the large asphericity in the supernova cores.
It is well known that the progenitors of
 core-collapse supernovae are a rapid rotator on the main sequence 
\cite{tass}. Typical rotational velocity on the equator are on the order
of $200$ km/s, which is the significant fraction of their breakup
rotational velocity \cite{fukuda}.
Recent theoretical studies suggest a fast 
rotating core prior to the collapse \cite{heger00}, although this is
not conclusive at all when the magnetic fields are taken into account 
\cite{heger03,heger04} (see Figure \ref{initial_ang}).
Neutron stars, which are created in the aftermath of 
the gravitational core collapse of massive stars at the end of their lives,
 receive the rapid rotation, which are
believed to be observed as pulsars. 
From the above facts, rotation seems naturally to be taken 
into account in order not only to clarify the explosion mechanism, 
but also to explain the observed properties of core-collapse supernovae.

 So far there have been some works devoted to the understanding of the effect of rotation upon the supernova explosion mechanism \cite{muhi,boden,  
sym, monch, yama94, fry00,fryer}. Among the studies, the systematic 
study of the rotational core-collapse, changing the initial angular momentum
distributions and the strength has been done 
(see, for example, \cite{yama94,kotakeaniso,ott}). 
In the following, we show how the hydrodynamics in rotational
core-collapse deviates from the one in spherical symmetry
 (section \ref{standardscenario}). 
%according to a recent study \cite{kotakeaniso}. 

\subsubsection{hydrodynamics in rotational core-collapse}
The story of rotating core is not greatly different from the standard
picture from the core bounce to the explosion as stated in section 
\ref{supernova_theory}. In addition to the gravity and pressure
gradients, one has to take into account the effect of centrifugal
forces.  
\begin{figure}
\begin{center}
\epsfxsize = 7 cm
\epsfbox{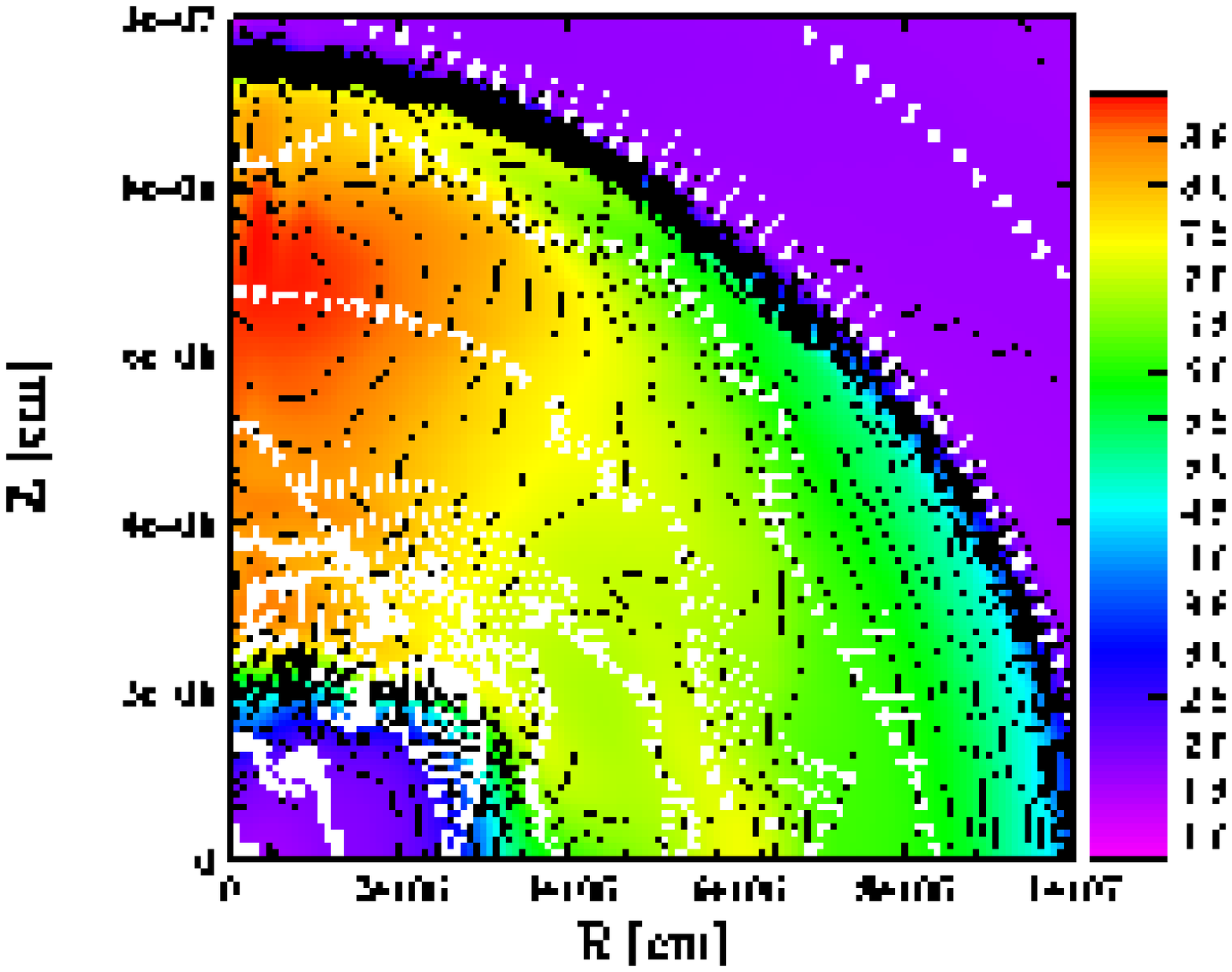}
\epsfxsize = 7 cm
\epsfbox{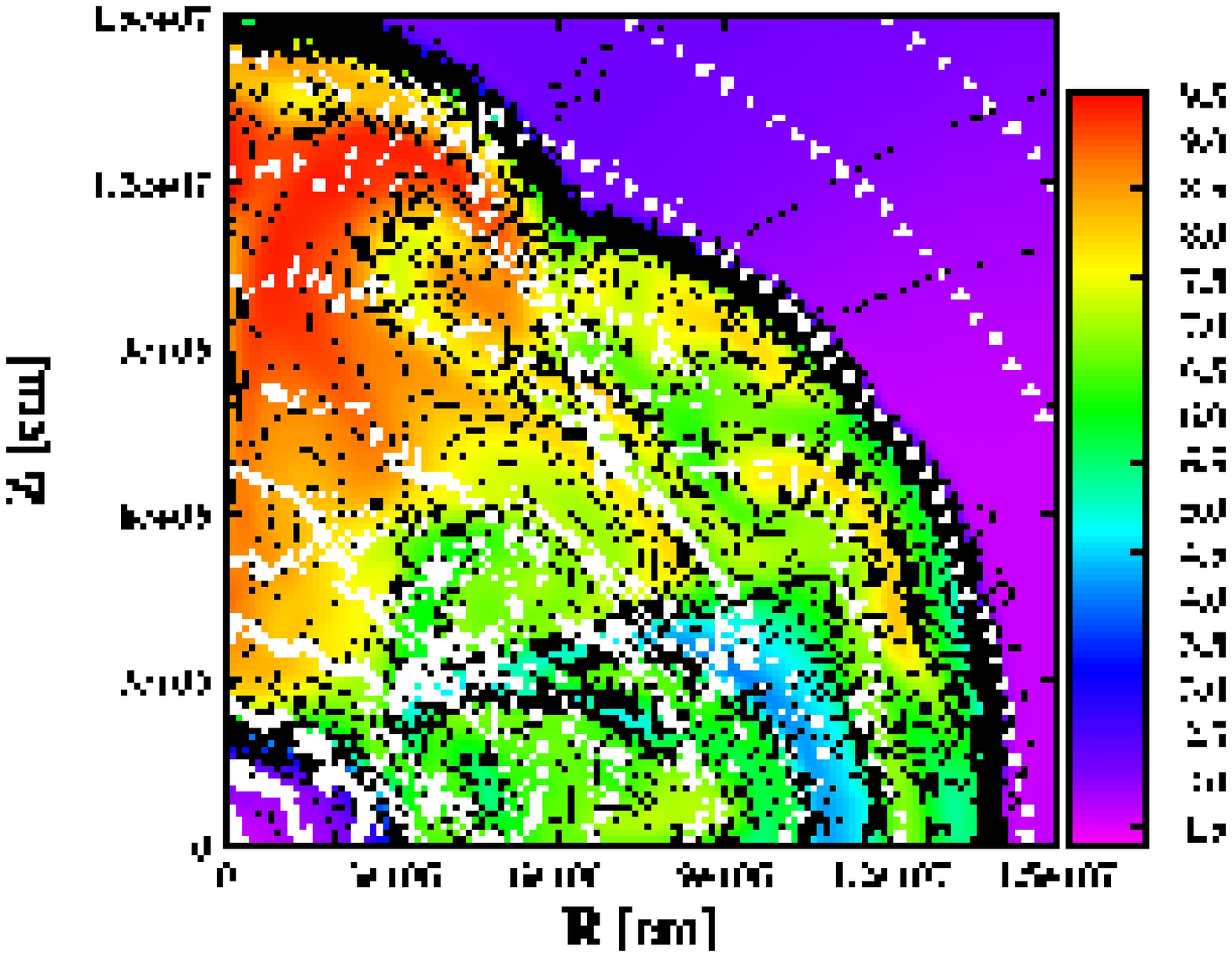}
\caption{Time evolution of the shock wave near core bounce for a typical 
rotational core-collapse model (we call it model A), whose initial angular momentum distribution is
 based on the rotational progenitor model by \cite{heger00}. The two
 snapshots show color-coded maps of entropy ($\rm{k_{B}}$) per nucleon
 together with the velocity fields. The left panel is for 2 msec after
 the core bounce when the shock wave begins to move upwards at the pole,
 the right panel is for 4 msec. In the case of weak differential
 rotation, it is found that the hot entropy blob is formed near the
 rotational axis and it moves upwards. Note the difference of the scale
 of the plots. These figures are taken from \cite{kotakeaniso}.}
\label{heger_1000}
\end{center}
\end{figure}
 In Figure \ref{heger_1000}, the time evolution of the shock wave in a rotational core-collapse simulation
\cite{kotakeaniso} is
shown. The initial angular velocity profile of this model is based on
the rotational progenitor model by \cite{heger00}, which is
approximately fitted by the following way, 
\begin{equation}
  \Omega(r) = \Omega_0 \times \frac{{R_{0}}^2}{r^2 + R_{0}^2},
\label{init_velo}
\end{equation}
where $\Omega(r)$ is an angular velocity, $r$ is a radius, and
$\Omega_0, R_{0}$ are model constants taking $\Omega_0 = 4~{\rm rad}~{\rm
s}^{-1}$ and $R_{0} = 1000~{\rm km}$ (see Figure \ref{initial_ang} and
A1000). As a
sideremark we note that $r$ can be also interpreted as the distance from the
rotational axis (the so-called cylindrical rotation). At first glance,
the rotation profile seems more natural due to the Poincar\'{e} and
Wavre theorem \cite{tass}, however, the shellular rotation 
profile, in which the angular velocity is constant with radius, is 
also pointed out to be natural because 
the horizontal turbulence is likely to be much stronger than the
 vertical one during stellar evolutions with rotation (for an elaborate description of the 
 presupernova models including rotation, the reader is referred to 
\cite{heger_phd}).  
Since even the-state-of-the art rotational progenitor models are
based on the spherically symmetric models, it may not be decided yet 
which rotation profile is correct. Considering the uncertainties of the
progenitor models,  the rotating 
pre-collapse models so far have been made by changing parametrically 
not only the rotational profile (shellular or cylindrical), but also 
the rotational velocity and the degree of the differential rotation,
and then adding them to
the spherically symmetric progenitor models.

Now let us return to the discuss of the model in Figure \ref{heger_1000} again.
At the central density of $2.3 \cdot 10^{14} \rm{g}\,\,\rm{cm}^{-3}$,
the model bounces at the pole first. The bounce epoch is slowed down
of the order of $10 {\rm ~msec}$ than the one 
in absence of rotation due to the centrifugal forces acting against the
gravitational pull. The occurrence of 
core bounce below the nuclear density ($\sim 3.0 \times
10^{14}\rm{g}\,\,\rm{cm}^{-3}$) is a general feature in case of
rotational core-collapse. This can be understood from the fact that
rotation acts like a gas with an adiabatic index of $\gamma = 5/3$. 
(see, for example, \cite{tass,monch,shapiro})
Due to this additional pressure support, whose gradient is steepest
along the rotational axis, the core bounce occurs at the subnuclear 
density at the pole first.  
%There is an another general feature that epoch of core bounce is slowed down  %These features can be also understood by the
%stabilizing effect of rotation. A critical condition against 
%the radial modes in rotating fluid is given by Ledoux, 
%\begin{equation}
%\gamma > \gamma_{\rm crit} = \frac{2}{3}\frac{2-5\beta}{1-2 \beta}.
%\end{equation} 
%Due to the conservation of angular momentum, the values of $\beta$
%becomes as high as $\sim 9 \%$. Since the adiabatic index is close to
%$4/3$ even
%From this analysis, one can expect that the even a
%small amount of rotation may be enough to stop the collapse and lead to
%the formation of stable rotating configurations below nuclear density.

\begin{figure}
\begin{center}
\epsfxsize=17.0cm
\epsfbox{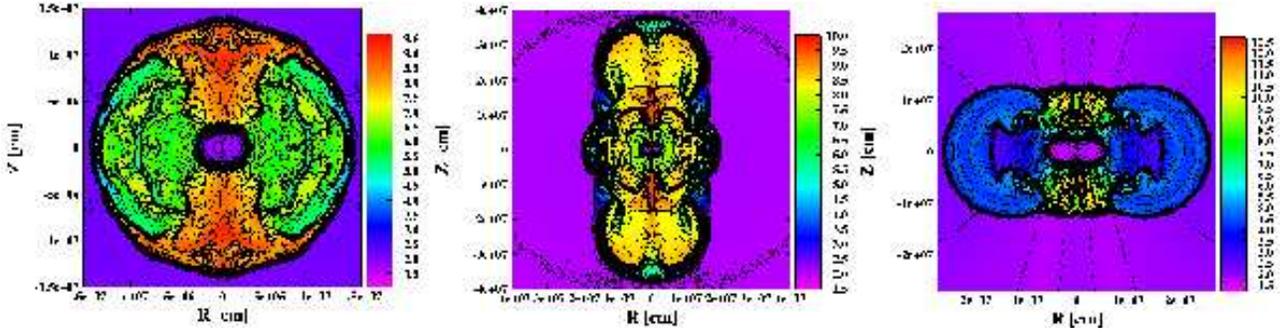}
\end{center}
\caption{Final profiles of the representative models. They show
 color-coded contour plots of entropy ($k_{\rm B}$) per nucleon.
The initial value of $T/|W|$ is $0.5 \%$ for the models of the left
 (model A) and
 central panels (model B), $1.5 \%$ for the model of the right panel (model C). Here
$T/|W|$ is the ratio of rotational to gravitational energy. The model
 of the central panel has stronger differential rotation than that of
 the left panel. The value of $R_{0}$ of model B (central panel) is
taken to be $100$ km while the initial value of $T/|W|$ is the same as
that of model A with $R_{0} = 1000~{\rm km}$.   
\label{fig1_IOP}}
\end{figure}

After core bounce, the shock wave begins to propagate a little bit faster along
the rotational axis supported by the buoyancy of the hot entropy blob
($S \geq 9.5 \rm{k_{\rm B}}/ \rm{nucleon}$) (see Figure
\ref{heger_1000}). At this time $R$, which is the aspect ratio of the
shock front, is 1.1, which is lowered to 1.0 by
the shock stagnation (see the right panel of Figure
\ref{fig1_IOP}). The initially prolate configuration ($R$ = 1.1) is
stretched in the direction of the equator after the shock is weakened at
the pole by the neutrino energy loss and ram pressure of the infalling
material and finally stalls in the iron core. Note that a large angular 
momentum tends to push the matter parallel to the equatorial plane. 

Depending on the total angular momentum and its distribution imposed
initially on the iron core, the effects of rotation on the core dynamics
have many variety.
In Figure \ref{fig1_IOP}, the entropy distributions at the shock-stall
for some representative models \cite{kotakeaniso} are presented.  
In fact, a variety of the final profiles is immediately seen in the figure.    
As the initial rotation rate becomes larger, 
the shape of the stalled shock wave
becomes more oblate due to the stronger centrifugal force 
(compare the left with right panel in Figure
\ref{fig1_IOP}). If the initial rotation rate is the same, 
the shape of the stalled shock is found to be elongated in the direction 
of the rotational axis as the differential rotation becomes stronger (compare
the left with the central panel of Figure
\ref{fig1_IOP}). 
This is related with the production of high entropy blobs by core
bounce. As mentioned above, the entropy blob formed near
the rotational axis floats up parallel to the axis and then stalls for
weak differential rotations. This makes the shock prolate at first. 
Then the matter distribution returns
to be spherical or oblate due to the centrifugal forces. On the other
hand, in the case of strong differential rotations, the shock wave 
formed first near the rotational axis hardly propagates and stalls very 
quickly. The high entropy blob begins to grow near the equatorial plane 
in this case. This then induces the flows towards the rotational axis. 
As a result, the final configuration becomes prolate. 

\begin{figure}
\begin{center}
\epsfxsize= 7.5 cm
\epsfbox{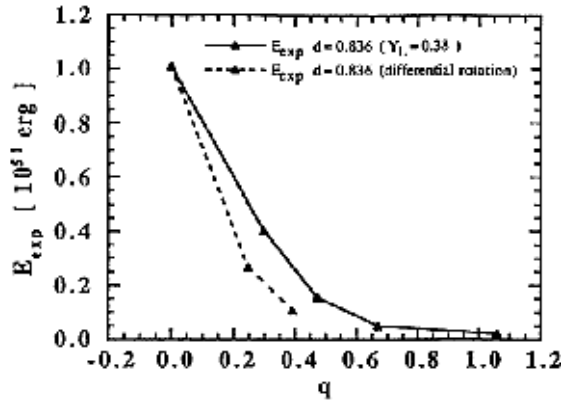}
\end{center}
\caption{Relation between the explosion energy and the initial 
total angular momentum taken from \cite{yama94}. $q$ in the $x$ axis means the normalized total angular momentum defined by $q \equiv J/(2GM/c^2$) with $J, M$ being the angular momentum and the mass of the core, respectively. 
The solid and dashed lines correspond to the uniform and differential rotation models, respectively. ``d=0.836'' in the figure indicates that in both cases the employed equation of state is the same (see \cite{yama94} for details). 
As the initial total angular momentum increases, the explosion energy monotonically decreases.
\label{yamasato94}
}
\end{figure}

When rotation is taken into account, the shock wave can generally 
reach further out than the one without  rotation ($R_{\rm stall} <
\sim 200 {\rm km}$). However, it was pointed that rotation does not good 
to the prompt explosion \cite{yama94}. 
This is because the centrifugal
force tends to halt the core collapse, which then reduces the conversion
of the gravitational energy to the kinetic energy as clearly seen from 
Figure \ref{yamasato94}. 

\subsubsection{anisotropic neutrino radiation in rotational
   core-collapse \label{anisoneu}}
Then the problem is whether rotation does good or harm to the neutrino-driven mechanism.
K. Sato's group in University of Tokyo has been paying attention
to the effect of rotation on the neutrino heating mechanism. First of
all,  Shimizu et al. (2001) \cite{shimi01} demonstrated that anisotropic neutrino
radiations induced by rotation may be able to enhance local heating
rates near the rotational axis and trigger globally asymmetric
explosions. The required anisotropy of the neutrino luminosity appears to
be not very large ($\sim  3\%$). 
In their study, the anisotropy of neutrino heating was given
 by hand and rotation was not taken into account, either. 
Kotake {\it et al.} (2003) \cite{kotakeaniso} demonstrated how large the anisotropy of
neutrino radiation could be, based on the two dimensional 
rotational core collapse simulations from the onset of gravitational
collapse of the core through the core bounce to the shock-stall 
. In their study,  a tabulated
EOS based on the relativistic mean field theory \cite{shen98} was
implemented and the electron captures and neutrino transport was
approximated by the so-called leakage scheme. They not only estimated
 the anisotropy of neutrino luminosity but also calculated local heating 
rates based on that. In the following, we state the main results shortly.

\begin{figure}
\begin{center}
\epsfxsize=7.5cm
\epsfbox{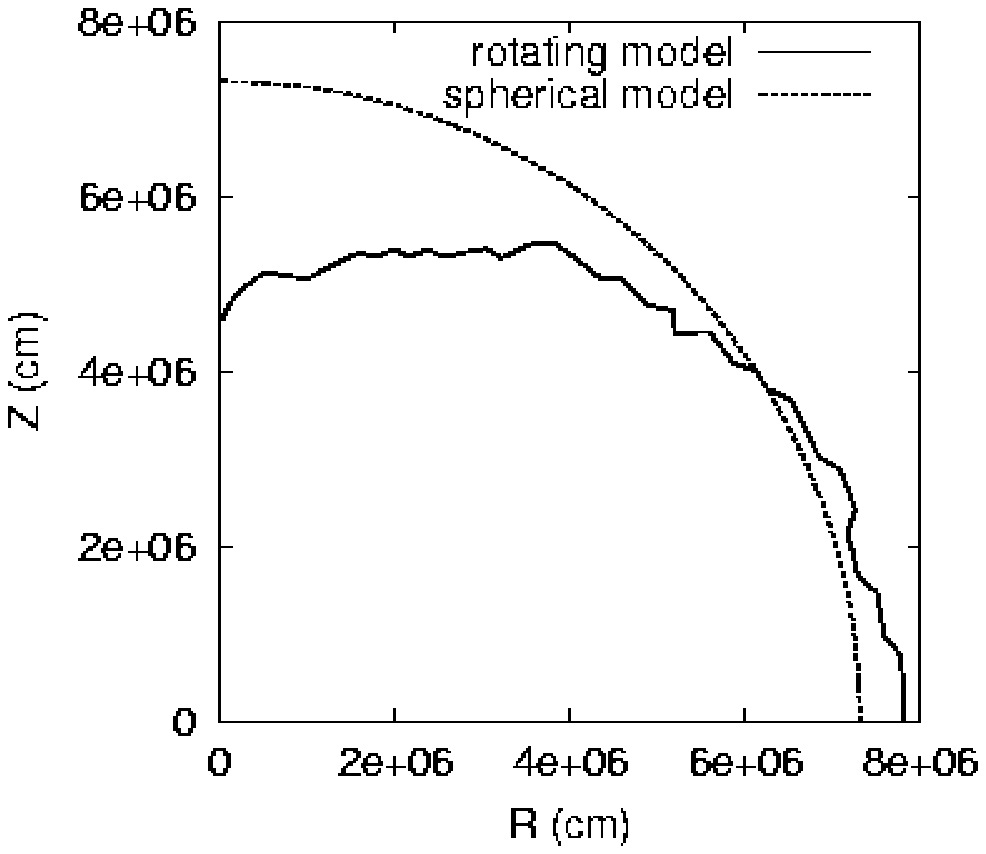}
\epsfxsize=7.5cm
\epsfbox{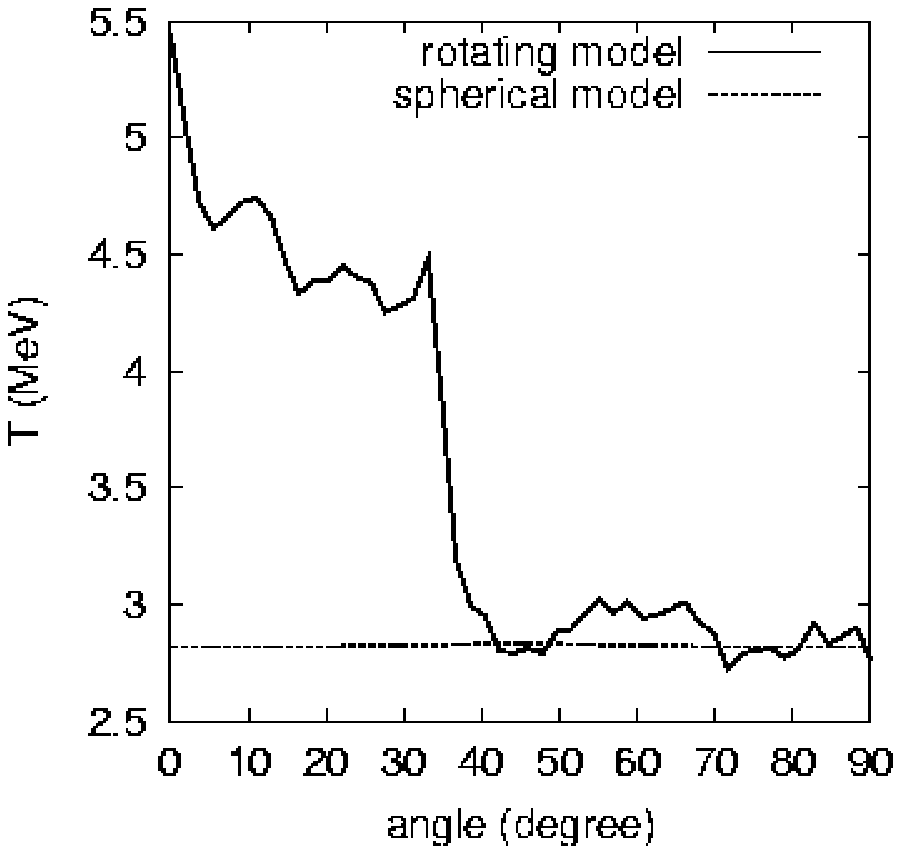}
\end{center}
\caption{Shapes of neutrino sphere (left panel), neutrino temperature
 vs. polar angle on the neutrino sphere (right panel) for 
 the rotating and the spherical models. These figures are taken from \cite{kotakeaniso} \label{fig2}}
\end{figure}

In the left panel of Figure \ref{fig2}, the neutrino spheres after the 
shock-stall ($\sim 50$ msec after core bounce) for the  
spherical and the rotating model (corresponding to model A in Figure
\ref{heger_1000}) are presented.
 For the rotating model, it is
found that the neutrino sphere forms deeper inside at the pole than for
the spherical model. This is a result of the fact that the density is
lower on the rotational axis in the rotation models than in the
spherical model because the matter tends to move away from the axis due
to the centrifugal force. The neutrino temperature profile on the neutrino 
sphere for the pair models is presented in the right panel of Figure 
\ref{fig2}. Note that 
the neutrino temperature is assumed to be equal to the
matter temperature. It is seen from the figure that the
 temperature varies with the polar angle for the rotating model.  
The neutrino temperature is higher at the pole for the rotating model
than for the spherical model. This can be understood from the fact that
the neutrino sphere is formed deeper inside for the rotational model
than for the spherical model, as mentioned above.

\begin{figure}
\begin{center}
\epsfxsize=15cm
\epsfbox{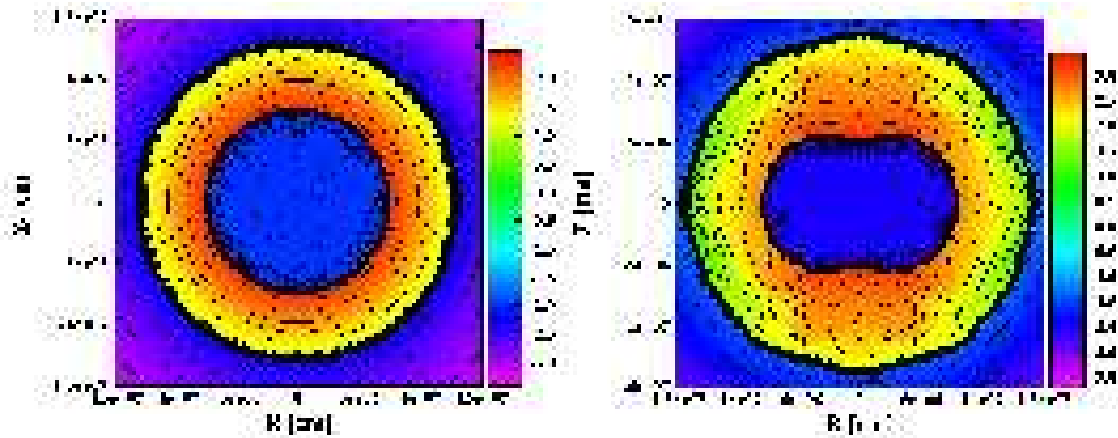}
\end{center}
\caption{Heating rates outside the neutrino sphere for the
 spherical model (right) and the rotating model (left). The color scale is 
  the logarithm of the heating rate
 ($\rm{MeV}~\rm{nucleon}^{-1}~\rm{s}^{-1}$). The neutrino sphere and the
 stalled shock are seen as the thick lines separating the bright color
 from the dark color region. Note that the value within the neutrino
 sphere is artificially modified to dark colors and has no physical
 meanings. These figures are taken from \cite{kotakeaniso}.
\label{fig3}}
\end{figure}

\begin{figure}
\begin{center}
\epsfxsize=7.5cm
\epsfbox{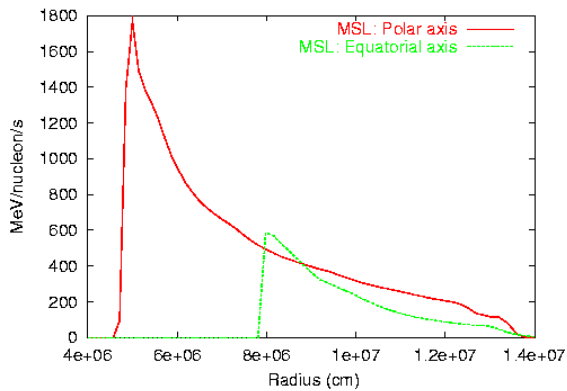}
\end{center}
\caption{The comparison of the heating rate ${Q^{+}}_{\nu}\,\,(\rm{MeV} /(\rm{nucleon} \cdot \rm{s}))$ along the rotational axis with that on the equator for the rotating model. The pole-to-equator ratio of the heating rate outside the neutrino sphere is about 3. These figures are taken from \cite{kotakeaniso}. 
\label{fig4_hikaku}}
\end{figure}

Based on the above results, the heating rates of the charged-current interaction: $\nu_e + n \rightarrow p + e^{-}$ outside 
the neutrino sphere were estimated. The neutrino emission from each
point on the neutrino sphere is assumed to be isotropic and take a
Fermi-Dirac distribution with a vanishing chemical potential. For the
details about the estimation, we refer readers to \cite{kotakeaniso}. In Figures  \ref{fig3} and \ref{fig4_hikaku}, the contour plots of the heating rate for the above two models and the neutrino heating rate along the rotational axis with that on the equatorial plane for the rotating model are presented, respectively. 
It is clearly seen from the figure that the neutrino heating occurs
anisotropically and is stronger near the rotational axis for the rotation
models. This is mainly because the neutrino temperatures at the rotational axis are higher than on the equatorial plane. In addition, the radius of the
 neutrino sphere tends to be smaller in the vicinity of the rotational
 axis. As a result, the solid angle of the neutrino sphere is larger
 seen from the rotational axis. These two effects make the neutrino
 heating near the rotational axis more efficient.

\begin{figure}
\begin{center}
\epsfxsize=7.5cm
\epsfbox{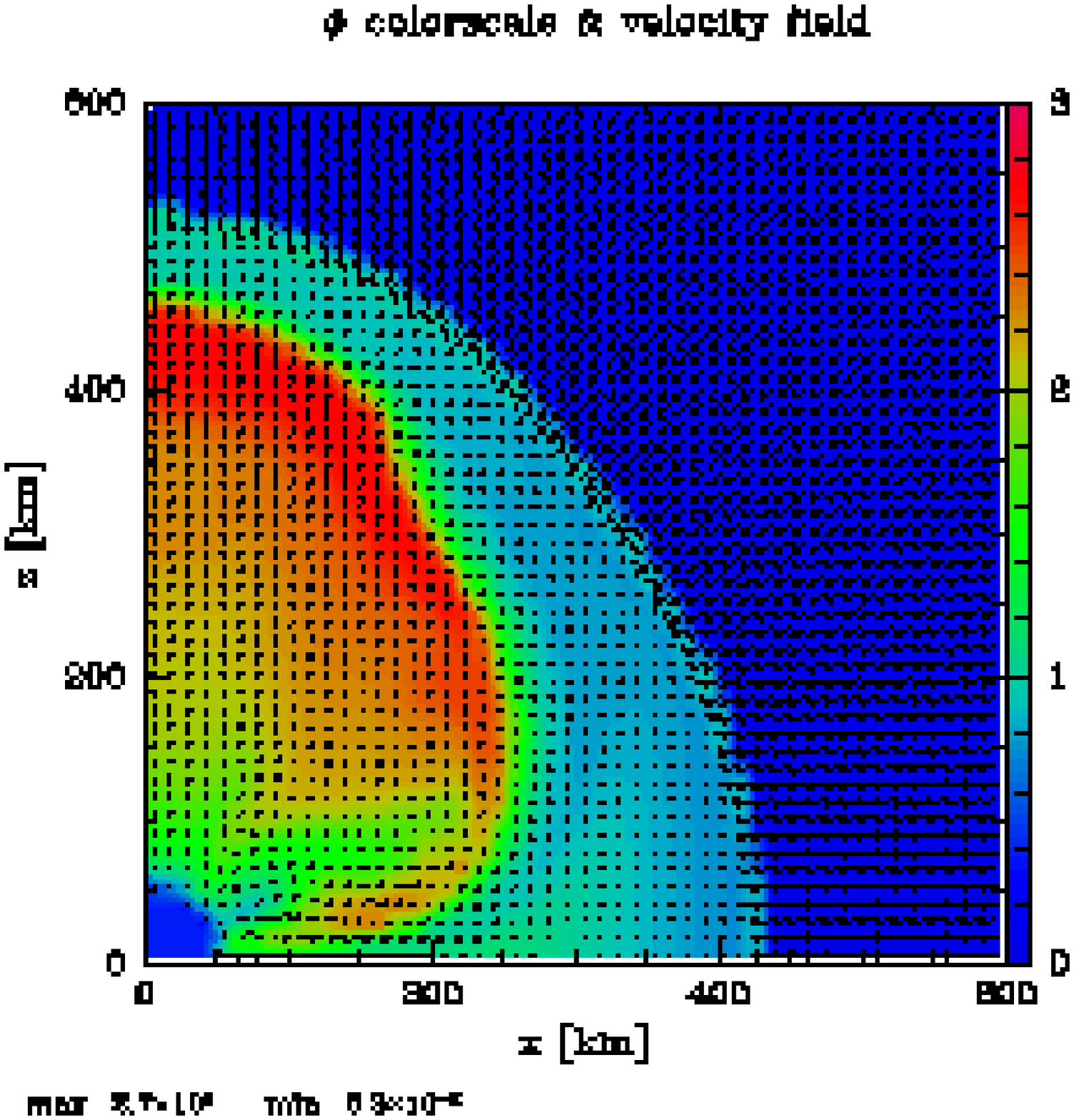}
\epsfxsize=7.5cm
\epsfbox{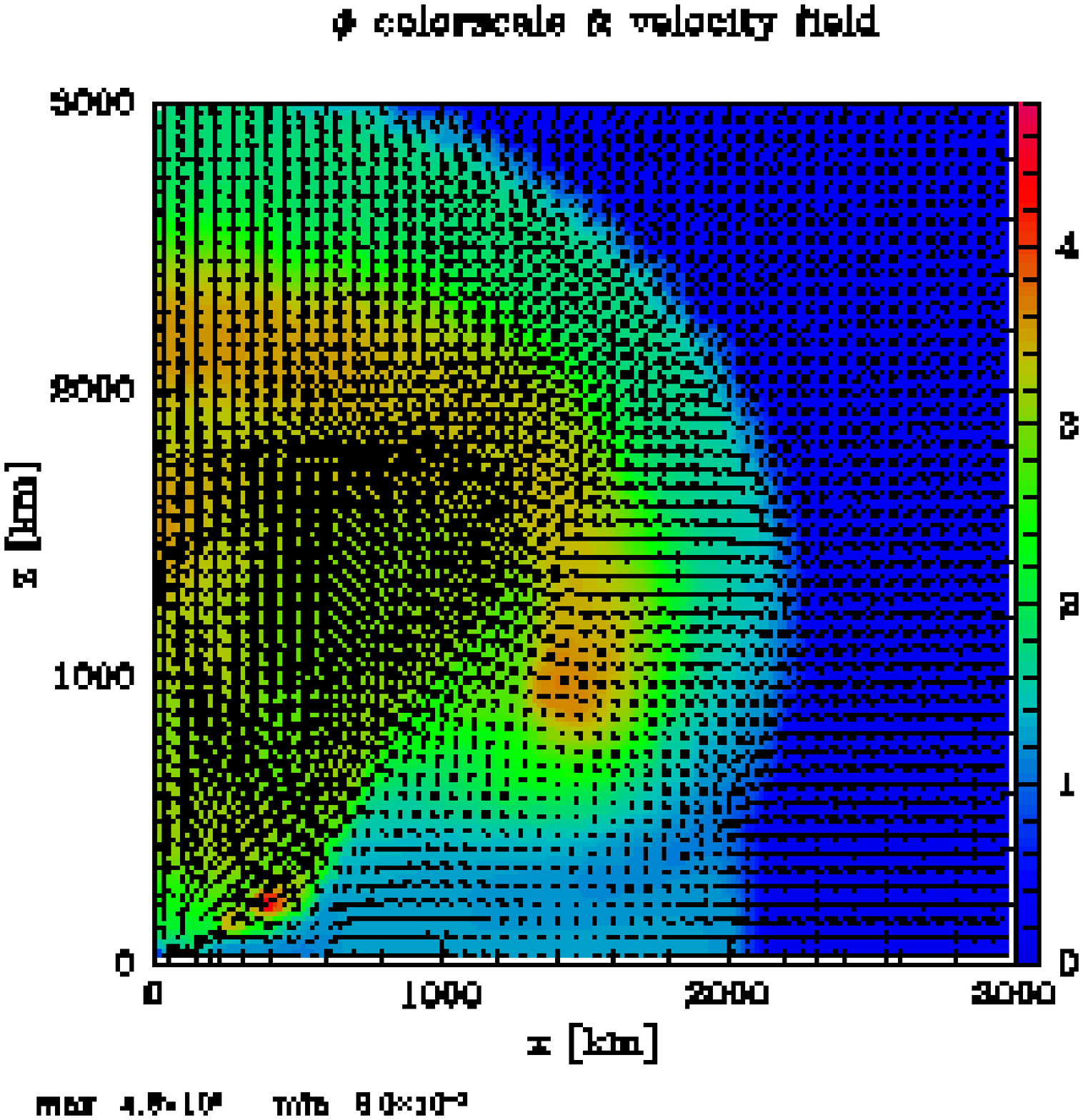}
\end{center}
\caption{Possible consequence of anisotropic neutrino radiation from the
 central protoneutron star, with the surface temperature of 4.5 MeV with
 10 $~\%$ neutrino flux enhancement along the rotational axis treated as
 the lightbulb approximation. The color map shows the dimensionless
 entropy distribution and the velocity fields (left $t=82~{\rm msec}$
 and right $t=244~{\rm msec}$ after the shock stall). This figure is
 taken from Madokoro {\it et al.} (2003) \cite{mado1}.}
\label{mado_fig}
\end{figure}     
\begin{figure}
\begin{center}
\epsfxsize=7.5cm
\epsfbox{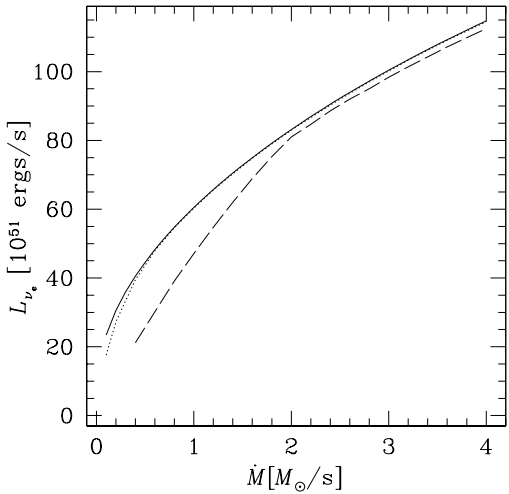}
\end{center}
\caption{Critical luminosity for the accretion shock revival as a function of the accretion rate. Solid line denotes the spherical model, and the other lines correspond to the rotating models with the different initial angular velocities ($\Omega = 0.03~\rm{rad}~\rm{s}^{-1}$; dotted line and 
$0.1~{\rm rad}{\rm s}^{-1}$; dashed line). It can be seen that the rotation can lower the critical luminosities for a give accretion rate. These figures are taken from \cite{yamayama}.}
\label{yamayama}
\end{figure}     
\begin{figure}
\begin{center}
\rotatebox{270}{\includegraphics[width=0.8\textwidth]{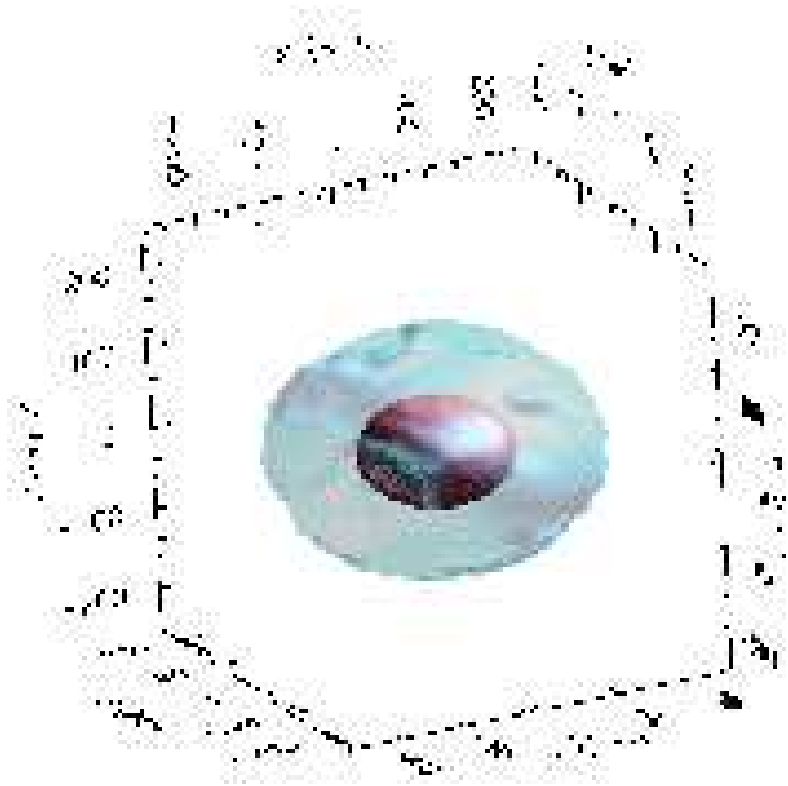}}
\end{center}
\caption{Density isosurface ($10^{11}~{\rm g}~{\rm cm}^{-3}$ (blue),
  $10^{12}~{\rm g}~{\rm cm}^{-3}$ (red)) implying the shapes of the
  neutrino spheres obtained in the 3D rotational core-collapse
  simulations by Fryer and Warren (2004) \cite{fryer} (at 45 msec after
  core bounce in model SN15BB-hr). It is shown that their shapes are
  deformed to be oblate and the aspect ratios of the neutrino spheres are
  about $\sim$ 2.}
\label{deformneutrino}
\end{figure}
\begin{figure}

\begin{center}
\epsfxsize = 7.0 cm
\epsfbox{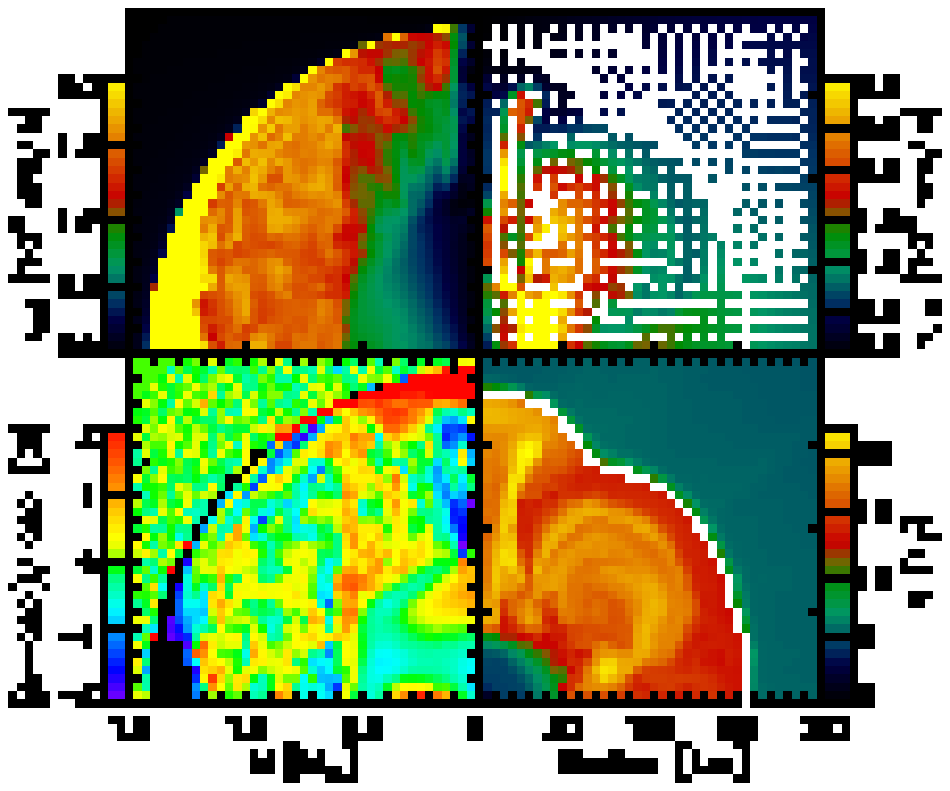}
\epsfxsize = 7.0 cm
\epsfbox{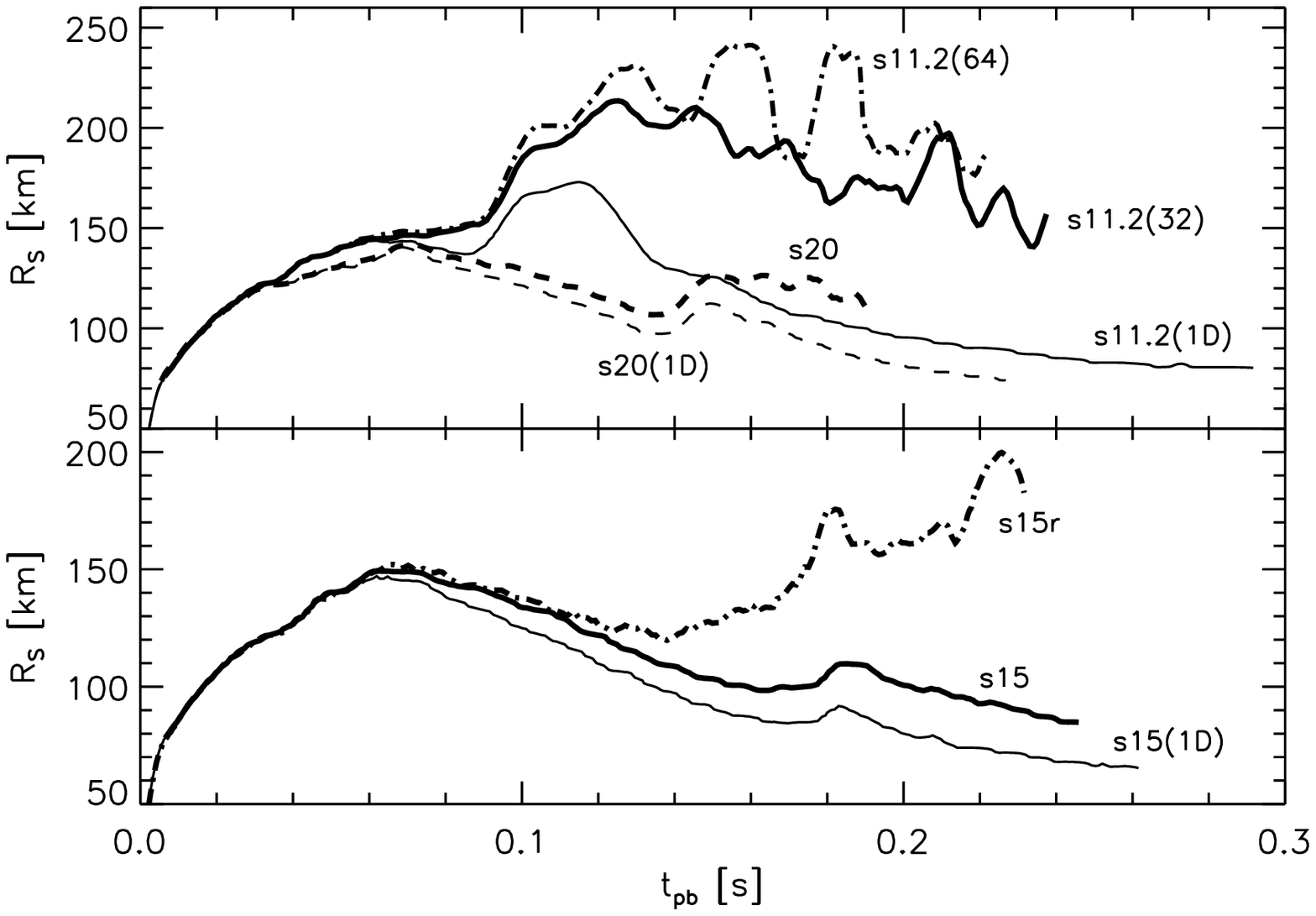}
\caption{Left is the snapshots of the stellar structure for a rotating
 model ($\Omega = 0.5~{\rm rad}~{\rm s}^{-1}$ imposed on $15M_{\odot}$
 progenitor model) at 198 msec after core bounce. The left panels show
 the rotational velocity (top) and the fluctuations of entropy (in
 percent) versus the enclosed mass, emphasizing the conditions inside
 the neutron star. The right panels display the rotational velocity
 (top) and the entropy as functions of radius. The arrow indicate the
 velocity field and the white line marks the shock front.
Right panel shows the shock radii ($R_s$) vs. postbounce time ($t_{\rm pb}$). The 2D models are compared to the corresponding 1D simulations (thin lines).
 These figures are taken from Buras {\it et al} (2003) \cite{buras}.}
\label{buras_fig}
\end{center}
\end{figure}     

\begin{figure}
\begin{center}
\epsfxsize = 12.0 cm
\epsfbox{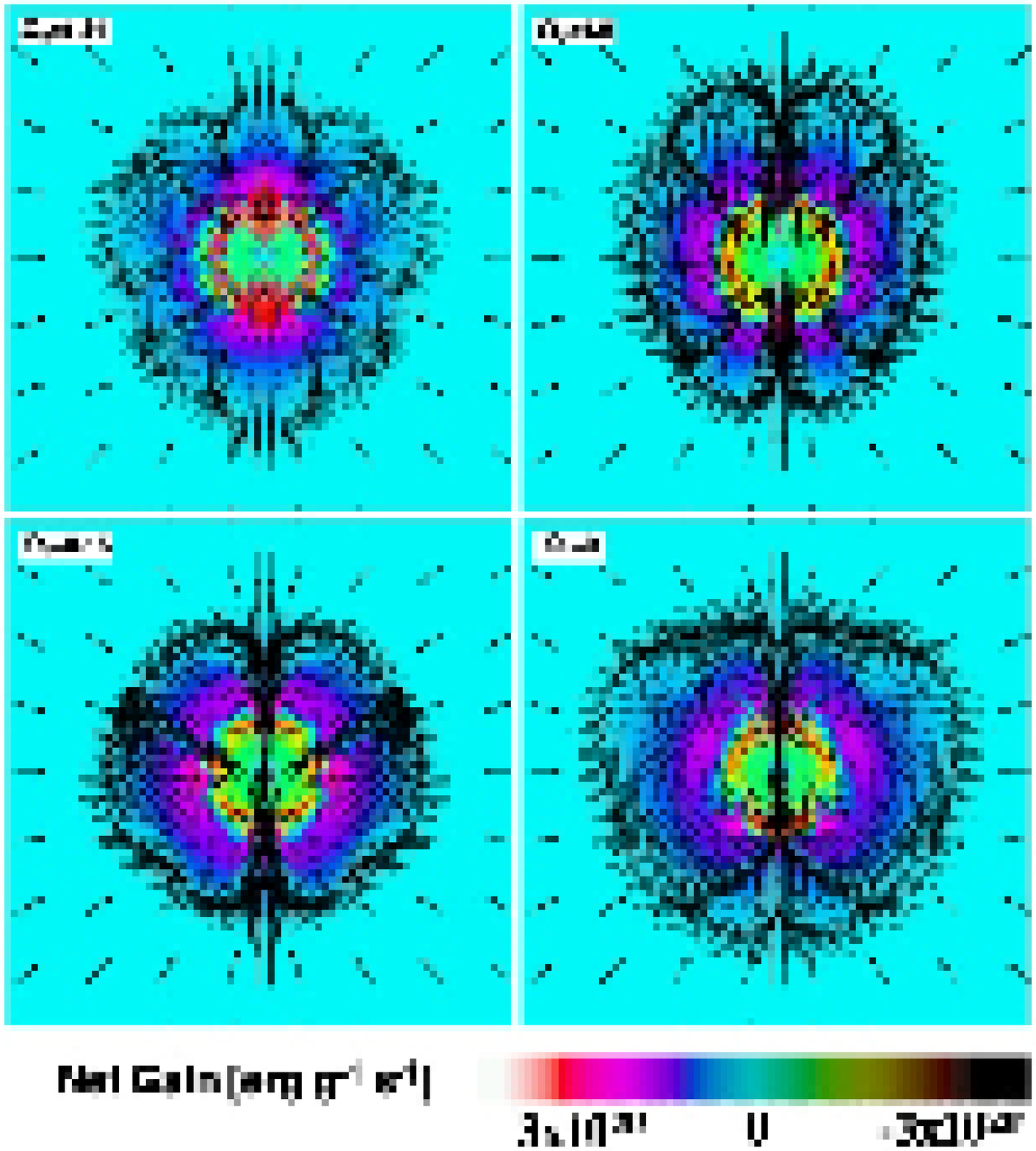}
\caption{Integrated net gain (in unit of ${\rm erg}~{\rm g}^{-1}~{\rm s}^{-1}$) in the rotational core-collapse simulations with multigroup neutrino transport with a flux limitter. $\Omega_0$ in each panel shows the initial angular velocity of the core. The inner region 600 km on a side is shown. With increasing initial angular velocity, the heating rate is shown to be more and more concentrated along the rotational axis. This figure is taken from Walder {\it et al} (2005) \cite{walder}.}
\label{walder_fig}
\end{center}
\end{figure}
   By performing the
 linear analysis for the convective stability in the corresponding models, 
it was found that the convective regions appear near the rotational axis 
(see \cite{kotakeaniso} and \cite{fry00}). 
This is because the gradient of the angular momentum is rather small near the axis and the stabilizing effect of
rotation is reduced there. The neutrino heating enhanced
near the rotational axis might lead to even stronger convection there
later on. Then, the outcome will be a jet-like explosion as considered in
 \cite{shimi01,mado1,mado2} (see Figure \ref{mado_fig}).  Here it should be mentioned 
that such jet like explosions play important roles in reproducing the
 synthesized abundance patterns of SN1987A \cite{nagataki1,nagataki2}.

Yamasaki and Yamada (2005) \cite{yamayama} recently reported the steady
accretion flows onto the protoneutron star with a standing shock, in
which they investigated how rotation affects the critical luminosity
required for the shock revival. Note that this study is the extension 
of the study by \cite{bur93}, in the sense that the former newly takes
into account the effect of rotation.  
For a given mass accretion rate, they found that rotation does lower the critical luminosity than the one in the case of the spherically symmetric mass accretion (see Figure \ref{yamayama}). This result is also in favor of rotation for producing the successful explosions. 

A series of SPH simulations in 2D or 3D of rotational core-collapse has been 
done by \cite{fry00,fryer}.  They referred
to the deformation of the neutrino sphere induced by rotation (see
figure \ref{deformneutrino}).
However, little effect of neutrino anisotropy on the
explosion was found. This is probably their models explode in the
prompt-shock timescale. 

Buras {\it et al.} (2003) \cite{buras} reported the core-collapse
simulations of a slowly rotating model, in which the state-of-the-art
neutrino reactions are included with the multigroup neutrino transport along
the radial rays. Although the shock wave reaches further out 
in the rotating model than the one in the spherically symmetric model,
the shock wave in the rotating model is shown to stall in the iron core 
(compare ``s15r'' (rotating model) and ``s15'' (spherical model) 
in the right-down panel of Figure \ref{buras_fig}).  

Very recently, Walder {\it et al} (2005) \cite{walder} reported the
two-dimensional rotating core-collapse simulations, in which the
multi-energy neutrino transport with the flux-limited diffusion
approximation was employed, and calculated the anisotropies of the
neutrino flux in the rotating cores (see Figure \ref{walder_fig}). 
The degree of the anisotropy
obtained in their simulations is almost the same with the aforementioned
results \cite{kotakeaniso} when the initial rotational velocity of the
core are close with each other. In addition, they pointed out that the
degree of the anisotropy becomes much smaller if the core rotates rather
slowly as suggested by the recent stellar evolution calculation
\cite{heger04} and thus concluded that the rotation-induced neutrino
heating anisotropy could not be a pivotal factor in the supernova
explosion mechanism. 

Here it should be mentioned that the lateral neutrino heating,
which is essential for the anisotropic neutrino radiation to work,
can neither treated very appropriately by the neutrino transport along the
radial rays \cite{buras} nor by the flux-limited diffusion approximation
\cite{walder}. Especially the diffusion approximation
could lead to the underestimation of the anisotropy ratios.
 The fully spatially multidimensional 
radiation-hydrodynamic 
simulations seem required to give us the answer whether the 
anisotropic neutrino radiation does really help the explosion. 

\begin{figure}
\begin{center}
\epsfxsize = 7.5 cm
\epsfbox{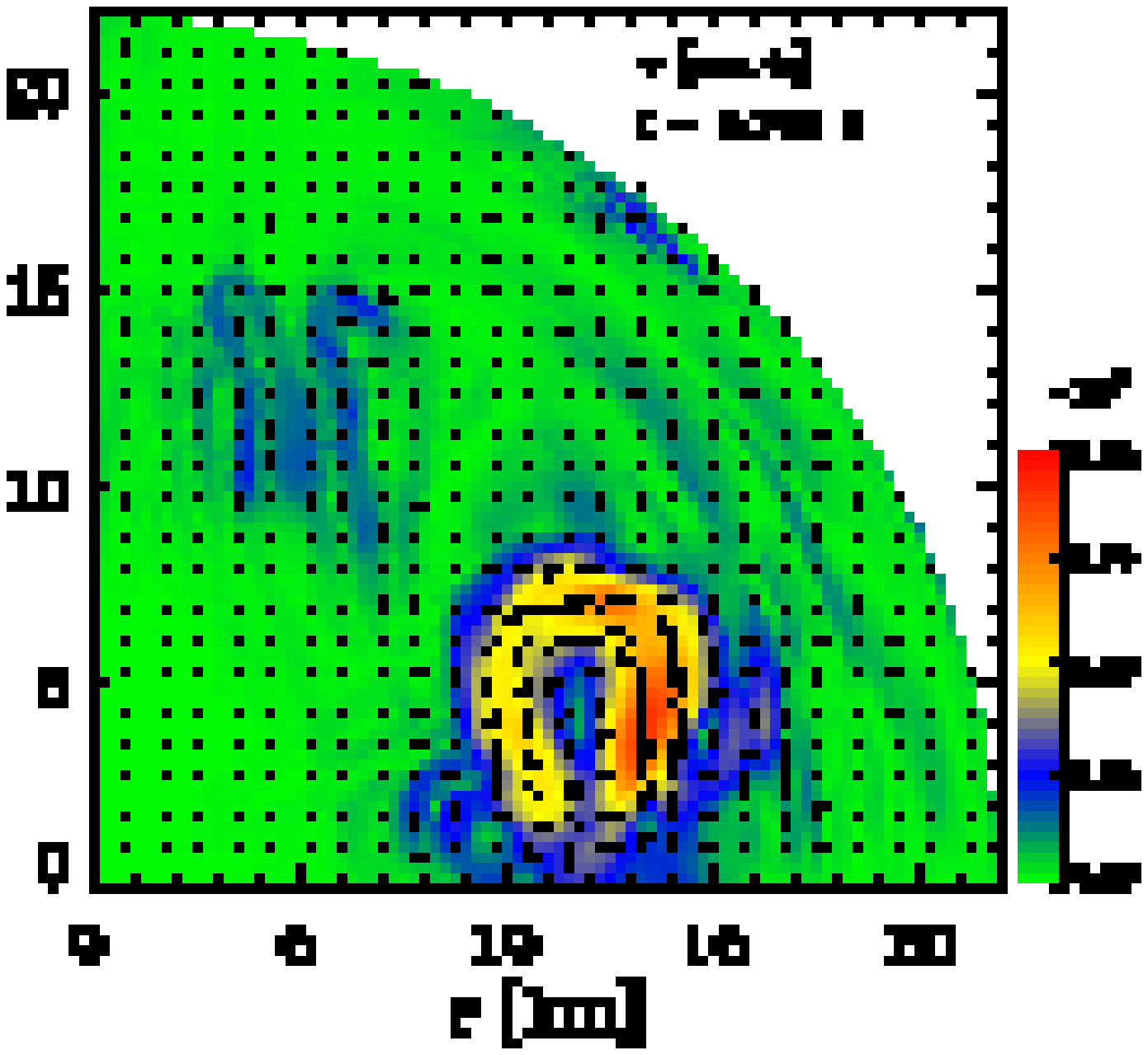}
\epsfxsize = 7.7 cm
\epsfbox{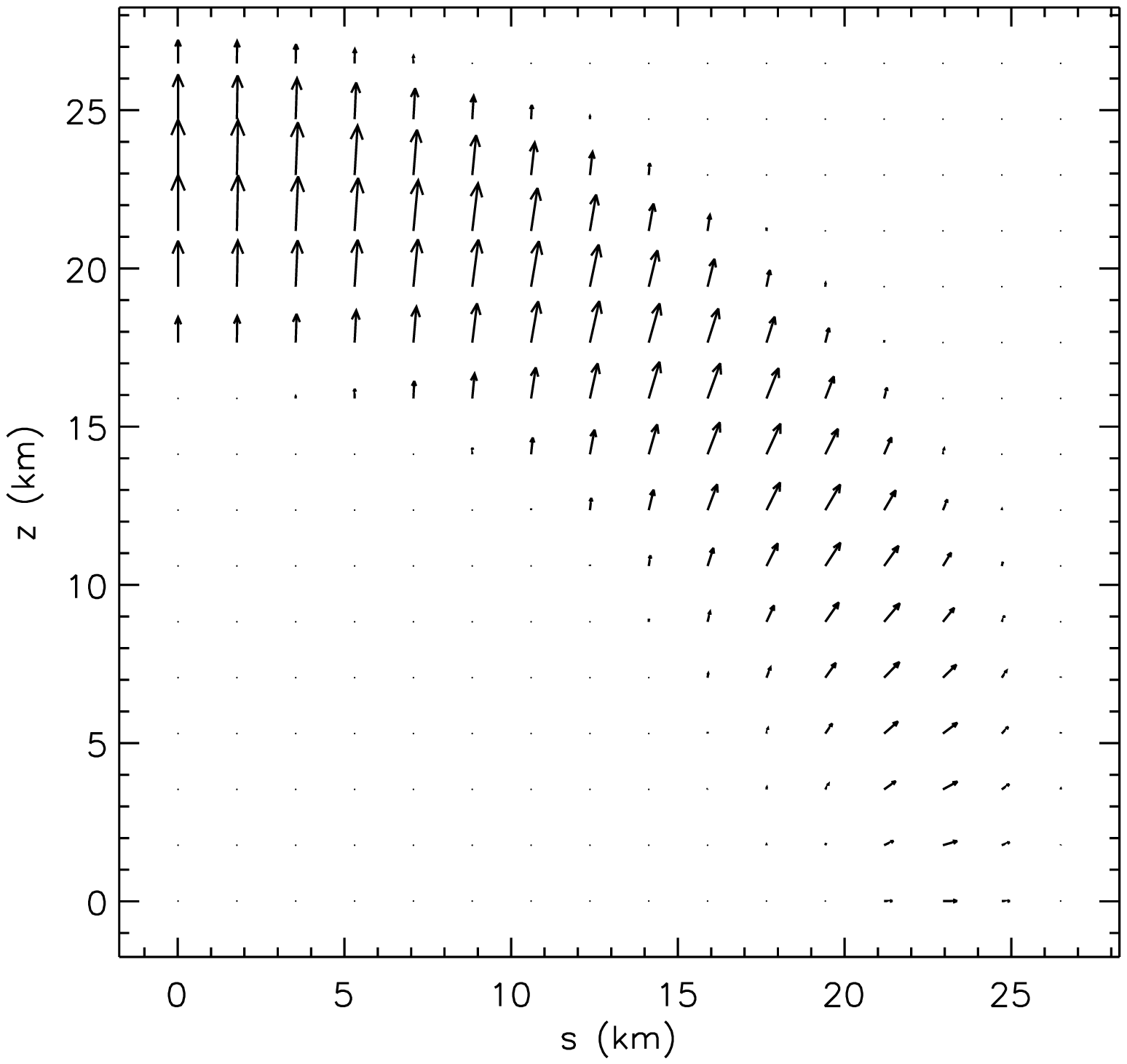}
\caption{Convection inside the rotating protoneutron stars.
In the left panel, the contour of the absolute value of the fluid
 velocity evolved about 750 msec after core bounce evolved by a
 numerical simulation \cite{jankeil} is shown. From the panel, the convection is shown to be 
suppressed near the rotational axis (vertical) and develop strongly near
 the equatorial plane. In the right panel, the direction, to which the
 convective motions are likely to occur at 500 msec after core bounce, is
 shown, which is obtained by a stability analysis \cite{miralles04}. The convection is more effective in the polar region. 
The left and right panel is
 taken from Janka \& Keil (1997) \cite{jankeil} and Miralles {\it et al}
 (2004) \cite{miralles04}, respectively. }
\label{rot_conv}
\end{center}
\end{figure}
\subsubsection{convection in rotating protoneutron stars}
As in the non-rotating cases, rotation should affect the convective
unstable regions in the protoneutron stars (PNSs). 
Janka and Keil (1997) \cite{jankeil} performed the two-dimensional
hydrodynamic simulation in the PNS treating the neutrino transport in
the grey flux-limited diffusion approximation and pointed out
that convections occurs only close to the equator (see the left panel
of Figure \ref{rot_conv}). On the other hand, 
 a recent study by Miralles {\it et al.} (2004) \cite{miralles04} suggests that the
convective unstable regions are formed along the direction of the
rotational axis preferentially (the right panel of Figure \ref{rot_conv}). 
Note that the computations by \cite{miralles04} 
are based on the linear stability and only valid in a steady-state 
situation, which is not always satisfied in the PNS. 
The results between the two studies, namely, the appearance
of the convective regions seems to change with the time evolution of the PNSs
 \cite{miralles04}. The effects of the heating anisotropy  
produced by the convections inside the rotating PNSs on the explosion 
mechanism are still an open question. 

As mentioned, the recent evolution models by 
\cite{heger04} suggest that the transport of 
angular momentum during the quasi-static evolutionary phase of the 
progenitor deprives the 
core of substantial fraction of its angular momentum, particularly when
the magnetic torque is taken into account \cite{spruit}. If this is really 
the case, the rotation will play no significant role in dynamics of 
core-collapse as shown by \cite{buras,muller03,walder}. 
%However, the evolution models are still 
%based on 1D calculations with some uncertainties in the treatment of 
%angular momentum transport and thus may not be the final answer yet.
\clearpage

\subsection{Roles of Magnetic Fields \label{magcurrent}}
Another possible cause for the asphericity of supernovae may be
magnetic fields. Soon after the discovery of pulsars, which are
the magnetized rotating neutron stars, the role
of rotation {\it plus} magnetic fields in the supernova explosion was
scrutinized \cite{ostriker,Bisno}. Because of the magnetic flux
conservations, a seed magnetic field in the stellar core can grow
significantly during core-collapse. In addition, winding of field lines
due to the differential rotation, which is natural after the collapse of
a spinning core, can further amplify the toroidal field component. If
the magnetic pressure becomes comparable to the thermal pressure,
magnetohydrodynamical forces can drive an explosion \cite{muller}, and 
accelerate axial jets \cite{meiner,sym,leblanc,ard}. Also magnetic buoyant
instabilities could produce mass motions along the rotational axis.
It should be mentioned that the necessary condition for the working of
the above mechanisms is that the core should be very strongly magnetized
initially. The magnetic field strength is required to be more than 
$\sim 10^{16}$G if the magnetic stress is to be comparable to the 
matter pressure in the supernova core after core bounce. Since the canonical value for the 
pulsar, $\sim 10^{12}$G, is negligibly small in terms of the above
effects on the dynamics of collapse, little attentions have been paid to
the magnetic supernovae. 

\begin{figure}
\begin{center}
\epsfxsize = 10.5 cm
\epsfbox{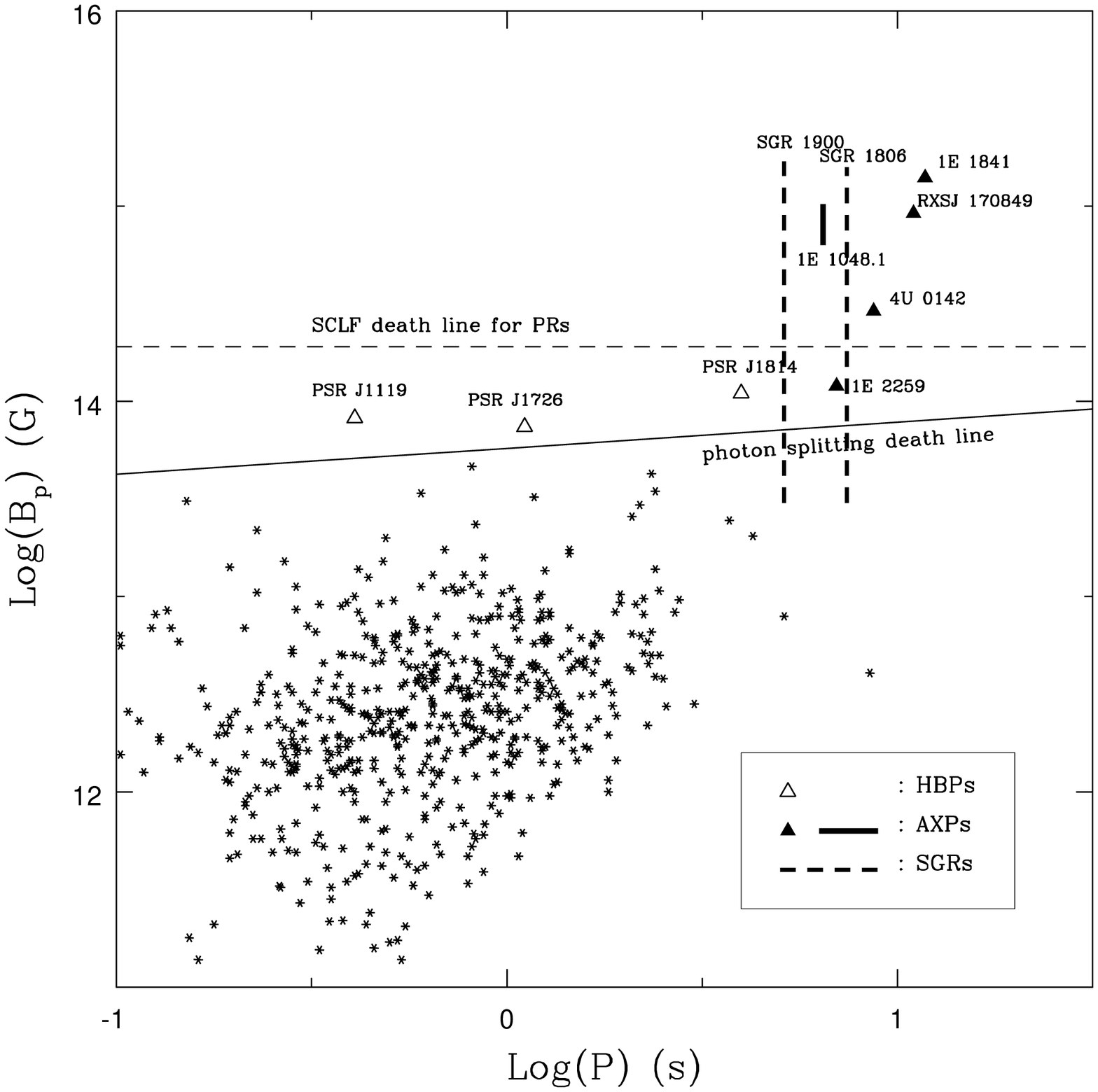}
\caption{$B_p - P$ diagram of radio pulsars and known magnetars taken
 from \cite{zhang}. The number of the magnetars discovered so far is about
10. See for details \cite{zhang}.}
\label{mag_fig}
\end{center}
\end{figure}

However, it has been
recently recognized these days that some neutron stars are indeed
strongly magnetized as high as $B \ge 10^{14}$ G.
 The strong dipolar magnetic fields at the surface can be estimated
 as follows,  
\begin{equation}
 B_{\rm dipole} \sim 3.0 \times 10^{14}\sqrt{\frac{P\dot{P}}{10^{-10}~{\rm s}}} ~[{\rm G}],
\end{equation}
 which is distinct from the normal radio pulsars by their long periods
 ($P$) and high-period derivatives ($\dot{P}$).
Some of them are soft gamma repeaters (SGRs) and others are anomalous 
X-ray pulsars (AXPs) or high magnetic field radio pulsars (HBPs) \cite{zhang,gusei}, which are collectively referred to as 
``magnetars'' \cite{duncan} (see Figure \ref{mag_fig} and \cite{woods}
 for review of the current observations). Although they are supposed to be a 
minor subgroup of neutron stars and the field strength might be lowered to
 the normal pulsars \cite{ibrahim}, these situations do revive the
study of the magnetic supernova again \cite{yama03,kotakemhd,takiwaki,sawai,kotakegwMHD}. 

\subsubsection{effect of magnetic fields on the prompt shock propagation}
As in the case of rotation \cite{yama94}, 
the first step to be investigated is the
effect of magnetic fields (in combination with rotation) on the dynamics 
of the prompt propagation of a shock wave \cite{yama03,takiwaki}.

The recent 2D magnetohydrodynamic simulations by Yamada \& Sawai \cite{yama03}
and Takiwaki {\it et al} \cite{takiwaki} showed that the combination of
rapid rotation, $T/|W| \ge 0.1 \%$ and strong poloidal magnetic field,
$E_{\rm m}/|W| \ge 0.1 \%$, leads in general to jet-like explosions. 
Here $T/|W|$ and $E_{\rm
m}/|W|$ represents the initial ratio of the rotational and magnetic
energies to the gravitational energy, respectively. $E_{\rm m}/|W| \ge
0.1 \%$ roughly means that the iron core has $B \geq 10^{12} ~{\rm G}$
before the onset of gravitational core-collapse. In the following, we
describe the magnetohydrodynamics features in such strongly 
magnetized and rapidly rotating models.

\begin{figure}
\begin{center}
\epsfxsize = 7 cm
\epsfbox{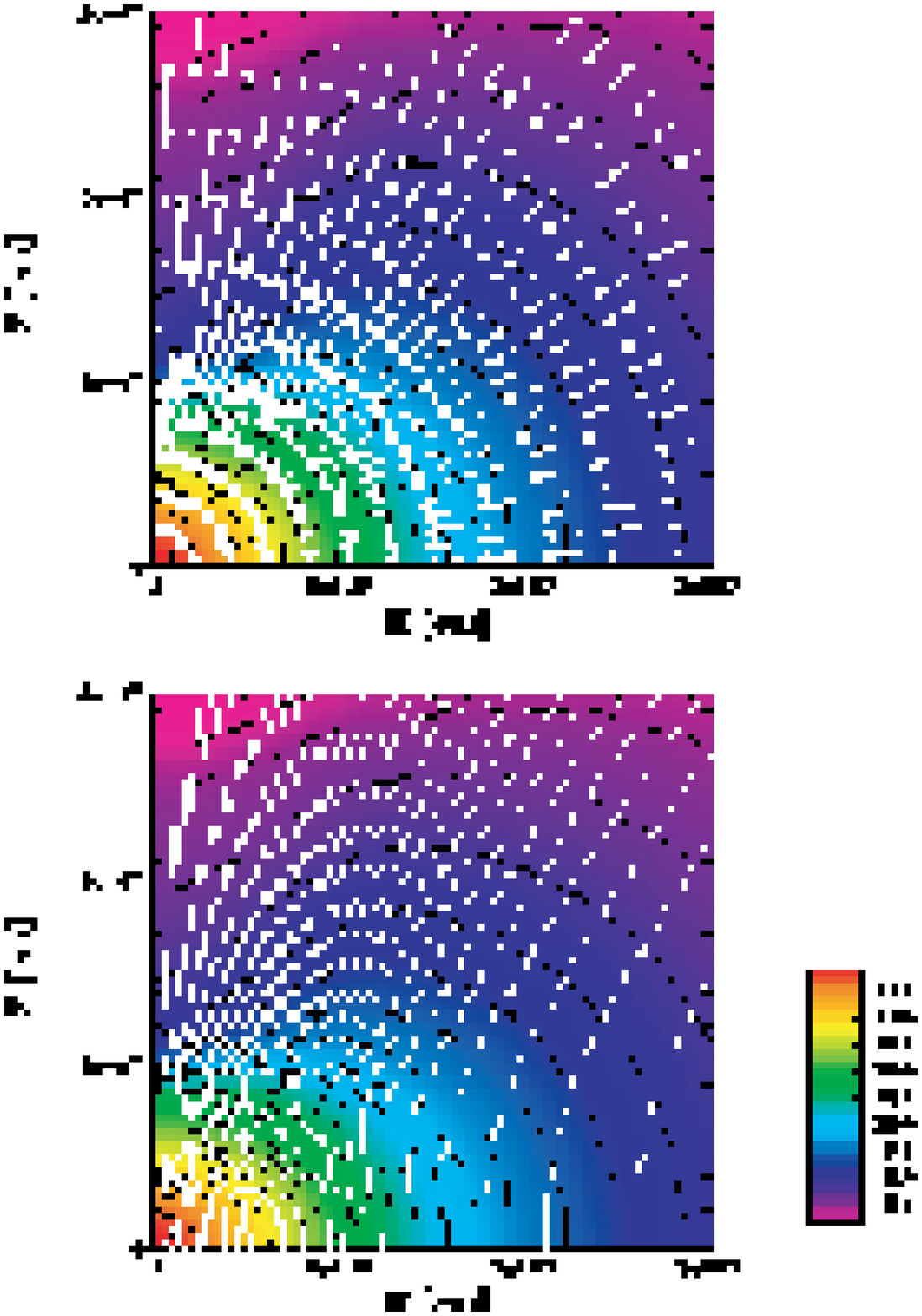}
\epsfxsize = 7 cm
\epsfbox{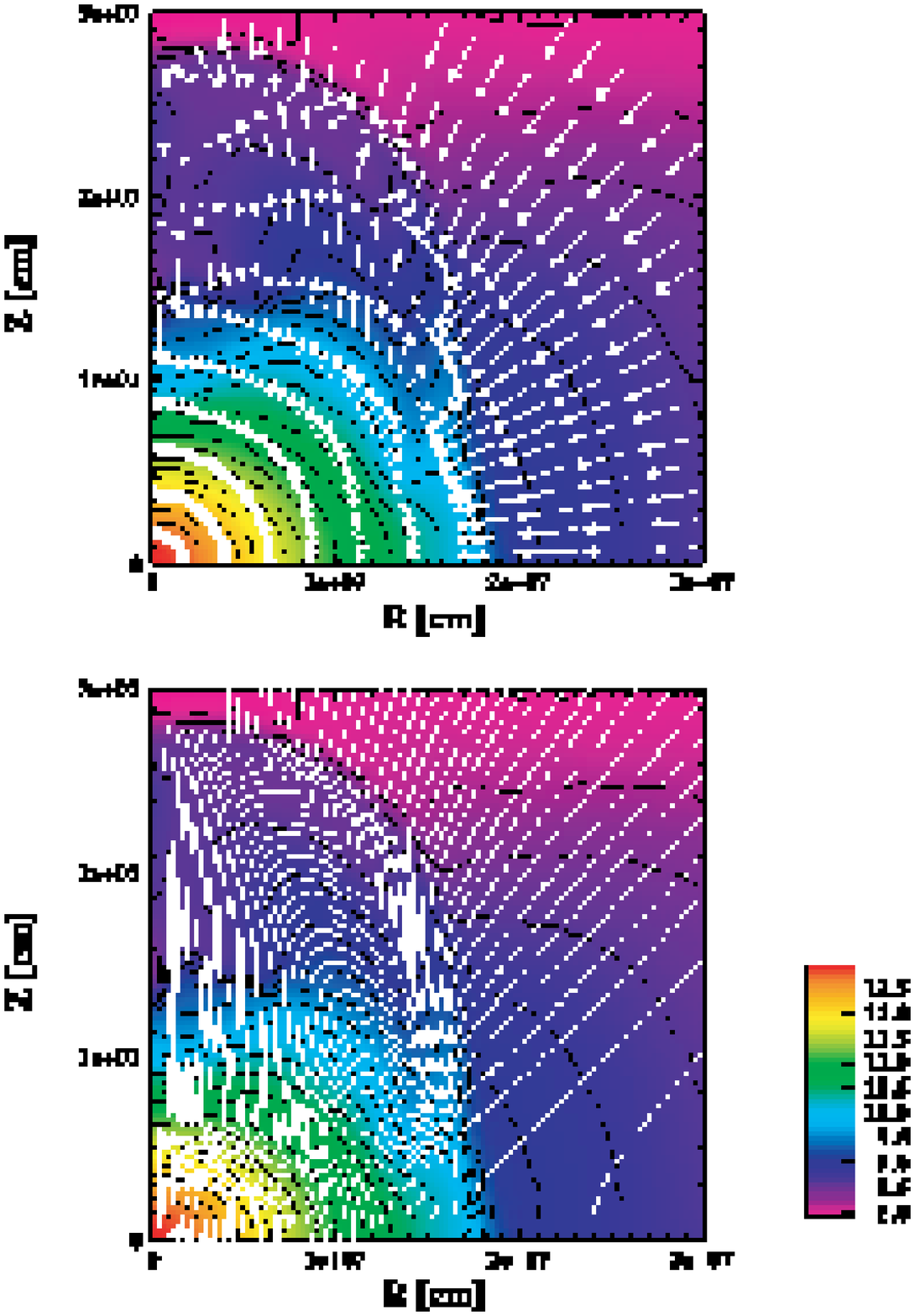}
\caption{Velocity fields (upper panels) and the poloidal magnetic fields
 (lower panels, white lines) on top of the density contours ($\log_{10} \rho$). 
Left and right is for core bounce and for 10 msec after core bounce.
These figures are taken from  Yamada and Sawai (2004) \cite{yama03}.}
\label{yama_mag}
\end{center}
\end{figure}

\begin{figure}
\begin{center}
\epsfxsize = 7.8 cm
\epsfbox{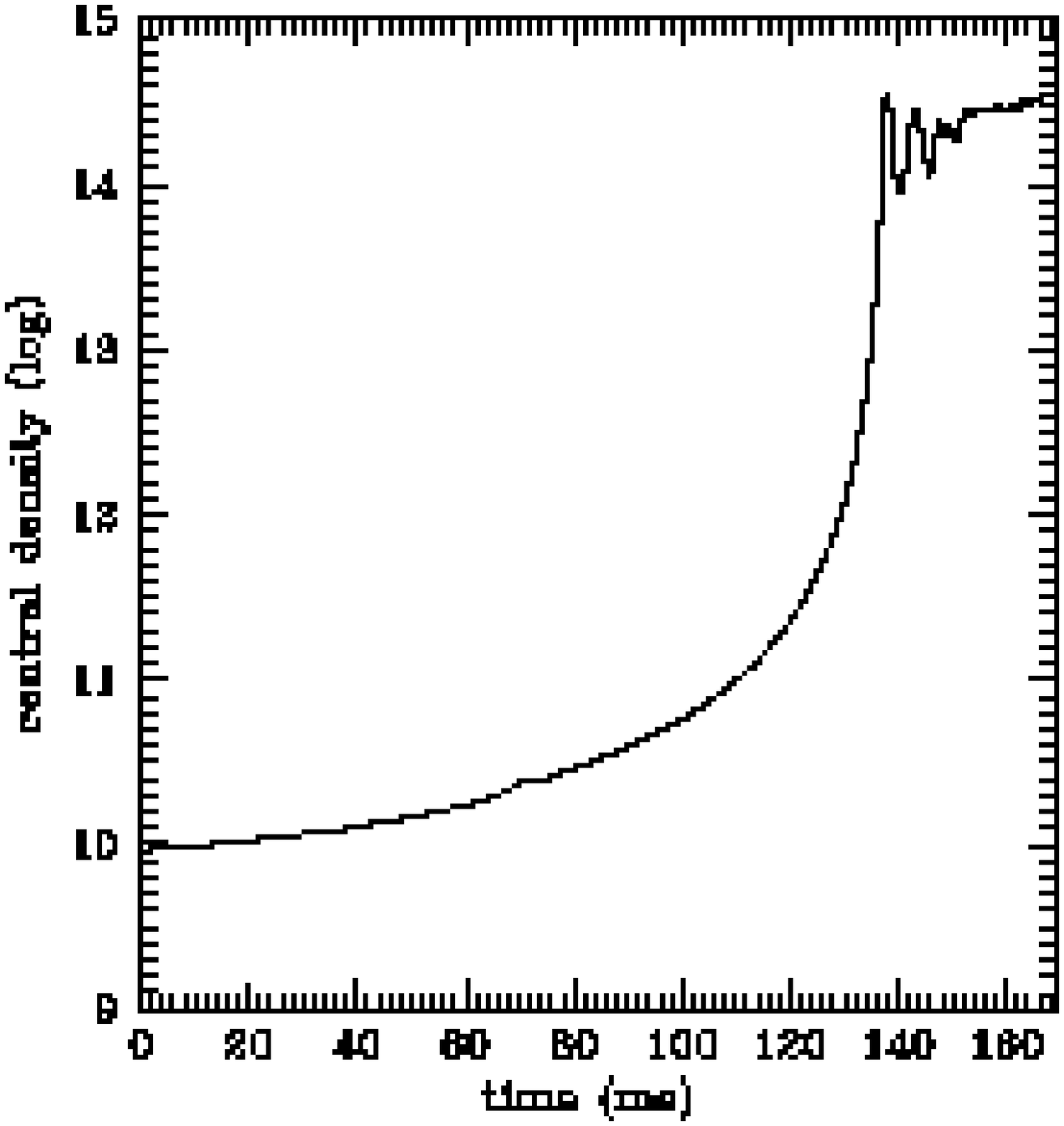}
\epsfxsize = 7. cm
\epsfbox{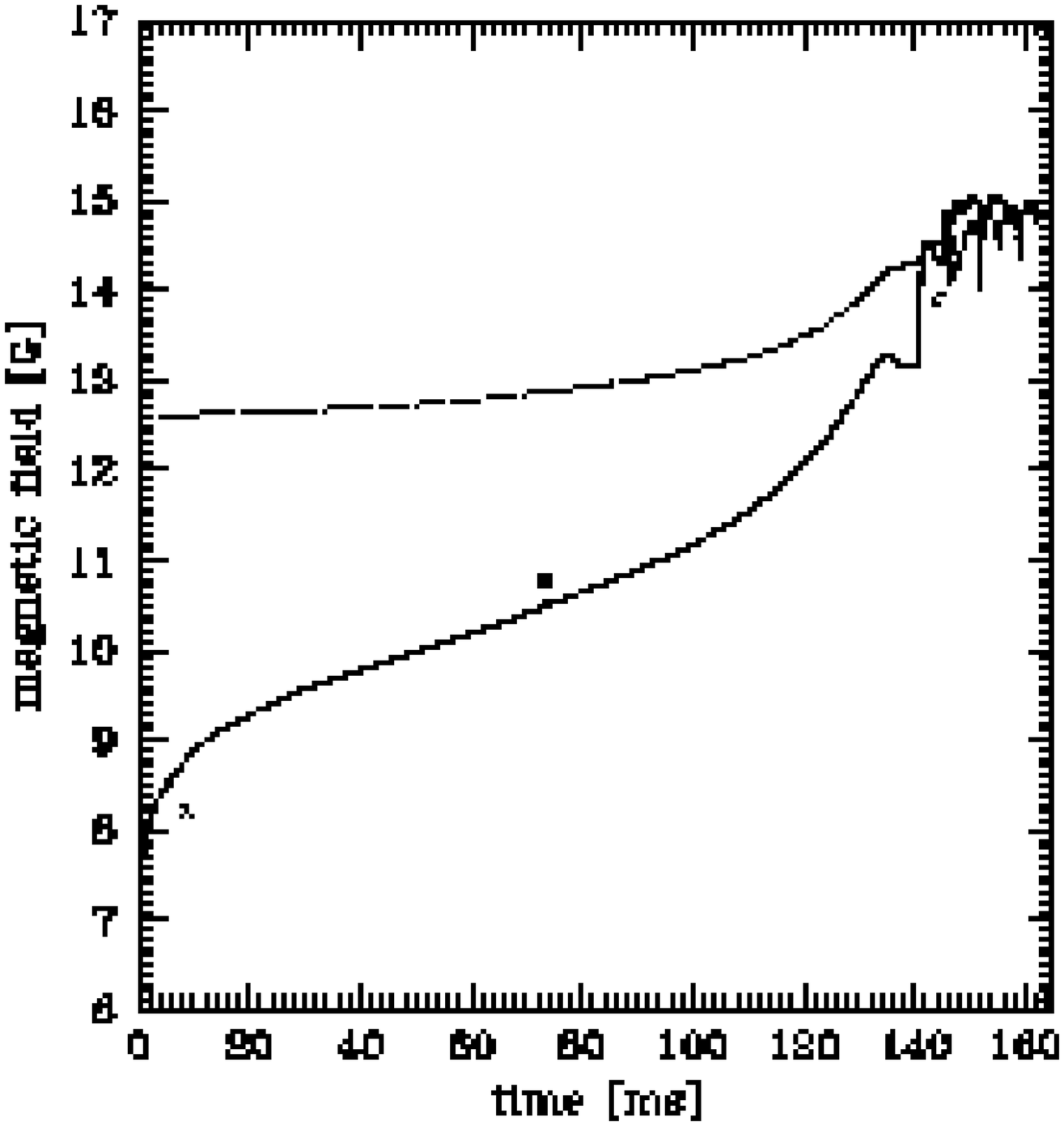}
\caption{Left panel is the time evolution of the central density and
 right is the time evolution of mean poloidal (dotted) and toroidal (solid) 
magnetic fields for a typical MHD model. Comparing the panels, one can
 see that ``C'' represents the time of core bounce. After core bounce,
 the toroidal component dominates over the poloidal one.
These figures are taken from  Sawai {\it et al.} (2005) \cite{sawai}.}
\label{sawai_fig}
\end{center}
\end{figure}

\begin{figure}
\begin{center}
\epsfxsize = 7 cm
\epsfbox{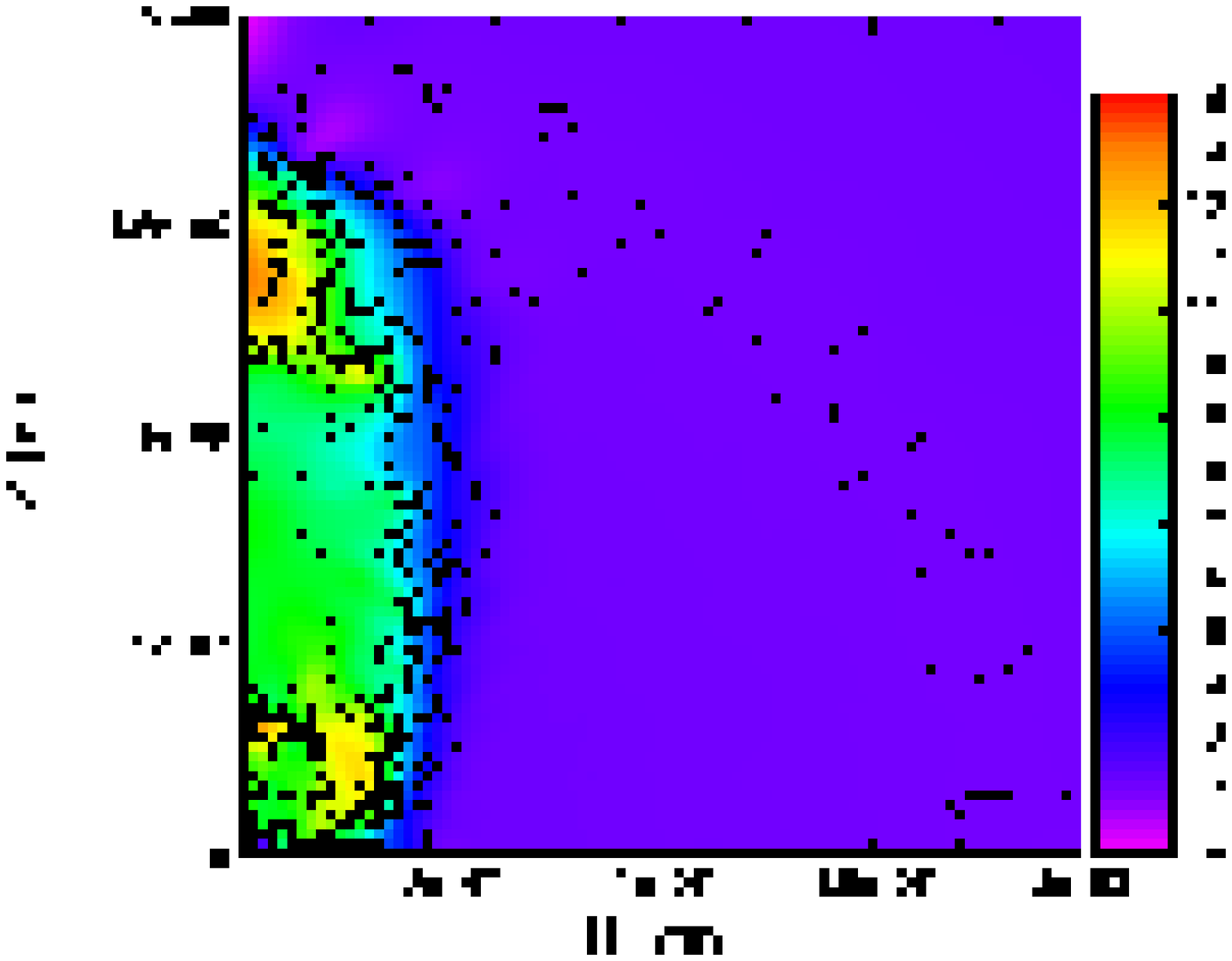}
\epsfxsize = 7 cm
\epsfbox{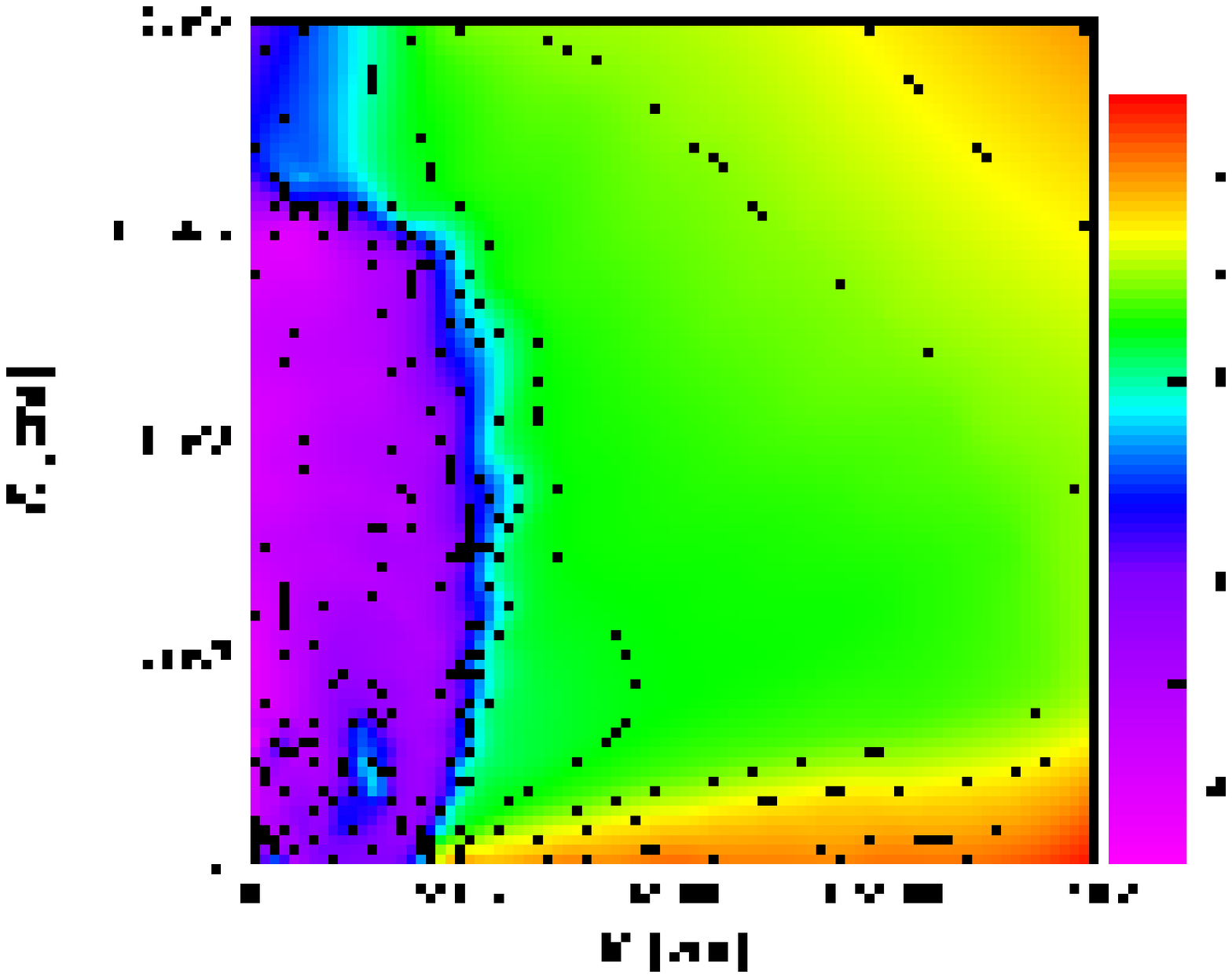}
\epsfxsize = 7 cm
\epsfbox{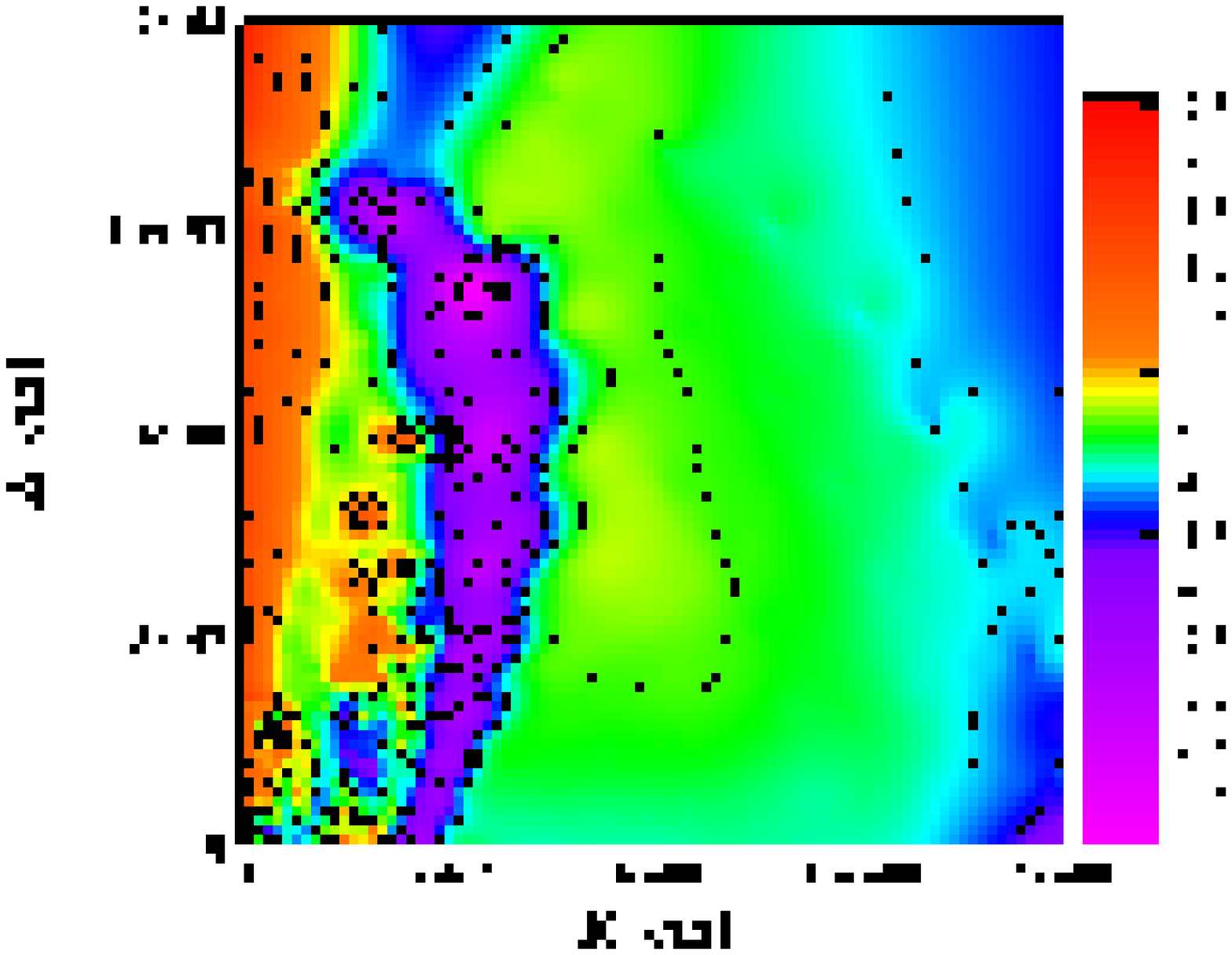}
\caption{A jet like explosion obtained in the rapidly rotating and strongly magnetized model studied in Takiwaki {\it et al} (2005) \cite{takiwaki}. Top left panel shows the contour of the entropy per nucleon near the shockbreak out of the iron core showing a jet like explosion. Top right panel shows the contour of the logarithm of the plasma beta, $\beta$, which is the ratio of the magnetic to matter pressure, representing that the shock wave is magneto-driven, since 
$\beta$ is greater than 1 just behind the shock wave. Bottom panel shows the contour of the ratio of the hoop stress to the gradient of the magnetic 
pressure, which demonstrates that the shock wave is collimated by the hoop 
stress near the rotational axis.}
\label{taki}
\end{center}
\end{figure}

The typical time evolutions are presented in Figure 
\ref{yama_mag}. The initial $T/|W|$ and $E_{\rm m}/|W|$ imposed
initially on the model is 1.5 $\%$ and 1.0 $\%$, respectively.
The initial field configuration is assumed to 
be parallel to the rotational axis with $B_{0} = 5.8 \times 10^{12}~{\rm
G}$. After core bounce, the fast magnetohydrodynamic shock is launched
in the direction of the rotational axis.
From Figure \ref{sawai_fig}, it can be seen that a seed magnetic field 
grow significantly during core-collapse due to the magnetic flux
conservation (see the right panel of Figure \ref{sawai_fig} from A to
C). Before core bounce (point C), the poloidal magnetic fields dominate
over the toroidal ones. After core bounce, winding of field lines
due to the differential rotation near the surface of the inner core can 
drastically amplify the toroidal components (see the right panel of
Figure \ref{sawai_fig} after point C).
In Figure \ref{taki}, the properties of the shock wave 
propagating further out from the central region are presented.
From the top left panel, the shock wave produced after core bounce becomes
magneto-driven, in the sense that the magnetic pressure becomes as
strong as the matter pressure behind the shock wave 
(see the top right panel of Figure \ref{taki}).  
Note that the magnetic pressure is much smaller than the matter pressure
in the unshocked inner core due to its high density, however, can be
much larger in the distant region from the center, because the matter
density drops much steeper than the magnetic fields which is almost
constant along the rotational axis. 
The magneto-driven shock wave is shown to be collimated along the
rotational axis because the 
the hoop stress $F_{\rm hoop} = \frac{B_{\phi}^2}{X}$, which collimates the shock wave, is dominant over the 
gradient of the magnetic pressure, $F_{\rm mag} =
\frac{1}{2}\frac{\partial B_{\phi}^2}{\partial X}$, acting to expand the
shock wave (see the bottom panel of Figure \ref{taki}). Here $X$
represents the distance from the rotational axis. 
These are the reasons for producing the jet-like explosion.

Furthermore it was pointed out that the explosion energy increases with the
initial magnetic fields strength, on the other hand, monotonically 
decreases with the the initial rotation rate \cite{takiwaki}. 
This is because the collimated shock wave
requires relatively lower energy to expel the matter near the rotational
axis as the initial field strength becomes larger. 
As a result, the stronger initial magnetic field is favorable to 
the robust explosion. 
Interestingly, the models with the
smallest magnetic fields studied in \cite{yama03,takiwaki} still 
produced the jet-like explosion although it takes longer for winding up 
the fields to launch the jet. This might suggest that even much smaller 
magnetic fields could be amplified in the collapsed core and play an 
important role for explosion.

\subsubsection{possible mechanisms for producing the pulsar-kicks}
 During core-collapse of such strongly magnetized models, 
the strength of the magnetic fields substantially
exceeds the QED critical value, $B_{\rm QED} = 4.4 \times 10^{13}$ (G),
above which the neutrino reactions are affected by
the parity-violating corrections to the weak interaction rates 
\cite{horowitz,arras1,arras2,ando}. 
Kotake {\it et al.} (2005) \cite{kotakemhd} estimated the  
corrections based on the results in the 2D MHD simulations and discussed 
its role for producing the pulsar kicks, which will be summarized
 shortly below (see
 also \cite{lai1,lai2,kusenko} for a review).

%%The physical origin of the pulsar kicks has been long controversial
%%among the unresolved mysteries in context of core-collapse supernovae. 
%As mentioned, recent analyses of 
% individual pulsar motions and observations of remnant associations 
% between supernovae and pulsars indicate that the neutron stars receive 
% large kick velocities at birth. The existence of pulsar kicks are also
% supported by the evolutionary studies of neutron star and black hole 
% binaries \cite{fryerkalogera,wex,mirabel} and by the
% detections of geodetic and orbital plane precessions in
% some binary pulsars \cite{cordes,kapsi}.
% The correlation between the direction of pulsar motions and the spin axis of 
% their supernovae could give us a clue to understand the kick mechanism,
% however,  it is statistically uncertain whether the spin-kick alignment
% is a generic feature of all pulsars. 

 \begin{figure}
\begin{center}
\epsfxsize=14.5cm
\epsfbox{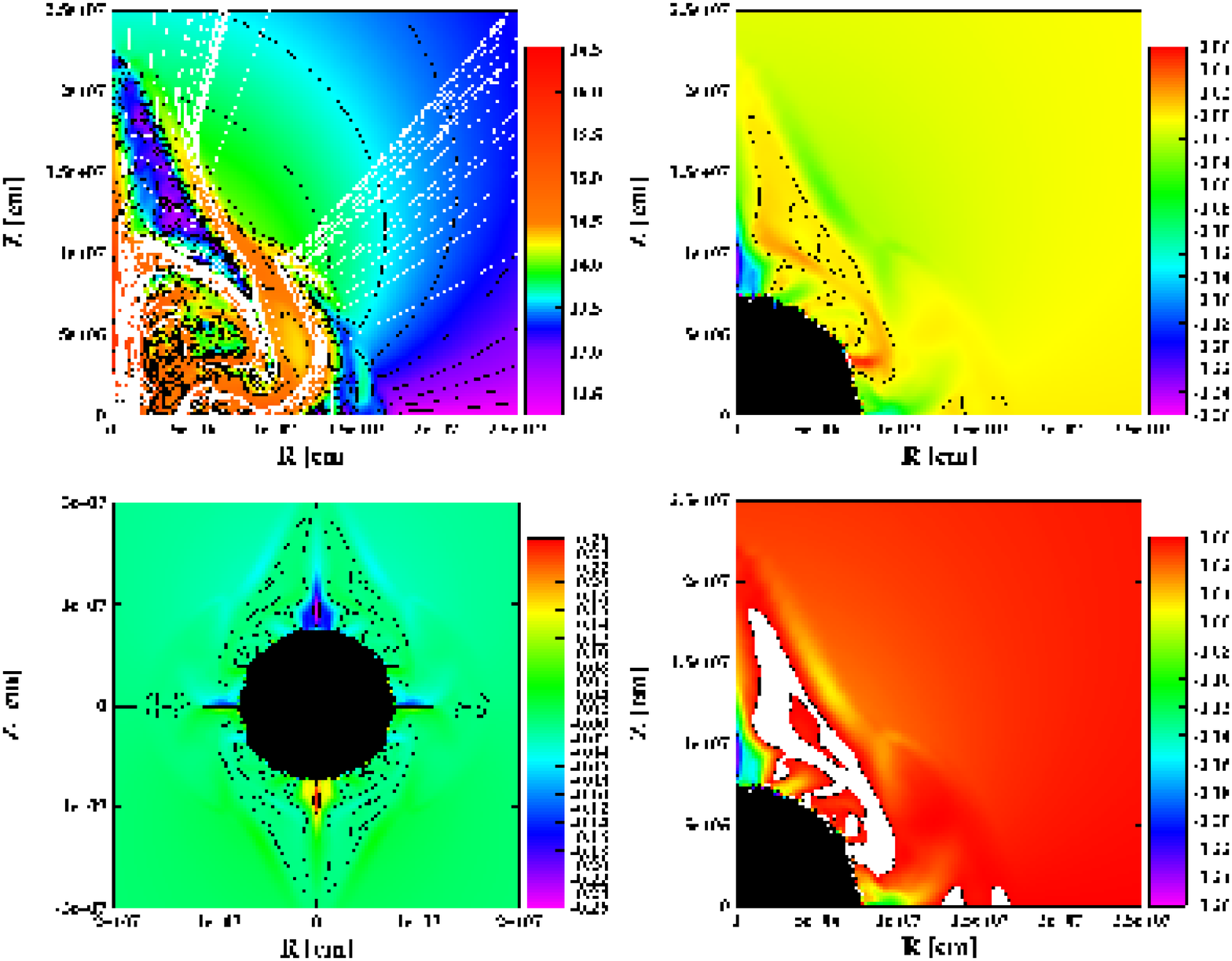}
\end{center}
\caption{Various quantities for a model with the strong magnetic fields and rapid rotation initially (see the context for the details of the model). 
Top left panel
 shows contour of the logarithm of the poloidal magnetic fields ($\log
 [B_{p}~{(\rm G})]$) with the magnetic field lines. Top right panel
 represents the ratio of the neutrino heating rate 
 corrected from the parity-violating effects, $\Delta Q_{\nu, \rm{B} \neq 0}^{+}$ to the heating rate without the corrections, $Q_{\nu, \rm{B} = 0}^{+}$. Note in the panel that the values of the color scale are expressed in percentage and that the central black region represents the region inside the neutrino sphere.
Bottom right panel is the same with the top right panel except
 that the bottom right panel only shows the regions with the negative
 values of the ratio. Thus the white region shows the regions with the
 positive values of the ratio. Bottom left panel shows the contour of
 the ratio in the whole region, which is prepared in order to see the
 global asymmetry of the heating induced by the strong magnetic fields. 
These figures are taken from Kotake {\it et al} (2005) \cite{kotakemhd}.}
\label{example_hikaku}
\end{figure}

%In addition, 
%most of the preceding studies relies on 
%the asymmetry of the neutrino emissions only by the magnetic fields 
%(see, however, \cite{janka99,duan04} for the simple models of atmosphere 
%in the supernova core).
%Kotake et al. (2005) \cite{kotakemhd} estimated the 
%corrections for discussing its role for producing the pulsar kicks.

%It is noted that rotation should also 
%contribute to produce the asymmetry of 
%the neutrino radiation \cite{kotakeaniso} as mentioned 
%in subsection \ref{roleofrotation}.
%  By performing a series of two-dimensional 
%simulations of magnetorotational core collapse of 15 $M_{\odot}$ 
%progenitor model \cite{ww:95}, in which we pay particular 
%attention to the models, which are assumed to have 
%the strong poloidal magnetic fields ($\sim 10^{12} $ G) 
%prior to core-collapse, we estimate   
%the parity-violating effects.

In the top left panel of Figure \ref{example_hikaku},
 the configuration of the poloidal magnetic fields after the shock stall
 ($\sim 50 $ msec after core bounce) is presented. For the initial
 condition for the model, the strong poloidal magnetic fields of
 $2\times 10^{12}$ G was imposed with the high angular velocity of 9 $\rm{rad}/ \rm{s}$  with a quadratic cutoff at 100 km radius in the iron core.  
It is seen from the panel that the poloidal magnetic fields are rather 
straight and parallel to the rotational axis in the regions near the
rotational axis, on the other hand, bent in a complex manner, in the other
central regions due to the convective motions after core bounce.
The top right panel of Figure \ref{example_hikaku} shows 
 the ratio of the neutrino heating rate 
 corrected from the parity-violating effects, $\Delta Q_{\nu, \rm{B} \neq 0}^{+}$ to the heating rate without the corrections, $Q_{\nu, \rm{B} = 0}^{+}$.  
%From the panel, one
% can see the shape of the neutrino sphere is rather spherical and the 
%resultant neutrino radiation becomes almost isotropic.
It is noted that the suppression or the enhancement of the heating rate 
through parity-violating effects is determined by the signs of the inner
product of $ {\mbox{\boldmath$\hat{n}$}} \cdot
 \mbox{\boldmath$\hat{\rm{B}}$}$, where  ${\mbox{\boldmath$\hat{n}$}}$,
$\mbox{\boldmath$\hat{\rm{B}}$}$ are the unit vector in the direction 
of the incoming neutrinos and along the magnetic field. If the product
 is positive (negative), then the corrections to the heating rates
 results in the suppression (enhancement) of the heating rate.
Reflecting the configurations of the magnetic fields,   
it is seen that the values of the ratio become
 negative in almost all the regions 
(see also the bottom right panel of Figure
 \ref{example_hikaku} for clarity). This means that the heating rate is 
reduced by the magnetic fields than that without. 
Especially, this tendency is most remarkable in the regions
 near the rotational axis and the surface of the neutrino sphere (see
 the regions colored by blue in the top right panel of Figure 
\ref{example_hikaku}). This is because the
magnetic fields are almost aligned and parallel to the rotational axis
 in the regions near the rotational axis.  In the bottom left panel of Figure 
\ref{example_hikaku}, the contour of the ratio in the 360 latitudinal
 degrees region of a star, is prepared in order to see the global asymmetry 
of the heating.  Since their simulations assumed the equatorial symmetry, 
the above features in the northern part of the star become reverse for 
the southern part. As a result, it was found that the heating rate is 
reduced about $\sim 0.5 \%$ in the vicinity of the north pole, on the other hand, enhanced about
$\sim 0.5 \%$ in the vicinity of the south pole in the computed model in
\cite{kotakemhd}. 
If the north/south asymmetry of the neutrino heating persists throughout 
the later phases, it is expected that the
pulsar is likely to be kicked toward the north pole.

\begin{figure}
\begin{center}
\epsfxsize=10cm
\epsfbox{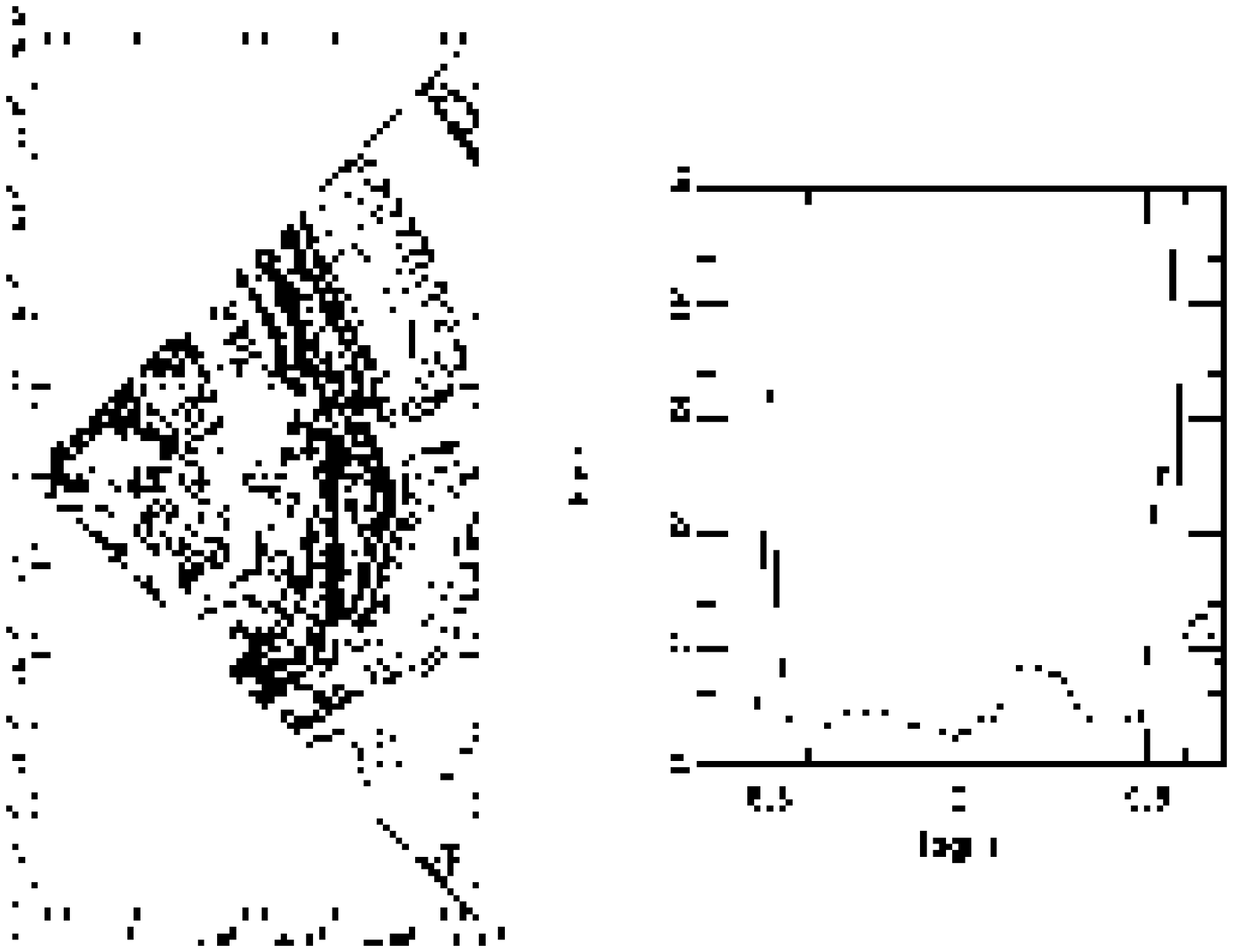}
\caption{Density inhomogenties (left) produced 
during oxygen shell burning and their root-mean-square azimuthal
 averages (right). These figures are taken from Bazan and Arnett (1998)
 \cite{bazan}.}
\label{bazan_fig}
\epsfxsize=10cm
\epsfbox{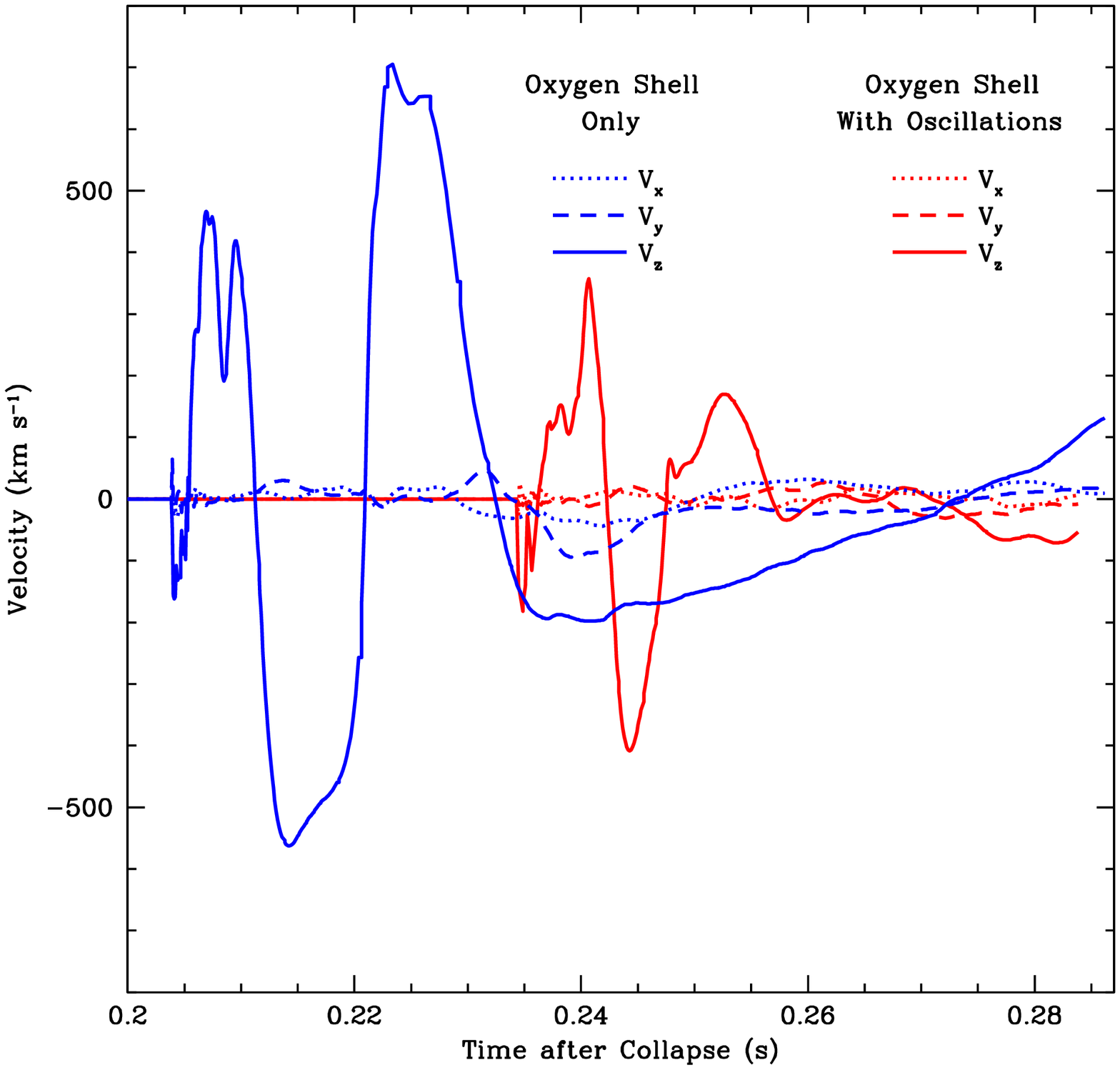}
\end{center}
\caption{Kick velocities ($x,y,z$ directions) of the neutron star 
due to the density inhomogenities prior to collapse. 
The inhomogenities were parametrically imposed in two ways, namely on 
the oxygen shell only and the core oscillation, with the density variations of $\leq 25 \%$.  After the oscillatory behaviors, which were pointed out to be due to the neutrino emission from material accreting onto the neutron star, the motions of the neutron star tend to damp in time with the kick velocity less than $200~{\rm km}~{\rm s}^{-1}$. This figure is taken from Fryer (2004) \cite{fryerkick}.}
\label{fryer_kick}
\end{figure}

\begin{figure}
\begin{center}
\epsfxsize=7cm
\epsfbox{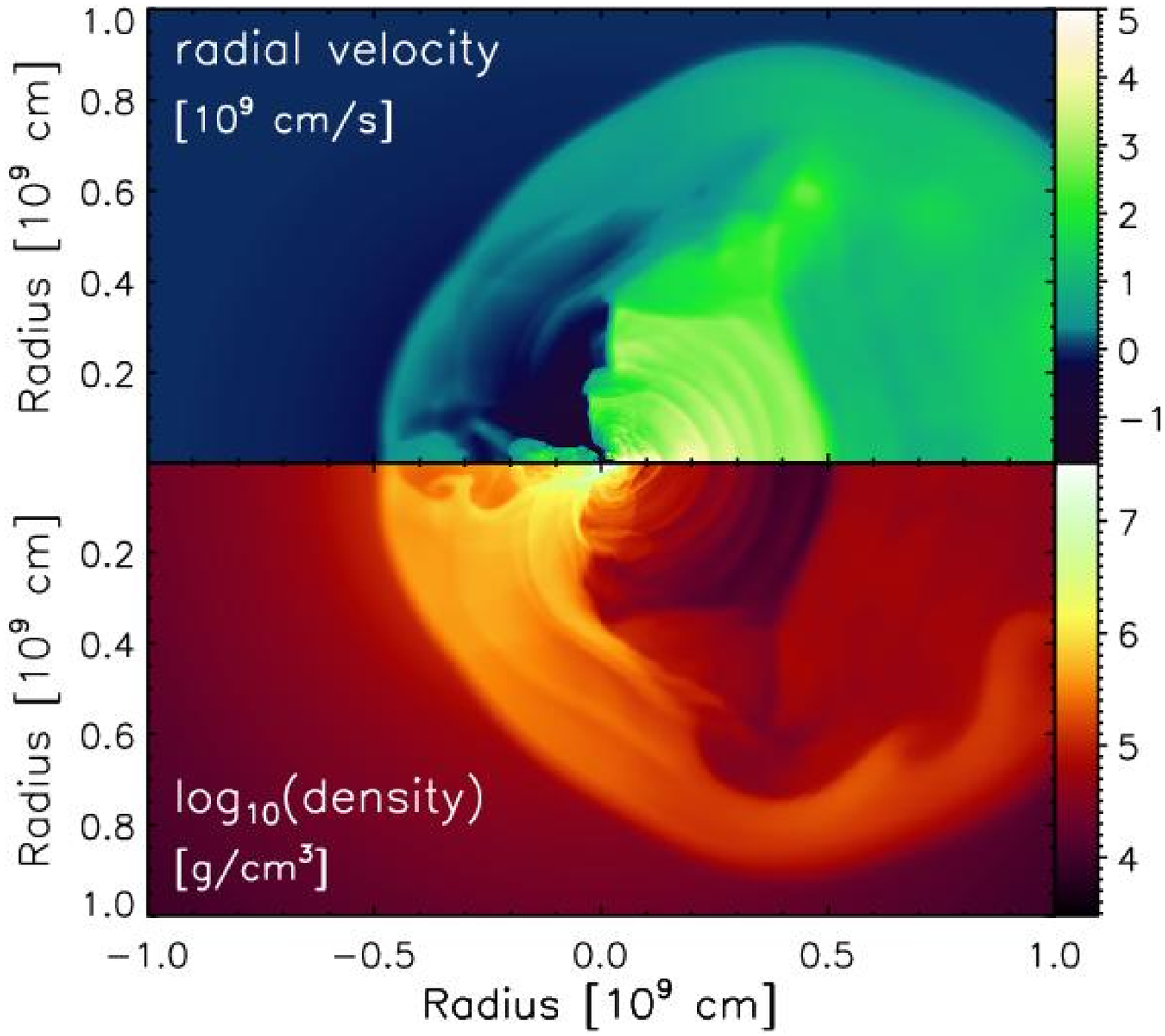}
\epsfxsize=8cm
\epsfbox{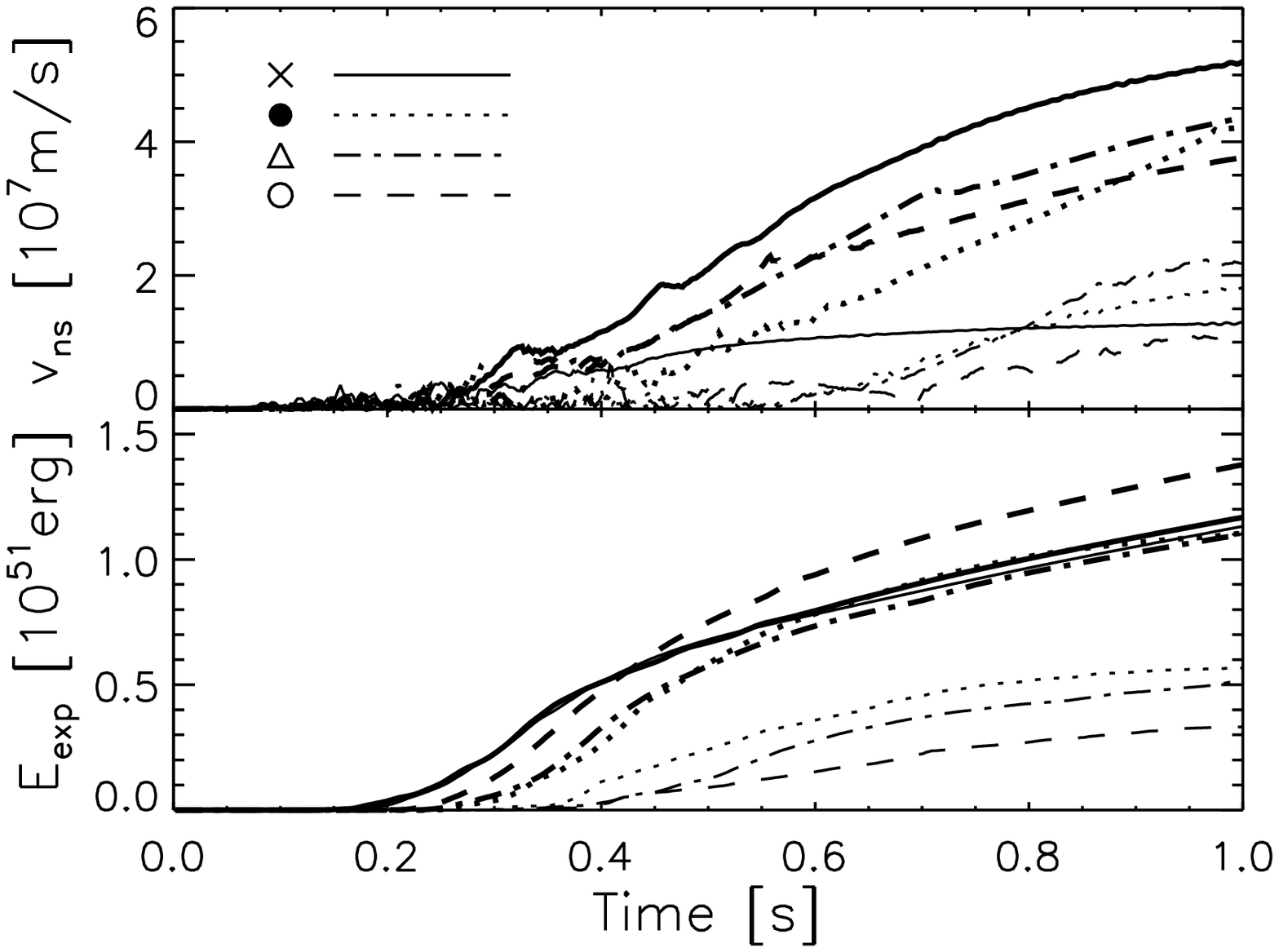}
\end{center}
\caption{The global asymmetry of $l =1$ mode and the kick velocities of the protoneutron star. Left panel shows the growth of $l=1 $
 at 1 s after core bounce. In the right panel, the velocities of the
 neutron star (top) and the explosion energies (bottom) as a function of
 time measured from core bounce are given for some representative
 models (see \cite{scheck} for details). At 1 s, it can be seen that the
 kick velocity becomes as high as $\sim 600$ km/s in one of the models
 wit the explosion energy of $\sim 1 \times 10^{51}$ erg. 
These figures are taken from Scheck {\it et al} (2004) \cite{scheck}.}
\label{scheckkick}
\end{figure}

There is an another class of the mechanism for producing the kicks,
which relies on the cause of the 
asymmetric explosion as a result of the global convective instabilities
\cite{burohey,goldreich,fryerkick} caused by the pre-collapse 
density inhomogeneities \cite{bazan} (see Figures \ref{bazan_fig}, \ref{fryer_kick})
or the local convective instabilities \cite{jankamueller94} 
formed after the onset of core-collapse. 
More recently, Scheck {\it et al} \cite{scheck}.
 recently pointed out that the random velocity
perturbation of $\sim 0.1$ \% added artificially at several
milliseconds after bounce could lead to the global asymmetry of
supernova explosion and cause the pulsar-kick (see Figure
\ref{scheckkick}).
While the orientation of pulsar-kicks  
is stochastic in their models, which is truly consistent with the
observation of the normal pulsars,
the asymmetry of the neutrino heating in the strong magnetic fields
mentioned above might predict the alignment of the magnetic axis and
the kick velocity in highly magnetized neutron stars.

%So far two major classes of mechanism for these kicks have been
%suggested (see \cite{lai2,lai1} for a review).
%More recently, Scheck {\it et al} reported that in the long
%duration explosions (more than a second after bounce), the
%neutrino-driven convections behind the expanding shock can lead the 
%global asymmetries, which accelerates the remnant neutron star to a 
%several hundreds of km/s \cite{scheck} (see Figure \ref{scheckkick}). 
% Further numerical investigations are required to see if this is really 
%the case. It is also interesting to study the
% effect of the anisotropic neutrino radiation in the strong magnetic
% field on the growth of the convective instability and/or MRI 
%in the later phases (Kotake {\it et al} in preparation 2005).

\subsubsection{effect of toroidal magnetic fields}
 While most of the magnetohydrodynamic (MHD) simulations in the context
 of core-collapse supernovae choose poloidal magnetic fields as 
initial conditions \cite{leblanc,Bisno,meiner,muller,sym,ard,yama03}, recent
 stellar evolution  calculations show that toroidal magnetic fields may be much stronger than poloidal ones
prior to collapse \cite{heger03,heger04,spruit}. Motivated by this situation,
Kotake {\it et al} (2004) investigated the models with predominantly
 toroidal magnetic fields,  changing the strength of rotation and the 
toroidal fields systematically. 

\begin{figure}
\begin{center}
\epsfxsize=15cm
\epsfbox{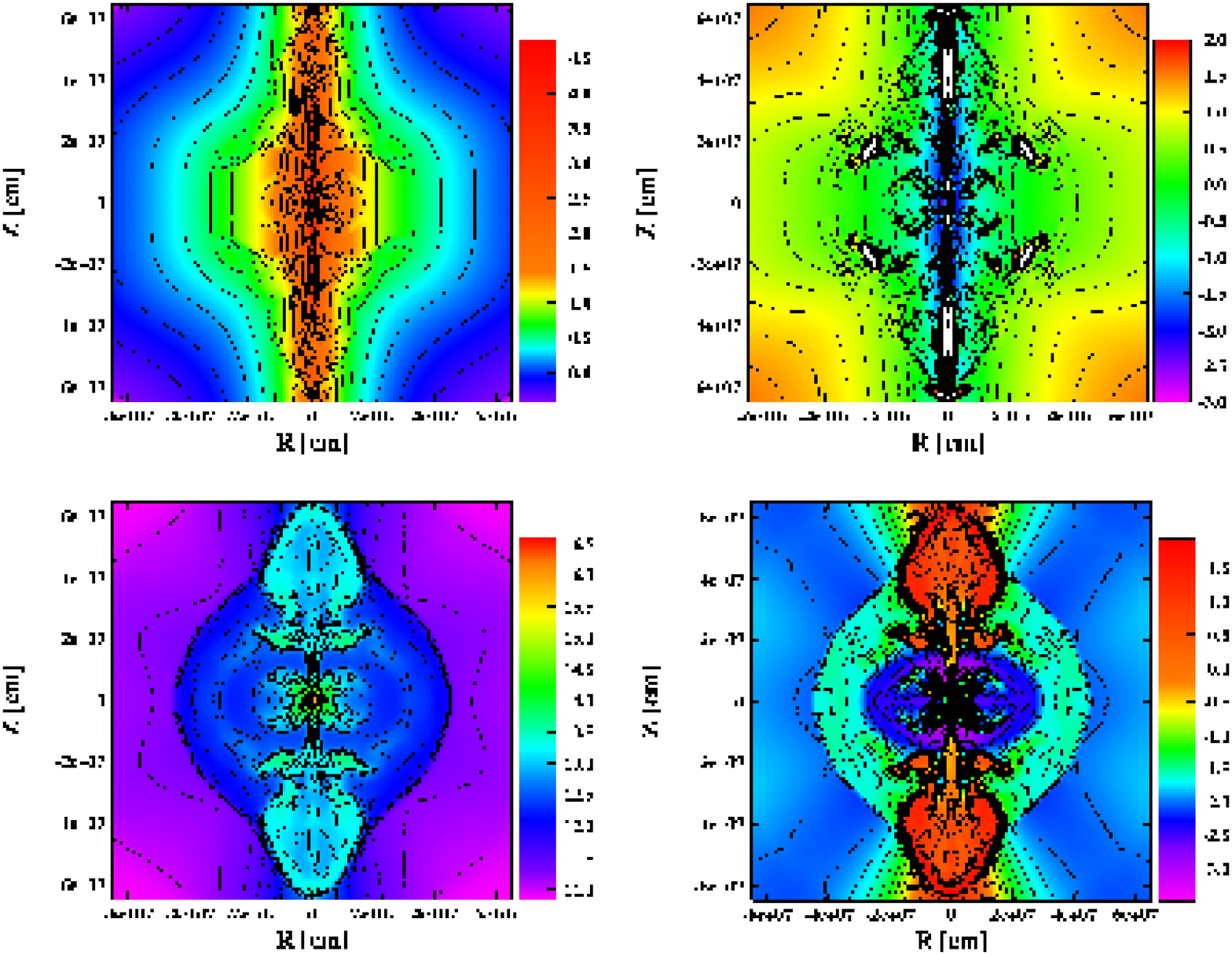}
\end{center}
\caption{Contour plots of various quantities for the model with the
 strongest magnetic field in the computations \cite{kotaketor}. The top left panel 
shows the logarithm of the
 angular velocity ($\rm{s}^{-1}$). The top right panel represents the
 magnetorotationally unstable regions to non-axisymmetric
 perturbations. The plot shows the growth time scale (s) of
 MRI. Note in this panel that the regions with white represent 
 stable regions against MRI. The bottom left panel shows 
 the logarithm of magnetic field strength (G). The bottom right
 panel displays the logarithm of the ratio of magnetic stress 
 to matter pressure in percentage.

\label{fig4}}
\end{figure}

The angular velocity profile for the model with the strongest toroidal 
magnetic fields in their simulations is given in the top left panel of Figure \ref{fig4}.
The initial values of $T/|W|$ and $E_{\rm m}/|W|$ are $0.5, 0.1 \%$, 
respectively, where $E_{\rm m}/|W|$ represents magnetic to
gravitational energy. In addition, the initial profiles of rotation and 
magnetic field are chosen to be cylindrical with strong differential
rotation for this model. 

In the top left panel, the negative gradient of angular
velocity, $d \Omega/ d X < 0$, can be found, where $\Omega$ is the angular
velocity and $X$ is the distance from the rotational axis. Such a region
is known to be unstable to non-axisymmetric perturbations
\cite{balbus1,akiyama}, as will be discussed in subsections of  
\ref{fond_MRI} and \ref{sub_MRI}. The characteristic time scale for the growth of the 
instability called the magnetorotational instability (MRI) is given 
as $\tau_{\rm MRI} = 4 \pi |d \Omega /d \log X|^{-1}$. 
The top right panel of Figure \ref{fig4} shows the contour of $\tau_{\rm
MRI}$ for the model. The typical time scale is found to be $\sim
O(10) $ ms near the rotational axis. This suggests that 
MRI induced by non-axisymmetric perturbations can grow on the
prompt shock time scale. 
The field strengths in the
protoneutron star become as high as $\sim 10^{16}$ G (see the bottom
left panel of Figure \ref{fig4}), and the ratio of 
magnetic stress to matter pressure gets as high as $0.9$ behind the
shock wave (see the bottom right panel of Figure \ref{fig4}). 

As for the anisotropic neutrino radiation observed in purely rotating
case (subsection \ref{anisoneu}), 
the feature is shown to be not changed significantly by the inclusion 
of very strong
toroidal magnetic fields ($\sim 10^{16}$ G in the protoneutron star).
Combined with the anisotropic neutrino radiation that heats matter
near the rotational axis more efficiently the growth of the instability 
is expected to further enhance heating near the axis. Furthermore, the
magnetic pressure behind the collimated shock wave is as strong as the
matter pressure in the vicinity of the rotational axis. 
From these results, one might speculate 
that the magnetar formation is accompanied by a jet-like explosion if 
it is formed in the magnetorotational collapse described above.

\clearpage
%%%%%%%%%%%%%%%%%%%%%%%%%%%%%%%%%%%%%%%%%%%%%%%%%%%%%%%%%%%%%%%%%%%%%%
\subsubsection{foundations of magnetorotational instability \label{fond_MRI}}
\begin{figure}
\begin{center}
\epsfxsize=14.5cm
\epsfbox{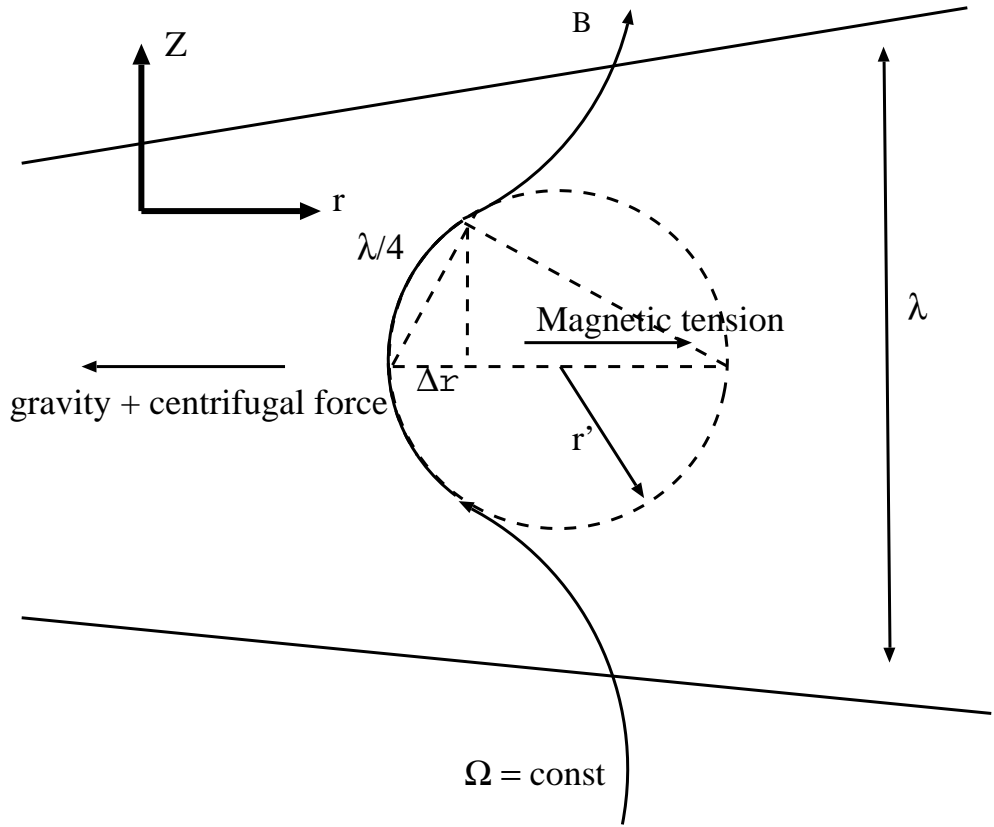}
\end{center}
\caption{Schematic figure prepared for the derivation of the growth 
condition of
 MRI. When the radial displacement of $\Delta r$ of the fluid element is
 imposed, the MRI develops if the sum of the gravity and the centrifugal force
 becomes larger than the magnetic tension (see text for details).}
\label{mri_schem}
\end{figure}     
So far in this section, we have concentrated on the models which have 
large magnetic fields prior to core-collapse.
The most mysterious is the origin of this large magnetic field.
The core might have strong magnetic fields prior to collapse already, 
although the evolution models indicate
quite the contrary \cite{heger04}. 
We may not need the strong
magnetic field initially if the magnetorotational instability (MRI)
sufficiently develops in the supernova core. 
This MRI was first discussed 
in the context of accretion disks by \cite{balbus1}.
% is introduced to the dynamics of supernova core by \cite{akiyama}. 
Before we mention the role of MRI in the supernova, we summarize, for convenience, the basic properties of MRI 
in the context of accretion disks around the black hole, which are 
common situations in the central part of the active galactic nuclei.

First of all, let us derive the growth condition for the MRI.
See a schematic Figure \ref{mri_schem}, describing the small
displacement of the magnetic fields in the accretion disk.
Before the displacement, the magnetic fields with strength $B$ 
are assumed to be uniform
and parallel to the rotational axis of the disk ($z$ axis). 
The displacement is assumed to occur at the radial distance $r$ from the
central region of density $\rho$. Furthermore, let us assume that the
disk is in a Kepler rotation, namely, the angular velocity of the disk
yields,
\begin{equation}
\Omega = \sqrt{\frac{GM}{r^3}},
\end{equation} 
where $M$ is the mass of the central object, presumably the black hole.

As in the Figure \ref{mri_schem}, let us displace the fluid element $\Delta
r$ inwards. Since the ideal MHD approximation can be well satisfied in the
accretion disks, the magnetic fields are frozen-in to the matter. Thus
the field line also moves $\Delta r$ inwards as in Figure
\ref{mri_schem}. Since the angular velocity of the magnetic field is
constant, the centrifugal force becomes smaller due to the displacement. 
The change of the centrifugal force can be written,
\begin{equation}
\delta F_{\rm rot} = (r - \Delta r) \rho \Omega^2.
\end{equation}
On
the other hand, the gravity in the radial direction increases,
\begin{equation}
\delta F_{\rm grav} = - \frac{\rho G M}{r^2}\Bigl(1 + 2\frac{\Delta r}{r}).
\end{equation} 
As a result, the net force in combination of the centrifugal force and
the gravity becomes,
\begin{equation}
\delta F_{\rm tot} = \delta F_{\rm rot} + \delta F_{\rm grav} = - 3 \rho \Omega^2 \Delta r,
\end{equation}
which acts to move the displaced field line inwards.   
If this force is stronger than the magnetic tension, which acts to put
the displacement back, the instability develops. The magnetic tension
is $B^2/r^{'}$, here $r^{'}$ is the curvature radius of the magnetic
field (see Figure \ref{mri_schem}). From a simple geometric calculation, 
one can obtain the magnetic tension,
\begin{equation}
\delta F_{\rm mag} = \frac{B^2}{r^{'}} = \frac{2 B^2 \Delta r}{(\lambda/4)^2},
\end{equation}
where $\lambda$ is the wavelength of the perturbation. Thus the
condition for the growth of the instability becomes,
\begin{equation}
\delta F_{\rm mag} + \delta F_{\rm top} = 
\Bigl(\frac{2 B^2 }{(\lambda/4)^2} - 3 \rho \Omega^2 \Bigr)\Delta r< 0.
\end{equation} 
 As a result, the wavelength of the perturbation satisfying the condition
 $\lambda > \lambda_c = 4 \sqrt{2/3}v_{\rm A}/\Omega$ becomes unstable,
 where $v_{A}$ is the Alfv\'{e}n velocity defined as $v_{\rm A} =
\frac{B}{\sqrt{4 \pi \rho}}$. This is the condition for the onset of the
 {\it MagnetoRotational Instability} (MRI).

\begin{figure}
\begin{center}
\epsfxsize=12.0cm
\epsfbox{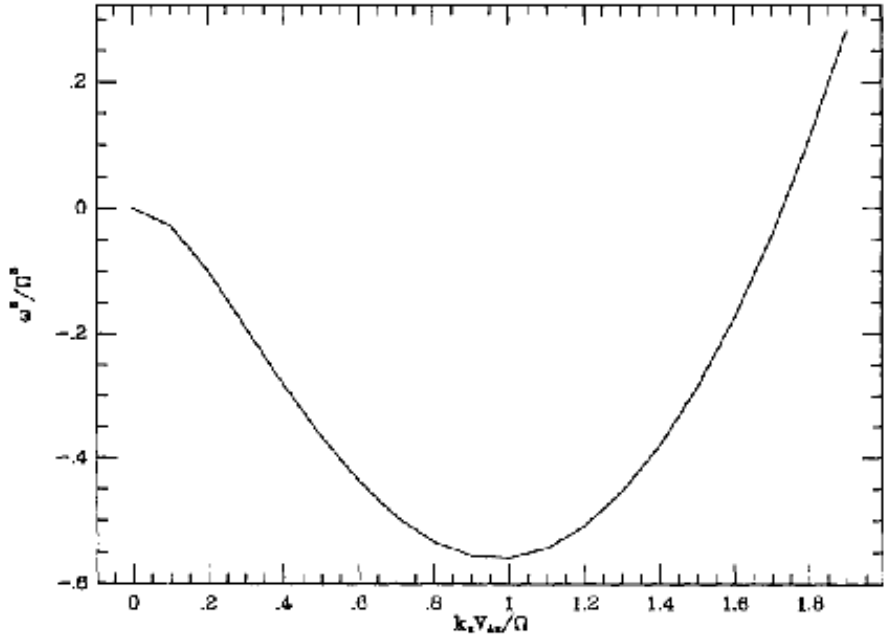}
\end{center}
\caption{The growth rate of MRI again axisymmetric perturbations in the incompressible and magnetized 
disk with a Kepler rotation. In the region with $\omega <0$, the MRI develops. 
This figure is taken from Balbus \& Hawley (1991) \cite{balbus2}.}
\label{bal_mri}
\end{figure}

\begin{figure}
\begin{center}
\epsfxsize=12.0cm
\epsfbox{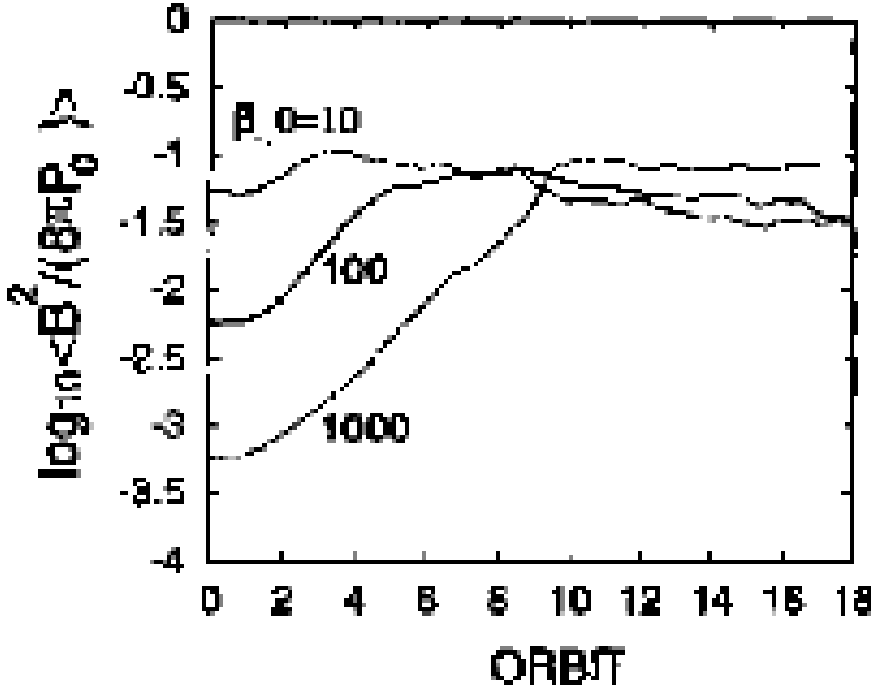}
\end{center}
\caption{The spatially averaged magnetic energy  
 normalized by the gas pressure as a function of the
 rotation period of the accretion disk obtained in the 3D simulations of
 accretion disk threaded by the toroidal magnetic fields. $\beta_0$
 represents the initial value of the plasma beta imposed on the disk. It
 can be seen that the magnetic energy increases exponentially regardless
 of the initial $\beta_0$ and reaches to the quasi-steady state with
 $P_{\rm gas}/P_{\rm mag} \sim 10$.
This figure is taken from Matsumoto {\it et al} (1999) \cite{matsumoto}.}
\label{matsu_mhd}
\end{figure}

Growth rate of the MRI can be roughly estimated as 
\begin{equation}
\eta_{\rm growth} \sim \frac{v_{\rm A}}{\lambda_{c}}.
\end{equation} 
To investigate the dependence of the growth rate on the wavelength of
the perturbation, one need more detailed analysis (see for a review 
\cite{balbus1}) than the one done above. In Figure \ref{bal_mri}, the
dispersion relation between the frequency ($\omega$) and the wavenumber
($k_{z}$ : along the unperturbed magnetic field) of perturbations is
shown. The above relation was found by the linear analysis by adding the  
axisymmetric perturbations to the Keplerian disk with uniform magnetic
fields parallel to the rotational axis and with the assumption that 
the fluid is incompressible (see for detail \cite{balbus2}). 
The oscillation, which is marginally stable ($\omega = 0$) in the absent
of the magnetic fields, is shown to be unstable ($\omega < 0$) in the
presence of the magnetic fields. The maximum growth rate (minimum of the
curve in Figure \ref{bal_mri}) is shown to be,
\begin{equation}
\eta_{\rm max} = \frac{3}{4}\Omega,
\label{eta_max}
\end{equation}
with
\begin{equation}
k_{z} v_{\rm A} = \sqrt{\frac{15}{16}} \Omega.
\end{equation}
Since this is very rapid indeed, the amplitude with the unstable mode 
 can be significantly amplified during several rotation
 periods. The MRI can amplify the fields even if the initial field is very weak. 
It should be noted that the growth rate is independent of the strength
of the magnetic field. Instead, the growth rate of MRI is
determined by the angular velocity (see Eq. (\ref{eta_max})).
This is the main characteristic of the MRI, which is distinct from other
magnetic instabilities such as Parker instability. 
As the MRI develops, 
 the initial poloidal magnetic fields are stretched to the toroidal
 ones, whose strength grows exponentially with time. It is noted that 
the MRI can also develop nonaxisymmetrically.
The evolution of the magnetic energy due to the nonaxisymmetric MRI in
the accretion disk threaded by the purely toroidal fields is shown in Figure
 \ref{matsu_mhd}.
The MRI and its associated angular momentum transport 
have been paid great attention in the community of the accretion disks
around the black hole.

\subsubsection{possibility of the growth of magnetorotational instability \label{sub_MRI} in supernovae}
\begin{figure}
\begin{center}
\epsfxsize=14.5cm
\epsfbox{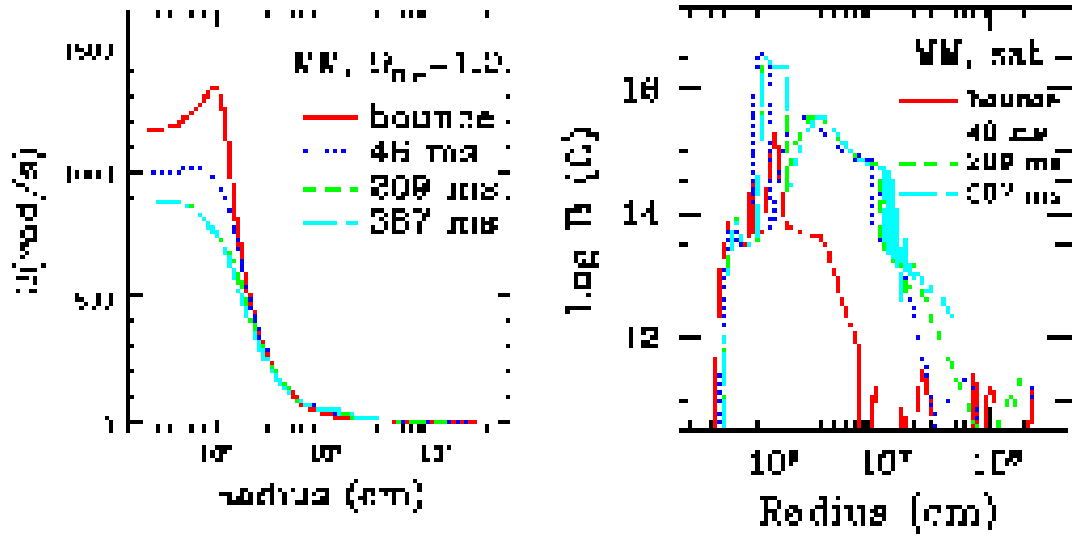}
\end{center}
\caption{The angular velocity (left panel) and the saturation magnetic
 fields (right panel) as a function of radius taken from Akiyama {\it et
 al} (2003) \cite{akiyama}. It is seen 
from the left panel that at core bounce, the angular velocity decreases
 sharply with the radius near $r \sim 10$ km, where is near the surface
 of the protoneutron star. This negative gradient of the angular
 velocity ($d\Omega / d r \le 0 $) is the criteria for the onset of the
 MRI (see \cite{balbus1} for its derivation). They pointed out that the magnetic fields could be amplified as high as $10^{16}$ G in the later phases as shown from the right panel.}
\label{akiyamasan}
\end{figure}     
%The most mysterious is the origin of this large magnetic field.
%The core might have strong magnetic fields prior to collapse already, 
%although the evolution models indicate
%quite the contrary \cite{heger04}. 
%We may not need the strong
%magnetic field initially if the magnetorotational instability (MRI)
%sufficiently develops. 
The MRI mentioned in the last subsection was first applied to the
dynamics of supernova core by Akiyama {\it et al} (2003) \cite{akiyama}. 
Compared with the linear growth of the toroidal
magnetic fields in case of field wrapping, the field is expected to grow 
exponentially due to MRI also in the supernova core. 
If it is true, we may need the only small seed magnetic fields. 
When MRI develops efficiently in the supernova core, the field might reach to the saturation
strength, which is estimated to be as high as 
\begin{eqnarray}
B_{\rm sat} &=& (4 \pi \rho)^{1/2} r \Omega  \nonumber \\
&\sim&  4 \times 10^{16} \Bigl(\frac{\rho}{1\times 10^{13}~{\rm g}~{\rm cm}^{-3}}\Bigr)^{1/2}
\Bigl(\frac{R}{10~{\rm km}}\Bigr)\Bigl(\frac{\Omega}{1000~{\rm rad}~{\rm s}^{-1}}\Bigr)~[{\rm G}],
\end{eqnarray}
 where we employ the typical values near the surface of the protoneutron star (see Figure \ref{akiyamasan}). 
Here the saturation field is determined by the condition that 
the toroidal component of the Alfv\'{e}n speed, $v_{\rm A} =
\frac{B}{\sqrt{4 \pi \rho}}$, comes into rough
equipartition with the rotation velocity, $v_{\rm rot} = r \Omega$, in
analogy with the case in the accretion disk \cite{balbus1}.

\begin{figure}
\begin{center}
\epsfxsize = 7 cm
\epsfbox{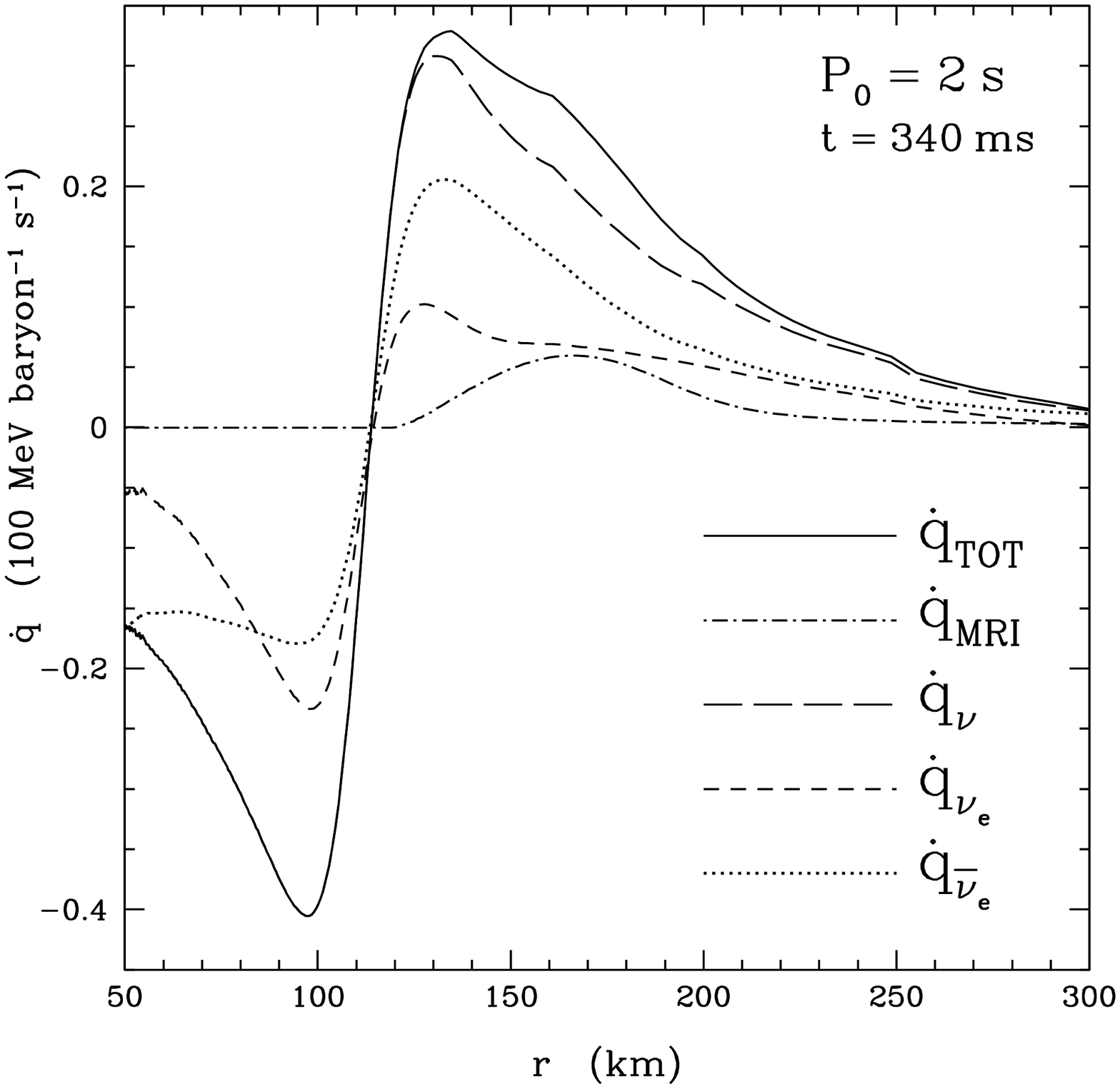}
\epsfxsize = 7 cm
\epsfbox{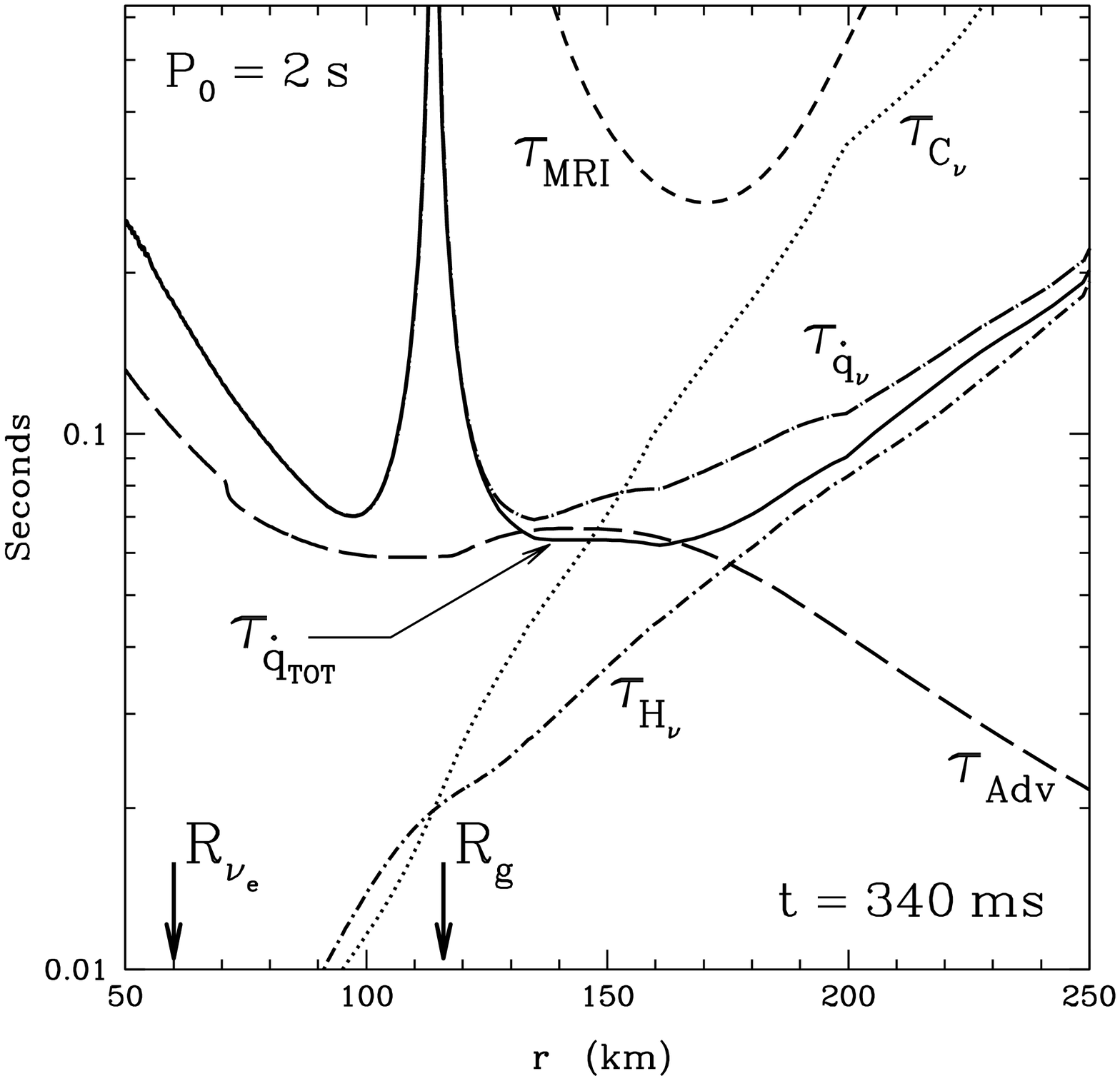}
\caption{Effect of viscous heating due to MRI on the explosion mechanism
 taken from Thompson et
 al. (2005) \cite{tomp04}. 
Left panel shows the energy deposition rate as a function of radius for
 a rotating model with the initial period of $2$ seconds, including
 viscous heating via the MRI with $\alpha = 0.1$. In this panel,
total energy deposition rate 
($\dot{q}_{\rm tot}=\dot{q}_{\nu} + \dot{q}_{\rm MRI}$,
 solid line), viscous energy deposition ($\dot{q}_{\rm MRI}$, dot-dashed
 line), and neutrino deposition rate from each species ($\nu_e$
 short-dashed line, $\bar{\nu}_e$ dotted line) are shown.
 The order
 of 10 \% increase of the heating rate 
due to the viscous heating above the gain radius ($R \sim 120 ~{\rm km}$)
 can be seen. This increase was pointed out to be sufficient to 
instigate the explosion. Right panel shows the various timescales of the
 same model in the left panel. $\tau_{H_{\nu}}$, $\tau_{C_{\nu}}$, 
$\tau_{\dot{q_{\nu}}}$, $\tau_{\rm adv}$, $\tau_{\rm MRI}$, and
 $\tau_{\dot{q_{\rm TOT}}}$, represent the timescales of neutrino
 heating, neutrino cooling, the advection of the infalling matter, 
the viscous heating of MRI, the total heating including the
 contributions of MRI, as a function of radius, respectively.
$R_{\nu_e}$ and $R_{\rm g}$ mark the position of the $\nu_e$
 neutrinosphere and the gain radius, respectively.
In the regions above the gain radius from 
$130 {\rm km}< R < \sim 170 ~{\rm km}$, the net heating time scale 
becomes shorter than the advection timescale of the infalling matter 
as a result of the viscous heating.  
Otherwise in the slowly rotating models $P_0 > 5 ~\rm{s}$, the explosions were
 not obtained in their computations, because there are no regions, in
 which the condition, $\tau_{\dot{q}_{\nu}} < \tau_{\rm adv}$, is satisfied.}
\label{tomp_vis}
\end{center}
\end{figure}

In such a case, the magnetic stresses generated by the MRI 
could be the origin of the viscous energy deposition \cite{tomp04}. 
By employing an $\alpha$ prescription for the
viscous dissipation, which is often used in the study of accretion disk
\cite{shakura}, the robust explosion
was pointed out to be obtained because the MRI heating aids the neutrino heating (see
Figure \ref{tomp_vis}).

Although there remains persistent concern of the treatment of the rotation and magnetic
fields in the above 1D model computations \cite{akiyama,tomp04}, 
the obtained implications seem to be important and should be examined by the multidimensional MHD 
simulations \cite{tomp04}. 
Currently the 3D MHD simulations 
have just begun to be 
investigated (see, for example \cite{liepen}). 
Very recently nucleosynthesis in the magnetic supernovae has been reported 
 \cite{nishimura}. Although the study of magnetic supernovae 
is still in its infancy with respect to the treatment of the neutrino physics,
this field seems to be blossoming with the recent developments both 
of the transport method \cite{buras,livne,walder} and  
the growing computing power. 

%In the left panel of Figure \ref{aniso_mag}, the distribution of the poloidal
%magnetic fields at $t \sim 20$ msec after core bounce is given. 
%The initially uniform magnetic field 
%is deformed to be dipole-like due to core contraction.
%Using this configuration of magnetic field as a background, 
%we estimate the neutrino heating rate by electron neutrino absorptions on
%neutron both with and without the
%parity-violating corrections and compare them by the same procedure as
%discussed earlier in this paper. As for the cross section with the
%parity-violating effects, we employ the formula by \cite{arras}.   
%The right panel of Figure \ref{aniso_mag} shows 
%the ratio of the heating rate 
%with the corrections to that without the corrections.
%It is found that the ratio is reduced by the corrections $\lesssim 0.1 \%$ 
%near the north pole, while it is enhanced 
%$\sim 0.1 \%$ in the vicinity of the south pole. 

\clearpage

\section{Gravitational Waves from Core-Collapse Supernovae \label{GW_sec}}

From this section, we review the studies about gravitational waves 
from core-collapse supernovae. As mentioned in section \ref{intro}, 
the detection of the gravitational signal is
important not only for the direct confirmation of general relativity but
also for the understanding of the explosion mechanism supernovae themselves. 
In combination with neutrino signals mentioned in section \ref{section:SN_nu_osc},
 the gravitational wave will enable us to see
directly the innermost part of an evolved star, where 
 the key physics related to the explosion mechanism such as the angular momentum distribution and the equation of state are veiled. 

In this section, we use the convention 
that Latin indices run from 1 to 3, Greek from 0 to 3, where 0 is
 the time component of four-vectors and that
partial derivatives $\partial/ \partial x^{\mu}$ of tensor
$T^{\alpha_1 \cdot\cdot\cdot \alpha_n}$ with respect to a coordinate
$x^{\alpha}$ are denoted by ${T^{\alpha_1\cdot\cdot\cdot \alpha_n}}_{,\alpha}$.

%\subsection{Emission mechanisms of Gravitational Waves in Core-Collapse Supernovae}
\subsection{Physical foundations} 
Before we discuss the gravitational waves from core-collapse supernovae,
we summarize the physical foundations of Gravitational Wave in {\it 
Einstein's theory of gravity}, for later convenience and completeness of
this review (see, also \cite{weinberg,thorne80,misner,takashi}). 

As well known, the {\it Einstein equations},
\begin{equation}
R_{\mu \nu} - \frac{1}{2}g_{\mu \nu} R = \frac{8\pi G}{c^4}T_{\mu \nu},
\label{einstein}
\end{equation}
 express the relation between matter distribution in spacetime,
 $T_{\mu \nu}$ on the right hand side representing the matter
 stress-energy tensor, and the curvature of spacetime on the left hand side, 
represented by the components of the Ricci tensor given by the contraction of $R_{\mu \nu} = {R^{\alpha}}_{\mu
\alpha \nu}$ of the Riemann curvature tensor, and the scalar curvature $R = {R^{\alpha}}_{\alpha}$. The Riemann
curvature tensor is connected to the metric $g_{\mu \nu}$ of spacetime
through the Chiristoffel symbols $\Gamma^{\alpha}_{\beta
\mu}$:
\begin{equation}
R^{\alpha}_{\beta \mu \nu} \equiv \Gamma^{\alpha}_{\beta \nu,\mu} -  \Gamma^{\alpha}_{\beta \mu,\nu} +  \Gamma^{\sigma}_{\beta \nu} \Gamma^{\alpha}_{\sigma \mu} -
 \Gamma^{\sigma}_{\beta \mu} \Gamma^{\alpha}_{\sigma \nu},
\end{equation}
with
\begin{equation}
\Gamma^{\alpha}_{\beta \mu} \equiv \frac{1}{2}g^{\alpha \sigma}(g_{\mu \sigma,
\beta} + g_{\sigma \beta,\mu} - g_{\beta \mu,\sigma}).
\end{equation}
If one defines the Einstein tensor $G_{\mu \nu}$ as
\begin{equation}
G_{\mu \nu} \equiv R_{\mu \nu } - \frac{1}{2}g_{\mu \nu} R,
\end{equation}
the Einstein equations can be compactly written,
\begin{equation}
G_{\mu \nu} = \frac{8\pi G}{c^4} T_{\mu \nu}.
\end{equation}
It is noted that 
the Einstein equations reduce to Poisson's equation, $\nabla^2 \phi = 
4 \pi G \rho$
in the Newtonian limit.
To solve the Einstein equations, which consists of 10 nonlinear partical 
derivative equations, has been a challenging problem.
So far only some exact solutions, such as Schwarzshild and Kerr
solutions etc, have been found in the very idealized 
physical situations. 

\subsubsection{weak-field approximation}
The gravitational waves far from the source can be characterized as
linear metric perturbations propagating on a flat backgroud.
Taking a first order perturbation from the Minkowskian metric:
$\eta_{\mu \nu} \equiv diag(1,-1,-1,-1)$, the metric $g_{\mu \nu}$ of 
spacetime can be decomposed as,
\begin{equation}
g_{\mu \nu} = n_{\mu \nu} + h_{\mu \nu},
\end{equation}
with the demand that 
\begin{equation}
|h_{\mu \nu}| \ll 1.
\label{verysmall}
\end{equation}
Roughly speaking, the weak field conditions can be satisfied when
$GM/(c^2 R) \ll 1$ with $M$ and $R$ being the characteristic mass and
the length scale of the system. If we take $M=M_{\odot}$, the
approximation is well valid in the distance $r \gg GM_{\odot}/c^2 \simeq 1.5~{\rm km}$.

To first order in $h$, the Ricci tensor becomes,
\begin{equation}
R_{\mu \nu} = \Gamma^{\lambda}_{\lambda \mu, \nu} - \Gamma^{\lambda}_{\mu \nu, \lambda} + O(h^2),
\label{ricci}
\end{equation}
and the affine connection is 
\begin{equation}
\Gamma^{\lambda}_{\mu \nu} = \frac{1}{2} \eta^{\lambda \rho}
(h_{\rho \nu, \mu} + h_{\rho \mu, \nu} - h_{\mu \nu, \rho}) +  O(h^2).
\label{chiri}
\end{equation}
Since we treat the linearized equations up to the order $O(h)$ in the
following, the raising and rising of all 
indices can be done using 
$\eta_{\mu \nu}$, not $g_{\mu \nu}$, that is,
\begin{equation}
\eta^{\mu\nu}h_{\nu\lambda} \equiv {h^{\mu}}_{\lambda},~\eta^{\lambda \rho} \frac{\partial}{\partial x^{\rho}} \equiv \frac{\partial}{\partial x_{\lambda}}, {\rm etc},
\end{equation}
because using $g_{\mu \nu}$ makes the equations non-linear again.

Introducing Eq. (\ref{chiri}) to (\ref{ricci}), one can obtain the first-order Ricci 
tensor,
\begin{equation}
R_{\mu\nu}^{(1)} = \frac{1}{2}(\Box h_{\mu\nu} - {h^{\lambda}}_{\nu, 
\lambda \mu} - {h^{\lambda}}_{\mu,\lambda\nu} + {h^{\lambda}}_{\lambda, \mu\nu}),
\end{equation}
here $\Box \equiv \eta^{\mu \nu}\frac{\partial^2}{\partial x^{\mu}
\partial x^{\nu}}$ is the d'Alembert operator. 
Then the Einstein field equations (Eq. (\ref{einstein})) read
\begin{equation}
\Box h_{\mu \nu} - {h^{\lambda}}_{\nu, 
\lambda \mu} - {h^{\lambda}}_{\mu,\lambda\nu} + {h^{\lambda}}_{\lambda, \mu\nu}= - \frac{16 \pi G}{c^4} T_{\mu \nu}.
\label{linearized}
\end{equation}
Since the above equations are invariant under arbitrary coordinate
transformations, the solutions cannot be determined uniquely.
Let us introduce the most general coordinate transformation in the
following form, 
\begin{equation}
x^{\mu} \Longrightarrow x^{'\mu} = x^{\mu} + \epsilon(x^{\mu}).
\label{coordinate}
\end{equation} 
Here it is noted that $\partial{\epsilon^{\mu}}/ \partial x^{\nu}$
should be at most of the same order of magnitude as $h_{\mu \nu}$ not
to violate the weak-field condition. Since the metric of spacetime in the new
coordinate is given by
\begin{equation}
g^{' \mu \nu} = \frac{\partial x^{'\mu}}{\partial x^{\lambda}}
\frac{\partial x^{'\nu}}{\partial {x^{\rho}}} g^{\lambda \rho},
\end{equation}
then 
\begin{equation}
h^{'\mu\nu} = h^{\mu \nu} - {\epsilon^{\mu}}_{,\lambda} \eta^{\lambda\nu} - {\epsilon^{\nu}}_{,\rho} \eta^{\rho\mu} = h^{\mu \nu } - \epsilon^{\mu,\nu}-\epsilon^{\nu,\mu}.
\label{hdash}
\end{equation}
One can easily check that the new $h^{'\mu \nu}$ satisfy the linearized
Einstein equations by introducing Eq. (\ref{hdash}) to Eq. (\ref{linearized}). 

The above property, namely gauge invariance of the field equations, is
a nuisance when one actually solves the field equations. To circumvent
this problem, one has only to choose the coordinate system. The most
familiar and convenient choice is to work in a harmonic coordinate, such
that,
\begin{equation}
g^{\mu \nu} {\Gamma_{\mu \nu}}^{\lambda} = 0.
\end{equation}
Using Eq .(\ref{chiri}), one can obtain equivalently,
\begin{equation}
{{h_{\nu}}^{\mu}}_{,\mu} = 
\frac{1}{2}{{h_{\mu}}^{\mu}}_{,\nu}.
\label{condi}
\end{equation}
If ${h^{\mu}}_{\nu}$ does not satisfy Eq (\ref{condi}), one can find
the new $h^{'\mu \nu}$ by performing the coordinate transformation
(Eq. (\ref{coordinate})) with
$\epsilon_{\nu}$ satisfying the condition,
\begin{equation}
\Box \epsilon_{\nu} = {{h_{\nu}}^{\mu}}_{,\mu} - \frac{1}{2}{{h_{\mu}}^{\mu}}_{,\nu}.
\end{equation} 
It should be noted that there still remains the freedom of the 
coordinate transformation. For example, perform the coordinate transformation
($x^{\mu} \rightarrow x^{' \mu} = x^{\mu} + {\epsilon}^{' \mu} (x)$) with
${\epsilon}^{'}_{\nu}$ satisfying the following condition,
\begin{equation}
\Box {\epsilon}^{'}_{\nu} = 0.
\label{remain}
\end{equation} 
Then the condition in Eq. (\ref{condi}) is indeed satisfied. 
We return to this problem soon in subsection \ref{vacuno}. 

Using the harmonic gauge condition (Eq. (\ref{condi})) 
in Eq. (\ref{linearized}), the field equations now read,
\begin{equation}
\Box {h}^{\mu \nu} = - \frac{16 \pi G}{c^4} T^{\mu \nu}.
\label{bert}
\end{equation}
Finally one can find the physical formal solution in a form of 
the time-retarded Green function,
\begin{equation}
h_{\mu \nu}(t,\mbox{\boldmath$x$}) 
= \frac{4G}{c^4}\int d^3 x^{'}~ \frac{T_{\mu \nu}(t-\frac{|\mbox{\boldmath$x$}-\mbox{\boldmath$x$}^{'}|}{c}, \mbox{\boldmath$x$}^{'})}{|\mbox{\boldmath$x$}-\mbox{\boldmath$x$}^{'}|}.
\label{retarded}
\end{equation}

\subsubsection{wave solutions in vacuum \label{vacuno}}
Now we move on to consider the solution of the linearized Einstein
equations in vacuum ($T_{\mu \nu} = 0$). 
Then Eq. (\ref{bert}) reduces to
\begin{equation}
\Box {h}_{\mu \nu} = 0.
\label{vacuum}
\end{equation}
The well-known plane wave solution is
\begin{equation}
{h}_{\mu \nu} = e_{\mu \nu} \exp(i k_{\lambda} x^{\lambda}).
\label{plane}
\end{equation}
Introducing the solution into Eq. (\ref{vacuum}) yields $ k_{\lambda}
k^{\lambda} =0$, which means that the gravitational waves travel along
the null geodesics at the speed of light. Since Eq. (\ref{plane}) should
satisfy the harmonic condition
(Eq. (\ref{condi})),
\begin{equation}
k_{\mu}{e_{\nu}}^{\mu} - \frac{1}{2}k_{\nu} {e_{\mu}}^{\mu} = 0.
\label{harmopol}
\end{equation} 
Let us express the remained freedom of the coordinate transformation
(Eq. (\ref{remain})) in the following form,
\begin{equation}
\epsilon^{\mu}(x) = i c^{\mu} \exp(i k_{\lambda} x^{\lambda}),
\label{epi}
\end{equation}
where $c^{\mu}$ is the constant vector.
Introducing Eq. (\ref{epi}) to Eq. (\ref{hdash}), one obtains
\begin{equation}
{e_{\mu\nu}}^{'} = e_{\mu\nu} + k_{\mu} c_{\nu} + k_{\nu} c_{\mu}.
\label{e_munu}
\end{equation}

For instance, let us consider a wave traveling in the $z$ direction
with wave vector,
\begin{equation}
k^1=k^2=0,~~k^0=k^3\equiv k >0.
\end{equation}
Using Eq. (\ref{e_munu}), we determine $c_{\mu}$ satisfying $e_{00}^{'},
e_{0i}^{'}=0$, namely
\begin{equation}
e_{00}^{'} = e_{00} + 2 k_0 c_0 = 0
\label{2-41}
\end{equation}
\begin{equation}
e_{0i}^{'}= e_{0i} + k_0 c_i + k_i c_0 = 0.
\label{2-42}
\end{equation}
Noting $k_0 = - k$, the component of $c_{\mu}$ becomes,
\begin{equation}
c_0 = \frac{e_{00}}{2 k},~c_1 = \frac{e_{01}}{k},~~c_2=\frac{e_{02}}{k},~~c_3 = \frac{e_{03}+kc_0}{k}.
\end{equation}
Since Eq. (\ref{harmopol}) is invariant under the gauge transformation
 (Eq. (\ref{e_munu})), we omit $'$ in the
 following. Using Eq. (\ref{2-41} and \ref{2-42}), 
Eq. (\ref{harmopol}) becomes,
\begin{equation}
k e_{3 \mu} - \frac{1}{2} k_{\mu}(e_{11} + e_{22} + e_{33}) =0.
\label{2-44}
\end{equation}
When we see the zero component of the above equation ($\mu = 0$),
\begin{equation}
e_{11} + e_{22} + e_{33} = 0,
\label{2-45}
\end{equation}
because
\begin{equation}
e_{30} = e_{03} = 0.
\end{equation}
Using Eq. (\ref{2-45}), Eq. (\ref{2-44}) means,
\begin{equation}
e_{3l} = 0.
\end{equation}
Taking $l=3$, Eq (\ref{2-45}) becomes
\begin{equation}
e_{11} + e_{22} =0.
\end{equation}
Finally, the $e_{\mu \nu}$ can be written in the following form,
\begin{equation}
                   \label{2-47} e_{\mu\nu} =
                    \left(
                     \begin{array}{cccc}
                      0 & 0 & 0& 0 \\
                      0 & e_{11} & e_{12} & 0 \\
		        0 & e_{12} & -e_{11} & 0 \\
                        0 & 0 & 0& 0 \\
                     \end{array}
                    \right)
\end{equation}
So far, the spatial component of $k^{\mu}$ is taken to be $z$ axis,
Eq. (\ref{2-47}) indicates that $e_{ij}$ with arbitrary $k^{\mu}$
satisfies,
\begin{equation}
e_{ij} \delta^{ij} = 0~({\rm Tranceless}) \nonumber,
\end{equation}  
\begin{equation}
e_{ij} k^{j} = 0~({\rm Transeverce}).
\end{equation}  
It is noted that the above choice of the gauge is the so-called
transverse-traceless (TT) gauge. We will use it in the following and denote it
by the superscript TT.
Since $e_{0\mu}$ was originately
correspond to the freedom of the coordinate choice, they have nothing to
do with the physical freedom. After all, the nonvanishing two components
of $e_{11}$ and $e_{22}$ remain. We call the true physical freedom as the gravitational wave.
 
%%%%%%%%%%%%%%%%%%%%%%%%%%%%%%%%%%%%%%%%%%%%%%%%%%%%%
\subsubsection{polarization of gravitational waves}
%%%%%%%%%%%%%%%%%%%%%%%%%%%%%%%%%%%%%%%%%%%%%%%%%%
\begin{figure}
\begin{center}
\epsfxsize=3cm
\epsfbox{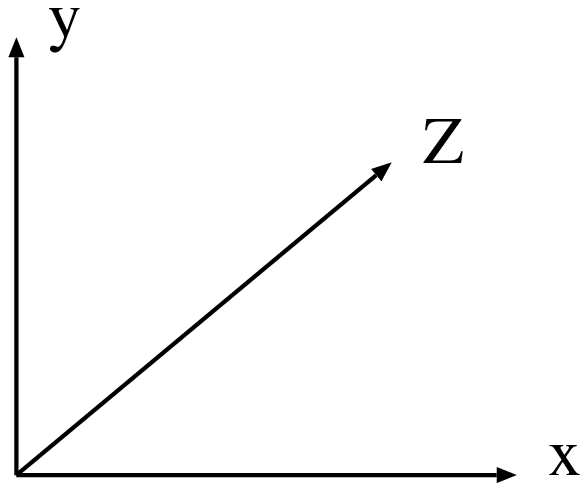}
\epsfxsize=8cm
\epsfbox{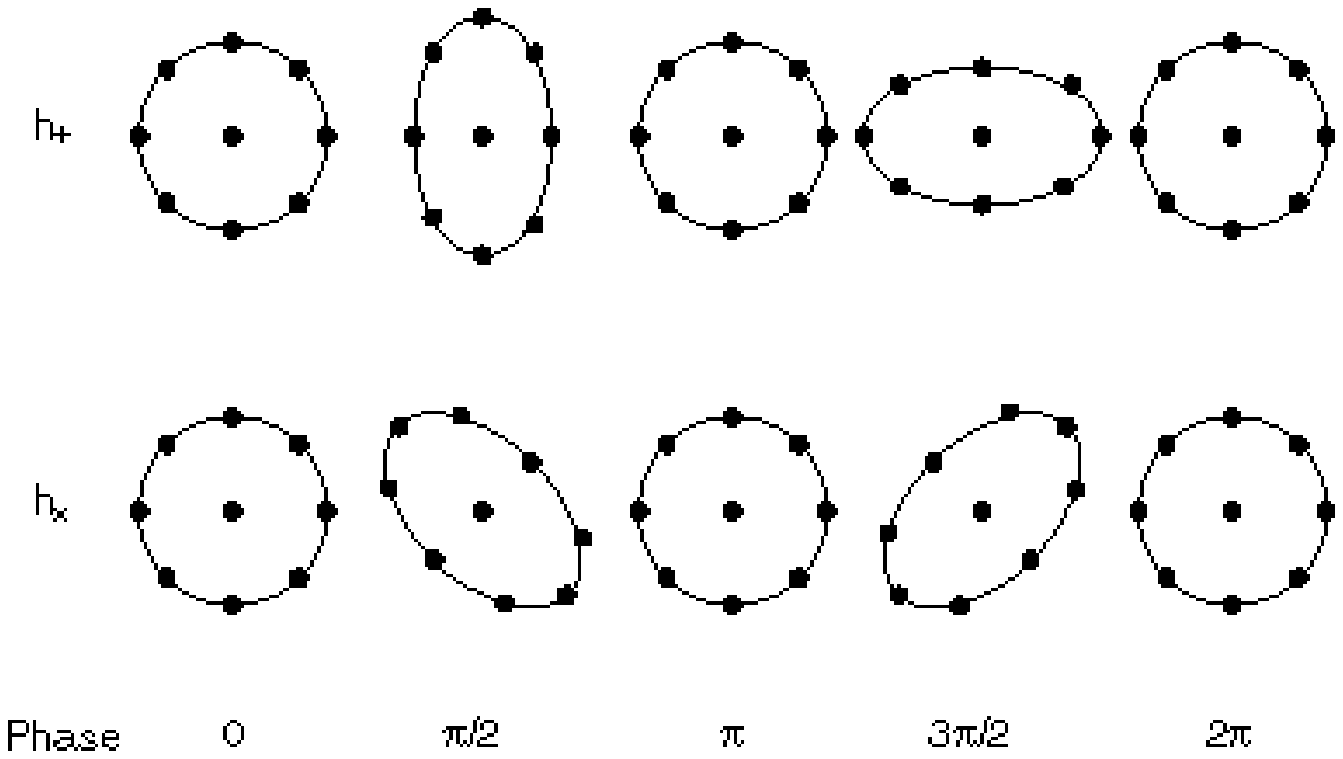}
\end{center}
\caption{Schematic figure describing the changes of the proper distance
 due to the plus $h_{+}$ and cross $h_{\times}$ modes of
 the gravitational wave, which propagates along the $+z$ axis. 
Note that horizontal and vertical directions
 express the $x$ and $y$ axis, respectively.
}
\label{pol_GW}
\end{figure}     
First of all, we pay attention to the two spatial components 
of the gravitational wave propagating $+z$ direction,
($h^{TT}_{\mu\nu}$ in Eq. (\ref{2-47})),
\begin{equation}
 \label{2-168}
h^{TT}_{ij} = 
                    \left(
                     \begin{array}{ccc}
                      h_{+} & h_{\times} & 0 \\
                      h_{\times} & - h_{+} & 0 \\
		      0 & 0 &  0 \\
                     \end{array}
                    \right).
\end{equation}
Here the corresponding metric, which expresses the superposition of the
two plane waves can be written in a form,
\begin{equation}
ds^2 = - dt^2 + (1 + h_{+})dx^2 + (1-h_{+})dy^2 + 2 h_{\times }dx dy + dz^2,
\label{metric}
\end{equation}
where $h_{+} = h_{+}(ct-z),~~h_{\times} = h_{\times}(ct-z)$. As clearly
seen from the metric, the gravitational wave changes the proper distance
in the plane ($x-y$ plane) perpendicular to the propagating direction
($z$ axis). When the gravitational wave with $h_{+}$ (plus mode)
propagates, the proper distance in the $x$ axis becomes longer (shorter)
and the one in the $y$ axis becomes shorter (longer) (see the top panel
of Figure \ref{pol_GW}).

Next, let us rotate the coordinate $\pi/4$ along the $+z$ axis, 
\begin{equation}
\label{2-169}
     \left(
                     \begin{array}{c}
                      x^{'} \\
                      y^{'}\\
                     \end{array}
                    \right)
   = 
                    \left(
                     \begin{array}{cc}
                       \cos \pi/4 & \sin \pi/4 \\
                      - \sin \pi/4  & \cos \pi/4\\
		         \end{array}
                    \right)
		     \left(
		      \begin{array}{c}
                      x \\
                      y \\
                     \end{array}
                    \right)
		     = 
		     \left(  
\begin{array}{c}
                      \frac{1}{\sqrt{2}}(x+y) \\
                       \frac{1}{\sqrt{2}}(-x+y) \\
                     \end{array}
                    \right).
\end{equation}
Then the metric in Eq. (\ref{metric}) reads,
\begin{equation}
ds^2 = - dt^2 + (1 + h_{\times})dx^{'2} + (1-h_{\times})dy^{'2} + 2 h_{+ }dx^{'} dy^{'} + dz^2.
\end{equation}
Thus one can see that the cross mode of the gravitational wave
$h_{\times}$ represents the change in the proper distance tilted $\pi/4$
with respect to the change formed by the plus mode (see the lower panel
of Figure \ref{pol_GW}).
In this way, gravitational waves act tidally, stretching and squeezing 
space in a quadrupole manner and thus object that they pass through.
%%%%%%%%%%%%%%%%%%%%%%%%%%%%%%%%%%%%%%%%%%%%%
\subsubsection{meaning of TT gauge}
%%%%%%%%%%%%%%%%%%%%%%%%%%%%%%%%%%
In this subsection, we discuss why TT gauge is important and how we can
extract the TT components from the arbitrary tensor $h_{ij}$.

First of all, we should be cautious that the solution of Eq. (\ref{vacuum}) is {\it not
only} the gravitational wave. For example, let us consider the static
solution around the mass point $M$ in the Newtonian limit,
\begin{equation}
g_{00} = - \Bigl(1 - \frac{2GM}{c^2r}\Bigr), ~g_{0j} =0,~g_{ij} = \delta_{ij}.
\end{equation} 
For the distant observer $r_0 \gg 2 GM/c^2$, the $00$ component of the
metric can be written,
\begin{equation}
g_{00} = - 1 + \epsilon \frac{r_0}{r},
\label{2-68}
\end{equation}
with $\epsilon$ being $2GM/c^2 r_0$. Comparing Eq. (\ref{2-68}) with
\begin{equation}
g_{\mu\nu} = \eta_{\mu\nu} + h_{\mu\nu},
\end{equation} 
one can notice that the solution includes a static
contribution. Furthermore, the freedom of the arbitrary coordinate
transformation could be remained. We should omit these components in
order to take the components of gravitational wave, which we are
interested in. In other words, it is necessary to take TT part from the
arbitrary $h_{ij}$. As a sideremark, let us summarize the procedure to do so in the
following \cite{takashi}.

First of all, we perform the infinitesimal coordinate transformation
(Eq. (\ref{hdash})) in order to satisfy the condition : ${h^{'}}^{0\mu}= 0$. The condition
is equivalently written,
\begin{equation}
0 = h^{'}_{00} = h_{00} - 2 \frac{\partial \epsilon_0}{\partial t}, 
\label{2-72}
\end{equation}
and 
\begin{equation}
0 = h^{'}_{0i} =  h_{0i} -  \frac{\partial \epsilon_i}{\partial t} - 
 \frac{\partial \epsilon_0}{\partial x^{i}}.
\label{2-73}
\end{equation}
From Eq (\ref{2-72}), $\epsilon_{0}$ can be determined and using it,
$\epsilon_i$ can be determined from Eq (\ref{2-73}). As a result, one
can confirm that ${h^{'}}^{0\mu}= 0$ can be really satisfied. Then the
problem is how one can find the TT components given the arbitrary tensor
$h_{ij}$. There is a general formula to do so, in which the arbitrary
tensor $h_{ij}$ can be decomposed as follows,
\begin{equation}
h_{ij} = h^{TT}_{ij} + (W_{i,j} + W_{j,i} - \frac{2}{3}\delta_{ij}{W^{l}}_{,l}) + \frac{\delta_{ij}}{3}{h^{l}}_{l}, 
\label{2-74}
\end{equation}
\begin{equation}
\delta^{lm} h^{TT}_{il,m}=0,~~\delta^{ij}h^{TT}_{ij} = 0,
\label{2-75}
\end{equation}
with $W_{i}$ is an arbitrary longitudinal vector. Define $\tilde{h}_{ij}
= h_{ij} - \delta_{ij}{h^{l}}_l/3$, then Eq. (\ref{2-74}) becomes
\begin{equation}
W_{i,j} + W_{j,i} - \frac{2}{3}\delta_{ij}{W^{l}}_{,l} = \tilde{h}_{ij} - h^{TT}_{ij}.
\label{2-76}
\end{equation} 
By taking divergence of Eq. (\ref{2-76}), Eq. (\ref{2-76}) becomes 
\begin{equation}
\triangle W_{i} + \frac{1}{3}({W^{l}}_{,l})_i = {\tilde{h}_{i,j}}^{j},
\label{2-77}
\end{equation}
because one can drop the TT components.
By taking divergence Eq. (\ref{2-77}) again, Eq. (\ref{2-77}) reads,
\begin{equation}
\triangle ({W^{l}}_{,l}) = \frac{3}{4}{\tilde{h}^{ij}}_{~,ji}.
\label{2-78}
\end{equation} 
To summarize, given $\tilde{h}_{ij}$, one can find $W^{l}_{,l}$ by
solving the Poisson equation of Eq. (\ref{2-78}). Then introducing the
solution of $W^{l}_{,l}$ into Eq. (\ref{2-77}), Eq. (\ref{2-77})
becomes,
\begin{equation}
\triangle W_{i}  = {\tilde{h}_{i,j}}^{j} - \frac{1}{3}({W^{l}}_{,l})_i.
\label{2-79}
\end{equation}
Solving the Poisson equation of  Eq. (\ref{2-79}), one can obtain
$W_{i}$. Then with Eq. (\ref{2-76}), TT component of $h_{ij}$ can be
determined.

 The above procedure for extracting TT components from the arbitrary
 tensor is a little bit complicated, but is far easier for the plane-wave
 solution in Eq. (\ref{plane}). This is because the spatial
 derivative can be replaced by $i \mbox{\boldmath$k$}$.
The projection tensor to change the arbitrary vector to transverse
 is given by,
\begin{equation}
P_{ij} = \delta_{ij} - n_i n_j,
\end{equation}
with
\begin{equation}
n_i = \frac{k_i}{k}.
\end{equation}
Thus the transverse component of $h_{ij}$ can be given by
\begin{equation}
h^{T}_{ij} = {P_i}^{l} {P_j}^m h_{lm}.
\end{equation}
Noting ${P_i}^{l} P_{lj} = P_{ij}$ and ${P^{l}}_{l} = 2$, the procedure
to make the tensor traceless can be given by,
\begin{equation}
h^{TT}_{ij} = {P_i}^{l} {P_j}^m h_{lm} - \frac{P_{ij}}{2}(P^{lm}h_{lm}).
\label{2-85}
\end{equation}
When the source of gravitational wave is far distant from us, the plane-wave
approximation in Eq. (\ref{plane}) is well satisfied. Thus we have only
to do the procedure in Eq. (\ref{2-85}), in order to extract the TT components.

\subsubsection{quadrupole formula}
In order to extract the components of gravitational wave, we should take TT components of  Eq. (\ref{retarded}). Let us write
Eq. (\ref{retarded}) again with using the geometrical unit ($G=c=1$) for simplicity,
\begin{equation}
h_{\mu\nu}(t,\mbox{\boldmath$x$}) = 4 \int d^3 x^{'}~
\frac{T_{\mu \nu}(t- |\mbox{\boldmath$x$}- \mbox{\boldmath$x$}^{'} |, \mbox{\boldmath$x$}^{'})}{|\mbox{\boldmath$x$}-\mbox{\boldmath$x$}^{'}|}.
\label{3-18}
\end{equation}
Integral with respect to $\mbox{\boldmath$x$}^{'}$ is performed in the
source volume. Even for a supernova in our galactic center 
($r = 10$~{\rm kpc}),  the distance to us ($r$) is far larger than
the scale of the source, 
\begin{equation}
r = |\mbox{\boldmath$x$}| \gg |\mbox{\boldmath$x$}^{'}| \sim L 
\sim O(10~{\rm km}).
\end{equation} 
Note that we set $L$ to be typical size of the inner core, because 
the gravitational wave is emitted most strongly at the epoch of core
bounce as will be mentioned later.

 Neglecting the terms higher than $O([L/r]^2)$, the denominator of 
Eq. (\ref{3-18}) becomes,
\begin{equation}
|\mbox{\boldmath$x$}-\mbox{\boldmath$x$}^{'}| \simeq 
r - \mbox{\boldmath$n$} \cdot \mbox{\boldmath$x$}^{'},
\end{equation} 
with
\begin{equation}
\mbox{\boldmath$n$} = \frac{\mbox{\boldmath$x$}}{r}.
\end{equation}
Taking TT of Eq. (\ref{3-18}), Eq. (\ref{3-18}) up to the lowest order
of $L/r$ becomes,
\begin{equation}
h^{TT}_{ij} = \frac{4}{r} 
\int d^3 x^{'}~
T^{TT}_{ij}(t- r + \mbox{\boldmath$n$} \cdot \mbox{\boldmath$x$}^{'},
 \mbox{\boldmath$x$}^{'}),
\label{3-24}
\end{equation}
here we remained the spatial components of $h_{\mu \nu}$ in order to
extract the gravitational wave. 
Let us expand Eq. (\ref{3-24}) as follows,
\begin{eqnarray}
T^{TT}_{ij}(t- r + \mbox{\boldmath$n$} \cdot \mbox{\boldmath$x$}^{'},\mbox{\boldmath$x$}^{'}) &=& \sum_{m=0}^{\infty} \frac{\partial}{\partial t^m} T^{TT}_{ij}(t- r,\mbox{\boldmath$x$}^{'}) 
\frac{(\mbox{\boldmath$n$} \cdot \mbox{\boldmath$x$}^{'})^{m}}{m!}
\label{3-25}
 \\ 
&=&T^{TT}_{ij}(t- r,\mbox{\boldmath$x$}^{'}) + \mbox{\boldmath$n$} \cdot \mbox{\boldmath$x$}^{'} \frac{\partial}{\partial t} T^{TT}_{ij}(t- r,\mbox{\boldmath$x$}^{'}) + \cdot \cdot \cdot.
\label{expand}
\end{eqnarray}
This expansion is allowed only when the motion of the source is much
slower than the speed of light.  Writing the
second term in Eq. (\ref{expand}) as follows,
\begin{equation}
\Bigl|\mbox{\boldmath$n$} \cdot \mbox{\boldmath$x$}^{'} \frac{\partial}{\partial t} T^{TT}_{ij}(t- r,\mbox{\boldmath$x$}^{'})\Bigr| \leq \frac{L}{c} 
\Bigl|\frac{\partial}{\partial t} T^{TT}_{ij}(t- r,\mbox{\boldmath$x$}^{'})\Bigr| \sim 
\frac{v}{c}\Bigl|T^{TT}_{ij}(t- r,\mbox{\boldmath$x$}^{'})\Bigr|,
\end{equation}
where $v$ is the typical velocity of the source, one can understand the
reason clearly. This approximation is often referred as the slow-motion 
approximation. We employ this in the following.

With Eq. (\ref{3-25}), Eq. (\ref{3-24}) simply reads,
\begin{equation}
h^{TT}_{ij} = \frac{2}{r} \sum_{m=0}^{\infty} n_{k_1}n_{k_2}\cdot\cdot\cdot
n_{k_m} {H^{TT}_{ij}}~^{k_1,k_2,,,,,k_m}(t-r),
\label{3-26}
\end{equation} 
here
\begin{equation}
{H_{ij}}~^{k_1,k_2,,,,,k_m} = \frac{2}{m!} \Bigl(\frac{\partial}{\partial t}\Bigr)^m \int T_{ij} x^{k_1} x^{k_2}\cdot \cdot \cdot x^{k_m} d^3 x.
\label{3-27}
\end{equation}

 Now let us take the lowest order ($m=0$) in Eq. (\ref{3-27}),
\begin{equation}
H_{ij} = 2 \int T_{ij}~d^3 x.
\label{3-28}
\end{equation}
Then we introduce the following identity,
\begin{equation}
(T^{\alpha \beta} x^{\mu} x^{\nu})_{, \alpha \beta} = (T^{\mu \beta} x^{\nu} + T^{\nu\beta}x^{\mu})_{,\beta} = 2 T_{\mu \nu}, 
\end{equation}
which can be readily proved using the energy-momentum conservation,
\begin{equation}
{T^{\mu\nu}}_{,\nu} = 0.
\end{equation}
Using the identity, the right hand side of Eq. (\ref{3-28}) becomes,
\begin{eqnarray}
 \int T^{ij}~d^3 x &=& \frac{1}{2}\Bigl(\int d^3 x (T^{00}x^i x^j)_{,00}   
+ 2 \int d^3 x (T^{k0}x^i x^j)_{,{k0}} + \int d^3 x(T^{kl}x^i x^j)_{,{kl}}
 \Bigr) \nonumber \\
&=&  \frac{1}{2}\Bigl(\int d^3 x (T^{00}x^i x^j)_{,{00}}   
+ 2 \int dS_k (T^{k0}x^i x^j)_{,{0}} + \int dS_k (T^{kl}x^i x^j)_{,{l}} \Bigr) \nonumber \\
&=& \frac{1}{2}\int d^3 x (T^{00}x^i x^j)_{,{00}} \nonumber \\
&=& \frac{1}{2}\frac{\partial^2}{\partial t^2} \int d^3x~\rho x^i x^j \equiv I_{ij},
\end{eqnarray}
here we used the Gauss's theorem from the second to the third column and we
 assumed the Newtonian perfect fluid, $T^{00}\sim \rho$. Note that in
 the final column, $I_{ij}$ represents the mass quadrupole moment.
Taking the TT part of $I_{ij}$ using Eq. (\ref{2-85}), one can find the
 so-called reduced mass quadrupole moment,
\begin{equation}
I^{TT}_{ij} = \int d^3 x~\rho(x^i x^j - \frac{1}{3}\delta_{ij}r^2) .
\end{equation}
Finally, we can reach to the quadrupole formula for the gravitational
waves,
\begin{equation}
h^{TT}_{ij}(t,\mbox{\boldmath$x$}) = \frac{2G}{c^4 r}\ddot{I}^{TT}_{ij} (t - \frac{r}{c}), ~~\Bigl(^{\cdot} \equiv \frac{\partial}{\partial t}\Bigr),
\label{SQF}
\end{equation}
where we have recovered $G$ and $c$.

\subsubsection{angular dependence of quadrupole formula}
In this subsection, we discuss the angular dependence of the
gravitational wave seen from the distant observer.

For the purpose, it is convenient to find the non-zero components of 
$h^{TT}_{ij}$ in the spherical coordinate, 
\begin{equation}
x  = r \sin \theta \cos \phi \nonumber, 
\end{equation}
\begin{equation}
y  = r \sin \theta \sin \phi \nonumber ,
\end{equation}
\begin{equation}
z  = r \cos \theta,
\end{equation}
not in the Cartesian coordinate.
We assume that the gravitational wave propagates along the $r$
direction. Due to the TT nature of the gravitational wave, $h_{rr}$,
$h_{r\theta}$, and $h_{r\phi}$ vanish. By performing a simple coordinate
transformation, for example for the $\theta$$\theta$ component,
\begin{equation}
h_{\theta \theta} = h^{TT}_{ij} \frac{\partial x^i}{\partial \theta}\frac{\partial x^j}{\partial \theta},
\end{equation} 
one can find the following nonzero components,
\begin{eqnarray}
h_{\theta\theta} &=& r^2\Bigl[(h^{Q}_{xx} - h^{Q}_{yy})\frac{(\cos^2 \theta + 1)}{4}\cos 2 \phi -  
\frac{h^{Q}_{xx} + h^{Q}_{yy} - 2 h^{Q}_{zz}}{4} \sin^2 \theta + 
 \nonumber \\ & &
h^{Q}_{xy}\frac{\cos^2\theta+1}{2}\sin2\phi-h^{Q}_{xz}\sin \theta\cos\theta\cos\phi-h^{Q}_{yz}\sin\theta\cos\theta\sin\phi\Bigr],
\label{3-54}
\end{eqnarray}
\begin{equation}
h_{\phi\phi} = - h_{\theta\theta}\sin^2 \theta,
\end{equation}
\begin{eqnarray}
\frac{h_{\theta \phi}}{r^2 \sin\theta} &=& - \frac{h^{Q}_{xx} - h^{Q}_{yy}}{2} \cos\theta\sin\phi + h^{Q}_{xy}\cos\theta\cos2\phi \nonumber \\
& &+ h^{Q}_{xz}\sin\theta \sin\phi - h^{Q}_{yz}\sin\theta \cos\phi,
\label{3-56}
\end{eqnarray}
where
\begin{equation}
h^{Q}_{ij} = \frac{2}{r}\ddot{I}^{TT}_{ij},
\end{equation}
represents the second time derivative of the reduced quadrupole moments
in the Cartesian coordinate.

Now let us define the two independent components as follows,
\begin{equation}
h_{+} = \frac{h_{\theta\theta}}{r^2},~~h_{\times} = \frac{h_{\theta\phi}}{r^2 
\sin\theta}.
\label{3-65}
\end{equation}
Then we perform a simple estimation for the gravitational wave emitted
from the rotating star in axisymmetry.

Given the density distribution of the rotating star of $\rho(R,Z)$ in
the cylindrical coordinate,
the quadrupole moments are
\begin{equation}
I_{xx} = \int \rho(R,Z)~R^2 \cos^2\phi~R~dR~dz~d\phi = \pi \int \rho(R,Z)~R^3 ~dR,
\end{equation}
\begin{equation}
I_{yy} = \int \rho(R,Z)~R^2 \sin^2\phi~R~dR~dz~d\phi = I_{xx},
\end{equation} 
\begin{equation}
I_{xy} = \int \rho(R,Z)~R^2 \sin\phi~\cos\phi~R~dR~dz~d\phi = 0,
\end{equation}
furthermore assuming the equatorial symmetry of the rotating star, the
remaining components are,
\begin{equation}
I_{zz} =  \int \rho(R,Z)~z^2 ~R~dR~dz~d\phi = 2 \pi \int \rho(R,Z)~z^2~R ~dR,
\end{equation}
\begin{equation}
I_{xz}=I_{yz}=0.
\end{equation}
As easily understood, if the star rotates stationally, no gravitational waves are
emitted because $\ddot{I}_{ij} =0$. The gravitational waves are emitted
when the rotating star contracts or expands dynamically, because the
time derivatives of the quadrupole moments have non-zero values. It
should be noted that the gravitational waves can be emitted from the
``axisymmetrically'' rotating stars, when the motion is dynamically
changing. 

From Eq. (\ref{3-54}) and (\ref{3-56}) with Eq. (\ref{3-65}), 
the gravitational waves are
\begin{equation}
h_{+} = - \frac{1}{r}(\ddot{I}_{xx} - \ddot{I}_{zz})\sin^2\theta,
\label{3-65}
\end{equation}
\begin{equation}
h_{\times} = 0.
\end{equation}
The gravitational waves are most strongly emitted in the direction
perpendicular to the rotational axis (Eq. (\ref{3-65})). Intuitively, it is natural because
the dynamical behavior of the rotating star can be seen most
drastically for the observer in the direction perpendicular to the
pole. On the
contrary, gravitational waves from the merging neutron stars are most
strongly emitted in the direction of the rotational axis.  
\subsubsection{quadrupole formula for supernovae}
To end this section, we introduce the quadrupole formula in a useful
form, which is often used for the computation of gravitational wave from core-collapse
supernovae.

First of all, let us define the tensor $f^{lm}_{ij}$,
\begin{equation}
\label{flm} 
f^{lm}_{ij} = r^2
\left(
                     \begin{array}{ccc}
                      0 & 0 & 0 \\
                      0 & W_{lm} & X_{lm} \\
		      0 & X_{lm} & -\sin^2 \theta W_{lm}
                     \end{array}
                    \right),
\end{equation}
with 
\begin{equation}
X_{lm} = 2 \frac{\partial}{\partial \phi}\Bigl(
\frac{\partial}{\partial \theta} - \cot \theta \Bigr) Y_{lm}(\theta,\phi),
\end{equation}
\begin{equation}
W_{lm} = \Bigl(\frac{\partial^2}{\partial \theta^2} - \cot \theta 
\frac{\partial}{\partial \theta} - \frac{1}{\sin^2 \theta}\frac{\partial^2}
{\partial \phi^2}\Bigr)Y_{lm}(\theta,\phi),
\end{equation}
where $Y_{lm}$ is the spherical harmonics.

After tedious calculations, one can check that $h^{TT}_{ij}$ can be
expressed by using $f^{lm}_{ij}$,
\begin{eqnarray}
h^{TT}_{ij} &=& \sqrt{\frac{32\pi}{15}}\frac{1}{8}(h^{Q}_{xx} - h^{Q}_{yy})
{\rm Re}(f^{22}_{ij}(\theta,\phi)) - \sqrt{\frac{16\pi}{3}}\frac{1}{24}
(h^{Q}_{xx} + h^{Q}_{yy} - 2 h^{Q}_{zz})
{\rm Re}(f^{20}_{ij}(\theta,\phi)) \nonumber \\ 
& & \sqrt{\frac{32\pi}{15}}\frac{1}{4}h^{Q}_{xy} 
{\rm Im}(f^{22}_{ij}(\theta,\phi)) - \sqrt{\frac{8\pi}{15}}\frac{1}{2}h^{Q}_{xz} 
{\rm Re}(f^{21}_{ij}(\theta,\phi))  \nonumber \\
& & - \sqrt{\frac{8\pi}{15}}\frac{1}{2}h^{Q}_{yz} 
{\rm Re}(f^{21}_{ij}(\theta,\phi)). 
\end{eqnarray}
More compactly, the above equation may be written,
\begin{equation}
h^{TT}_{ij} = \sum_{m=-2}^{m=2}A_{2m}f^{2m}_{ij}.
\end{equation}
Although we have so far considered the gravitational wave up to the
quadrupole, the gravitational wave in all order is shown to be
expressed by the multiple expansions in the following form,
\begin{equation}
h^{TT}_{ij} = \sum_{l=2}^{\infty}\sum_{m=-l}^{l}\frac{1}{r}
\Bigl(\frac{d^l}{dt^l}I^{lm}(t-r)f^{lm}_{ij} + \frac{d^l}{dt^l}S^{lm}(t-r)d^{lm}_{ij}\Bigr),
\end{equation}
where 
\begin{equation}
I^{lm} = \frac{16 \pi}{(2l+1)!!}\Bigl(\frac{(l+1)(l+2)}{2(l-1)l}\Bigr)^{1/2} 
\int T_{00} Y^{lm*} r^{l}~d^3 x,
\label{Ilm} 
\end{equation}
represents the mass quadrupole, 
\begin{equation}
S^{lm} = - \frac{32 \pi}{(2l+1)!!}\Bigl(\frac{(l+2)(2l+1)}{2(l-1)(l+1)}\Bigr)^{1/2} \int \epsilon_{jpq} x_{q} (-T_{0q}) Y_j^{l-1,lm*} r^{l-1}~d^3 x
\end{equation}
represents the mass-current quadrupole with  $Y_j^{l-1,lm*}$ being the
pure orbital spherical harmonics, and  $f^{lm}_{ij},d^{lm}_{ij}$
are the pure-spin harmonics (see \cite{thorne80} for a complete description). Note that $n!! =
n\cdot(n-2)\cdot\cdot\cdot 1$.

In case of axisymmetry ($m=0$), the gravitational wave up to the
quadrupole ($l=2$) becomes,
\begin{equation}
h^{TT}_{ij} = \frac{1}{r}
\frac{d^2}{dt^2}I^{20}(t-r)f^{20}_{ij}.
\label{final-3}
\end{equation}
Noting in Eq. (\ref{Ilm}) that $Y_{20} = \sqrt{\frac{5}{16\pi}}(3 \cos^2
\theta - 1)$ and $T_{00}= \rho$, Eq. (\ref{Ilm}) reads
\begin{equation}
I^{20} = \frac{G}{c^4}\frac{32 \pi^{3/2}}{\sqrt{15}} \int_{0}^{1} d\mu
\int_{0}^{\infty} dr \rho(\frac{3}{2}\mu^2 - \frac{1}{2})r^4,
\end{equation}
where $\mu = \cos \theta$.
And $f^{20}_{ij}$ becomes,
\begin{equation}
\label{flm} 
f^{20}_{ij} = \frac{1}{8}\sqrt{\frac{15}{\pi}}
\left(
                     \begin{array}{ccc}
                      0 & 0 & 0 \\
                      0 & \sin^2\theta & 0 \\
		      0 & 0 & -\sin^2 \theta \\
                     \end{array}
                    \right).
\end{equation}
Finally one can obtain the nonvanishing component in the following, 
\begin{equation}
h_{+} \equiv h^{TT}_{\theta\theta} = - h^{TT}_{\phi\phi} = \frac{1}{8}\sqrt{\frac{15}{\pi}} 
\sin^2 \theta \frac{A^{E2}_{20}}{r}, 
\label{101}
\end{equation}
where
\begin{equation}
A^{E2}_{20} \equiv \frac{d^2}{dt^2} I^{20}.
\label{A}
\end{equation}
It is noted that the numerical treatment of the second time derivatives
in Eq. (\ref{A}) are
formidable. Using the hydrodynamic equations of perfect fluid, the time
derivatives can be eliminated. For example, we take the first time
derivative of $A^{E2}_{20}$,
\begin{equation}
N^{E2}_{20} \equiv  \frac{\partial}{\partial t}A^{E2}_{20} = \frac{G}{c^4}\frac{32 \pi^{3/2}}{\sqrt{15}} \int_{0}^{1} d\mu
\int_{0}^{\infty} dr \Bigl[\frac{\partial}{\partial t}\rho\Bigr](\frac{3}{2}\mu^2 - \frac{1}{2})r^4.
\label{NE2}
\end{equation}
Using the equation of mass conservation expressed in the spherical coordinates,
\begin{eqnarray}
0 &=&\frac{\partial \rho}{\partial t} + \frac{\partial}{\partial x^{k}}(\rho v_{k}) = \frac{\partial \rho}{\partial t} + \frac{1}{r^2}\frac{\partial}{\partial r}(r^2 \rho v_{r}) +  
\frac{1}{r\sin \theta}\frac{\partial}{\partial \theta}
(\sin \theta \rho v_{\theta}) \nonumber \\
&=& \frac{\partial \rho}{\partial t} + \frac{1}{r^2}\frac{\partial}{\partial r}(r^2 \rho v_{r}) - 
\frac{1}{r}\frac{\partial}{\partial \mu}
(\sqrt{1- \mu^2}\rho v_{\theta}),
\end{eqnarray}
with $\mu = \cos \theta$ and introducing $\frac{\partial \rho}{\partial t}$ to Eq. (\ref{NE2})
yields,
\begin{equation}
N^{E2}_{20} = \frac{G}{c^4} \frac{32 \pi^{3/2}}{\sqrt{15 }} \int_{0}^{1} 
\int_{0}^{\infty}  r^3 \,dr \,d\mu \,\rho [ {v_r} ( 3 \mu^2 -1) - 3 v_{\theta} \mu \sqrt{1-\mu^2}].
\end{equation}

In this way taking one more time derivative using the Euler equations,
the well known form of $A^{E2}_{20}$ in the literature can be obtained,
\begin{eqnarray}
 A_{20}^{\rm{E} 2} &=& \frac{G}{c^4} \frac{32 \pi^{3/2}}{\sqrt{15 }} \int_{0}^{1} 
\int_{0}^{\infty}  r^2 \,dr \,d\mu \,\rho [ {v_r}^2 ( 3 \mu^2 -1) + {v_{\theta}}^2 ( 2 - 3 \mu^2)
 - {v_{\phi}}^{2} \nonumber \\ 
& & - 6 v_{r} v_{\theta} \,\mu \sqrt{1-\mu^2} 
 - r \partial_{r} \Phi (3 \mu^2 -1) + 3 \partial_{\theta} \Phi \,\mu
\sqrt{1-\mu^2}], 
\label{103}
\end{eqnarray}
where $ \partial_r = \partial/ \partial r,\,\,\partial_{\theta} =
\partial/ \partial \theta$.
%% koko
With Eq. (\ref{101}) and Eq. (\ref{103}), one can extract the
gravitational waves from the numerical simulations assuming axisymmetry.

Now that we have mentioned the physical foundations, we move now on to 
the discussion of the gravitational waves in core-collapse supernovae 
from the next subsection. The readers, who are interested in 
the detection techniques of gravitational waves by interferometric
detectors, please see, for example \cite{hugh} for a review. 
\clearpage

\subsection{Gravitational waves at core bounce }
If the gravitational collapse of the supernova core proceeds 
spherically, no gravitational waves can be emitted. 
The gravitational core-collapse should proceed
 aspherically and dynamically for the emissions of the gravitational waves.  

As mentioned in subsection \ref{roleofrotation}, stars are generally rotating. 
This stellar rotation has been long supposed to play an
important role in the gravitational waves from core collapse
supernovae. The large-scale
asphericities at core bounce induced by rotation 
can convert the part of the gravitational energy 
into the form of the gravitational waves. In this section, we review the 
gravitational waves emitted at core bounce in rotating supernovae
 (see also \cite{new} for a review).
%In the later phases after core bounce, two other sources have been 
%considered to be the sources of gravitational waves, namely convective
%motions and anisotropic neutrino emission in the core. We review and
%discuss them in subsection \ref{aniso_grav}.

\subsubsection{characteristic properties  \label{gw_bounce}}

First of all, we make an order-of-magnitude estimate of the
amplitude and frequency of the gravitational waves emitted at core bounce for
later convenience.

A characteristic amplitude of gravitational waves at core bounce can be approximately 
estimated with the help of the standard quadrupole formula (see Eq. (\ref{SQF}), for example, \cite{shapiro}) as follows,
\begin{eqnarray}
h &=& \frac{2G}{c^4 D} \ddot{I_{ij}} \sim \frac{2G}{c^4 D} \frac{M R^2}{T_{\rm dyn}^2} \epsilon \nonumber \\
&\sim& 
\frac{300 {\rm cm}}{D} \epsilon \Bigl(\frac{M}{M_{\odot}}\Bigr)
 \Bigl(\frac{R}{10~{\rm km}}\Bigr)^2 \Bigl(\frac{T_{\rm dyn}}{1~{\rm ms}}\Bigr)^{-2} 
\nonumber \\ &\sim& 
10^{-20}\epsilon \Bigl(\frac{10~{\rm kpc}}{D}\Bigr) \Bigl(\frac{R}{10~{\rm km}}\Bigr)^2 \Bigl(\frac{T_{\rm dyn}}{1~{\rm ms}}\Bigr)^{-2},
\label{simple_estimate}
\end{eqnarray}
where $D$ is the distance to the source, $\ddot{I_{ij}}$ is the second
time derivative of the quadrupole moment of $I_{ij}$, $M$, $R$, and
$T_{\rm dyn}$
represents the typical mass and radius of the inner core and the
dynamical timescale at core bounce, respectively. We assume that the
supernova occurs at our galactic center at the distance of $10~{\rm
kpc}$. $\epsilon$ is a parameter, representing the degree of the
nonsphericity as well as the degree of compaction, which
may be optimistically estimated to be the order of 10 $\%$ in rapidly
rotating supernova cores.

In addition, the typical frequency of the gravitational waves, which can
be approximately estimated by the inverse of the dynamical timescale, is expected to be
in the following range, 
\begin{equation}
\nu_{\rm GW} \sim \frac{1}{T_{\rm dyn}} \simeq O(100~{\rm Hz}) \sim  1~{\rm kHz}. 
\end{equation}
We will later see that these values have the right order of magnitude.
In the following, we review the study of gravitational wave at core bounce 
in rotating core-collapse supernovae.

\subsubsection{waveforms in rotating core-collapse supernovae}
Realistic progenitor models \cite{heger00,heger04,ww:95} (in section \ref{pre}), which 
current researchers in this area can obtain, had been hard to access for those in the early studies. 
Some were forced to assume the collapse of oblate spheroids with
 pressureless dust \cite{thuan}, others tried to include the effect of the
 internal pressure additionally, by which closer situations at core
 bounce were examined \cite{shapiro1,shapiro2,shapiro3}. 
Although the obtained waveforms and the
 amplitudes of the GWs are quantitatively different from the ones 
in the current numerical studies, these pioneering studies were
 valuable in the sense that they constructed the formalism of
 the standard quadrupole formula still used by current studies 
\cite{turner1979}, and that 
they obtained the qualitative understanding of the effect of initial
 angular momentum of the stars and the stiffness of equations of state
 on the waveforms near core bounce \cite{shapiro3}. 

\begin{figure}[H]
\begin{center}
\epsfxsize=5cm
\epsfbox{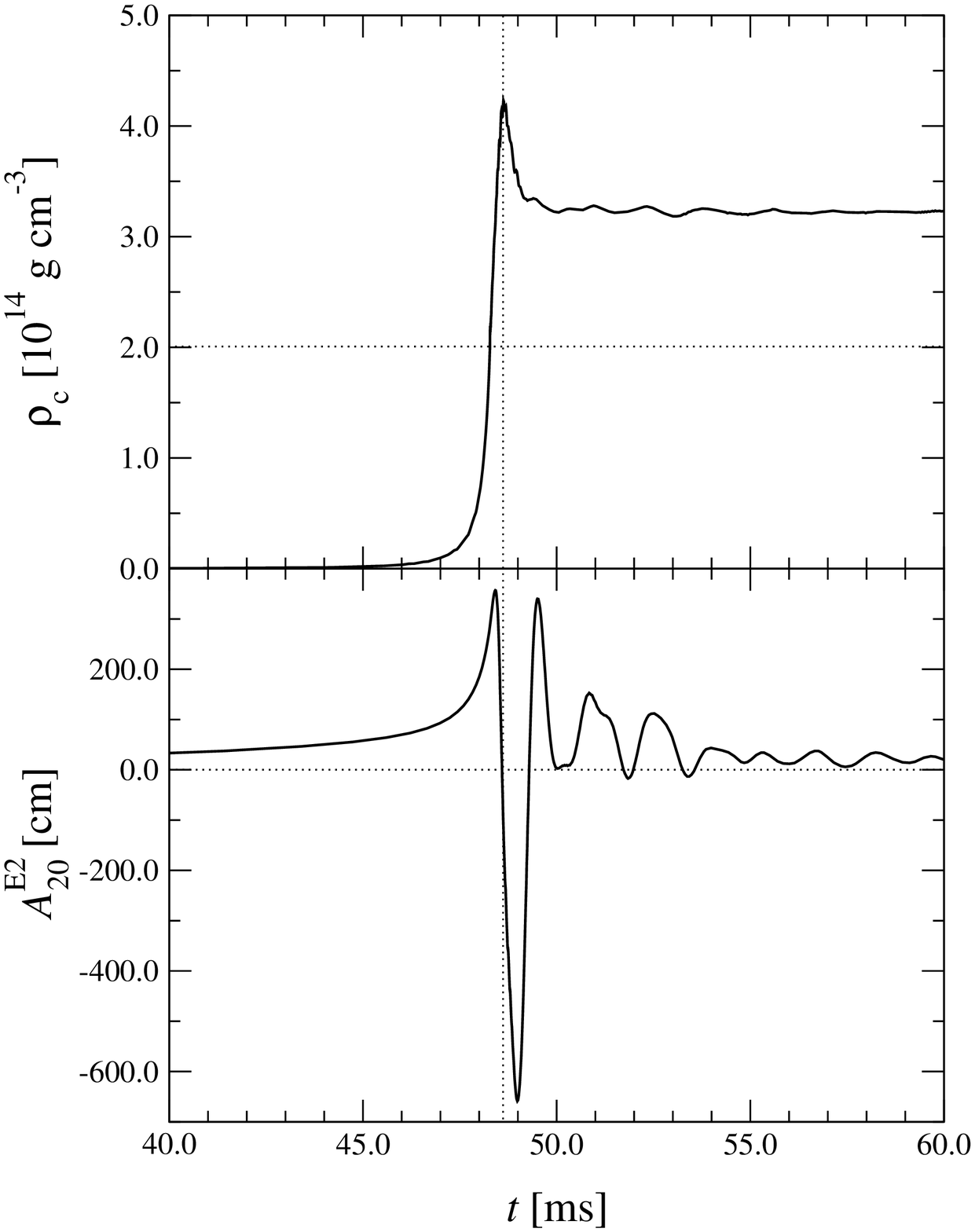}
\epsfxsize=5cm
\epsfbox{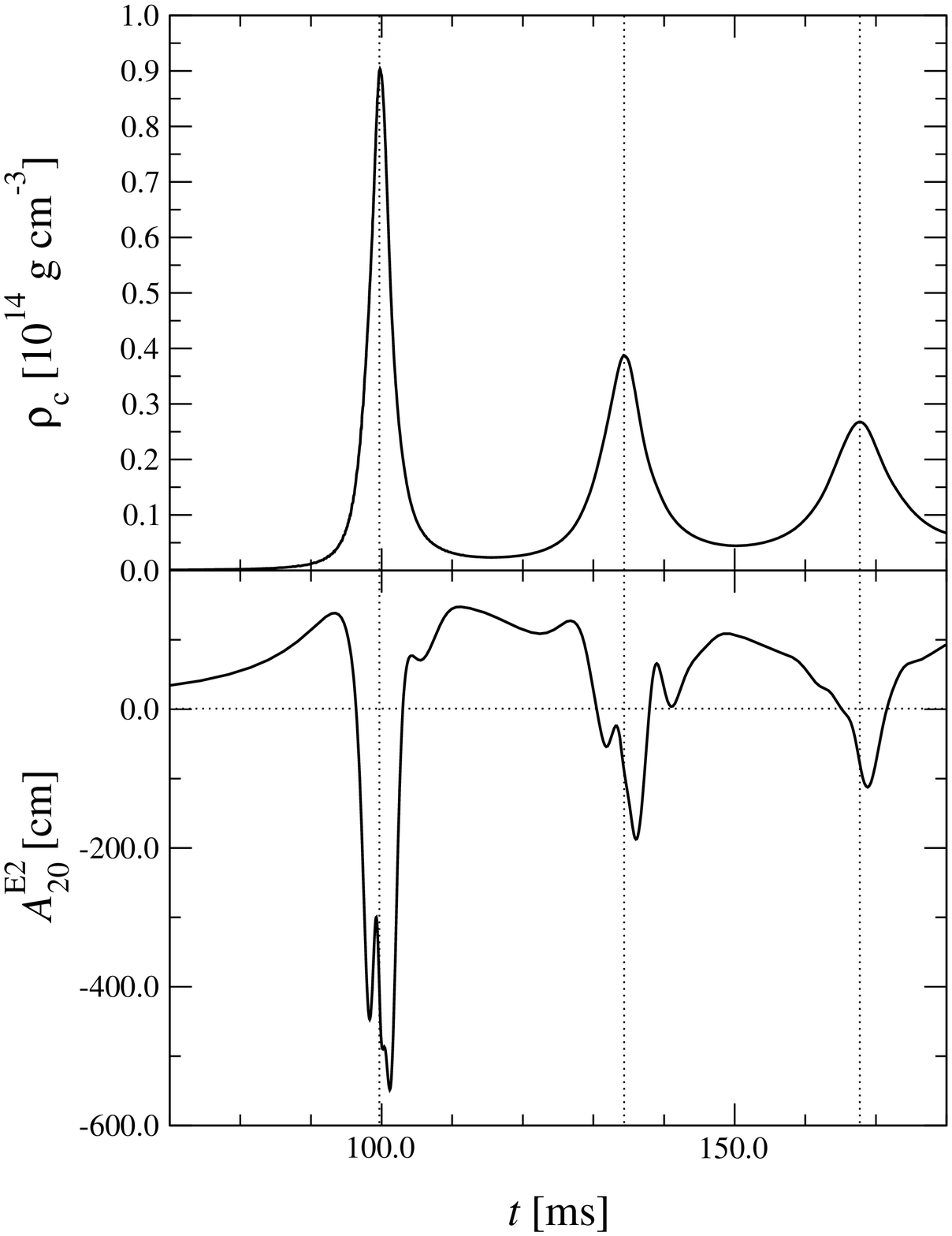}
\epsfxsize=5cm
\epsfbox{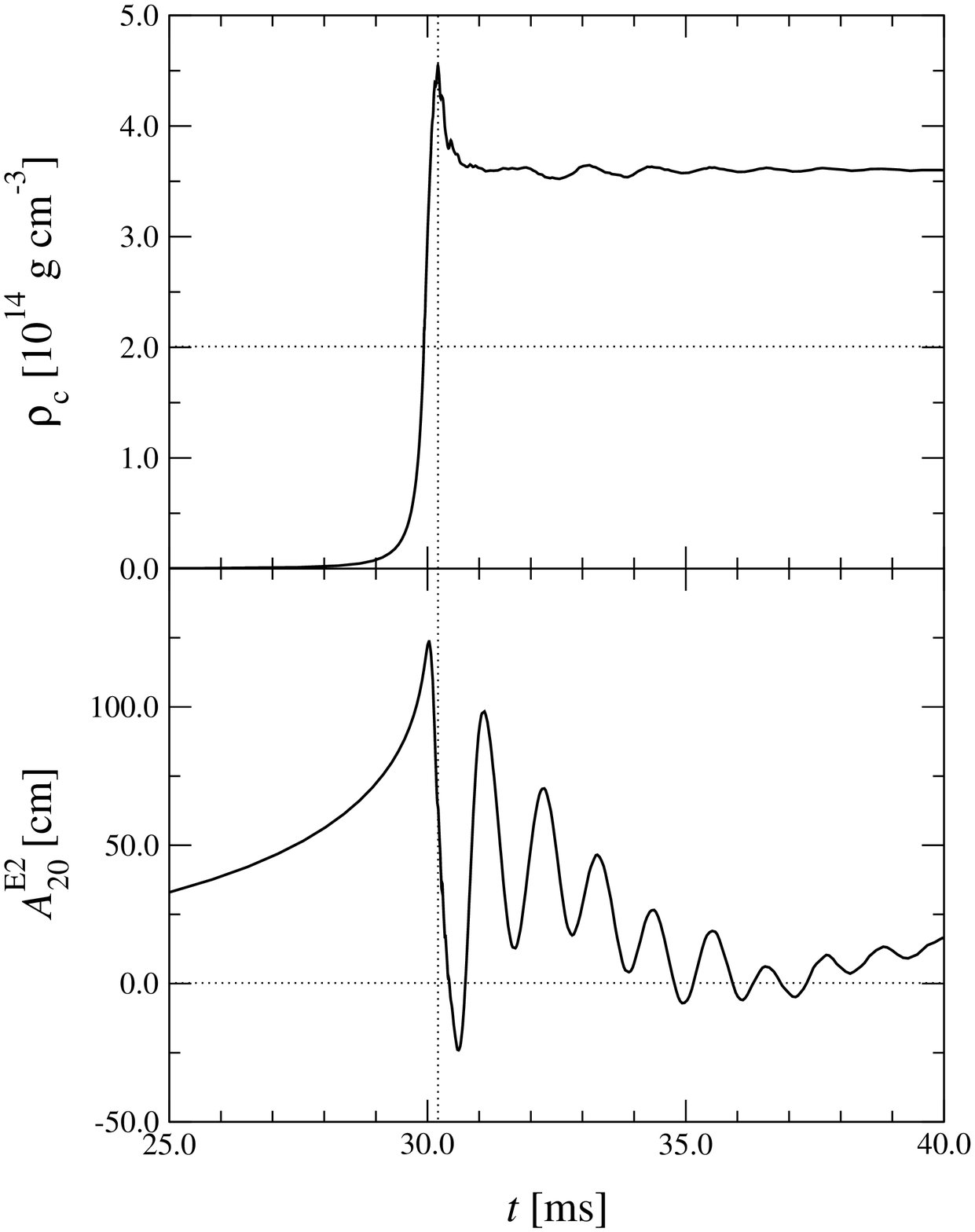}
\end{center}
\caption{Typical waveforms of gravitational waves (bottom) with
 the time evolution of the central densities (top) panels taken from
 \cite{dimmel}. Bottom left, middle, right panels correspond to 
type I, II, and III waveforms, respectively. The vertical dotted lines 
represent the epoch of core bounce.
The peak spike and the subsequent spikes represent the gravitational
 waves emitted at core bounce and at the oscillation of the inner core 
produced by its inertia after core bounce.
}
\label{ZM}
\end{figure}     

In the early 1980's, E. M\"{u}ller in the Max-Planck Institute for
Astrophysics (MPA) performed two-dimensional (2D) axisymmetric core 
collapse simulations with better iron core models and
calculated the quadrupole GW emission  \cite{mueller1982}. 
As for the microphysics, he employed a finite-temperature 
equation of state, however excluded the treatment of neutrino energy
loss for simplicity. 
Due to the poor computational intensity at that time,  
only a small set of models could be investigated. However,
it was found that differential rotation enhanced the efficiency of the
GW emission. Afterward, developing the 2D hydrodynamic code employed in
\cite{mueller1982},  
M\"{o}nchmeyer {\it et al} accounted for electron capture and treated neutrino transfer
 by making use of a
leakage scheme for simplicity \cite{moenchgw}.  By computing four
models changing the initial angular momentum parametrically, 
they categorized the obtained shapes of the waveforms 
into two distinct classes. The waveform categorized as Type I 
is distinguished by a large amplitude peak at core bounce and
subsequent damping ring-down oscillations. This waveform is obtained 
when the initial angular momentum is small, which leads to core bounce
near at nuclear density. A typical waveform for a Type I is shown in the
left panel of Figure \ref{ZM}. They also found the waveform identified 
as Type II, which shows a several distinct peaks caused by multiple
bounce (see the middle panel of Figure \ref{ZM} for an example of a Type
II waveform). From their study, it was found that the gravitational signals at
core bounce are largest with amplitudes less than $\sim 10^{-20}$ for a
source at the distance of 10 kpc in the frequency range of $5\times 10^2
- 10^3$ Hz. One may find these values are roughly in agreement with the
one obtained in the simple order-of-the-magnitude estimation in Eq. 
(\ref{simple_estimate}). Note here that the waveforms shown in Figure \ref{ZM} are from
the study of Dimmelmeier {\it et al} \cite{dimmel} discussed later. 

Since a clear criteria determining the types of the waveforms might not
be obtained by the study of M\"{o}nchmeyer {\it et al}. due to their limited
models, Zwerger {\it et al.} simulated the collapse of
a large number of the initial models (78 models) with varying amounts of 
the initial rotation rates, the degree of differential rotation, and the
stiffness of equation of state \cite{zweg}. In order to make this large survey
possible, they employed a simplified equation of state and did not take
into account electron capture and neutrino transport. Their initial
models were constructed in a rotational equilibrium by the method of
\cite{eriguchi}. In contrast, all the preceding
studies, with the exception of \cite{bonazzola}, 
constructed the initial models just by adding the angular 
momentum to the spherically symmetric progenitor models by hand. 
In the study by \cite{zweg}, the
rotational equilibrium was produced by a polytropic equation of state
with the initial adiabatic index of $\Gamma_r = 4/3$, and then, the core-collapse was 
initiated by dropping the adiabatic index down to 4/3, varing the values
from $\Gamma_r = $1.28 to 1.325. In addition to the cold part, the
equation of state  consists of the thermal part and 
the stiff part in order to take into
account the shock heating and the repulsive action of nuclear forces, 
respectively. 

With these computations, they found that the type of the
waveforms was mainly determined by the stiffness of the cold part of 
equation of state, $\Gamma_r$. The type I and II was obtained for the
models with relatively softer ($\Gamma_r \leq \sim 1.31$) and 
stiffer equations of state ($1.32 \leq \sim \Gamma_r $),
respectively. In addition, they found a smooth transition from type II
to type I while fixing other parameters, such as the initial rotation rate
and the degree of differential rotation. They explained the
 cause of the transition as follows. 
As the value of $\Gamma_r$ becomes smaller, the
core-collapse is enhanced. This results in the core bounce at the higher
density. Since the typical interval between the multiple bounces should be an
order of the dynamical timescale $t_{\rm dyn} \sim 1/ \sqrt{G\rho}$, the 
higher density at bounce results in a shorter interval between the
subsequent bounces. This makes the transition to type II to type I.
As for the degree of the differential rotation, they did not find a
large effect on the transition of the waveforms. In addition to the
above waveforms, it is noted that they observed an another class of the
waveform, the so-called type III (the right panel of Figure \ref{ZM})
for their models with the extremely lower values of $\Gamma_r = 1.28$.
Between the initial models constructed in rotational equilibrium and
those not in rotational equilibrium, they found no significant changes in the waveforms.
Employing the extensive sets of the initial models, they pointed out that 
the maximum amplitudes of the GWs were in the range of $4\times 10^{-22}
\leq \sim h \leq \sim 4 \times 10^{-20}$ for a source at the distance
of 10 kpc with the typical frequencies between 500 to 1000 Hz.

More recently, Kotake {\it et al} (2003) 
calculated the waveforms by performing 2D rotational core-collapse
simulations, in which they employed
 a realistic equation of state (EOS) and took into account electron
 captures and neutrino transport by the so-called leakage scheme 
\cite{kotakegw,kotakegwMHD}. Employing the-state-of-the-art equation of
state, it was found that the typical frequencies and the
amplitudes of the gravitational waves are consistent with the previous
studies qualitatively and quantitatively. 
Furthermore they pointed out
the importance of detecting the second peaks of the gravitational waves,
because they will give us the information as to the angular momentum
distribution of evolved massive stars. This is explained below.

\begin{figure}
\begin{center}
\epsfxsize=7.5cm
\epsfbox{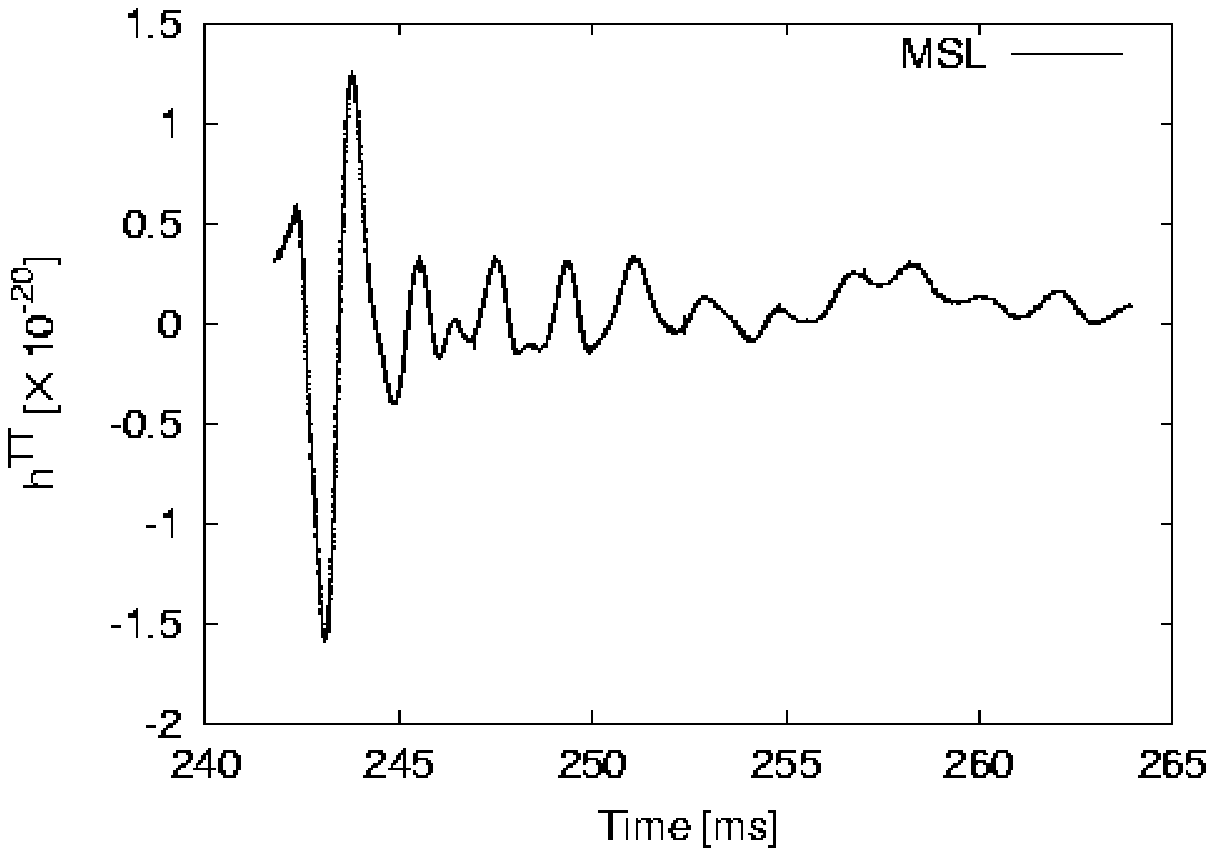}
\epsfxsize=7.5cm
\epsfbox{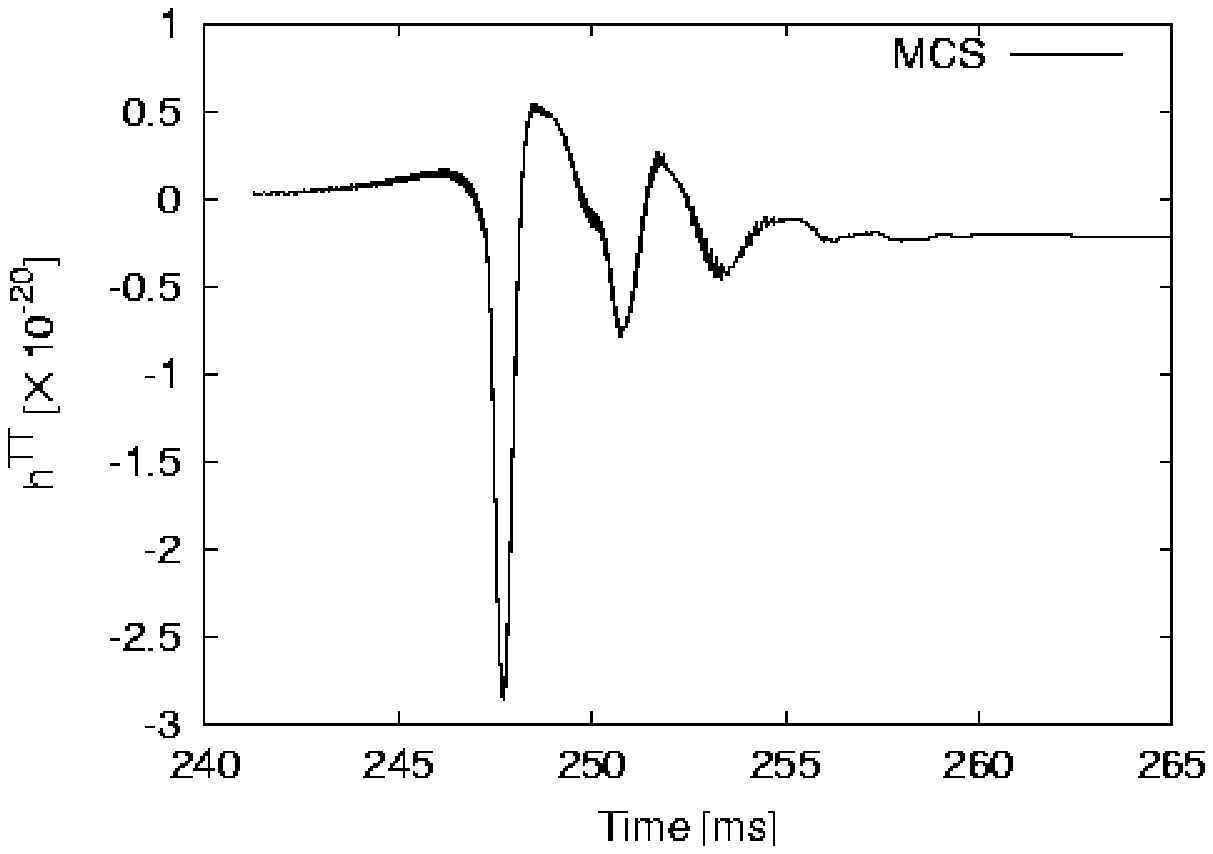}
\end{center}
\caption{Time evolutions of the amplitude of gravitational wave for
 the representative models taken from \cite{kotakegw}. Note that the distance of the source is
 assumed to be located at the distance of 10 kpc. \label{fig5_gw}}
\end{figure}

\begin{figure}
\begin{center}
\epsfxsize=12.5cm
\epsfbox{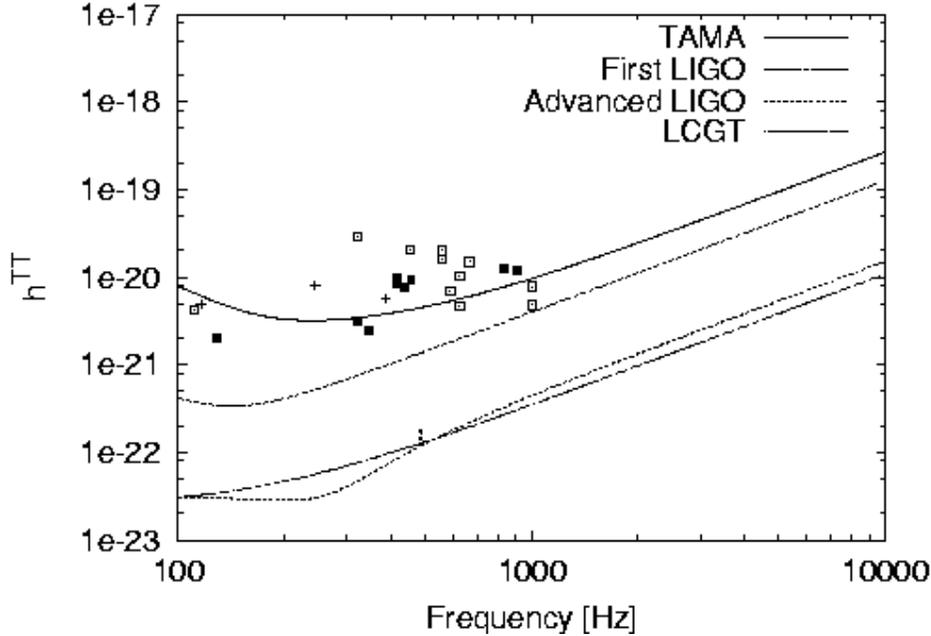}
\end{center}
\caption{Detection limits of TAMA \cite{tama}, first LIGO
 \cite{firstligo}, advanced LIGO \cite{advancedligo}, and LCGT \cite{lcgt} with
 the amplitudes from numerical simulations \cite{kotakegw}. The open squares
 represent the maximum amplitudes for all the models, while 
 the pluses and the closed squares stand for the amplitudes of the second
 peaks for the models with strong differential rotation and cylindrical
 rotation law (negative sign of the peak) and for the other models
 (positive sign of the peak), respectively.  
Note that the source is assumed
 to be located at the distance of 10 kpc. \label{fig9}}
\end{figure}

The waveform for a Heger's $15M_{\odot}$ rotational progenitor model 
studied in \cite{kotakegw} 
is given in the left panel of Figure \ref{fig5_gw}, which can be
categorized into the type I waveform. 
While in the right panel of Figure \ref{fig5_gw}, 
the waveform (type II) is given for the model, which has a cylindrical 
rotation law with the strong differential rotation, in which the initial
angular velocity is assumed to yield to a quadratic cutoff at 100 km radius.
Comparing the panels, one can see that the sign of the second peak is
 negative in the left panel, while it is positive
 for the model in the right panel. Note that
the second peak is defined to be the place where the absolute amplitude 
is the second largest. 
They found that this characteristics 
that the sign of the second peak is negative is common to the models 
with the strong differential rotation and the cylindrical rotation law. 
The absolute amplitudes of the peak and the second peak studied are shown 
in Figure \ref{fig9}. In addition to the first peak, the second peaks
are also shown to be 
within the detection limit of the first LIGO for a source within 10
kpc. It seems quite possible 
for the detectors of next generation such as the advanced LIGO and LCGT
to detect the difference of the sign.  
Therefore, it may be possible to obtain in this way 
the otherwise  inaccessible information about the angular momentum 
distribution of evolved massive stars.

%The maximum amplitudes for all the models studied in Kotake {\it et al.}
%(2003) range  $5\times 10^{-21} \leq
%h^{\rm TT} \leq 3\times 10^{-20}$ for a source at the distance of 10 kpc.
%This is almost the same as obtained in the preceding rotational
%core-collapse simulations.

\begin{figure}
\begin{center}
\epsfxsize = 12.5 cm
\epsfbox{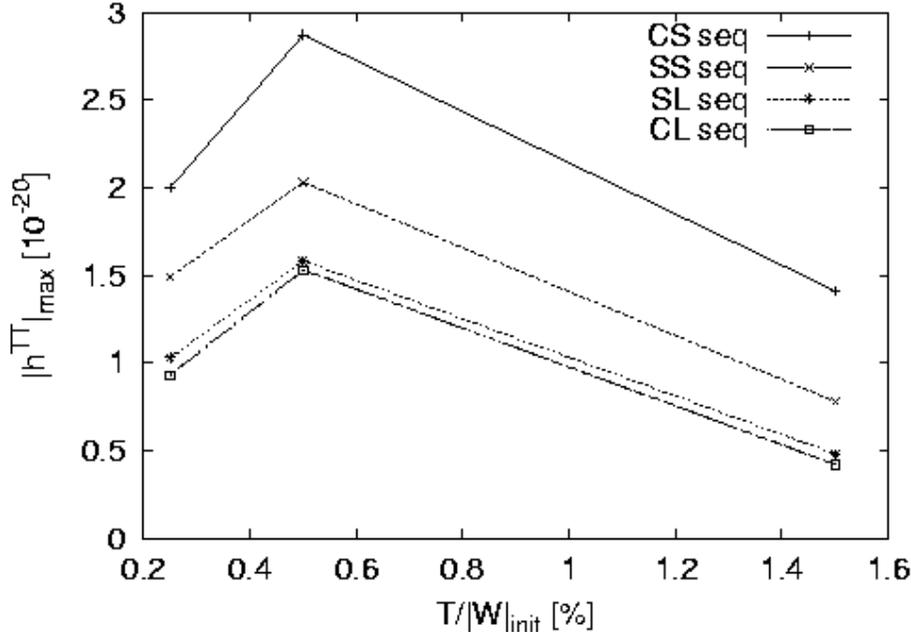}
\caption{Relations between $T/|W|_{\rm init}$ and the peak amplitude
 $|h^{\rm TT}|_{\rm max}$ for all the models \cite{kotakegw}.
In the figure, ``CS, SS, SL, CL seq'' represent the model sequences whose
 initial rotation profiles differ (see \cite{kotakegw} for details).
Note that the distance of the source is 
 assumed to be located at the distance of 10 kpc. }
\label{fig6}
\end{center}
\end{figure}

They also discussed the relation between the maximum amplitudes of
gravitational wave and the initial $T/|W|$, which is the ratio of the
rotational to the gravitational energy. 
From Figure \ref{fig6}, it was found that the largest amplitude is obtained 
for the moderate initial rotation rate (i.e., $T/|W|_{\rm init} = 0.5
\%$) when one fixes the initial rotation law and the degree of differential
rotation. This is understood as follows. The amplitude of gravitational
wave is roughly proportional to the inverse square of the typical
dynamical scale, $t_{\rm dyn}$. Since  
$t_{\rm dyn}$ is proportional to the inverse square root of the central
density $\rho$, the amplitude is proportional to the density. As a
result, the amplitude becomes smaller as the initial rotation rates
become larger because the density decreases then. 
On the other hand, the amplitude is proportional to the 
quadrupole moment, which becomes larger in turn as the
total angular momentum increases. This is because stronger
centrifugal forces not only make the mass of the inner core larger, but
also deform it. The amplitude of gravitational wave is determined by the
competition of these factors. As a result, the amplitudes is found to become
maximal for moderate initial rotation rates. This is also noticed by the
earlier work by Yamada and Sato (1995) \cite{ys}.
 Combining the result by Finn \cite{finn}, who 
pointed out by a perturbation technique that 
the amplitudes of the GWs are proportional to the square of
the angular momentum $h \sim J^2$ in slowly rotating cases, 
the relation between the initial angular momentum and the peak
amplitudes up to the rapidly rotating cases could be understood.

\subsubsection{effects of magnetic fields}
In addition to rotation discussed so far, 
Kotake {\it et al} (2004) investigated the effect of magnetic fields on
the gravitational signals \cite{kotakegwMHD}.

They extended the quadrupole formulae (Eq. (\ref{SQF})) in a form
including the contributions from the electromagnetic fields. 
To begin with, $I^{20}$ in Eq. (\ref{A}) should be replaced by
$M_{20}^{\rm{E} 2}$ as follows, 
\begin{equation}
A_{20}^{\rm{E} 2} = \frac{d^2}{dt^2} M_{20}^{\rm{E} 2}, 
\end{equation}
where the mass quadrupole formula is given as  
\begin{equation}
M_{20}^{\rm{E} 2} = \frac{G}{c^4}\frac{32 \pi^{3/2}}{\sqrt{15}}\int_{0}^1 d\mu
\int_{0}^{\infty} dr~\rho_{*} \Bigl(\frac{3}{2} \mu^2 - \frac{1}{2}\Bigr) r^4, 
\end{equation}
where $\rho_{*}$ represents the total energy density including the
contribution from the magnetic field,
\begin{equation}
\rho_{*} = \rho + \frac{B^2}{8 \pi c^2}.
\label{modrho}
\end{equation}
By a straightforward, however tedious, calculation to replace the time
derivatives by the spatial derivatives applying the continuity equation, 
the equation of motion, and the
induction equation, 
\begin{equation}
\frac{\partial \mbox{\boldmath$B$}}{\partial t} = \nabla \times ( \mbox{\boldmath$v$}\times \mbox{\boldmath$B$}),
\end{equation}
noting the divergence-free constraint
($\nabla \cdot \mbox{\boldmath$B$} = 0$),  $A_{20}^{\rm{E} 2}$ can be
transformed into the following form,
\begin{equation}
 A_{20}^{\rm{E} 2} \equiv  {A_{20}^{\rm{E} 2}}_{\rm, quad} 
+  {A_{20}^{\rm{E} 2}}_{\rm ,Mag},
\end{equation}
where ${A_{20}^{\rm{E} 2}}_{\rm, quad}$ is the contribution from the matter:
\begin{eqnarray}
 {A_{20}^{\rm{E} 2}}_{\rm,quad} &=& \frac{G}{c^4} \frac{32 \pi^{3/2}}{\sqrt{15 }} 
\Biggl(\int_{0}^{1}d\mu 
\int_{0}^{\infty}  r^2  \,dr \,\rho [ {v_r}^2 ( 3 \mu^2 -1) + {v_{\theta}}^2 ( 2 - 3 \mu^2)
 - {v_{\phi}}^{2} - 6 v_{r} v_{\theta} \nonumber \\
& & \,\mu \sqrt{1-\mu^2}
- r \partial_{r} \Phi (3 \mu^2 -1) + 3 \partial_{\theta} \Phi \,\mu
\sqrt{1-\mu^2}]
 - \nonumber\\
& & \int_{0}^{1} d\mu
\int_{0}^{\infty}  r^3  \,dr [q_{r}(3\mu^2 -1) - 3~q_{\theta}~\mu 
\sqrt{1 - \mu^2}]
\Biggr), 
\label{quad}
\end{eqnarray}
%${A_{20}^{\rm{E} 2}}_{\rm, AV}$ is the contribution from the viscosity:
% \begin{eqnarray}
%{A_{20}^{\rm{E} 2}}_{\rm ,AV}= \frac{G}{c^4} \frac{32 \pi^{3/2}}{\sqrt{15 }} \%int_{0}^{1} d\mu
%\int_{0}^{\infty}  r^3  \,dr [q_{r}(3\mu^2 -1) - 3~q_{\theta}~\mu 
%\sqrt{1 - \mu^2}],
%\end{eqnarray}
${A_{20}^{\rm{E} 2}}_{\rm Mag} \equiv {A_{20}^{\rm{E} 2}}_{j \times B} +
{A_{20}^{\rm{E} 2}}_{\rho_{\rm m}}$ is the contribution from the magnetic field:
\begin{eqnarray}
{A_{20}^{\rm{E} 2}}_{j\times B} &=& 
\frac{G}{c^4} \frac{32 \pi^{3/2}}{\sqrt{15}} 
\int_{0}^{1} d\mu \int_{0}^{\infty} r^3 \,dr  \Bigl[(3 \mu^2 - 1)~ 
\frac{1}{c}~(\mbox{\boldmath$j$}\times \mbox{\boldmath$B$})_{r} - 
\nonumber \\
& & 3 \mu \sqrt{1 - \mu^2}~\frac{1}{c}~(\mbox{\boldmath$j$} \times \mbox{\boldmath$B$})_{\theta}\Bigl], 
\label{jB}
\end{eqnarray}
\begin{eqnarray}
{A_{20}^{\rm{E} 2}}_{\rho_{\rm m}} &=&\frac{G}{c^4} \frac{32 \pi^{3/2}}{\sqrt{15}}  \int_{0}^{1}d\mu \int_{0}^{\infty} \,dr ~\frac{1}{8 \pi c}
\nonumber \\
& & \frac{d}{dt}\Biggl[
\frac{\partial}{\partial \theta}[B_{r} r^3 (3~\mu^2 - 1)]E_{\phi} - 
\frac{\partial}{\partial r}[B_{\theta} r^3(3\mu^2 -1)]rE_{\phi} + 
\nonumber\\
&+& 
\frac{\partial}{\partial r}[B_{\phi}   r^3(3 \mu^2 - 1)]rE_{\theta}
- \frac{1}{\sin \theta}\frac{\partial}{\partial \theta}[B_{\phi}\sin \theta
r^3 (3 \mu^2 - 1)]E_{r}\Biggr]. 
\label{rhom}
\end{eqnarray}
${A_{20}^{\rm{E} 2}}_{j \times B},
{A_{20}^{\rm{E} 2}}_{\rho_{\rm m}}$ represent the contribution from
${j \times B}$ part and from the time
derivatives of the energy density of electro-magnetic fields, respectively.
Only the first time derivative of the magnetic fields is remained, because this
is the leading order and the numerical treatments of the second time
derivatives are formidable. Then the total gravitational amplitude can
be written as follows,
\begin{equation}
h^{\rm TT} \equiv h^{\rm TT}_{\rm quad} + h^{\rm TT}_{j \times B} + h^{TT}_{\rho_{\rm m}}, 
\label{tot}
\end{equation}
where the quantities of the right hand of the equation are defined by
Eqs. (\ref{101}), (\ref{quad}), (\ref{jB}), and (\ref{rhom}). 
Note that $q_{r}$ and $q_{\theta}$ 
in Eq. (\ref{quad})
represents the gravitational waves contributed from the artificial 
viscosity (see e.g., \cite{moenchgw}). 
When one uses an artificial viscosity of von
Neumann and Richtmyer, which is a most popular one, 
the concrete form of $q_{i}$ is,
\begin{equation}
q_{i} = \nabla_{i}~[l^2~\rho~(\nabla \cdot \mbox{\boldmath$v$})^2],
\label{av}
\end{equation}
where $i = r, \theta$ with $l$ defining the dissipation length.
Using above quadrupole formula including contributions from the electromagnetic
fields \cite{yama03, kotakegwMHD}, they calculated the waveforms by
performing the 2D magnetohydrodynamic core-collapse simulations 
\cite{kotakegwMHD}. 

With these computations, they found that the amplitude is affected in 
the strongly magnetized models whose initial $E_{\rm m}/|W|$
is greater than 0.1 \%, where $E_{\rm m}/|W|$ represents the magnetic to
the gravitational energy. This is natural because the 
amplitude contributed from the electromagnetic fields should be an order of
\begin{equation}
R_{\rm mag} = \frac{{B_{\rm c}^2}/{8 \pi }}{\rho_{\rm c}c^2} \sim ~10 ~\%
~\Bigl(\frac{B_{\rm c}}{\rm{several}~\times 10^{17}~\rm{G}}\Bigr)^2 
\Bigl(\frac{\rho_{\rm c}}{10^{13}~ \rm{g}~\rm{cm}^{-3}}\Bigr)^{-1},
\end{equation}
with $B_{\rm c},\rho_{\rm c}$ being the central magnetic field and the
central density near core bounce, respectively.
Thus, strongly magnetized models, whose central magnetic fields at core
bounce become as high as  
$ \sim 10^{17}$ G, can affect the amplitude. 
\begin{figure}
\begin{center}
\epsfxsize=12.5cm
\epsfbox{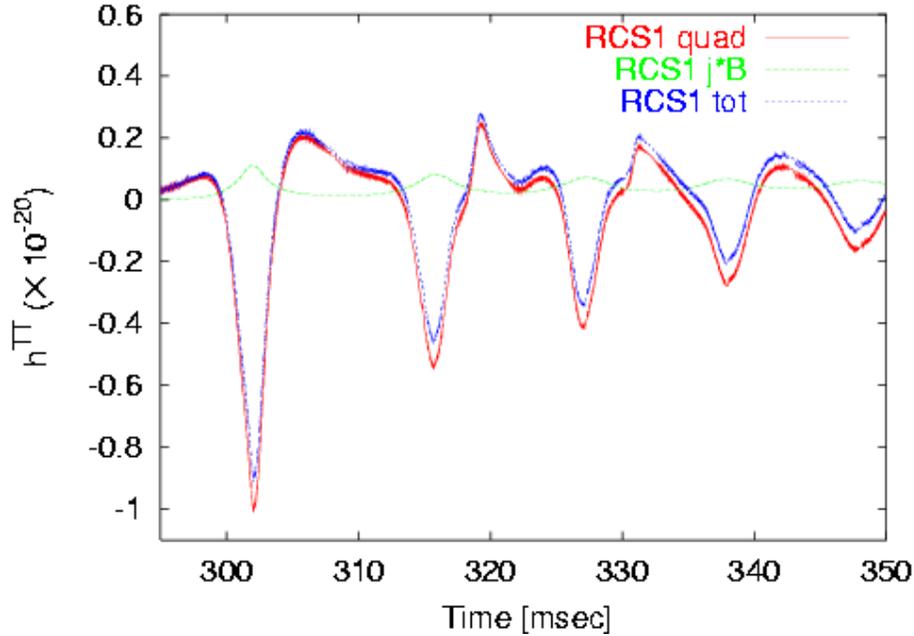}
\end{center}
\caption{Waveforms for a model with the strongest magnetic field
 obtained in the magnetohydrodynamic core-collapse simulation \cite{kotakegwMHD}. In the
 figure, ``quad'',''$j \times B$''  represent the
 contributions from the mass quadrupole moment and from the electromagnetic field, 
respectively (see Eq. (\ref{tot})). The total amplitude is denoted as ``tot''. Note that the source is assumed 
to be located at the distance of 10 kpc.\label{kotake_gw_mag}}
\end{figure}

It was furthermore found that the contribution of the electromagnetic
fields changes in the
opposite phase to the matter contribution (see Figure \ref{kotake_gw_mag}). 
Together with a slight offset of the electromagnetic part, the negative part of the amplitude 
becomes less negative, while the positive part becomes more positive
(see Figure \ref{kotake_gw_mag}).  
As a result, the peak amplitude at core bounce is found to be 
lowered by $\sim 10$ \%
They confirmed that the amplitudes of second peaks and the difference of its sign, from which
 one may know the information of the angular momentum of the core as
 mentioned, are still within the
 detection limit of the first LIGO for the galactic supernova, although the
 characteristics of second peaks are reduced by the incursion of the
 strong magnetic fields.

\subsubsection{effects of realistic equations of state}
We turn to the effect of the equation of state (EOS)
on the gravitational signals. Needless to say, EOS is an important
microphysical ingredient for determining the dynamics of core collapse and, 
eventually, the gravitational wave amplitude.  As a realistic EOS,
Lattimer-Swesty (LS) EOS \cite{Lat91} has been used in recent papers
discussing gravitational radiations from the rotational core collapse
\cite{ott,muller03}. It has been difficult to investigate the effect of EOS's on 
the gravitational signals because available EOS's
based on different nuclear models are limited. Recently, a new complete
EOS for supernova simulations has become available
\cite{shen98,sumi_prep}. The EOS is based on the relativistic
mean field (RMF) theory combined with the Thomas-Fermi approach.

\begin{figure}
\begin{center}
\epsfxsize=7.5cm
\epsfbox{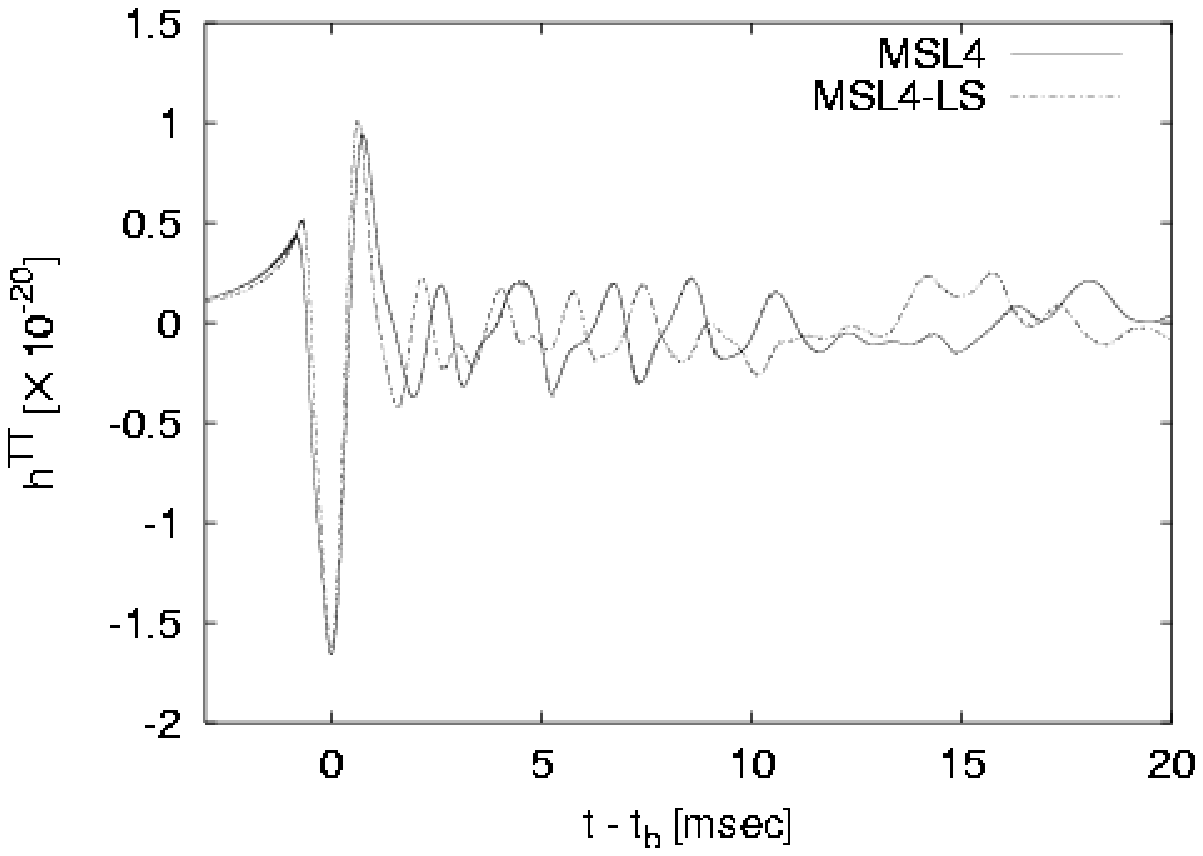}
\epsfxsize=7.5cm
\epsfbox{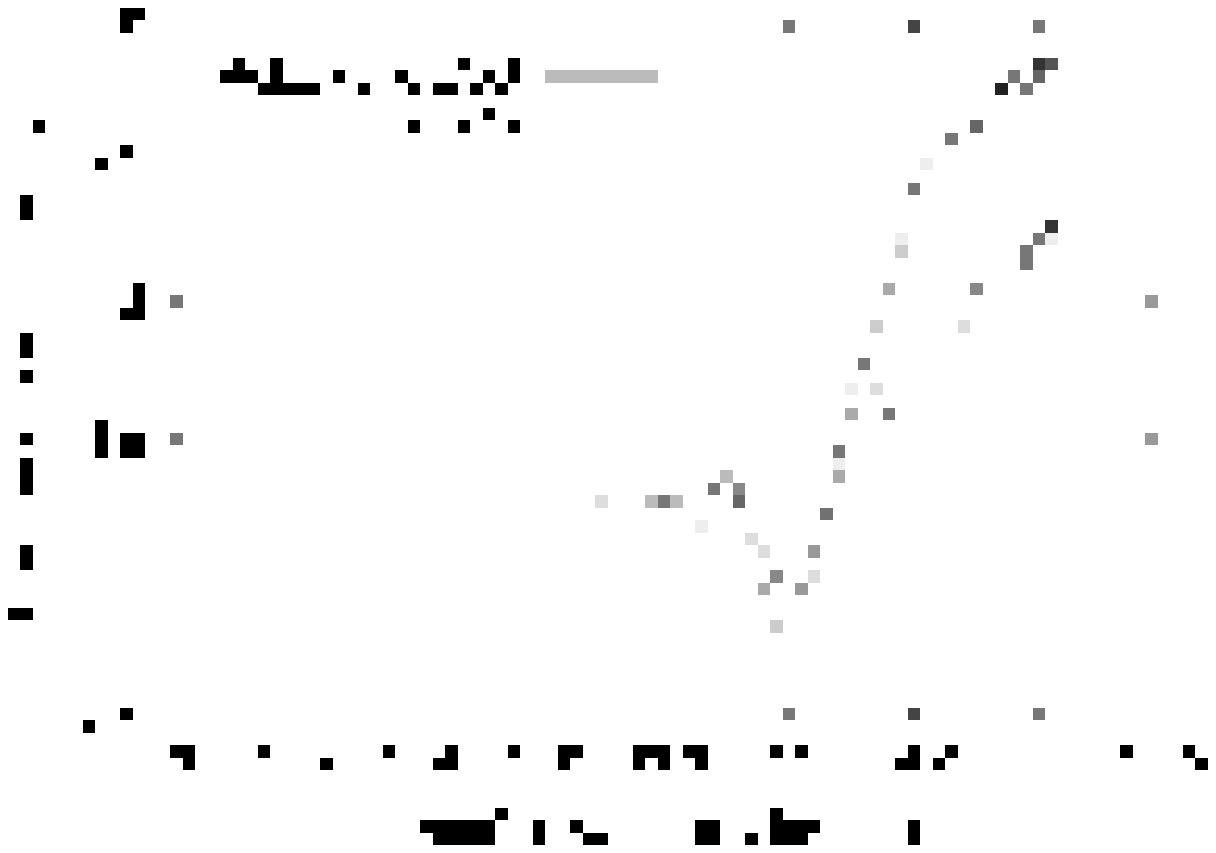}
\end{center}
\caption{Waveforms (left panel) for the models with the relativistic EOS
 (the solid line labeled as MSL4) and with LS EOS (the dashed line
 labeled as MSL4-LS) and the relation between the central density and
 the effective adiabatic index $\gamma$ near core bounce (right
 panel).These figures are taken from \cite{kotakegwMHD}. \label{fig7}}
\end{figure}

 By implementing these two realistic EOS's, 
Kotake {\it et al.} (2004) looked into the difference of the
gravitational wave signals \cite{kotakegwMHD}. The left panel of Figure \ref{fig7} shows
the waveforms for the models with the relativistic EOS (model MSL4)
or the LS EOS (model MSL4-LS). The maximum
amplitudes for the two models do not differ significantly (see the left
panel of Figure \ref{fig7}). The important
difference of the two EOS's is the stiffness. As seen from
the right panel of Figure \ref{fig7}, LS EOS is softer than the
relativistic EOS, which makes the central density larger at core bounce
and thus results in the shorter time interval between the
subsequent bounces.
On the other hands, softer EOS results in the smaller lepton fraction 
in the inner core, which reduces the mass quadrupole moments at core bounce. 
By the competition of
these factors (see Eq. (\ref{simple_estimate})), the maximum amplitude 
remains almost the same between the two realistic EOS's, while the
typical frequencies of the gravitational wave become slightly higher for
the softer equation of state. Furthermore it was found that the
aforementioned type III waveform observed in a very soft EOS polytropic 
equation of state \cite{zweg} does not appear when the realistic equations of state are employed.  

\subsubsection{non-axisymmetric simulations}
\begin{figure}
\begin{center}
\epsfxsize = 7.5 cm
\epsfbox{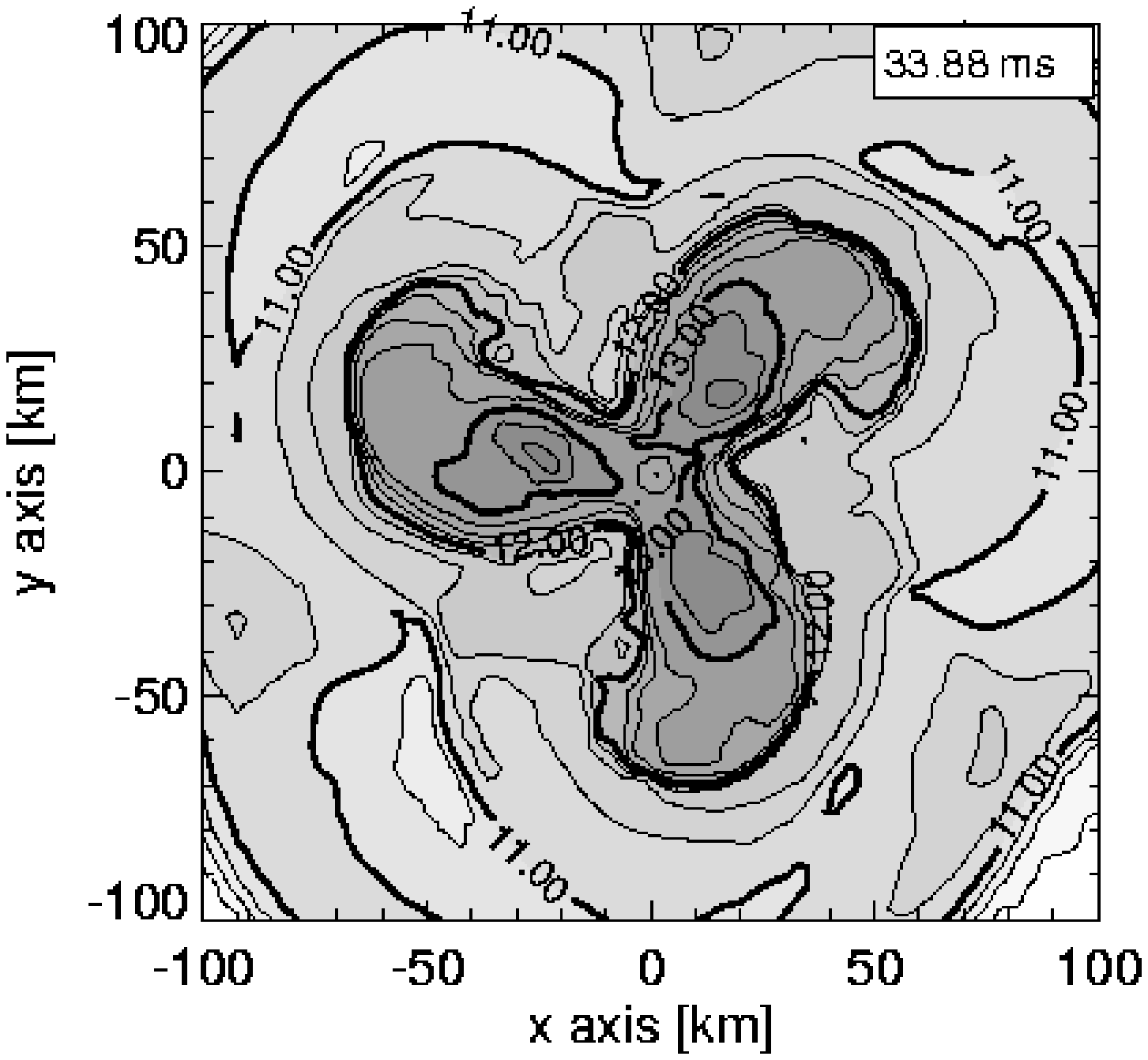}
\epsfxsize = 7.5 cm
\epsfbox{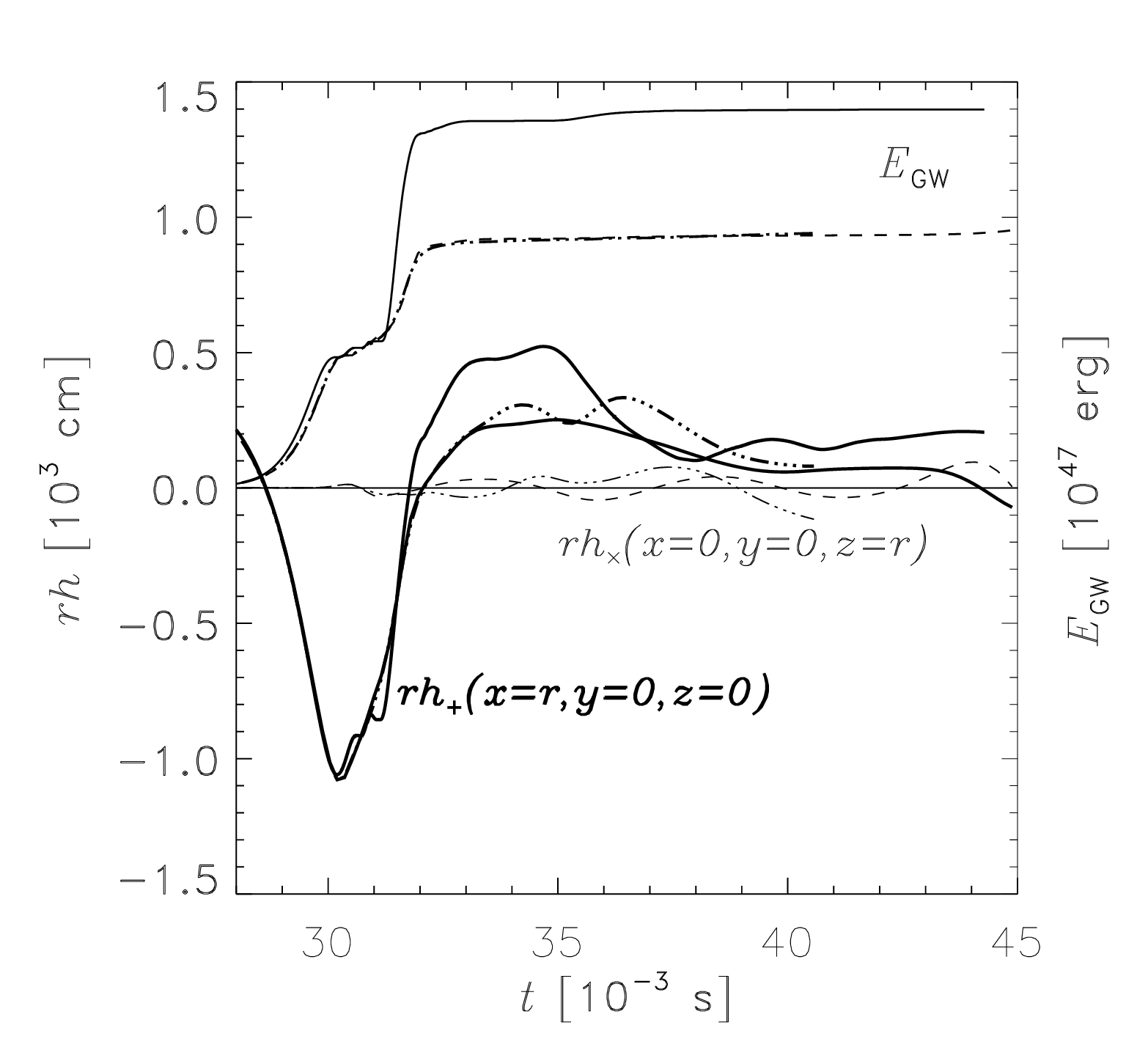}
\caption{Gravitational radiation in a 3D model taken from
 \cite{rampp}. Left panels shows the contour of density in the
 equatorial plane after about 3 msec after core-bounce showing 
the growth of the three arms. From the left panel, it can be seen that 
the difference of gravitational amplitudes between the 3D models 
(dashed and dashed-dotted lines) and the axisymmetric models (solid
 line) at core-bounce (30 msec) are too small to see by eye. Cross mode
 of the gravitational waves $h_{\times}$, which is of genuine 3D origin begins to
 grow after core bounce ($t \ge 33$ msec), however, does not grow later
 on.
(see \cite{rampp} for details).}
\label{rampp_clump}
\end{center}
\end{figure}

In addition to the above 2D simulations, several 3D simulations have been
computed. The first 3D hydrodynamic core-collapse
simulations well beyond the core bounce was performed by \cite{rampp}.
The initial condition for their study was based on the configuration  
at several milliseconds before core bounce in the rapidly rotating 
2D models of Zwerger {\it
et al} \cite{zweg}. In addition to the configuration, they imposed low
mode ($m = 3$) density perturbation and followed the growth of the
instability, where $m$ denotes the azimuthal quantum number. They observed the three clumps merged into a bar-like
structure due to the growth of the non-axisymmetric instability (see
the left panel of Figure \ref{rampp_clump}). In
fact, their models are rapid rotator whose value of $T/|W|$ at core
bounce exceeds the 
critical value,
 beyond which MacLaurin spheroids become dynamically unstable 
against tri-axial perturbations ($T/|W| > T/|W|_{\rm dyn} \simeq 27.0 \% $). 
However, they found that the maximum amplitudes of the gravitational
waves were only $\sim 2 \%$ different from the 2D cases by Zwerger {\it
et al.}\cite{zweg} (see the
right panel of Figure \ref{rampp_clump}).

3D rotational core-collapse simulations of Fryer and his collaborators \cite{fryer,fryerkick}
 seem in favor of the above result. They have investigated whether
the core fragmentation, and thus, the significant deviations 
of the gravitational radiation from 2D studies happen or not. As for the
 numerical computations, they have performed 3D smoothed
particle hydrodynamic (SPH) simulations, with a realistic equation of
 state and the flux-limited diffusion approximation method for neutrino transfer. As for the initial model, they employed an
 rapidly rotating model with the initial value of 
$T/|W|_{\rm initial}$ of
$\sim 3$ \%.
As a result, they also found that no fragmentation, although near core bounce 
the value of $T/|W| $ approaches to a critical value of $\sim $14 \%, 
beyond which the secular instability due to the non-axisymmetric
perturbations sets in. Since the maximum value of 
$T/|W|_{\rm initial}$ predicted by the recent evolution models are $
 \leq \sim 0.5 \%$ \cite{heger00}, they concluded that the fragmentation or
dynamical bar instabilities are unlikely to occur with any of the
currently-produced supernova progenitors.

\begin{figure}
\begin{center}
\epsfxsize=8cm
\epsfbox{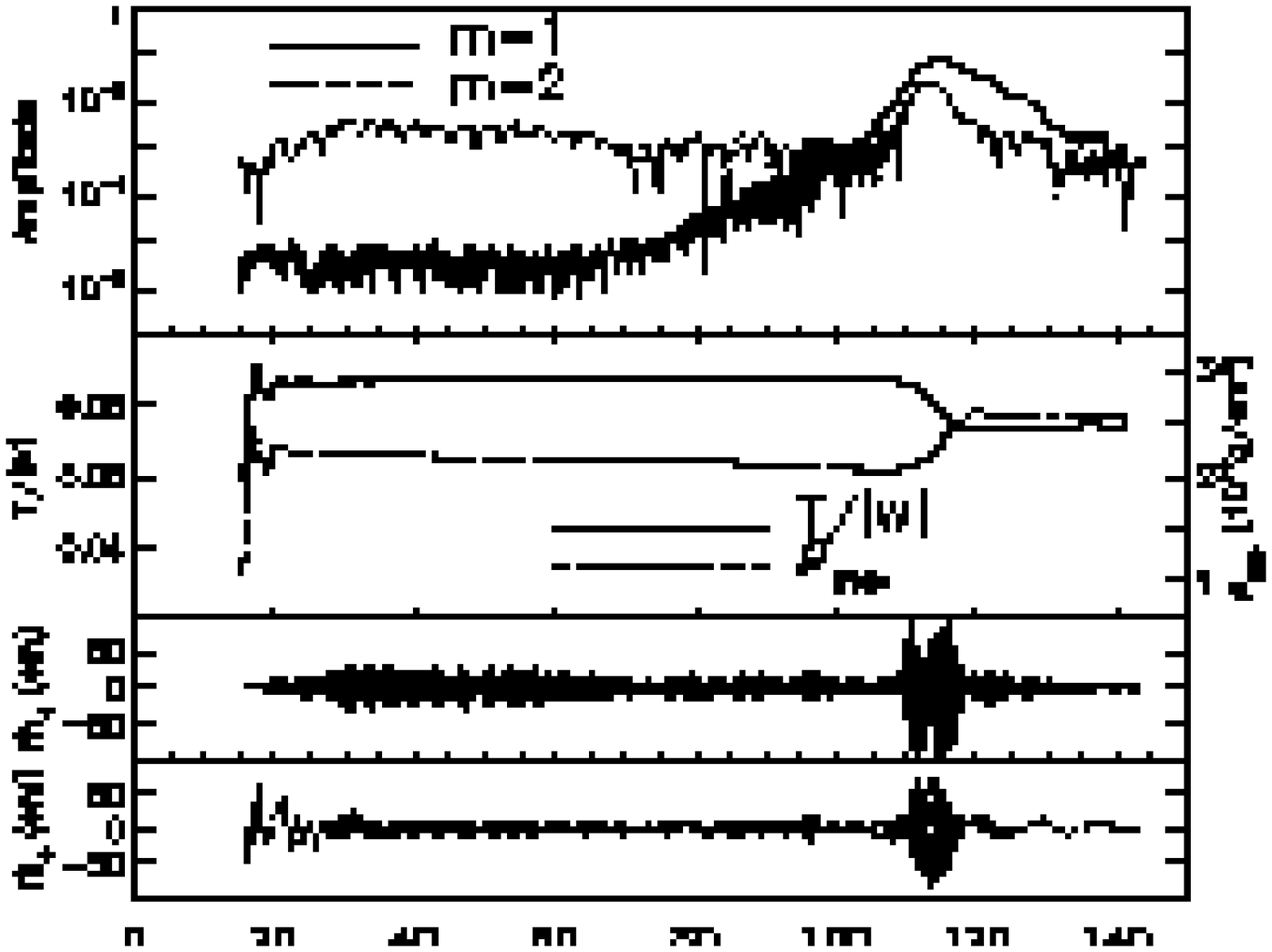}
\caption{Time evolution of various quantities in a 3D model with
 $T/|W|_{\rm init} = 0.2 \%$ calculated
 by Ott {\it et al}. (2005) \cite{ott_one_arm}. Time is measured from the epoch of core bounce $t_b$. Top panel shows that the amplitude of the $m = 1$ mode precedes
 that of the $m=2$. Middle panel shows the time evolution of $T/|W|$ and
 the core's maximum density. It is shown from the panel that after the epoch of $t - t_b \sim 100$ ms, when the $m=2$
 mode begins to be amplified, the transfer of the angular momentum
 becomes active which results in the increase of the maximum density and
 the decrease of the $T/|W|$. The bottom panel shows the gravitational
 strain at the distance to the source $r$ as viewed down the rotational
 axis (solid curve) and as viewed along the equatorial plane (dotted curve). One can
 see that the waveform traces the time evolution of the $m=2$ mode. Note in
 the panel that $rh = 100$ cm corresponds to $h \sim 3 \times 10^{-21}$ for a
 galactic supernova.       }
\label{ott_one_arm}
\end{center}
\end{figure}

While the past studies terminated the 3D simulations at several tens
milliseconds after core bounce,  Ott {\it et al} (2005) 
investigated the growth of the non-axisymmetric structure until the rather later phases ($\ge 100$ msec) after core
bounce \cite{ott_one_arm}. 
They found 
that the growth of the $m=1$ mode, 
the so-called one-armed instability \cite{centrella,shibata_one_1,shibata_one_2,saijo}, 
precedes the growth of the bar-mode ($m=2$) instability (see Figure
\ref{ott_one_arm}), where $m$ denotes the azimuthal quantum number. Since the criterion of
the growth of the one-armed instability is lower than that of the
bar-mode instability \cite{centrella, shibata_one_1,shibata_one_2,saijo}, they pointed out that the initial rotation rate
required for the sufficient gravitational radiation from a galactic
supernova enough to be detected by the future detectors
 can be as small
as $T/|W|_{init} = 0.2 \%$, which is much smaller than the one previously 
assumed for igniting the growth of the bar-mode instability.    
\subsubsection{general relativistic studies}
\begin{figure}
\begin{center}
\epsfxsize = 9 cm
\epsfbox{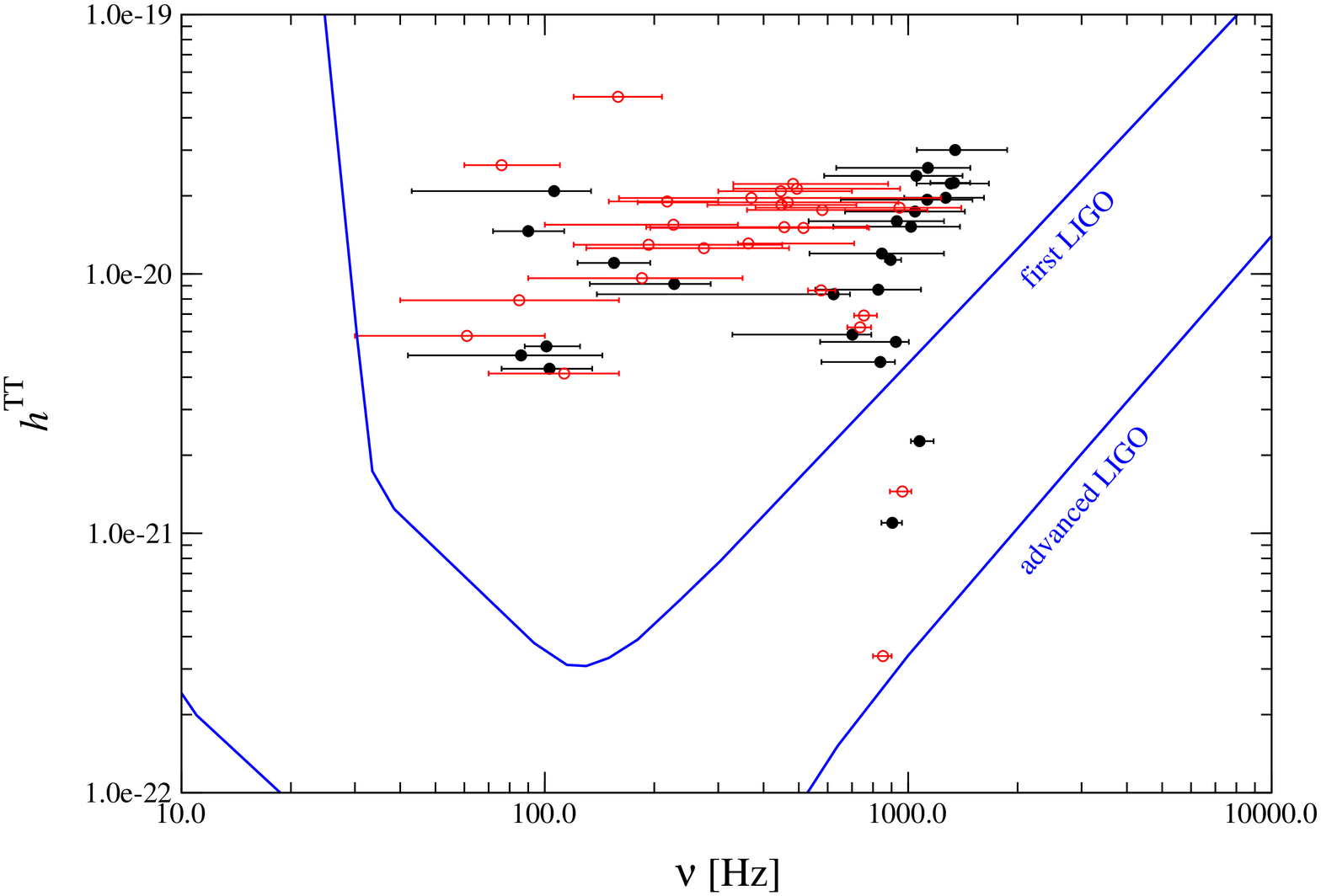}
  \caption{~Prospects of detection of the gravitational wave signal
    from axisymmetric rotational supernova core collapse in relativistic
    (black filled circles) and Newtonian (red unfilled circles)
    gravity studied in Dimmelmeier {\it et al} \cite{dimmel}. The figure gives the (dimensionless) gravitational wave
    amplitude $ h^{\rm TT} $ and the frequency range for all 26
    models. For a source at a distance of 10~kpc the signals of all
    models are above the burst sensitivity of the LIGO~I detector
    (except for some low amplitude, high frequency models), and well
    above that of the LIGO~II interferometer. It can be seen that the
 typical frequencies of the gravitational waves are blue-shifted when
 relativistic effects are taken into account. This figure is taken from
 Dimmelmeier {\it et al} \cite{dimmel}.}
  \label{blue_shift}
\end{center}
\end{figure}

All the above computations employ the Newton gravity.
In the following, we give a brief description of general relativistic (GR)
studies. The study by \cite{dimmel} may be one of the representative 
GR studies for computing the GWs in stellar-collapse in the sense that 
they compared the properties of the GWs obtained by Newtonian and GR 
simulations systematically. As in the work of \cite{zweg}, they characterized the
model difference by the degree of differential rotation, initial
rotation rates, and adiabatic indices of equation of state and computed
26 initial models. The conformally flat (CF) metric was used to approximate
the space time geometry in their GR hydrodynamic simulations. As well known,
the CF approximation gives the exact solution of Einstein's equation in
spherical symmetry. The approximation may not be so bad unless the
configurations are extremely deviated from the spherical symmetry.
However, they were forced to employ the quadrupole formula for
extracting the GWs because CF approximation eliminates the GW emission
from the spacetime. With these computations,
it was found that the qualitative 
features of the GWs obtained in the Newtonian studies of \cite{zweg}
are almost true for their studies. Quantitatively, however, it was
pointed out that relativistic
effects make the central density at core bounce much higher 
than that in the Newtonian gravity. This is simply due to the
enhancement of the gravity due to the GR effect. As for the peak amplitudes
at core bounce, no significant differences between the GR and Newtonian
case were found, while the typical frequencies at the peak amplitudes
are blue-shifted for the GR models. This may be because the higher
central density makes the timescale at core bounce shorter, which leads
to the higher frequency (see Figure \ref{blue_shift}).

\begin{figure}
\begin{center}
\epsfxsize=6.5cm
\epsfbox{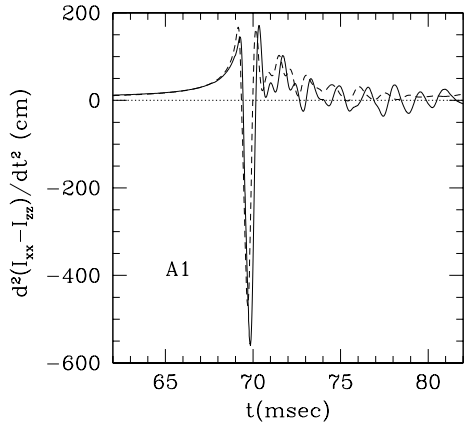}
\end{center}
\caption{Comparison of the waveforms between the fully general
 relativistic (solid line) and conformally flat approximation (dotted
 line) calculations. This
 figure is taken from Shibata and Sekiguchi (2004) \cite{shibaseki} \label{shibaseki}.}
\end{figure}

Fully general relativistic collapse simulations from core-collapse to
the formation of a neutron star have been performed by the group of 
M. Shibata \cite{shibaseki} (see the reference, therein). It is mentioned
that not only in their studies but also in other GR studies, the polytropic equations of state 
are employed in order to reduce the computational costs required 
for including a realistic equation of state. With these computations,
it was found that not only the evolution of central density 
during core-collapse, bounce, and the formation of PNS, but also the
waveforms are qualitatively in good agreement with those in the study of 
\cite{dimmel}, except for a factor of $\sim $ 2 difference of the
amplitudes in the ring-down phase (see Figure
\ref{shibaseki}). 
As a result, they concluded that the approximated method by 
\cite{dimmel} is appropriate for following the axisymmetric stellar
core-collapse associated with the formation of the neutron stars and 
for estimating the emitted gravitational waves. Needless to say,
the full GR calculations are indispensable for estimating the
gravitational waves from the core-collapse of very massive stars associated with the 
formation of the black hole.

\begin{figure}
\begin{center}
\epsfxsize = 7cm
\epsfbox{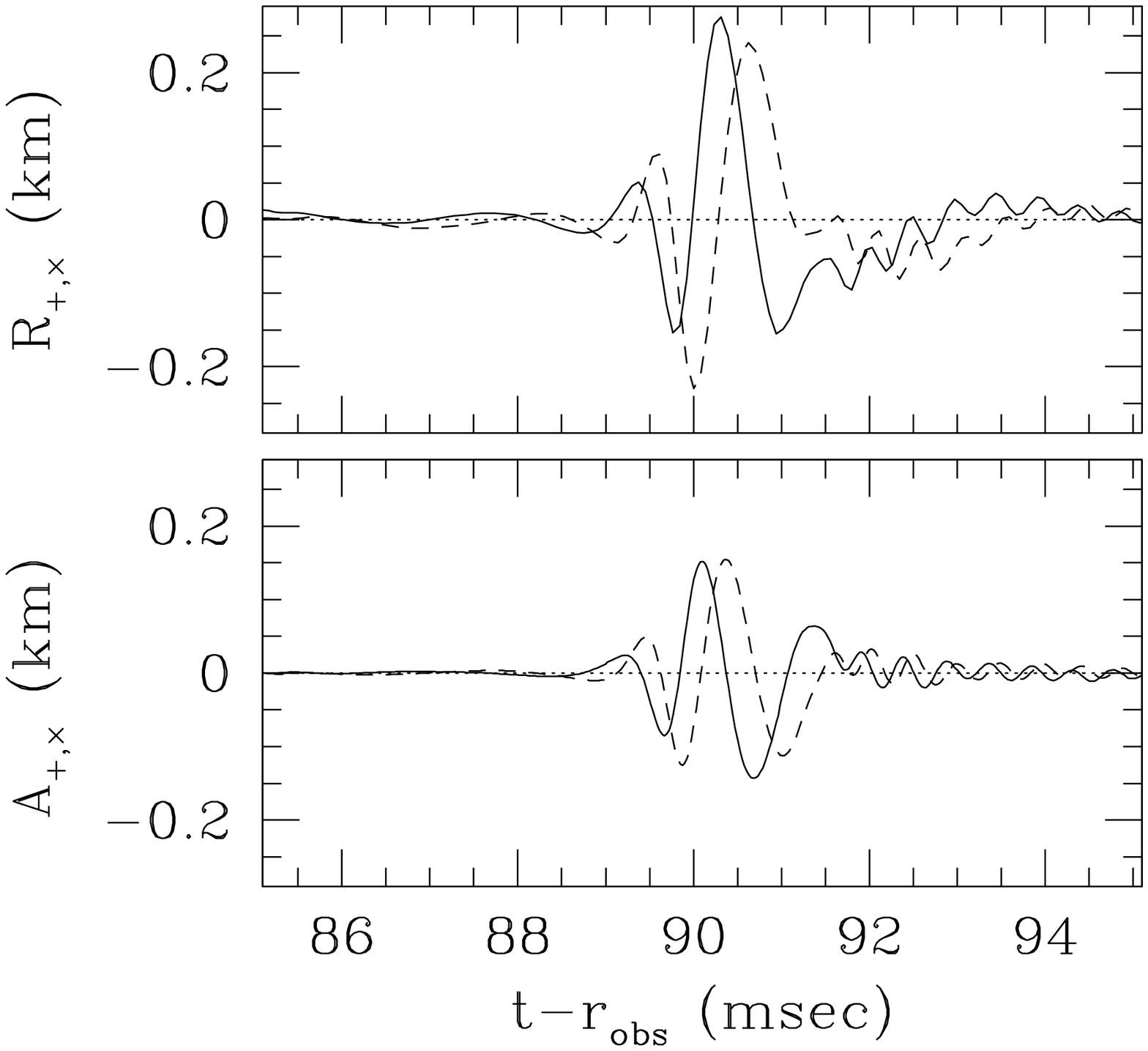}
\epsfxsize = 7cm
\epsfbox{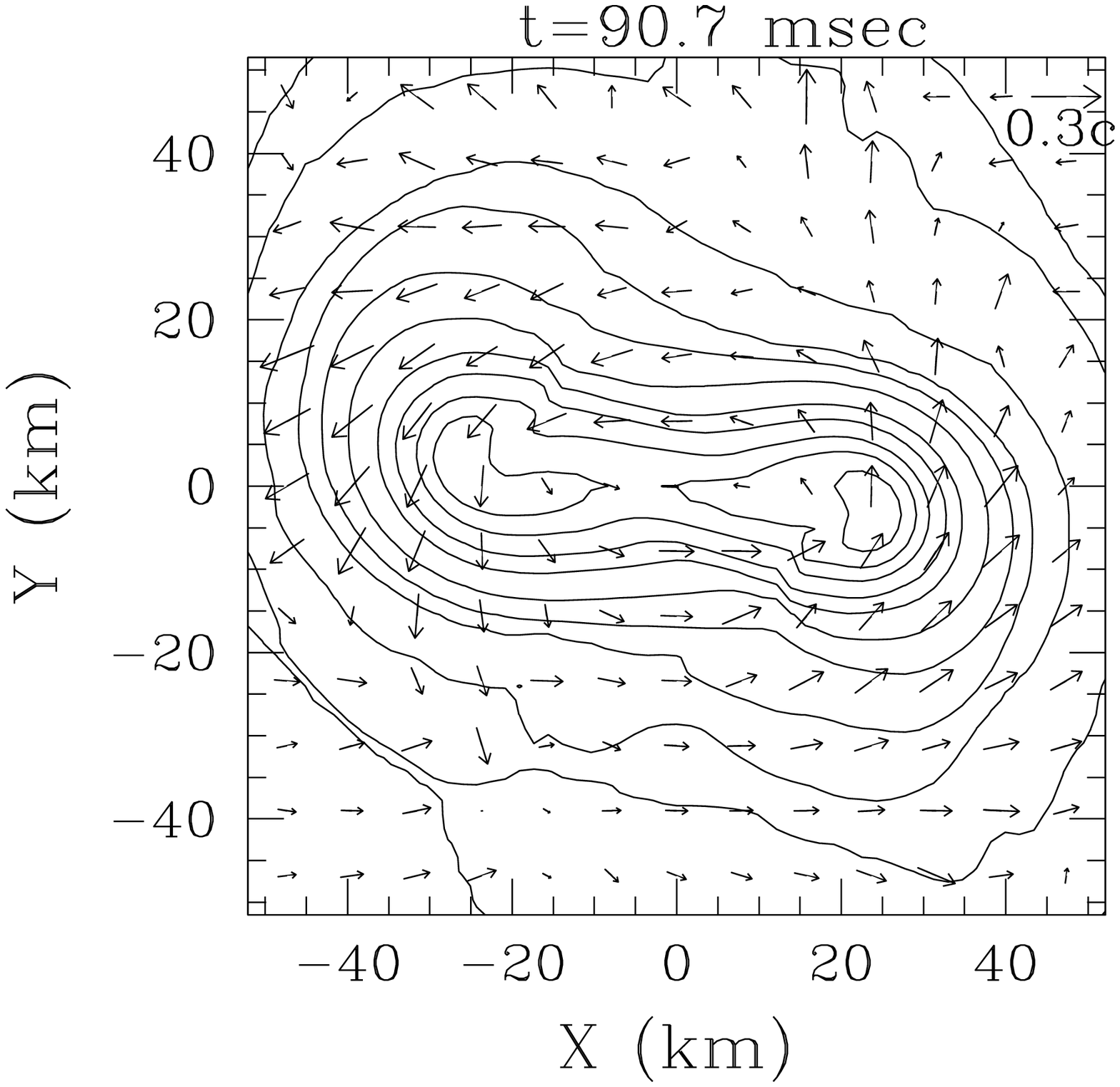}
\caption{Results of the first 3D full GR calculations taken from Shibata
 and Sekiguchi (2005) \cite{shiba_onearm}. Left panel shows the gravitational waveform of
 the model with $T/|W|_{\rm init} \sim 1.8 \%$ with the strong differential
 rotation (model M7c2 in \cite{shiba_onearm}). $R_{+,\times}$ is the
 amplitude computed by the gauge-invariant extraction method, and 
 $A_{+,\times}$ is the one by the quadrupole formula (see \cite{shiba_onearm} for details). Right panel
 shows the snapshot of the contour of the
 density on the equatorial plane at $t = 90.7$ ms with the vector fields
 (arrows), showing the bar-mode
 instability does develop, leading to the significant change in the
 amplitudes near the corresponding time. The maximum amplitude 
in the right panel can be translated into a dimensionless strain of 
$h \sim 10^{-18}$ for a galactic
 supernova with frequencies about 1 kHz. }
\label{shiba_barmode}
\end{center}
\end{figure}
The numerics of the general relativistic studies have seen
major progress recently. Dimmelmeier {\it et al.} recently succeeded 
in extending their 2D GR code to 3D with the CF approximation \cite{dimmel04}.
 Shibata and Sekiguchi (2005) have performed the fully general
 relativistic 3D simulations and pointed out that the amplitudes of the
 gravitational waves can be enhanced by a factor of $10$ than the
 axisymmetric collapse due to the
 growth of the dynamical bar-mode instabilities, when the core initially
 rotates very differentially with rapid rotation of $1 \% \le
 T/|W|_{\rm init} \le 2 \% $
 \cite{shiba_onearm} (see Figure \ref{shiba_barmode}). They discussed that the enhancement of the
 self-gravity due to the general relativistic effects results in a more
 efficient spin-up of the core, and thus, leading to the growth of the
 instability, which would be underestimated in the study of Rampp {\it
 et al} mentioned above. Very recently, fully general relativistic and
 magnetohydrodynamic simulations have also been reported to be
 practicable \cite{duez1,duez2}. 
The vary wide varieties of relativistic astrophysical
events are expected to be investigated by the new-coming 3D GR studies.

\clearpage
\subsection{Gravitational waves from convection and
  anisotropic neutrino radiation \label{aniso_grav}}

All the studies, which we reviewed so far, paid attention to the gravitational 
signals produced near core bounce due to the large-scale aspherical
motions of matter induced by core's rotation without/with magnetic fields. In addition, 
two other sources of the GW emissions 
have been considered to be important in the
later phases after core bounce, namely convective motions and anisotropic
neutrino radiation, both of which can contribute to the non-spherical
part of the energy momentum tensor of the Einstein equations. While
gravitational waves from convective motions are originated 
from the aspherical motions of matter as well as the ones at core
bounce, gravitational waves from neutrinos have some different features,
which will be explained in the next section. 

\subsubsection{foundation of gravitational waves from neutrinos} 
As mentioned, the gravitational-wave signals at core bounce are emitted as
bursts in which the wave amplitude rises from zero at core bounce,
oscillates for several cycles and then approaches to zero (see for
example Figure \ref{ZM}) due to the hydrodynamic motions of the central
core. In addition, there is another class of gravitational wave, that is, 
{\it bursts with memory}, in which the wave amplitude rises from zero and 
then after the neutrino bursts settles into a non-zero final value
\cite{braginskii}. The gravitational waves from anisotropic neutrino
radiation from core-collapse supernovae is categorized to this class,
which has been originately pointed out in late 1970's by \cite{epstein_gw,turner1979}.
The detectability of such effect was discussed by 
\cite{braginskii} through the ground-based laser
interferometers. According to \cite{epstein_gw}, we summarize the
formulation of gravitational waves from neutrinos in the
following, which will be useful in the later discussions.

\begin{figure}
\begin{center}
\epsfxsize = 7 cm
\epsfbox{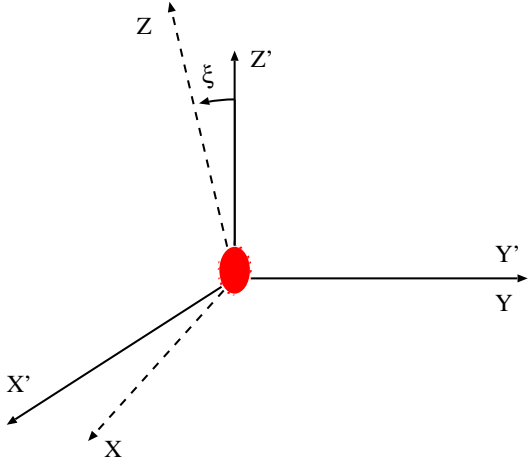}
\vspace{0.5 cm}
\caption{Source coordinate system $(x^{'},y^{'},z^{'})$ and observer
 coordinate system  $(x,y,z)$. The observer resides at the
 distant point on the $z$-axis. The viewing angle is denoted by $\xi$
 which is the angle between $z$ and $z^{'}$ axis. The $z^{'}$ axis coincides with the
 symmetry axis of the source, presumably the rotational axis. Central
 red region indicates the anisotropic neutrino radiation from a supernova.}
\label{gw_sch_fig}
\end{center}
\end{figure}

At first, one should define a concrete form of an energy-momentum tensor
of the neutrino radiation field to compute the gravitational waves.  Then the following form is naturally assumed,
\begin{equation}
T^{ij}(t, \mbox{\boldmath$x$}) = n^{i}n^{j}r^{-2}L_{\nu}(t-r)f(\Omega,t-r),
\label{gw_aniso_1}
\end{equation}   
where $\mbox{\boldmath$n$}= \mbox{\boldmath$x$}/r,~r =
|\mbox{\boldmath$x$}|,~
f(\Omega,t)\geq 0,$ and $\int f(\Omega,t) d\Omega =1$. This source
tensor represents radiation fields of neutrinos being released at the speed of light from the
point $\mbox{\boldmath$x$} = 0$ to an observer at a distance of $r$. 
The functions $L_{\nu} (t)$ and
$f(\Omega, t)$ are the rate of energy loss and 
the angular distribution of neutrino radiation, respectively, at time $t$. Given the
energy-momentum tensor, one can calculate the transverse-traceless (TT)
gravitational field from the corresponding source as follows,
\begin{equation}
h_{\nu, TT}^{ij} = 4\int T_{TT}^{ij}(t-|\mbox{\boldmath$x$}-
\mbox{\boldmath$x$}^{'}|, \mbox{\boldmath$x$}^{'})|\mbox{\boldmath$x$}-
\mbox{\boldmath$x$}^{'}|^{-1}~d^3 x^{'}
\label{gw_aniso_2}
\end{equation} 
Note that dashed $(^{'})$ variables represent the quantities observed in
a coordinate frame $(x^{'},~y^{'},~z^{'})$, on the other hand,
non-dashed variables represent the quantities observed in the observer's
frame $(x,~y,~z)$ (see Figure \ref{gw_sch_fig}). 
For convenience, we write Eq. (\ref{gw_aniso_1}) as follows,
\begin{equation}
T^{ij}(t, \mbox{\boldmath$x$}) = n^{i}n^{j}r^{-2}\int_{-\infty}^{\infty}
L_{\nu}(t^{'})f(\Omega^{'},t^{'})~\delta(t-t^{'}-r)dt^{'}.
\end{equation}
Introducing Eq. (\ref{gw_aniso_1}) to (\ref{gw_aniso_2}) and performing
the integration with respect to $r^{'}$ and we may obtain,
\begin{equation}
h^{ij}_{\nu, TT}(t, \mbox{\boldmath$x$}) =  4 \int_{-\infty}^{t-r}
 \int_{4 \pi} 
\frac{(n^{i}n^{j})_{ TT} L_{\nu}(t^{'})f(\Omega^{'},t^{'})}
{t -t^{'} - r\cos \theta^{'}}d\Omega^{'}~dt^{'}.
\label{gw_aniso_3}
\end{equation}
In addition, we make use of the approximation that the gravitational
wave signal measured by an observer at time $t$ is caused by radiation
emitted at time $t^{'} = t - r$. Hence we take $t - t^{'} = const =
r$, which means that only a neutrino pulse itself is assumed to cause a
gravitational wave signal. This procedure is equivalent to eliminate the 
non-zero value of $h_{\nu}$ in case of isotropic neutrino radiation 
appeared in Eq. (\ref{gw_aniso_3}). We verify this in the later section.

%Next we proceed to calculate $(n^{i}n^{j})_{\rm TT}$.
In the geometrical setup shown in Figure \ref{gw_sch_fig}, we assume
that the $z$-axis lies on the ($x^{'},z^{'}$) plane for convenience. In
this case, the two polarization states of gravitational waves satisfying
the transverse-traceless conditions become 
\begin{equation}
 h^{+}_{\nu,\rm TT} \equiv h^{xx}_{\nu} = - h^{yy}_{\nu},
\end{equation}
and 
\begin{equation}
 h^{\times}_{\nu,\rm TT} \equiv h^{xy}_{\nu} =  h^{yx}_{\nu},
\end{equation}
in the observer coordinates. It is noted that the sum of the squared
amplitudes $|h_{+}|^2 + |h_{\times}|^2$ is invariant under the rotation
about the $z$-axis.
Using the following relations between the two coordinates,
\begin{equation}
\sin \theta^{'} \cos \phi^{'} = \sin \theta \cos \phi \cos \xi + \cos \theta \sin \xi,
\end{equation}
\begin{equation}
\sin \theta^{'} \sin \phi^{'} = \sin \theta \sin \phi,
\end{equation}
\begin{equation}
\cos \theta^{'} = - \sin \theta \cos \phi \sin \xi + \cos \theta \cos \xi,
\end{equation}
 one can obtain the following,
 \begin{equation}
h_{\nu, +}^{\rm TT} = \frac{2}{r}\int_{-\infty}^{t - \frac{R}{c}}dt^{'}
\int_{4 \pi} d\Omega^{'} (1 + \cos \theta)~\cos2\phi~
 L_{\nu}(t^{'})f(\Omega^{'},t^{'}),
\end{equation}
while the counter part of the amplitude, $h_{\times}$, is obtained just
replacing $\cos 2 \phi$ with $\sin 2 \phi$, which immediately becomes
zero by integrating over the angle $\phi$ due to the axisymmetric source
we consider here. 

In the above equation, it should be noted that $\theta$ and $\phi$ are
required to be expressed in terms of the angles $\vartheta^{'},
~\phi^{'}$ with respect to the source coordinate valuables,
and the viewing angle of $\xi$. In the following, we consider
two cases, in which the observer is situated parallel to the $z^{'}$
axis ($\xi = 0$) or perpendicular to the $z^{'}$ axis ($\xi
= \pi/2$). In the former case, one easily obtains,
\begin{equation}
h_{\nu, \rm p}^{\rm TT} = \frac{2}{r}\int_{-\infty}^{t - \frac{R}{c}}dt^{'}
\int_{4 \pi} d\Omega^{'} (1 + \cos \vartheta^{'})~\cos2\varphi^{'}~
 L_{\nu}(t^{'})f(\Omega^{'},t^{'}),
\end{equation} 
which becomes zero in case of the axisymmetric radiation source. 
Here the subscript $_{\rm p}$ suggests that the observer is situated in
the polar axis relative to the source coordinate frame.
In the latter case, the observer is positioned perpendicular to the
source's $z^{'}$ axis (seen from the $e$quator), and the field becomes,
\begin{equation}
h_{\nu, \rm e}^{\rm TT} = \frac{2}{r}\int_{-\infty}^{t - \frac{R}{c}}dt^{'}
\int_{4 \pi} d\Omega^{'} \Psi(\vartheta^{'}, \varphi^{'})~
 L_{\nu}(t^{'})f(\Omega^{'},t^{'}),
\label{polar}
\end{equation} 
where,
\begin{eqnarray}
\Psi({\vartheta^{'},\varphi^{'}}) &=& (1 + \sin\vartheta^{'} \cos\varphi^{'}) 
\frac{\cos^2 \vartheta^{'}- \sin^2 \vartheta^{'}\sin^2\varphi^{'}}
{\cos^2 \vartheta^{'}+ \sin^2 \vartheta^{'}\sin^2\varphi^{'}}.
\label{equator}
\end{eqnarray}
 Since the amplitudes from neutrinos become largest seen from the
 equatorial plane of the source, we regard the Eq. (\ref{polar})
 as the base formula to compute the amplitudes from neutrinos.
 we use the  
gravitational waves from neutrinos are anti-beaming 

Next, we make an order-of-magnitude estimate of the
amplitude of the gravitational waves for neutrinos \cite{epstein_gw,mull_mem}.
From Eq. (\ref{polar}), one can estimate the amplitudes, $h_{\nu}$, as follows,
\begin{equation}
h_{\nu} = \frac{2 G}{c^4 D} \int_{- \infty}^{t - R/c} dt~L_{\nu}(t^{'})\cdot 
\alpha(t'), 
\end{equation}
 where $D$ is the distance to the source and $\alpha(t^{'})$ is 
the time-dependent anisotropy parameter,
\begin{equation}
\alpha(t^{'}) = \int_{4 \pi} d\Omega^{'} \Psi(\vartheta^{'}, \varphi^{'})~
 f(\Omega^{'},t^{'}),
\end{equation} 
representing the degree of the deviation of the neutrino emission
 from spherical symmetry. 
  Inserting typical values into the above
 equation, one can find the typical amplitudes,
\begin{equation}
h_{\nu} \sim 1.6 \times 10^{2} {\rm cm}~\frac{1}{D}
\Bigl(\frac{\alpha}{0.1}\Bigr)
 \Bigl(\frac{L_{\nu}}{10^{52}~{\rm erg}~{\rm s}^{-1}}\Bigr)
 \Bigl(\frac{\Delta t}{1 ~{\rm s}}\Bigr),
\label{estim}
\end{equation}
where we take an emission time of $\Delta t = 1$ s assuming constant
radiation and the optimistic
values of $\alpha \sim 0.1$ suggested from the numerical results
\cite{mull_mem}. Thus it is expected that the gravitational-wave 
amplitude from neutrinos can be larger than the one emitted at core 
bounce in rotational core-collapse (see Eq. (\ref{simple_estimate})). 
The typical frequency of the gravitational waves from neutrinos is
expected to be lower by an order of magnitude than the one at the core
bounce in rotational-core collapse, because the dynamical scale is not
determined at the central core $\rho \sim 10^{14} ~{\rm g}~{\rm cm}^{-3}$ 
but at the neutrinosphere $\rho \sim 10^{12}~{\rm g}~{\rm cm}^{-3}$.
%may be roughly estimated by the inverse of the dynamical timescale 
%near the neutrinosphere $\tau_{\nu} \sim 4 ~{\rm ms}
%~\rho_{\nu}/(1\times 10^{12}~{\rm g}~{\rm cm}^{-3})$, 
%\begin{equation}
%\nu_{\nu} \le \frac{1}{\tau_{\nu}} \sim O(10) {\rm Hz}.
%\end{equation} 

It is noted that the gravitational memory stems from the change in the
transverse-traceless part of the Coulomb-type ($\propto 1/r$), and
thus, can appear in other astrophysical events.
Recently, the memory effect 
generated by a point particle whose velocity 
changes via gravitational interactions with other objects is studied 
\cite{Segalis}. 
Further, the memory in the context of jets in gamma-ray
bursts is studied, which predicts that such gravitational waves are
likely to be detected by the space-based laser 
interferometers such as LISA and DECIGO/BBO \cite{Sago,hiramatsu}.

Now we return to mention the gravitational waves from neutrinos and
convections from core-collapse supernovae in the following sections.

\subsubsection{gravitational waves from convections and neutrinos in non-rotating stars}
\begin{figure}
\begin{center}
%\epsfbox{figure_kotake/BF2D_yevar.ps}
\epsfxsize = 7 cm
\epsfbox{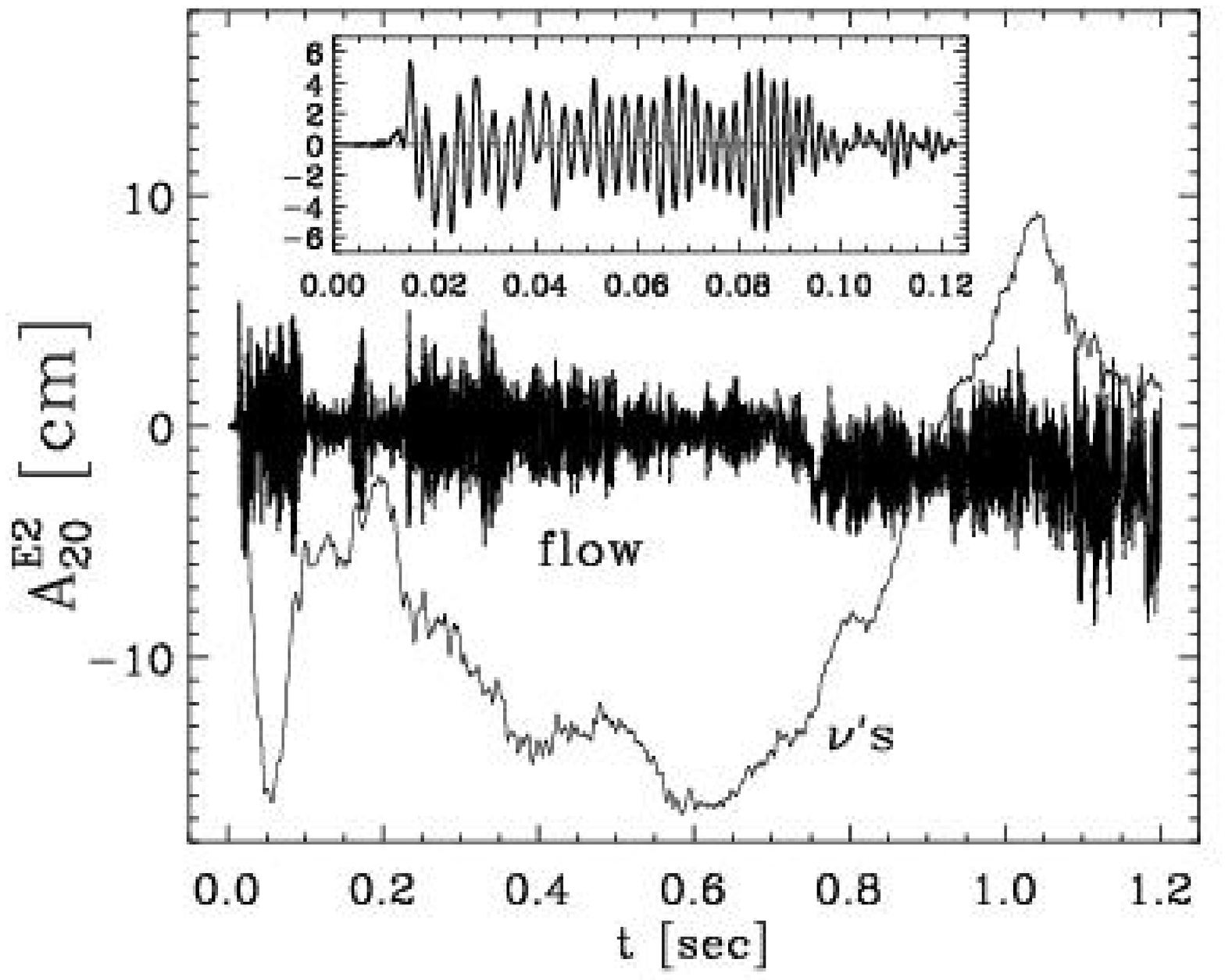}
\epsfxsize = 7 cm
\epsfbox{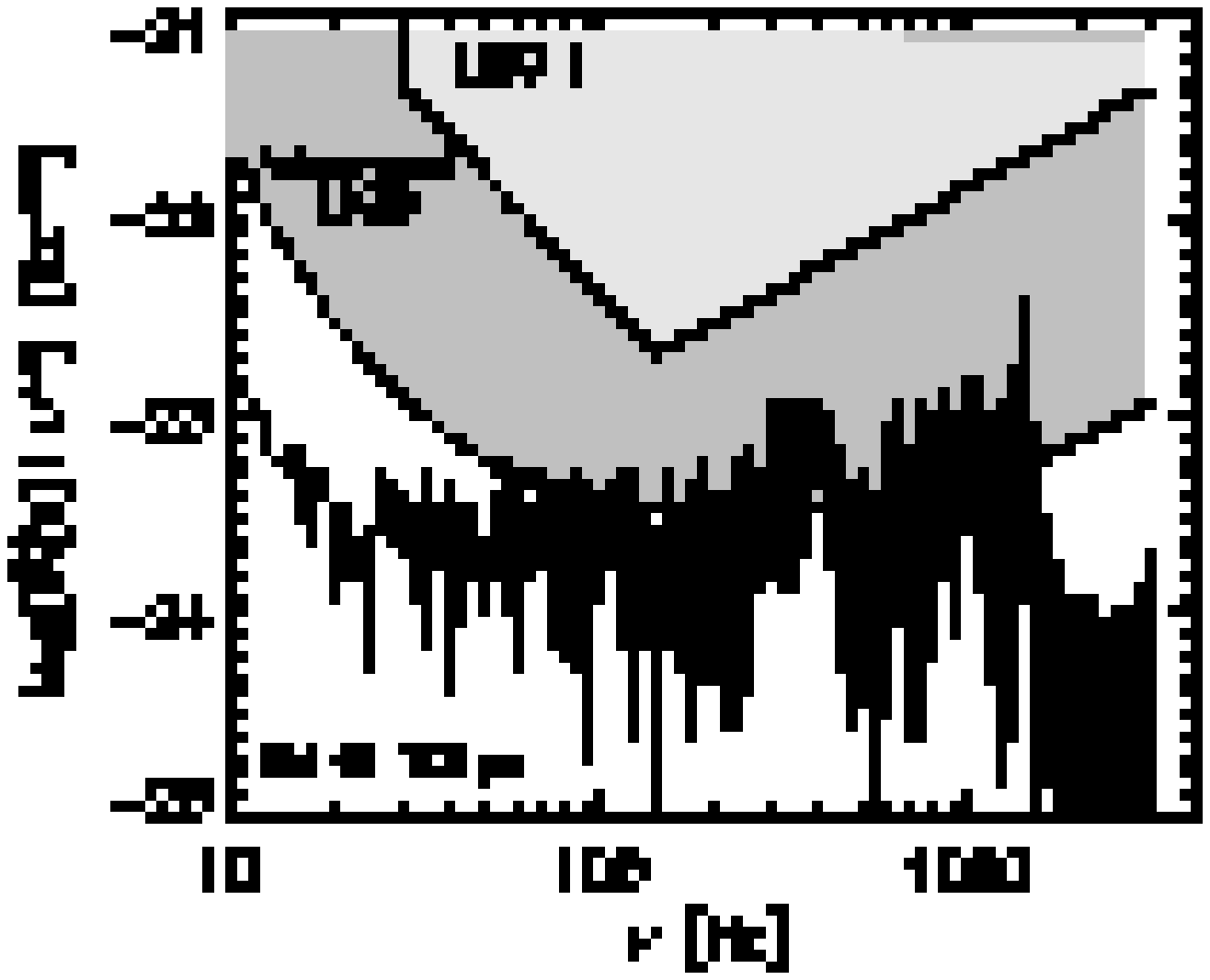} 
\vspace{0.5 cm}
\caption{Computations of the gravitational wave 
in the non-rotating protoneutron star (PNS) taken from \cite{muller03}.
The left panel
 shows the waveform due to the convective motion (thick line labeled as 
``flow'') and the anisotropic
 neutrino radiation (thin line labeled as ``$\nu'{\rm s}$''). The insert shows an
 enlargement of the signal until $\sim 100$ ms from the start of 
the simulation. 
The right panel shows the total gravitational spectrum
 with the sensitivity curves for some detectors. The source is located
 at a distance of 10 kpc.
These figures are taken from M\"{u}ller {\it et al} (2003) \cite{muller03}. }
\label{figmueller1}
\end{center}
\end{figure}

 As mentioned in section 
\ref{convec_chap2}, the convections are likely to occur in the
protoneutron stars and in the hot bubbles regions. 
In Figure \ref{figmueller1},
a typical GW waveform due to the convective motions and the associated
anisotropic neutrino radiation 
inside the non-rotating protoneutron star (PNS) is presented.
From the left panel, the time interval of the each 
GW signal from convections (thick line) 
is found to be very short with an order of milliseconds, while the
waveform associated with the neutrinos (thin line) shows much less time structure. 
The short interval of the GWs from matter
 reflects the timescale of the convective motion inside the
PNS, which may be roughly estimated as follows, 
$t_{\rm conv} \leq \sim R_{\rm PNS}/v_{\rm conv} \sim {O \rm (ms)} 
~(R_{\rm PNS}/ 20 {\rm km})/(v_{\rm conv}/1\times 10^{9}~{\rm cm}~ 
{\rm s}^{-1})$ with $R_{\rm PNS}$  and $v_{\rm conv}$ being the 
size of the PNS and the typical velocity of
the convective motions. Since the amplitudes both from matter and
neutrinos result from the small-scale motions induced
mainly by the negative gradient of the lepton fraction, the amplitudes
become much smaller than the ones at core bounce in rotational
core-collapse (typically $\le 1/10$). 
Due to the smaller amplitudes and the higher
frequencies of the emitted gravitational waves, both unlike the ones 
obtained at core bounce in rotational core-collapse, 
they are marginally within the detection limits 
for the laser interferometer in the next 
generation (see the right panel of Figure \ref{figmueller1}).
 
\begin{figure}
\begin{center}
\epsfxsize=7.5cm
\epsfbox{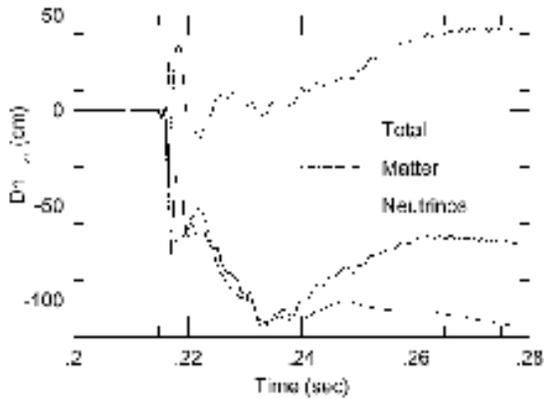}
\end{center}
\caption{The gravitational wave strain, $h_{\rm zz}^{TT}$ times the
 distance to the supernova, D, versus time. Core bounce is at 0,215
 seconds. Each line shows the corresponding contributions to the
 gravitational waves. The contributions of
matter motion and neutrinos to the GW amplitude can be seen  
of opposite sign at core bounce. Only for the first 20 ms after core bounce, 
the gravitational waves from neutrinos are shown to dominate over the
 mass motions. This figure is taken from Burrows \& Hayes \cite{burohey}.}
\label{figburohey}
\end{figure}

Not only inside the PNS, but also outside the PNS, the convections are 
likely to be induced, as stated earlier (see section \ref{convec_chap2}). 
Burrows \& Hayes (1996) \cite{burohey} performed 2D simulations, 
in which the density of the precollapse core was artificially reduced 15
\% within $20$ degree of the pole, and
demonstrated how the initial density inhomogeneity affects the 
gravitational waves both from the convective motions and the 
anisotropic neutrino radiation. The large density inhomogeneity assumed
in their work was predicted by the stellar evolution calculations, 
pointing out that they could be formed and amplified during silicon and oxygen 
burning stages \cite{bazan,goldreich}.
 The obtained properties of the waveform is presented in Figure
 \ref{figburohey}. They discussed that the total amplitude could be 
detected by the advanced LIGO, with a signal-to-noise ratio of 10, 
for a supernovae at a distance of 10 kpc.
% Furthermore, they pointed out that the initial
%large density asymmetries present prior to collapse can be the source of
%the larger proper motions observed in pulsars. In fact, a recoil
%velocity of the central neutron star becomes as high as $\sim 530~{\rm
%km}~{\rm s}^{-1}$.  

\begin{figure}
\begin{center}
\epsfxsize=7cm
\epsfbox{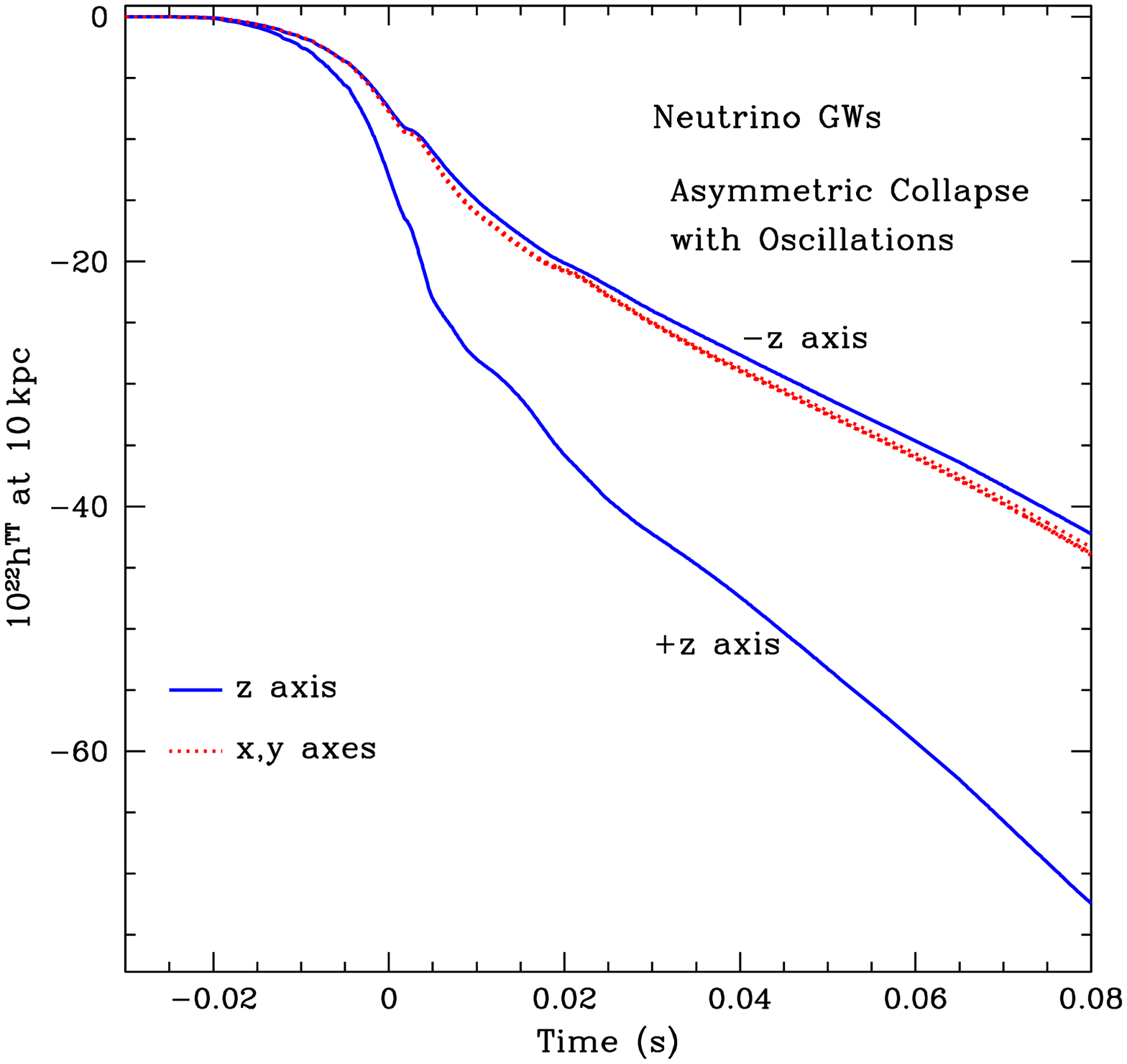}
\epsfxsize=7cm
\epsfbox{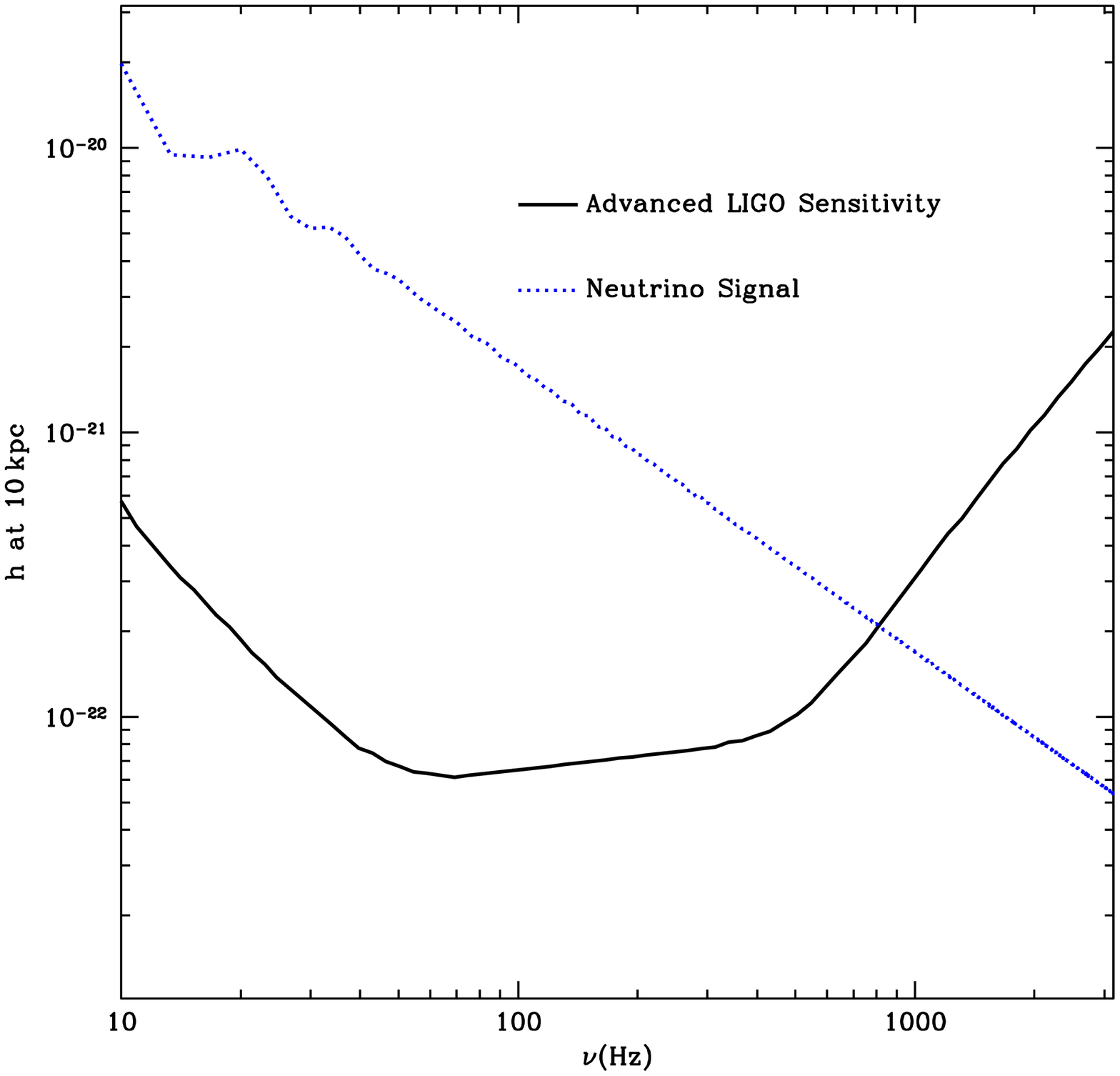}
\caption{Gravitational from anisotropic neutrino radiation as a function of time and observer location (left) and its spectrum (right), obtained in the 3D simulations \cite{fryergw}. In the model,  a 25 $\%$ core oscillation perturbation is assumed in the iron core of $15 M_{\odot}$ progenitor star for producing the asphericity of the neutrino radiation field. From the right panel, it can be seen that the gravitational wave from the neutrinos peaks at lower frequencies and within the detection limits for the Advanced LIGO for the galactic supernova at the distance of $10$ kpc. These figures are taken from Fryer {\it et al} \cite{fryergw}.}
\label{fryergw04}
\end{center}
\end{figure}

With almost the same motivation for the investigation, 
Fryer {\it et al} performed 3D SPH simulations of the
inhomogeneous core-collapse and discussed the waveforms \cite{fryer,fryerkick}.
The computed signatures of the gravitational waves are consistent with
the study of \cite{burohey}. 
They discussed that such gravitational waves are within the detection limits 
for the advanced LIGO for the galactic
supernova (see Figure \ref{fryergw04}). As shown by the simple
order-of-magnitude estimates (Eq. (\ref{estim})), it is seen from the right
panel that the peak amplitude form neutrinos becomes 
as high as the one at core bounce 
($h \sim 10^{-20}$) with the relatively lower peak frequency.  
%As a sideremark, the estimated recoil
%velocity of the protoneutron star is less than
%$\sim 200~{\rm km}~{\rm s}^{-1}$. This discrepancy may be due to 
%the problem of the fixing inner boundary. In the study of \cite{burohey}, the
%region inside a radius of 15 km was followed in 1D taken to be fixed
%boundary to reduce the computational burden, on the other hand, 
%such assumption was not taken in the study of \cite{fryer}.  

Both in the above studies of \cite{burohey} and \cite{fryer,fryerkick}, 
a large density inhomogeneity ($\sim O(10) \%$ fluctuations in the
density) prior to core-collapse is assumed in their initial conditions.
On the other hand, it is noted that a recent study pointed out by the
linear stability analysis in the cores of supernova progenitor stars
that the timescale for the growth of the nuclear burning 
(the so-called $\epsilon$ mechanism) is much longer than the time until the
commencement of core-collapse, hence such a large inhomogeneity may not 
develop \cite{murphy}.    
\begin{figure}
\begin{center}
\epsfxsize= 5cm
\epsfbox{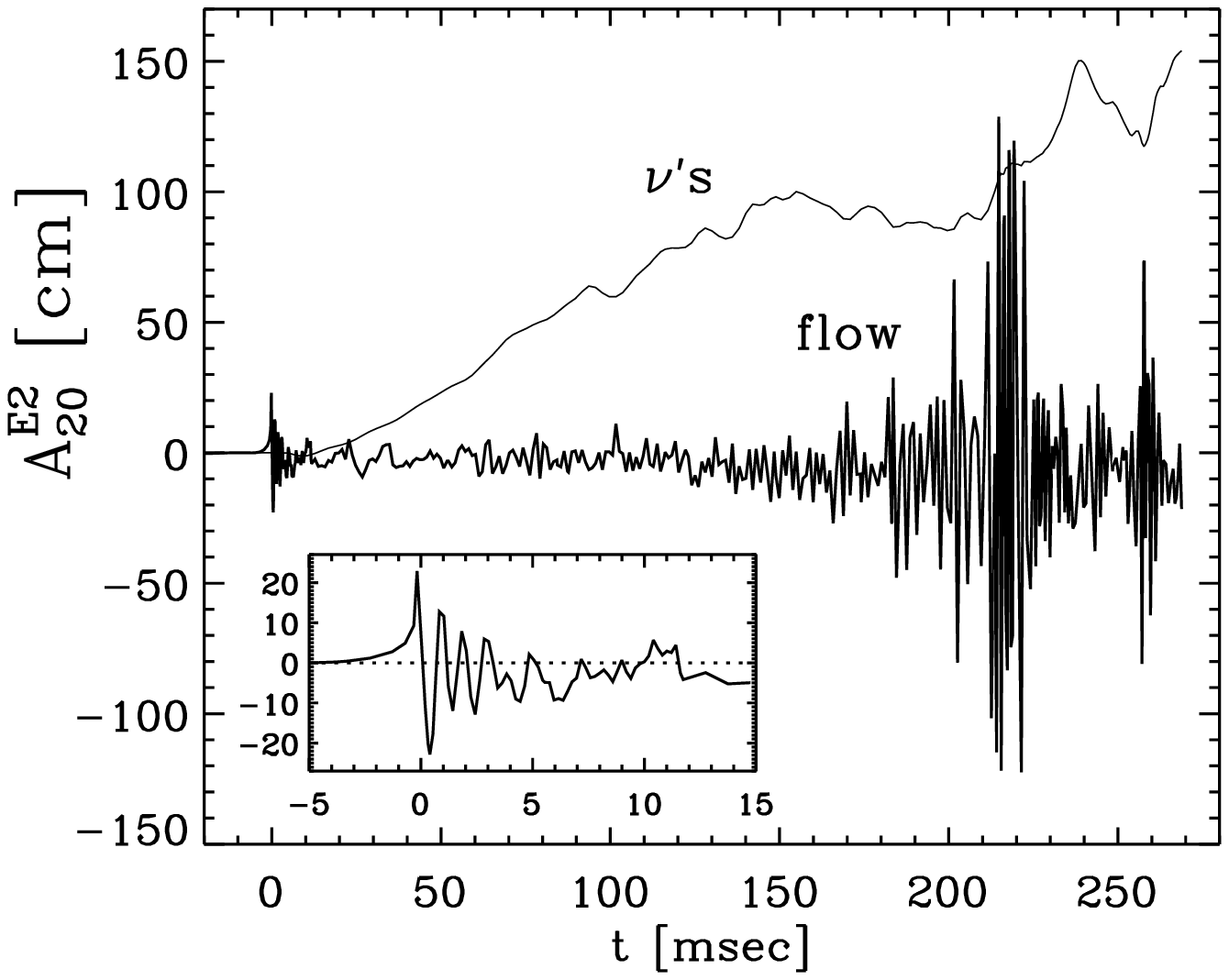}
\epsfxsize=5cm
\epsfbox{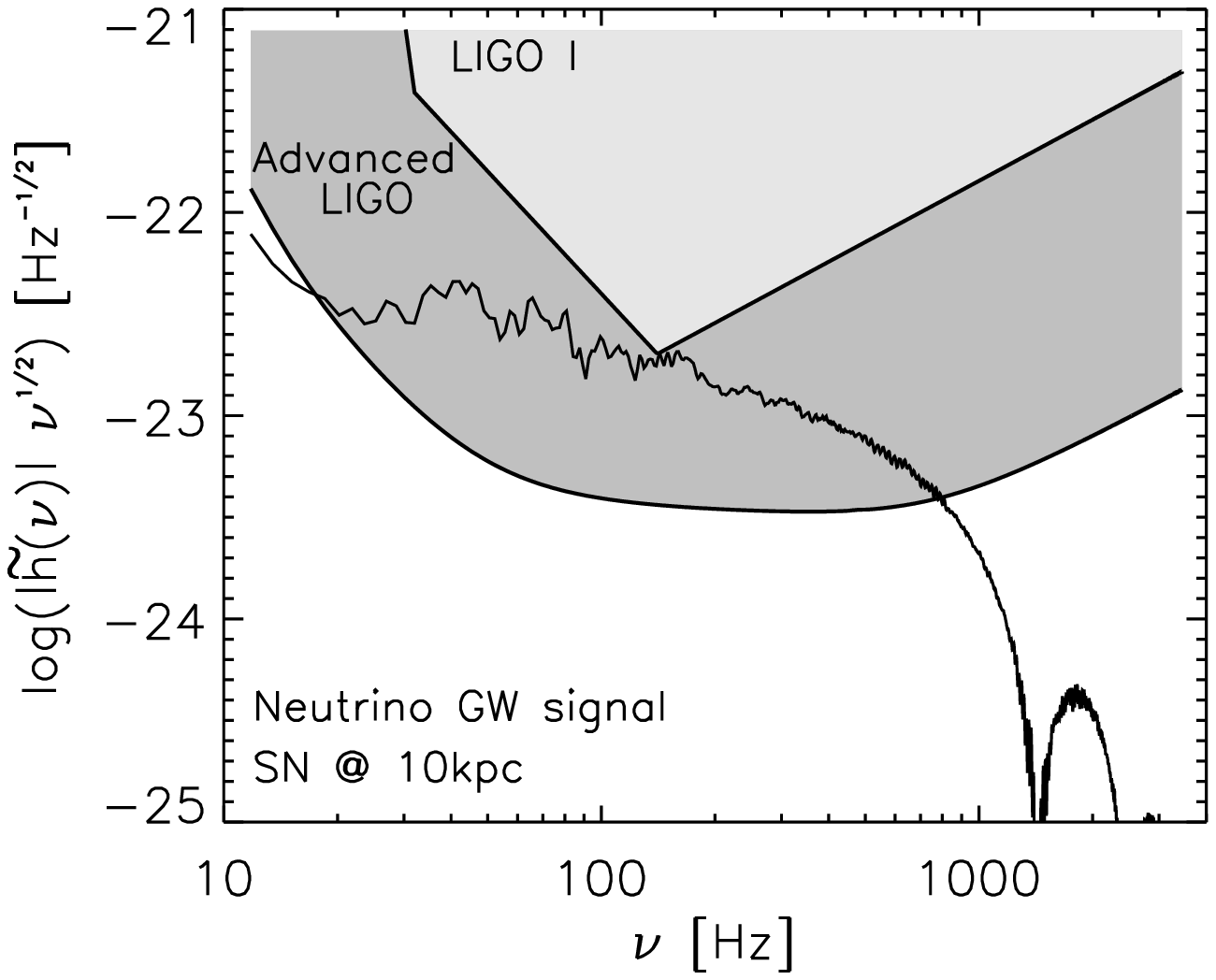}
\epsfxsize=5cm
\epsfbox{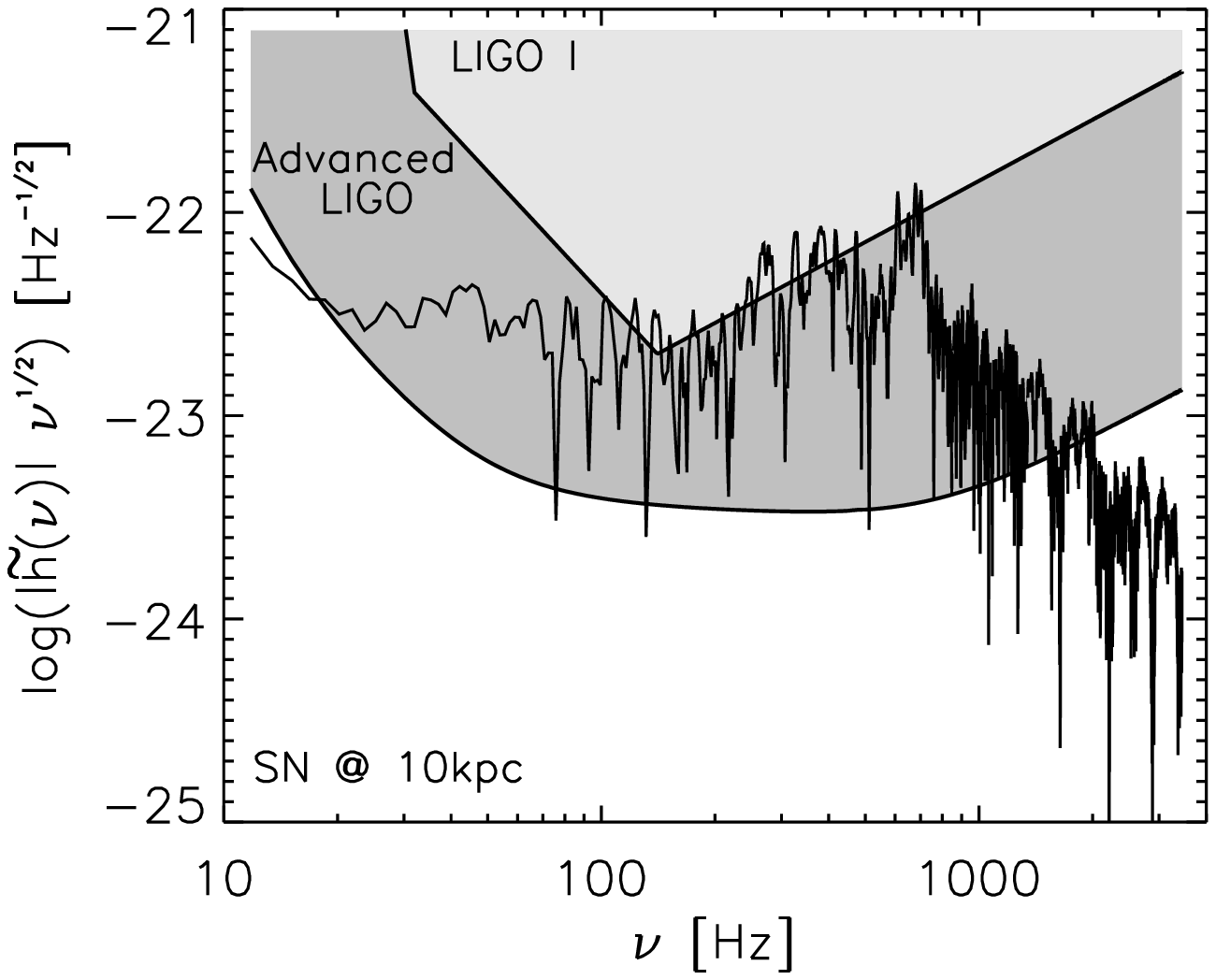}
\end{center}
\caption{Computations of the GW in rotational core collapse obtained by
 the state-of-the-art simulations ($\Omega = 0.5~{\rm rad}~{\rm s}^{-1}$
 is imposed on $15M_{\odot}$ progenitor model) \cite{muller03}.
In the left panel the waveforms contributed from neutrinos and the
 matter flows are shown. It is shown that the gravitational waves from
 neutrinos dominate over the ones from the flows almost always. 
The middle panel shows the spectrum of gravitational waves contributed
 from neutrinos. The right panel shows the total gravitational spectrum
 with the sensitivity curves for some detectors. Comparing the middle
 with the right panel, one can see that the gravitational waves in the lower
 frequency ($\leq 100$ Hz) are dominated by the neutrinos. 
These figures are taken from M\"{u}ller et al. (2003) \cite{muller03}.
}
\label{figmueller2}
\end{figure}
\subsubsection{gravitational waves from anisotropic neutrino radiation in rotating stars}
Another possibility to induce the
anisotropy of neutrino emissions is the stellar rotation.  
Recently, M\"{u}ller et al (2004) \cite{muller03} performed 
the rotational core collapse simulations employing the 
 elaborate neutrino transport with the detailed microphysics and 
calculated the gravitational waves from neutrinos. 
They found that the gravitational waves from the neutrinos 
grows due to convections and dominate over those of the matter 
at core bounce (see Figure \ref{figmueller2}). 

In their study, the initial models were limited to
the rather slower rotating cases based on a recent stellar evolution
models, in which the magnetic braking is taken into account 
\cite{heger03,heger04}. 
Recently, 
Kotake {\it et al} (2005) \cite{kotake_aniso_gw}
calculated the
waveforms from neutrinos employing a series of more rapidly rotating
models with changing the rotational profiles and the degree of 
differential rotation parametrically in order to see the effects of
anisotropy of the neutrino radiation induced dominantly by rapid rotation.

\begin{figure}[hbt]
\begin{center}
\epsfxsize=7.5cm
\epsfbox{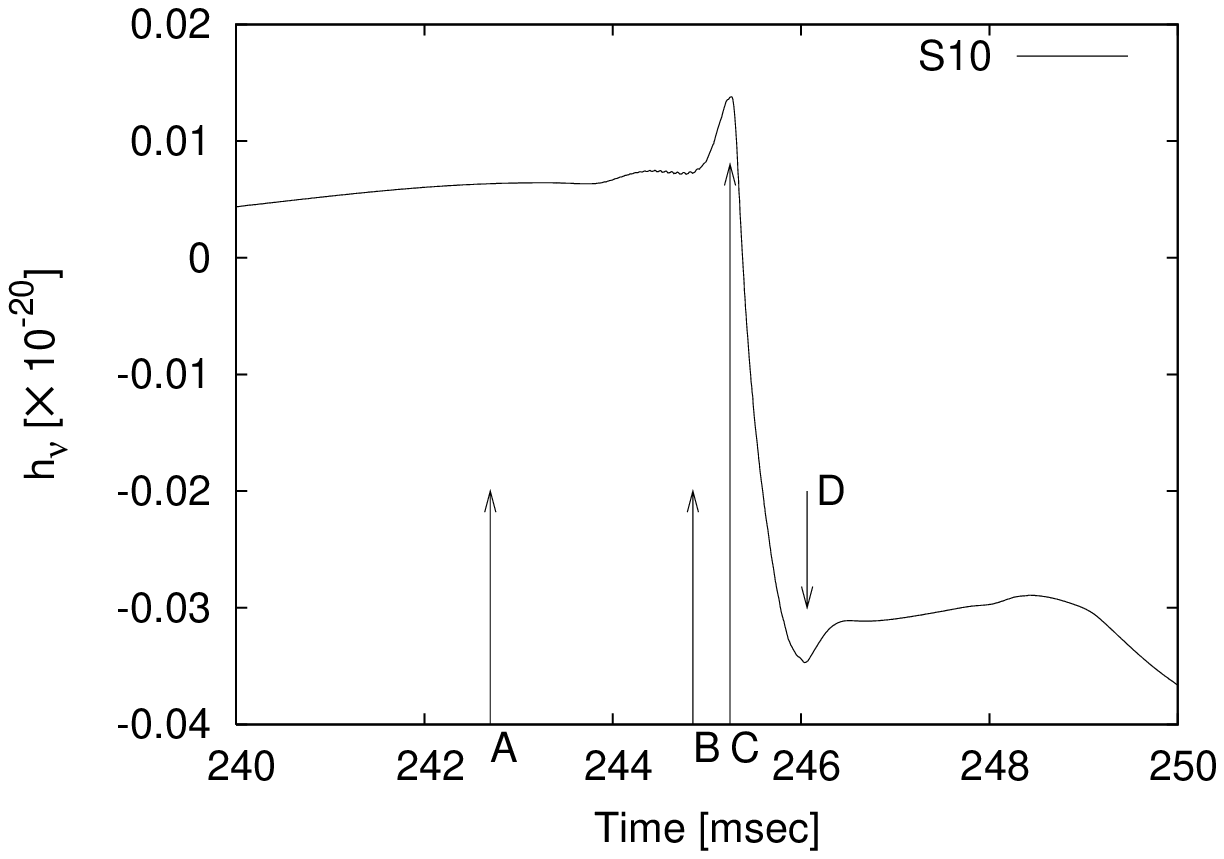}
\epsfxsize=7.5cm
\epsfbox{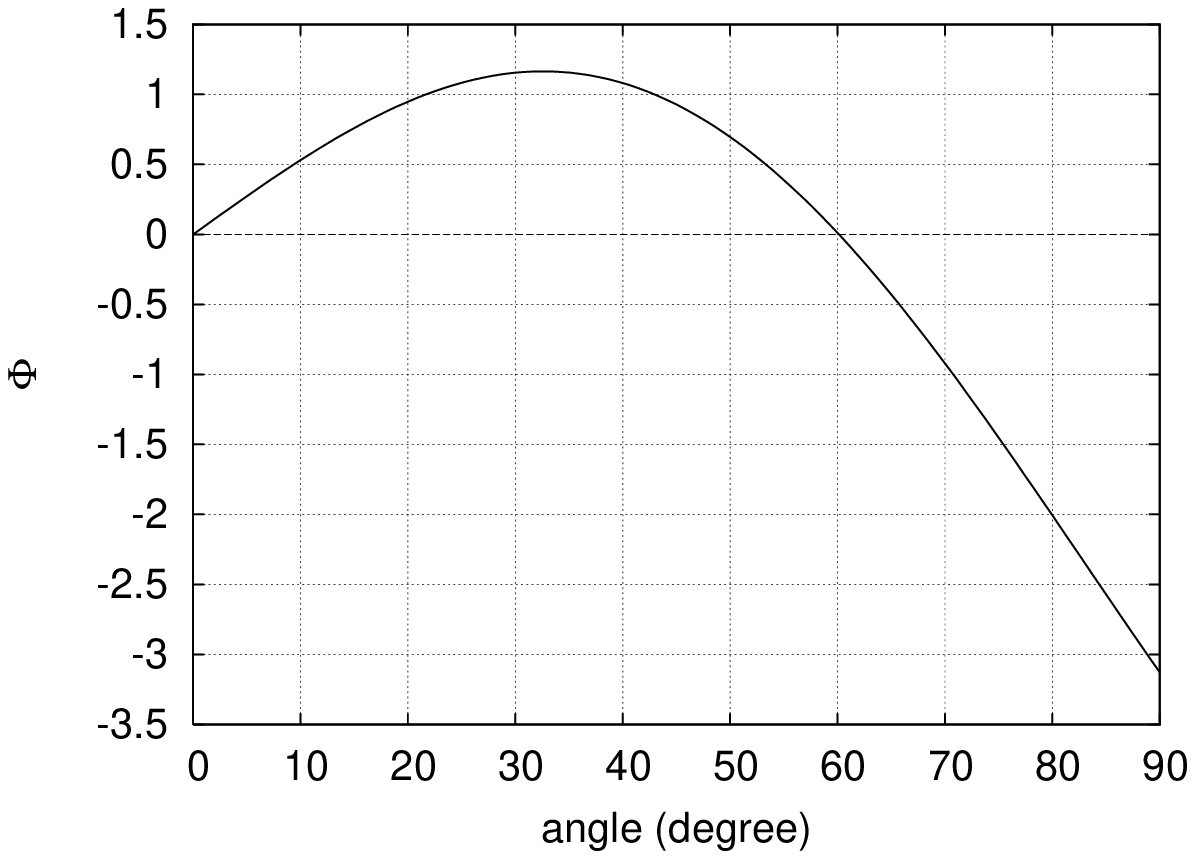}
\caption{Left panel is the waveform due to the anisotropic neutrino radiation for model S10.
In the panel, points A, B, C, and D represents the epoch of core
 bounce ($t = 242.7~{\rm ms}$), the onset of the first neutronization ($t = 244.8 ~{\rm ms}$), the peak of the
 neutronization, which corresponds to the onset of the second
 neutronization ($t = 245.3~ {\rm ms}$), and the offset of the
 second neutronization ($t = 246.1~{\rm ms}$), respectively. Right panel
 represents the angular dependence of $\Phi$ in Eq. (\ref{tt}). Note that the angle is measured
 from the rotational axis. These figures are
 taken from \cite{kotake_aniso_gw}.}
\label{fig0}
\end{center}
\end{figure}
\begin{figure}
\begin{center}
\epsfxsize=7.5cm
\epsfbox{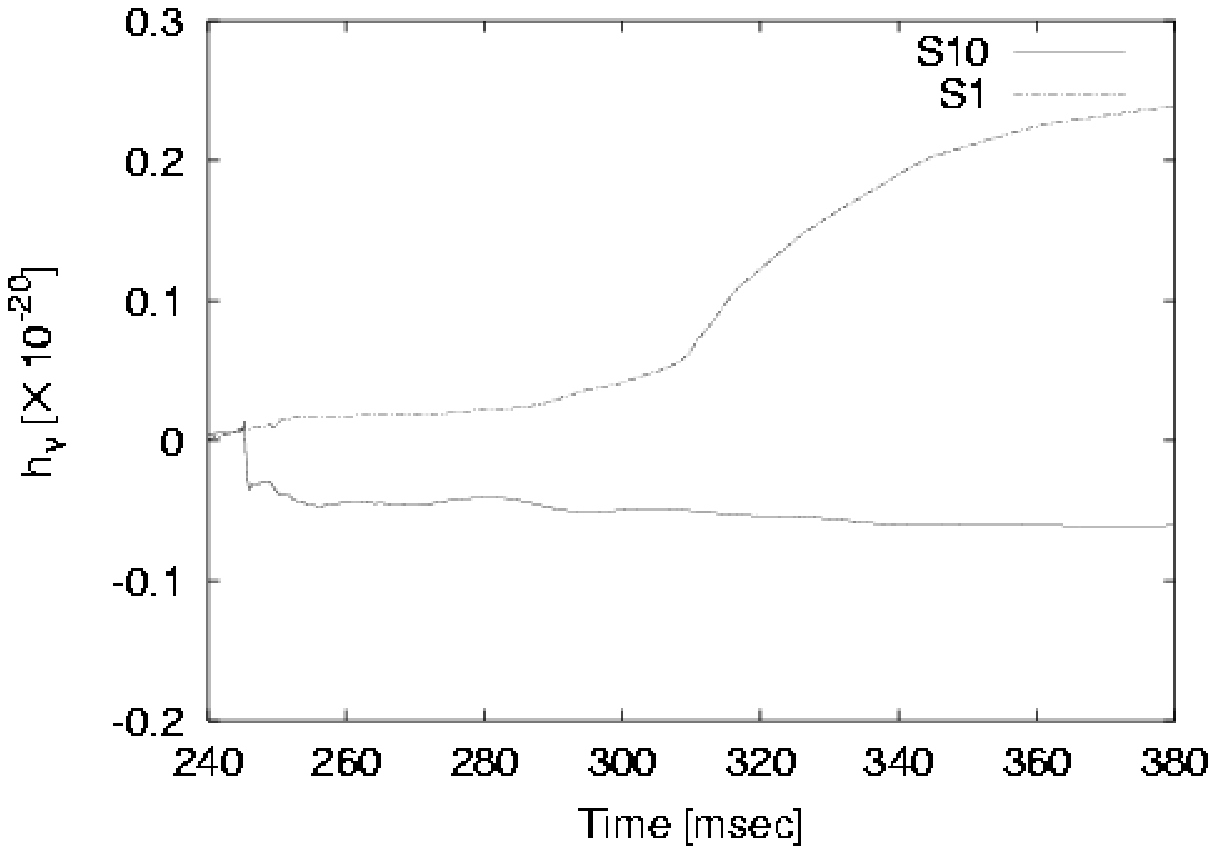}
\epsfxsize=7.5cm
\epsfbox{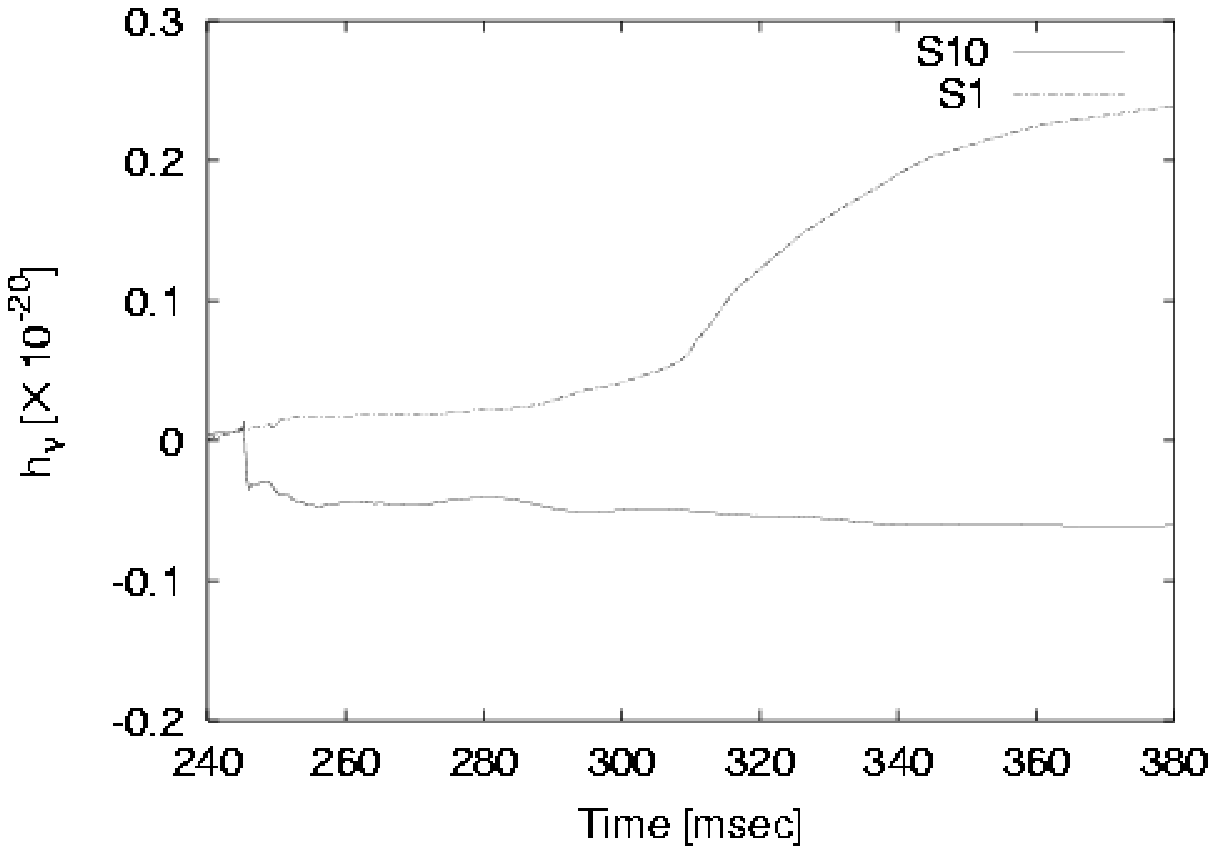}
\caption{Waveforms due to the anisotropic neutrino radiations for some
 representative models with the shell-type (left panel) and 
the cylindrical rotation (right panel) profiles. Note that the source is
 assumed to be located at the distance of 10 kpc. These figures are
 taken from \cite{kotake_aniso_gw}.}
\label{fig1_agw}
\end{center}
\end{figure}
%\subsection{Properties of waveforms from anisotropic neutrino radiation}
%First of all, we take model S10 as an example
%model, because the model is based on the recent stellar evolution
%calculation \cite{heger00}. 
In the left panel of Figure \ref{fig0}, the waveform for a typical model
studied in \cite{kotake_aniso_gw} near core bounce is given. Note that the model is based on the 
recent stellar evolution calculation while excluding the effects of
the magnetic fields \cite{heger00}.
About 2.1 ms after core bounce at point A in the panel, 
the amplitude begins to rise more steeply  than before 
(see point B in the figure).  This epoch corresponds to the so-called 
neutronization, which occurs when the shock wave goes over the neutrino sphere.
%At this moment, the trapped electron neutrinos inside the neutrino
%sphere decouple from the matter and propagate ahead of the shock waves.
Here it should be noted that the neutronization occurs anisotropically in their
rotating models. Due to the non-sphericity of the shapes of the shock
 wave and the neutrino sphere, the neutronization can occur more than
 once while it occurs only once for spherical models.
In which direction the 
neutronization occurs is determined by the shapes of the deformed
 neutrino sphere and the anisotropically propagating shock wave.
The first neutronization (at point B in the panel) occurs near the 
pole at the radius of  $\sim 20~{\rm km}$ along the rotational axis. 
The shape of the shock wave formed by core bounce is prolate 
initially because the bounce occurs near the rotational axis. 
This prolate shock wave crosses the neutrino sphere, whose shape 
is deformed to be oblate due to rotation. 
Since the area of surface, where the first neutronization occurs, 
is small, the neutrino luminosity is relatively low at this epoch
 ($L_{\nu} \sim 10^{52}~\rm{erg}~\rm{s}^{-1}$). 
About 0.5 ms after the first neutronization at point B in the left panel
of Figure \ref{fig0}, the amplitude shows sudden fall from $t = 245.3$
ms (point C) to $t = 246.1$ ms (point D). 
 After the first neutronization occurs at the pole, 
the shock wave in the polar region is weakened by the neutrino energy loss and ram pressure of the infalling material.
Then, the subsequent shock wave is formed in the vicinity of the equatorial plane and
 begins to move rather parallel to the plane. Since the first shock
 wave almost stalls along the rotational axis during a several
 millisecond after the first neutronization, the shape of the shock wave
 becomes rather oblate due to the propagation of the second shock wave
 rather parallel to the equatorial plane. 
As this oblate shock wave propagates along
the equatorial plane, it crosses the oblate neutrino sphere. At this
moment, the second neutronization occurs, where its luminosity becomes
maximum, whose value reaches as high as $\lesssim 10^{54} {~\rm erg}
~{\rm s}^{-1}$.

They explained the signatures of the gravitational waves at the two
epochs of the neutronization as follows. 
At the first neutronization, the neutrino emissions are 
concentrated near the rotational axis. For convenience to understand the
relation between the direction of neutrino emissions and the resultant
properties of the waveforms, we analytically 
integrate over $\varphi^{'}$ in Eq. (\ref{polar}) and 
obtain the formula in the closed form as follows,
\begin{equation}
h^{\rm TT}_{\nu} = \frac{8G}{c^4 R} \int_{-\infty}^{t - R/c} dt^{'}
\int_{0}^{\pi/2}~d\theta^{'}~\Phi(\theta^{'})~\frac{dL_{\nu}(\theta^{'},t^{'})}{d\Omega^{'}},
\label{tt}
\end{equation} 
where $\Phi (\theta)$ is the latitudinal angle dependent function,
\begin{eqnarray}
\Phi({\vartheta^{'}}) &=& \sin \vartheta^{'} \Bigl(-  \pi +  \int_{0}^{2\pi} d\varphi^{'}\frac{1 + \sin\vartheta^{'}\cos\varphi^{'}}{1 + \tan^2\vartheta^{'}
\sin^2\varphi^{'}}\Bigr),
\label{equator} 
\end{eqnarray}
(see
the right panel of Figure \ref{fig0}) and
${dL_{\nu}(\theta,t)}/{d\Omega}$ is the latitudinal angle dependent
neutrino luminosity. From the formula, one can readily see 
that no gravitational waves can be emitted if the neutrino emissions are spherical.

From the feature of $\Phi$ seen in the panel, the enhancement of the neutrino
radiation near the rotational axis does not work sufficiently to change
the amplitude although the amplitude shows a rise at the
first neutronization due to the non-zero contributions of $\Phi$
and the neutrino luminosity near the rotational axis (see from point B to C in Figure \ref{fig0}).
On the other hand, at the second neutronization, the neutrino
emissions are concentrated in the vicinity of the
equatorial plane. The value of the
function $\Phi$ is negative ($\theta = \pi/2$) and its absolute value is large 
compared to the value near the rotational axis ($\theta \sim 0$) (see
the right panel of Figure \ref{fig0}). In addition, the neutrino luminosity is much
larger at the second neutronization. This is because the area of the
surface at the second neutronization is quite larger than that at the 
first neutronization near the pole, since
the oblate shock wave crosses the oblate neutrino sphere. As a result,
the amplitude of the gravitational waves from the neutrinos
shows a steep fall at the second neutronization. 

%\subsection{Effect of differential rotation}

Moreover they demonstrated that the differential rotation mainly
determines the waveforms. In Figure \ref{fig1_agw}, the waveforms $h_{\nu}$ from near core
bounce up to the final time of their simulation are presented for the
representative models with the 
shell-type (left panel) and cylindrical rotation profiles (right panel), 
respectively. It can be seen from the figures that the amplitude due to
the neutrinos in the later times becomes much larger for the
models with the stronger differential rotation (compare S10 (weakest
differential rotation with initial angular velocity cut of $1000$ km)
with S1 (strongest differential rotation with initial angular velocity
cut of $100$ km), CS10
(weakest one) with CS1 (strongest one), respectively). 
Simultaneously, it is found that this feature is regardless of the 
rotational profiles (compare the
left (shell-type rotation) with the right panels (cylindrical rotation) 
in Figure \ref{fig1_agw}). As the differential rotation becomes stronger,
the shape of the shock wave becomes more prolate, which makes the
neutrino emission more stronger in the vicinity of the rotational axis.
This makes the gravitational amplitudes from the neutrinos more larger 
(see \cite{kotake_aniso_gw} for the more detailed explanations).

\begin{figure}[hbtp]
\begin{center}
\epsfxsize=7.5cm
\epsfbox{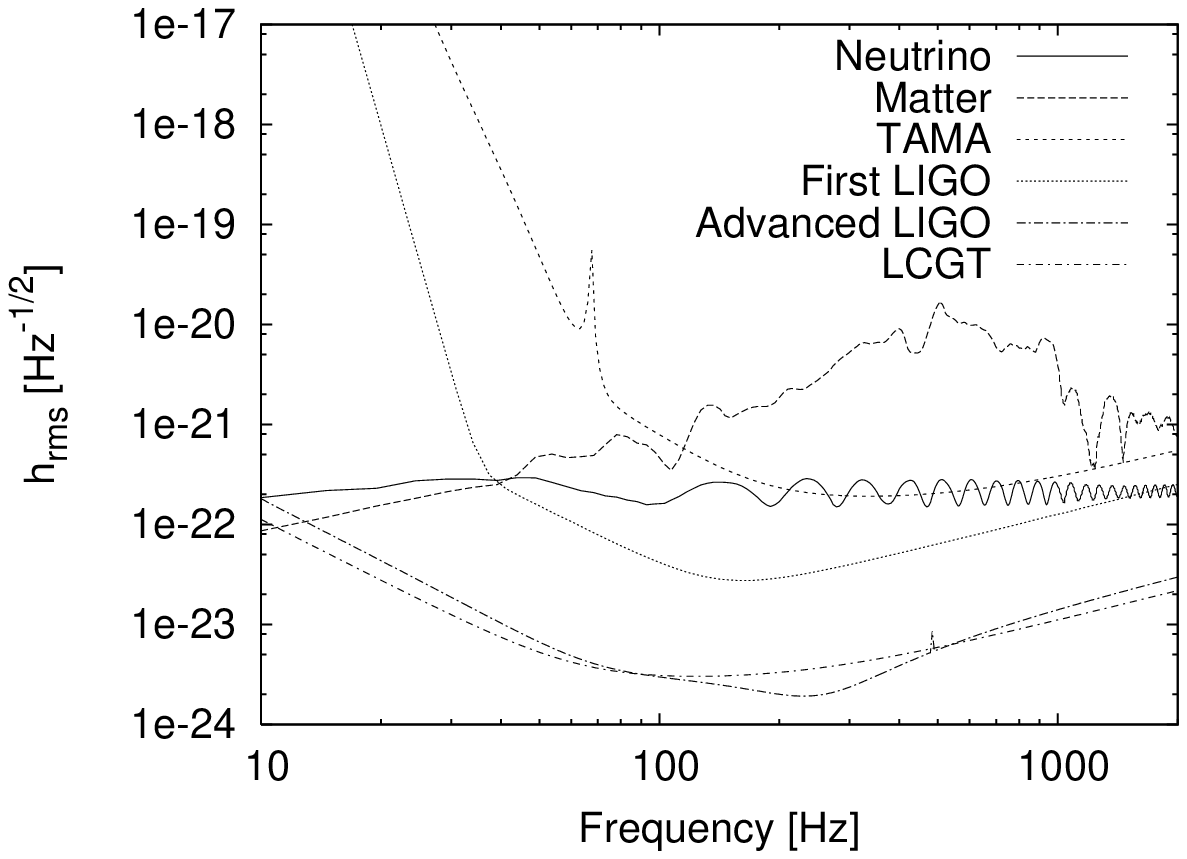}
\epsfxsize=7.5cm
\epsfbox{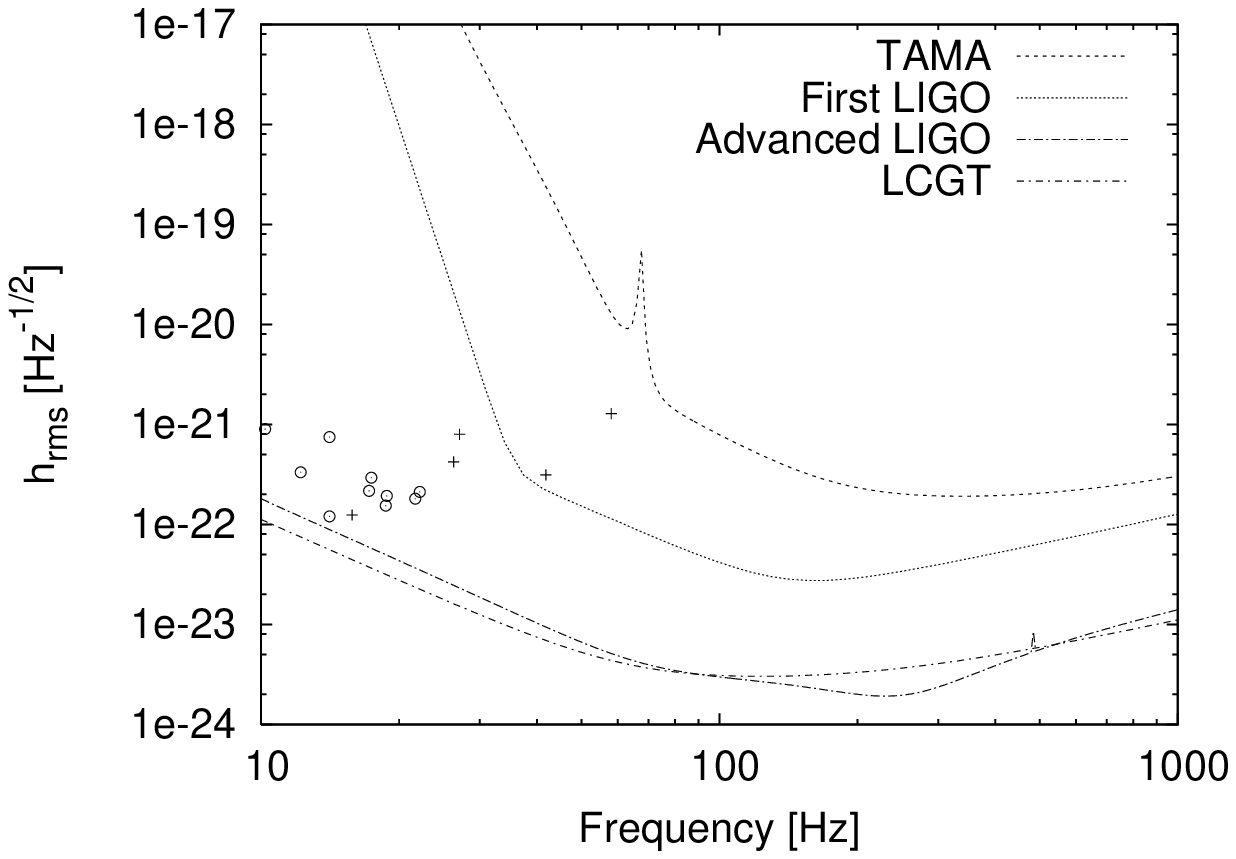}
\caption{Detection limits of TAMA, first LIGO, advanced LIGO, and Large-scale
 Cryogenic Gravitational wave Telescope (LCGT) with
 the expected gravitational wave spectrum obtained from the 
numerical simulations. The left panel shows the gravitational-wave
 spectrum contributed from neutrinos (solid) and from the matter (dashed)
in a rotating model with $\Omega = 4~{\rm rad}~{s}^{-1}$ imposed
 initially on a 15 $M_{\odot}$ progenitor model. 
 In the right panel, the open circles and the pluses 
 represent the amplitudes of $h_{\nu, {\rm eq}}$ with the characteristic
 frequencies of $\nu_{\rm eq}$ for the models with the cylindrical and 
the shell-type rotation profiles, respectively. Under the frequency
 of $\nu_{\rm eq}$, the gravitational waves from the neutrinos dominate
 over those from the matter contributions. 
From the panel, it is
 seen that the gravitational waves from neutrinos dominate over the ones
 from the matter in a lower frequency ($f \leq 100$ Hz).
Note that the source is assumed to be located at the distance of 10 kpc.
These figures are
 taken from \cite{kotake_aniso_gw}.}
\label{S}
\end{center}
\end{figure}

%\subsection{Detectability}
In the left panel of Figures \ref{S}, the root mean square (rms) 
sensitivity curves with the rms gravitational wave spectra 
for the standard model is given. From the panel, 
it can be seen that the gravitational waves due
to the anisotropic neutrino radiation dominate over those due to the 
mass motions at the frequency lower than several 10 Hz.   
In the 21 models computed, it is found that the values 
of $\nu_{\rm eq}$ ranges from 8  to 58.1 Hz with the rms amplitude 
ranging from $7.28 \times 10^{-23}$ to $1.28 \times 10^{-21}$ and 
that the values of $h_{\nu, \rm eq}$ generally become larger for the
stronger differential rotation models. Here $\nu_{\rm eq}$ represents a
characteristic frequency, below which the
gravitational waves from the neutrinos dominate over those from the mass 
motions, and the corresponding rms gravitational wave amplitude is named
as $h_{\nu,\rm eq}$. In the right panel of
Figure \ref{S}, the values of  $h_{\nu, \rm eq}$ with $\nu_{\rm eq}$ for all the models are plotted with the sensitivity curves of laser interferometers. 
It can be seen that the gravitational waves from the neutrinos are
the detection limits of the detectors in
the next generation such as LCGT and advanced LIGO.
Thus, it is suggested that the detection of the gravitational wave 
at the low frequency range ($\lesssim 100$ Hz) becomes more hopeful due to
the contributions from the anisotropic neutrino radiation.

\clearpage

\section{Concluding Remarks}
The aim of writing this article was to provide an overview of what we
currently know about the explosion mechanism, neutrinos, and
gravitational waves in core-collapse supernovae.
As we have discussed, much progress has been made in each topics.

Recently, multidimensional studies and simulations of core-collapse
supernovae have come into blossom again since 1990's when the direct
observations of global asymmetry in SN 1987A were reported. The
current trend might be ascribed to the fact that spherically symmetric 
supernova simulations have not yet produced explosions, albeit with the 
probably ``ultimate'' sophistication of the neutrino-transport method
including the state-of-the-art microphysics. A step beyond the spherical
models is in steady progress. It seems most natural and convincing 
to investigate the effect of the asphericity on the neutrino heating 
mechanism. Many ingredients to produce asymmetry in the supernova core
 have been considered, e.g., stellar rotation, magnetic fields, convection,
 and the standing accretion shock instability.
In order to see the real outcome of them, we should be able to perform 
at least two-dimensional fully angle-dependent radiation-hydrodynamic
calculations.
%, which is still computationally intensive. 
In fact, several groups are really pursuing it with the use of advanced
numerical techniques. 

Understanding the explosion mechanism of
core-collapse supernovae is
important not only for itself but also for the theoretical
understanding for the other (astro)physical relevance, 
such as neutrino and gravitational-wave
emissions. Conversely, we are now being able to understand 
them from the observations. In fact, neutrino and 
gravitational-wave astronomy are now becoming reality. 

Neutrino is a powerful tool to probe deep inside of the supernova
while we can see just its surface by electromagnetic waves.
In fact, supernova gave the first stage for the Neutrino Astronomy
when we observed neutrinos from SN1987A.
Neutrinos reflect the physical state of the core and even the density
structure of the mantle if we consider neutrino oscillation.
Actually neutrino oscillation is a necessary physics when we want
to extract information on supernova from observed neutrinos
because neutrino oscillation changes the neutrino spectra.
However, this cannot be done so easily because there are some
ambiguities in one of neutrino oscillation parameters, $\theta_{13}$,
and mass hierarchy, although we have other parameters with high
accuracies. Conversely, this means that supernova can be a unique
laboratory which probes the unknown neutrino parameters which
can not be studied by other experiments such as solar, atmospheric,
accelerator and reactor experiments.

Gravitational wave is also a powerful tool, which gives us the information
deep inside the supernova core. If the supernova core rotates rapidly, 
the burst-like gravitational waves are emitted at core bounce, 
and their amplitudes are as strong as the currently running 
interferometric observatories could detect for a galactic supernova.
While this is not the case, gravitational waves emitted in the later
phase due to the convective motions and the anisotropic neutrino
emissions will be most promising for the detections. Especially, the gravitational signal
from neutrinos have different features from the other, in the sense that
they have {\it memory}, the detection may need somewhat different
technique. If we could detect the gravitational waves from anisotropic
neutrino emissions
simultaneously with neutrinos themselves, 
it will bring about the
great progress in the understanding of the explosion mechanism, because 
the anisotropy should play a important role.
Moreover, the mutual
understanding of the explosion mechanism, the supernova neutrinos, and
the gravitational waves, which we reviewed somewhat separately in
this article, will be greatly progressed.

Moreover, supernova study is indispensable for the understanding of some hot astrophysical issues, such as the 
central engine of gamma-ray bursts and the nucleosynthesis in the explosion
of population III stars.  Further
 study will have a great impact not only on disclosing the mechanism of such 
astrophysical phenomena, but also on unveiling the fundamental properties
of particle physics.

\clearpage

\section*{Acknowledgments}
We are happy to thank Shoichi Yamada for helpful conversations.
K.K. is thankful to K. Numata and M. Ando for helpful discussions and to
A. Kotake for continuing encouragements.
K.T. would like to thank J. R. Wilson and A. Serenelli for informative
discussions and M. Takahashi for a great help.
It is pleasure to thank Alex Heger, Anthony Mezzacappa, Hideki Madokoro,
Kohsuke Sumiyoshi, Robert Buras, Adam Burrows, Matthias Liebend\"{o}rfer, Hans Thomas
Janka, Todd Thompson, Christian Ott, Shizuka Akiyama, John Blondin, Tatsuya Yamasaki, Rolf Walder,
Ewald M\"{u}ller, Chris Fryer, Leonard Scheck, James R Wilson,
Aldo Serenelli, Shoichi Yamada, Tomoya
Takiwaki, Hidetomo Sawai, and Yudai Suwa for helpful conversations
and/or permission to reprint figures from their published works.
Finally, we would like to gratefully thank to the late Professor
John Bahcall, who contributed not only
to the neutrino astrophysics, which we mention in this article, but also
to many areas of astrophysics including the study of dark matter in the 
universe, quasar properties, galactic structure, 
and the identification of the first neutron star companion.
Without his great contributions, the neutrino astrophysics could never be
developed as it is today.

This work was supported in part by the Japan Society for
Promotion of Science (JSPS) Research Fellowships (K.K., K.T.) and a
Grant-in-Aid for Scientific Research from the Ministry of
Education, Science and Culture of Japan through No.S 14102004,
No. 14079202, and No. 14740166.


\clearpage

\section*{References}
\begin{thebibliography}{}
% aaa 
\bibitem{Achkar95}
Achkar B {\it et al} 1995
Search for neutrino oscillations at 15, 40 and 95 meters from a nuclear power reactor at Bugey
{\it Nucl. Phys. B} {\bf 434} 503-532

\bibitem{Aglietta92}
Aglietta M {\it et al} 1992 
THE MOST POWERFUL SCINTILLATOR SUPERNOVAE DETECTOR: LVD
{\it Il Nuovo Cimento A} {\bf 105} 1793-1804

\bibitem{AkhmedovLunardiniSmirnov02}
Akhmedov E Kh {\it et al} 2002
Supernova neutrinos: difference of $\nu_{\mu} - {\nu}_{\tau}$ fluxes and conversion effects
{\it Nucl. Phys. B} {\bf 643} 339-366
\bibitem{akiyama} Akiyama S {\it et al} 2003 The Magnetorotational Instability in Core-Collapse Supernova Explosions {\it Astrophys. J.} {\bf 584} 954 - 70
\bibitem{Alekseev87}
Alekseev E N {\it et al} 1987
POSSIBLE DETECTION OF A NEUTRINO SIGNAL ON 23 FEBRUARY 1987
AT THE BAKSAN UNDERGROUND SCINTILLATION TELESCOPE OF THE INSTITUTE OF NUCLEAR RESEARCH
{\it JETP Lett.} {\bf 45} 589-592

\bibitem{AMANDAHP}
AMANDA Collaboration, web page, http://amanda.wisc.edu/

\bibitem{AMANDA02}
AMANDA Collaboration 2002 {\it Mod. Phys. Lett. A} {\bf 17} 2019

\bibitem{tama} Ando M and TAMA collaboration 2002 {\it Class.\ Quantum
	Grav.} {\bf 19} 1409
\bibitem{AndoSato02}
Ando S and Sato K 2002
Determining the Supernova Direction by its Neutrinos
{\it Prog. Theor. Phys.} {\bf 107} 957

\bibitem{ando} Ando S 2003 Asymmetric neutrino emission due to
	neutrino-nucleon scatterings in supernova magnetic fields
 {\it Phys. Rev. D} {\bf 68} 063002
\bibitem{ArafuneFukugita87}
Arafune J and Fukugita M 1987 
PHYSICAL IMPLICATIONS OF THE KAMIOKA OBSERVATION OF NEUTRINOS FROM SUPERNOVA SN1987A.
\bibitem{ard} Ardeljan N V {\it et al} 2000 Nonstationary magnetorotational processes in a rotating magnetized cloud {\it Astron. Astrophys.} {\bf 355} 1181 - 90
\bibitem{arnet1}Arnett W D 1983 Neutrino escape, nuclear dissociation,
	and core collapse and/or explosion {\it Astrophys. J.} {\bf 263} L55-L57

{\it Phys. Rev. Lett.} {\bf 59} 367

\bibitem{ArnettRosner87}
Arnett W and Rosner J 1987 
NEUTRINO MASS LIMITS FROM SN1987A
{\it Phys. Rev. Lett.} {\bf 58} 1906

\bibitem{arras1} Arras P and Lai D 1999 Can Parity Violation in Neutrino Transport Lead to Pulsar Kicks? {\it Astrophys. J.} {\bf 519} 745 - 9

\bibitem{arras2} Arras P and Lai D 1999 Neutrino-nucleon interactions in
	magnetized neutron-star matter: The effects of parity violation
	{\it Phys. Rev. D} {\bf 60} 043001-1 - 28
\bibitem{arzoumanian} 
Arzoumanian Z {\it et al} 2002 The Velocity Distribution of Isolated
	Radio Pulsars {\it Astrophys. J.} {\bf 568} 289 - 301 
% bbb
\bibitem{label1} Baade W and Zwicky F 1934 Remarks on Super-Novae and
 Cosmic Rays {\it Phys. Rev.} {\bf 46} 76-77 
\bibitem{BahcallHP}
J. N. Bahcall, web page, http://www.sns.ias.edu/~jnb/

\bibitem{BahcallText}
Bahcall J N 1989 {\it Neutrino Astrophysics} (Cambridge University Press).

\bibitem{Bahcall04}
Bahcall J N 2004
Solar Models and Solar Neutrinos: Current Status
hep-ph/0412068

\bibitem{BahcallBasu05}
Bahcall J N {\it et al} 2005 
Helioseismological Implications of Recent Solar Abundance Determinations
{\it Astrophys. J} {\bf 618} 1049-1056

\bibitem{BahcallGlashow87}
Bahcall J N and  Glashow S L 1987 
UPPER LIMIT ON THE MASS OF THE ELECTRON-NEUTRINO
{\it Nature} {\bf 326} 476

\bibitem{BahcallSerenelliBasu05}
Bahcall J N {\it et al} 2005 
New solar opacities, abundances, helioseismology, and neutrino fluxes
{\it Astrophys. J} {\bf 621} L85-L88

\bibitem{BahcallSpiegelPress}
Bahcall J N {\it et al} 
{\em SN1987A in The Large Magellanic Cloud},
Proceedings of the Fourth George Mason Astronomy Workshop,
Fairfax, Virginia, ed. M. Kafatos, (Cambridge Univ. Press), p. 172.

\bibitem{balbus1} Balbus S A and Hawley J F 1998 Instability, turbulence, and enhanced transport in accretion disks Reviews of Modern Physics
 {\bf 70} 1-53
\bibitem{balbus2} Balbus S A and Hawley J F 1991 A powerful local shear instability in weakly magnetized disks. I - Linear analysis. II - Nonlinear evolution
{\it Astrophys. J.} {\bf 376} 214 - 33

\bibitem{Bandyopadhyay03}
Bandyopadhyay A {\it et al} 2003 
Prospects of probing $\theta_{13}$ and neutrino mass hierarchy by Supernova Neutrinos in KamLAND
hep-ph/0312315

\bibitem{Battistoni03}
Battistoni G {\it et al} 2003 
The FLUKA atmospheric neutrino flux calculation
{\it Astrop. Phys.} {\bf 19} 269-290


 \bibitem{baron} Baron E {\it et al} 1985 Type II supernovae in
	12$M_{\odot}$ and 15$M_{\odot}$ stars: The equation of state and general
	relativity {\it Phys. Rev. Lett.} {\bf 55} 126-9 
\bibitem{bazan} Bazan G and Arnett D 1998 Two-dimensional Hydrodynamics of Pre--Core Collapse: Oxygen Shell Burning {\it Astrophys. J.} {\bf 496} 316 - 32

\bibitem{BeacomVogel99}
Beacom J F and Vogel P 1999 
Can a supernova be located by its neutrinos?
{\it Phys. Rev. D} {\bf 60} 033007
\bibitem{bethegain} Bethe H A and Wilson J R 1985 Revival of a stalled
	supernova shock by neutrino heating {\it Astrophy. J.} {\bf
	295} 14-23
\bibitem{Bethe86}
Bethe H A 1986 
A POSSIBLE EXPLANATION OF THE SOLAR NEUTRINO PUZZLE
{\it Phys. Rev. Lett.} {\bf 56} 1305
\bibitem{betherev} Bethe H A 1990 Supernova mechanisms {\it
	Rev. Mod. Phys.} {\bf 62} 801-866 

\bibitem{Bionta87}
Bionta R M {\it et al} 1987 
OBSERVATION OF A NEUTRINO BURST IN COINCIDENCE WITH SUPERNOVA SN1987A IN THE LARGE MAGELLANIC CLOUD
{\it Phys. Rev. Lett.} {\bf 58} 1494-1496
\bibitem{Bisno}Bisnovatyi-Kogan G S and
		       Ruzmaikin A A 1976 The accretion of matter by a
		       collapsing star in the presence of a magnetic
		       field. II - Selfconsistent stationary picture
		       {\it Astrophys. Space. Science} {\bf 42} 401 - 24
\bibitem{blo03} 
Blondin J M {\it et al} 2003 Stability of Standing Accretion Shocks,
	with an Eye toward Core-Collapse Supernovae {\it Astrophys. J.}
	{\bf 584} 971 -80 
\bibitem{Boehm00a}
Boehm F {\it et al} 2000 
Search for Neutrino Oscillations at the Palo Verde Nuclear Reactors
{\it Phys. Rev. Lett.} {\bf 84} 3764-3767

\bibitem{Boehm00b}
Boehm F {\it et al} 2000 
Results from the Palo Verde Neutrino Oscillation Experiment
{\it Phys. Rev. D} {\bf 62} 072002

\bibitem{Bratton88}
Bratton C B {\it et al} 1988 
ANGULAR DISTRIBUTION OF EVENTS FROM SN1987A
{\it Phys. Rev. D} {\bf 37} 3361-3363
\bibitem{braginskii} Braginskii V B \& Thorne K S 1987
	Gravitational-wave bursts with memory and experimental prospects
	{\it Nat} {\bf
	327} 123 - 5
\bibitem{boden} Bodenheimer P and Woosley S E 1983 A two-dimensional supernova model with rotation and nuclear burning {\it Astrophys. J.} {\bf 269} 281 - 91

 \bibitem{bonazzola} Bonazzola S and Marck J A 1993 Efficiency of
	gravitational radiation from axisymmetric and 3 D stellar
	collapse. I - Polytropic case {\it Astron. Astrophys.} {\bf 267}
	623 - 33

\bibitem{buras} Buras R {\it et al} 2003 Improved Models of Stellar Core
	Collapse and Still No Explosions: What Is Missing? \PRL {\bf 90}
	241101-1 - 4

\bibitem{burasbrem} Buras R {\it et al} 2003 Electron Neutrino Pair
	Annihilation: A New Source for Muon and Tau Neutrinos in
	Supernovae {\it Astrophys. J.} {\bf 587} 320-6

\bibitem{Burrows92}
Burrows A 1992 
The future of supernova neutrino detection
{\it Phys. Rev. D} {\bf 45} 3361-3385
\bibitem{burro93} Burrows A {\it et al} 1993 A Convective Trigger for Supernova Explosion
  {\it Astrophys. J.} {\bf 418} L33 - 5 

\bibitem{bur93} Burrows A and Goshy J
    1993 A Theory of Supernova Explosions {\it Astrophys. J. Lett.} {\bf
    416} 75-78


\bibitem{buroheyfrex} Burrows A {\it et al} 1995 On the Nature of
	Core-Collapse Supernova Explosions {\it Astrophys. J.} {\bf 435} 830-50


\bibitem{burohey} Burrows A and Hayes J 1996 Pulsar Recoil and Gravitational Radiation Due to Asymmetrical Stellar Collapse and Explosion \PRL {\bf 76} 352 -5 
\bibitem{bursaw1} Burrows A and 
Sawyer R F 1998 Effects of correlations on neutrino opacities in nuclear
	matter \PR {\it C} {\bf 58} 554 - 571
\bibitem{bursaw2} Burrows A and 
Sawyer R F 1999 Many-body corrections to charged-current neutrino
	absorption rates in nuclear matter \PR {\it C} {\bf 59} 510 -4

\bibitem{thomp_detailed} Burrows A {\it et al} 2004 Neutrino Opacities
in Nuclear Matter {\it Nuc. Phys. A.} in press
\bibitem{bruenn85} Bruenn S W 1985 Stellar core collapse - Numerical
	model and infall epoch {\it Astrophys. J. Suppl.}
	{\bf 58} 771-841
\bibitem{bruenn87}Bruenn S W 1987 Neutrinos from SN1987A and current
	models of stellar-core collapse \PRL {\bf 59} 938-941
\bibitem{bruennhax} Bruenn S W and Haxton W C 1991 Neutrino-nucleus
	interactions in core-collapse supernovae {\it Astrophys. J.}
	{\bf 376} 678-700.
\bibitem{bruenn92}Bruenn S W 1992 in Proceeding of First Symposium on
Nuclear Physics in Universe, in press

\bibitem{bruennmezza94} Bruenn S W {\it et al} 1994 Prompt convection in core
	collapse supernovae {\it Astrophys. J.} {\bf 433} L45 - 8 

\bibitem{bruenn_ion}  Bruenn S W and Mezzacappa A 1997 Ion screening
	effects and stellar collapse {\it Phys. Rev. D} {\bf 56} 7529 - 47

 
% ccc
\bibitem{cardall05} Cardall C Y 2005 Supernova neutrino challenges
	astro-ph/0502232
\bibitem{carter} Carter G W and Prakash M 2002 
The quenching of the axial coupling in nuclear and neutron-star matter
	{\it Phys. Lett. B} {\bf 525} 249-254 
\bibitem{CCFR95}
CCFR Collaboration 1995 
Limits on $\nu_\mu(\overline{\nu}_\mu)\to\nu_\tau(\overline{\nu}_\tau)$ 
and $\nu_\mu(\overline{\nu}_\mu)\to\nu_e(\overline{\nu}_e)$ Oscillations 
from a Precision Measurement of Neutrino-Nucleon Neutral Current Interactions
{\it Phys. Rev. Lett.} {\bf 75} 3993-3996

\bibitem{CCFR97}
CCFR Collaboration 1997 
A High Statistics Search for muon-neutrino(anti-muon-neutrino) $\rightarrow$
electron-neutrino(anti-electron-neutrino) Oscillations in the Small Mixing Angle Regime
{\it Phys. Rev. Lett.} {\bf 78} 2912-2915
\bibitem{centrella} Centrella J M {\it et al} 2001 Dynamical Rotational
	Instability at Low T/W  {\it Astrophys. J.} {\bf 550} L193 - L196 
\bibitem{chandra} Chandrasekar S 1938 {\it An introduction to the study
	of stellar structure} University of Chicago Press reissued by
	Dover Press
\bibitem{CHOOZ98}
CHOOZ Collaboration 1998 
Initial Results from the CHOOZ Long Baseline Reactor Neutrino Oscillation Experiment
{\it Phys. Lett. B} {\bf 420} 397-404

\bibitem{CHOOZ99}
CHOOZ Collaboration 1999
Limits on neutrino oscillations from the CHOOZ experiment
{\it Phys. Lett. B} {\bf 466} 415

\bibitem{christlieb} Christlieb N {\it et al} 2002 A stellar relic from
	the early Milky Way {\it Natur} {\bf 419} 904 - 6
\bibitem{Cirelli04}
M. Cirelli,2004 
Sterile Neutrinos in astrophysical and cosmological sauce
astro-ph/0410122
\bibitem{Cleveland98}
Cleveland B T {\it et al} 1998 
MEASUREMENT OF THE SOLAR ELECTRON NEUTRINO FLUX WITH THE HOMESTAKE CHLORINE DETECTOR
{\it Astrophys. J.} {\bf 496} 505-526
\bibitem{colgate}  Colgate S A and White R H 1966 The Hydrodynamic
	Behavior of Supernovae Explosions {\it Astrophys. J.} {\bf 143}
	626-81

\bibitem{cordes}  Cordes J M {\it et al} 1990 Polarization of the binary radio pulsar 1913 + 16 - Constraints on geodetic precession{\it Astrophys. J.} {\bf 349} 546 - 52  

% ddd
\bibitem{Davis68}
Davis R {\it et al} 1968 
SEARCH FOR NEUTRINOS FROM THE SUN
{\it Phys. Rev. Lett.} {\bf 20} 1205-1209
\bibitem{DigheKachelriessRaffeltTomas04}
Dighe A S {\it et al} 2004 
Signatures of supernova neutrino oscillations in the Earth mantle and core
{\it JCAP 0401} 004

\bibitem{DigheKeilRaffelt03a}
Dighe A S {\it et al} 2003 
Detecting the Neutrino Mass Hierarchy with a Supernova at IceCube
{\it JCAP 0306} 005

\bibitem{DigheKeilRaffelt03b}
Dighe A S {\it et al} 2003 
Identifying Earth matter effects on supernova neutrinos at a single detector
{\it JCAP 0306} 006

\bibitem{DigheSmirnov00}
Dighe A S and Smirnov A Yu 2000 
Identifying the neutrino mass spectrum from a supernova neutrino burst
{\it Phys. Rev. D} {\bf 62} 033007

\bibitem{dimmel}Dimmelmeier H {\it et al} 2002 Relativistic simulations of rotational core collapse II. Collapse dynamics and gravitational radiation
	{\it Astron.~Astrophys.} {\bf 393} 523 - 42

\bibitem{dimmel04} 
Dimmelmeier H {\it et al} astro-ph/0407174 
\bibitem{duez1} Duez M D {\it et al} 2005 Relativistic
	Magnetohydrodynamics In Dynamical Spacetimes: Numerical Methods
	And Tests astro-ph/0503420
\bibitem{duan04} Duan H and  Qian Y Z 2004
 {\it Phys. Rev. D} {\bf 69} 123004-1 -16

\bibitem{duez2} Duez M D {\it et al} 2005 Excitation Of MHD Modes With
	Gravitational Waves: A Testbed For Numerical Codes astro-ph/0503421 

\bibitem{duncan} Duncan R C and Thompson C 1992 Formation of very strongly magnetized neutron stars - Implications for gamma-ray bursts {\it Astrophys. J. Lett.} {\bf 392} L9 -13

\bibitem{DziewonskiAnderson81}
Dziewonski A M and Anderson D L 1981 
PRELIMINARY REFERENCE EARTH MODEL
{\it Phys. Earth. Planet. Inter.} {\bf 25} 297-356
% eee
\bibitem{epsteinconv} Epstein R I 1979 Lepton-driven convection in
	supernovae {\it
	Mon. Not. Roy. Aca. P.} {\bf 188} 305 -25
\bibitem{epstein_gw} Epstein R {1978} The generation of gravitational
	radiation by escaping supernova neutrinos  {\it
	Astrophys. J.} {\bf 223} 1037-1045
\bibitem{eriguchi} Eriguchi Y and M\"{u}ller E 1985 A general
	computational method for obtaining equilibria of
	self-gravitating and rotating gases {\it Astron. Astrophys.}
	{\bf 146} 260 - 8
 
% fff
\bibitem{FeruglioStrumiaVissani02}
Feruglio F {\it et al} 2002 
Neutrino oscillations and signals in beta and 0nu2beta experiments
{\it Nucl. Phys. B} {\bf 637} 345-377

\bibitem{FeruglioStrumiaVissani03}
Feruglio F {\it et al} 2003 
Neutrino oscillations and signals in beta and 0nu2beta experiments
{\it Nucl. Phys. B} {\bf 659} 359-362
\bibitem{finn} Finn L S 1991 New York 
Academy Sciences Annals {\bf 631} 156
\bibitem{Fogli03}
Fogli G L {\it et al} 2003 
Solar neutrino oscillation parameters after first KamLAND results
{\it Phys. Rev. D} {\bf 67} 073002

\bibitem{Fogli05}
Fogli G L {\it et al} 2005 
Probing supernova shock waves and neutrino flavor transitions in next-generation 
water-Cherenkov detectors
{\it JCAP} {\bf 0504} 002
 
\bibitem{fog1} Foglizzo T  2001 Entropic-acoustic instability of shocked
	Bondi accretion I. What does perturbed Bondi accretion sound
	like ? {\it Astron. Astrophys.} {\bf 368} 311-24 

\bibitem{fog2} Foglizzo T  2002 Non-radial instabilities of isothermal Bondi accretion with a shock: Vortical-acoustic cycle vs. post-shock acceleratio
 {\it Astron. Astrophys.} {\bf 392} 353-68

\bibitem{frebel} Frebel A {\it et al} 2005 Nucleosynthetic signatures of
	the first stars  {\it Natur} {\bf 434}
	871 - 3
\bibitem{fried} Freedman D Z 1974 Coherent effects of a weak neutral
	current {\it Phys. Rev. D} {\bf 9} 1389-92
\bibitem{fryerkalogera} Fryer C and 
Kalogera V 1997 Double Neutron Star Systems and Natal Neutron Star Kicks {\it Astrophys. J.} {\bf 489} 244 - 53
\bibitem{fryer03} Fryer C L and Warren M S 2002
Modeling Core-Collapse Supernovae in Three Dimensions {\it Astrophys. J.} {\bf 574} L65-8
\bibitem{fry00} Fryer C L and Heger A 2000 Core-Collapse Simulations of Rotating Stars {\it Astrophys. J.} 541, 1033 -50
\bibitem{fryer} Fryer C L and Warren M S 2004 
The Collapse of Rotating Massive Stars in Three Dimensions 
{\it Astrophys. J.} {\bf 601} 391 - 404 
\bibitem{fryerkick} Fryer C L 2004 Neutron Star Kicks from Asymmetric
	Collapse {\it Astrophys. J. Lett.} {\bf 601} 
L175 - 8
\bibitem{fryergw} Fryer C L {\it et al} 2004 Gravitational Waves from Stellar Collapse: Correlations to Explosion Asymmetries  {\it Astrophys. J.} {\bf 609} 288-300
\bibitem{fuller1} Fuller G M {\it et al} 1980 Stellar weak-interaction
	rates for sd-shell nuclei. I - Nuclear matrix element
	systematics with application to Al-26 and selected nuclei of
	importance to the supernova problem {\it Astrophys. J. Suppl.}
	{\bf 42} 447-473

\bibitem{fuller2} Fuller G M {\it et al} 1982 Stellar weak interaction rates for intermediate mass nuclei. III - Rate tables for the free nucleons and nuclei with A = 21 to A = 60 {\it Astrophys. J. Suppl.}
	{\bf 48} 279 - 319

\bibitem{fuller3} Fuller G M {\it et al} 1982 
Stellar weak interaction rates for intermediate-mass nuclei. II - A = 21
	to A = 60 {\it Astrophys. J. }
	{\bf 252} 715 - 70
\bibitem{fesen} Fesen R A 1996 An Optical Survey of Outlying Ejecta in
	Cassiopeia A: Evidence for a Turbulent, Asymmetric Explosion
	{\it Astrophys. J. Supple.} {\bf 133} 161-86  
\bibitem{fukuda} Fukuda I 1982 A statistical study of rotational
	velocities of the stars {\it Pub. Astron. Soc. Pac.}{\bf 94}
	271-284
% ggg
\bibitem{GiacomelliGiorgini05}
Giacomelli G and Giorgini M 2005
Atmospheric neutrino oscillations
hep-ex/0504002.

\bibitem{Gil-BotellaRubbia03}
Gil-Botella I and Rubbia A 2003 
Oscillation effects on supernova neutrino rates and spectra and detection of the shock 
breakout in a liquid Argon TPC
{\it JCAP} {\bf 0310} 009

\bibitem{weber}Goldreich P and Weber S V 1980 {\it Astrophys. J.} {\bf 238} 991-997
Homologously collapsing stellar cores
\bibitem{goldreich} Goldreich P {\it et al} 1996 {\it Unresolved Problems in
			      Astrophysics} (Princeton University
			      Press, Princeton)
\bibitem{Gonzalez03}
Gonzalez-Garcia M C and Pena-Garay C 2003 
Three-Neutrino Mixing after the First Results from K2K and KamLAND
{\it Phys. Rev. D} {\bf 68} 093003

\bibitem{Goswami03}
Goswami S 2003 
Solar Neutrino Experiments: An Overview
hep-ph/0303075.

\bibitem{gusei} 
Guseinov O H {\it et al} 2003 On period and burst histories of AXPs and SGRs and the possible evolution of these objects on the P - Pdot diagram 
{\it Inter. J. Mod. Phys. D}  {\bf 12} 1 

% hhh
\bibitem{HalzenJacobsenZas96}
Halzen F {\it et al} 1996 
Ultra-Transparent Antarctic Ice as a Supernova Detector
{\it Phys. Rev. D} {\bf 53} 7359-7361

\bibitem{hamuy} Hamuy M 2004 Review on the Observed and Physical
	Properties of Core Collapse Supernovae {\it Stellar Collapse}
	(Dordrecht: Kluwer Academic Press) 39-61
\bibitem{hannestad} Hannestad S and 
Raffelt G 1998 Supernova Neutrino Opacity from Nucleon-Nucleon
	Bremsstrahlung and Related Processes {\it Astrophys. J.} {\bf 507} 339-352
\bibitem{haxton} Haxton W C 1998 Neutrino Heating in Supernovae \PRL
	{\bf 60} 1999-2002

\bibitem{heger00} Heger A {\it et al} 2000
Presupernova Evolution of Rotating Massive Stars. I. Numerical Method and Evolution of the Internal Stellar Structure {\it Astrophys. J.}{\bf 528} 368 -96

\bibitem{heger03} Heger A {\it et al} 2003 Presupernova Evolution of
	Rotating Massive Stars and the Rotation Rate of Pulsars {\it Stellar rotation} Proceeding
	of IAU symposium No. 215 astro-ph/0301374 

\bibitem{heger04} Heger A {\it et al} 2004
Presupernova Evolution of Differentially Rotating Massive Stars Including Magnetic Fields astro-ph/0409422

\bibitem{hegershinka} Heger A {\it et al} 2003 How Massive Single Stars
	End Their Life  {\it Astrophy. J.} {\bf
	591} 288-300

\bibitem{heger_prl}
Heger A {\it et al} 2001 Presupernova Collapse Models with Improved
	Weak-Interaction Rates \PRL {\bf 86} 1678 - 1681

\bibitem{heger_weak} Heger A {\it et al} 2001 Presupernova Evolution
	with Improved Rates for Weak Interactions {\it Astrophys. J.} {\bf 560} 307 - 325

\bibitem{heger_phd} Heger A 1998 The presupernova Evolution of Rotating
	Massive Stars PhD Thesis Max-Planck-Institute

\bibitem{helfand} Helfand D J {\it et al} 2001 Vela Pulsar and Its Synchrotron Nebula {\it Astrophys. J.} {\bf 556} 380 - 91

\bibitem{hera92} Herant M {\it et al} 1992 Postcollapse hydrodynamics of
	SN 1987A - Two-dimensional simulations of the early evolution
	{\it Astrophys. J.} {\bf 395} 642 - 53 
\bibitem{hera} Herant M {\it et al} 1994  Inside the supernova: A
	powerful convective engine {\it Astrophys. J.} {\bf 435} 339 - 61 
\bibitem{hill1} Hillebrandt W 1982 An exploding 10 solar mass star - A
	model for the Crab supernova {\it Astron. Astrophys.} {\bf 110}
	L3-L6
\bibitem{hillnomo} Hillebrandt W {\it et al}1984
 Supernova explosions of massive stars - The mass range 8 to 10 solar
	masses {\it Astron. Astrophys.} {\bf 133}
	175-84
\bibitem{hillenie} Hillebrandt W and Niemeyer J C 2000 
Type IA Supernova Explosion Models {\it Ann. Rev. Astron. Astrophys.}
	{\bf 38} 191-230

\bibitem{Hirata87}
Hirata K {\it et al}1987 
OBSERVATION OF A NEUTRINO BURST FROM THE SUPERNOVA SN1987A
{\it Phys. Rev. Lett.} {\bf 58} 1490-1493

\bibitem{Hirata88}
Hirata K S {\it et al} 1988 
OBSERVATION IN THE KAMIOKANDE-II DETECTOR OF THE NEUTRINO BURST FROM SUPERNOVA SN1987A
{\it Phys. Rev. D} {\bf 38} 448-458
\bibitem{hiramatsu} Hiramatsu T {\it et al} 2005 Gravitational Wave
	Background from Neutrino-Driven Gamma-Ray Bursts submitted to {\it
	Mon. Not. Roy. Astr. S.}
\bibitem{hix} Hix W R {\it et al} 2003 Consequences of Nuclear Electron
	Capture in Core Collapse Supernovae \PRL {\bf 91} 201102-1 - 4

\bibitem{hoflich} Hoflich P {\it et al} 1991 Asphericity Effects in Scattering
	Dominated Photospheres  {\it Astrophys. J.} {\bf 246}  481-9 
\bibitem{hoefrev}
H\"{o}flich P et al 2004 Asymmetric supernova explosions {\it Stellar Collapse}
	(Dordrecht: Kluwer Academic Press) 237-58
\bibitem{Honda01}
Honda M {\it et al} 2001 
Comparison of 3-Dimensional and 1-Dimensional Schemes in the calculation of Atmospheric Neutrinos
{\it Phys. Rev. D} {\bf 64} 053011

\bibitem{Honda04}
Honda M {\it et al} 2004 
A New calculation of the atmospheric neutrino flux in a 3-dimensional scheme
{\it Phys. Rev. D} {\bf 70} 043008

\bibitem{horowitz_ion} Horowitz C J 1997 Neutrino trapping in a
	supernova and the screening of weak neutral currents {\it Phys. Rev. D}
{\bf 55} 4577-4581
\bibitem{horomag} Horowitz C J 2002 Weak magnetism for antineutrinos in supernovae \PR {\it D}, 
{\bf 65} 043001-1 - 12 


\bibitem{horowitz} Horowitz C J and Li G 1998 Cumulative Parity Violation in Supernovae \PRL {\bf 80} 3694 - 97
\bibitem{horonuc} Horowitz C J {\it et al} 2004 Nonuniform neutron-rich
	matter and coherent neutrino scattering \PR{\it C} {\bf 70}
	065806-1 - 15
\bibitem{hugh} Hughes S A {\it et al} 2001
New physics and astronomy with the new gravitational-wave observatories 
 Proceedings of the 2001 Snowmass Meeting 
 astro-ph/0110349
% iii
\bibitem{ibrahim} Ibrahim A I {\it et al} 2003 New Evidence of
	Proton-Cyclotron Resonance in a Magnetar Strength Field from SGR
	1806-20 {\it Astrophys. J. Lett.} {\bf 584} L17-L22
\bibitem{IceCubeHP}
IceCube Collaboration, web page, http://icecube.wisc.edu/

\bibitem{IceCube04}
IceCube Collaboration 2004 
Sensitivity of the IceCube Detector to Astrophysical Sources of High Energy Muon Neutrinos
{\it Astropart. Phys.} {\bf 20} 507-532

\bibitem{ito_ion} Itoh N {\it et al} 2004 Ion-Ion Correlation Effect on
	the Neutrino-Nucleus Scattering in Supernova Cores  {\it
	Astrophys. J.} {\bf 611} 1041 - 1044
\bibitem{iwamoto} Iwamoto N {\it et al} 2005 The first chemical
	enrichment in the universe and the formation of hyper metal-poor
	stars {\it Natur} in press
% jjj
\bibitem{jankeil} Janka H T and Keil W 1997 Perspective of Core-Collapse 
beyond SN 1987A {\it Proc. of the
			 Colloquium in Honor of Prof. G. Tammann} 
astro-ph/9709012
\bibitem{JankaHillebrandt89}
Janda H T and Hillebrandt W 1989 {\it Astron. Astrophys.} {\bf 224} 49
\bibitem{janka99} Janka H T and Raffelt G G 1999 {\it Phys. Rev. D} {\bf
	59} 023005-1 -8 
\bibitem{jankashock} Janka H T 2001 Conditions for shock revival by
	neutrino heating in core-collapse supernovae {\it
	Astron. Astrophys.} {\bf 368} 527-60 
\bibitem{jankamueller94} Janka H T and Mueller E 1994 Neutrino heating,
	convection, and the mechanism of Type-II supernova explosions {\it
	Astron. Astrophys.} {\bf 306} 167-98
\bibitem{janka04manu}  Janka {\it et al} 2004  Explosion Mechanism of
	Massive Stars {\it Stellar Collapse}
	(Dordrecht: Kluwer Academic Press) 65-93
\bibitem{jankapro04}
Janka H T {\it et al}  2004 Supernova Asymmetries and Pulsar Kicks --
	Views on Controversial Issues  arXiv:astro-ph/0408439
% kkk
\bibitem{K2K01}
K2K Collaboration 2001 
Detection of Accelerator-Produced Neutrinos at a Distance of 250 km
{\it Phys. Lett. B} {\bf 511} 178-184

\bibitem{K2K03}
K2K Collaboration 2003 
Indications of Neutrino Oscillation in a 250 km Long-baseline Experiment
{\it Phys. Rev. Lett.} {\bf 90} 041801

\bibitem{K2K04}
K2K Collaboration 2004 
Search for Electron Neutrino Appearance in a 250 km Long-baseline Experiment
{\it Phys. Rev. Lett.} {\bf 93} 051801

\bibitem{K2K05}
K2K Collaboration 2005 
Evidence for muon neutrino oscillation in an accelerator-based experiment
{\it Phys. Rev. Lett.} {\bf 94} 081802 
\bibitem{Kachelriess05}
Kachelriess M {\it et al} 2005 
Exploiting the neutronization burst of a galactic supernova
{\it Phys. Rev. D} {\bf 71} 063003
\bibitem{KamLANDHP}
KamLAND Collaboration, web page, http://www.awa.tohoku.ac.jp/KamLAND/index.html

\bibitem{KamLAND03}
KamLAND Collaboration 2003 
First Results from KamLAND: Evidence for Reactor Anti-Neutrino Disappearance
{\it Phys. Rev. Lett.} {\bf 90} 021802

\bibitem{KamLAND04}
KamLAND Collaboration 2004 
A High Sensitivity Search for $\bar{\nu}_{e}$'s from the Sun and Other Sources at KamLAND
{\it Phys. Rev. Lett.} {\bf 92} 071301

\bibitem{KamLAND05}
KamLAND Collaboration 2005 
Measurement of Neutrino Oscillation with KamLAND: Evidence of Spectral Distortion
{\it Phys. Rev. Lett.} {\bf 94} 081801
\bibitem{KARMEN98}
KARMEN Collaboration 1998 
Measurement of the energy spectrum of $\nu_e$ from muon decay and implications 
for the Lorentz structure of the weak interaction
{\it Phys. Rev. Lett.} {\bf 81} 520-523

\bibitem{KARMEN02}
KARMEN Collaboration 2002 
Upper limits for neutrino oscillations muon-antineutrino to electron-antineutrino from muon decay at rest
{\it Phys. Rev. D} {\bf 65} 112001
\bibitem{kapsi} Kapsi V M {\it et al} 1996 {\it Nature} {\bf 381} 584
\bibitem{Keil03}
Keil M T PhD thesis TU M\"unchen 2003
Supernova Neutrino Spectra and Applications to Flavor Oscillations
astro-ph/0308228
\bibitem{KeilRaffeltJanka03}
Keil M T {\it et al} 2003 
Monte Carlo Study of Supernova Neutrino Spectra Formation
{\it Astrophys. J} {\bf 590} 971-991

\bibitem{keil} Keil W {\it et al} 1996 Ledoux Convection in Protoneutron
	Stars---A Clue to Supernova Nucleosynthesis? {\it
	Astrophys. J. Lett.} {\bf 473} L111-4 
\bibitem{KolbStebbinsTurner87}
Kolb E {\it et al} 1987 
HOW RELIABLE ARE NEUTRINO MASS LIMITS DERIVED FROM SN1987A?
{\it Phys. Rev. D} {\bf 35} 3598
\bibitem{kotakegw} Kotake K {\it al} 2003 Gravitational radiation from
	axisymmetric rotational core collapse {\it Phys. Rev. D.} {\bf 68} 044023
\bibitem{kotakeaniso} Kotake K {\it et al} 2003 Anisotropic Neutrino Radiation in Rotational Core Collapse
 {\it Astrophys. J.} {\bf 595} 304 - 16 
\bibitem{kotake_aniso_gw} Kotake {\it et al} 2005 Gravitational Waves
	from Anisotropic Neutrino Radiation in Rotational Core-Collapse
	submitted to {\it Phys. Rev. D}
\bibitem{kotaketor} Kotake K {\it et al} 2004 Magnetorotational Effects
	on Anisotropic Neutrino Emission and Convection in Core Collapse
	Supernovae  {\it Astrophys. J.} {\bf 608} 391 - 404  

\bibitem{kotakemhd} Kotake 
K {\it et al} 2005 North-South Neutrino Heating Asymmetry in Strongly Magnetized and Rotating Stellar Cores {\it Astrophys. J.} {\bf 618} 474 - 84 
\bibitem{kotakegwMHD} Kotake K {\it et al} 2004 Gravitational radiation
	from rotational core collapse: Effects of magnetic fields and
	realistic equations of state {\it Phys. Rev. D} {\bf 12}
	124004-1 - 11
\bibitem{KraussTremaine88}
Krauss L M and Tremaine S 1988 
Test of the Weak Equivalence Principle for Neutrinos and Photons
{\it Phys. Rev. Lett.} {\bf 60} 176-177

\bibitem{KuoPantaleone88}
Kuo T K and Pantaleone J 1988 
Supernova neutrinos and their oscillations
{\it Phys. Rev. D} {\bf 37} 298-304

\bibitem{KuoPantaleone89}
Kuo T K and Pantaleone J 1989 
NEUTRINO OSCILLATIONS IN MATTER
{\it Rev. Mod. Phys.} {\bf 61} 937

\bibitem{kolbe03} Kolbe E {\it et al} 2003 Neutrino-nucleus reactions
	and nuclear matter {\it J. Phys. J: Nucl. Part. Phys.} {\bf 29}
	2569-596
\bibitem{kusenko} Alexander K 2004 Pulsar Kicks from Neutrino
	Oscillations {\it Int. J. Mod. Phys. D} {\bf 13} 2065-2084
% lll
\bibitem{lai1} Lai D 2004 Neutron Star Kicks and Supernova Asymmetry {\it 3D Signatures of Stellar Explosion, a workshop honoring
			      J.C. Wheeler's 60th
			      Birthday} astro-ph/0312542
\bibitem{lai2} Lai D {\it et al}~2001 Pulsar Jets: Implications for Neutron Star Kicks and Initial Spins {\it Astrophys. J.} {\bf 549} 1111 - 8 

\bibitem{Lat91} Lattimer J and Douglas Swesty F 1991 A generalized
	equation of state for hot, dense matter {\it Nuc. Phys. A} 
{\bf 535} 331-376

\bibitem{lcgt} LCGT Collaboration {\it Int. J. Mod. Phys. D.} {\bf{5}} 557



\bibitem{lmp1} Langanke K and Mart\'{i}nez-Pinedo G 1999 Supernova electron
	capture rates on odd-odd nuclei {\it Phys. Lett. B} {\it 453}  187 - 93

\bibitem{lmp2} Langanke K and Mart\'{i}nez-Pinedo G 2000 Shell-model
	calculations of stellar weak interaction rates: II. Weak rates
	for nuclei in the mass range /A=45-65 in supernovae environments
	{\it Nuc. Phys. A} {\bf 673}. 481 - 508
\bibitem{langanke04} Langanke K {\it et al} 2003 Consequences of Nuclear
	Electron Capture in Core Collapse Supernovae \PRL {\bf 90}
	241102-1 - 4
\bibitem{langankeprl04} Langanke K {\it et al} 2004 Supernova Inelastic
	Neutrino-Nucleus Cross Sections from High-Resolution Electron
	Scattering Experiments and Shell-Model Calculations \PRL {\bf
	93} 202501-1 - 4
\bibitem{langankerev} Langanke K and Mart\'{i}nez-Pinedo G 2003 Nuclear
	weak-interaction processes in stars  {\it
	Rev. Mod. Phys.}
	{\bf 75}. 819 - 62 
\bibitem{lazzati} Lazzati D 2004 Gamma-Ray Burst Progenitors Confront
Observations  Xth Marcel Grossmann Meeting on General Relativity, Rio de
Janeiro, Brazil, July 2003
\bibitem{leblanc} LeBlanc J M and Wilson J R 1970 A Numerical Example of the Collapse of a Rotating Magnetized Star {\it Astrophys. J.} {\bf 161} 541 - 52
\bibitem{LEP03}
LEP Collaborations 2003 
A Combination of Preliminary Electroweak Measurements and Constraints on the Standard Model
hep-ex/0312023.

\bibitem{Liebendorfer01a}
Liebend\"orfer M {\it et al} 2001 
Conservative general relativistic radiation hydrodynamics in spherical symmetry and comoving coordinates
{\it Phys. Rev. D} {\bf 63} 104003
%\bibitem{Liebendorfer01b}
%Liebend\"orfer M {\it et al} 2001 
%Probing the gravitational well:No supernova explosion in spherical symmetry with 
%general relativistic Boltzmann neutrino transport
%{\it Phys. Rev. D} {\bf 63} 103004.
%\bibitem{Liebendorfer05}
%Liebend\"orfer M {\it et al} 2005
%Supernova Simulations with Boltzmann Neutrino Transport: A Comparison of Methods 
%{\it Astrophys. J.} {\bf 620} 840-860

\bibitem{Lieben01} Liebend\"orfer M {\it et al} 2001 Probing the gravitational
	well: No supernova explosion in spherical symmetry with general
	relativistic Boltzmann neutrino transport {\it Phys. Rev. D}
	{\bf 63} 103004-1 - 13 
\bibitem{Liebendoerfer_et_al_04} Liebend\"orfer M {\it et al}  2004 A Finite Difference Representation of Neutrino Radiation Hydrodynamics in Spherically Symmetric General Relativistic Spacetime {\it Astrophys. J. Suppl.} {\bf 150} 263 -316
\bibitem{liepen} Liebend\"orfer M {\it et al} 2004, submitted to {\it 
Nucl. Phys. A} astro-ph/0408161

\bibitem{Lieben03} Liebend\"orfer,
		       M {\it et al} 2005 Supernova Simulations with Boltzmann
	Neutrino Transport: A Comparison of Methods {\it Astrophys. J.}
	{\bf 620} 840-60


\bibitem{livne}Livne E {\it et al} 2005 Two-dimensional, Time-dependent,
	Multigroup, Multiangle Radiation Hydrodynamics Test Simulation
	in the Core-Collapse Supernova Context {\it Astrophys. J.}
	{\it 609} 277-287
\bibitem{leonard} Leonard D C {\it et al} Is It Round?
	Spectropolarimetry of the Type II-p Supernova 1999EM
 {\it Astrophys. J.} {\bf 553} 861-85

\bibitem{lorimer} Lorimer 
D R {\it et al}1997 Pulsar statistics - IV. Pulsar velocities {\it
	Mon. Not. Roy. Aca. P.} {\bf 289} 592 - 604 
\bibitem{Longo88}
Longo M J 1988
New Precision Tests of the Einstein Equivalence Principle from Sn1987a
{\it Phys. Rev. Lett.} {\bf 60} 173-175

\bibitem{LoredoLamb02}
Loredo T J and Lamb D Q 2002 
Bayesian analysis of neutrinos observed from supernova SN 1987A
{\it Phys. Rev. D} {\bf 65} 063002
\bibitem{LSND01a}
LSND Collaboration 2001  
Measurements of Charged Current Reactions of $\nu_e$ on $^{12}C$
{\it Phys. Rev. C} {\bf 64} 065501

\bibitem{LSND01b}
LSND Collaboration 2001 
Measurement of electron-neutrino electron elastic scattering
{\it Phys. Rev. D} {\bf 63} 112001

\bibitem{LSND01c}
LSND Collaboration 2001 
Evidence for neutrino oscillations from the observation of nu -bare appearance in a nu -bar¦Ì beam
{\it Phys. Rev. D} {\bf 64} 112007

\bibitem{LunardiniSmirnov01}
Lunardini C and Smirnov A Yu 2001 
Supernova neutrinos: Earth matter effects and neutrino mass spectrum
{\it Nucl. Phys. B} {\bf 616} 307-348

\bibitem{LunardiniSmirnov03}
Lunardini C and Smirnov A Yu 2003 
Probing the neutrino mass hierarchy and the 13-mixing with supernovae
{\it JCAP 0306} 009

\bibitem{LunardiniSmirnov04}
Lunardini C and Smirnov A Yu 2004 
Neutrinos from SN1987A: flavor conversion and interpretation of results
{\it Astropart. Phys.} {\bf 21} 703-720

\bibitem{lyne} Lyne A G and Lorimer
			      D R 1994 High Birth Velocities of Radio
	Pulsars {\it Nature} {\bf 369} 127 
% mmm
\bibitem{MACRO01}
MACRO Collaboration 2001 
Matter Effects in Upward-Going Muons and Sterile Neutrino Oscillations
{\it Phys. Lett. B} {\bf 517} 59-66

\bibitem{MACRO03}
MACRO Collaboration 2003 
Atmospheric neutrino oscillations from upward throughgoing muon multiple scattering in MACRO
{\it Phys. Lett. B} {\bf 566} 35-44

\bibitem{MakiNakagawaSakata62}
Maki Z {\it et al} 1962 
REMARKS ON THE UNIFIED MODEL OF ELEMENTARY PARTICLES
{\it Prog. Theor. Phys.} {\bf 28} 870

\bibitem{Maltoni03}
Maltoni M {\it et al} 2003
Status of three-neutrino oscillations after the SNO-salt data
{\it Phys. Rev. D} {\bf 68} 113010

\bibitem{Maltoni04}
Maltoni M {\it et al} 2004 
Status of global fits to neutrino oscillations
{\it New J. Phys.} {\bf 6} 122

\bibitem{MayleWilsonSchramm87}
Mayle R {\it et al} 1987 
NEUTRINOS FROM GRAVITATIONAL COLLAPSE
{\it Astrophys. J} {\bf 318} 288-306
\bibitem{macfad} Woosley S E and MacFadyen A I 1999 Central engines for
gamma-ray bursts {\it Astron. Astrophys.} {\bf 138} 499-502

\bibitem{mado1} Madokoro H {\it et al} 2003 Global Anisotropy versus Small-Scale Fluctuations in Neutrino Flux in Core-Collapse Supernova Explosions  {\bf 592}  1035-41
\bibitem{mado2} Madokoro H {\it et al} 2004 Importance of Prolate Neutrino Radiation in Core-Collapse Supernovae: The Reason for the Prolate Geometry of SN1987A? {\it Pub. Astron. Soc. J.} {\bf 56} 663-9
\bibitem{marek_ion} Marek A {\it et al} 2005 On ion-ion correlation
	effects during stellar core collapse submitted to {\it
	Astron. Astrophys.} astro-ph/0504291
\bibitem{matsumoto} Matsumoto R and Shibata K  1999 Global three-dimensional
MHD simulations of accretion disks and jet formation in AGNS {\it
Advances in Space Research}  {\bf 23} 1109 - 13 
\bibitem{mazzali} Mazzali P A {\it et al} 2003 The Type Ic Hypernova SN
2003dh/GRB 030329 {\it Astrophys. J. } {\bf 599} L95 - L98  
\bibitem{MikheyevSmirnov85}
Mikheyev S P and Smirnov A Yu 1985 {\it Yad. Fiz.} {\bf 42} 1441

\bibitem{MikheyevSmirnov86a}
Mikheyev S P and Smirnov A Yu 1986 {\it Nuovo Cim. C} {\bf 9} 17

\bibitem{MikheyevSmirnov86b}
Mikheyev S P and Smirnov A Yu 1986 {\it Sov. Phys. JETP} {\bf 64} 4

\bibitem{MikheyevSmirnov89}
Mikheyev S P and Smirnov A Yu 1989 
RESONANT NEUTRINO OSCILLATIONS IN MATTER
{\it Prog. Part. Nucl. Phys.} {\bf 23} 41-136

\bibitem{MiniBooNE00}
MiniBooNE Collaboration 2000 
MiniBooNE: Status of the Booster Neutrino Experiment
{\it Nucl. Phys. Proc. Suppl.} {\bf 91} 210-215

\bibitem{MiniBooNE04}
MiniBooNE Collaboration 2004 
MiniBooNE and Sterile Neutrinos
hep-ex/0407027.
\bibitem{misner} Misner, C.~W., Thorne, K.~S., \&
			      Wheeler,~J.~A. 1973, Gravitation (San
			      Francisco, U.~S.~A: Freeman)
\bibitem{murphy} Murphy J W {\it et al} 2005 Pulsational Analysis of the
	Cores of Massive Stars and Its Relevance to Pulsar Kicks  {\it
	Astrophys. J.} {\bf
	615} 460 - 474
\bibitem{mezza98} Mezzacappa A {\it et al} 1998 An Investigation of
	Neutrino-driven Convection and the Core Collapse Supernova
	Mechanism Using Multigroup Neutrino Transport {\it Astrophys. J}
	{\bf 495} 911-26
\bibitem{mezza1998} Mezzacappa A {\it et al} 1998 The interplay between
	proto-neutron star convection and  neutrino transport in
	core-collapse supernovae {\bf 493} 848-62 

\bibitem{monch} M\"{o}nchmeyer R M and M$\ddot{\rm u}$ller E 1989, in NATO ASI Series, Timing Neutron Stars, ed. H. \"{O}gelman \& E.P.J van der Heuvel (New York: ASI) 
\bibitem{moenchgw} M\"{o}nchmeyer R {\it et al} 1991 Gravitational waves
	from the collapse of rotating stellar cores {\it
	Astron. Astrophys.} {\bf 246} 417 - 40


\bibitem{muller} M\"{u}ller E and 
Hillebrandt W 1979 A magnetohydrodynamical supernova model {\it
Astron. Astrophys.} {\bf 80} 147 - 54

\bibitem{muhi} M\"{u}ller E and Hillebrandt W 1981 The collapse of rotating stellar cores {\it Astron. Astrophys.} {\bf 103} 358 - 66

\bibitem{mueller1982} M\"{u}ller E 1982 Gravitational radiation from
	collapsing rotating stellar cores {\it Astron. Astrophys.} {\bf
	114} 53 - 9
\bibitem{mull_mem} M\"{u}ller E and Janka H T 1997 Gravitational
	radiation from convective instabilities in type II supernova
	explosions {\it Astron.~Astrophys.} {\bf 317} 140 - 63
\bibitem{muller03} M{\" u}ller E {\it et al} 2004 Toward Gravitational Wave Signals from Realistic Core-Collapse Supernova Models 
 2004 {\it Astrophys. J.} {\bf 603}  221 - 30 


\bibitem{mirabel} Mirabel I F {\it et al} 2002 The runaway black hole GRO J1655-40{\it Astron. Astrophys.} 
{\bf 395} 595 - 9 
\bibitem{mendez} Mendez M {\it et al} 1988 SN 1987A - A linear
	polarimetric study {\it Astrophys. J.} {\bf 334}  295 - 307 
\bibitem{miralles04} Miralles J A {\it et al} 2004 Anisotropic convection in rotating proto-neutron stars {\it Astron. Astrophys.} {\bf 420} 245 - 9 
\bibitem{meiner} Meier D L {\it et al} 1976 Magnetohydrodynamic
	phenomena in collapsing stellar cores {\it Astrophys. J.} {\bf
	204 } 869 - 78
% nnn
\bibitem{takashi} Nakamura T{\it et al} 1998 Kyoto University
press ISBN4-87698-032-2
\bibitem{nagataki1}
Nagataki S {\it et al} 1997 Explosive Nucleosynthesis in
	Axisymmetrically Deformed Type II Supernovae {\it Astrophys. J.}
	{\bf 486} 1026 - 35

\bibitem{nagataki2} Nagataki S 2000 Effects of Jetlike Explosion in SN
	1987A {\it Astrophys. J. Supple.} {\bf 127} 141 - 57 
\bibitem{new} New K S 2003 Gravitational Waves from Gravitational Collapse  {\it Living Reviews in Relativity} {\bf 6} 2 - 71
\bibitem{nishimura} Nishimura S {\it et al} 2005 R-Process Nucleosynthesis in MHD Explosions of Core-Collapse Supernovae 
submitted to {\it Astrophys. J.}
\bibitem{nometal94}
Nomoto K et al 1994
\emph{Supernovae, Les Houches Session LIV},
(Amsterdam: Elsevier/North-Holland) 489
\bibitem{NOMAD01}
NOMAD Collaboration 2001 
Final NOMAD results on $nu_mu->nu_tau$ and $nu_e->nu_tau$ oscillations including a new search 
for $nu_tau$ appearance using hadronic tau decays
{\it Nucl. Phys. B} {\bf 611} 3-39

\bibitem{NOMAD03}
NOMAD Collaboration 2003 
Search for nu(mu)-->nu(e) Oscillations in the NOMAD Experiment
{\it Phys. Lett. B} {\bf 570} 19-31

\bibitem{NuTeV02}
NuTeV Collaboration 2002 
Search for ¦Ìe and ¦Ìe Oscillations at NuTeV
{\it Phys. Rev. Lett.} {\bf 89} 011804

\bibitem{RamppJanka00}
Rampp M and Janka H -Th 2000 
Spherically Symmetric Simulation with Boltzmann Neutrino Transport of Core Collapse 
and Post-Bounce Evolution of a 15 Solar Mass Star
{\it Astrophys. J} {\bf 539} L33-L36

% ooo
\bibitem{ostriker} Ostriker J P and Gunn J E 1971 Do Pulsars Make Supernovae?
 {\it Astrophys. J. Lett.} {\bf 164} L95 - 104 
\bibitem{ott} Ott, 
C D {\it et al} 2004 Gravitational Waves from Axisymmetric, Rotating
	Stellar Core Collapse {\it Astrophys. J.} {\bf 600} 834 - 64

\bibitem{ott_one_arm}   Ott C D {\it et al} (2005)
One-armed Spiral Instability in a Slowly Rotating, Post-Bounce Supernova
	Core {\it Astrophys. J. Lett.} in press  astro-ph/0503187 
% ppp
\bibitem{pavlov} 
Pavlov G G {\it et al} 2001 Variability of the Vela Pulsar Wind Nebula Observed with Chandra {\it Astrophys. J. Lett.} {\bf 554} L189 - 92 

\bibitem{plait} Plait P C {\it et al} 1995 HST observations of the ring
	around SN 1987A {\it Astrophys. J.} {\bf 439} 730-751 
\bibitem{piran} Piran T 2004 The physics of gamma-ray bursts {\it
Rev. Modern Phys} {\bf 76} 1143-1210 
% qqq

% rrr
\bibitem{rampp} Rampp M {\it et al} 
 1998 Simulations of non-axisymmetric rotational core collapse {\it Astron. Astrophys.} {\bf 332} 969 - 83 
\bibitem{ramp} Rampp M and Janka H T  2000 Spherically Symmetric
	Simulation with Boltzmann Neutrino Transport of Core Collapse
	and Postbounce Evolution of a 15 $M_{\odot}$ Star {\it
	Astrophys. J.}{\bf 539} L33-6
\bibitem{RamppJanka02}
Rampp M and Janka H -Th 2002 
Radiation hydrodynamics with neutrinos: Variable Eddington factor method for 
core-collapse supernova simulations
{\it Astron. Astrophys.} {\bf 396} 361

\bibitem{rampp_buras} Rampp M {\it et al} 2002 Core-collapse supernova
	simulations: Variations of the input physics {\it Proceedings of
	the 11th Workshop on "Nuclear Astrophysics"} astro-ph/0203493
\bibitem{reddy} 
Reddy S {\it et al.} 1999 Effects of strong and electromagnetic
	correlations on neutrino interactions in dense matter \PR {\it
	C} {\bf 59} 2888 - 918  
% sss

\bibitem{sawai} Sawai H {\it et al} 2005 The Core-Collapse Supernova
	with "Non-Uniform" Magnetic Fields  {\it Astrophys. J.} in press
\bibitem{scheck} Scheck L {\it et al} 2004 Pulsar Recoil by Large-Scale
	Anisotropies in Supernova Explosions \PRL {\bf 92} 011103-1 -4 
\bibitem{shakura} Shakura N I and Sunyaev R A 1973 Black holes in binary
	systems. Observational appearance. {\it Astron. Astrophys.} {\bf
	24} 337 - 55 
\bibitem{shapiro} Shapiro S L and Teukolsky S A 1983 {\it Blach Holes,
	White Dwarfs, and Neutron Stars} (John Wiley \& Sons)
\bibitem{shimi01} Shimizu T M {\it et al} 2001 
        {\it Astrophys. J.} {\bf 552} 756
\bibitem{spruit} Spruit H C  2002  Dynamo action by differential rotation in a stably stratified stellar interior {\it Astron. Astrophys.} {\bf 381} 923 -32 
\bibitem{sym} Symbalisty E 1984 Magnetorotational iron core collapse
{\it Astrophys. J.} {\bf 285} 729 -46
\bibitem{Sago} Sago N {\it et al} 2004 Gravitational wave memory of
	gamma-ray burst jets {\it Phys. Rev. D} {\bf 70} 104012
\bibitem{saijo} Saijo M {\it et al} 2003 One-armed Spiral Instability in
	Differentially Rotating Stars {\it Astrophys. J.} {\bf 595} 352 - 64
\bibitem{Segalis}
  Segalis E B \& Ori A 2001 Emission of gravitational radiation from
	ultrarelativistic sources {\it Phys. Rev. D} {\bf 64} 064018
\bibitem{Satotrap1} Sato K 1975 Neutrino Degeneracy in Supernova Cores
	and Neutral Current of Weak Interaction {\it Prog. Theor. Phys.}
	{\bf 53} 595-7

\bibitem{Satotrap2} Sato K 1975 Supernova explosion and neutral currents
	of weak interaction {\it Prog. Theor. Phys.}
	{\bf 54} 1325-38

\bibitem{SatoSuzuki87}
Sato K and Suzuki H 1987 
Analysis of neutrino burst from the supernova 1987A in the Large Magellanic Cloud
{\it Phys. Rev. Lett.} {\bf 58} 2722-2725
\bibitem{schneider} Schneider R {\it et al} 2003 Low-mass relics of
	early star formation {\it Natur} {\bf 422}
	869 - 71
\bibitem{SchiratoFuller02}
Schirato R C and Fuller G M 2002 
Connection between supernova shocks, flavor transformation, and the neutrino signal
astro-ph/0205390 
\bibitem{Selvi03}
Selvi M {\it et al} 2003 
Study of the effect of neutrino oscillation on the supernova neutrino signal with the LVD detector
hep-ph/0307287


\bibitem{shapiro1} Saenz R A and Shapiro S L 1978 Gravitational
	radiation from stellar collapse - Ellipsoidal models {\it
	Astrophys. J.} {\bf 221} 286 - 303 


\bibitem{shapiro2} Saenz R A and Shapiro S L 1979 Gravitational and
	neutrino radiation from stellar core collapse Improved
	ellipsoidal model calculations {\it Astrophys. J.} {\bf 229}
	1107 - 25 

\bibitem{shapiro3} Saenz R A and Shapiro S L 1981  Gravitational
	radiation from stellar core collapse. III - Damped ellipsoidal
	oscillations {\it Astrophys. J.} {\bf 244} 1033 - 8 

\bibitem{shen98} Shen H 1998 Relativistic equation of state of nuclear
	matter for supernova and neutron star {\it Nuc. Phys. A} 
{\bf 637} 435-50
\bibitem{shibata_one_1} Shibata M {\it et al} 2002 Dynamical instability
	of differentially rotating stars {\it Mon. Not. Roy. Astr. S.}
	{\bf 334}  L27 - L31
\bibitem{shibata_one_2} Shibata M {\it et al} 2003 Dynamical bar-mode
	instability of differentially rotating stars: effects of
	equations of state and velocity profiles  {\it
	Mon. Not. Roy. Astr. S.} {\bf 343}. 619 - 26

\bibitem{shibaseki} Shibata M and Sekiguchi Y 2004 Gravitational waves
from axisymmetric rotating stellar core collapse to a neutron star in
full general relativity {\it Phys. Rev. D} {\bf 69} 084024-1 - 16

\bibitem{shiba_onearm} Shibata M and Sekiguchi Y 2005 Three-dimensional
	simulations of stellar core collapse in full general relativity:
	Nonaxisymmetric dynamical instabilities {\it Phys. Rev. D} {\bf 71}
024014
\bibitem{SmirnovSpergelBahcall94}
Smirnov A Yu {\it et al} 1994 
Is Large Lepton Mixing Excluded?
{\it Phys. Rev. D} {\bf 49} 1389-1397

\bibitem{SNOHP}
SNO Collaboration, web page, http://eta.physics.uoguelph.ca/sno/

\bibitem{SNO00}
SNO Collaboration 2000 
The Sudbury Neutrino Observatory
{\it Nucl. Instrum. Meth. A} {\bf 449} 172-207

\bibitem{SNO01}
SNO Collaboration 2001 
Measurement of the rate of $nu_e + d --> p + p + e^-$ interactions produced by 8B solar neutrinos 
at the Sudbury Neutrino Observatory
{\it Phys. Rev. Lett.} {\bf 87} 071301

\bibitem{SNO02}
SNO Collaboration 2002 
Direct Evidence for Neutrino Flavor Transformation from Neutral-Current Interactions 
in the Sudbury Neutrino Observatory
{\it Phys. Rev. Lett.} {\bf 89} 011301

\bibitem{SNO04}
SNO Collaboration 2004 
Measurement of the Total Active 8B Solar Neutrino Flux at the Sudbury Neutrino Observatory 
with Enhanced Neutral Current Sensitivity
{\it Phys. Rev. Lett.} {\bf 92} 181301

\bibitem{SNO05}
SNO Collaboration 2005 
Electron Energy Spectra, Fluxes, and Day-Night Asymmetries of $^{8}$B 
Solar Neutrinos from the 391-Day Salt Phase SNO Data Set
nucl-ex/0502021.
\bibitem{Strumia04}
Strumia A 2004 
Searches for sterile neutrinos (and other light particles)
hep-ph/0407132

\bibitem{StrumiaVissani05}
Strumia A and Vissani F 2005
Implications of neutrino data circa 2005
hep-ph/0503246
\bibitem{SumiyoshiSuzukiToki95}
Sumiyoshi K {\it et al} 1995 
Influence of the symmetry energy on the birth of neutron stars and supernova neutrinos
{Astron. Astrophys.} {\bf 303} 475
\bibitem{sumi} Sumiyoshi K {\it et al} 2001 r-Process in Prompt
	Supernova Explosions Revisited {\it Astrophys. J.} {\bf 562}
	880-8
\bibitem{sumi_prep} Simiyoshi K {\it et al }2004 Properties of a relativistic
	equation of state for collapse-driven supernovae {\it Nuc. Phys. A} 
{\bf 730} 227-51

\bibitem{sumi_shock} Simiyoshi K {\it et al }2005 Postbounce evolution
of core-collapse supernovae: Long-term effects of equation of state,
{\it Astrophys. J.} in press
\bibitem{sumi_private} Simiyoshi K 2005 in private communication
\bibitem{SKHP}
Super-Kamiokande Collaboration web page,
http://www-sk.icrr.u-tokyo.ac.jp/
\bibitem{SKsolar98}
Super-Kamiokande Collaboration 1998 
Measurements of the Solar Neutrino Flux from Super-Kamiokande's First 300 Days
{\it Phys. Rev. Lett.} {\bf 81} 1158-1162
Erratum-ibid. {\bf 81} 4279

\bibitem{SKatm98}
Super-Kamiokande Collaboration 1998 
Evidence for oscillation of atmospheric neutrinos
{\it Phys. Rev. Lett.} {\bf 81} 1562-1567

\bibitem{SK99}
Super-Kamiokande Collaboration 1999 
Calibration of Super-Kamiokande Using an Electron Linac
{\it Nucl. Instrum. Meth. A} {\bf 421} 113-129

\bibitem{SKsolar99}
Super-Kamiokande Collaboration 1999 
Measurement of the solar neutrino energy spectrum using neutrino-electron scattering
{\it Phys. Rev. Lett.} {\bf 82} 2430-2434

\bibitem{SKsolar01}
Super-Kamiokande Collaboration 2001 
Solar 8B and hep Neutrino Measurements from 1258 Days of Super-Kamiokande Data
{\it Phys. Rev. Lett.} {\bf 86}  5651-5655

\bibitem{SKsolar02}
Super-Kamiokande Collaboration 2002 
Determination of Solar Neutrino Oscillation Parameters using 1496 Days of Super-Kamiokande-I Data
{\it Phys. Lett. B} {\bf 539} 179-187

\bibitem{SKatm04}
Super-Kamiokande Collaboration 2004 
Evidence for an oscillatory signature in atmospheric neutrino oscillation
{\it Phys. Rev. Lett.} {\bf 93} 101801

\bibitem{SKsolar04}
Super-Kamiokande Collaboration 2004 
Precise Measurement of the Solar Neutrino Day/Night and Seasonal Variation in Super-Kamiokande-I
{\it Phys. Rev. D} {\bf 69} 011104

\bibitem{SKatm05}
Super-Kamiokande Collaboration 2005 
A Measurement of Atmospheric Neutrino Oscillation Parameters by Super-Kamiokande I
hep-ex/0501064

\bibitem{Soudan03}
Soudan 2 Collaboration 2003 
Observation of Atmospheric Neutrino Oscillations in Soudan 2
{\it Phys. Rev. D} {\bf 68} 113004

\bibitem{SSC}
Supernova Science Center, web page, http://www.supersci.org/

\bibitem{Suzuki93}
Suzuki S in {\em Proc. of the International Symposium on Neutrino Astrophysics:
Frontiers of Neutrino Astrophysics}, edited by Y. Suzuki and K. Nakamura,
(Universal Academy Press Inc., Tokyo, 1993), number 5 in Frontiers Science Series, p. 219.
\bibitem{suzuki} Suzuki H 1994 Supernova neutrinos {\it Physics and
	Astrophysics of Neutrinos} (Springer-Verlag) 763-847
% ttt
\bibitem{KTearth02}
Takahashi K and Sato K 2002 
Earth effects on supernova neutrinos and their implications for neutrino parameters
{\it Phys. Rev. D} {\bf 66} 033006 hep-ph/0110105 

\bibitem{KT03}
Takahashi K and Sato K 2003 
Effects of neutrino oscillation on supernova neutrino: inverted mass hierarchy
{\it Prog. Theor. Phys.} {\bf 109} 919-931 hep-ph/0205070 

\bibitem{KTmass03}
Takahashi K {\it et al} 2003 
Supernova Neutrinos, Neutrino Oscillations, and the Mass of the Progenitor Star
{\it Phys. Rev. D} {\bf 68} 113009 hep-ph/0306056 

\bibitem{KTshock03}
Takahashi K {\it et al} 2003 
Shock propagation and neutrino oscillation in supernova
{\it Astropart. Phys.} {\bf 20} 189-193 astro-ph/0212195 

\bibitem{KTearth01}
Takahashi K {\it et al} 2001 
The Earth effects on the supernova neutrino spectra
{\it Phys. Lett. B} {\bf 510} 189-196 hep-ph/0012354

\bibitem{KT01}
Takahashi K {\it et al} 2001 
Effects of Neutrino Oscillation on the Supernova Neutrino Spectrum
{\it Phys. Rev. D} {\bf 64} 093004 hep-ph/0105204 

\bibitem{takiwaki} Takiwaki T {\it et al} 2005 Magneto-driven Shock
	Waves in Core-Collapse Supernovae {\it Astrophys. J.} {\bf 616} 
	1086 - 94 
\bibitem{tass} Tassoul J L 1978 {\it Theory of Rotating
			      Stars} (Princeton: Princeton Univ. Press)
\bibitem{thuan} Thuan T X and Ostriker J P 1974 Gravitational Radiation
	from Stellar Collapse {\it Astrophys. J. Lett.} {\bf 191} L105 -7
\bibitem{turner1979} Turner M S and Wagoner R V 1979 {\it 
			      Gravitational Radiation} (Cambridge
			      Univ. Press, Cambridge)
\bibitem{thorne80} Thorne K S 1980 Multipole expansions of
gravitational radiation {\it Review of Modern Physics}
{\bf 52} 299 - 338


\bibitem{tomp} Thompson  
T A {\it et al} 2003 Shock Breakout in Core-Collapse Supernovae and Its Neutrino Signature {\it Astrophys. J.} {\bf 592} 434


\bibitem{tomp04} 
Thompson T A {\it et al} Viscosity and Rotation in Core-Collapse Supernovae 
2005 {\bf 620} 861 - 77
\bibitem{firstligo} Thorne K S 1995 {Gravitational Waves. In {\it
	Proceedings of the Snowmass 95 Summer Study on Particle and
	Nuclear Astrophysics and Cosmology,} World  Scientific,
	pp. 398-425 }
\bibitem{timmes} Timmes F X {\it et al} 1996 The Neutron Star and Black
	Hole Initial Mass Function {\it Astrophy. J.} {\bf
	457} 834 - 43
\bibitem{turner1979} Turner M S 1978 Gravitational radiation from
	supernova neutrino bursts {\it Nat} {\bf 274} 565 - 6
\bibitem{Tomas03}
Tomas R {\it et al} 2003 
Supernova pointing with low- and high-energy neutrino detectors
{\it Phys. Rev. D} {\bf 68} 093013

\bibitem{Tomas04}
Tomas R {\it et al} 2004 
Neutrino signatures of supernova shock and reverse shock propagation
{\it JCAP 0409} 015
\bibitem{ThompsonBurrowsPinto03}
Thompson T A {\it et al} 2003
Shock Breakout in Core-Collapse Supernovae and Its Neutrino Signature
{\it Astrophys. J} {\bf 592} 434

\bibitem{TimmesWoosleyWeaver96}
Timmes F X {\it et al} 1996 
The Neutron Star and Black Hole Initial Mass Function
{\it Astrophys. J} {\bf 457} 834

\bibitem{TotaniSatoDalhedWilson98}
Totani T {\it et al} 1998 
Future Detection of Supernova Neutrino Burst and Explosion Mechanism
{\it Astrophys. J} {\bf 496} 216-225
% uuu
\bibitem{umeda} Umeda H \& Nomoto K 2003 First-generation
	black-hole-forming supernovae and the metal abundance pattern of
	a very iron-poor star {\it Natur} {\bf 422} 871 - 3
% vvv

% www
\bibitem{walder} Walder R {\it et al} 2004 Anisotropies in the Neutrino Fluxes and Heating Profiles in Two-dimensional, Time-dependent, Multi-group Radiation Hydrodynamics Simulations of Rotating Core-Collapse Supernovae {\it Astrophys. J.} in press
\bibitem{wang96} 
Wang L {\it et al} 1996 Broadband Polarimetry of Supernovae: SN 1994D,
	SN 1994Y, SN 1994ae, SN 1995D, and SN 1995H {\it Astrophys. J.}
	{\bf 467} 435-45  

\bibitem{wang01} 
Wang L {\it et al} 2001 Bipolar Supernova Explosions {\it Astrophys. J.} {\bf 550} 1030-5 


\bibitem{wang02} Wang L {\it et al} 2002 The Axisymmetric Ejecta of Supernova 1987A\ 2002 {\it Astrophys. J.} {\bf 579} 671-7 


\bibitem{gensan1} Watanabe G {\it et al} 2003 Structure of cold nuclear
	matter at subnuclear densities by quantum molecular dynamics
	{\bf 68} 035806-1 -20
\bibitem{gensan2} Watanabe G {\it et al} 2005 Simulation of Transitions
	between ``Pasta'' Phases in Dense Matter \PRL {\bf 94}	031101-1
	- 4
\bibitem{weinberg} Weinberg S 1972 Gravitation and Cosmology (John Wiley \& Sons) 
\bibitem{weiss} Weisskopf M C {\it et al} 2000 Discovery of Spatial and
Spectral Structure in the X-Ray Emission from the Crab Nebula {\it
Astrophys. J.} {\bf 536} L81-L84.
\bibitem{wex} Wex N {\it et al} 2000 Constraints on Supernova Kicks from the Double Neutron Star System PSR B1913+16 {\it Astrophys. J.} {\bf 528} 401 - 9

\bibitem{willi} Willingale R {\it et al} 2002 X-ray spectral imaging and
	Doppler mapping of Cassiopeia A {\it Astron. Astrophys.} {\bf 381} 1039-48
\bibitem{WilsonMayleWoosleyWeaver86}
Wilson J R {\it et al} 1986 {\it Ann. NY Acad. Sci} {\bf 470} 267

\bibitem{Wolfenstein78}
Wolfenstein L 1978 
Neutrino oscillations in matter
{\it Phys. Rev. D} {\bf 17} 2369-2374

\bibitem{Wolfenstein79}
Wolfenstein L 1978 
Neutrino oscillations and stellar collapse
{\it Phys. Rev. D} {\bf 20} 2634-2635

\bibitem{WoosleyHegerWeaver02}
Woosley S E {\it et al} 2002
The evolution and explosion of massive stars
{\it Rev. Mod. Phys.} {\bf 74} 1015

\bibitem{WoosleyWeaver95}
Woosley S E and Weaver T A 1995 
THE EVOLUTION AND EXPLOSION OF MASSIVE STARS. 2. EXPLOSIVE HYDRODYNAMICS AND NUCLEOSYNTHESIS
{\it Astrophys. J. Suppl.} {\bf 101} 181-235


\bibitem{advancedligo}
	Weinstein A 2002 {\it Class.\ Quantum
	Grav.} {\bf{19}} 1575
\bibitem{wilson1985} Wilson J R 1985 {\it Numerical Astrophysics}
	(Boston:Jones \& Barlett)

\bibitem{wilsonmayle88} Wilson J R and Mayle R W 1988 Convection in core collapse supernovae {\it
	Phys. Rep.} {\bf 163} 63-78 

\bibitem{wilsonmayle93} Wilson J R and Mayle R W 1993 Report on the
	progress of supernova research by the Livermore group {\it
	Phys. Rep.} {\bf 227} 97-111 
\bibitem{woods} Woods P M and Thompson C 2004 Soft Gamma Repeaters and
	Anomalous X-ray Pulsars: Magnetar Candidates "Compact Stellar
	X-ray Sources", eds. W.H.G. Lewin and M. van der Klis (astro-ph/0406133)
\bibitem{woosrevshinka} {Woosley} S E and {Weaver} T A 1986
The physics of supernova explosions {\it Ann. Rev. Astron. Astrophys.}
	{\bf 24} 205 - 253
\bibitem{ww:95}
{Woosley} S E and {Weaver} T A 1995 The Evolution and Explosion of
	Massive Stars. II. Explosive Hydrodynamics and Nucleosynthesis
	{\it Astrophys. J. Suppl.} {\bf 101} 181 - 230
\bibitem{wooshinka} Woosley S E {\it et al} 2002 The evolution and
	explosion of massive stars {\it Rev. Mod. Phys.} {\bf 74} 1015-71

% xxx

% yyy
\bibitem{yahil_lattimer}
Yahil A and Lattimer J M 1992 in Supernova: A survey of Current
Research, eds M.J. Ress and R.J. Stoneham (Dordrecht:Reidel)
\bibitem{yama94} Yamada S and Sato K 1994 Numerical study of rotating core collapse in supernova explosions {\it Astrophys. J.} {\bf 434} 268 -76 
\bibitem{ys} Yamada S and Sato K 1995 Gravitational Radiation from
	Rotational Collapse of a Supernova Core {\it Astrophys. J.} {\bf
	450} 245 - 52
\bibitem{yama03} Yamada S and Sawai H 2004 Numerical Study on the Rotational Collapse of Strongly Magnetized Cores of Massive Stars {\it Astrophys. J.} {\bf 608} 907 - 24  
\bibitem{yamatoki} Yamada S and Toki H 2000 Neutrino-nucleon reaction
	rates in the supernova core in the relativistic random phase
	approximation \PR{\it C} {\bf 61} 015803-1 - 16 
\bibitem{yamayama} Yamasaki T and Yamada S 2005 Effects of rotation on the revival of a stalled shock in supernova explosions submitted to {\it Astrophys. J.} astro-ph 0412625
% zzz
\bibitem{zhang} Zhang B and Harding A K 2000 High Magnetic Field Pulsars and Magnetars: A Unified Picture {\it Astrophys. J. Lett.} {\bf 535} L51 - 54
\bibitem{zweg} Zwerger T and M\"{u}ller E 1997 Dynamics and
	gravitational wave signature of axisymmetric rotational core
	collapse. {\it Astron. Astrophys.} {\bf 320} 209 -27













 

























%\include{reference_kotake}
%\include{reference_ktaro}
\end{thebibliography}
\end{document}